\documentclass[rmp,aps,twocolumn,floatfix,citeautoscript]{revtex4}
\usepackage{epsfig}
\usepackage{amsmath}
\usepackage{amsfonts}
\usepackage{bm}
\usepackage{euscript}
\usepackage{graphicx} 
\usepackage{wasysym}
\usepackage{color}
\usepackage{subfigure}
\usepackage{hyperref}

\newcommand{\mathnotation}[2]{\newcommand{#1}{\ensuremath{#2}}}

\newcommand{\Order}[1]{\mathcal{O}\!\l(#1\r)}

\newcommand{\R}{\mathbb{R}}
\newcommand{\Z}{\mathbb{Z}}

\newcommand{\eps}{\epsilon}

\newcommand{\jac}{\nabla}

\newcommand{\xvec}[1]{\mbox{\boldmath$#1$}}
\newcommand{\hspf}{,\hspace{0.5cm}}
\newcommand{\wt}[1]{\widetilde{#1}}
\newcommand{\PA}{\xvec{\Phi}_\mathcal{A}}
\newcommand{\PB}{\xvec{\Phi}_\mathcal{B}}
\newcommand{\PC}{\xvec{\Phi}_\mathcal{C}}
\newcommand{\PX}{\xvec{\Phi}}
\newcommand{\M}[2]{\mathcal{#1}^{\mbox{\tiny (#2)}}}

\mathnotation{\A}{\mathcal{A}}
\mathnotation{\B}{\mathcal{B}}
\mathnotation{\C}{\mathcal{C}}

\mathnotation{\Ax}{\mathcal{A}}
\mathnotation{\Bx}{\mathcal{B}}
\mathnotation{\Cx}{\mathcal{C}}

\graphicspath{{Figures/}}

\renewcommand{\l}{\left}               
\renewcommand{\r}{\right}              
\mathnotation{\ee}{{\mathrm e}}        
\mathnotation{\dint}{\,{\mathrm{d}}}   
\renewcommand{\time}{t}                
\mathnotation{\xc}{\theta}             
\mathnotation{\yc}{y}                  
\mathnotation{\xcp}{\bar\xc}           
\mathnotation{\ycp}{\bar\yc}           
\mathnotation{\xcsep}{\xc_{\mathrm{s}}}
\mathnotation{\ycsep}{\yc_{\text{s}}}  
\mathnotation{\uc}{u}                  
\mathnotation{\vc}{v}                  
\mathnotation{\Uc}{\Omega}             
\mathnotation{\Xc}{\Theta}             
\mathnotation{\Yc}{Y}                  
\mathnotation{\Xcp}{\overline{\Xc}}    
\mathnotation{\Ycp}{\overline{\Yc}}    
\mathnotation{\T}{T}                   
\mathnotation{\decfp}{\mu}             
\mathnotation{\decvar}{\alpha}         
\mathnotation{\lyap}{\lambda}          
\mathnotation{\gap}{d}                 
\mathnotation{\dotgap}{\skew{8}\dot{d}}
\mathnotation{\lB}{\ell_{\mathrm{B}}}  
\mathnotation{\tB}{\time_{\mathrm{B}}} 
\mathnotation{\Aw}{\mathcal{A}_{\mathrm{w}}}
\mathnotation{\Sgen}{S}                
\mathnotation{\Fgen}{F}                
\mathnotation{\aaa}{a}                 
\mathnotation{\Ham}{H}                 

\newcommand{\bk}{\boldsymbol{k}}

\newcommand{\bu}{\boldsymbol{u}}
\newcommand{\bx}{\boldsymbol{x}}

\newcommand{\bnabla}{\boldsymbol{\nabla}}

\newcommand{\pd}[2]{\frac{\partial #1}{\partial #2}}
\newcommand{\td}[2]{\frac{d #1}{d #2}}
\newcommand{\intd}{\; \textrm{d} }

\renewcommand{\l}{\left}
\renewcommand{\r}{\right}

\renewcommand{\A}{\uc_1}
\renewcommand{\epsilon}{\varepsilon}

\begin{document}

\title{Frontiers of chaotic advection}

\author{Hassan Aref}
\affiliation{Engineering Science \& Mechanics, Virginia Tech, Blacksburg, Virginia 24061, USA}
\author{John R. Blake}
\affiliation{School of Mathematics, University of Birmingham, Edgbaston, Birmingham, UK}
\author{Marko Budi\v{s}i\'{c}}
\affiliation{
Department of Mathematics and Clarkson Center for Complex Systems Science (C3S2), 
Clarkson University, Potsdam, New York 13699, USA}
\author{Silvana S. S. Cardoso}
\affiliation{Department of Chemical Engineering and Biotechnology, University of Cambridge,  Cambridge, UK}
\author{Julyan H. E. Cartwright}
\affiliation{Instituto Andaluz de Ciencias de la Tierra, CSIC--Universidad de Granada and  Instituto Carlos I de F\'{\i}sica Te\'orica y Computacional, Universidad de Granada, Granada, Spain}
\author{Herman J. H. Clercx}
\affiliation{Fluid Dynamics Laboratory, Department of Applied Physics  and JM Burgerscentrum, Eindhoven University of Technology, 
Eindhoven, The Netherlands}
\author{Kamal El Omari}
\affiliation{ Universit\'e Pau \& Pays Adour, Laboratoire des Sciences de l'Ing\'enieur Appliqu\'ees \`a la M\'ecanique et au G\'enie \'Electrique --- F\'ed\'eration IPRA, EA4581, Pau, France}
\author{Ulrike Feudel}
\affiliation{Institut f\"ur Chemie und Biologie des Meeres, Carl von Ossietzky Universit\"at Oldenburg, 
Oldenburg, Germany}
\author{Ramin Golestanian}
\affiliation{Rudolf Peierls Centre for Theoretical Physics, University of Oxford, Oxford,
UK}
\author{Emmanuelle Gouillart}
\affiliation{Surface du Verre et Interfaces, UMR 125 CNRS/Saint-Gobain, 
Aubervilliers, France}
\author{GertJan F. van Heijst}
\affiliation{Fluid Dynamics Laboratory,
Department of  Applied Physics  and JM Burgerscentrum,
Eindhoven University of Technology, 
5600 MB Eindhoven, The Netherlands}
\author{Tatyana S. Krasnopolskaya}
\affiliation{
Institute of
Hydromechanics, National Academy of Sciences of Ukraine, 
Kiev, Ukraine}
\author{Yves Le Guer}
\affiliation{Universit\'e Pau \& Pays Adour, Laboratoire des Sciences de l'Ing\'enieur Appliqu\'ees \`a la M\'ecanique et au G\'enie \'Electrique --- F\'ed\'eration IPRA, EA4581, Pau, France}
\author{Robert S. MacKay}
\affiliation{Mathematics Institute, University of Warwick, Coventry, 
UK}
\author{Vyacheslav V. Meleshko}
\affiliation{Department  of Theoretical and Applied Mechanics, 
Kiev National Taras Shevchenko University, 
Kiev, Ukraine}
\author{Guy Metcalfe}
\affiliation{School of Mathematical Sciences and {MAXIMA} Academy for
 Cross \& Interdisciplinary Mathematical Applications, Monash
 University, Clayton, Australia}
\author{Igor Mezi\'c}
\affiliation{Departments of Mechanical Engineering and Mathematics, University of California, Santa Barbara, 
California 93106, USA}
\author{Alessandro P. S. de Moura}
\affiliation{Institute for Complex Systems and Mathematical Biology, 
King's College, University of Aberdeen, Aberdeen, 
UK}
\author{Oreste Piro}
\affiliation{Department of Physics, University of the Balearic Islands, 
Palma de Mallorca, Spain}
\author{Michel F. M. Speetjens}
\affiliation{Department of Mechanical Engineering and JM Burgerscentrum, Eindhoven University of Technology, 
Eindhoven, The Netherlands}
\author{Rob Sturman}
\affiliation{Department of Applied Mathematics, University of Leeds, Leeds, 
UK}
\author{Jean-Luc Thiffeault}
\affiliation{Department of Mathematics, University of Wisconsin --  Madison,  Madison, Wisconsin 53706,
USA}
\author{Idan Tuval}
\affiliation{
Mediterranean Institute for Advanced Studies (CSIC-UIB), 
Esporles, Spain}

\date{\today: Version 4.1 --- the premium edition}

\begin{abstract}
This work reviews the present position of and surveys future perspectives in the physics of chaotic advection: the field that emerged three decades ago at the intersection of fluid mechanics and nonlinear dynamics, which encompasses a range of applications with length scales ranging from micrometers to hundreds of kilometers, including systems as diverse as mixing and thermal processing of viscous fluids, microfluidics, biological flows, and oceanographic and atmospheric flows.
\end{abstract}

\maketitle
\enlargethispage{\baselineskip}
\tableofcontents

\section{Introduction}\label{introduction}

\begin{quote}
Since things in motion sooner catch the eye than what not stirs. \\
Shakespeare, \emph{Troilus and Cressida}, \\act III scene 3
\end{quote}

A dynamical process like stirring a fluid holds a great deal more interest than a static system for physicists just as it does for everyone else. Stirring and mixing of scalars (additives, nutrients, heat, $\ldots$) by fluid flows is the common denominator in a wide variety of natural and industrial fluid systems of size extending from micrometers to hundreds of kilometers. Industrial examples range from mixing in the rapidly expanding field of micro-fluidics, encompassing applications as diverse as micro-electronics cooling, micro-reactors, ``labs-on-a-chip'' for molecular analysis and biotechnology and ``smart pills'' for targeted drug delivery, up to the mixing and thermal processing of viscous fluids with compact processing equipment. Examples in nature include magma transport in the Earth's mantle and dispersion of hydrocarbons within fractured rock, as well as gas exchange in lung alveoli and the distribution of blood-borne pathogens, and large-scale dispersion of pollutants in Earth's atmosphere and oceans.

Given its ubiquity in industry and nature, insight into the mechanisms underlying mixing, and ways purposefully to employ and control them, have great scientific, technological and social relevance, and are imperative for further development of fluid-processing technologies in, especially, micro-fluidics applications and process engineering. Although this insight remains incomplete to date, important physical and mathematical approaches for the analysis and understanding of mixing in laminar flows have become available during the last three decades. Moreover, both the advancement of measurement technologies to investigate mixing in laboratory set-ups --- such as laser-induced fluorescence, particle-tracking velocimetry, micro-particle image velocimetry, and so on --- and the rapid development of passive and active mixing elements for micro-fluidic devices, open the perspective for quantitative mixing studies for industrial and micro-fluidic applications.  Key challenges for advancing this field may occur to the reader during the course of this article; we will defer our thoughts on these to the discussion in Section~\ref{perspectives}.

The context of our review is the dynamical-systems and mathematical-physics perspective on fluid transport and mixing. 
Early works on the subject from the 1960s are owed to \textcite{arnold1965} and \textcite{henon1966}; the 1980s brought \textcite{arter1983} and \textcite{Aref1984}.
Three decades have passed since \textcite{Aref1982,Aref1984,Aref02} introduced the term \emph{chaotic advection}\footnote{An older term, \emph{Lagrangian turbulence}, is sometimes used as a synonym for chaotic advection, but is also applied to Lagrangian aspects of turbulent flows in general, so we prefer the term \emph{chaotic advection}.}, and over twenty-five years since a textbook on the field appeared \cite{Ottino1989}. During that time a great deal of research has been done. It is time to summarize thirty years of chaotic advection.
 Our review focuses on theoretical and mathematical-physics concepts and numerical approaches of stirring and mixing in viscous fluids. 
Its scope should be seen within the broader context of mixing in fluid flows and of earlier reviews.
Thus we complement the textbook and  review of  \textcite{Ottino1989,Ottino90} with developments from the last two decades. We have a more physical approach than the strongly mathematical perspective of \textcite{arnold1992}  (see also \textcite{Arnold1998}), which reviews mathematical concepts of transport for hydrodynamics. 
\textcite{mezic2013analysis} discussed the Koopman operator approach and 
\textcite{Haller2015} examined Lagrangian coherent structures; for this reason these two topics are reviewed herein only to a limited extent. 
And, as this is a review paper and not a book, we do not document the ever growing field of applications in exhaustive detail, but we provide some illustrative examples. 
Phenomena such as chemical reactions in flows, transport of active matter, aggregation processes, droplet dynamics and granular flow (see, for example, \textcite{Ottino2000}) are outside the scope of the present review and are touched upon only when they affect our core concern.
The 2000 Laporte Prize lecture \cite{Aref02} links chaotic advection with dynamical systems theory, which is the central theme of this review; its basic ideas are our starting point.

The number of papers in chaotic advection is now so numerous that anything approaching a complete coverage of the field is impossible. Here we present a view arising from a group of researchers predominantly working in the field of stirring and mixing in viscous fluids from a theoretical, numerical and/or mathematical-physics point of view, although experimental excursions are not excluded. Some of the choices to illustrate concepts are unavoidably colored by the interests of the researchers involved in this review, and should not be considered as exhaustive. 

We begin with an informal definition of chaotic advection.  We then look at the differences between open and bounded flows; in Section~\ref{open} we look at unbounded flows and in Section~\ref{closed}, we discuss ideas on the role of walls. The new frontier is, undoubtedly, three-dimensional (3D) unsteady flow, treated in Section~\ref{3D}. There is ongoing interest in elucidating chaotic advection from numerical and experimental data that we discuss in Section~\ref{data}. In  Section~\ref{turbulent} we discuss what is meant by a laminar or a turbulent flow;  these terms are frequently used in this discipline and yet are open to a range of interpretations. In Section~\ref{quality+measures} we examine the mixed state itself; there are interesting aspects of the mixed state, in particular the overarching problem of the quality of mixing and mixing measures, that are well worth looking at in more geometrical and topological detail.  ``Chaotic advection plus'',  treated in Section~\ref{advection_plus}, is a more and more active area; the increasing range of application of these ideas is most encouraging.  Lastly, in Section~\ref{perspectives}, we conclude with perspectives on these frontiers of chaotic advection.

\subsection{Synopsis of key concepts}\label{key}

At its simplest we may consider a flowing fluid to consist of only fluid
particles, aggregations of material elements small enough to satisfy
the requirements to treat the fluid as a continuum.  Each fluid
particle --- if a particle is conceptually or actually marked,
often called a tracer --- is denoted by its position $\xvec{x}$ and
moves passively with the fluid velocity $\xvec{u}$ according to
the kinematic equation
\begin{equation} 
\label{E2'} 
\frac{d\xvec{x}}{dt}=\xvec{u},\quad \xvec{x}(0)=\xvec{x}_0.
\end{equation} 
Deceptively simple, the kinematic equation can be taken as the
elementary definition of velocity or as defining a
dynamical system in which a given velocity field generates so-called Lagrangian
trajectories for the fluid particles.  Indeed, Eq.~\eqref{E2'} has the
formal solution $\xvec{x}(t)=\xvec{\Phi}_t(\xvec{x}_0)$ describing the
Lagrangian trajectory of a tracer released at $\xvec{x}_0$ with a
corresponding Poincar\'{e} map defined by
$\xvec{x}_{k+1}=\xvec{\Phi}(\xvec{x}_k)$, where
$\xvec{x}_k=\xvec{x}(kT)$ is the tracer position after $k$ periods of
a time-periodic flow.  
The velocity $\xvec{u}$ is often derived from the (steady)
Navier--Stokes and continuity equations
\begin{equation}
\label{E1'}
Re~\xvec{u}\cdot\nabla \xvec{u} =-\nabla p
+\nabla^{2}\xvec{u},\quad\xvec{\nabla}\cdot\xvec{u}=0,
\end{equation}
here given in non-dimensional form for incompressible fluids,
with $Re=UL/\nu$ the Reynolds number.

A special, but important case occurs for  two-dimensional (2D), divergence-free flows
where the velocity $\xvec{u}$ can be derived from a so-called stream function
$\Psi(x,y)$ and the kinematic equation written in terms of the stream
function as
\begin{equation}
\label{kinematic_Hamiltonian}
\frac{d x}{dt} = \frac{\partial\Psi}{\partial y}, \quad
\frac{d y}{dt} = -\frac{\partial\Psi}{\partial x}.
\end{equation}
Equation~\eqref{kinematic_Hamiltonian} is identical to Hamilton's equations
of motion for a one-degree-of-freedom dynamical system with the
identification of the position coordinates $x$ and $y$ respectively as
canonical position and momentum coordinates along with the
identification of $\Psi$ with the Hamiltonian function.  This crucial
insight has allowed over a century of theoretical developments from
Hamiltonian mechanics to be brought to bear on 2D flow problems and
links 2D fluid flow to many other areas of physics.

Under quite general conditions some trajectories or sets of trajectories advected by
Eq.~\eqref{E2'} form barriers to material transport, manifolds such as material lines or sheets that are invariant under the flow. These barriers are persistent, and even in
transient flows are long-lived.  In informal terms, these material
curves or sheets are the Lagrangian coherent structures (LCS; see Section~\ref{sec:LCS}) whose
material lines attract or repel the neighboring material.
LCS can both facilitate and retard transport fluxes; they organize and
mediate all transport and interaction of matter and energy in the
flow.  Finding, classifying and manipulating these structures plays the central role in both analysis of and design with
flows. These organizing structures can meander wildly throughout
the space of interest, and the wild meanders associated with chaotic dynamics give the title
\emph{chaotic advection} to this entire field of study.

\subsection{A definition of chaotic advection}

In many applications one wants to maximize the rate of mixing of a fluid. In the simplest setting, this means that we want to reduce as much as possible the time it takes for molecular diffusion to homogenize an initially inhomogeneous distribution of a scalar tracer. If there is no advection, molecular diffusion by itself takes a very long time to achieve homogeneity, even in quite small containers. So we use advection to accelerate this process. The classical and more well-known way to do so is through turbulence: by imposing a high Reynolds number in a 3D flow, we trigger the formation of a Kolmogorov energy cascade \cite{k41a,k41b,frisch1996turbulence,tritton,kundu}  whereby energy flows from large to small scales. This energy cascade is mirrored by a corresponding cascade in any scalar field advected along with the flow, whose distribution develops in this process small-scale structures, which are then rapidly homogenized by molecular diffusion. From the point of view of mixing, such turbulence is therefore a way to create --- quickly --- small-scale structures in the spatial distribution of advected fields, resulting in their being smoothed by diffusion.

Chaotic advection \cite{Aref1984} is a different way to generate small-scale structures in the spatial distribution of advected fields, by using the stretching and folding property of chaotic flows. This chaotic dynamics quickly evolves any smooth initial distribution into a complex pattern of filaments or sheets --- depending on the dimensionality of the system --- which tends exponentially fast to a geometric pattern with a fractal structure. Owing to the stretching, the length scales of the structures in the contracting directions decrease exponentially fast, and when they become small enough, they are smoothed out by diffusion. This is a purely kinematic effect, which does not need high Reynolds numbers, and exists even in time-dependent 2D Stokes flows. 
Chaotic advection can thus be defined as \emph{the creation of small scales in a flow by its chaotic dynamics}.
Mixing by chaotic advection has the advantages over turbulence that it does not require the larger input of energy needed to maintain the Kolmogorov cascade that turbulent mixing does; and it can be set up in situations  --- such as microfluidics  --- in which a high Reynolds number is not an option.

\subsection{Stirring and mixing}

The terminology regarding mixing and stirring is not always consistent in the literature. We suggest the following (although it is impossible to be taxative, as both words are so embedded in common usage). Stirring is advective redistribution --- i.e., purely kinematic transport --- and mixing is stirring together with diffusive effects. 
To add molecular diffusion to the mathematical conception of advection laid out in Section~\ref{key} , one has 
the advection--diffusion equation
\begin{equation}
\label{eqn:ADE}
\frac{\partial\phi}{\partial t} + 
\mathbf{v}\cdot\nabla\phi=\frac{1}{Pe}\nabla^2\phi
\end{equation}
with appropriate boundary and initial conditions, where $\phi$ is the
scalar concentration and the P\'{e}clet number $Pe$ balances diffusive and
advective time-scales.  
The natural scale for mixing is the Batchelor scale \cite{Batchelor1959}
\begin{equation}
\label{eq:batchelor-scale}
\lB = \sqrt{\frac{\kappa}{\lambda}},
\end{equation} 
where~$\kappa$ is the molecular
diffusivity and $\lambda$ is the
Lyapunov exponent of the flow. At scales smaller than
$\lB$, diffusion smooths out concentration gradients and mixing is
achieved at the molecular scale.

The sections of this review concentrate upon stirring --- Sections~\ref{open}, \ref{closed},  \ref{3D}, \ref{data},  \ref{turbulent}
--- and mixing ---  Sections~\ref{advection_plus}, \ref{perspectives}.
Section~\ref{quality+measures} is the bridge between the two parts.

\section{Unbounded flows}
\label{open}

\begin{figure}[tb]
\begin{center}
\includegraphics[width = 0.8\columnwidth]{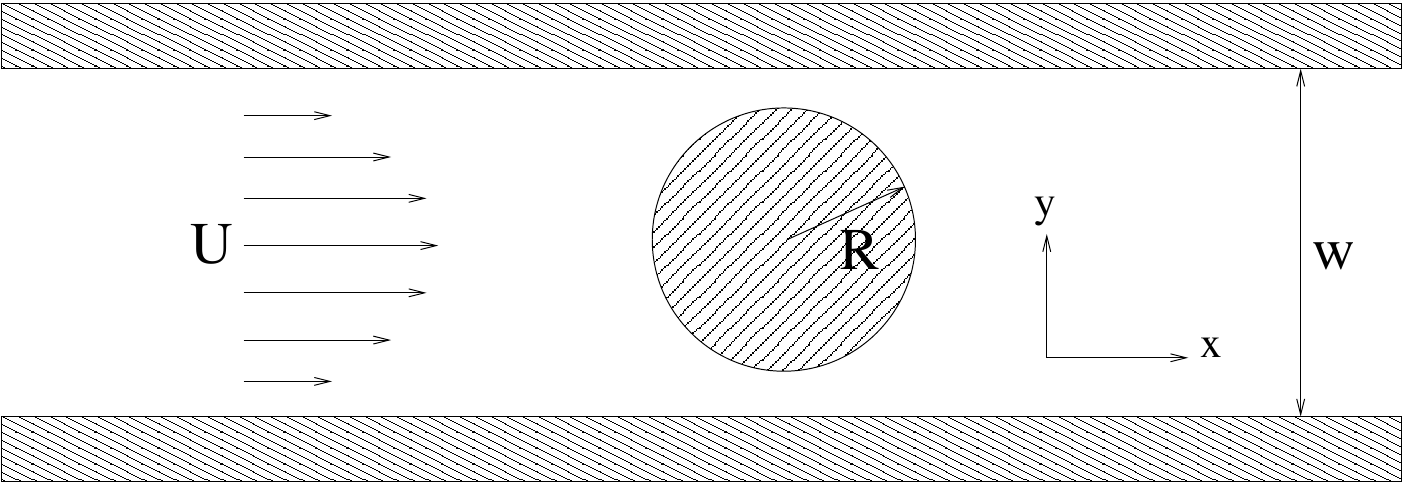}
\end{center}
\caption{Set-up for an open flow in a channel with an obstacle.}
\label{fig:channel_flow}
\end{figure}

Chaos in open flows manifests itself
through the appearance of fractal structures in the advection of an
initially smooth distribution of passive
tracers~\cite{Tel-et-al-05}. These fractal patterns arise as a direct
result of the existence of a non-attracting chaotic set in the
advection dynamics~\cite{cscatbook}.

Let us begin by considering a 2D channel flow
past an obstacle (Fig.~\ref{fig:channel_flow}). Fluid particles
come from an inflow region, may stay in the wake of the obstacle for
some time, and then leave through the outflow. If we consider a
limited region around the obstacle as our \emph{observation region}
$R$, most fluid particles stay only a finite time in $R$, before
escaping to the outflow region \cite{Jung-et-al-93}. The dynamics of
advection in an open flow is therefore \emph{transient}.
The transient nature of
the dynamics makes mixing in open flows qualitatively different from
the closed flow case discussed in Section~\ref{closed}: the very definition of mixing and its
mathematical formulation are different.

\subsection{The chaotic saddle and its invariant sets}\label{saddle}

\begin{figure}[tb]
\begin{center}
\includegraphics[width = 0.8\columnwidth]{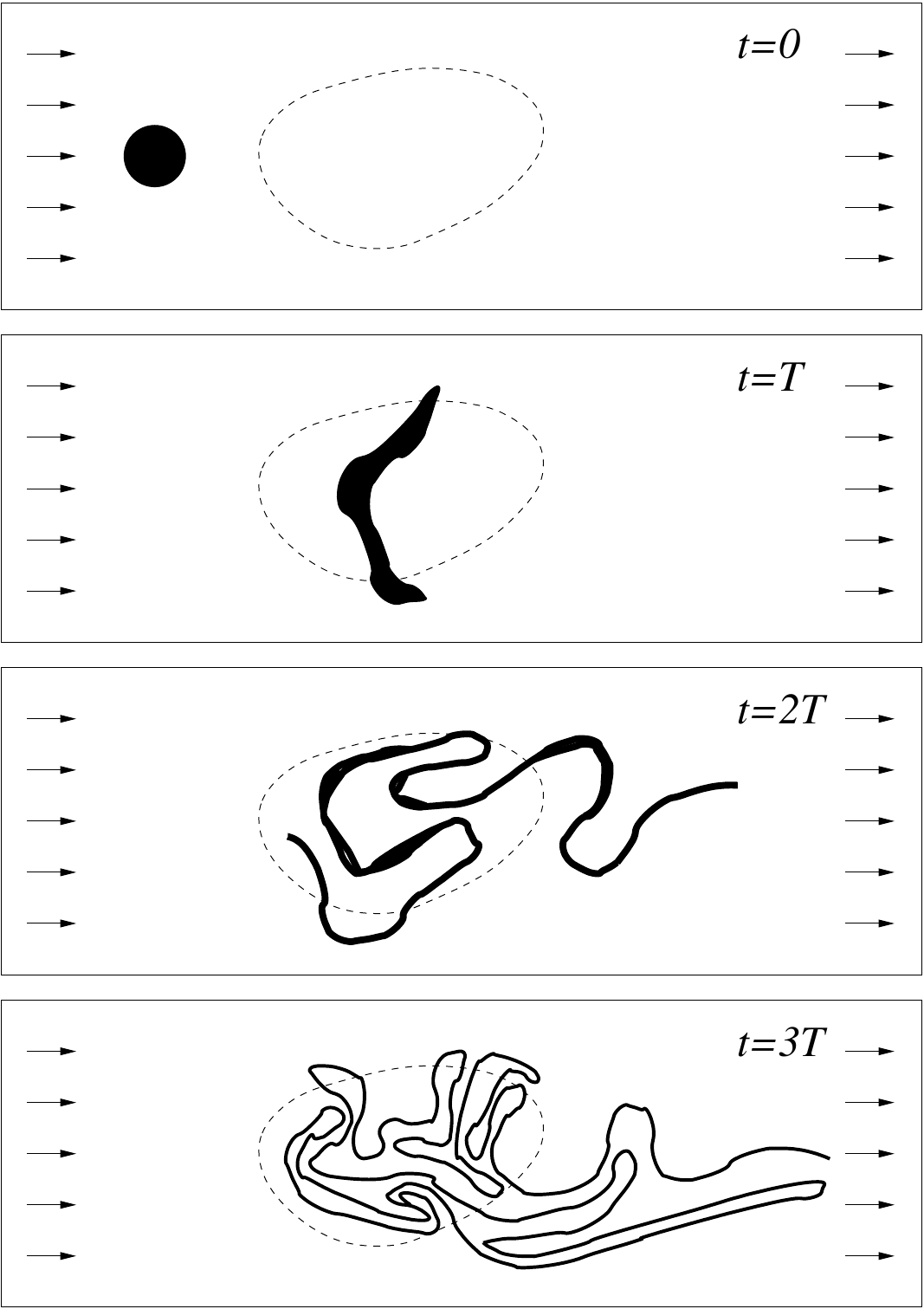}
\end{center}
\caption{Sketch of a dye droplet reaching the mixing region --- dashed line --- of
  an open flow displaying chaotic advection. After some time, the
  remaining dye approaches the unstable manifold of the chaotic
  saddle. Reprinted from \textcite{Tel-et-al-05} with permission from Elsevier.}
\label{fig:filament_formation}
\end{figure}

\begin{figure*}[tb]
\centering
\begin{tabular}{cccc}
  \includegraphics[width=0.2\textwidth]{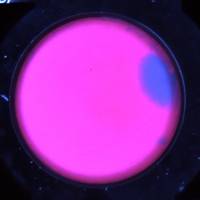} &
  \includegraphics[width=0.2\textwidth]{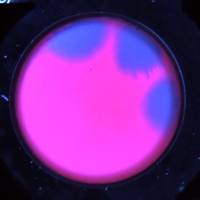} &
  \includegraphics[width=0.2\textwidth]{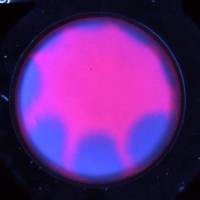} &
  \includegraphics[width=0.2\textwidth]{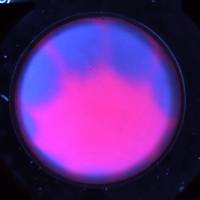} \\
  \includegraphics[width=0.2\textwidth]{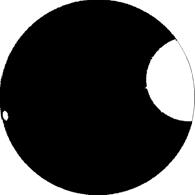} &
  \includegraphics[width=0.2\textwidth]{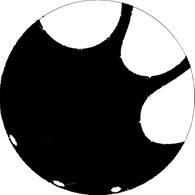} &
  \includegraphics[width=0.2\textwidth]{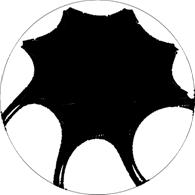} &
  \includegraphics[width=0.2\textwidth]{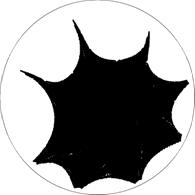} \\
  1 & 3 & 8 & 13 
\end{tabular}
\caption{Filamentary manifolds evolving and thinning in a Hele-Shaw flow experiment (top) and computer simulation (bottom) with an opposite inflow and outflow creating a dipole that is on for a scaled time $\tau$, after which the dipole is reoriented by an angle $\Theta$. Number of iterations are shown at the bottom for $\tau = 0.1$ and $\Theta = \pi/4$.   
 Reprinted from \textcite{Metcalfe_PiP_2010,Metcalfe_ECC_2010} with permission from the Royal Society.
  }
\label{fig:expt}
\end{figure*}

In open chaotic flows, the perpetual alternation of stretching and
folding results in filaments that grow ever thinner because of the
fluid's escape, as depicted in Fig.~\ref{fig:filament_formation}. As
they become thinner, they grow longer and more convoluted.  This
happens because of the stretching and folding properties of the
dynamics.  To understand the consequences of this stretching and
folding, let us take an initial region of the flow, located in a place
where mixing occurs; for example, immediately behind the wake of an
obstacle.  If the flow is chaotic, this initial region will be
repeatedly stretched and bent back on itself.  After some time, some
of the fluid in our original region has escaped, but some has come back
and intersects the original region.  This intersection is the
fundamental characteristic of a Smale horseshoe \cite{smale1967,smale1998,shub2005}, and it immediately
follows from Smale's results that there are infinitely many unstable
periodic orbits in the intersection region, as well as an uncountable
infinity of aperiodic orbits.  These orbits are collectively referred
to as the \emph{chaotic saddle}.  The set of initial conditions giving
rise to trajectories that approach one of these orbits asymptotically
consists of orbits that never escape --- they are trapped --- since they
converge to orbits bound to a compact region of space: the
intersection region previously discussed.  

Open flows are therefore characterized by a transient dynamics of most
fluid particles.  More precisely, if one randomly picks an initial condition
in the inflow region, the corresponding trajectory will
escape to the outflow region with probability 1. In other words, the
set of initial conditions which stay trapped forever in an open flow
has Lebesgue measure zero.

Although they have zero measure in phase space, these trapped orbits
are very important for open flows, because they govern the long-time
advection dynamics: those orbits that take a long time to escape
correspond to initial conditions lying close to the trapped
trajectories. If the advective dynamics of the open flow is chaotic,
each trapped trajectory converges asymptotically as
$t\rightarrow\infty$ to one of the orbits in the 
chaotic saddle~\cite{cscatbook}; orbits in the chaotic saddle do not go to the outflow
region for $t\rightarrow\infty$, and they do not go to the inflow
region for $t\rightarrow-\infty$. So these orbits lie on a confined
portion of space, where mixing takes place in open flows; we
refer to this region as the \emph{mixing region} from now on. In the
example of a flow past an obstacle, the mixing region and the chaotic
saddle are located in the wake of the obstacle, and the mixing region typically extends
for no more than a few times the length of the
obstacle~\cite{Jung-et-al-93}.

The set of trapped trajectories corresponds to 
the stable manifold of the chaotic saddle
\cite{cscatbook}. Conversely, the set of orbits that converge to the
chaotic saddle in backward time (i.e., for $t\rightarrow-\infty$) is
the saddle's unstable manifold.  Both the stable and unstable
manifolds have important physical interpretations for the advection
dynamics. Long-lived orbits are close
to the stable manifold.  The physical meaning of the unstable manifold
comes from the fact that those trajectories that stay a long time in
the mixing region, that is, those lying close to the stable manifold in the
inflow region, will trace out the unstable manifold on their way out
towards the outflow region.  An initial blob of dye, or anything else 
that passively follows the flow, 
is repeatedly stretched and folded by the flow in the mixing region,
generating a convoluted filamentary structure that converges to the
unstable manifold; a sketch of this process is shown
in Fig.~\ref{fig:filament_formation}.  As a consequence, the unstable
manifold can be observed directly in experiments by following
a dye as it is advected in the fluid \cite{sommerer}. Once the bulk of the dye has
escaped, what still remains in the observation region shadows the
unstable manifold. 

An experimental example is
shown in the top line of Fig.~\ref{fig:expt} where a potential flow
dipole in a Hele-Shaw cell is periodically reoriented
\cite{Metcalfe_PiP_2010,Metcalfe_ECC_2010}; the bottom line shows a
computed version of the same flow.  The initial condition of pink
fluid is removed by in-flowing blue fluid.  As the flow proceeds, the
persistent pink lines thin but never leave the cell; these are
examples of filamentary manifolds.  The large pink blob is an island
that incoming fluid also never displaces.

\subsection{Partial mixing, fractals and fractal dimensions}

The repeated stretching and folding of fluid elements caused by the
presence of the chaotic saddle results in mixing.
In open flows
this kinematic mechanism only has limited time to act because
of the transient nature of the advection dynamics.  One can therefore
say that open chaotic flows induce partial mixing: an initial blob of
dye is deformed into a set of very thin and long filaments along the
unstable manifold of the chaotic saddle.

\begin{figure}[tb]
\begin{center}
\includegraphics[width = \columnwidth]{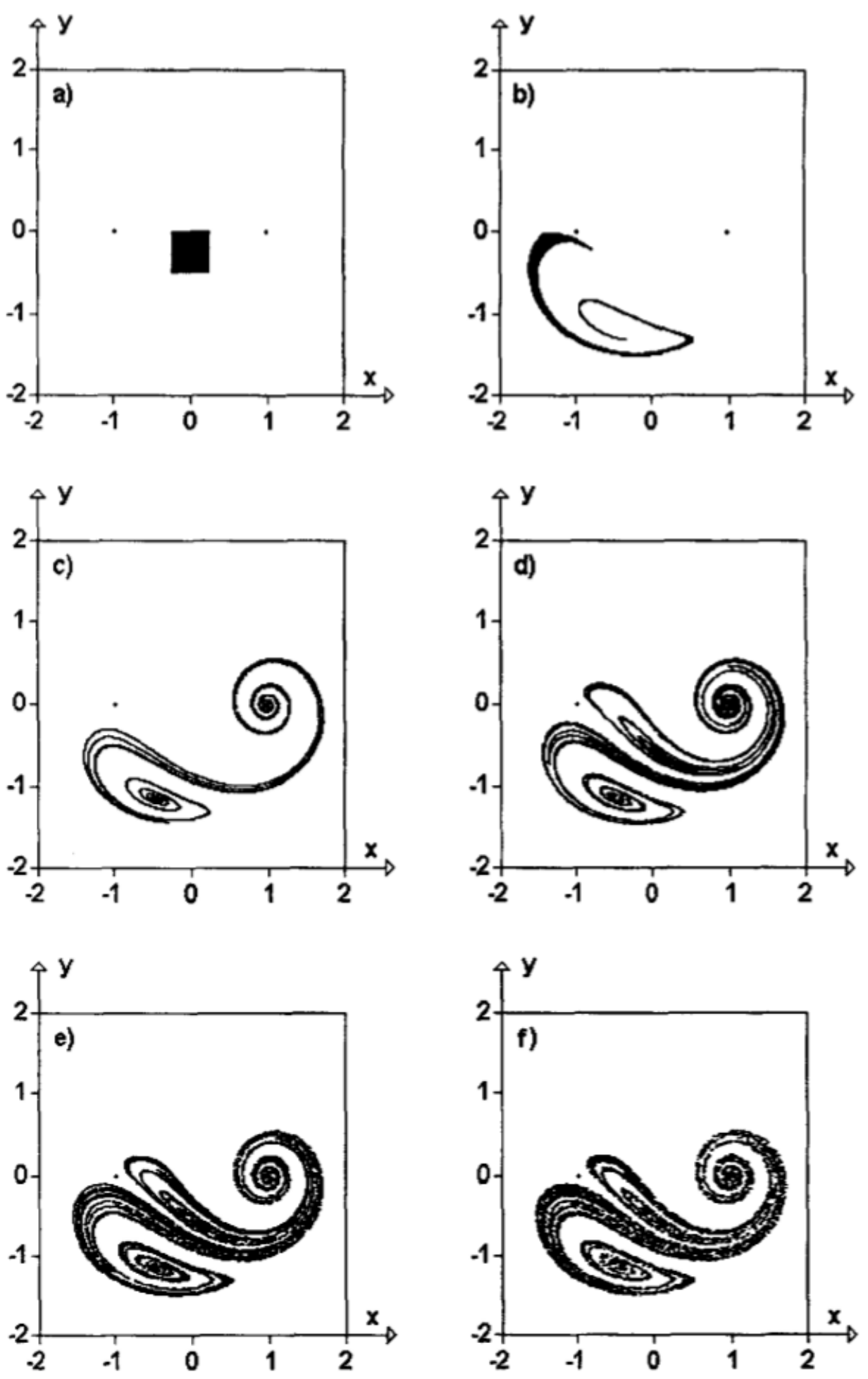}
\end{center}
\caption{Blinking vortex-sink flow: Evolution of a set of particles.  Reprinted from \textcite{PHR,Tel-et-al-05} with permission from Elsevier.}
\label{fig:filaments}
\end{figure}

We illustrate this in Fig.~\ref{fig:filaments} for the blinking vortex-sink flow, an
idealized periodic open chaotic flow used to study the dynamics of chaos
and mixing in open flows \cite{PHR}.  It consists of two sinks that are
alternately open and closed. One sink is open for half the period,
while the other is closed, and then the first sink closes while
the other one opens for the remaining half period, and so on
cyclically. This is a generalization of the famous blinking vortex
system introduced by \textcite{Aref1984},  the difference being that the vortices are
also sinks, which creates an escape and turns the dynamics into an
open flow.
A fluid particle in blinking vortex-sink flow follows a trajectory determined by the
equations of motion
\begin{equation}
\dot{r} =-C/r, \quad \dot{\varphi}=K/r^2.
\end{equation}
Solving these with initial conditions
$r_0$ and $\varphi_0$, we get
\begin{equation}
\label{singlevortexmotion}
r(t) = (r_0^2-2Ct)^{1/2}, \quad
\varphi (t) = \varphi_0 -\frac{K}{C} \ln{\frac{r(t)}{r_0}}.
\end{equation}
Without loss of generality, we choose the positions of the
vortices at $x=\pm a$, $y=0$, where $a$ is a parameter of the system.
Since we have an analytical expression for the motion of fluid
particles for each of the half-periods, we may find an
expression for the new position $\xvec{r}_{n+1}$ after one period as a
function of the position $\xvec{r}_n$ at the beginning of the
period. This is best done using a complex representation for the
position of a fluid particle, $z=x+iy$. The mapping from the initial
position $z_n$ to the new one $z_{n+1}$ is then given by
\begin{eqnarray}\label{eq:blinkingvortexflow}
z_{n+1/2}  &=&  \left( z_n+a\right) \left( 1-\frac{CT}{\left| z_n+
a\right| ^2}\right) ^{1/2-iK/(2C)}-a ; \nonumber \\
z_{n+1}  &=&  \left( z_{n+1/2}-a\right) \left( 1-\frac{CT}{ \left|
z_{n+1/2}- 
a\right| ^2}\right) ^{1/2-iK/(2C)} \nonumber \\
&&+a . 
\end{eqnarray}
Here $z_{n+1/2}$ is an intermediate variable representing the
particle's position after the first half-period.

One can see in Fig.~\ref{fig:filaments} that the filaments are
arranged in intricate layers, such that if one zooms in around a given
filament, the nearby filaments are oriented along roughly the same
direction. 
In other words, in the case of 2D flows, due to stretching and folding one finds, in a small volume, an infinite number of sections of the manifold, lined up in parallel, and densely packed in the perpendicular direction. More precisely, the unstable manifold is
locally the direct product of a Cantor set and a 1D
smooth curve \cite{cscatbook}.  This means that at any given point the
asymptotic distribution of any tracer advected by the flow varies
smoothly in the direction along the unstable manifold at that point,
while it varies wildly in directions transversal to the unstable
manifold \cite{Tel-et-al-05}.  This is a defining feature of the
SRB (Sinai--Ruelle--Bowen) measures \cite{Ott}, which
describe the natural probability distributions of transient chaotic
systems.  Although we have been focusing on the unstable manifold, the
same properties are shared by the stable manifold as well.

From the point of view of mixing, it is clear from
Fig.~\ref{fig:filaments}, and from the discussion above on the
structure of the unstable manifolds, that there is efficient mixing in
the directions locally transversal to the unstable manifold, but no
mixing happens in the direction of the unstable manifold.  
Contrast this to the case of closed flows (Section~\ref{closed}), where the
unstable manifold is space-filling, and mixing will eventually take
place along all directions, given enough time for the system to
evolve.

How does one quantify the amount of mixing in an open flow? Looking at
Fig.~\ref{fig:filaments}, one would intuitively want to measure mixing
by how much area in the picture is occupied by points lying close
to both the black and white regions.  But what does ``close to''
mean?  We have stated that the natural scale for mixing 
is the Batchelor scale, Eq.~\eqref{eq:batchelor-scale}.
The unstable manifold separates the black and white regions in the
limit $t\rightarrow\infty$.  Let us thus define $A(\eps)$ to be the
area of the set $S$ of points such that their distance from the
unstable manifold is smaller than $\eps$, restricted to a finite
observation region $R$ containing the chaotic saddle.  $A$ can be
estimated by covering $R$ with a grid of size $\eps$ and counting the
number $N(\eps)$ of grid elements that intersect with the unstable
manifold.  $A$ is then given by $A(\eps) = N(\eps)\eps^2$.  The
way $N(\eps)$ scales with $\eps$ is governed by the fractal
  dimension $D$ of the unstable manifold \cite{Ott,falconer}:
\begin{equation}
N(\eps) \sim \eps^{-D}.
\end{equation}
The area $A(\eps)$ therefore scales with $\eps$ as
\begin{equation}
A(\eps) \sim \eps^{2-D}.
\end{equation}
For a regular, non-chaotic 2D flow, the unstable set is a simple
1D curve, and thus $D=1$.  In this case $A$ is
proportional to $\eps$.  If the flow displays chaotic advection, $D$
satisfies $1<D\leq2$, and $A$ decreases sub-linearly with $\eps$.  This
means that in a chaotic flow, the mixing area $A(\eps)$ decays
very slowly with $\eps$. Using $\eps=\lB$, 
from
Eq.~\eqref{eq:batchelor-scale} we see that this results in a
slow decay of $A$ with the diffusivity.
This slow decrease of $A(\epsilon)$ with $\epsilon$ has many consequences for the dynamics of processes taking place in the flow. These include a singular increase in the rate of chemical reactions in open chaotic flows \cite{Tel-et-al-05}, and an anomalous scaling in the collision rate of particles \cite{moura2011}.

We have concentrated on the meaning of the fractal dimension of
the unstable manifold.  The fractal dimension of the stable manifold
also has a physical meaning: it is a measure of the sensitivity of the
dynamics of fluid particles in open flows to the initial
conditions~\cite{Grebogi1983}.  Let $p(\eps)$ be the probability that
the trajectories corresponding to two initial conditions separated by
a small distance $\eps$ eventually separate before escaping, so that
they escape following completely different paths.  $p(\eps)$ can be
numerically calculated by choosing a large number of pairs of points
located randomly in space, and following their trajectories until they
escape. If, for example, they escape in different cycles (assuming that
the flow is time-periodic), we consider them to have separated.  The
initial conditions within a distance $\eps$ of the stable manifold are
at risk of separating, and thus $p(\eps)$ is proportional to
$A(\eps)$, and scales as
\begin{equation}
\label{def_dim}
p(\eps) \sim \eps^{2-D}.
\end{equation}
$p(\eps)$ can be considered as a measure of the uncertainty in the
prediction of the ultimate fate of the trajectory of a given fluid
particle, when its initial condition is given with an experimental
error of size $\eps$.  Decreasing $\eps$ means an increase in
accuracy in the determination of the initial condition.  For
non-chaotic flows, $D=1$, and therefore $p(\eps)\sim\eps$;
decreasing $\eps$ by a factor of 10 would decrease the uncertainty
by the same factor, as one might expect.  If the flow
displays chaotic advection, however, $p(\eps)$ does not scale
linearly with $\eps$, and the uncertainty decreases more slowly.
For the case of $D=1.9$, for example, it would take a decrease of
ten orders of magnitude in $\eps$ to reduce $p(\eps)$ by a factor
of 10.

\subsection{Hyperbolicity and the Grassberger--Kantz relation}

Open hyperbolic systems have
exponential decay: if we keep track of the time
evolution of a typical area of flow, the amount $Q(t)$ of this initial
area still remaining in the mixing region at time $t$ decays
exponentially with $t$ for large $t$: $Q(t)\sim \exp{(-\kappa_e
  t)}$. $\kappa_e$ is the escape rate of the flow. 
  It satisfies
$\kappa_e<\lambda$, where $\lambda$ is the chaotic saddle's Lyapunov
exponent.  The fractal dimension
$D$ of the unstable manifold, the Lyapunov exponent $\lambda$ and the
escape rate $\kappa_e$ are related by the Grassberger--Kantz
formula~\cite{Grassberger1985}:
\begin{equation}
D = 2-\frac{\kappa_e}{\lambda}.   \label{grasskantz}
\end{equation}
More rigorously, we should have $D_1$, the information
  dimension \cite{falconer}, instead of the box-counting dimension $D$ in
  Eq.~\eqref{grasskantz}, but since $D$ and $D_1$ are almost always very close
for open flows, this approximation is valid in most cases.

\subsection{Robustness of the chaotic saddle}

In the previous discussion, and in most of what follows in this Section, we
concentrate on the case of 2D flows.  Furthermore, we
have concentrated on the motion of fluid particles, that is, of
passive tracers that assume exactly the velocity of the surrounding
fluid.  The fractal structure of the chaotic saddle and its associated
invariant manifolds persist, however, in the case of actual,
finite-sized particles, which have inertia and whose velocities do not
coincide with that of the fluid's velocity field
\cite{Vilela2006,Vilela2007,Springer2010}.  There are some
considerable differences between the dynamics of fluid particles and
that of inertial particles, in particular the possibility of the
appearance of attractors in the latter case
\cite{Cartwright2002_1,Benczik2002,Motter2003,Springer2010}.  But even when the global
dynamics has attractors, chaotic saddles are still present, and the
system is still governed by fractal structures in phase space
connected to a chaotic saddle, as in the simpler case of passive
advection.

The same overall picture remains valid for 3D systems
\cite{Cartwright1996,Tuval2004,Moura2004b}; in this case, the stable
and unstable manifolds are a fractal set of sheets, instead of
segments.  Periodicity is also not required for the existence of the
chaotic saddle: aperiodic and random flows can also result in
well-defined fractal structures in phase space
\cite{Karolyi2004,Rodrigues2010}.

The conclusion is that the concepts of chaotic saddle and its stable
and unstable manifolds are remarkably robust, and are not consequences
of over-simplified models of flows.

\subsection{Transport barriers and KAM islands: the effective dimension}
\label{kam}\label{KAM}

In discussions about chaotic open flows and the chaotic saddle it is
often assumed, sometimes tacitly, that the dynamics is hyperbolic.
The reason is partly that the hyperbolic case is more tractable, and
there are more rigorous results available. However,
non-hyperbolicity occurs in many important cases, and is to be
expected in many very general scenarios in fluid dynamics.  For
example, it can be shown that the dynamics of 2D advection of a flow
past an obstacle becomes chaotic immediately after the transition of
the flow from stationary to time-dependent, as the Reynolds number is
increased beyond a critical value; furthermore, the dynamics is
non-hyperbolic for a range of Reynolds numbers past the transition
point, independently of the shape of the obstacle or the particular
features of the flow \cite{Biemond2008}.  Many other systems of
interest are non-hyperbolic, so it is imperative to understand
the mixing dynamics in the non-hyperbolic case.

\begin{figure}[tb]
\begin{center}
\includegraphics[width = \columnwidth]{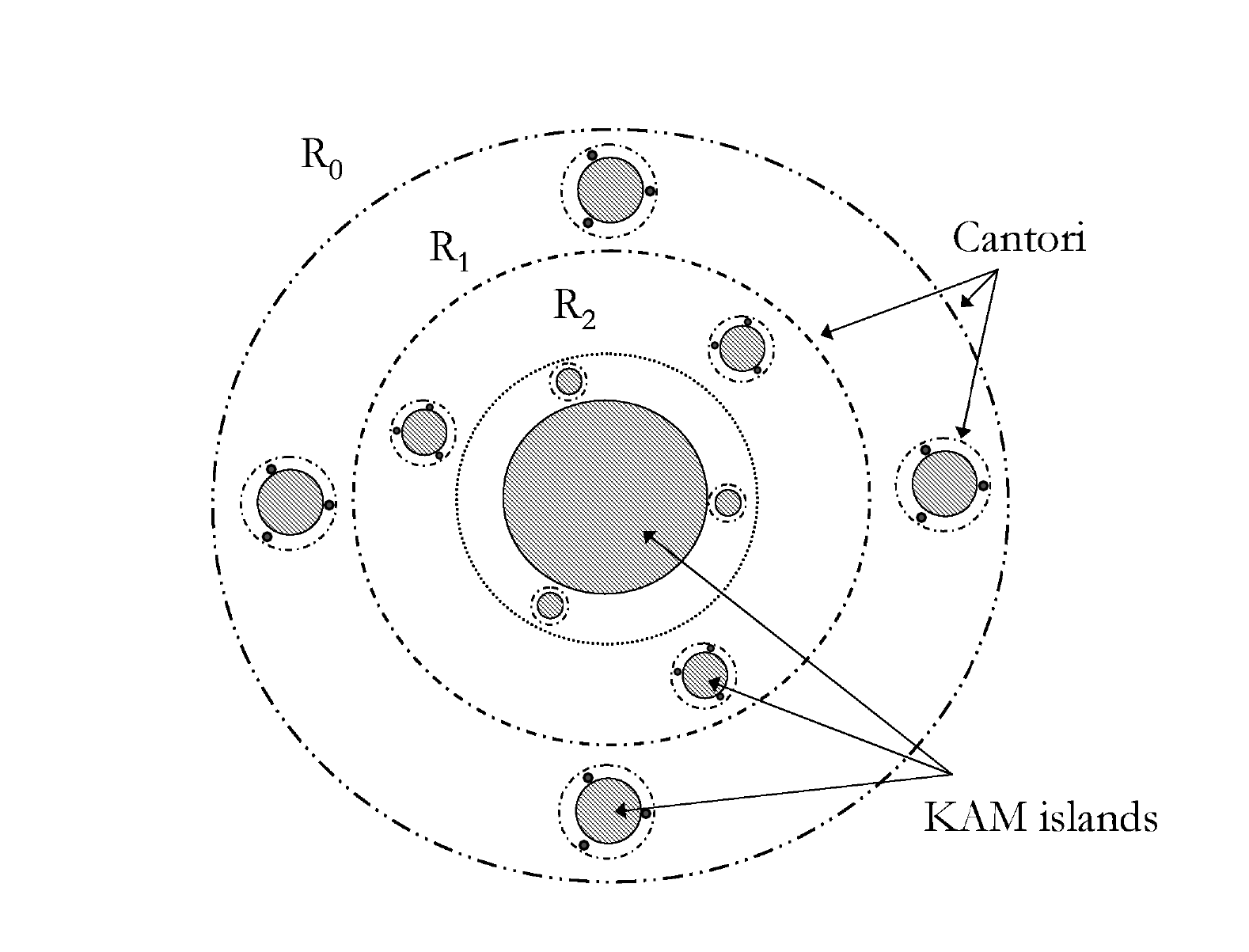}
\end{center}
\caption{KAM tori and cantori: hierarchical structure. Solid circles indicate KAM tori and cantori are
  represented by the dotted lines. Reprinted from \textcite{Tel-et-al-05} with permission from Elsevier.} \label{fig:cantori}
\end{figure}

Non-hyperbolicity is manifested through the appearance of stable
orbits in space. These orbits are surrounded by stable KAM
(Kolmogorov--Arnol'd--Moser) islands \cite{Ham_chaos}.  KAM vortices
are well known in closed flow (Section~\ref{closed}), and they have been extensively studied
in that context.  What is perhaps less well known is that they can
also appear in open flows,  for example in the flow of Fig.~\ref{fig:expt}, 
and when they do, they play a crucial role
in the mixing dynamics. They have been observed in geophysical 2D
flows, such as the stratospheric polar vortex, which plays a crucial
role in the process of ozone depletion \cite{vortex_atm}; and also in
ocean circulation patterns \cite{Abraham-98,Abrah1,Abrah2}. The
islands form a fractal hierarchical structure, with large islands
being surrounded by smaller islands, and these in turn are surrounded
by even smaller islands, and so on (Fig.~\ref{fig:cantori}).  The
presence of KAM islands means that there is a finite volume of initial
conditions in the mixing region whose orbits do not escape,
corresponding to those initial conditions lying in the islands.
Moreover, fluid particles with initial conditions outside the
interaction region cannot enter the islands.  As a result, the set of
initial conditions outside the mixing region whose trajectories end up
trapped there still has zero measure, as in the hyperbolic
case. However, the islands have deep consequences for the transient
dynamics, resulting in important differences between the hyperbolic
and non-hyperbolic cases.

The transport of fluid in the vicinity of the islands is dominated by
cantori, which are remnants of broken up KAM tori. Cantori are
invariant sets of the dynamics, as are KAM islands; but in
contrast to those, fluid particles can cross from one side of a
cantorus to the other  \cite{MacKay:1984bz,Ham_chaos}. However, it typically takes very long times to
do so, and as a consequence the cantori act as partial transport
barriers. The overall picture of non-hyperbolic transport is sketched
in Fig.~\ref{fig:cantori}.

\begin{figure}[tb]
\begin{center}
\includegraphics[width = \columnwidth]{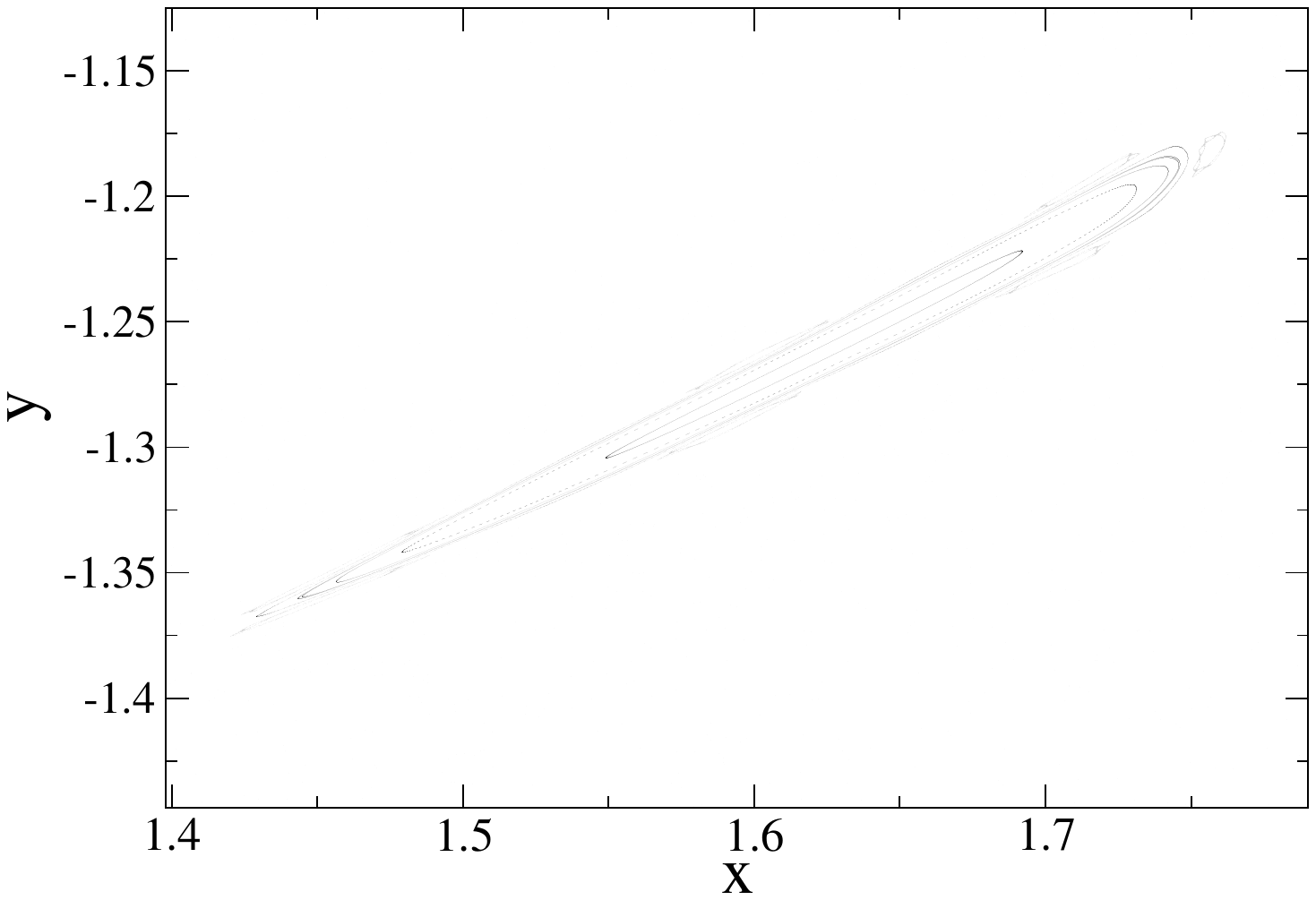}
\includegraphics[width = \columnwidth]{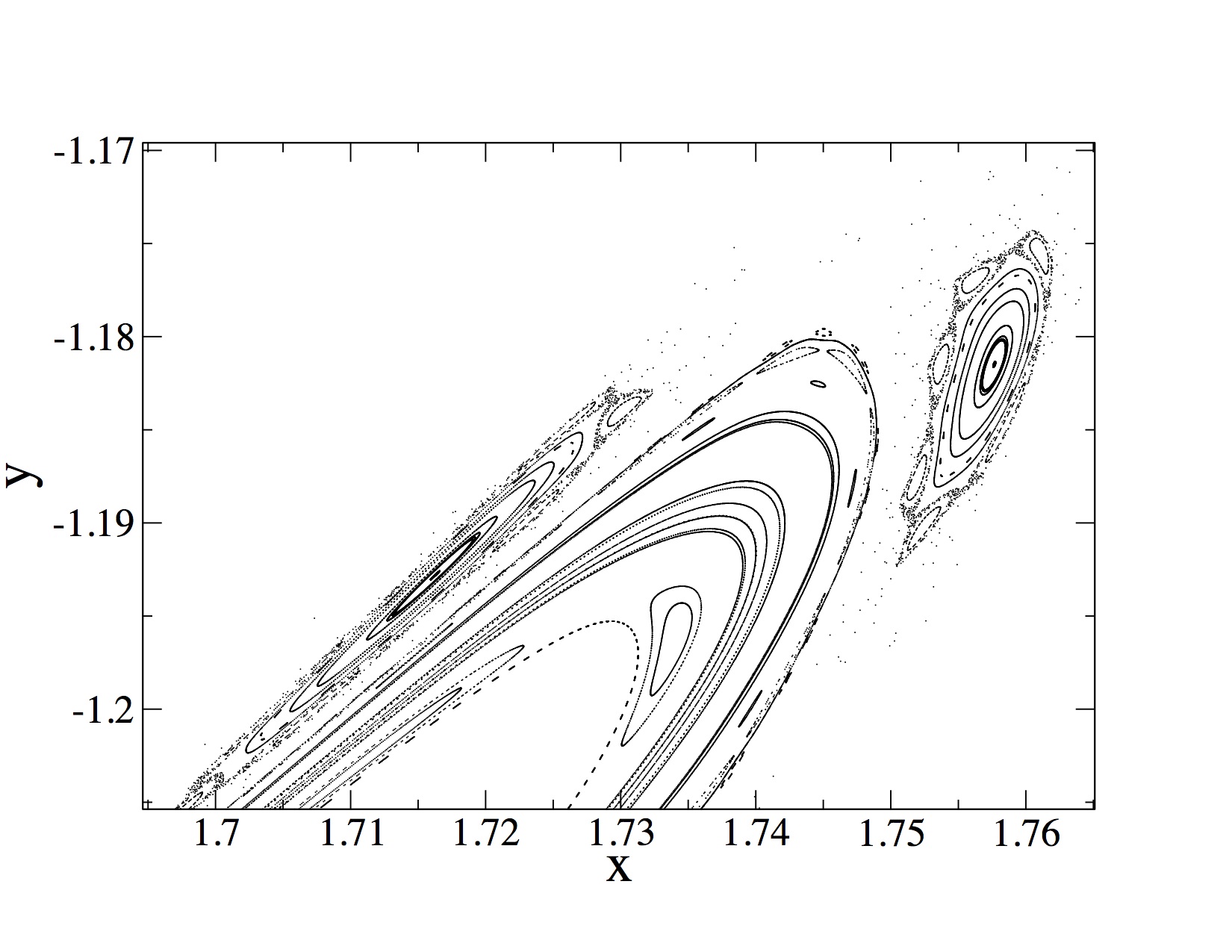}
\end{center}
\caption{
Blinking vortex-sink flow (Eq.~\eqref{eq:blinkingvortexflow}): Poincar\'e map with
$C=T=a=1$, $K=18$, parameters for which the flow is
  non-hyperbolic. The map shows the orbits of a few fluid particles,
  with positions taken at discrete times, at the beginning of every
  period of the flow. The picture on the bottom is a magnification of a
  small region of the top image, and shows the self-similar
  structure of the KAM islands.
  } \label{fig:blink_islands}
\end{figure}

Figure~\ref{fig:blink_islands} shows Poincar\'e sections for a
flow simulation with non-hyperbolic advection dynamics. The magnification shows
the striking self-similar organization of the islands.  The effect of
cantori on the advection dynamics can be seen in the cloud of points
surrounding the sub-islands on the upper right and to the left of the
main island in the magnified figure. These points are snapshots taken
at the start of every period of a single orbit that meanders inside
the cantorus surrounding these islands. This orbit eventually escapes
after thousands of cycles. Another cantorus can just be seen
surrounding the main island. These cantori are in turn surrounded by a
larger cantorus encircling the whole structure, which is apparent from
the higher density of points in the region around the complex of
islands in the bottom figure of Fig.~\ref{fig:blink_islands}. An experimental
example showing KAM islands in an open flow motivated by mixers in the
food industry was investigated by \textcite{Gouillart2009}.

The partition of space by the KAM islands and cantori into distinct
domains separated by transport barriers has no counterpart in
hyperbolic systems, and is the cause of the profound differences in
the dynamics of hyperbolic and non-hyperbolic flows.  A direct
consequence of the self-similar structure of the transport barriers
depicted in Fig.~\ref{fig:cantori} is the phenomenon known as
stickiness: in non-hyperbolic flows, many trajectories spend
extremely long times inside cantori, leading to very long typical
escape times compared to hyperbolic dynamics.  Once inside, an orbit may
enter an inner cantorus located within another cantorus, and so on to
arbitrarily high levels in the cantorus hierarchy.  So once a fluid
particle is inside a cantorus, it will wander within a fractal
labyrinth from which escape is likely to take a very long time. 
Note that the preceding discussion holds for 2D systems only. However, with more degrees of freedom there is also the possibility of ArnolÕd diffusion \cite{Arnold1964}.

Even in non-hyperbolic flows it is still true that fluid particles
with initial conditions outside of KAM islands will eventually escape
with 100\% probability: the component of the chaotic saddle outside
the islands has zero measure.  But stickiness makes escape
sub-exponential, in marked contrast with hyperbolic flows. 
In non-hyperbolic flows, the number
$N(t)$ of particles with initial conditions chosen randomly
in a region with no intersection with KAM islands that have not
escaped up to time $t$ follows a power law \cite{Meiss1985}:
\begin{equation}
\label{power_law_escape}
N(t) \sim t^{-\gamma},
\end{equation}
with $\gamma>0$.

It has been shown that a direct consequence of the slower escape
dynamics described by Eq.~\eqref{power_law_escape} is that the fractal
dimension $D$ of the stable (and unstable) manifold is equal to the
dimension of the embedding space, $D=2$ \cite{Lau}.  From the
interpretation of the fractal dimension as a measure of uncertainty of
transient systems, expressed mathematically by Eq.~\eqref{def_dim},
the fact that $D$ assumes the maximum possible value in non-hyperbolic
systems suggests that these systems have an extreme sensitivity to
initial conditions.  Indeed, the exponent in Eq.~\eqref{def_dim}
vanishes for $D=2$, which means that the ``uncertainty probability''
$p(\eps)$ decreases more slowly than a power law for small
$\eps$.
The fact that $D=2$ in non-hyperbolic open flows suggests that
predicting asymptotic properties of trajectories in these systems is
an almost impossible task.  The reason for this unpredictability is
the very long time it takes initial conditions inside cantori to
escape: two initially close trajectories will have much more time
to spend in the mixing region to separate and follow independent paths
before they escape.  Figures~\ref{fig:cantori} and
\ref{fig:blink_islands} also suggest that the unpredictability is
greater for initial conditions located in deeper levels of the
cantorus hierarchy, as they have longer escape times.

\begin{figure}[tb]
\begin{center}
\includegraphics[width = 0.8\columnwidth]{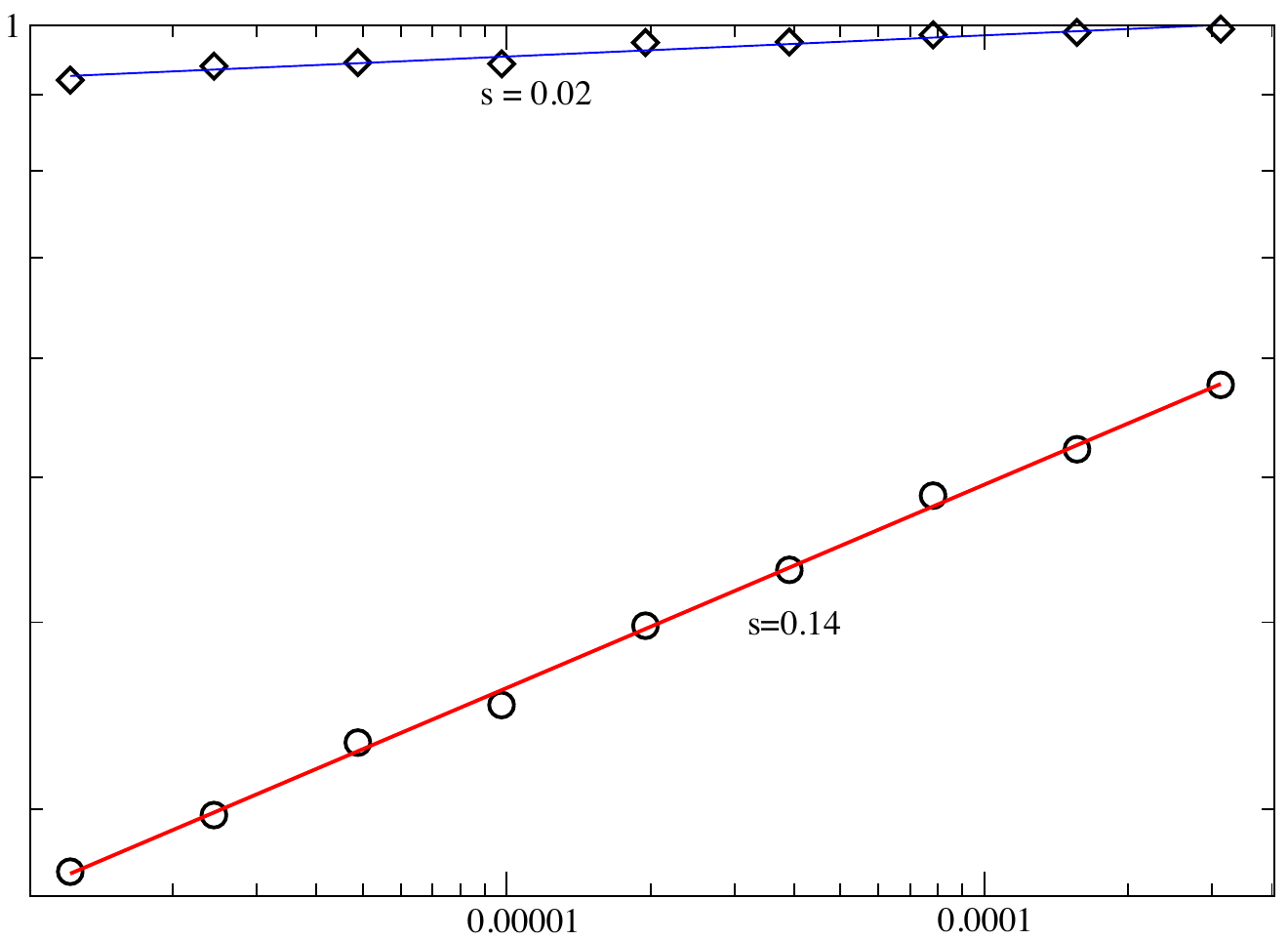}
\end{center}
\caption{Blinking vortex-sink flow: Fraction of uncertain pairs as a function of separation
  $\epsilon$ between points in a pair, with $\eta=1$, $\xi=18$; see Fig.~\ref{fig:filaments}.  Initial
  conditions for the bottom points are taken in the outermost
  cantorus, on the segment $x_0=1.75$, $y_0\in(-1.18,-1.175)$; for the
  top points, initial conditions are in an inner cantorus, on $x_0=1.75$,
  $y_0\in(-1.193,-1.192)$. The numbers beside each line are the
  slope coefficients obtained from fitting,
  $f(\epsilon)\sim\epsilon^s$.
  } \label{fig:uncplot_nhyp}
  \end{figure}

To investigate the sensitivity to initial conditions in different
areas of space in Fig.~\ref{fig:blink_islands} one may calculate
numerically the fraction $f(\eps)$ of $\eps$-separated pairs of points
whose escape times differ by one period or more.  For a sufficiently
large sample, we expect $f(\eps)\propto p(\eps)$. The result, for initial conditions taken
in two different cantori, is
plotted in Fig.~\ref{fig:uncplot_nhyp}.
Figure~\ref{fig:uncplot_nhyp} seems to go against the claim that $D=2$
for non-hyperbolic systems, since this would predict that the plot of
$f(\eps)$ versus $\eps$ should be a line with zero slope.  But in
non-hyperbolic systems, the $\eps\rightarrow 0$ limit in
Eq.~\eqref{def_dim} converges sub-logarithmically with $\eps$
\cite{Lau}.  This extremely slow convergence means that reaching this
limit usually requires values of $\eps$ so small they are not
physically meaningful.  Any model of a physical system has a lower
scale below which the model is no longer valid; for example, the size
of advected particles or the finite resolution of our measurements.
This implies that the dimension that is physically relevant for
realistic systems is not the mathematical definition Eq.~\eqref{def_dim}
with its unreachable limit, but is given instead by an effective
  dimension $D_{\mathrm{eff}}$ \cite{Moura2004,Motter2005}, defined
as an approximation of the fractal dimension for a finite range of
$\eps$:
\begin{equation}
  D_{\mathrm{eff}}(\eps) = 2 - \frac{d\ln
f(\eps)}{d\ln\eps}\approx\mbox{const.} \quad \mbox{for~}
\eps_1 < \eps < \eps_2,
\label{effdim}
\end{equation}
valid in a range $(\eps_1,\eps_2)$, with $\eps_1 \ll
\eps_2$.  $D_{\mathrm{eff}}$ satisfies
$D_{\mathrm{eff}}(\eps)\rightarrow 2$ as $\eps\rightarrow 0$,
in accordance with Eq.~\eqref{def_dim}.
From Eq.~\eqref{effdim}, the results in Fig.~\ref{fig:uncplot_nhyp}
can be interpreted as yielding the effective fractal dimensions of the
stable and unstable manifolds for two different locations in space:
$D_{\mathrm{eff}}=1.86$ inside the outermost cantorus, and
$D_{\mathrm{eff}}=1.98$ inside one of the inner cantori.  The
effective dimension therefore depends on the position in
non-hyperbolic systems, in contrast to the actual fractal dimension,
which is 2 anywhere.  The greater escape time in the inner cantori means
that the invariant manifolds of the chaotic saddle have more time
to be stretched and folded and distorted by advection, hence the
greater effective dimension.

Because of time-reversal symmetry, the stable and unstable manifolds
have the same fractal dimensions, and also the same effective
fractal dimensions.  We argued above that the fractal dimension of the
unstable manifold is a measure of lower-scale mixing efficiency for
open flows.  This means that the fluid in regions of space surrounded
by cantori will be extremely well mixed, and the efficiency of mixing
increases as we go deeper into the cantorus structure, and reaches the
maximum limit of $D_{\mathrm{eff}}\rightarrow 2$ for regions buried
deep within the cantori.  This picture is somewhat at odds with an
idea prevalent in this field that KAM islands are obstacles to mixing.
That view is justified in closed flows (Section~\ref{closed}), where one  wants
to mix the fluid homogeneously throughout the container; this is
not possible in the presence of KAM islands. In open flows,
however, the fluid to be mixed usually comes from the inflow region,
and thus from outside the KAM islands, and so this is not an issue if
the material to be mixed is injected into the flow outside the
islands. For open flows, one wants the flow to be well mixed by the
time it reaches the outflow region.  The cantori surrounding KAM
islands greatly enhance this kind of mixing, by causing fluid to spend
very long times within themselves.  This comes at a cost: the time it
takes for any given piece of fluid to escape a cantorus to the outflow is very
much increased by the stickiness.  If one has a continuum input of dye
or other material one wants to mix, however, this may not be relevant
in practice.  All this suggests that in open flows the best strategy
to achieve optimal mixing would be to inject material inside
the cantori, but still outside the islands.

\section{The role of walls}\label{closed}

Many studies of mixing over the years have used maps of flows in periodic domains (cat map, standard map, etc) to great effect to generate insight into the evolution of chaotic dynamics.  However, in actual containers the solid boundaries throw up several new effects whose consequences for mixing are not confined to thin boundary layers but penetrate into the bulk of the flow.
Mixing follows a different dynamics when the flow is confined to a closed space. As discussed, in general stirring induces chaotic advection in the flow, causing fluid elements to be repeatedly stretched and folded, generating over time a very fine pattern of thin filaments with a complex structure. In a closed container, these filaments eventually spread throughout the available space, as long as the flow has no prominent regular islands, and total mixing is achieved in the asymptotic limit of $t\rightarrow\infty$. The physically relevant questions are then related to the time-scale over which mixing is achieved, and how the system approaches the limit of becoming perfectly mixed.

\subsection{Things are not always exponential}

The prototype of chaotic mixing is as follows: stirring a fluid
promotes chaotic advection, which leads to an exponential stretching
of fluid elements.  These fluid elements carry some concentration of a
substance to be mixed, and as they are stretched gradients of
concentration increase exponentially.  This allows molecular diffusion
to act efficiently, and the uniformization of the concentration
proceeds at a much faster rate than it would have in the absence of
stirring \cite{Eckart1948, Welander1954, Batchelor1959}.  Typically,
this decay is exponential in time; the decay constant is not, however,
simply the average rate of stretching (infinite-time Lyapunov
exponent), but is obtained from the distribution of finite-time
Lyapunov exponents in a nontrivial manner \cite{Antonsen1995,
  Antonsen1996, Balkovsky1999, Falkovich2001, ThiffeaultAosta2004}.
This is the `local' picture of chaotic mixing; in some cases it
must be supplemented by a more global approach, where one analyses the
advection--diffusion operator \cite{Pierrehumbert1994,
  Fereday2002, Wonhas2002, Pikovsky2003, Fereday2004, Thiffeault2003d,
  Haynes2005}.  However, whether the decay of concentration
is locally or globally controlled, the rate is still exponential.

This exponential-decay framework is helpful, but it is complicated by
the presence of walls.  In this case, several authors \cite{JTA89,Jones1994,
  Chertkov2003b, Lebedev2004, Schekochihin2004, Salman2007,
  Popovych2007, Chernykh2008, MacKay_CCT2007, Boffetta2009,
  Zaggout2012} have suggested that the no-slip boundary condition and
the presence of separatrices on the walls slow down mixing: the decay
is power law rather than exponential.  (This is connected to a
breakdown of hyperbolicity.)  Recent
experiments \cite{Gouillart2007,Gouillart2008,Gouillart2009,Gouillart2010}
have confirmed this hypothesis, and also showed that for a significant period of
time the rate of decay of variance is dramatically reduced, even away
from the walls, due to the entrainment of unmixed material into the
central mixing region.

In this section we describe the limiting effect of boundaries on
chaotic mixing, following \citet{Gouillart2007,Gouillart2008}.  We
then explain how creating closed orbits near the wall alleviates the
problem somewhat, by `shielding' the central mixing region from the
detrimental effect of walls \cite{Gouillart2010b,Thiffeault2011c}.
We end by exploring mixing by non-reciprocal contractible loops of the wall's positions; a class of protocols directly related to the concept of geometric phases \cite{arrieta2015}.

\subsection{Passive scalar near a wall}

\begin{figure}[tb]
  \subfigure[]{
    \includegraphics[width=.46\columnwidth]{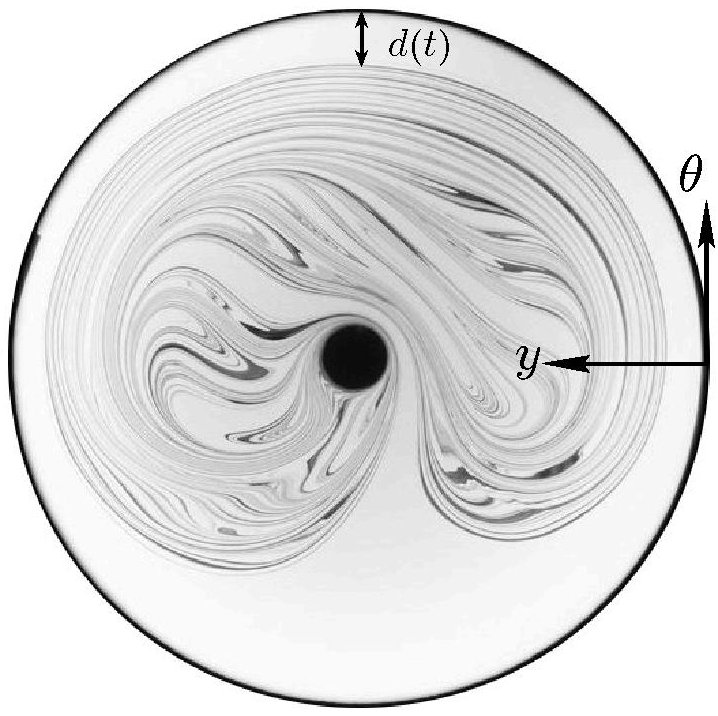}
    \label{fig:fig8exp}
  }\hspace{.2em}
  \subfigure[]{
    \includegraphics[width=.46\columnwidth]{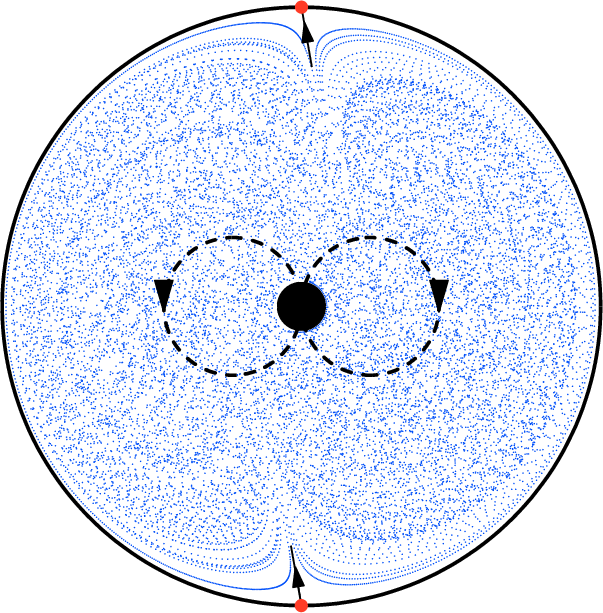}
    \label{fig:eightpoinc_wout=0}
  }
  \caption{Rod mixer: (a) Experiment with a figure-of-eight stirring protocol,
    showing an advected blob of dye (India ink) in sugar syrup.  The
    coordinate system used here is also indicated, as is the
    distance $\gap(\time)$ between the dark mixing pattern and the
    wall. (b) Numerical Poincar\'e section, showing the two
    fixed points and their
    separatrices. Reprinted with permission from \citet{Thiffeault2011c}. Copyright (2011) the American Physical Society.
    }
\end{figure}

Consider the experiment shown in Fig.~\ref{fig:fig8exp}:
a dark blob of ink in a light fluid has been stretched and folded
repeatedly by the periodic movement of a rod.  The movement of the rod
defines a figure-of-eight stirring protocol with period $T$, shown as
a dashed line in Fig.~\ref{fig:eightpoinc_wout=0}.  3D
effects are negligible, and the fluid  flow can be treated as
a Stokes flow. The mixing pattern has a kidney shape, and it slowly
grows and approaches the wall.  The distance of closest approach at
the top is $\gap(\time)$, where~$\time$ is time.

Inside the central mixing region, we assume the action of the flow is
that of a simple chaotic mixer.  By this we mean that fluid elements
are stretched, on average, at a given rate~$\lambda$
(the Lyapunov exponent).  Hence, after a
time~$\time$ a blob of initial size~$\delta$ will have
length~$\delta\,\ee^{\lambda\time}$.  However, because of diffusion,
its width will stabilize at an equilibrium between compression and
diffusion at a
scale $\lB$,  the Batchelor length, Eq.~\eqref{eq:batchelor-scale}
\cite{Batchelor1959,Balkovsky1999,ThiffeaultAosta2004}.

 We emphasize that the flow in Fig.~\ref{fig:fig8exp} is globally
chaotic in the sense that it does not possess visible islands, as
evidenced by the numerical Poincar\'e section in
Fig.~\ref{fig:eightpoinc_wout=0}.  The chaotic region extends all the
way to the wall, but there are clearly two special points at
the wall, at the top and bottom --- shown as dots in
Fig.~\ref{fig:eightpoinc_wout=0} --- corresponding to separatrices.
They are associated with the stable (top) and unstable (bottom)
manifolds of two distinguished non-hyperbolic (parabolic) fixed points at the
wall. 

Each period, the pattern gets progressively closer to the wall.
Assuming molecular diffusion can be neglected, because of area
preservation some white fluid must have entered the central mixing
region.  It does so in the form of white strips, visible as
layers inside the pattern of Fig.~\ref{fig:fig8exp}.  If we assume
that the mixing pattern grows uniformly along the periphery of the
wall, we can write the width~$\Delta(\time)$ of a strip injected at
period~$\time/\T$ as
\begin{equation}
  \Delta(\time) = \gap(\time) - \gap(\time+\T) \simeq
  -\T\,\dotgap(\time) \ge 0,
\end{equation}
where we also assumed that~$\gap(\time)$ changes slowly in time.

Now, if a white strip is injected at time~$\tau$, how long does it
persist before it is wiped out by diffusion?  The answer is the
solution~$\time$ to the equation
\begin{equation}
  \Delta(\tau)\,\ee^{-\lambda(\time-\tau)} = \lB\,.
  \label{eq:age}
\end{equation} 
This means that the strip initially had
width~$\Delta(\tau)$ when it was injected, it gets compressed by the
flow in the central mixing region by a
factor~$\exp{(-\lambda(\time-\tau))}$ depending on its
age,~$\time-\tau$, and once it is compressed to the Batchelor
length~$\lB$ it quickly diffuses away.  Thus, we can
solve Eq.~\eqref{eq:age} to find the age the strip has when it gets wiped
out by diffusion,
\begin{equation}
  \time-\tau = \lambda^{-1}\log(\Delta(\tau)/\lB).
  \label{eq:age2}
\end{equation}

Eventually, at time~$\tB$, any newly-injected filament will have width
equal to the Batchelor length.  This occurs when
\begin{equation}
  \Delta(\tB) = \lB,
  \label{eq:tB}
\end{equation}
which can be solved for~$\tB$ given a form for~$\Delta(\time)$.  After
this time it makes no sense to speak of newly-injected filaments as
`white,' since they are already dominated by diffusion at their birth.
Hence, the description we present here is valid only for times earlier
than~$\tB$, but late enough that the edge of the mixing pattern has
reached the vicinity of the wall.

In their experiments, \citet{Gouillart2007,Gouillart2008} measured the
intensity of pixels in the central mixing region.  They observed
for \hbox{$\T \ll \time \lesssim \tB$} that the concentration variance
is dominated by the proportion of strips in the central region that
are still white at that time.  Because of area conservation, the total
area of injected white material that is still visible at time~$\time$
is proportional to
\begin{equation}
  \Aw(\time) = \gap(\tau(\time)) - \gap(\time),
  \label{eq:Aw}
\end{equation}
where we use Eq.~\eqref{eq:age2} to solve for~$\tau(\time)$, the injection
time of the oldest strip that is still white at time~$\time$.  Hence,
the goal is to estimate~$\Aw(\time)$ for times~\hbox{$T \ll \time
  \lesssim \tB$}, since~$\Aw$ is directly proportional to the
concentration variance.  To do this we need~$\tau(\time)$, which
requires specifying~$\Delta(\time)$.  We now look at three possible forms,
corresponding to a free-slip wall, a no-slip wall, and a moving
no-slip wall.

\subsection{Exponential approach to a free-slip wall}

Consider first the case where~$\gap(\time)=\gap(0)\,\ee^{-\mu\time}$
for some positive constant~$\mu$.  We have~$\Delta(\time)=-\T\dotgap =
\mu\T\gap(0)\,\ee^{-\mu\time} = \Delta(0)\,\ee^{-\mu\time}$.
From Eq.~\eqref{eq:tB}, we have~$\tB=\mu^{-1}\log(\Delta(0)/\lB)$, and
from Eq.~\eqref{eq:age2},
\begin{equation}
  \time - \tau = \frac{\mu}{\lambda-\mu}\,(\tB - \time).
  \label{eq:exptau}
\end{equation}
By assumption, $\tau<\time<\tB$, so for consistency we
require~$\mu<\lambda$, i.e., the rate of approach toward the wall is
slower than the natural decay rate of the chaotic mixer. The area of
white material in the mixing region is then obtained
from Eq.~\eqref{eq:Aw}:
\begin{equation}
  \Aw(\time) = \gap(0)\,\ee^{-\mu\time}
  \l(\exp\l(\frac{\mu^2}{\lambda-\mu}(\tB-\time)\r) - 1\r),
\end{equation}
which in the wall-dominated regime ($\lambda/\mu\gg 1$) can be
approximated by
\begin{equation}
  \Aw(\time) \sim
  \gap(0)\,\lambda^{-1}\mu^2\,(\tB-\time)\,\ee^{-\mu\time},
  \qquad \time \lesssim \tB.
  \label{eq:Awexp}
\end{equation}
The decay rate of the white area is completely dominated by the
walls.  The central mixing process is potentially more efficient
($\lambda>\mu$), but it is starved by the boundaries.

If~$\mu>\lambda$, we have $\time>\tB$ in Eq.~\eqref{eq:exptau}, since
newly injected strips reach the Batchelor length before strips
that were injected previously.  This violates our assumptions, and we
conclude that in that case the white strips can be neglected; the
decay rate of the concentration variance is then given by the natural
decay rate~$\lambda$.

As an example of an exponential approach to the wall, consider the
velocity field near a free-slip boundary,
\begin{subequations}
\begin{align}
  \uc(\xc,\yc) &= \uc_0(\xc) + \Order{\yc},\\
  \vc(\xc,\yc) &= -\uc_0'(\xc)\yc + \Order{\yc^2},
\end{align}%
\label{eq:slipvel}%
\end{subequations}
which satisfies the incompressibility constraint.  Here~$\uc$ is the direction parallel to the wall, and~$\vc$ is perpendicular to the wall.  The perpendicular distance from the wall is~$\yc$, and $\xc$ is a angle around the circular boundary.
(Since the dynamics
near the wall are slow, we can use a steady flow here to model the
time-$\T$ Poincar\'e map.)  A separatrix is a distinguished streamline that ends at the boundary at some position~$\xc=\xcsep$.   Along a separatrix at~$\xc=\xcsep$, we
have~$\uc_0(\xcsep)=0$ since the velocity field changes sign.  The
rate of approach along the separatrix is thus given by~$\dotgap =
\vc(\xcsep,\gap) = -\uc_0'(\xcsep)\gap$, so that~$\mu=\uc_0'(\xcsep)$.
Hence, if~$\uc_0'(\xcsep)>\lambda$ the rate of decay of concentration
variance will not be limited by wall effects.

\subsection{Algebraic approach to a no-slip wall}

If the fluid at the wall is subject to no-slip boundary conditions,
the Taylor expansion Eq.~\eqref{eq:slipvel} is modified to become
\begin{subequations}
\begin{align}
  \uc(\xc,\yc) &= \uc_1(\xc)\yc + \Order{\yc^2},\\
  \vc(\xc,\yc) &= -\tfrac12\uc_1'(\xc)\yc^2 + \Order{\yc^3}.
\end{align}%
\label{eq:noslipvel}%
\end{subequations}
The rate of approach along the separatrix at~$\xc=\xcsep$ is given
by~$\dotgap = \vc(\xcsep,\gap) = -\tfrac12\uc_1'(\xcsep)\gap^2$, with
asymptotic solution~$\gap(\time)\sim 2/(\uc_1'(\xcsep)\time)$,
for~$\gap(0)\uc_1'(\xcsep)\time\gg1$.  This is independent of the initial
condition~$\gap(0)$: asymptotically, a fluid particle forgets its
initial position; this explains why material lines bunch up against
each other faster than they approach the wall, as reflected by the
front in the upper part of Fig.~\ref{fig:fig8exp}.  The total area of
remaining white strips at time~$\time$ as given by Eq.~\eqref{eq:Aw} is
proportional to
\begin{equation}
  \Aw(\time) = \frac{2}{\uc_1'(\xcsep)\tau} - \frac{2}{\uc_1'(\xcsep)\time}
  = \frac{2}{\uc_1'(\xcsep)}\,\frac{\time-\tau}{\tau\time}\,.
  \label{eq:Awalg0}
\end{equation}
The width of injected strips is~$\Delta(\time) = -\T\dotgap =
2\T/(\uc_1'(\xcsep)\time^2)$.  Equation~\eqref{eq:age2} cannot be
solved exactly, but since~$\tau(\time)$ is algebraic its right-hand
side is not large, implying that~$\time/\tau\simeq
1$ for large~$\time$.  We can thus replace~$\tau$ by~$\time$
in Eq.~\eqref{eq:age2} and the denominator of Eq.~\eqref{eq:Awalg0}, and find
\begin{equation}
  \Aw(\time) \simeq \frac{2}{\uc_1'(\xcsep)}\,
  \frac{\log(\Delta(\time)/\lB)}{\lambda\,\time^2}\,,\quad
  \frac{1}{(\gap(0)\uc_1'(\xcsep))} \ll \time \ll \tB.
  \label{eq:Awalg}
\end{equation}
Compare this to the exponential case of Eq.~\eqref{eq:Awexp}: the decay
of concentration variance is now algebraic ($1/\time^2$), with a
logarithmic correction.  The form Eq.~\eqref{eq:Awalg} has been verified
in experiments and using a simple map
model~\cite{Gouillart2007,Gouillart2008}.

\subsection{Dynamics near a moving no-slip wall}

Now consider the case of a rotating wall, where we add a constant
speed~$\Uc>0$ to the velocity~$\uc$ in Eq.~\eqref{eq:noslipvel}.  Again we
look for fixed points: all the parabolic fixed points on the wall have
disappeared, as well as the two separatrices.  Since~$\A(\xc)$ is
continuous, has two zeros, and~$\A'(0)>0$, $\A(\xc)$ must have a
minimum at some angle~$\xc_*$, where~$\A'(\xc_*)=0$ and
hence~\hbox{$\vc(\xc_*,\yc)=0$} for all~$\yc$.  Enforcing that the
along-wall velocity also vanish, there will be a fixed point at~$\yc_* =
-\Uc/\A(\xc_*)$.  Now we look at the linearized dynamics near the
fixed point.  Let~$(\xc,\yc) = (\xc_* + \Xc,-\Uc/\A(\xc_*) + \Yc)$;
then
\begin{subequations}
\begin{align}
  \dot\Xc &= \A(\xc_*)\Yc + \Order{\Xc^2,\Yc^2,\Xc\Yc},\\
  \dot\Yc &= -\tfrac12\A''(\xc_*)\yc_*^2\,\Xc + \Order{\Xc^2,\Yc^2,\Xc\Yc}.
\end{align}%
\end{subequations}%
The linearized motion thus has eigenvalues~$\lambda_\pm = \pm\lambda
= \pm\sqrt{-{\A''(\xc_*)}/{2\A(\xc_*)}}\,U$, where the argument in
the square root is non-negative since~$\A(\xc_*)<0$
and~$\A''(\xc_*)\ge0$.  For~$\A''(\xc_*)>0$ and~$\Uc>0$, this is a
hyperbolic fixed point, and the approach along its stable manifold is
given by~$\Yc(\time) \sim \Yc_0 \,\exp(-\lambda\,\time)$
for~$(\Xc_0,\Yc_0)$ initially on the stable manifold.  Compare this to
the~$1/\time$ approach for a fixed wall: the approach to the fixed
point is now exponential, at a rate proportional to the speed of
rotation of the wall.  One expects that this exponential decay will
dominate if it is slower than the mixing rate in the bulk.  Otherwise,
if~$\lambda$ is large enough, then the rate of mixing in the bulk
dominates.

\begin{figure}[tb]
\begin{center}
\subfigure[]{
  \includegraphics[width=.4825\columnwidth]{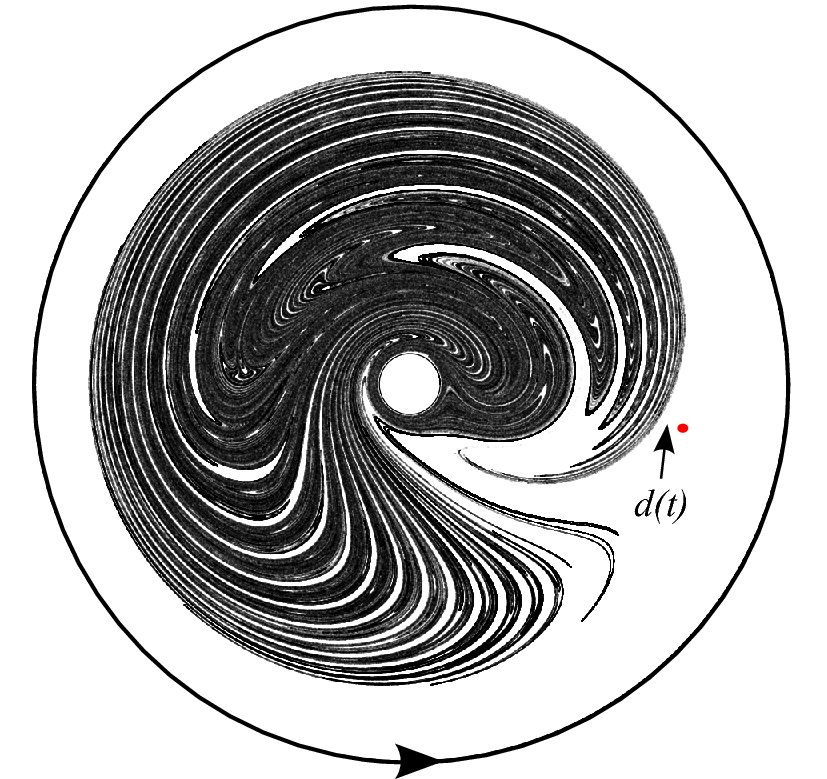}
  \label{fig:fig8rot}
}\hspace{.2em}
\subfigure[]{
  \includegraphics[width=.445\columnwidth]{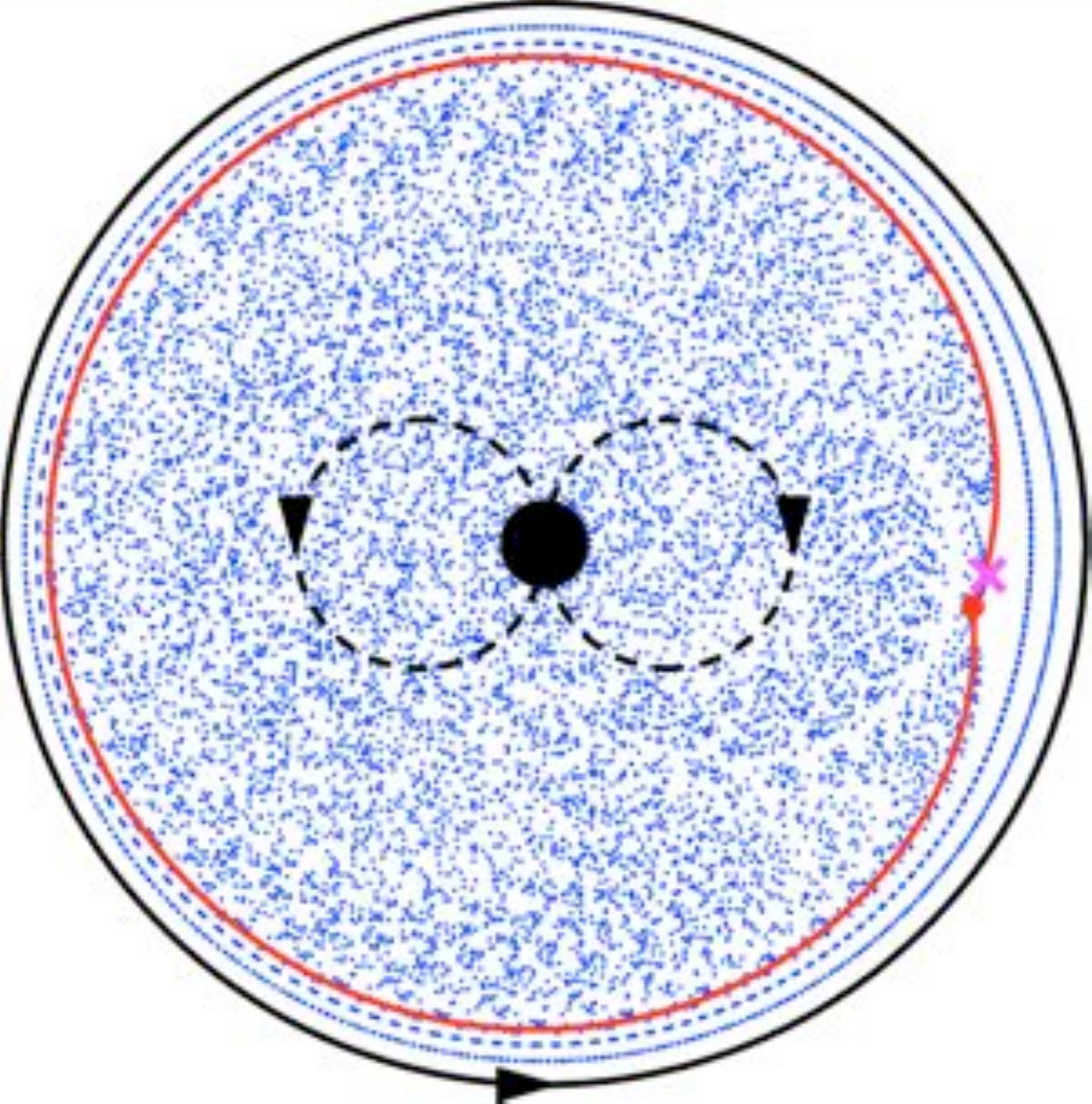}
  \label{fig:eightpoinc_wout=0p2}
}
\end{center}
\caption{Rod mixer: (a) Numerical simulation of dye advection for a wall rotating
  at velocity~$\Uc\T=0.2$, with the rod moving in a figure-of-eight
  pattern. (b) Poincar\'e section, which shows a large chaotic region
  and closed orbits near the wall, with a separatrix in between.
  Reprinted with permission from \citet{Thiffeault2011c}. Copyright (2011) the American Physical Society.
  }
\end{figure}

Figure~\ref{fig:fig8rot} shows a numerical simulation of the flow
pattern for a wall rotating at a rate~$\Uc\T=0.2$.  The hyperbolic
fixed point is indicated by a dot, as is the distance~$\gap(\time)$
between the mixing pattern and the hyperbolic point.  Note the unmixed
region between the rotating wall and the mixing pattern.
Figure~\ref{fig:eightpoinc_wout=0p2} is a Poincar\'e section that
shows the presence of the unmixed region, which consists of closed
orbits.  Numerical simulations have confirmed
that the decay rate of a passive scalar in the central region is
indeed exponential, so the rotating wall can help recover exponential
mixing \cite{Thiffeault2011c}.  However, the price to pay is that there is now an unmixed
region surrounding the region of good mixing.  Whether this is a price
worth paying depends on the specific application.

\begin{figure}[tb]
\begin{center}
\subfigure[]{
  \includegraphics[width=.46\columnwidth]{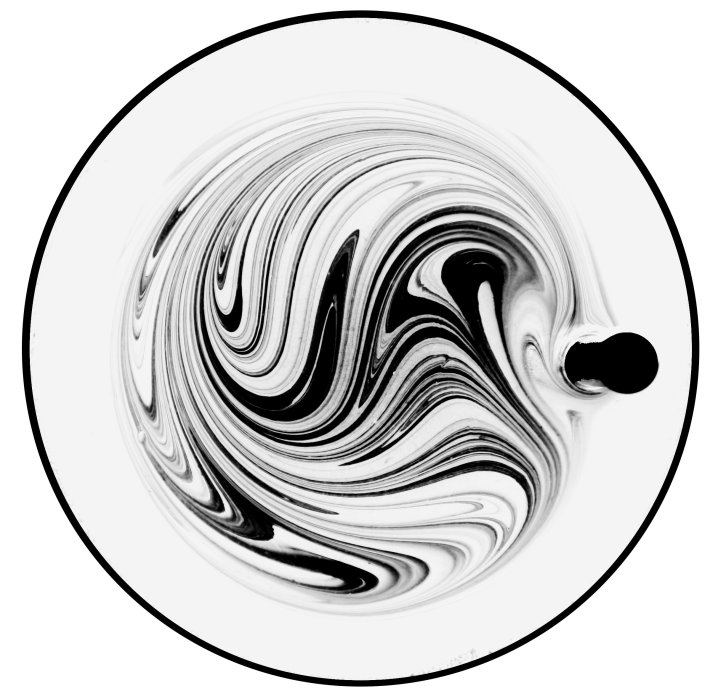}
  \label{fig:trm1}
}\hspace{.2em}
\subfigure[]{
  \includegraphics[width=.46\columnwidth]{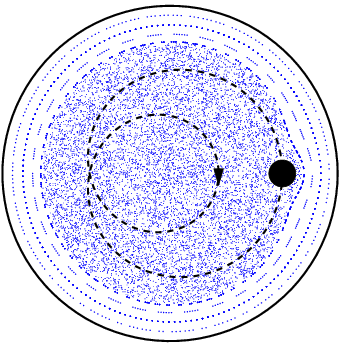}
  \label{fig:trmpoinc}
}
\end{center}
\caption{Rod mixer: The `epitrochoid' stirring protocol.  (a) Experiment; (b)
  Poincar\'e section, also showing the rod's trajectory.  Closed
  orbits are present near the wall, even though the wall is
  fixed. 
  Reprinted with permission from \citet{Thiffeault2011c}. Copyright (2011) the American Physical Society.
 }
\label{fig:trm}
\end{figure}

Another strategy to mimic a moving wall, and thus recover
exponential mixing, is to move the rod in a looping `epitrochoid' motion, shown
in Fig.~\ref{fig:trm}.  This motion creates closed trajectories near
the wall, as is evident in the Poincar\'e section, 
Fig.~\ref{fig:trmpoinc}.  \citet{Thiffeault2011c} have verified
experimentally that the decay of the passive scalar in this case is
indeed exponential, for the same reason as for the moving wall.  However,
the analysis of the near-wall map for this system is more complicated
than for a moving wall and has not been carried out.
3D effects also remain to be investigated: these could
hold some surprises, since the nature of separatrices at the wall is
potentially much richer.

\subsection{Geometric mixing}

\begin{figure*}[tb]
\begin{center}
\includegraphics[width=0.8\textwidth,clip=true]{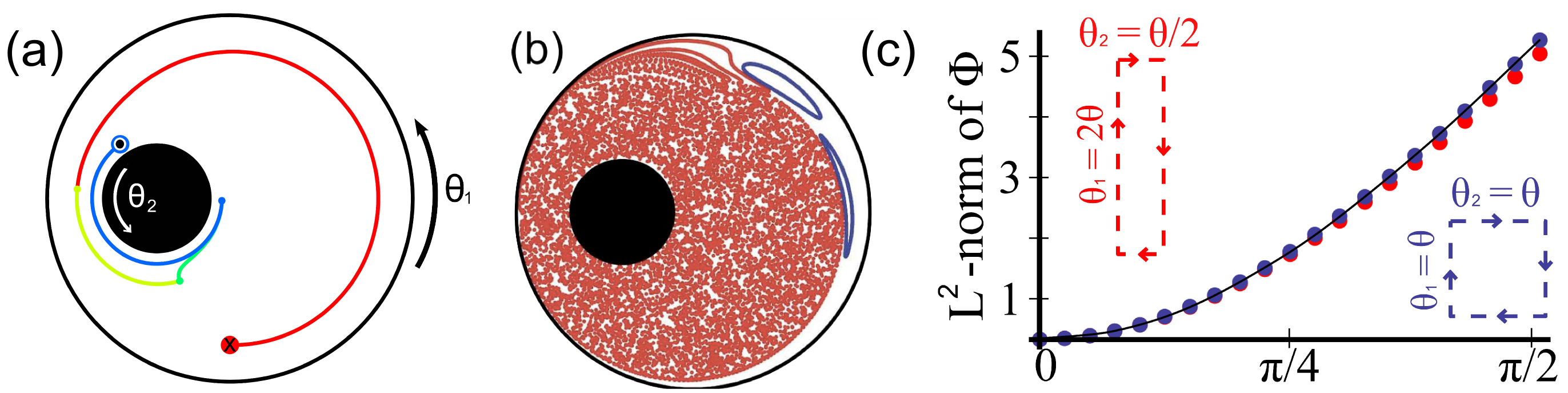}
\end{center}
\caption{\label{geom_mixing_fig}(a) A finite-area non-reciprocal contractible loop. The journal-bearing flow with cylinder radii $R_1 = 1.0$, $R_2 = 0.3$ and eccentricity $\varepsilon = 0.4$, taken around a closed square parameter loop with  $\theta_1 =  \theta_2 = \theta = 2\pi$ radians. The four segments of the loop are plotted in different colors/shades to enable their contributions to the particle motion to be seen. A trajectory beginning at $(0.0, 0.8)$ is shown. (b) Poincare maps demonstrate geometric mixing for this flow for the same cylinder radii, eccentricity and displacements as in (a). Chaotic trajectories are marked in a different shade/color to regular ones. $10 000$ iterations of the parameter loop are shown. (c ) The $L^{2}$-norm of the geometric phase grows quadratically with $\theta$ for loops with small area. Two distinct loops with equal area are shown. Reprinted from~\citet{arrieta2015}.}
\end{figure*} 

Protocols aimed for efficient mixing of fluids heavily rely upon the generation of a chaotic kinematic template. In most cases, mixing protocols can be designed free of major geometrical constraints. In particular, no limitation is usually imposed upon the relative displacement of boundaries. But, can we achieve fluid mixing if we limit the allowed motion of boundaries to the subset returning to their original position after each iteration of the protocol? The answer to the above questions is related to the concept of geometric phases \cite{shapere,shapere2}: the failure of system variables to return to their original values after a closed circuit in the parameters.

In the zero Reynolds number limit, fluid inertia is negligible, fluid flow is reversible, and an inversion of the movement of the walls leads, up to perturbations owing to particle diffusion, to unmixing, as \textcite{taylor_film} and \textcite{heller} demonstrated. This would seem to preclude the use of reciprocating motion to stir fluid at low Reynolds numbers; it would appear to lead to perpetual cycles of mixing and unmixing. But, is that always the case? Can cyclic changes in the shape of the containers lead to efficient mixing?

The well-known 2D mixer based on the journal-bearing flow \cite{aref1986,chaiken,Ottino1989,tabor} may be used as an example of how nonreciprocal cycling of the deformable boundaries of a container can be used as a tool for fluid mixing at low Reynolds number  \cite{arrieta2015}; Fig. \ref{geom_mixing_fig}. Considering as parameters in this device the positions of the outer and inner cylindrical walls of the container specified respectively with the angles $\theta_1$  and  $\theta_2$  from a given starting point, a geometric phase might arise from driving this system around a loop in the parameter space. In a Heller-Taylor-type unmixing demonstration the parameter loop is very simple:  $\theta_1$  first increases a certain amount and then decreases the same amount while $\theta_2$ remains fixed. This loop encloses no area, and reversibility ensures that the phase is zero. To obtain a finite-area non-reciprocal contractible loop we can, for instance, rotate first one cylinder, then the other, then reverse the first, and finally reverse the other. If we now perform a parameter loop by the sequence of rotations detailed above, we arrive back at our starting point from the point of view of the positions of the two cylinders, so it is, perhaps, surprising that the fluid inside does not return to its initial state. We illustrate the presence of this geometric phase in Fig. \ref{geom_mixing_fig}(a) in which an example of the trajectory of a fluid particle is shown as the walls are driven through a non-reciprocal contractible loop. 

The long-term fluid dynamics elicited by a repeated realization of the same contractible non-reciprocal loop is shown in Fig. \ref{geom_mixing_fig}(b). A single fluid particle has covered most of the area available to it between the two cylinders. This is fluid mixing induced entirely by a geometric phase; we may call it geometric mixing. Geometric mixing therefore creates chaotic advection. The geometric phase scales with the area of the parameter loop and it is independent, at least for small enough loops, of the specifics of the trajectory in parameter space (Fig.~\ref{geom_mixing_fig}(c)).
 
The example above illustrates a general class of protocols in which mixing arises as a consequence of a geometric phase induced by a contractible non-reciprocal cycle in the parameters defining the shape of the container. It turns out that the mixing efficiency estimated from the stretching of material lines is roughly proportional to the geometric phase. Mixing in the corresponding flows can be also considered as the result of chaos arising in the mapping describing the motion of fluid elements during one cycle. When the cycle is reciprocal, this map is the identity and a small departure from reciprocity corresponds to a small departure from the identity map. Hence, the problem of mixing by nonreciprocal cycles is closely related to the class of dynamical systems constituted by perturbations of the identity \cite{arrieta2015}. The structure of chaos in this class of dynamics has been greatly overlooked in the literature, which points to a much needed revisiting of this associated problem.

\section{The new frontier: 3D unsteady flow}\label{3D}

Most results to date of chaotic advection, indeed of dynamical systems in general, have been found by examination of 2D maps or flows where stable elliptic fixed points and the stable and unstable 1D manifolds of hyperbolic points define just a few Lagrangian coherent structures that control all of the transport behavior.  However, in 3D there is an explosion of complexity in the number of possible Lagrangian structures and connections between them.  This is due both to impossibility of the existence of any stable fixed points and to hyperbolic manifolds existing as both sheets and curves.  It is still an open question --- especially in experiment --- how these structures fit together to control mixing rates and the distribution of material and energy in a 3D stirred flow.

\subsection{Motivation and background}
\label{Background}

Coherence and, intimately related to that, invariance are key notions in the investigation and classification of transport phenomena in (laminar) fluid flows. These notions can in general be defined in several ways (refer, e.g., to Section~\ref{data}). The discussion within this section adopts the Lagrangian perspective of organization of fluid trajectories into coherent structures collectively defining the flow topology that geometrically determine the advective transport of material. It is important to note in this context that basically {\em any} group or union of fluid trajectories constitutes a material entity (Section~\ref{sec:LCS}), suggesting a certain arbitrariness in the definition of coherent structures. An example of an (in general) non-unique entity is a stream surface in 3D steady flows: any material line advected unobstructed by the flow describes a stream surface. Consider to this end Poiseuille flows, where advection of any family of closed material curves released at the inlet yields a valid foliation into stream surfaces (Section~\ref{LaminarFlow}).
Connection with other properties or entities of the system renders coherent structures in the web of Lagrangian fluid trajectories unique. Structures tied directly to properties of the kinematic equation governing Lagrangian motion are, arguably, the most fundamental kind and include entities such as separatrices due to discrete symmetries, families of invariant surfaces due to continuous symmetries, 2D manifolds and tubes associated with closed streamlines/periodic lines and 1D/2D manifolds associated with isolated stagnation/periodic points. Such structures constitute elements of the flow topology or, equivalently, the ``ergodic partition'' (Section~\ref{visualization}). Coherent structures may also be defined indirectly as, for instance, material entities distinguished by Lyapunov exponents (Section~\ref{sec:LCS}), topological deformation of enclosing material curves (Section~\ref{sec:braids}) or leakage from Eulerian regions.  
However, the discussion below concerns the coherent structures directly formed by the Lagrangian fluid trajectories.

Well-known examples of coherent structures of the ``direct'' kind are the KAM islands and unstable and stable manifolds of
hyperbolic periodic points that constitute the flow topologies of 2D time-periodic flows in bounded domains
(Section~\ref{closed}). 
Lagrangian transport in other flow configurations has received considerably
less attention and remains the subject of ongoing investigations.
This section concerns one such class of configurations: 3D unsteady flows. 

Current insight into the fundamentals of Lagrangian transport in 3D (un)steady flows is to a great extent based on kinematic properties of divergence-free vector fields and volume-preserving maps. This encompasses any incompressible unsteady flow, $\xvec{\nabla}\cdot\xvec{u}=0$, as well as any compressible steady flow $\xvec{\nabla}\cdot(\rho\xvec{u})=\xvec{\nabla}\cdot\xvec{u}'=0$ where $\xvec{u}' = \rho \xvec{u}$. 
Groundbreaking progress has come from the 3D extensions of the KAM theorem \cite{cheng90,Mezic1994,broer96} and of the Poincar\'e--Birkhoff theorem \cite{cheng90b} that respectively
describe the fate of non-resonant invariant tori and resonant trajectories in 3D volume-preserving maps. Existence of 3D counterparts to these theorems was first hypothesized on the basis of a classification of
volume-preserving maps by the number of action variables \cite{Feingold1987,Feingold1988,Feingold1988b,Feingold1989}. Further important results include generic reductions in flow complexity by symmetries \cite{Mezic1994,Haller1998}, the formation of invariant manifolds of various topologies due to constants of motion \cite{Mezic1994,Haller1998,gomez02,Mullowney2005}, local and global breakdown of invariant manifolds by resonances \cite{Mezic2001,Feingold1988,Cartwright1994,Vainchtein2006,Vainchtein2007,Meiss2012} and universal properties of the Lagrangian transport between flow regions \cite{Mackay1994,Lomeli2009}.
These phenomena require in principle only satisfaction of continuity
and compliance with certain kinematic conditions. However, whether a real fluid flow indeed admits the latter conditions --- and the associated Lagrangian dynamics --- depends essentially on momentum conservation.

Consider the 3D steady momentum equation once again
\begin{eqnarray}
\rho\xvec{u}\cdot\xvec{\nabla}\xvec{u}=-\xvec{\nabla}p + \mu\xvec{\nabla}^2\xvec{u},
\label{NS0}
\end{eqnarray}
where $\xvec{u}$, as before, is the velocity, $p$ the pressure, $\rho$ the density and $\mu$ the dynamic viscosity.
We recast this for the present
discussion into the alternative form
\begin{eqnarray}
\rho\xvec{w}\times\xvec{u}=-\xvec{\nabla}\zeta + \mu\xvec{f},\quad \xvec{f}=\xvec{\nabla}\times\xvec{w},
\label{NS1}
\end{eqnarray}
with $\xvec{w} \equiv \xvec{\nabla}\times \xvec{u}$ the vorticity, $\xvec{w}\times\xvec{u}$ the Lamb vector and $\xvec{f}$ the flexion field, representing inertia and viscous forces, respectively, and 
\begin{equation}
\zeta=p + \rho g z + e_k 
\label{bernoulli}
\end{equation}
the Bernoulli function \cite{Yanna1998}. 
Here $e_k=\rho\xvec{u}\cdot\xvec{u}/2$ the kinetic energy, and gravity is defined as $\xvec{g}=-g\xvec{e}_z$. This form of the momentum equation directly reveals that it reduces from the 3D steady Navier-Stokes equation to the 3D steady Euler equation for both inviscid ($\mu=0$) and flexion-free ($\xvec{f}=\xvec{0}$) flows. 3D steady Euler flows are special in that universal conditions for (the absence of) chaos can be formulated on the basis of momentum conservation. It is well-known that they admit chaos upon satisfying the Beltrami condition $\xvec{w}\times\xvec{u}=\xvec{0}$; in all other cases they yield $\xvec{u}\cdot\xvec{\nabla}\zeta=\xvec{0}$ and
possess invariant manifolds defined by level sets of $\zeta$ (Section~\ref{LaminarFlow}).
Moreover, these invariant manifolds are diffeomorphic either to cylinders or tori \cite{Arnold1998,Mezic1994}. Hence, the Lagrangian dynamics as described above can happen only in Euler flows (locally) meeting the Beltrami condition. Thus the latter facilitates (yet not per se {\it causes}) certain kinematic events. (Consult \textcite{Arnold1998} for further properties of 3D steady Euler flows.)
The ABC (Arnol'd--Beltrami--Childress) flow, for example, always satisfies the Beltrami condition and is the archetypal flow for many studies on 3D chaotic advection \cite{Dombre1986,Feingold1988,Cartwright1994,Haller2001}.

Irrotational flexion fields ($\xvec{\nabla}\times\xvec{f}=\xvec{0}$) imply $\xvec{f}=-\xvec{\nabla}\sigma$, with $\sigma$ the flexion potential, reducing Eq.~\eqref{NS1} essentially to an Euler form $\rho\xvec{w}\times\xvec{u}=-\xvec{\nabla}\zeta'$, with $\zeta'=\zeta+\mu\sigma$ \cite{Yanna1998}. Here the Beltrami condition again determines the Lagrangian dynamics. Thus flexion-free flows, though strictly incorporating viscous effects, behave effectively as inviscid flows. It must be stressed that Stokes flows, though governed by $\mu\xvec{f}=\xvec{\nabla}p$ and thus also meeting $\xvec{\nabla}\times\xvec{f}=\xvec{0}$, are excluded from this behavior. Here the flexion potential equals $\sigma = -p/\mu+c$, with $c$ an arbitrary constant, implying $\zeta'=c$ and, in consequence, conservation of $\zeta'$ tells us nothing. Hence, akin to a Beltrami flow, $\zeta'$ is {\it not} a useful constant of motion. This point exposes an intriguing contrast: Euler flows satisfy $\xvec{\nabla}\zeta'=\xvec{0}$ only in the exceptional Beltrami case; Stokes flows, on the other hand, {\it invariably} satisfy $\xvec{\nabla}\zeta'=\xvec{0}$. Thus Euler flows generically are non-chaotic, while 3D Stokes flows have no obstacle to chaos. This observation has the fundamental implication that, lacking a universal dynamical restriction, Stokes flows can be integrable only due to symmetries.

Realistic 3D steady flows typically have significant inertia and viscosity, implying a rotational flexion field ($\mu\xvec{\nabla}\times\xvec{f}=\xvec{\nabla}\times(\rho\xvec{w}\times\xvec{u})\neq\xvec{0}$), meaning that
in general they are devoid of constants of motion: $\xvec{u}\cdot\xvec{\nabla}\zeta =\mu\xvec{u}\cdot\xvec{f}\neq\xvec{0}$ (refer
to \textcite{Kozlov1993} for a rigorous discussion). Here, similarly to Stokes flows, absence of a universal dynamical mechanism means
integrability can ensue only from symmetries. These must yield a flexion field perpendicular to $\xvec{u}$. It is important to note in relation
to Euler flows that realistic fluid flows admit 3D chaos without satisfying the Beltrami condition. Thus Beltrami flows (e.g., the widely-used
ABC flow) may be too restrictive for general studies on 3D chaotic advection in realistic fluid flows \cite{mezic2002extension}.

Laminar flows are for increasing $Re$ progressively better described by the Euler limit of the momentum equation and, given that the Beltrami
condition is exceptional, therefore typically tend to become integrable. (It must be stressed that laminar flow is assumed at all times here.)
Significant viscous effects, for example due to (local) breakdown of symmetries or boundary layers, (locally) disrupt the integrability of the Euler approximation and thus promote, or at least facilitate, chaotic advection in high-$Re$ (yet laminar) 3D steady flows \cite{Yanna1998,mezic2001chaotic}. Conversely, increasing $Re$ tends to augment the Euler-flow region and, in consequence, to suppress 3D chaos. Realistic high-$Re$ laminar flows thus generically lean towards a non-chaotic bulk flow; chaos, if occurring, tends to be confined to boundary layers and certain localized areas with symmetry breakdown.

Flows with low $Re$ have significant viscous effects throughout the entire flow domain and, contrary to Euler flows and high-$Re$ flows, in
principle admit global chaos. Here chaos (or absence thereof) is intimately related to symmetries. The linearity of the momentum equation in the Stokes limit ($Re=0$) causes symmetries in geometry and boundary conditions to be imparted on the flow. Hence, Stokes flows, akin to Euler flows, often are integrable yet due to different mechanisms. Consider, for example, 3D lid-driven cavity flow inside cubic and cylindrical domains; here symmetries result in closed streamlines in the Stokes limit \cite{Shankar1997,Shankar2000}. Nonlinearity due to fluid inertia ($Re>0$) or asymmetry in geometry and/or flow forcing are necessary ingredients for chaos to occur in 3D steady viscous flows \cite{Bajer1990,Bajer1992,Shankar1998,Shankar2000}.
This discloses a remarkable difference with the high-$Re$ regime in that here increasing $Re$  promotes rather than suppresses chaos. This implies an essentially nonlinear dependence of the chaotic Lagrangian dynamics on $Re$. However, the complete story of the routes between the generically integrable states in the Stokes and Euler limits remains unexplored to date.

The terrain of 3D unsteady flows is even less charted than that of their steady counterparts. Here the LHS of the momentum equation Eq.~\eqref{NS1} becomes
augmented by an unsteady term $\rho\partial\xvec{u}/\partial t$ and in principle any flow --- including non-Beltrami Euler flow --- is non-integrable and admits 3D chaos. Integrability thus, reminiscent of 3D steady viscous flows, seems to hinge entirely on symmetries and linearity and for unsteady flows can in all likelihood be expected only in the Stokes limit. Studies on Lagrangian dynamics in realistic 3D unsteady fluid flows have to date been few and far between and restricted to time-periodic flows constructed by systematic reorientation of piecewise steady flows: the bi-axial unsteady spherical Couette flow \cite{Cartwright1995,Cartwright1996}; the cubic lid-driven cavity \cite{anderson99,anderson2006}; and the cylindrical lid-driven cavity \cite{Malyuga2002,Michel2004,Pouransari2010}.

The discussion hereafter concentrates on 3D unsteady flows and exemplifies typical behavior by way of the above-mentioned cylindrical lid-driven cavity. This system possesses a rich dynamics and thus enables a good demonstration of what may happen in this class of flows. Two topics are considered that, as in the simpler flow systems discussed above, are key to Lagrangian dynamics and 3D chaos. First, the role of symmetries in the integrability of the Stokes limit (Section~\ref{NonInertial}). Second, the breakdown of this integrability by fluid inertia (Section~\ref{Inertial}). These phenomena are examined in terms of the formation of coherent structures and the associated freedom of motion for tracers. The discussion below on the cylindrical lid-driven cavity in essence concerns an overview and recapitulation of the main results of the separate studies in \textcite{Malyuga2002,Michel2004,Speetjens06b,MichelChaos,Pouransari2010,Speetjens2013}.

\subsubsection{3D square cylinder flow}
\label{ModelFlow}

\begin{figure}
 \subfigure[]{
\includegraphics[width=0.48\columnwidth]{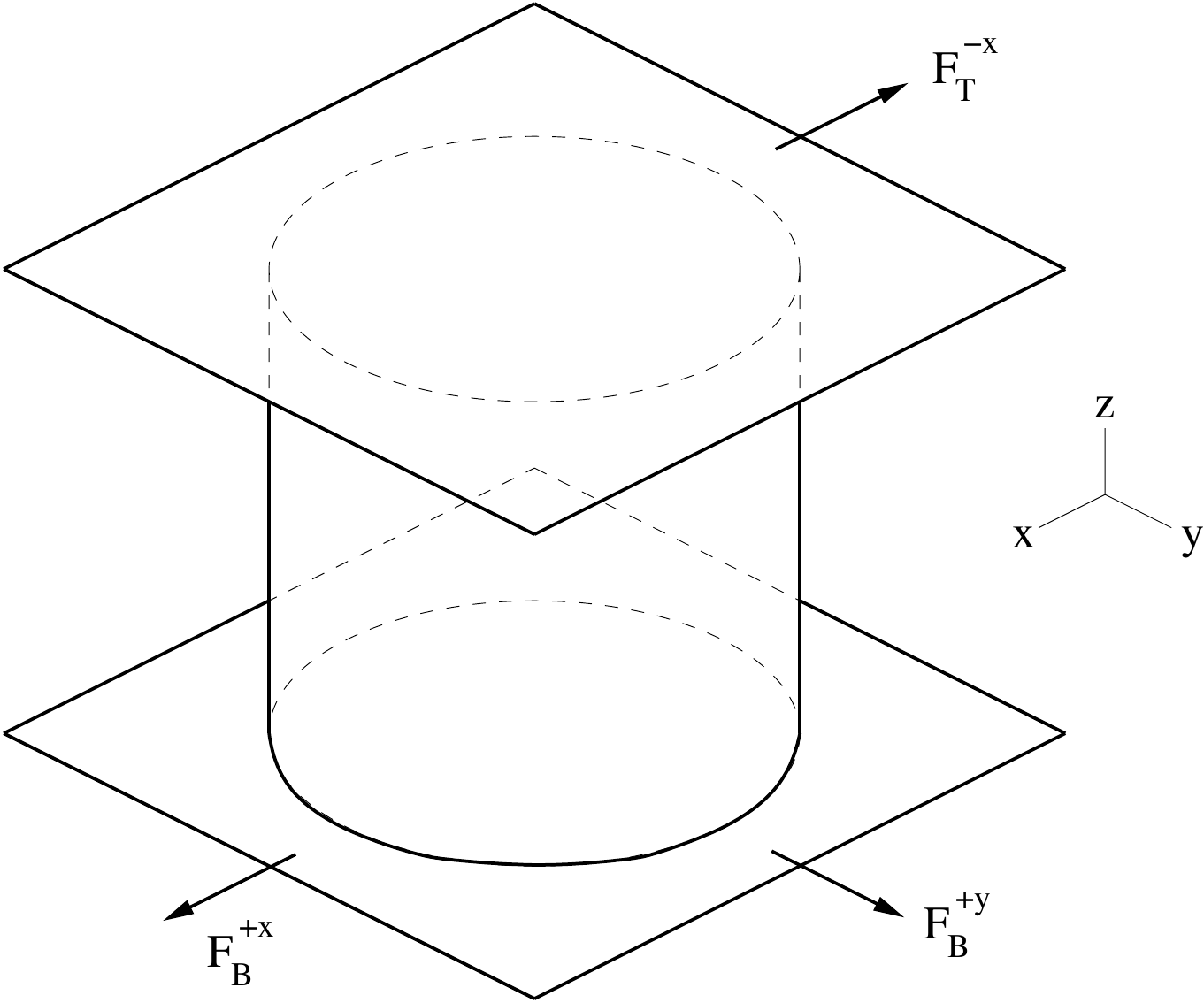}}
 \subfigure[]{\includegraphics[width=0.48\columnwidth]{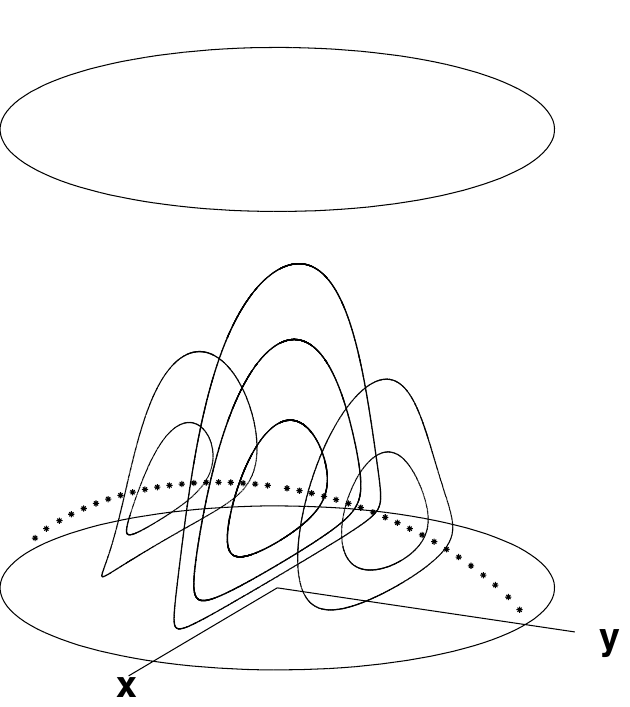}}
\caption{3D square cylinder flow: Non-dimensional flow configuration. (a) Flow domain and forcing. (b) Streamline portrait of base flow $\xvec{F}^{\mbox{\tiny\it +x}}_B$.
Reprinted with permission from \textcite{Michel2004}. Copyright (2004) Cambridge University Press.
}
\label{Configuration}
\end{figure}

Lagrangian features of 3D unsteady flows are exemplified by way of a simple yet realistic fluid flow: the time-periodic flow inside a 3D square cylinder $[r,\theta,z]=[0,1]\times[0,2\pi]\times[-1,1]$ \cite{Malyuga2002,Michel2004}.
The fluid is set in motion via time-periodic repetition of a sequence of piecewise steady translations (``forcing steps'') of the end-walls with unit
velocity $U=1$ and relative wall displacement $D=L/R$ ($L$ and $R$ are physical wall displacement and cylinder radius, respectively) by prescribed forcing
protocols. Figure~\ref{Configuration}(a) shows a schematic of the flow configuration; forcing protocols are specified below and are composed of the forcing
steps indicated by the arrows. Highly-viscous flow conditions are assumed such that transients during switching between forcing
steps are negligible (i.e., $T_\nu/T_{step}\ll 1$, with $T_\nu=R^2/\nu$ the viscous time-scale and $T_{step}$ the duration of one forcing step). Under this premise the internal
flow consists of piecewise steady flows that are each governed by the non-dimensional steady Navier--Stokes and continuity equations, Eq.~\eqref{E1'},
$
Re~\xvec{u}\cdot\nabla \xvec{u} =-\nabla p +\nabla^{2}\xvec{u}$,
$\xvec{\nabla}\cdot\xvec{u}=0.
$
Non-dimensionalization follows from substitution of the scaling
$\xvec{x} = R\xvec{x}'$, $\xvec{u} = U\xvec{u}'$ and $p=Pp'$ in Eq.~\eqref{NS0}, with primes indicating dimensionless variables (omitted above
for brevity). The characteristic pressure is given by $P=\mu U/R^2$ and ensues from assuming laminar flow conditions dominated by a force
balance between viscous forces and pressure gradient. Thus the Reynolds number appears before the inertial term and parameterizes
perturbation of the Stokes limit.

The motion of passive tracers is governed by the kinematic equation, Eq.~\eqref{E2'},
with formal solution $\xvec{x}(t)=\xvec{\Phi}_t(x_0)$ describing the Lagrangian trajectory of a tracer released at $\xvec{x}_0$. The
corresponding Poincar\'{e} map (which has also been referred to as  a \emph{Liouvillian map} in the case of 3D volume-preserving flows\footnote{The present class of divergence-free flows ($\xvec{\nabla}\cdot\xvec{u}=0$) may in the literature alternatively be
denoted ``volume-preserving flows'' or ``solenoidal flows''.} \cite{Cartwright1994,Cartwright1995,Cartwright1996}) is defined by $\xvec{x}_{k+1}=\xvec{\Phi}(\xvec{x}_k)$, where
$\xvec{x}_k=\xvec{x}(kT)$ is the tracer position after $k$ periods of the time-periodic forcing protocol.
The following forcing protocols --- denoted protocols $\mathcal{A}$, $\mathcal{B}$, and $\mathcal{C}$ hereafter --- are considered:
\begin{equation}
\PA = \xvec{F}^{\mbox{\tiny\it +y}}_B\xvec{F}^{\mbox{\tiny\it +x}}_B\hspf
\PB = \xvec{F}^{\mbox{\tiny\it -x}}_T\xvec{F}^{\mbox{\tiny\it +x}}_B\hspf
\PC = \xvec{F}^{\mbox{\tiny\it +y}}_B\xvec{F}^{\mbox{\tiny\it -x}}_T\xvec{F}^{\mbox{\tiny\it +x}}_B,
\label{Protocols}
\end{equation}
with subscripts in the forcing steps referring to the top ($T$) and bottom ($B$) end-walls and superscripts indicating the
translation direction (Fig.~\ref{Configuration}(a)). All forcing steps are transformations of the base flow $\xvec{F}^{\mbox{\tiny\it +x}}_B$ according to
\begin{equation}
\xvec{F}^{\mbox{\tiny\it +y}}_B=\mathcal{F}_{\pi/2}(\xvec{F}^{\mbox{\tiny\it +x}}_B),\quad
\xvec{F}^{\mbox{\tiny\it -x}}_T=S_z\mathcal{F}_{\pi}(\xvec{F}^{\mbox{\tiny\it +x}}_B),\quad
\label{Protocols2}
\end{equation}
with $\mathcal{F}_\alpha:\theta\rightarrow\theta + \alpha$ and $S_z : (x,y,z) \rightarrow (x,y,-z)$. Furthermore, the relative displacement is fixed at $D=5$, leaving only $Re$
as a control parameter for each forcing protocol. Maps $\xvec{\Phi}_{\mathcal{A},\mathcal{B},\mathcal{C}}$ as well as the underlying base flow $\xvec{F}^{\mbox{\tiny\it +x}}_B$
each exhibit particular 3D dynamics and thus serve to demonstrate fundamental aspects of 3D flows. Results have been obtained via numerical simulations \cite{Pouransari2010}.

\subsubsection{Coherent structures in 3D systems}
\label{coherent3D}

Coherent structures in the flow topology are spatial entities in the web of Lagrangian fluid trajectories that exhibit a certain invariance to the mapping $\PX$. Four kinds --- based on classifications in \textcite{GuckenheimerHolmes1983,Feingold1988} --- can be distinguished in 3D time-periodic systems, defined by
\begin{equation}
\M{X}{k}=\PX^k(\M{X}{k}),
\label{Structures}
\end{equation}
with $\M{X}{k}$ constituting periodic points ($\M{P}{k}$), periodic lines ($\M{L}{k}$), invariant curves ($\M{C}{k}$) and invariant surfaces
($\M{S}{k}$) of order $k$ (i.e., invariant with respect to $k$ forcing cycles). Note that periodic lines consist of periodic points, meaning that each constituent point is invariant; invariant curves and surfaces are only invariant as an entire entity. Brouwer's fixed-point theorem states that any continuous mapping of a convex space\footnote{A space is termed convex if for any pair
of points within the space, any point on the line joining them is also within the space. The present cylindrical domain is such a convex
space.} onto itself has at least one fixed point \cite{Zeidler2012,Michel2004}. This puts forward periodic points and associated coherent structures
as the most fundamental building blocks of 3D flow topologies. Periodic points and lines fall within one of the following categories: node-type and focus-type
periodic points and elliptic and hyperbolic periodic lines \cite{Malyuga2002}. (Periodic lines admit segmentation into elliptic and hyperbolic
parts.) Isolated periodic points and hyperbolic and elliptic lines imply pairs of stable ($W^s$) and unstable ($W^u$) manifolds, arising
as surface-curve pairs ($W^{s,u}_{2D}$,$W^{u,s}_{1D}$) for points and as surface-surface pairs
($W^{s,u}_{2D}$,$W^{u,s}_{2D}$) for lines. Elliptic lines form the center of concentric tubes. The 1D manifolds of
isolated periodic points define invariant curves $\M{C}{k}$; 2D manifolds and elliptic tubes
define invariant surfaces $\M{S}{k}$. Period-1 structures are the most important for the flow topology, as they determine the global organization. Higher-order
structures are embedded within lower-order ones and thus concern ever smaller features. The discussion below thus
is restricted to period-1 structures.

\subsection{Degrees of integrability in 3D unsteady Stokes flows}\label{NonInertial}

Flows often accommodate symmetries due to the geometry of the flow domain and the mathematical structure of the governing
conservation laws. Such symmetries, if present, play a central role in the formation of coherent structures and, inherently, in the spatial confinement of tracer motion. Symmetries in fact are the only mechanism that may accomplish integrability in 3D viscous flows (Section~\ref{Background}). In 2D time-periodic flows this typically results in symmetry groups of coherent structures or physical separation of flow
regions by symmetry axes \cite{Franjione1989,Ottino1994,Meleshko1996}. In 3D time-periodic flows this may furthermore
suppress truly 3D dynamics \cite{Feingold1988,Mezic1994,Haller1998,Malyuga2002,Michel2004}. Such manifestations of symmetries are demonstrated below for the time-periodic cylinder flow in the non-inertial limit, $Re=0$. The impact of fluid inertia is examined in Section~\ref{Inertial}.

\begin{figure}[tb]
 \subfigure[]{
\includegraphics[width=0.48\columnwidth,height=0.51\columnwidth]{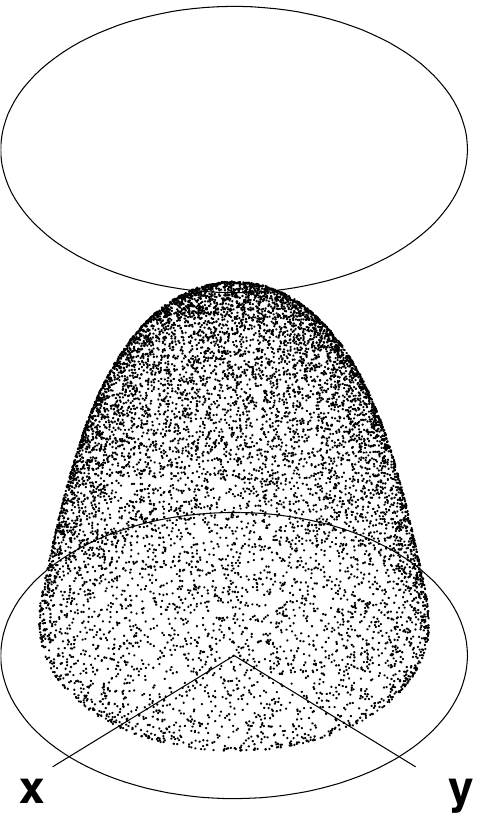}}
 \subfigure[]{
\includegraphics[width=0.48\columnwidth,height=0.51\columnwidth]{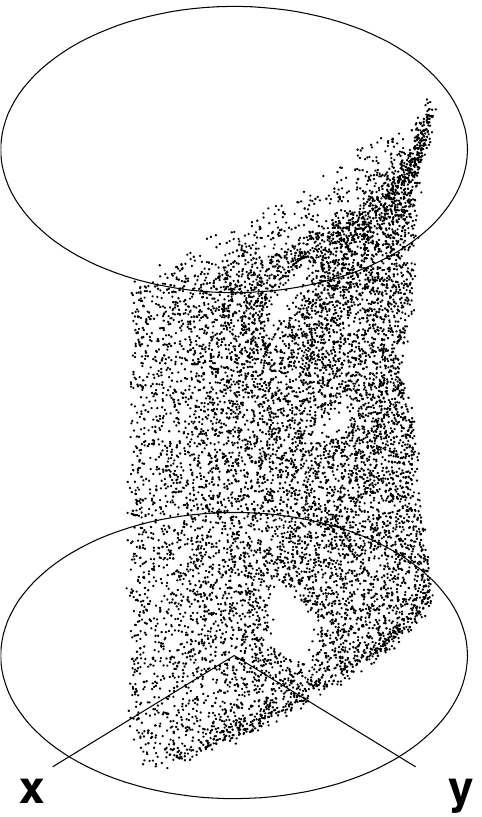}}
 \subfigure[]{
\includegraphics[width=0.48\columnwidth,height=0.51\columnwidth]{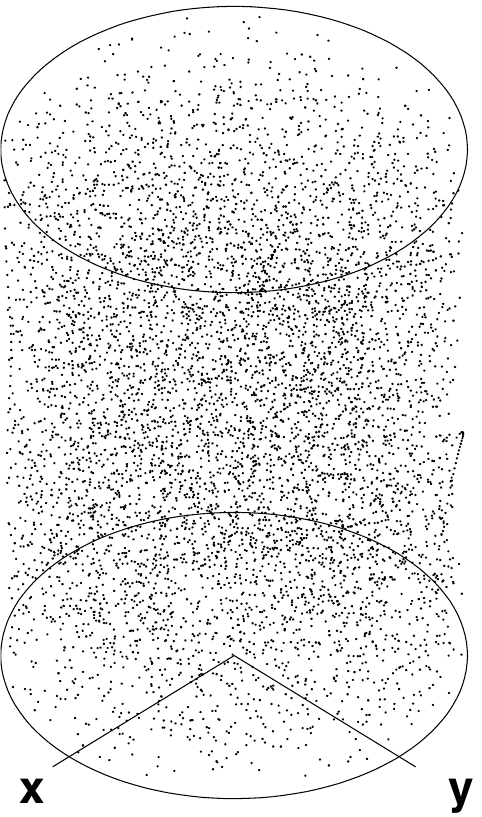}}
\caption{3D square cylinder flow:  Degrees of integrability of 3D unsteady Stokes flows demonstrated by Poincar\'{e} sections of
single tracers for the time-periodic forcing protocols according to Eq.~\eqref{Protocols}. (a) Protocol \Ax; (b) Protocol \Bx; (c) Protocol \Cx. 
Reprinted with permission from \textcite{Michel2004}. Copyright (2004) Cambridge University Press.
}
\label{long_term}
\end{figure}

The flow field \xvec{u} governed by Eq.~\eqref{NS0} collapses to the simple form
\begin{eqnarray}
u_{r}(\xvec{x}) = \mathrm{u}_{r}(r,z) \cos \theta,\quad
u_{\theta}(\xvec{x}) = \mathrm{u}_{\theta}(r,z) \sin \theta,\quad \nonumber \\
u_{z}(\xvec{x}) = \mathrm{u}_{z}(r,z) \cos \theta,
\label{exact}
\end{eqnarray}
in the non-inertial limit. (The italic $u$'s refer to the actual velocity components of 
the 3D velocity; the roman $\mathrm{u}$'s correspond with the part of each component that depends 
on $r$ and $z$.) This implies closed streamlines in the base flow $\xvec{F}^{\mbox{\tiny\it +x}}_B$ that are symmetric about the planes $x=0$ and $y=0$  (representing reflections); Fig.~\ref{Configuration}(b) \cite{Shankar1997} and, inextricably connected with that, two constants of motion of the generic form
\begin{eqnarray}
F_1(\xvec{x})=f_1(r,z)\hspf F_2(\xvec{x})=f_2(r,z)\sin\theta,
\label{com1}
\end{eqnarray}
satisfying $dF/dt=\xvec{u}\cdot\nabla F=0$. (An analytical expression for $F_1$ is given in \citet{Malyuga2002}.)
The properties of Eq.~\eqref{com1} have essential ramifications for the flow topologies of the forcing protocols of Eq.~\eqref{Protocols}. Figure~\ref{long_term} offers some first insight into the dynamics
by way of the Poincar\'{e} sections of a single passive tracer. Tracers released under Protocol $\mathcal{A}$ (Fig.~\ref{long_term}(a)) are confined to invariant spheroidal surfaces on which they perform effectively 2D (chaotic) dynamics. This occurrence of chaos on a submanifold of co-dimension one is an essentially 3D phenomenon; refer, e.g., to \citet{gomez02,Mullowney2005,Mullowney2008,meier07,sturman08} for dynamically similar systems. Protocol $\mathcal{B}$ (Fig.~\ref{long_term}(b)) restricts tracers to a quasi-2D (chaotic) motion within thin shells parallel to the $yz$-plane. Truly 3D (chaotic) dynamics covering the entire flow domain occurs only for Protocol $\mathcal{C}$ (Fig.~\ref{long_term}(c)). These dramatic differences in dynamics signify the presence of geometric restrictions on the tracer motion, akin to the KAM islands and cantori of 2D flows (Section~\ref{KAM}), in Protocols $\mathcal{A}$ and \Bx. This is a direct consequence of symmetries, as we shall elaborate below.

The above observations furthermore demonstrate that the quality \emph{integrability} takes on a subtler meaning in 3D flows, in that various degrees of
integrability --- and, inherently, spatial confinement --- can be distinguished, ranging from restriction to closed trajectories (base flow; Fig.~\ref{Configuration}(b)) to
global 3D chaotic advection (Protocol \Cx; Fig.~\ref{long_term}(c)). The classification of 3D time-periodic fluid flows introduced by \textcite{Cartwright1996} may be understood in terms of this notion of degrees of integrability. The cylinder flow in its Stokes limit encompasses all degrees of integrability in 3D time-periodic systems.

\subsubsection{2D (chaotic) dynamics within invariant manifolds}\label{ProtoA}
\label{ChaosInManifolds}

\begin{figure}
 \subfigure[]{
\includegraphics[width=0.55\columnwidth,height=0.57\columnwidth]{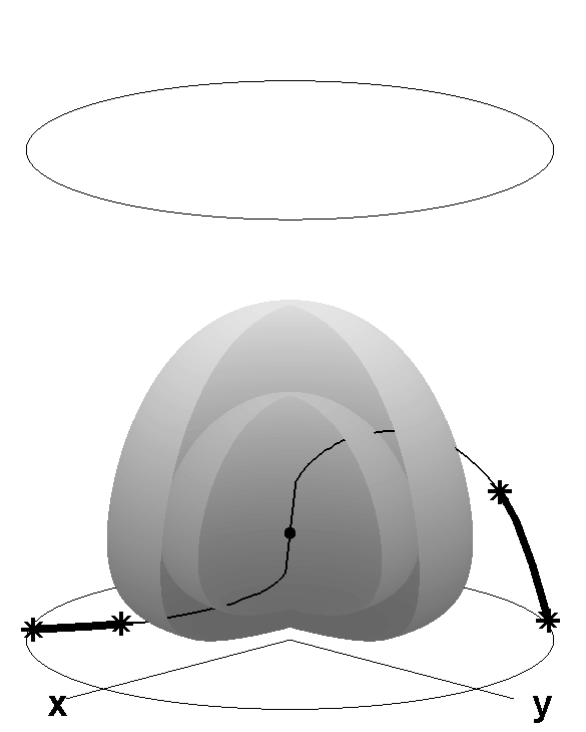}} \\
 \subfigure[]{\includegraphics[width=0.48\columnwidth,height=0.51\columnwidth]{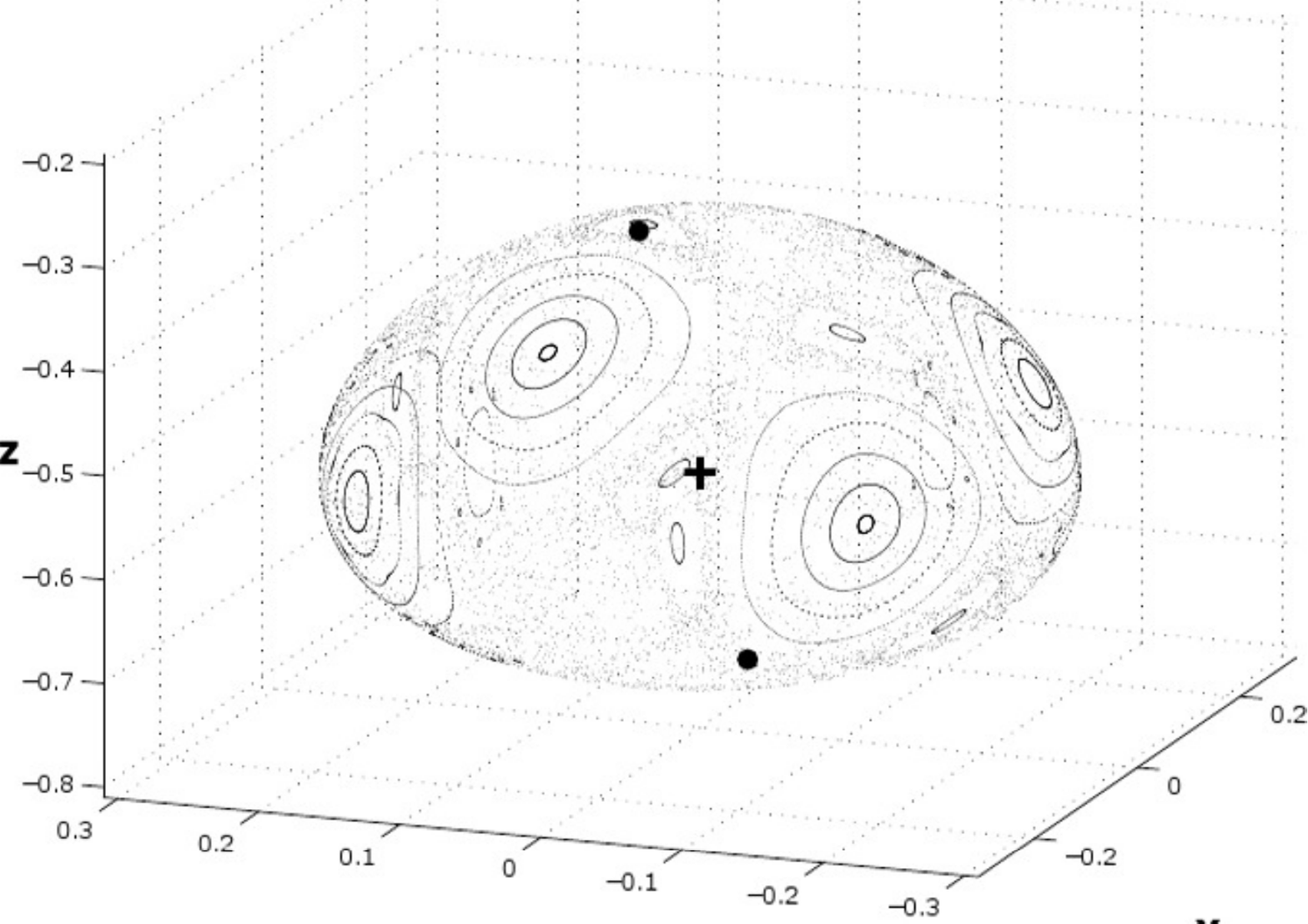}}
 \subfigure[]{\includegraphics[width=0.48\columnwidth,height=0.51\columnwidth]{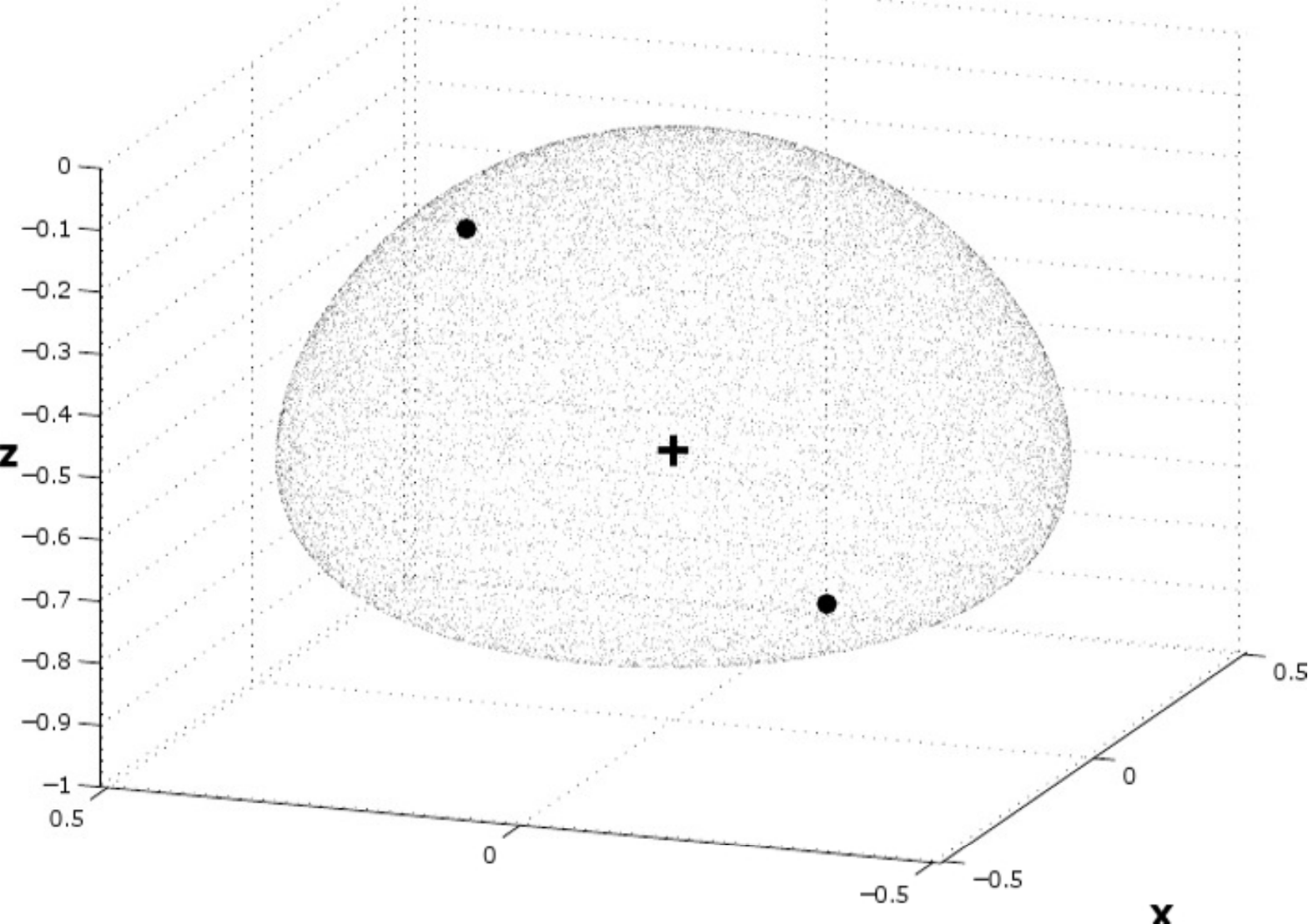}}
\caption{3D square cylinder flow: Formation of spheroidal invariant surfaces and intra-surface Hamiltonian dynamics in 3D unsteady Stokes flows demonstrated by Protocol $\mathcal{A}$. Heavy/normal sections of the period-1 line indicate elliptic/hyperbolic segments. Intra-surface dynamics are visualized by Poincar\'{e} sections of a ring of tracers.
(a) Spheroids and period-1 line;
(b, c).
Reprinted from \textcite{Speetjens06b} with the permission of AIP Publishing.
}
\label{ProtoA1}
\end{figure}

The restriction of tracers in Protocol $\mathcal{A}$ to invariant surfaces arises from a hidden axi-symmetry in the base flow that is retained by any forcing protocol involving reorientations of only one end-wall. According to Eq.~\eqref{com1}, constant of motion $F_1$  is invariant under the continuous transformation ${\mathcal{F}}_{\alpha}:\theta \rightarrow \theta + \alpha$, with $0\leq\alpha\leq 2\pi$, i.e., $\mathcal{F}_\alpha(F_1)=F_1$. The level sets of $F_1$ are defined by the surfaces of revolution of the trajectories $dr/dz=\mathrm{u}_{r}(r,z)/\mathrm{u}_{z}(r,z)=g(r,z)$ in the $rz$-plane \cite{Speetjens06b}. Figure~\ref{ProtoA1}(a) shows members of the infinite family of concentric spheroidal surfaces thus formed. Their emergence is entirely consistent with the generic property that a continuous symmetry in a bounded 3D steady flow --- here the base flow --- partitions the flow topology into a finite number of families of nested invariant tori or spheroids (see Theorem 4.1 in \textcite{Mezic1994}). Furthermore, spheroidal invariant surfaces imply
closed streamlines \cite{Mezic1994}. This explains the flow topology of the base flow in its Stokes limit (Fig.~\ref{Configuration}(b)).

Further organization of the flow topology of Protocol $\mathcal{A}$ results from discrete symmetries arising from the base flow. Transformations Eq.~\eqref{Protocols2} through $\PA$ following Eq.~\eqref{Protocols} translate into
\begin{eqnarray}
\PA = S_1 \PA^{-1} S_1,\quad
\PA = \wt{S} \PA \wt{S},
\label{sym_proto_a}
\end{eqnarray}
with $S_1 : (x,y,z) \rightarrow (-y,-x,z)$, $\wt{S}=S_2 \xvec{F}^{\mbox{\tiny\it +x}}_B$ and $S_2 : (x,y,z) \rightarrow (y,x,z)$. (Note that these
symmetry operators are consistent with the above continuous axi-symmetry by acting only within a given invariant
spheroid.) The time-reversal reflectional symmetry $S_1$ has the fundamental consequence that the flow must possess at
least one period-1 line $\M{L}{1}$, viz.\ within the symmetry plane $I_1=S_1(I_1)$ (plane $y=-x$; \textcite{Michel2004}). Coexistence of $S_1$ with $\wt{S}$ dictates that $\M{L}{1}$ be invariant to both discrete symmetries, i.e.,
\begin{eqnarray}
\M{L}{1}=S_1(\M{L}{1})=\wt{S}(\M{L}{1})= S_1\wt{S}(\M{L}{1})= \wt{S}S_1(\M{L}{1}),
\label{SymGroupA2}
\end{eqnarray}
meaning they essentially shape the period-1 line and its associated structures. The curve in Fig.~\ref{ProtoA1}(a) outlines the period-1 line for $D=5$, where heavy and normal parts indicate elliptic and hyperbolic segments, respectively. The hyperbolic segment is invariant to $\wt{S}$; left and right elliptic segments form symmetry pairs related via $\wt{S}$. Higher-order periodic lines are subject to a similar organization \cite{Speetjens06b}. Experimental validation of periodic lines and their fundamental link with symmetries is discussed in \textcite{Znaien2012}.

The intra-surface topologies within the invariant spheroids are organized by the periodic points defined by their
intersection with the periodic lines. The segmentation into elliptic and hyperbolic parts results in multiple kinds of
intra-surface topologies. Figure~\ref{ProtoA1}(b) shows an invariant spheroid that intersects with the hyperbolic
segment of the above period-1 line and exposes a topology that consists of two pairs of period-2 islands arranged around
the two hyperbolic period-1 points (dots) and enveloped by their heteroclinically-interacting manifolds (not shown). The
stable $(W^s_i)$ and unstable $(W^u_i)$ manifolds of each period-1 point relate via $W^u_i=S_1(W^s_i)$; manifolds of the
period-1 points interrelate via $W^u_1=\wt{S}(W^u_2)$ and $W^s_1=\wt{S}(W^s_2)$. (Note that the intra-surface manifolds merge
into 2D manifolds $W^{s,u}_{2D}$ in 3D space with the same symmetry properties.) The two pairs of period-2 islands
correspond to two pairs of period-2 elliptic points (facing points make one pair) that are on elliptic segments of two
period-2 lines intersecting the invariant surface. These entities relate, in a similar manner as their period-1
counterparts, via symmetries $S_1$ and $\wt{S}$. Figure~\ref{ProtoA1}(c) shows an invariant spheroid that
intersects with hyperbolic segments of said periodic lines, leading to fully-chaotic intra-surface dynamics.

Periodic points within invariant spheroids belonging to periodic lines imply the map $\PA$ is locally area-preserving
in their proximity~\cite{gomez02}. This has the important implication that the tracer motion
within each spheroid is essentially similar to that in 2D area-preserving maps. This underlies the 2D Hamiltonian
intra-surface topologies in Fig.~\ref{ProtoA1} with generic composition according to Section~\ref{KAM};
compare Fig.~\ref{ProtoA1}(b) with Fig.~\ref{fig:blink_islands}. Imperative for the rich intra-surface
dynamics here as well as in other 3D systems (see e.g., \textcite{gomez02,Mullowney2005,Mullowney2008,meier07,sturman08}) is the existence of multiple isolated periodic points of different type. This, through Brouwer's fixed point theorem, suggests that convexity of the invariant surfaces is a necessary (if not sufficient) condition. Absence of such complex dynamics in the classical case of invariant tori, which are non-convex and generically accommodate a dense winding of a single trajectory, supports this conjecture. However, conclusive establishment of the conditions that admit intra-surface chaos remains outstanding.

\subsubsection{Quasi-2D (chaotic) dynamics within subregions}\label{ProtoB}
\label{Quasi2DChaos}

An immediate consequence of including both top and bottom walls in $\PB$ is the vanishing of constant of motion $F_1$ and the associated continuous axi-symmetry. Thus here tracers are, in contrast to Protocol \Ax, no longer restricted to invariant
surfaces. Protocol $\mathcal{B}$  nonetheless accommodates discrete symmetries, reading
\begin{eqnarray}
\PB = \bar{S} \PB^{-1} \bar{S},\quad
\PB = \bar{S}' \PB^{-1} \bar{S}',\quad
\PB = S_y \PB S_y,
\label{sym_proto_b}
\end{eqnarray}
with $S_y : (x,y,z) \rightarrow (x,-y,z)$, $\bar{S}=S_x \xvec{F}^{\mbox{\tiny\it +x}}_B$,
$S_x : (x,y,z) \rightarrow (-x,y,z)$, $\bar{S}'=S_z\bar{S}S_z$ and $S_z$ as before. Note that symmetries
$\bar{S}$ and $\bar{S}$ --- and corresponding symmetry planes $\bar{I}=\bar{S}(\bar{I})$ and $\bar{I}'=S_z(\bar{I})$ ---
are conjugate in that they relate via $S_z$; the latter is a time-reversal symmetry hidden in $\bar{S}$ and $\bar{S}'$: $\PB = S_z \PB^{-1} S_z$.
Time-reversal symmetry again implies at least one period-1 line $\M{L}{1}$ within the corresponding symmetry plane.
However, coexistence of two such symmetries imposes an additional restriction compared to Protocol $\mathcal{A}$ in that period-1
lines identify with the intersections $\M{L}{1}\in\bar{I}\cap\bar{I}'$ of the conjugate symmetry planes so as to
belong simultaneously to both \cite{Michel2004}. The symmetry of $\bar{I}$ and $\bar{I}'$ about $z=0$ implies organization of period-1 lines into the group
\begin{eqnarray}
\mathcal{M}_B=\{
\{\M{L}{1}_{z,1},\M{L}{1}_{z,2},\dots,\M{L}{1}_{z,n}\},
\{\M{L}{1}_{1},S_z(\M{L}{1}_{1})\}, \nonumber \\
\{\M{L}{1}_{2},S_z(\M{L}{k}_{2})\},\dots,
\{\M{L}{1}_{m},S_z(\M{L}{1}_{m})\}
\},
\label{SymGroupB3}
\end{eqnarray}
with $\M{L}{k}_{z,i}=S_z(\M{L}{k}_{z,i})$ ($i\in[1,n]$) period-1 lines within $z=0$ and
$\{\M{L}{k}_{i},S_z(\M{L}{k}_{i})\}$ ($i\in[1,m]$) symmetry pairs about $z=0$. Note that all period-1 lines
possess the self-symmetry $\M{L}{k}_{z,i}=S_y(\M{L}{k}_{z,i})$ and  $\M{L}{k}_{i}=S_y(\M{L}{k}_{i})$ about the plane
$y=0$. Thus here a symmetry {\it group} of period-1 lines forms; for Protocol $\mathcal{A}$ a
{\it single} period-1 line consisting of symmetric segments forms.

\begin{figure}[tb]
 \subfigure[]{
\includegraphics[width=0.49\columnwidth,height=0.57\columnwidth]{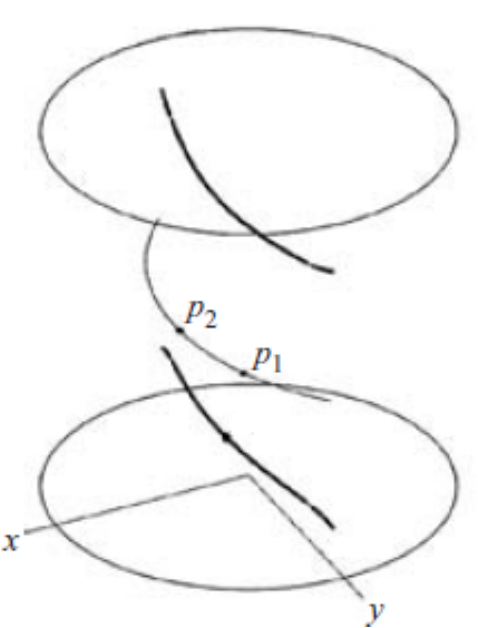}} \\
 \subfigure[]{\includegraphics[width=0.51\columnwidth,height=0.55\columnwidth]{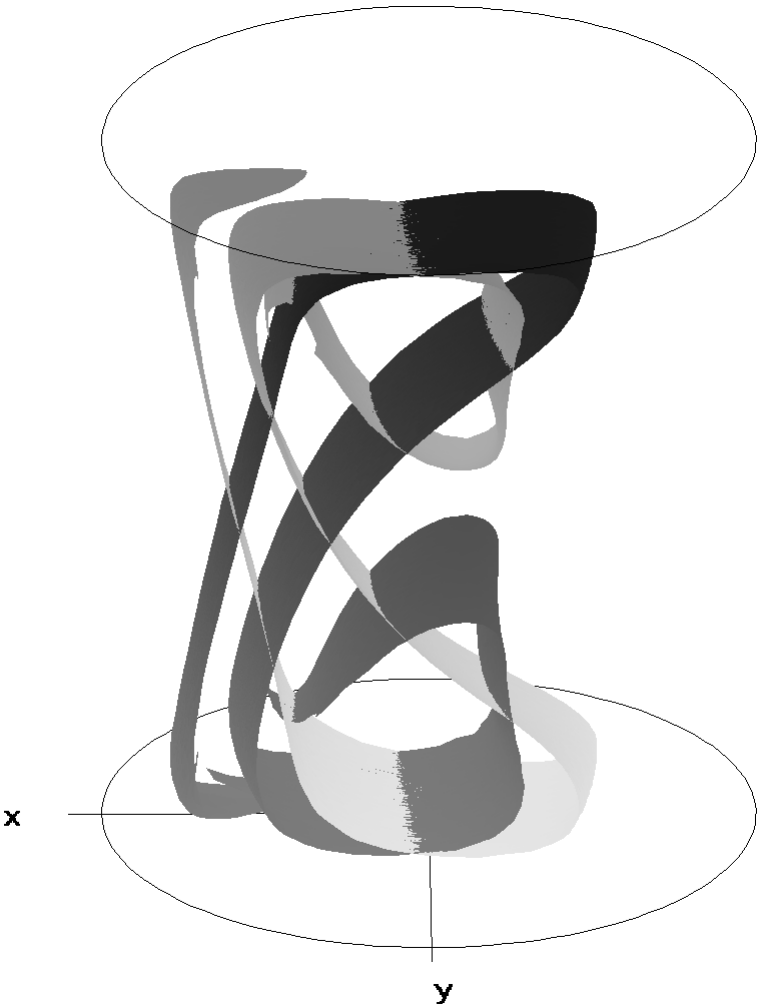}}
 \subfigure[]{\includegraphics[width=0.47\columnwidth,height=0.52\columnwidth]{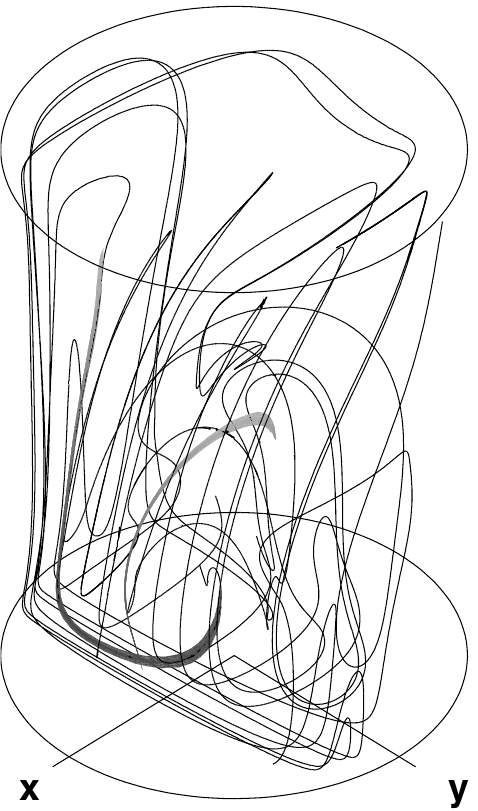}}
\caption{3D square cylinder flow: Formation of periodic lines and isolated periodic points and associated manifolds in 3D unsteady Stokes flows demonstrated by Protocols \B~and \C. (a) Period-1 lines for Protocol \Bx. (b) Manifolds of segment $p_1-p_2$. (c)  Manifold pairs for Protocol \Cx.
(a) Reprinted with permission from \textcite{Michel2004},  copyright (2004) Cambridge University Press; 
(b) from \textcite{michelthesis}.}
\label{ProtoBandC1}
\end{figure}

Mass conservation implies at least one period-1 line that, for given symmetries, must sit in the plane $z=0$, meaning
that $\M{L}{k}_{z,1}$ always exists. Figure~\ref{ProtoBandC1}(a) shows a typical symmetry group
$\mathcal{M}_B=\{\M{L}{k}_{z,1},\M{L}{k}_{1},S_z(\M{L}{k}_{1})\}$ of period-1 lines, with $\M{L}{k}_{z,1}$ fully
hyperbolic and the conjugate pair $\{\M{L}{k}_{1},S_z(\M{L}{k}_{1})\}$ fully elliptic. The stable/unstable
manifolds $W^{s/u}_{2D}$ of line segment $p_1-p_2$ of $\M{L}{k}_{z,1}$, relating via $W^u_{2D}=S_z(W^s_{2D})$, are shown
in Fig.~\ref{ProtoBandC1}(b) and envelop the elliptic region comprising concentric tubes --- the 3D counterparts to KAM
islands --- centered on the elliptic lines (not shown). The manifolds extend primarily in the direction normal to
$\M{L}{k}_{z,1}$ (i.e., parallel to the $yz$-plane) and exhibit only marginal $y$-wise variation. Moreover, they exhibit
transversal interaction, which is a fingerprint of chaotic dynamics in 2D systems. This manifold
behavior causes the quasi-2D chaotic tracer motion within a thin layer normal to the period-1 lines, as illustrated in
Fig.~\ref{long_term}(b). Tracers released near the elliptic segments exhibit similar behavior by remaining
confined to thin slices of elliptic tubes (not shown). Hence, tracer dynamics within each layer is of
a basically Hamiltonian nature and is thus intimately related to the intra-surface dynamics of Protocol~\Ax. A primary difference
with the latter is that tracers are not strictly confined to an invariant surface. Whether these less restrictive
conditions may be of any consequence is an open question. Protocols $\mathcal{A}$ and $\mathcal{B}$ thus reveal that time-reversal
symmetries, through their link with periodic lines, imply effectively (quasi-)2D dynamics. This suppression of truly
3D dynamics is an essentially 3D manifestation of this kind of symmetries; their role in 2D systems primarily concerns
formation of symmetry groups.

\subsubsection{Unrestricted 3D (chaotic) dynamics}\label{ProtoC}

Inclusion of a third forcing step results in a flow that is devoid of global symmetries. This in principle paves the way to 3D chaotic advection for viscous flows. In particular the absence of time-reversal symmetries is of fundamental consequence in that periodic lines must thus no longer be present. However, the current flow must, according to Brouwer's fixed-point theorem, accommodate at least one isolated period-1 point (Section~\ref{coherent3D}). Two node-type period-1 points indeed exist, and have associated manifold pairs ($W^{u}_{2D}$,$W^{s}_{1D}$) with essentially 3D foliations that densely fill the entire flow domain (Fig.~\ref{ProtoBandC1}(c)). (The 2D manifolds have a dominant stretching direction and thus assume a particular shape.) Here the stable and unstable manifolds, in contrast with those associated with periodic lines, are not related via a time-reversal symmetry. This asymmetry in time results in essentially 3D transport and is a key element in the truly 3D chaotic dynamics demonstrated in Fig.~\ref{long_term}(c). Moreover,
the role of manifolds is fundamentally different here compared with Protocols $\mathcal{A}$ and \Bx. The stability properties of the manifolds admit only homo-/hetero-clinic $W_{2D}$--$W_{1D}$ interactions; any interactions between the 2D manifolds are impossible on grounds of identical stability (Fig.~\ref{ProtoA_Re=100}(a), for example, shows a homoclinic transversal $W^{s}_{2D}$--$W^{u}_{1D}$ interaction for Protocol $\mathcal{A}$ at $Re=100$.) Furthermore, interactions between 2D manifolds of isolated periodic points can generically happen only through either merger into heteroclinic surfaces or formation of a heteroclinic orbit connecting both points by transversal intersection. 3D counterparts to transversal manifold interaction as, for example,  shown in Fig.~\ref{ProtoBandC1}(b) for Protocol~$\mathcal{B}$ are non-existent for isolated periodic points.\footnote{The periodic points by definition are asymptotic limits of any transversal intersections. Topological consistency then admits two situations: (i) periodic points are part of the intersection (i.e., the heteroclinic orbit); (ii) intersections are isolated points ($W_{2D}$--$W_{1D}$ interactions).}
This suggests that, contrary to the (quasi-)2D dynamics associated with periodic lines, the intrinsic hyperbolicity of isolated period-1 points generically suffices for 3D chaotic advection to occur \cite{Michel2004}. Formation of transport barriers by heteroclinic merger of 2D manifolds is the sole non-chaotic case. However, that is an atypical case in 3D unsteady flows.

\subsection{Breakdown of invariant manifolds by fluid inertia}\label{Inertial}
\label{Inertia0}

\label{Inertia1}

Fluid inertia, $Re>0$, has a strong impact upon transport properties. It eliminates any symmetry that hinges upon linearity of the Stokes limit of the momentum equation and may thus facilitate (if not directly cause) 3D chaotic advection \cite{Bajer1992,Cartwright1996}. In our example, the cylinder flow, inertia introduces a secondary circulation to the base flow $\xvec{F}^{\mbox{\tiny\it +x}}_B$ transverse to its primary circulation in the non-inertial limit (Fig.~\ref{BaseFlowInertial}(a)). This breaks the symmetry about
$x=0$ and causes the closed streamlines to become non-closed and wrapped around invariant tori defined by the level sets of a new constant of motion $\bar{F}$ that emerges in favor of constants of motion $F_{1,2}$ \cite{Cartwright1996}. (The symmetry of the base flow about $y=0$ is preserved for $Re>0$.) These tori, in turn,
undergo a progressive disintegration into tori with winding numbers\footnote{The winding number or rotation number $W$ represents the number of revolutions around the axis of rotation required for completing a full loop on a closed trajectory. The
closed streamlines in Fig.~\ref{Configuration}(b), for example, have $W=1$.} $W>1$ (island chains in
Fig.~\ref{BaseFlowInertial}(b)) and chaotic seas (Fig.~\ref{BaseFlowInertial}(c)) with increasing $Re$, according
to the Hamiltonian response scenario for 2D time-periodic systems to perturbations and
described by the KAM and Poincar\'{e}--Birkhoff theorems \cite{Ottino1989,Cartwright1996}. The cross-sections
in Fig.~\ref{BaseFlowInertial} are dynamically equivalent to said 2D systems \cite{Cartwright1996,Bajer1994}; compare their structure with Fig.~\ref{fig:cantori}.

\subsubsection{Response of invariant tori}
\label{Inertia2}

\begin{figure}[tb]
 \subfigure[]{\includegraphics[width=0.49\columnwidth,height=0.55\columnwidth]{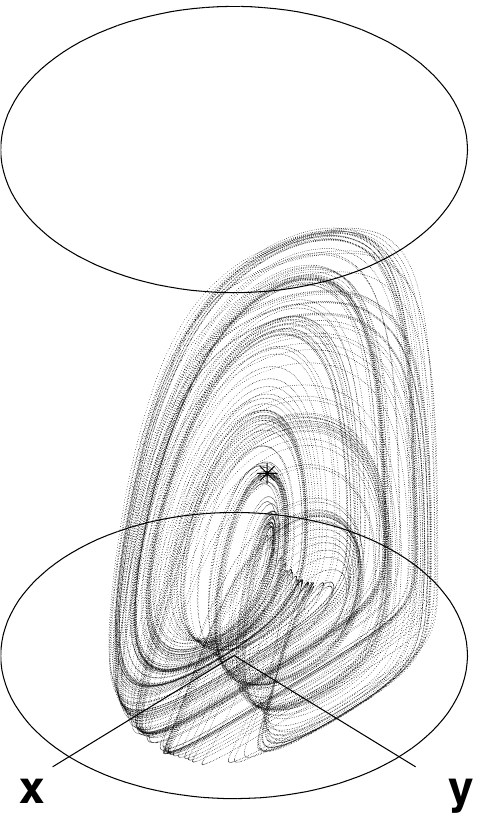}} \\
 \subfigure[]{\includegraphics[width=0.48\columnwidth,height=0.55\columnwidth]{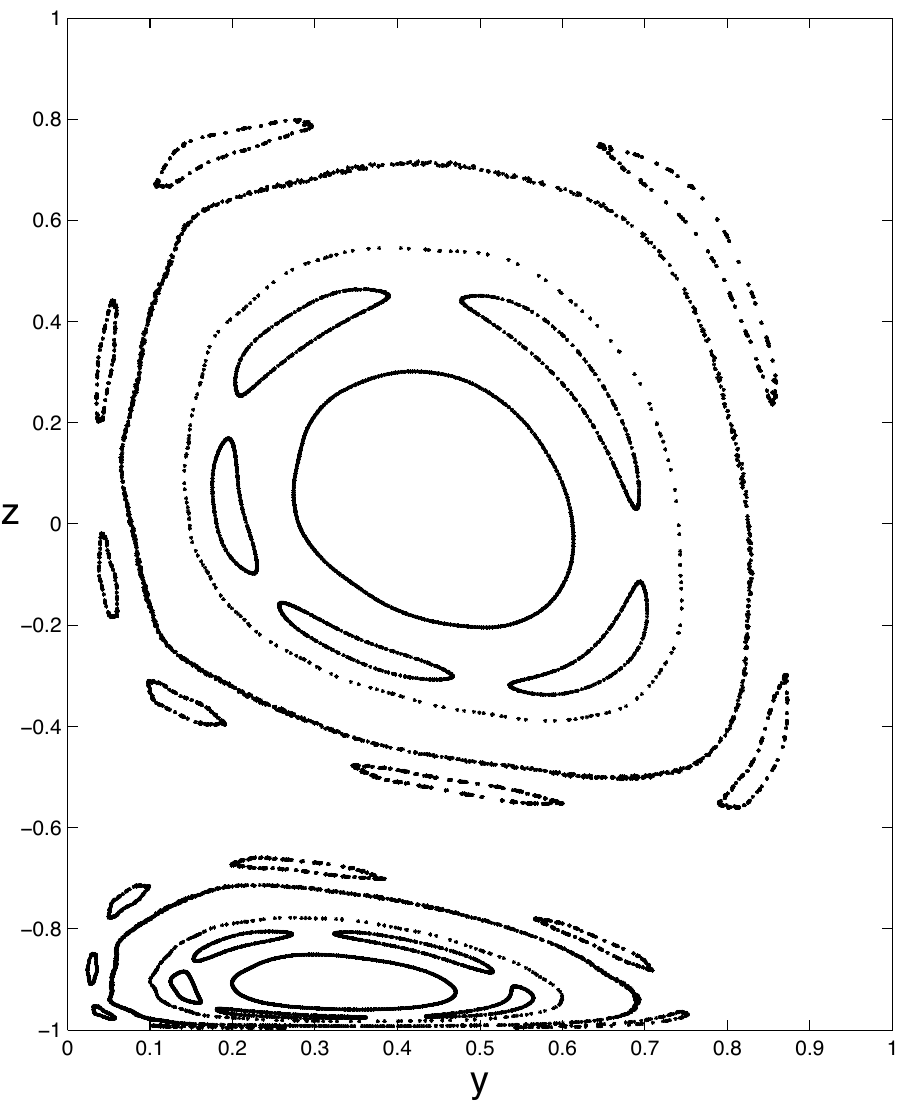}}
 \subfigure[]{\includegraphics[width=0.48\columnwidth,height=0.55\columnwidth]{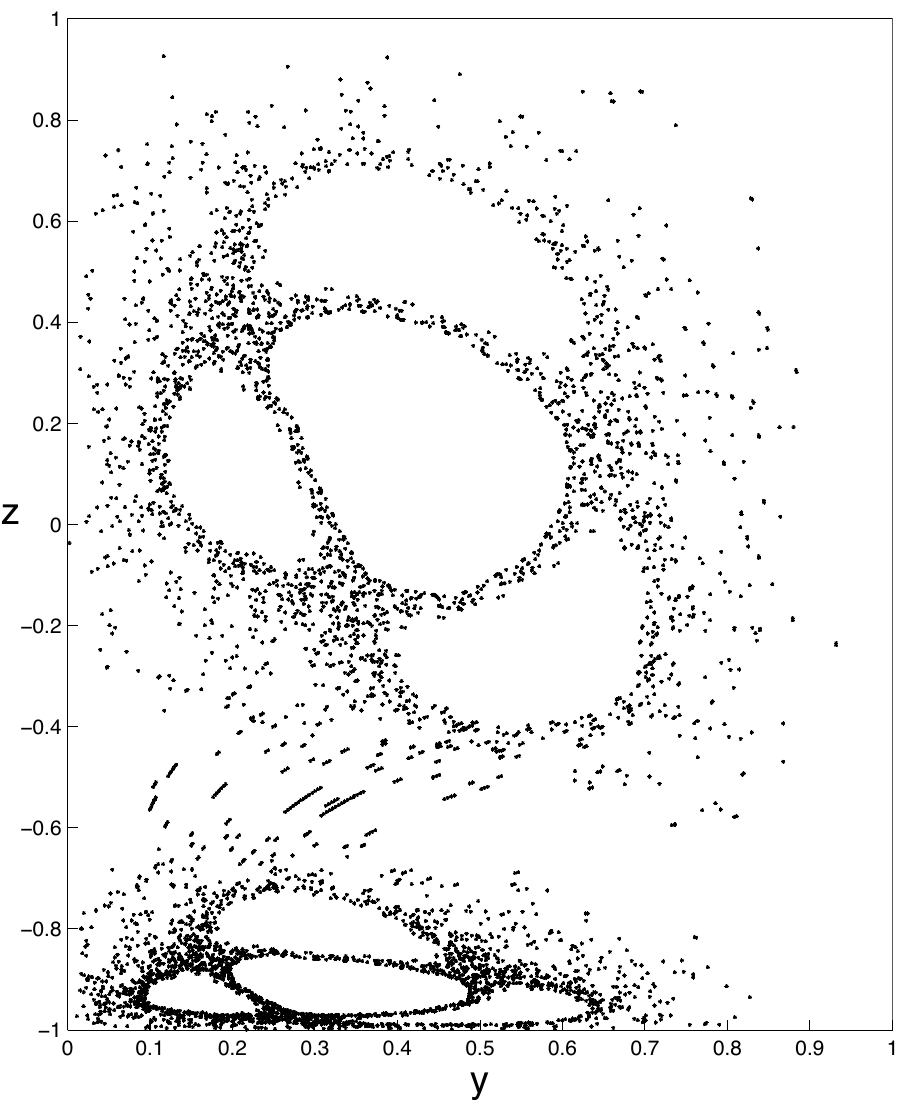}}
\caption{3D square cylinder flow: Formation and subsequent breakdown of invariant tori due to fluid inertia demonstrated by the
base flow $\xvec{F}^{\mbox{\tiny\it +x}}_B$. (a) single streamline for $Re=100$; (b) and (c): cross-sections
of the streamline pattern with plane $x=0.12$ at $Re=50$ and $Re=100$, respectively.
From \textcite{michelthesis}.}
\label{BaseFlowInertial}
\end{figure}

Toroidal invariant surfaces in the Poincar\'{e} maps of 3D time-periodic flows exhibit similar behavior in that they also partly survive (inertia-induced) perturbations; this results in Poincar\'{e} sections of composition similar to the 3D streamline portrait in Fig.~\ref{BaseFlowInertial} yet typically comprising various arrangements of multiple families of tori \cite{Cartwright1994,Dombre1986,Feingold1987,Feingold1988}. This survival of invariant tori is described by a 3D counterpart to the KAM theorem \cite{cheng90,Mezic1994,broer96}. Invariant tori accommodating resonant (i.e., closed) trajectories with $W>1$ disintegrate into tori with $W>1$ in a way similar to the 3D steady case (Fig.~\ref{BaseFlowInertial}(b)) \cite{Cartwright1996}. The fate of such resonant trajectories is described by the 3D counterpart to the Poincar\'e--Birkhoff theorem \cite{cheng90b}.

A special case exists in perturbed time-periodic systems near the limiting case $W=1$. Trajectories in the Poincar\'{e} map in general wrap themselves densely around invariant tori in a manner akin to in the perturbed base flow (Fig.~\ref{BaseFlowInertial}). However,
so-called resonant sheets, material surfaces in the unperturbed flow composed of period-$p$ points with $p\geq 1$, interrupt this
process and, if occurring, cause local defects in the tori by which tracers can jump randomly between their intact sections. This behavior
is termed resonance-induced dispersion (or resonance-induced diffusion) and facilitates global tracer distribution \cite{Feingold1988,Piro1988,Cartwright1994,Cartwright1995,Cartwright1996}. Resonance-induced dispersion is extremely slow compared to truly 3D chaotic advection, since tracers are confined to segments of tori in between jumps. Tracers nonetheless describe
space-filling trajectories that visit the entire domain in the course of time. The dynamics within the jump zones is governed by isolated periodic points that remain from the (originally entirely periodic) resonance sheets upon perturbation. The associated manifold pairs ($W^{s,u}_{2D}$,$W^{u,s}_{1D}$) intersect these zones transversally and thus enable the jumping between segments of tori that underlies resonance-induced dispersion \cite{Mezic2001}.

\subsubsection{Response of invariant spheroids}
\label{Inertia3}

Studies on the responses of invariant surfaces to perturbations are almost exclusively restricted to tori. However, the classification theorem for closed surfaces states that any orientable closed surface, which includes level sets of a constant of motion in bounded flows, is topologically equivalent to a sphere or a connected sum of tori \cite{Alexandroff1961}. This puts forward invariant spheroids, besides invariant tori, as a second fundamental form of invariant surfaces relevant in the present context \cite{michelthesis}; recall in this regard that tori and spheroids are also
the key invariant surfaces in 3D steady flows; Section~\ref{ProtoA}. Essentially different dynamics upon perturbation are likely
on grounds of fundamental topological properties. Tori are doubly-connected non-convex manifolds; spheroids are singly-connected convex manifolds. Tori accommodate trajectories that are either closed or densely wrapped; spheroids typically accommodate chaotic seas and elliptic islands. Model flows offer a way to investigate the response of invariant spheroids to inertial perturbations under realistic conditions. This may contribute to a more complete picture of the fate of invariant surfaces subject to inertial perturbations.

\begin{figure}[tb]
 \subfigure[]{\includegraphics[width=0.48\columnwidth,height=0.55\columnwidth]{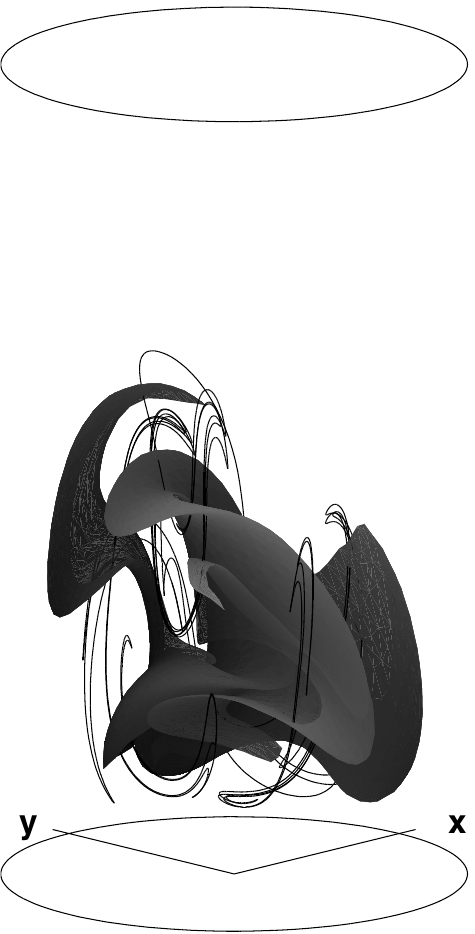}}
 \subfigure[]{\includegraphics[width=0.48\columnwidth,height=0.55\columnwidth]{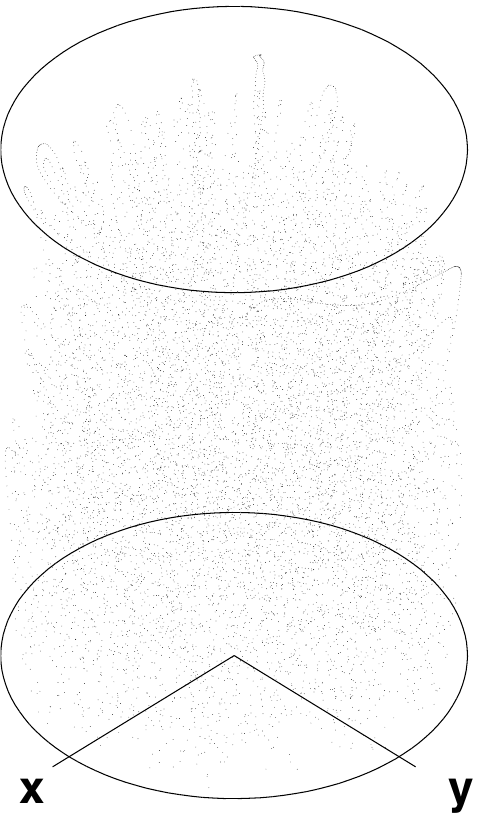}}
\caption{3D square cylinder flow:  Emergence of 3D global chaos and isolated periodic points with associated manifolds due to inertia-induced breakdown of spheroidal invariant surfaces in 3D unsteady flows demonstrated by Protocol $\mathcal{A}$ at $Re=100$.
(a) Manifold pair ($W_{2D}^s,W_{1D}^u$) of the isolated periodic point;
(b) Poincar\'{e} section of a single tracer.
(a) From \textcite{michelthesis};
(b) reprinted from \textcite{Speetjens06b} with the permission of AIP Publishing.
}
\label{ProtoA_Re=100}
\end{figure}

The effect of fluid inertia upon invariant spheroids may be demonstrated for Protocol \Ax; this is representative of generic forcing protocols with such a topology \cite{Pouransari2010}. Inertia breaks both the time-reversal reflectional symmetry $S_1$ in Eq.~\eqref{sym_proto_a} and the continuous axi-symmetry due to constant of motion $F_1$; only symmetry $\wt{S}$ is preserved for $Re>0$ \cite{Speetjens06b}. This causes the period-1 line $\M{L}{1}$ (Fig.~\ref{ProtoA1}(a)) to give way to a focus-type isolated period-1 point with a ($W^{s}_{2D}$,$W^{u}_{1D}$) manifold pair that, for sufficiently high $Re$, completely destroys the invariant spheroids and, in consequence, yields 3D chaotic tracer motion. Figure~\ref{ProtoA_Re=100} demonstrates this process for $Re=100$, where, consistent with permissible manifold interactions of isolated periodic points (Section~\ref{ProtoC}), homoclinic transversal $W^{s}_{2D}$--$W^{u}_{1D}$ interaction occurs. However, even minute departures from the non-inertial limit may change the flow topology drastically. This is discussed below.

\begin{figure}[tb]
 \subfigure[]{\includegraphics[width=0.48\columnwidth]{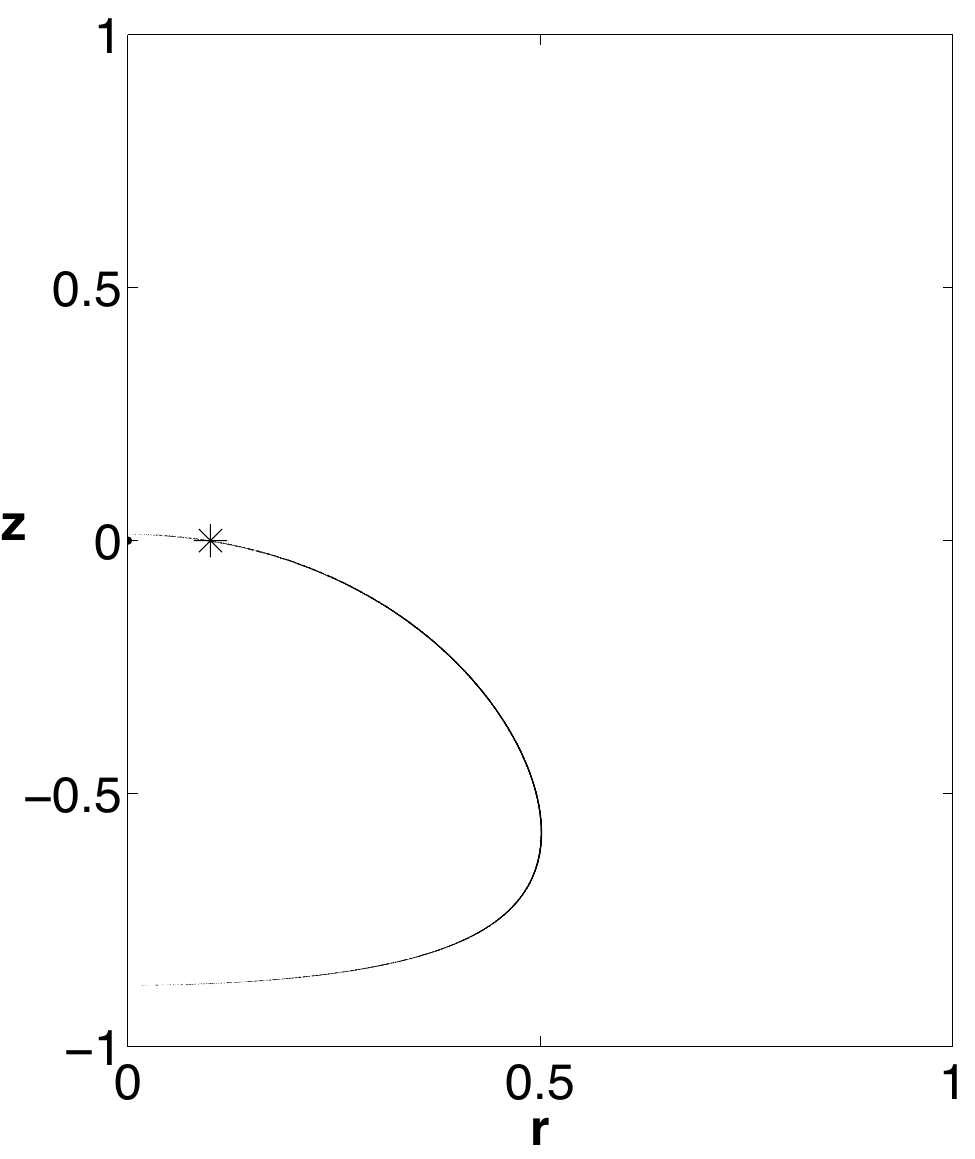}}
 \subfigure[]{\includegraphics[width=0.48\columnwidth]{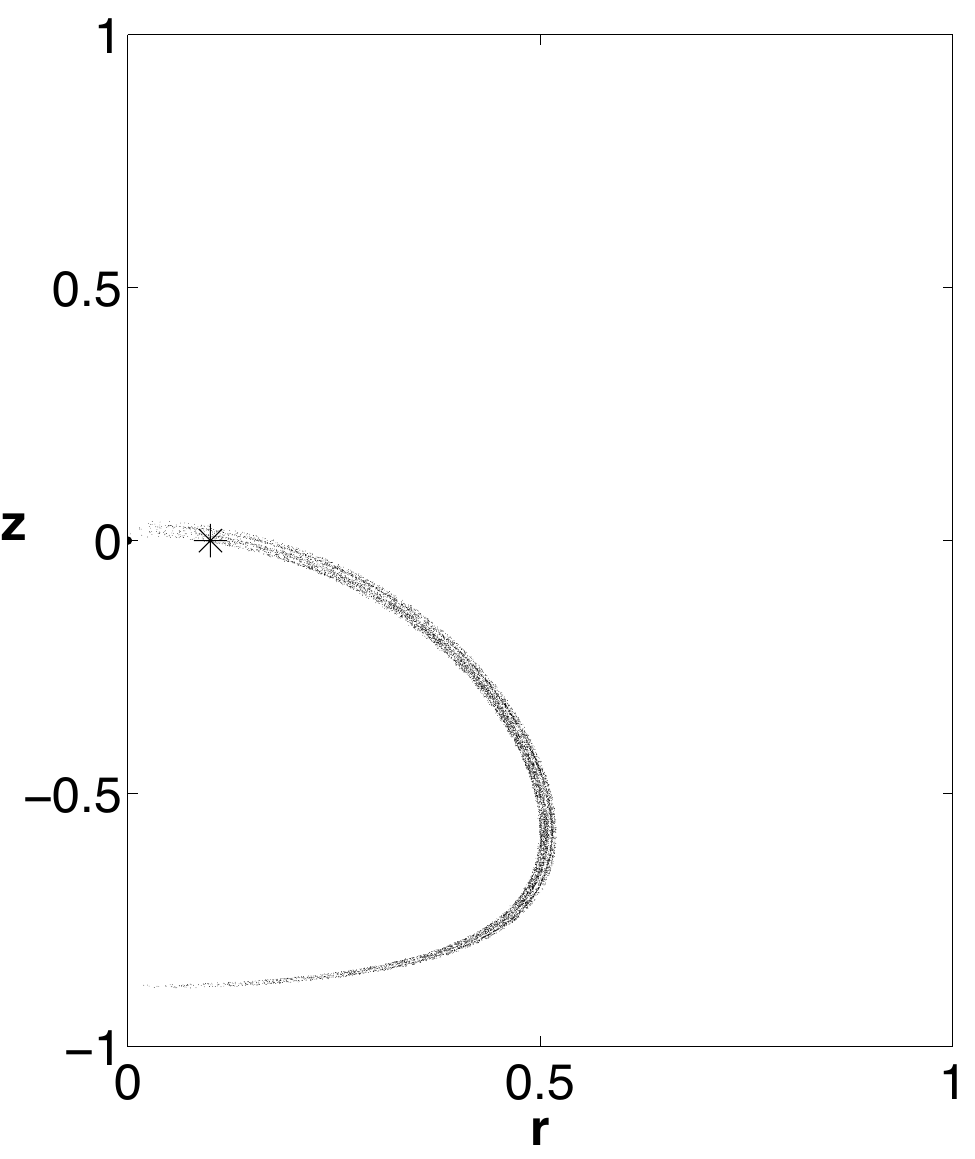}}
 \subfigure[]{\includegraphics[width=0.48\columnwidth]{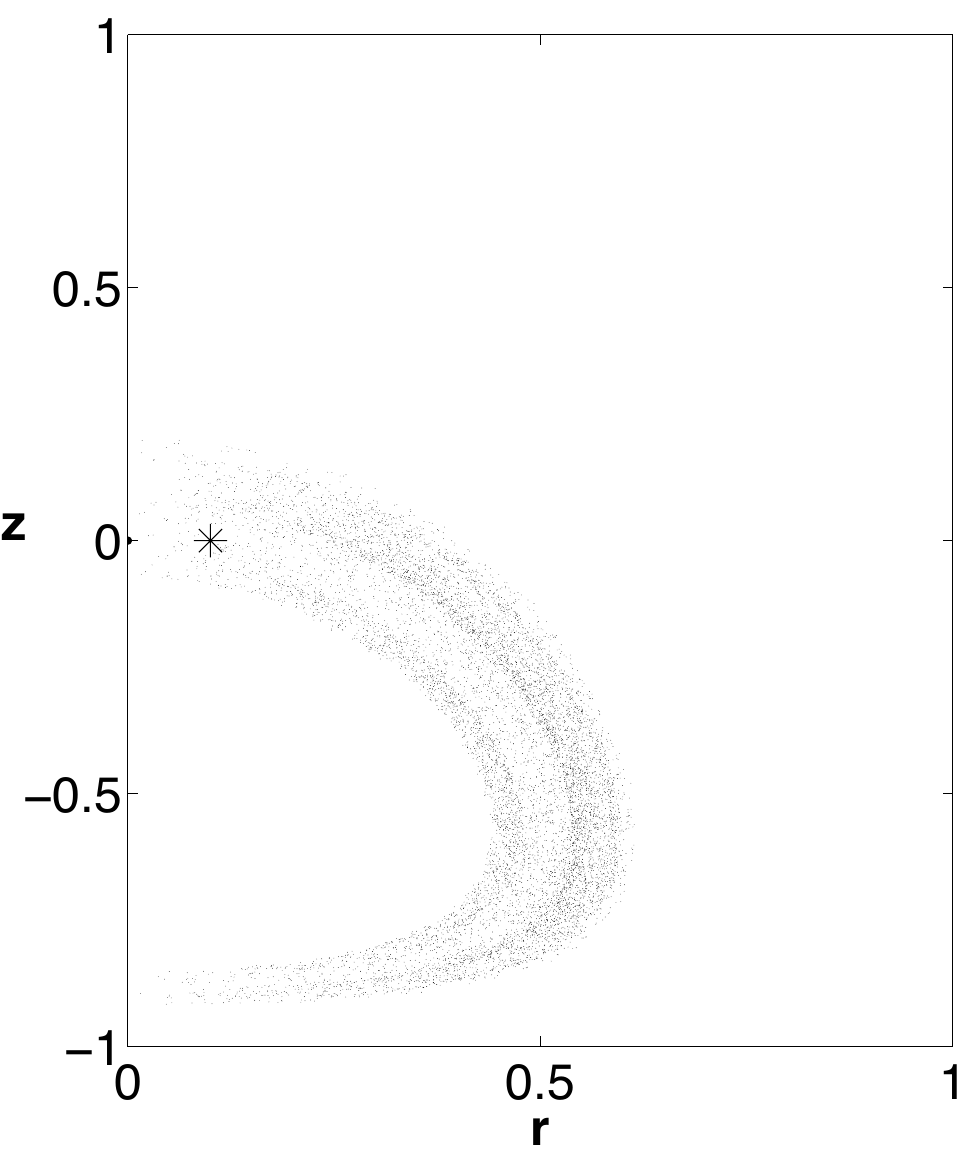}}
\caption{3D square cylinder flow: Inertia-induced drifting of tracers transverse to invariant spheroids in
3D unsteady flows  demonstrated by the Poincar\'{e} section of a single tracer (10\,000 periods) for Protocol \Ax. The asterisk denotes the initial tracer position. (a) $Re=0$; (b) $Re=0.1$; (c) $Re=1$. 
Reprinted from \textcite{Speetjens06b} with the permission of AIP Publishing.
}
\label{AdiabaticShells}
\end{figure}

\begin{figure}[tb]
 \subfigure[]{\includegraphics[width=0.48\columnwidth]{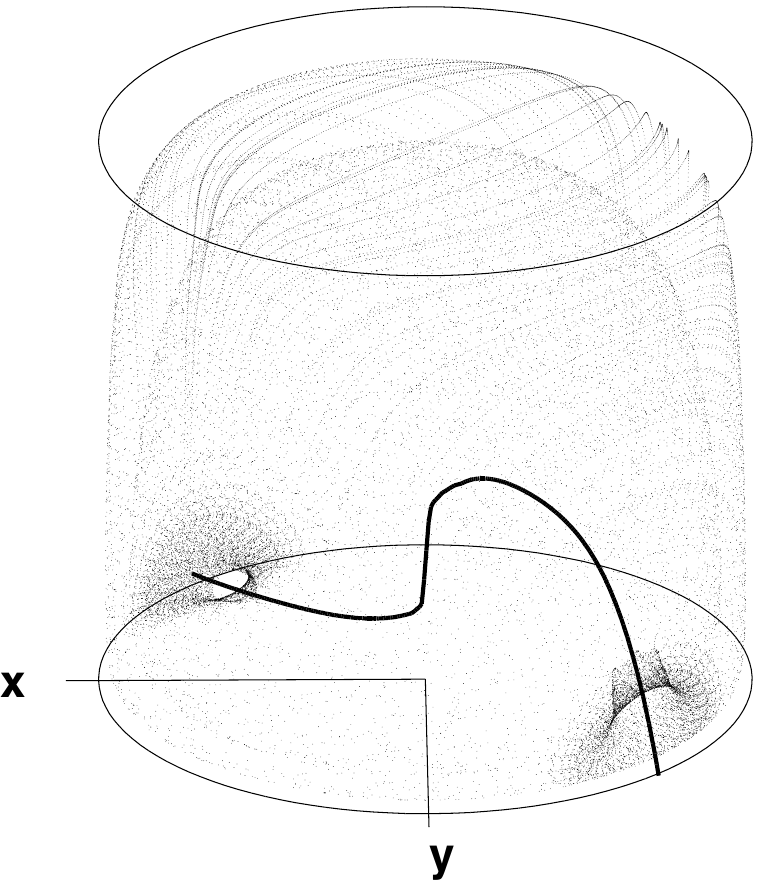}}
 \subfigure[]{\includegraphics[width=0.48\columnwidth]{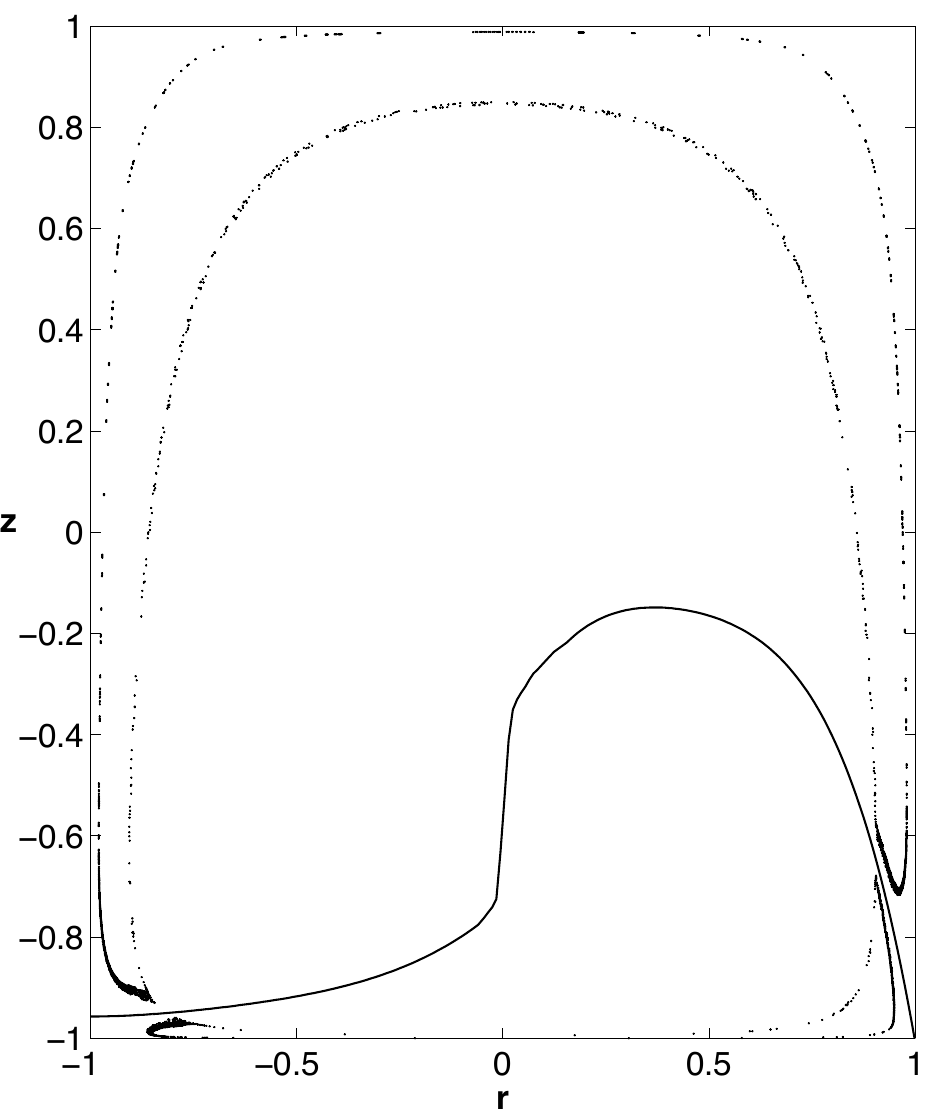}}
\caption{3D square cylinder flow:  Formation of adiabatic structures by resonance-induced merger of adiabatic shells and elliptic tubes in 3D unsteady flows
for small departures from the non-inertial limit  demonstrated for Protocol \Ax.  $Re=0.1$;
(a) Perspective view; (b) Slice centered upon symmetry plane $y=-x$.
Reprinted from \textcite{Speetjens06b} with the permission of AIP Publishing.
}
\label{ClosedStructures}
\end{figure}

\begin{figure}[tb]
 \subfigure[]{\includegraphics[width=0.48\columnwidth]{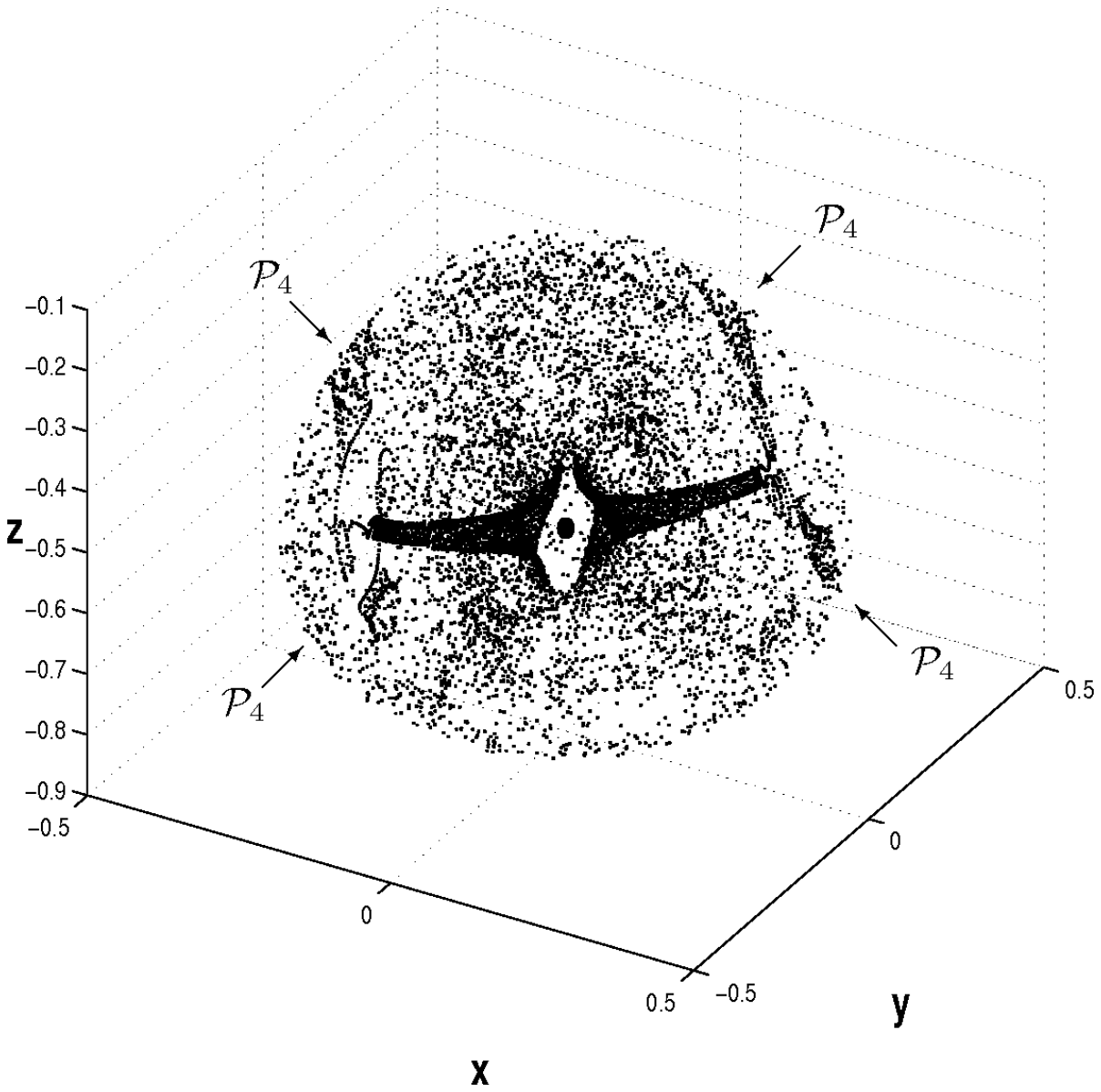}}
 \subfigure[]{\includegraphics[width=0.48\columnwidth]{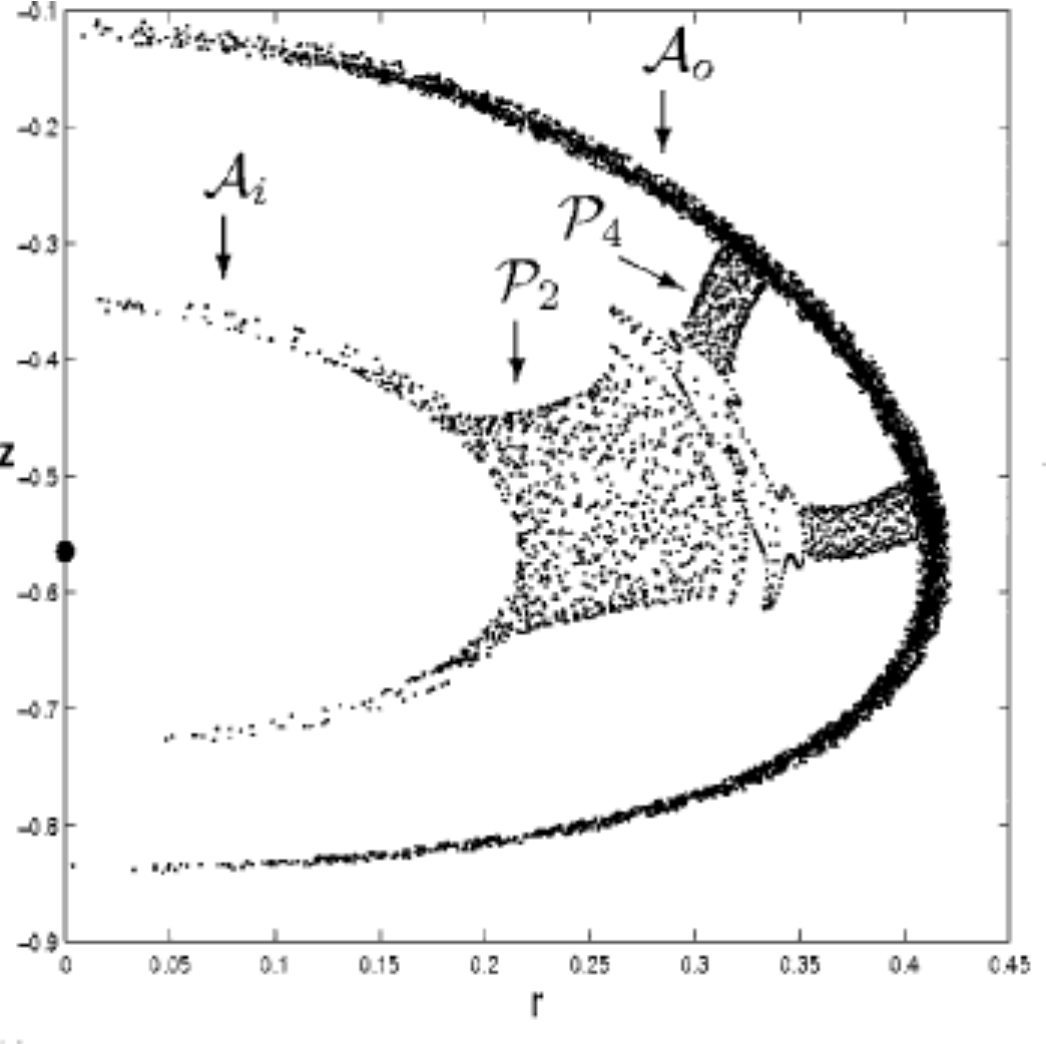}}
\caption{3D square cylinder flow:  Formation of leaky adiabatic structures by resonance-induced merger in 3D unsteady flows for small departures from the non-inertial limit demonstrated for Protocol \Ax. $Re=0.1$; (a) Perspective view; (b) $rz$-projection.
Reprinted from \textcite{Speetjens06b} with the permission of AIP Publishing.
}
\label{OpenStructures}
\end{figure}

Secondary circulation causes progressive drifting of tracers transverse to the invariant spheroids that grows stronger
with increasing $Re$. This is demonstrated in Fig.~\ref{AdiabaticShells} by means of the $rz$-projection of a Poincar\'{e}
section of a single tracer for $10^4$ forcing periods. The drifting tracers remain confined within thin shells centered upon the
invariant spheroids for $Re\lesssim 0.1$ (Figs~\ref{AdiabaticShells}(a,b)) for time spans of $\mathcal{O}(2\times10^4)$ periods due to the averaging out of transverse excursions (``averaging principle'' \cite{Arnold1978}). Thus invariant spheroids survive in an approximate way as so-called adiabatic shells \cite{Speetjens06b}. However, this survival occurs only for regions with chaotic intra-surface dynamics.
This puts forth the rather intriguing notion that intra-surface chaos promotes the persistence of partial transport
barriers, or equivalently, that 2D chaos suppresses the onset of 3D chaos. The formation of a complete adiabatic shell in
Fig.~\ref{AdiabaticShells}(b) thus signifies an underlying invariant spheroid with fully-chaotic tracer
motion. Chaotic and non-chaotic regions of invariant spheroids (e.g., shown in Fig.~\ref{ProtoA1}(b)) transform into incomplete adiabatic shells and elliptic tubes, respectively, that
merge into intricate adiabatic structures by a mechanism termed resonance-induced merger
\cite{Speetjens06b,MichelChaos}. Figure~\ref{ClosedStructures} shows an adiabatic structure formed by
 resonance-induced merger, comprising an inner and outer adiabatic shell, connected via an elliptic tube on each elliptic
segment of the period-1 line for $Re=0.1$. Both tubes, similarly to the underlying elliptic segments of the
period-1 line of the non-inertial limit, form a symmetry pair related via $\wt{S}$.  Resonance-induced merger thus results in a family of nested closed adiabatic structures that are topologically equivalent to
tori. A similar adiabatic structure due to  resonance-induced merger is shown in Fig.~\ref{OpenStructures}, but with two differences
compared to the above: (i) connecting tubes undergo a bifurcation from period-2 ($\mathcal{P}_2$) to period-4
($\mathcal{P}_4$) structures; (ii) inner ($\mathcal{A}_i$) and outer ($\mathcal{A}_o$) shells are ``leaky.'' This
leakiness causes tracer exchange with a chaotic environment (not shown) and thus sets up a net tracer circulation through the structure from outer to inner shell. This temporary tracer entrapment within a coherent structure (residence time $\mathcal{O}(10^3)$
periods) bears a certain resemblance to prolonged confinement in the chaotic-saddle region (Section~\ref{saddle}) and stickiness
near cantori (Section~\ref{KAM}) in 2D (open) flows.

The above exposes a fundamental difference in response scenarios of invariant tori and spheroids. The case of tori seems
fairly well understood, although a complete rigorous description of the breakdown and survival mechanisms remains
outstanding. The dynamics of perturbed invariant spheroids, on the other hand, remains largely an open problem.
Fully-chaotic spheroids survive weak inertia as complete adiabatic shells and constitute transport barriers akin to those of the non-inertial limit.
Non-chaotic regions on sub-families of invariant spheroids have fundamental ramifications by causing the formation of intricate adiabatic
structures through resonance-induced merger. Its occurrence for a wide range of forcing protocols \cite{Speetjens06b,MichelChaos,Pouransari2010} and different
systems \cite{Moharana2013} suggests that resonance-induced merger is a universal phenomenon and is part of an essentially 3D route to chaos.
The study in \textcite{Wu2014} supports this assertion by demonstrating that resonance-induced merger can in fact be triggered by weak perturbations of
arbitrary nature and provides the first experimental evidence of its physical existence.

Resonances as a key mechanism in response scenarios are the common denominator for tori and spheroids. However, their
emergence and manifestation are very different. Known resonances on tori in time-periodic systems include isolated closed trajectories
and, in case of the very dense windings underlying resonance-induced dispersion, resonant sheets: material surfaces of period-$p$ points. The former cause disintegration of the entire torus in question; the latter break-up into disconnected segments. Whether
further types may exist is open. Topological considerations suggest strings of isolated period-$p$ points as special
case of isolated closed trajectories (i.e.,  a localized counterpart to resonant sheets.) Resonances on spheroids emerge as periodic points belonging to periodic lines in the 3D domain. Resonance-induced merger seems to rely in particular on isolated degenerate periodic points connecting elliptic and hyperbolic segments of periodic lines. Moreover, instead of causing disintegration and chaos, resonance-induced merger reorganizes and reorders the flow topology through the formation of
distinct coherent structures. Resonance-induced dispersion  and resonance-induced merger nonetheless exhibit a remarkable similarity that hints at a comparable mechanism. Tracer switching effectuated by the isolated periodic points inside the perturbed resonant sheets during resonance-induced dispersion, demonstrated in \textcite{Mezic2001} and supported by experiments in \textcite{Solomon2003}, bears a great resemblance to the migration of tracers from an adiabatic shell into an elliptic tube in the case of resonance-induced merger. 
Further explorations are needed  and it may be of interest to compare other studies on resonance phenomena \cite{Litvak2002,Vainchtein2006}.

\section{Chaotic advection from data}\label{data}

Accurate prediction of material transport in geophysical flows is growing in importance in the aftermath of large-scale catastrophic events like the volcanic eruption of Eyjafjallaj{\"o}kull (Iceland, 2010), the Deepwater Horizon oil spill (Gulf of Mexico, 2010) [see e.g., \textcite{Domingos2013}], and the nuclear disaster in Fukushima (Japan, 2011) [see e.g., \textcite{Domingos2015}]. The main avenues for the material transport have been connected to the skeleton of chaotic advection, and detecting these routes from measured data in (near) real time has become one of the major goals at the intersection of dynamical systems and fluid dynamics. Most approaches use data in the form of numerical vector fields, interpolated and extrapolated from telemetry, e.g., from high-frequency radar and satellite altimetry, although Lagrangian sensor data are also available and studied. The approaches reviewed here are aimed at defining and computing skeletons of chaotic advection directly from available data, without intermediary equation-based models.

\subsection{Lagrangian coherent structures}
\label{sec:LCS}

\begin{figure}
  \centering
  \includegraphics[width=0.8\columnwidth]{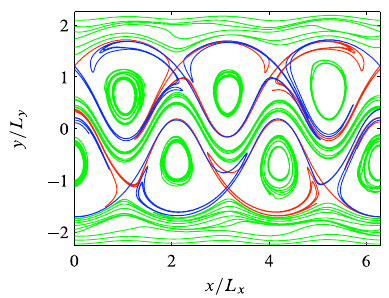}
  \caption{Elliptic and forward- and backward- hyperbolic Lagrangian
    coherent structures (LCS) in a time-periodic Bickley jet, detected
    based on the variational re-definition of LCS. 
    Reprinted from \textcite{haller2012} with permission from Elsevier.
    \label{fig:LCS-bickley} }
\end{figure}

As we have seen, invariant manifolds play the central role in analysis of time-invariant velocity fields as they act as barriers for material transport. If in addition an invariant manifold attracts or repels nearby material, it organizes  behavior of the flow in its vicinity. Since invariant manifolds align locally with eigenvectors of the linearized velocity field, methods based on flow linearization are able to compute accurately routes for material transport in time-invariant flows. For time-varying flows, however, manifolds that align with eigenvectors of linearized flow do not act as barriers to material transport. Therefore, a different strategy is needed to find lower-dimensional structures that organize the flow.

All material lines advected by the flow are barriers to material transport; in informal terms, Lagrangian coherent structures  are those material lines that strongly attract or repel the neighboring material. The initial operational definition specified attracting (respectively, repelling) LCS as material lines that are linearly stable in forward (respectively, backward) time for a longer period than any of their neighbors \cite{haller_finding_2000, Haller:2000us, Haller:2001ed, Haller2001, Haller:2002bf}.

LCS were initially approximated by ridges of finite-time Lyapunov exponent (FTLE) and finite-size Lyapunov exponent (FSLE) fields, which were assumed to exist near true LCS; see \textcite{Shadden:2005vn, Joseph:2002vi, haller_lagrangian_2011, Tallapragada:2013bh}. FTLEs and FSLEs measure the rate of separation between nearby trajectories, which is associated with the linear stability of those trajectories. Since FTLE or FSLE fields are relatively easy to compute and understand, they were quickly adopted as proxies to the routes of material transport in geophysical studies, e.g., in  \textcite{Rypina:2007ev,Shadden:2009cn, Bettencourt:2012dv}, but also in analyses of mixing, e.g., \textcite{lukens_using_2010}. Post-processing and use of FTLE or FSLE fields remains a challenge as they may indicate the presence of LCS where there are none (\textcite{karrasch_finite-size_2013}). In certain cases, FTLE or FSLE analysis is justified only in a qualitative, but not quantitative, sense; see \textcite{BozorgMagham2013}. Nevertheless, the ease of interpretation of FTLE fields means that they remain an active area of research in experimental studies, e.g., \textcite{Raben2013}.

\begin{figure}
\includegraphics[width=\columnwidth]{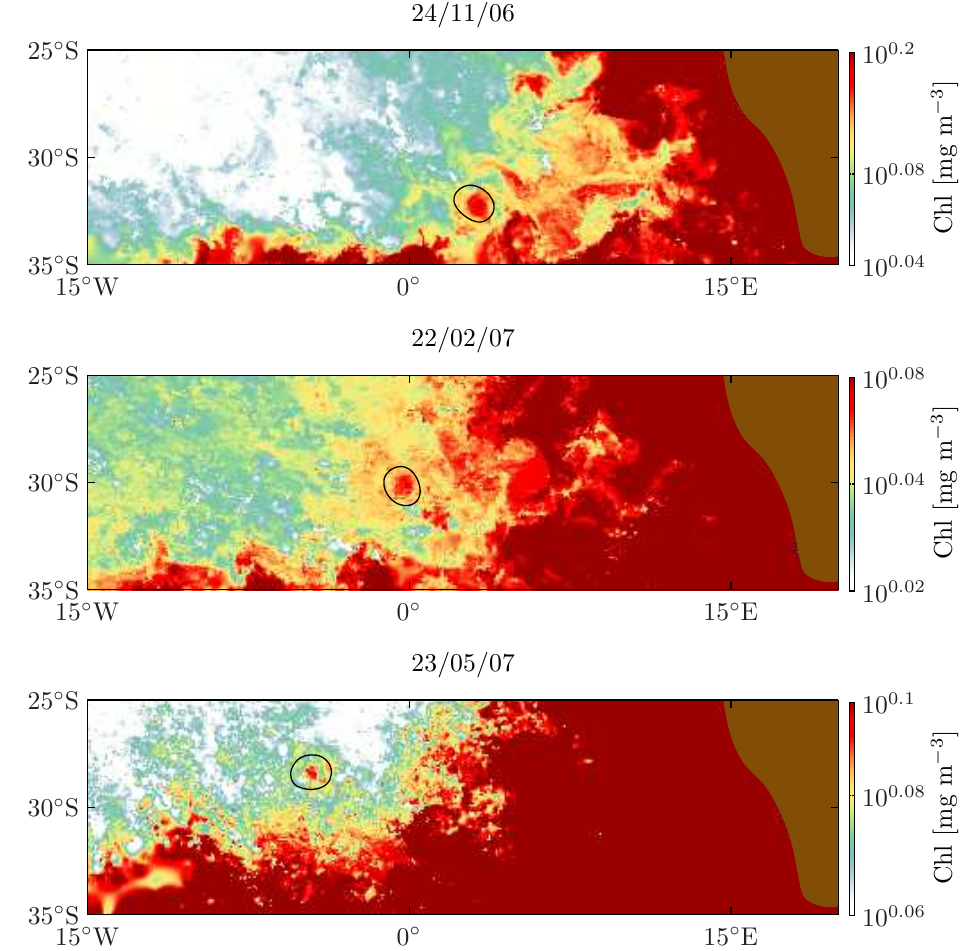}
\caption{A satellite image showing chlorophyll being carried by an
  Agulhas ring away from the southern tip of Africa (bottom-right in
  the image). Superimposed in black is an elliptic Lagrangian coherent
  structure  corresponding to the ring. \textcopyright American Meteorological Society.  Used with permission. From
  \textcite{beron-vera_objective_2013}. \label{fig:LCS-agulhas} }
\end{figure}

Lagrangian coherent structures have recently been re-defined with the
requirements that the defined structures be objective with respect to
the reference frame, observable over finite-time periods, coherent
under Lagrangian transport, and spatially continuous, as detailed in
\textcite{Haller2015}.  Under these requirements, a variational
principle for LCS has been formulated, which is then used as an
operational procedure for calculating the LCS \cite{Haller:2011kr,
  haller2012, Blazevski:2013ws}.  According to the variational
definition, LCS are shadowed by minimal geodesics induced by the
Riemannian metric of the Cauchy--Green strain tensor
\(C_{t}(x) = \jac\Phi_{t}^{\top}\jac\Phi_{t}\),
i.e., the ``absolute-value-squared'' of the Jacobian matrix
\(\jac \Phi_{t}\)
of the flow map.  Fig.~\ref{fig:LCS-bickley} shows an example
  of the invariant manifolds found in this manner in a jet flow. The
theory presently covers a broad array of LCS in 2D and 3D velocity
fields, classified according to the type of deformation they exert on
the surrounding material \cite{Haller2015}.

Compared to FTLE or FSLE computations, the variational definition results in a computationally more involved procedure. The algorithm has been made more accessible through a freely-available LCS toolbox for MATLAB.\footnote{\url{http://georgehaller.com/software/software.html}} As an example of the use of this approach, we highlight detection of an elliptic LCS that matches a physically-observed phenomenon known as the Agulhas ring, associated with the transport of cold water from the southern tip of Africa \cite{beron-vera_objective_2013};  Fig.~\ref{fig:LCS-agulhas}.
A recent detailed review of the Lagrangian coherent structure theory \cite{Haller2015} provides further references to studies that have applied variational LCS to analysis of geophysical flows, nowcasting and forecasting, as well as discussions of numerical considerations associated with the technique.

\subsection{Coherent sets}\label{sec:coherent-sets}

\begin{figure}
\includegraphics[width=\columnwidth]{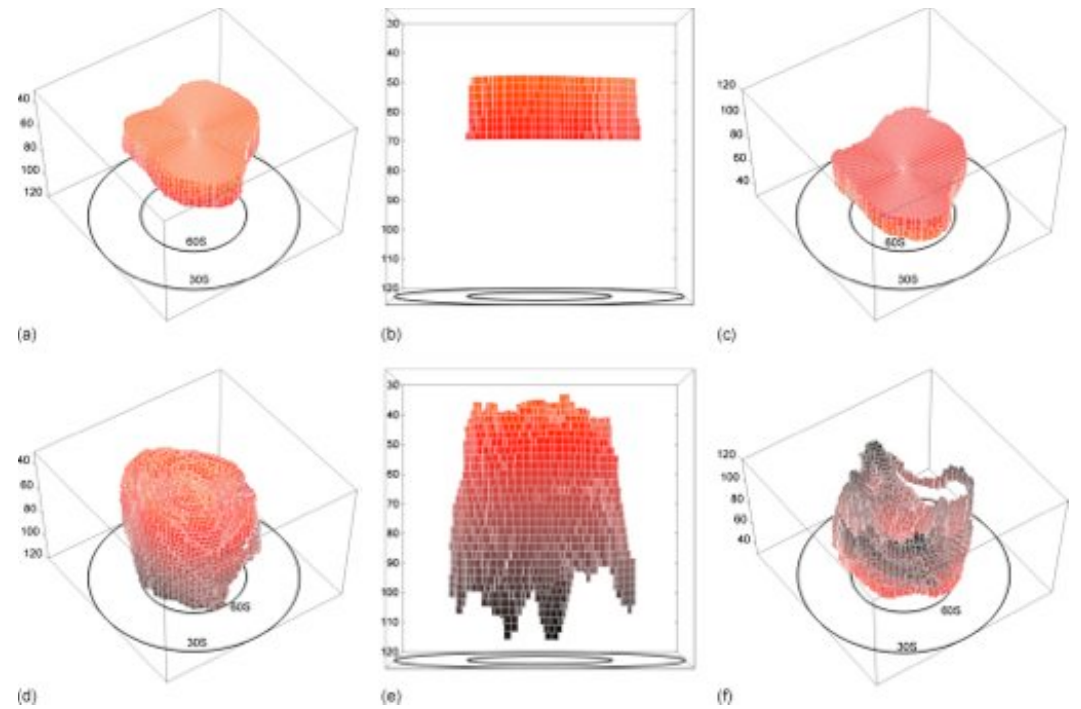}
\caption{Coherent set of the Antarctic atmospheric vortex, visualized
  using the Perron--Frobenius technique. Images within each row are different
  views of the same set; rows are visualizations for data recorded seven days
  apart. Reprinted from \textcite{Froyland:2010jo} with the permission of AIP Publishing.
}
\label{fig:coherent_sets}
\end{figure}

Instead of looking at barriers to transport, as is the case with Lagrangian coherent structures, a skeleton of chaotic advection can be proposed to be a collection of regions from which tracer material does not leak out. Such collections are routes through which the material travels, therefore serving as a road-map for transport.

For a tracer distributed in the set \(A\) at initial time \(t_{0}\), \(\Phi_{t}(A, t_{0})\) represents the set containing the tracer after it is advected by the flow over the time interval \(t\). The sets  \(C_{0}\) (source) and  \(C_{t}\) (target) are a \emph{coherent pair} if the tracer placed in the source fills the target entirely, without leaking, that is
\begin{equation}
  \label{eq:coherent-pair}
  \mu[ \Phi_{t}(C_{0},t_{0}) \cap C_{t} ]  \approx \mu[C_{t}],
\end{equation}
where the size of sets is measured by an application-relevant measure \(\mu\), e.g., fluid volume. Finding pairs of coherent sets can be formulated as a variational problem based on the linear Perron--Frobenius transfer operator. The transfer operator acts on tracer distributions \(\lambda\) by composing them with the flow map \([P_{t_{0},t} \lambda](A) := \lambda[ \Phi_{-t}(A,t_{0}) ]\). A solution to the search for coherent pairs is conveniently given by the eigenfunctions of the transfer operator. Similar methods were previously used  successfully in the reconstruction of invariant measures of dynamical systems in \textcite{Dellnitz:1999tr,Dellnitz:2002wma}, almost-invariant sets in time-independent dynamical systems in \textcite{Froyland:2003jj, Froyland:2009ti}, and coherent pairs in time-dependent flows in \textcite{Froyland:2010jo,froyland_coherent_2010}.

Numerical approaches predominantly rely on Ulam approximations of transfer operators, which discretize the fluid domain into smaller cells. Launching a large number of Lagrangian particles from each set, the transfer operator is approximated by a large, but sparse, Markov chain transition matrix. Discretization cells corresponding to non-zero elements of eigenvectors of the Ulam matrix form sets that approximate coherent pairs, as explained in the review by \textcite{Froyland:2013}  (see Fig.~\ref{fig:coherent_sets}).

In the context of fluid flows, the transfer operator approach has been applied to geophysical transport: e.g., atmospheric flows in \textcite{santitissadeekorn_optimally_2010}, and oceanic flows in \textcite{Froyland:2012fo,Sebille2012}. Case studies comparing Lagrangian coherent structures and coherent sets \cite{Froyland:2009ti,Tallapragada:2013bh} show that both approaches capture similar features in flows, which is explained by \textcite{froyland_finite-time_2012}. A further connection between coherent sets and braid-theoretic studies of mixing (see Section~\ref{sec:braids}) can be found in \textcite{Grover:2012iy}. That study points out that while material in coherent sets might not be well mixed with the surrounding fluid, the coherent sets may act as ``ghost'' rods that stir the fluid around them, promoting efficient mixing in the rest of the fluid.

\subsection{Mesochronic analysis}
\label{sec:mesochronic}

\begin{figure}
\includegraphics[width=\columnwidth]{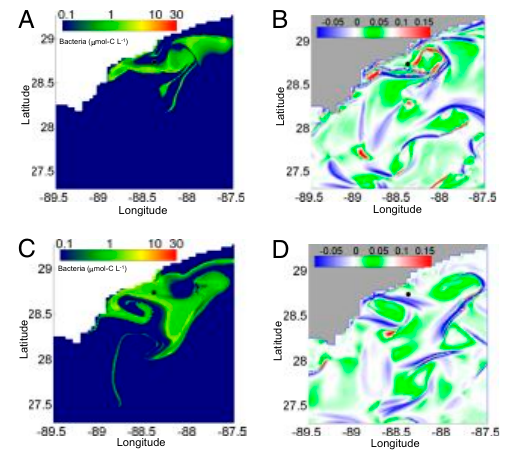}
\caption{Satellite images of bacterial abundance in the Gulf of Mexico (left column) and computed mesohyperbolic regions (right column) show that finite-time deformation computed from time-averaged velocity indicates chaotic advection activity.
Reprinted with permission from \textcite{valentine2012}, copyright (2012) National Academy of Sciences.
}\label{fig:mesochronic}
\end{figure}

Given any observable function, i.e., a field over states, \(f(x_{0},t_{0})\), we can generate its time-averaged, or \emph{mesochronic}, counterpart \(\wt{f}_{t}(x_{0},t_{0})\) by evolving a trajectory \(x_{t}(x_{0},t_{0})\) from the initial condition \((x_{0},t_{0})\) and computing the Lagrangian average of \(f\) along it:
\begin{equation}
  \label{eq:mesochronic-observable}
  \wt{f}_{t}(x,t_{0}) := \frac{1}{t}\int_{t_{0}}^{t_{0} + t} f(x_{\tau}(x,t_{0}), \tau) d\tau.
\end{equation}
It is a well-known result in ergodic theory that such functions will be constant over sets invariant under dynamics when \(t\to\infty\), regardless of \(f\) initially chosen. This result was exploited in \textcite{mezic1994thesis} and later in \textcite{Mezic1999,Levnajic:2010gq,Budisic2012} to construct an ergodic partition for closed, autonomous and periodic dynamical systems (see also Section~\ref{visualization}).

In data-based systems, the averaging period \(t\) is always finite, but dynamically-relevant information can still be extracted from \(\wt{f}\), in particular if the observable \(f\) is chosen to relate to the velocity vector field \(\mathbf{u}\) of the flow. (These methods are referred to as \emph{parcel schemes} in \textcite{samelson_lagrangian_2013}, and as \emph{Lagrangian descriptors} in \textcite{mancho_lagrangian_2013}.) The first attempt at doing so in the context of chaotic advection was by \textcite{Malhotra:1998tk}, who computed so-called \emph{patchiness} plots by averaging the magnitude of the velocity field \(\lVert \mathbf{u} \rVert\). The boundaries in patchiness plots were recognized by \textcite{Poje:1999wf} to correspond to certain analogs of invariant manifolds in 2D turbulence (see \textcite{Haller:1998cx}). The same quantity was again studied by
  \textcite{mancho_lagrangian_2013,madrid_distinguished_2009} and shown to align with ridges of the finite-time Lyapunov exponent field. Instead of the norm of velocity, \textcite{Haller:2003cf} computed time-averages of scalar quantities related to the Jacobian matrix \(\jac\mathbf{u}\), and showed that it is possible to produce a finer characterization of deformation in a flow which, again, agreed with proxies of Lagrangian coherent structures.

Instead of averaging scalars related to the velocity field, \textcite{Mezic:2010kh} computed time-averages of the velocity vector field itself, producing the so-called \emph{mesochronic velocity}. These averages  \(\wt{\mathbf{u}}_{t}\), computed using Eq.~\eqref{eq:mesochronic-observable}, are directly related to the flow map \(\Phi_{t}\) as the average velocity is just the displacement caused by dynamics, divided by the duration of the motion:
\begin{equation}
  \wt{\mathbf{u}}_{t}(x,t_{0}) = (\Phi_{t}(x,t_{0}) - x)/t.
  \label{eq:mesochronic-flow-connection}
\end{equation}
The mesochronic velocity Jacobian matrix \(\jac \wt{\mathbf{u}}_{t}\) is therefore directly related to the Jacobian matrix of the flow map \(\jac \Phi_{t}\). Since the spatial derivative \(\jac\) and the averaging integral \(t^{-1}\int_{t_{0}}^{t_{0}+t} d\tau\)  do not commute, this analysis is different from the analysis in \textcite{Haller:2003cf} mentioned earlier.
Unlike maximal finite-time Lyapunov exponents, \(t^{-1}\log \lVert\jac \Phi_{t}\rVert\), that measure the \emph{magnitude} of deformation that the material experiences, the scalar field \(\det \jac \wt{\mathbf{u}}_{t}\) uncovers the \emph{character} of deformation for planar incompressible flows, e.g., strain (mesohyperbolicity), or rotation (mesoellipticity). Partitioning the state space based on values of \(\det \jac \wt{\mathbf{u}}_{t}\) is a direct generalization of the Okubo--Weiss deformation criterion \cite{Okubo:1970tr,Weiss:1991kv} from infinitesimal to finite advection times. Note that \textcite{Greene:1979jr,Greene:1968ua} previously used a similar quantity to predict the order of destruction of KAM tori in perturbed Hamiltonian maps, the phenomenon shown to be relevant to mixing in Section~\ref{Inertial}. The mesochronic calculations were further generalized to 3D flows in \textcite{budisic2016}.

The analysis of the mesochronic Jacobian has been applied to the
prediction of the oil-slick transport in the aftermath of the Deepwater Horizon spill.  \textcite{Mezic:2010kh} showed that regions of
hyperbolicity correspond to jets that dispersed the slick, while
elliptical zones correspond to eddies in which the slick accumulated.
In a followup paper, the technique contributed towards resolving the problem of "missing oil", indicating the correct locations and volumes of the oil post-spill \cite{valentine2012} (see Fig.~\ref{fig:mesochronic}).

\subsection{Braids of Lagrangian trajectories}
\label{sec:braids}

\begin{figure}
\includegraphics[width=\columnwidth]{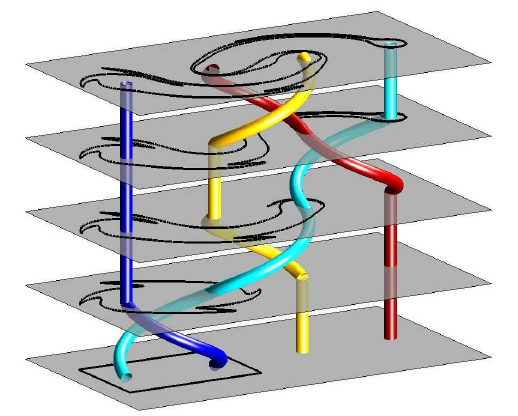}
\caption{A braid constructed from the motion of stirrers in a vat, along with a sample material line deformed by the motion. Reprinted from \textcite{thiffeault2006topology}.
}\label{fig:braids-schematic}
\end{figure}

Although flows are most commonly analyzed through their vector fields, in certain situations data available about the flow are too sparse to allow interpolation of the (continuous) vector field. An example is the Argo Program\footnote{\url{http://www.argo.ucsd.edu}} that measures  global oceans using sensors relaying their positions, temperature, and salinity as they are advected by ocean currents. In other cases, recorded data are sparse because detailed data would be too complex to use and act upon. For example, industrial stirring devices often use stirring rods following simple protocols that induce complicated flows surrounding them (Section~\ref{sec:heat_transfer}). In both of these instances, braid-based approaches are able to estimate properties of the flow from only a few Lagrangian trajectories, whether sensor or stirring-rod paths, significantly reducing the need for detailed measurements.

\begin{figure}
\includegraphics[width=0.5\columnwidth]{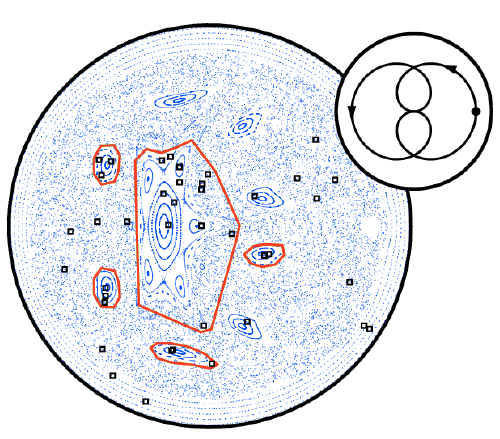}
\caption{An approximation to the boundary of a coherent structure in a stirred vat of fluid estimated from only 40 sampled trajectories; compare Lagrangian coherent structure computation that would require knowledge of the entire vector field. 
Reprinted from \textcite{Allshouse:2012kc} with permission from Elsevier.
}\label{fig:braids-coherence}
\end{figure}

A seminal paper by \textcite{boyland2000topological} (anticipated a decade earlier by \textcite{mackay1990_knot})
takes Lagrangian trajectories of a time-varying planar flow and treats them as entangled strands in the 3D space--time domain, resulting in an approach termed \emph{braid dynamics}. The tangle of trajectories is represented by a braid, which is a sequence of symbols encoding only relative positions of trajectories and their exchanges. Braids can be interpreted as reduced models of flows, acting on closed material lines tightened around trajectories to form so-called loops, which are deformed by interchanges of trajectories.

Regardless of the number and locations of trajectories forming the braid, topological entropy of the braid is always the lower bound for the topological entropy of the flow, which, in turn, is connected to mixing rates (see Section~\ref{quality+measures}). Such lower bounds have been used to optimize stirring protocols used for mixing in viscous fluids, see \textcite{boyland2000topological,thiffeault2006topology,Finn2011}  (see Fig.~\ref{fig:braids-schematic}). As an analysis procedure, the entropy of the flow can be bounded either by computing braids of individual trajectories, as in \textcite{Thiffeault2005,Thiffeault2010}, or braids of coherent structures that act as ``ghost rods'', stirring the surrounding fluid, as in \textcite{Gouillart2006,Grover:2012iy,Stremler:2011hu,Tumasz2013}.

The topological entropy of a braid is the fastest rate at which a loop can
be deformed under a braid.  The slowest rates of deformation are also
significant, as minimally-growing loops can enclose trajectories that
remain trapped together in a coherent structure
\cite{Allshouse:2012kc}.  Such loops then provide approximations to
boundaries of coherent structures, which may be computed from several
trajectories alone, instead of from the fully-resolved velocity
field; see Fig.~\ref{fig:braids-coherence}.

The braid approach allows one to process extremely sparse data of planar flows in a computationally efficient manner and to obtain estimates of mixing rates and coherent structures. The computations involved in braid dynamics may be unfamiliar to researchers in fluid dynamics. However, they can readily be performed with the help of freely-available software\footnote{MATLAB toolbox \texttt{braidlab}, see \textcite{Thiffeault2014v3}}. Unfortunately, the braid approach does not so far extend to 3D fluid flows.

\section{Laminar versus turbulent flows}\label{turbulent}

A fundamental dichotomy between laminar and turbulent flows
has been signalled in the introduction and at points throughout this review, and exists in the fluid dynamics literature more generally.
We believe, however, that a more refined categorization of fluid flows is required; below we provide a roadmap for rethinking this question, based on what has been learned in the last several decades through studies of chaotic advection.

\subsection{Laminar flows}\label{LaminarFlow}

The basic image of laminar flow is that the flow takes place in layers.  Thus a good initial definition of laminar flow is that the velocity field has a codimension-1 foliation whose leaves are invariant.  A  codimension-1 foliation of an $n$-dimensional manifold is a decomposition into injectively immersed submanifolds\footnote{An injectively immersed submanifold of a manifold $M$ is the image of a one-to-one differentiable map $f:N \to M$ from a reference manifold $N$ into $M$ for which the derivative $Df$ has maximal rank everywhere.}
called  leaves such that in a neighborhood of any point the decomposition is diffeomorphic to the decomposition of $\R^n$ into the planes $x_n=$ constant, where $x_n$ is the last coordinate in $\R^n$.
Our discussion will progress from steady flows to time-dependent flows.

All steady mass-preserving flows\footnote{Mass-preservation allows more generality than volume-preservation, e.g.~compressible flow, and gives the same results, suitably interpreted.  Thus there is a preserved stream function $\Psi$, such that the velocity is given by $v_x = \frac{1}{\rho}\frac{\partial \Psi}{\partial y}, v_y = -\frac{1}{\rho}\frac{\partial \Psi}{\partial x}$, where $\rho dx\wedge dy$ is the preserved mass-form.} in 2D are laminar, or almost so.  The leaves can be taken to be the level sets of the stream function.  Yet this example already shows that the definition needs generalizing to allow finitely many singular leaves of lower dimension.  The problem is that although the generic level set is a smooth 1D submanifold, for critical values of the stream function the level set may be a single point or contain saddle points or worse.  There are various ways of prescribing which forms of singular leaf are allowed and how the leaves fit together around it.  This is especially well worked out in the theory of pseudo-Anosov maps (for references aimed at chaotic advection, see \textcite{boyland2000topological,M01}), but there are various not-quite-equivalent formulations, so we refrain from being prescriptive.

The case of 2D steady mass-preserving flows illustrates another feature of laminar flows that may arise.  If the 2D manifold is not simply connected then for a vector field $v$ preserving a non-degenerate 2-form\footnote{An $n$-form is an antisymmetric $n$-linear functional of $n$ tangent vectors at each point.  If $n$ is the dimension of the manifold, the $n$-form is {\em non-degenerate} if at every point there is an $n$-tuple of tangent vectors on which it is non-zero.}
$\omega$ (formulating mass-preservation in a way that is more general than volume-preservation) there is not necessarily a single-valued stream function, but the flow still preserves a closed 1-form\footnote{The operator $i_v$ is the contraction with the vector field $v$ as the first argument; thus for all tangent vectors $\xi$, $\alpha(\xi) = \omega(v,\xi)$. An $n$-form $\alpha$ is closed if its exterior derivative $d\alpha$ is zero; equivalently if the integral of $\alpha$ over the boundary of any $(n+1)$-dimensional ball is zero, which is true here because $v$ preserves $\omega$.}
 $\alpha = i_v\omega$.
For example, the vector field $\dot{x}=1,\dot{y}=V$ (constant) on the 2-torus $\R^2/\Z^2$ preserves area $\omega = dx\wedge dy$ and the 1-form $\alpha(\xi) = \omega(v,\xi) = \xi_y-V\xi_x$ for tangent vectors $\xi = (\xi_x,\xi_y)$, which is $d\Psi(\xi)$ for $\Psi = y-Vx$ on $\R^2$, but $\Psi$ is multivalued\footnote{A closed 1-form is called {\em exact} if it can be written as $d\Psi$ for some single-valued function $\Psi$, which is then a genuine stream function.}   
on $\T^2$.  Then the level sets of $\Psi$ are replaced by the integral submanifolds of the 1-form, which in general are only injectively immersed submanifolds rather than true submanifolds (e.g., $V$ irrational in the example).
Indeed in this example they wind round $\T^2$ densely.  It is density of invariant leaves that gives a laminar flow the potential for chaotic mixing, in contrast to those with a single-valued stream function.  This simple example does not have chaotic mixing but we will show examples in Section~\ref{sec:turb} that do.

Moving to steady 3D flows, the prime example of Poiseuille flow in a circular pipe is laminar: the obvious foliation by concentric cylinders works provided we allow the singular 1D leaf down the axis.  But this example also allows many other foliations with invariant leaves.  Choose any foliation of an initial cross-section and let its leaves flow with the vector field to give a foliation of the whole domain.  So there is a high degree of non-uniqueness in the foliation.  If one adds swirl this still works.

Any steady 3D mass-preserving flow with a continuous symmetry is laminar \cite{Mezic1994,Haller1998}, because it has a stream function whose level sets make the leaves.  Here is an outline proof:  Given a 3-form $\omega$ representing mass density, a velocity field $v$ preserving $\omega$ and a vector field $u$ representing a symmetry (i.e.,~also preserving mass and commuting with $v$) and independent of $v$ modulo a set of zero volume, let $\gamma = i_v i_u \omega$.  It is a 1-form and a little calculation shows that it is closed ($d\gamma=0$).  Thus the field of planes defined by $\gamma=0$ is integrable to a foliation. 
Its leaves are invariant because $i_v i_u \omega (v) = 0$ by anti-symmetry of $\omega$. The constructed foliation is also invariant under the symmetry.

What about dynamically defined classes of steady flow, like Euler flows or Stokes flows?  For steady Euler flows the Bernoulli function (Eq.~\eqref{bernoulli})  is conserved so its level sets provide a foliation with invariant leaves, unless the Bernoulli function is constant. In the latter case $\textrm{curl}\,v = \kappa' v$ for some function $\kappa'$ of position, and $\kappa'$ is conserved so its level sets provide a foliation with invariant leaves, unless $\kappa'$ is constant.  The latter case gives the Beltrami flows, for which a complete understanding is still lacking (\textcite{Arnold1998}, Ch II.1).  
Is every 3D Stokes flow laminar in this sense?  Following \textcite{boyland2000topological}, we feel that this is probably true, because they minimize dissipation rate, which should make them analogous to Thurston's simplest representatives of isotopy classes of surface homeomorphisms, each of which has an invariant foliation (two in the case of pseudo-Anosov components), but we are not aware of a definitive answer.

Are there 3D steady mass-preserving flows which are {\em not} laminar?  One might be tempted to say that any chaotic flow would give an example, because most candidate leaves grow exponentially under the flow, but in fact the nicest forms of chaotic flow possess invariant foliations.  Specifically, each uniformly hyperbolic 3D mass-preserving steady vector field has a forward contracting foliation and a backward contracting foliation (with no singularities, even).  An example is Arnol'd's fast dynamo flow (see, e.g., Ch II.5.E of \textcite{Arnold1998}), the vector field $(0,0,1)$ on the manifold made by identifying horizontally opposite sides of a cube by translation and the top to the bottom by the matrix $\mathbf{A}=\left[\begin{array}{cc}
2 & 1 \\
1 & 1
\end{array}\right]$.  The foliation by the surfaces $\gamma y -x=C$ constant, where $\gamma = (1+\sqrt{5})/2$, is invariant.  
To interpret this correctly, one has to identify the values $C+1, C-\gamma$ and $\gamma^2C$ with $C$.  Thus each leaf is dense in the manifold. The foliation $y+\gamma x = $ constant is also invariant. The flow is chaotic in many good senses: in particular, displacements in the direction $(\gamma,1,0)$ grow exponentially at rate $2\log\gamma$.  Although it looks artificial, versions of this flow can be made in 3D containers in Euclidean space \cite{MacKay_CCT2007}.  

A genuine obstacle to laminar flow is elliptic periodic orbits.  If a leaf passes through an elliptic periodic orbit then after one period the leaf comes back rotated by an amount not equal to $0$ or $\pi$, contradicting the condition for a foliation.  One can allow a finite number of elliptic periodic orbits by our decision above to allow finitely many singularities in the foliation, but it would probably be a mistake to allow infinitely many, or at least to allow singularities to have accumulation points.  Thus any mass-preserving steady 3D flow with a set of elliptic periodic orbits with an accumulation point is not laminar.  Generically, within the class of smooth-enough mass-preserving steady 3D flows, if there is one elliptic periodic orbit then it is an accumulation point of others. So, except for the highly exceptional integrable cases (i.e., those with a stream function), and the hyperbolic cases, all the rest are very likely not laminar, unless those arising dynamically are a very special subset (as we guess may be the case for Stokes flows).

A generalization of the concept of invariant foliation that may be useful is that of measurable partition \cite{rokhlin1960new,mezic1994thesis}.
In this case, the elements of the partition are not necessarily submanifolds and can even be fractal sets. If a flow possesses an invariant, nontrivial\footnote{A trivial measurable partition of a set $A$ consists of the set itself and the empty set.} 
measurable partition it could be considered laminar.
Nontriviality could be too weak though, because a flow on an annulus that mixes the top half of the annulus and the bottom part of the annulus separately would have only two, thick laminae. Recently, the concept of ergodic quotient has been introduced for flows in bounded domains \cite{Levnajic:2010gq,Budisic:2012woa}. Roughly speaking, the ergodic quotient space is obtained by identifying each subset of the physical space that the bounded fluid flow ``samples" well (precisely speaking, time-averages of continuous functions are equal to space-averages with respect to a measure defined on that subset; e.g.~the invariant subset might be a torus, the flow might induce a dense winding on the torus and the measure is the surface area on the torus, and then the ergodic quotient is a single point). The ergodic quotient of a flow on the annulus that we described above would consist of two points, representing the top and bottom part of the annulus. However, a shear flow in such an annulus --- produced e.g., by the outside circle of the annulus rotating and the inside also rotating at smaller angular velocity, as in Couette flow --- would have ergodic quotient that is a closed interval, and is thus 1D. We could then say that the flow is $d$-laminar with (possibly nonuniform) thickness of the laminae, provided the fractal dimension of the nontrivial ergodic quotient is $1-d$.
In the case of $d=0,$ we retain the label of ``laminar" instead of ``$0$-laminar".

Let us move now to time-dependent flows.  For these flows it is essential to consider the vector field as living in a domain of space-time rather than just space.  So an $n$-D time-dependent flow corresponds to a vector field on an $(n+1)$-D manifold, where the $t$-component is just 1.  Our definition proposes that the flow is laminar if there is an $n$-D foliation of this $(n+1)$-D space-time, with invariant leaves.
Let us think about this for $n=2$. As for Poiseuille flow, any foliation of the initial domain at $t=0$ is transported to a foliation of the whole space-time with invariant leaves.  But the leaves may become more and more contorted at large times.  So we put a restriction on the types of foliation that we allow, namely the diffeomorphisms that map the local decompositions onto the planes $x_n=$ constant should be bounded in $C^2$. 
In the case of periodic time-dependence, we would furthermore want the foliation to have the same period in time.  Thus, because of elliptic islands, the beloved example of blinking vortex flow \cite{Aref1984} is almost certainly not laminar.

\subsection{Turbulent flows}
\label{sec:turb}

A good starting point for considering turbulent flow is the viewpoint of \textcite{Taylor1954}, that a flow is turbulent if the velocity auto-correlation integral in time converges.  To define this, a (possibly time-dependent) velocity field $v$ in a subset of Euclidean space induces a flow map $\phi_{t,s}$ from time $s$ to time $t$.  Let
$$C_{ij}(\tau;t) = \int v_i(\phi_{t+\tau,t}(x),t+\tau) v_j(x,t) \mu_t(dx)$$
where $\mu_t$ represents the mass density at position $x$ and time $t$.  Say the flow is {\em turbulent} if
$$\int_0^\infty C_{ij}(\tau;t) d\tau < \infty$$
for each $t$.  Probably if it holds for one value of $t$ then it holds for all, though we have not seen this proved.

Note, however, that Taylor's condition can hold for some steady flows, contradicting the commonly held notion that turbulent flows must be time-dependent.  For example, modify the vector field in Arnol'd's fast dynamo flow to $(0,0,g(x,y,z))$ for some function $g>0$ such that $g(x,y,1) = g(A(x,y),0)$ and the ratio of period to number of revolutions in $z$ is different for two periodic orbits.  Then the flow preserves the modified mass $g^{-1} dx\wedge dy\wedge dz$ and by Anosov's alternative \cite{Anosov1967} it is mixing in the ergodic theorist's sense, i.e., $C(\tau) = \int \psi(\phi_\tau x)\chi(x)\ d\mu(x) \to 0$ as $\tau \to \infty$ for all pairs of functions $\psi \in L^\infty$, $\chi \in L^1$.  A stronger non-triviality condition on $g$ probably makes $\int_0^\infty C(\tau) d \tau$ converge.\footnote{Taylor's definition of turbulence is not immediately applicable here because Arnol'd's flow does not live on a subset of Euclidean space, so the meaning of multiplying components of a vector at different points is unclear. However, one can probably turn it into an example in a bounded container with stress-free boundaries for which the velocity autocorrelation function does make sense and is integrable, so this example shows that turbulence might not require time dependence.} Note that this flow preserves the same foliation as for the case $g=1$, so is also laminar.

On the other hand, Taylor's condition never holds for steady flows with no-slip boundaries.  The issue is that the volume remaining in a neighborhood of the boundary for time at least $\tau$ is at least $C/\tau$ for some $C$ (depending on the neighborhood), and thus the velocity autocorrelation integral diverges \cite{Jones1994}.  The effect of walls has been addressed further in Section~\ref{closed}.

Also, there are flows that are neither turbulent nor laminar.  Take for example a steady 3D flow with infinitely many elliptic periodic orbits.  As we have discussed in the previous subsection it is not laminar.  Yet if the elliptic periodic orbits are surrounded by a set of positive volume of invariant tori, as given generically by KAM theory for smooth enough mass-preserving vector fields, then the velocity auto-correlation integral diverges.
 
Note that Taylor's notion is in essence a Lagrangian notion, since it requires integration of physical quantities along particle paths. 
It can be easily reformulated in the context of the Koopman operator \cite{mezic2013analysis}, requiring that the Koopman operator spectrum of the solution of the dynamical evolution equation (e.g., the Navier--Stokes equation) with appropriate initial and boundary conditions does not have any eigenvalues and associated eigenfunctions, except for trivial (constant) ones \cite{arnol?d1968ergodic}. In other words, the flow generated by the solution has the mixing property. An alternative, Eulerian view would be to require integrability over time of the Eulerian velocity autocorrelation function that is given by 
$$C_{ij}(\tau;t) = \int v_i(x,t+\tau) v_j(x,t) dt$$ for every $x$. This, in fact, can be related to properties of another 
Koopman operator, the one associated with the velocity phase space. In particular, the property would be implied
if the related Navier--Stokes attractor has the mixing property. Nevertheless, even this is not enough to capture the full idea of what most people want to mean by ``turbulence": the definition does not say
anything about the spatial distribution of eddies and their size structure; a very important notion in the turbulent energy transfer context
\cite{frisch1996turbulence}. Specifically, a flow with a very simple spatial dependence --- with all the velocity vectors 
pointing in one direction --- but complex temporal dependence could have the mixing property on the strange attractor in
Navier--Stokes phase space, so be turbulent in the sense set out here, but clearly does not have the energy-cascade property of turbulence. Such a flow would also be laminar in the sense of our previous discussion.

\subsection{Synthesis}

Is laminar versus turbulent flow really a dichotomy?  The answer is clearly no. We have shown that with common definitions there are flows that are both laminar and turbulent and there are flows that are neither. And we have not even addressed the case of flows that one might want to say are laminar in some regions and turbulent in others. The variety of flows that exist is not captured by the current simplistic laminar versus turbulent classification.  More work needs to be done and we hope that this discussion might provide some ideas.

\section{The structure of the mixed state: Quality and measures of mixing}\label{quality+measures}

\begin{figure}[tb]
  \centering
  \includegraphics[width=0.6\columnwidth]{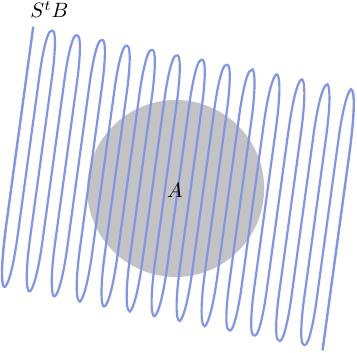}
  \caption{An advected patch $S^t B$ that has undergone strong
    mixing. At late times the patch covers an arbitrary reference
    patch $A$.
    \label{fig:mixpatch}}
\end{figure}

At the heart of all applications of chaotic advection lies the desire to exploit the inherent complexity in deterministic chaotic dynamics for fluid mixing. There are several different mathematical diagnostics for revealing this complexity, for example recurrence, transitivity, ergodicity, entropy; which is the most relevant will depend on the particular application. The most natural one of concern for mixing applications is arguably the measure-theoretic notion of (strong) mixing: a measure preserving (invertible) transformation  $f: M \to M$ is {\em strong mixing}\footnote{Typically in ergodic theory the short form \emph{mixing} is used as a synonym for \emph{strong mixing}, but here we use the full term to avoid confusion both with the everyday sense of the term mixing and with the fluid dynamical definition we put forward in Section~\ref{introduction}.} if for any two  measurable sets $A, \, B \subset M$,
\begin{equation}\label{eq:mixing}
\lim_{n \to \infty} \mu (A \cap f^n (B) ) = \mu (A) \mu (B)/ \mu(M).
\end{equation}
We may understand the meaning of this expression as follows: in a container $M$ of fluid, if $B$, a region originally occupied by a tracer, is mixed to become $S^t B$ (in continuous time, or equivalently $f^n(B)$ in discrete time), as depicted in Fig.~\ref{fig:mixpatch}, then for any arbitrary part $A$ of the container, the
amount of tracer in $A$ after mixing is
$\mu (A \cap f^n (B) ) /  \mu (A)$ and in the limit, $n \to \infty$, this should be $\mu (B)/ \mu(M)$. 
 For incompressible fluids the measure $\mu$ can be assumed to be Lebesgue, that is, area or volume and we may normalize the
 the volume of the domain to $\mu(M)=1$.  This definition effectively says that any two sets become asymptotically independent of each other for a strong mixing system, as illustrated in Fig.~\ref{fig:mixpatch}. Strong mixing implies the weaker property of weak mixing,
\begin{equation}\label{eq:weak}
\lim_{n \to \infty} \frac{1}{n} \sum_{k=0}^{n-1}  \left|  \mu(A \cap f^n (B) ) - \mu(A)\mu(B) \right| = 0.	
\end{equation}
In turn weak mixing implies ergodicity, which can be defined by
\begin{equation}\label{eq:ergodic}
\lim_{n \to \infty} \frac{1}{n} \sum_{k=0}^{n-1}  \mu(A \cap f^n (B) ) = \mu(A)\mu(B) .	
\end{equation}
Note that for each of these properties, we consider the quantity $\mu (A \cap f^n (B))$ in relation to $\mu(A)\mu(B)$. Setting $\mu (A \cap f^n (B)) - \mu(A)\mu(B) \equiv \xi_{f,n}$ we see that $f$ is strong mixing if $\xi_{f,n}$ converges to 0 as $n \to \infty$, weak mixing if $|\xi_{f,n}|$ converges to 0 in the sense of Ces\`aro summation, and ergodic if $\xi_{f,n}$ converges to 0 in the Ces\`aro sense. The hierarchy is strict: strong mixing implies weak mixing, which itself implies ergodicity. The converse is not true; an irrational rotation is an example of an ergodic system that is not weak mixing, while an interval exchange transformation can be weak mixing but not strong mixing.

Mathematical models that underpin the basics of fluid mixing devices can be proven to exhibit this behavior. For example, the cat map $\mathbf{A}$ --- an integral part of the Arnol'd fast dynamo model discussed in Section~\ref{turbulent} --- a paradigmatic model of repeated shearing in orthogonal directions, can be shown to be strong mixing, and to have positive topological entropy (equal to the logarithm of its largest eigenvalue). Another less simple yet analysable example is the map of \citet{Cerbelli2005}, which is strong mixing yet shows multifractal structure (some aspects of this case are treated further in \citet{MacKay2006}). In practice however,  it may be implausible to expect the strong-mixing property Eq.~(\ref{eq:mixing}), and impossible to verify, since to do so requires computing over all possible regions $A$ and $B$. These issues, coupled with the ubiquity of mixing problems in applications, have led to reformulations of Eq.~\eqref{eq:mixing}. For example, in section \ref{sec:mixnorms} we discuss {\em weak convergence}, and a method to establish this condition, while in section \ref{rates} we describe the {\em decay of correlations} of observable functions, which is used in ergodic theory to estimate rates of mixing. These two notions are closely related, but we will describe each in terminology and notation most closely associated with their own bodies of literature. 

First we observe that taking a physical measurement such as concentration  involves a choice of scale. In sections \ref{sec:mixquality}, \ref{sec:gibbs} and \ref{sec:second} we take a new look at classical ideas of measuring the quality of a mixture, in which the choice of scale provides details of interest about a mixing process. Section \ref{sec:mixnorms} contrasts this with the recent concept of multiscale measures for mixing.

The quality or `goodness' of mixing may be quantified by using
statistical quantities like the \emph{coarse-grained density} and the
entropy. These quantities may be employed for this purpose if at
any moment in time the distribution of the (marked) fluid to be
mixed in the ambient fluid domain is exactly known. We restrict
ourselves here to 2D incompressible flow, focussing on
the problem of a blob of colored fluid introduced into a fluid
in which some specified flow is present. We will use the concept
of the coarse-grained density of the distribution --- introduced by
\textcite{Gibbs1902} --- as a basic measure of the three criteria of the
mixed state: the {\em averaged square density} \cite{Welander1955},
the {\em entropy} \cite{Gibbs1902} and the {\em intensity of
segregation} \cite{Danckwerts1952}. By using these criteria we can
estimate the time necessary for  the mixed state to become uniform
within some specified range, for a given volume element size (the
`grain'). It is
important to note that the three criteria are not independent
and that they are statistical measures of the first order. 
For a more   complete description of a mixture, we  also
use the \emph{scale of segregation} \cite{Danckwerts1952},   
which is a
statistical measure of the second order and 
represents an average of the size of the clumps of the mixed
component. The first-order statistics provide information
related to the coarse-grained density distribution calculated in each 
grain at a time. And the second-order statistics gives  information about 
the correlations of this density in two different grains at the same moment of time.

\subsection{Definitions of mixing quality}\label{sec:mixquality}

We first illustrate the concept of coarse-grained density as a quantification of mixing in an example problem of 2D stirring in a square cavity. The cavity is covered by a grid of $N_{\delta}$ square cells of side $\delta$ and area $S_{\delta} = \delta^2$, and initially contains a square black blob in its center. The area of the cavity is $S=N_{\delta}S_{\delta}$, while the blob's area is $S_b$, which is conserved under area-preserving stirring. We denote the area of the blob inside cell $n$ by $S_b^{(n)}$, and the proportion of black $D_n = S_b^{(n)}/S_{\delta}$, which may be considered a probability density. Calculating the average of $D_n$ by summing over cells we obtain
\[
 \langle D \rangle =\frac{1}{N_{\delta}} \sum_{n=1}^{N_{\delta}}
D_n=\frac{1}{N_{\delta}S_{\delta}} \sum_{n=1}^{N_{\delta}}
S_b^{(n)} = \frac {S_b}{S},
\]
the ratio of the  area $S_b$  of the
colored matter and the total area $S$ of the cavity. This value
does not change in the course of the stirring and is the mean or
uniform density $\langle D \rangle$ of the colored blob in the
cavity. (The angle brackets here and later denote an average over the
cavity.) However, using the square density defined by
$
D_n^2 = ({S_b^{(n)}} / {S _{\delta}})^2
$
 and averaging over the
area of the cavity, keeping the cell area constant as before,
 we get the inequality:
\begin{equation}
 \langle D^2 \rangle =
\frac{1}{N_{\delta}} \sum_{n=1}^{N_{\delta}} D_n^2 = \frac{1}{S}
\sum_{n=1}^{N _ {\delta}} D_n S_b^ {(n)}
 \leq \frac{S_b}{S}
\label{D^2}
\end{equation}
because $D_n \leq 1$.

\begin{figure}[tb]
\centering
\subfigure[$ \langle D \rangle$ = $ \langle D^2 \rangle=1/4$]{
\includegraphics[width=0.45\columnwidth]{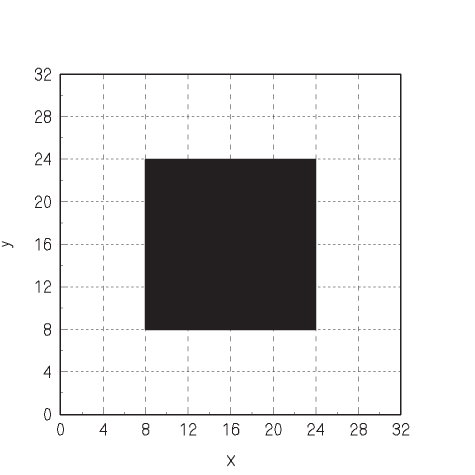}
\label{ris1a}
}
\subfigure[$ \langle D \rangle$ = $ \langle D^2 \rangle=1/4$]{
\includegraphics[width=0.45\columnwidth]{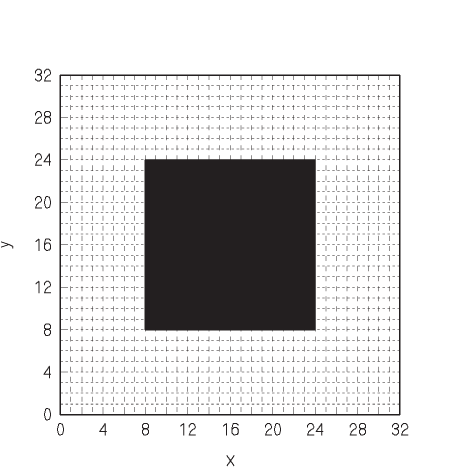}
\label{ris1b}
}
\subfigure[$ \langle D \rangle=1/4$; $ \langle D^2 \rangle=35/256<1/4$]{
\includegraphics[width=0.45\columnwidth]{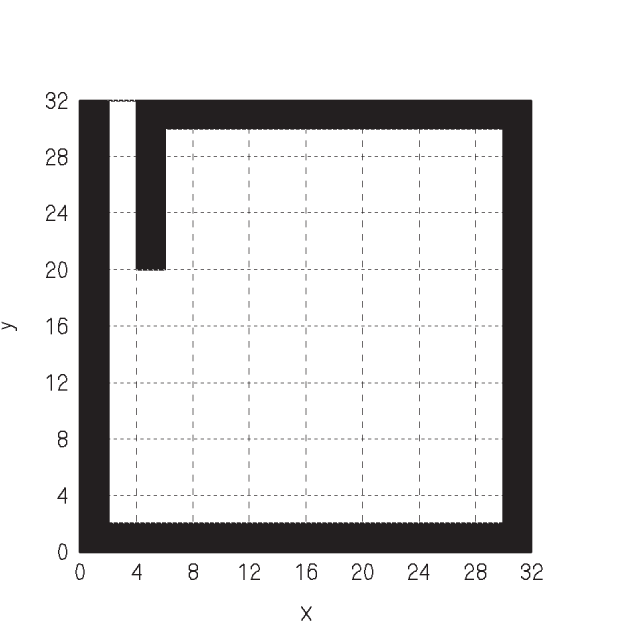}
\label{ris2a}
}
\subfigure[$ \langle D \rangle$ = $ \langle D^2 \rangle=1/4$]{
\includegraphics[width=0.45\columnwidth]{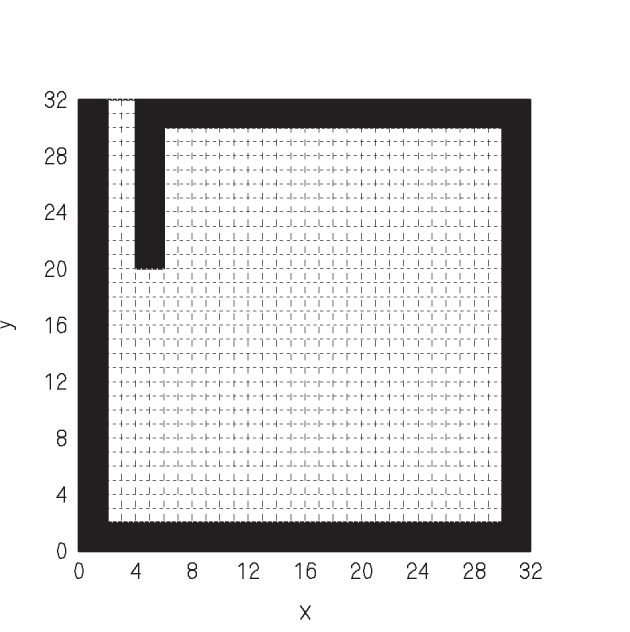}
\label{ris2b}
}
\subfigure[$ \langle D \rangle=1/4$; $ \langle D^2 \rangle=3/40<35/256<1/4$]{
\includegraphics[width=0.45\columnwidth]{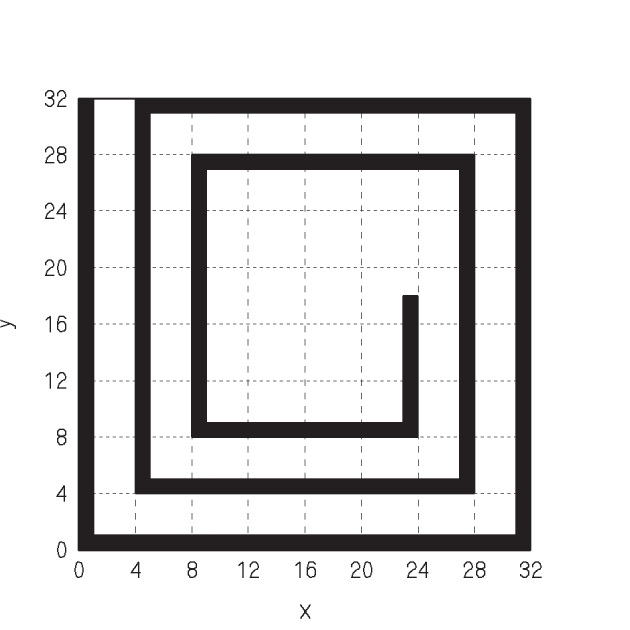}
\label{ris3a}
}
\subfigure[$ \langle D \rangle$ = $ \langle D^2 \rangle=1/4$]{
\includegraphics[width=0.45\columnwidth]{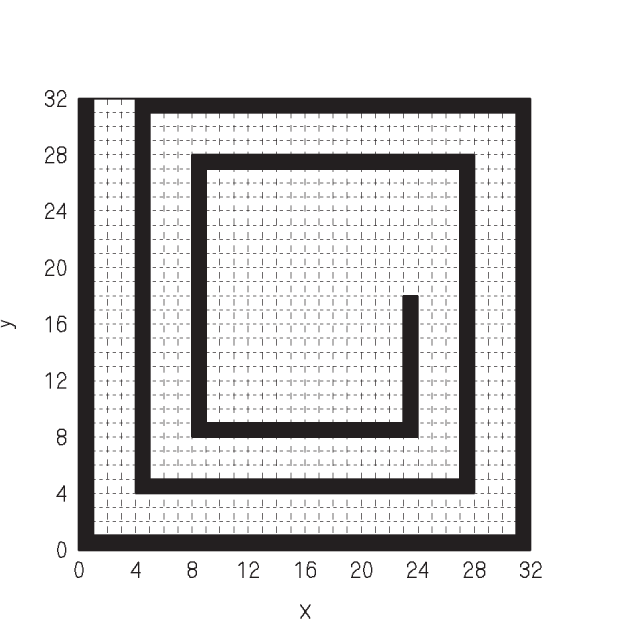}
\label{ris3b}
}
\caption{Square cavity with black blob initially in the center, and after subsequent stirring. The left column has grid of cell size $\delta=4$, while the right column has $\delta=1$. Calculations make explicit the dependence of $ \langle D \rangle$ and $ \langle D^2 \rangle$ on scale. Reprinted from \textcite{Krasnopolskaya2009} with permission from Springer.
}
\label{mkh1}
\end{figure}

Figure~\ref{mkh1} shows a square cavity with a central black blob for two different grid sizes. Calculations given in detail in \textcite{Krasnopolskaya2009} show how the averages $ \langle D \rangle$ and $ \langle D^2 \rangle$ depend on the cell size. In each of Figs~\ref{ris1a} and \ref{ris1b} grid cells are either completely filled or completely empty, giving $D_n = 1$ or $0$ for each $n$. Thus we have $ \langle D \rangle$ = $ \langle D^2 \rangle=1/4$. Figures \ref{ris2a} and \ref{ris2b} show the blob after stirring has taken place. Now in Fig.~\ref{ris2a} $D_n$ can take values 0, 1, 1/2 or 3/4, and we have $ \langle D^2 \rangle=35/256<\langle D \rangle=1/4$, while in Fig.~\ref{ris2b} all cells are either completely filled or empty, so that  $ \langle D \rangle$ = $ \langle D^2 \rangle=1/4$. Similarly in Figs~\ref{ris3a} and \ref{ris3b}, after further stirring, we have  $ \langle D^2 \rangle=3/40<\langle D \rangle=1/4$ when $\delta=4$, and $\langle D \rangle$ = $ \langle D^2 \rangle=1/4$ when $\delta=1$. Thus for $\delta=4$, $\langle D^2 \rangle$ decreases with the thinning of the black striations, while if $\delta=1$, all cells are filled or empty, and $\langle D^2 \rangle$ does not change.  In terms of statistical mechanics, $D_n$ is the 
coarse-grained density, which is different from the 
fine-grained density $f_d$ of
the infinitesimal, super-differential elements $dS_f$, which are always assumed to be small compared to the width of the area of the colored matter. Moreover, $dS_f$ is always so small that it either is located inside the colored matter and $f_d =  1$  or it is outside  and $f_d =0$   \cite{Gibbs1902}.

It was shown by Gibbs, for the special case of mixing two fluids approaching the perfectly mixed state, that the final state
of mixing is characterized by a minimum statistical square
density, i.e., 
\begin{equation}
{\langle (D_n - \langle D \rangle{} )^{2}\rangle} \equiv {\langle
D^2 \rangle}-{\langle D \rangle}^2\rightarrow 0. \label{den}
\end{equation}
 Thus, going
to a uniform mixture in time, the mean square density
 $ {\langle D^2 \rangle}$
will approach its minimum ${\langle D \rangle}^{2}$.  The rate of
decrease of these values is not only time dependent but also
depends on the cell sizes.

It is also possible to use the analogy of entropy, i.e., $-
D_n\,\log{D_n}$, instead of $D_n^2$ as a statistical measure. If
each cell is empty or completely filled, the entropy $-D_n\,\log{D_n}$  equals zero.  The
entropy measure changes only in those
  boxes
where $0<D_n<1$. Moreover, for  $0<D_n<1$,  $-\log{D_n}$ is always
positive, so the more boxes (partially) covered by  colored material the larger
is
 $-\sum_{n=1}^{N_{\delta}} D_n \log {D_n}$.
As a result, for a  good mixing process, the  entropy of the
mixture
$
e = -\langle D\log {D} \rangle
$
will grow in time to its maximum. The entropy measure is not  independent of the square density
measure; both of them  have first-order statistics (one element
of area at a time).

\textcite{Danckwerts1952} defined two properties that are useful in
evaluating the quality of mixing with diffusion and chemical
reactions: the scale of segregation $L_C$ (the measure of the second order statistics) and the
intensity of segregation $I_C$ (the measure of the first order statistics). The scale of segregation is a
measure of the size of
  clumps in a mixture, while the intensity of segregation
refers to the variance in  composition. For the intensity of
segregation Danckwerts introduced the formula
\begin{equation}
I_C=\frac{ \int_{S} (C - {\langle C \rangle} )^{2}\,{d}S}
{{\langle C \rangle} (1-{\langle C \rangle} )S}= \frac{\langle (C
- {\langle C \rangle})^{2}\rangle} {{\langle C \rangle}(1-{\langle
C \rangle})}
\end{equation}
where $C$ is the local concentration, which is in Gibbs' and
Welander's definitions equal to the fine-grained density $f_d$. 
It is easy to see that for the
fine-grained density $\langle(f_d- \langle f_d\rangle)^{2}\rangle
= \langle f_d \rangle - {\langle f_d \rangle}^{2}$, and  in that
case $I_C$ always equals to one. Consequently,
 for mixing without diffusion and chemical
reactions the intensity of segregation $I_C$ is not decreasing,
but equals a constant, initial value.

Therefore, one may suggest  a modification of the intensity of
segregation by making
Gibbs' mean square density Eq.~(\ref{den}) dimensionless
 by dividing by
${\langle D \rangle} (1-{\langle D \rangle})$,
\begin{equation}
I =\frac{\langle (D_n - {\langle D \rangle})^{2}\rangle} {{\langle
D \rangle}(1-{\langle D \rangle})}. \label{inten}
\end{equation}
For  good mixing $\langle (D_n-{\langle D \rangle})^2 \rangle$
 tends to zero, which means that $I$ also tends to zero.
(This definition of $I$ is different from a similar measure
$I_O$ proposed by \textcite{Ottino1989}, who defined $I_O$   as the square
root of the  mean square density divided by ${\langle D
\rangle}^2$, so $I_O^2=\langle (D_n-{\langle D
\rangle})^2\rangle/{\langle D \rangle}^2$.)  For the calculation of
$I$ as proposed in Eq.~(\ref{inten}) and what is in fact the
coarse-grained modification of $I_C$, it is necessary to adopt  an
additional assumption when considering mixing by a set of $N$
points: it is   assumed  that each  of the $N$ points (which
together represent the colored blob) carries   a small undeformed
area equal to $S_b/N$, which cannot be a correct approximation for
continuous mixing flow with large stretching and folding.

The scale of segregation $L_C$ was defined by  \textcite{Danckwerts1952}
by means of the correlation function
\begin{equation}
K_C(\mbox{\boldmath$ \eta$})=\langle (C_{1}- {\langle C \rangle})
(C_{2}-{\langle C \rangle}) \rangle,
\label{cor}
\end{equation}
which shows how the concentration fluctuations $C-{\langle C
\rangle} $ at points 1 and 2, separated by the  vector \boldmath$
\eta $, \unboldmath differ from each other. The normalized
correlation function is called the correlation coefficient
\begin{equation}
\rho_c(\mbox{\boldmath$ \eta$})= \frac{\langle (C_{1}- {\langle C
\rangle})(C_{2}-{\langle C \rangle})\rangle} {\langle (C-{\langle
C \rangle})^{2}\rangle}.
\end{equation}
It is obvious that $\rho(\xvec{0})  =1$. When
$|$\boldmath $\eta$\unboldmath $|$ exceeds a certain value, the
relationship between the concentrations at points 1 and 2 may
become random  when $K_C$(\boldmath$\eta$)  \unboldmath   is equal
to zero. If a mixture consists of clumps,
$|$\boldmath$\eta$\unboldmath$|$ at which $K_C$(\boldmath$\eta$)
 \unboldmath is equal to zero (say,
$|$\boldmath$\eta$\unboldmath$|=\xi$) is approximately the average
clump size in the direction \boldmath$ \eta $. \unboldmath More
precisely, the average clump radius   in the direction of
\boldmath$ \eta $ \unboldmath is
$
L_C(\mbox{\boldmath$ \eta$})=\int_{0}^{\xi}\rho_c(\mbox{\boldmath$
\eta$})d | \mbox{\boldmath$\eta$}|.
$

The mixing  patterns which we are discussing do not consist
of
 a random distribution of clumps,
but of layered structures. However, the coarse-grained
representations
 of these patterns may look like a collection of clumps.
If we indicate cells for which the density
$D_n$ is larger than
 ${\langle D \rangle}$ with black,
  the cells for which $D_n$ equals
   ${\langle D \rangle}$ with gray, and   cells where $0 \leq D_n<
{\langle D \rangle}$ with white, then such a representation has
the appearance of a collection of white and black clumps with gray
clumps in between them as a transition. Moreover, with
 the coarse-grained correlation function
defined as
$
K(\mbox{\boldmath$ \eta$})=\langle (D_{1}- {\langle D \rangle})
(D_{2}-{\langle D \rangle}) \rangle,
$
 (where
$D_1$ and $D_2$ correspond to coarse-grained density in the boxes
1 and 2 separated by vector  \boldmath$ \eta $\unboldmath) the
short-term regularity
 (when $K>0$)  in
the interval $(0,\xi)$
  gives  important information about  the mixture
pattern and can be examined.
 Short-term regularity means that on average in two boxes
at any distance $|$\boldmath$\eta$\unboldmath$|<\xi$ the
fluctuations $D_n-{\langle D \rangle}$
  have the same sign (i.e., the same color) and thus $K>0$.
For $|$\boldmath$\eta$\unboldmath$|=\xi$ the fluctuations become
uncorrelated and therefore $K=0$.
 Thus, the distance
$|$\boldmath$\eta$\unboldmath$|=\xi$
 in the direction
\boldmath $ \eta $ \unboldmath is related to the average clump
size in this direction, and the value of the scale of segregation
\begin{equation}
L(\mbox{\boldmath$ \eta$})=\int_{0}^{\xi} \frac{\langle (D_{1}-
{\langle D \rangle})(D_{2}-{\langle D \rangle})\rangle}
{\langle(D-{\langle D \rangle})^{2}\rangle}\, {d}
|\mbox{\boldmath$\eta$}| \label{scale}
\end{equation}
gives  the average radius of the  clump. Complementary to
intensity of segregation $I$, the scale of segregation $L$ can be
used as a measure of clump sizes of the coarse-grained description
of mixing patterns.
 The dynamics of such scales should reflect the changes of sizes of
unmixed  regions, where $D_n$ is always larger than ${\langle D
\rangle}$.

\subsection{Gibbs' example of mixing}\label{sec:gibbs}

\begin{figure}[tb]
\centering
\includegraphics[width=0.6\columnwidth]{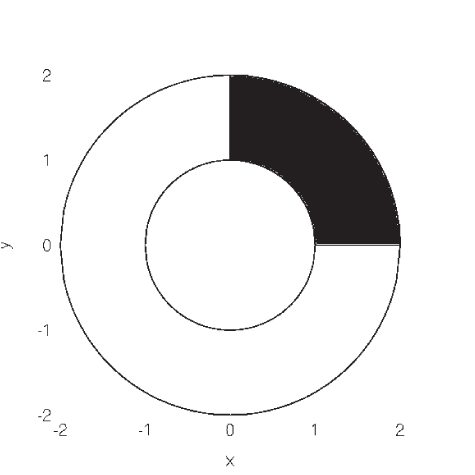}
\caption{Gibbs' example: The geometry of the Gibbs problem,
with viscous fluid occupying the annular region between two
co-axial cylinders The outer cylinder is rotated at a constant
speed ($V_2$), while the inner cylinder is kept at rest. Initially,
the dyed fluid (black) occupies a $90^{\circ}$ sector.}
\label{mkh4}
\end{figure}

In his example of fluid stirring, \textcite{Gibbs1902} described the case
of the advection of colored fluid in the annular domain between two
infinite coaxial cylinders rotating at different speeds.
 The cross-sectional area between the cylinders is shown
 in Fig.~\ref{mkh4}, with the dyed (black) fluid occupying a $90^{\circ}$ sector
 between the cylinders.
The flow is driven by the rotation of  the outer cylinder: the
tangential  velocity is $V _ {2} \neq 0 $ at $r=2$, while the
inner cylinder is kept fixed ($V _ {1}=0 $ at $r=1$). This forcing
drives a purely azimuthal flow, with the azimuthal velocity
component given by \cite{Krasnopolskaya2004,Krasnopolskaya2009}
\begin {equation}
u _ {\theta} = Ar+ {B \over r},\quad A={V_{2}b \over{b^2-a^2 }},
\quad B=-{V_{2}a^2b \over{b^2-a^2 }}.
\label {vel1}
\end {equation}
 For the case under consideration $a=1$ and $b=2$, therefore, $A=2/3V_{2}$ and $B=-2/3V_{2}$.
The advection equations are
$ {dr}/{dt} = u_r=0$, 
$ r{d\theta}/{dt} = u _ {\theta} $,
and with initial conditions
$r=r _ {in}, \; \theta =\theta _ {in} $ at $t=0$,
these equations describe the motion of a passive Lagrangian particle in a position   $(r, \theta) $ at time $t$ in the known  Eulerian velocity field \boldmath$
v $ \unboldmath $=$ $(u_r,u _ {\theta})$.

\begin{figure}[tb]
\centering
 \subfigure[]{\includegraphics[width=0.6\columnwidth]{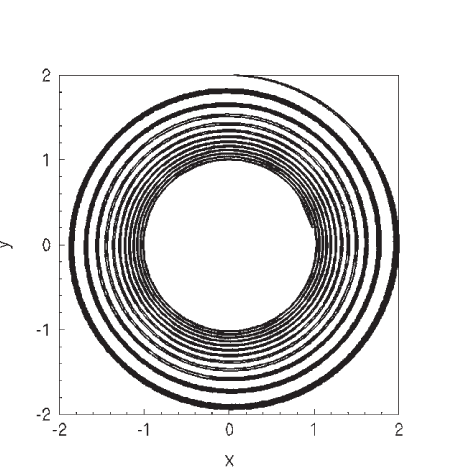}}
 \subfigure[]{\includegraphics[width=0.6\columnwidth]{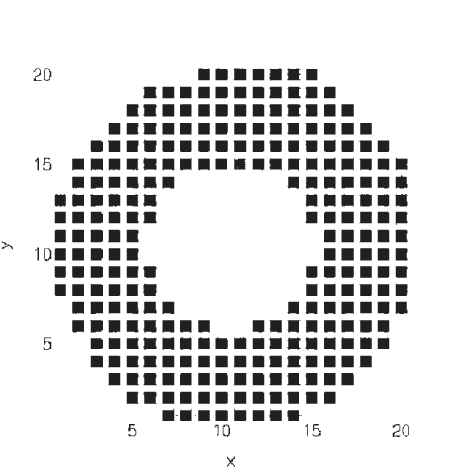}}
 \subfigure[]{\includegraphics[width=0.6\columnwidth]{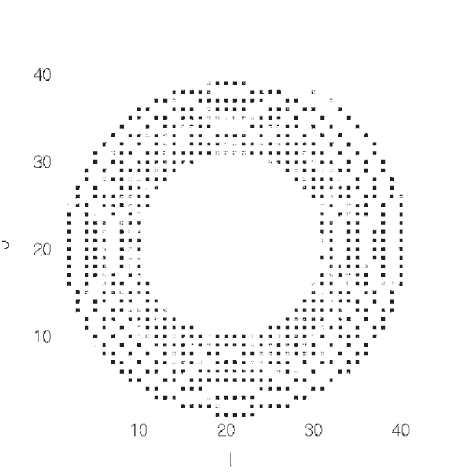}}
\caption{Gibbs' example: The evolution of the colored fluid (a)
after 12 revolutions of the outer cylinder and the corresponding
square density distribution ($ {\langle D^2 \rangle} $) for (b)
$\delta=0.2$   and (c) $\delta=0.1$. Reprinted from \textcite{Krasnopolskaya2009} with permission from Springer.
} \label{mkh5}
\end{figure}

The advection of the colored fluid may be studied by using the contour
tracking method \cite{Krasnopolskaya1996}, in which
the interface between colored and uncolored fluid is covered by a number
of passive tracer particles. After 12 complete revolutions of the outer
cylinder the colored fluid occupies a spiral-shaped region as shown
in Fig.~\ref{mkh5}(a). For continued rotation of the outer
cylinder, the number of windings of the spiral pattern increases
steadily: for an infinite number of revolutions the number of
windings would be infinite and the mixture uniform.

We now apply the square density and intensity of
segregation measures  to quantify the mixing quality. Figure~\ref{mkh5}(b)
replicates Fig.~\ref{mkh5}(a), but now in the form
of the square density ($D_n$) distribution for a cell size
$\delta=0.2$. A similar pattern is shown in Fig.~\ref{mkh5}(c),
but for $\delta=0.1$. Comparison of the latter two graphics shows
that for a coarser grid (larger cells) the mixture looks uniform,
while for a finer grid (smaller cells) the mixing is non-uniform,
with empty (white) cells occurring in between the black cells.

\begin{figure}[tb]
\centering
\includegraphics[width=\columnwidth]{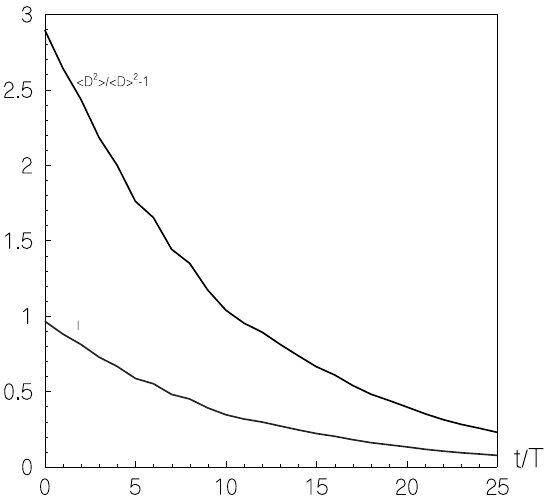}
\caption{GibbsÕ example: Calculated evolution of the intensity of
segregation $I$ and the square density $ {\langle D^2
\rangle}/{\langle D \rangle}^2 -1 $ as a function of time t/T,
with $T$ the time of a half revolution of the outer cylinder. Reprinted from \textcite{Krasnopolskaya2009} with permission from Springer.
}
\label{mkh6}
\end{figure}

The evolution of the intensity of segregation $I$ 
as a function of $t/T$ (with $T$ the time of a half revolution of the outer cylinder) as calculated
for $\delta=0.1$ and the
 square density $ {\langle D^2
\rangle}/{\langle D \rangle}^2 -1 $   is shown in
Fig.~\ref{mkh6}. It is observed  that the mixture becomes more
uniform with increasing number of revolutions. In this case the
intensity of segregation $I$ is more convenient as a criterion
because it changes between two fixed values: it decreases from $1$
(initial state) to $0$ (complete mixing).

\subsection{A second example: mixing in a wedge cavity}\label{sec:second}

\subsubsection{Definition of the problem}

\begin{figure}[tb]
\centering
\includegraphics*[width=\columnwidth]{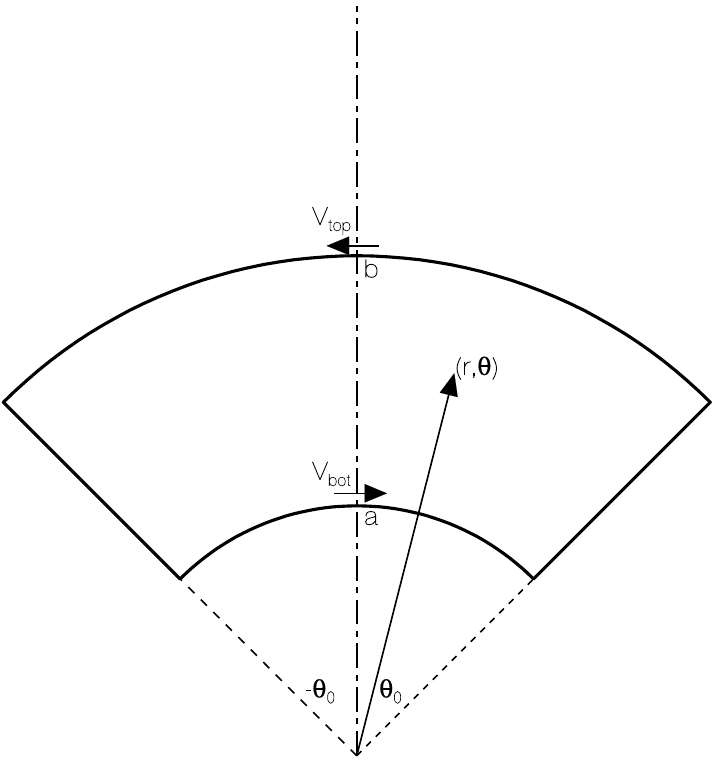}
\caption{Wedge cavity flow: geometry.
}
\label{mkh7}
\end{figure}

As a second example, we consider a 2D creeping flow
of an incompressible viscous fluid in an annular wedge cavity,
$a\leq{r}\leq{b}$, $|\theta|\leq{\theta_{0}}$ (Fig.~\ref{mkh7}),
driven by periodically time-dependent tangential velocities
 $V_{bot}(t)$ and $V_{top}(t)$
at the curved bottom and top boundaries $r=a$ and $r=b$,
respectively. The side walls,  $a\leq{r}\leq{b}$, $|\theta| =
\theta_{0}$ are fixed. We consider a discontinuous mixing protocol
with the bottom and top walls alternatingly rotating over an angle
$\Theta$ in clockwise and counterclockwise directions,
respectively. More specifically, we consider the case
$
V_{bot}(t) =  {2a \Theta / T}$,  $V_{top}(t) = 0$,
for $kT < t \leq \left( k + {1 / 2} \right) T;
$
$
V_{bot}(t) = 0$,  $V_{top}(t) = - {2b \Theta / T};$
for 
$
 \left( k + {1 / 2} \right) T < t \leq (k + 1)
T$, 
where $k=0,1,2,\ldots$.   $\Theta$ is the angle of wall rotation  and
$T$ is the  period of the wall motion.

The radial  and azimuthal  velocity components  $u_r$ and
$u_{\theta}$ can be expressed by means of the stream function
$\Psi(r,\theta,t)$ as
$
u_{r}={1 / r}\,{\partial{\Psi} / \partial{\theta}},
$
$
u_{\theta}=-{\partial {\Psi}/ \partial r}.
$
For a quasi-stationary creeping flow in the Stokes
regime 
the stream function $\Psi$
satisfies the biharmonic equation
$
\nabla^{2}\nabla^{2}\Psi = 0,
$
with $\nabla^2$ the Laplace operator 
and the boundary conditions
$
\Psi= 0,$ $\partial\Psi/\partial r=-V_{bot}$ at $r=a,$ $|\theta| \leq \theta_{0}$;
$
\Psi= 0,$ $\partial\Psi/\partial r=-V_{top}$ at $r=b,$ $|\theta| \leq \theta_{0}$;
$
\Psi= 0,$ $\partial\Psi/\partial {\theta}=0$ at $a\leq{r}\leq{b}$, $|\theta|=\theta_{0}.
$ 
Therefore,
we have the classical biharmonic problem  for the stream function
$\Psi$
with prescribed values of this function
and its outward normal derivative at the boundary. 
This wedge-cavity flow problem was solved analytically by
\textcite{Krasnopolskaya1996}. Their analytical solution was used
for the numerical evolution of the interface line between the
marker fluid and the ambient fluid, which was carried out by the
`dynamical' contour tracking algorithm.

\subsubsection{Wedge cavity flow}

The results discussed here correspond to a typical wedge cavity
with sector angle  $\theta_0=\pi/4$ and radius ratio $b/a=2$ (see
Fig.~\ref{mkh7}). Using the dimensionless parameter
$H={\Theta}/{\theta_0}$ and a
 fixed value for the period $T_p$, the discontinuous  mixing
 protocol is completely defined.
 The value of $H$
 is proportional to the displacement of
 the top and bottom
walls during one period. Multiplied by the number of periods $M$,
 the value $HMT_p$  serves as a
 measure of the  energy supplied during the mixing process.
In what follows we restrict to the  cases $H=2$ and $4$.

\begin{figure}[tb]
\centering
\includegraphics*[width=\columnwidth]{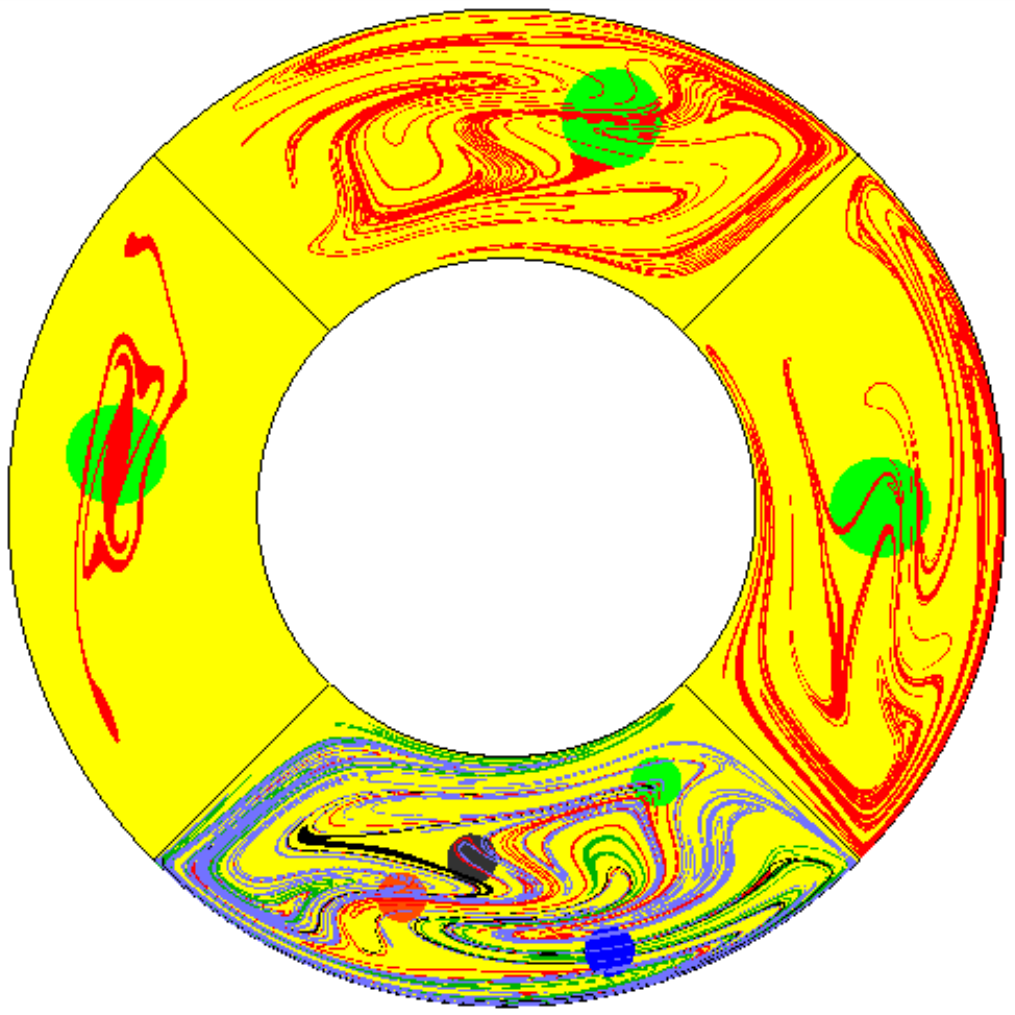}
\caption{Wedge cavity flow: Mixing patterns obtained in the wedge-cavity flow
for different initial positions of the tracer blobs (colored/shaded
circles) and different wall displacements $H$. Upper, lower, and
right cavities show the results for the same protocol $H = 4$ and $M
= 12$ for different initial positions of the blob. Upper: blob
initially placed around a single hyperbolic point; right: for a
slightly different initial position; lower: blob split up into
four smaller blobs, located in different positions. Left cavity:
initial blow positioned around an elliptic point of period-1, for
$H = 2$. In all cases the energy input was $3HM$. 
Adapted from \textcite{Krasnopolskaya1999} with permission from Elsevier.
}
\label{mkh8}
\end{figure}

Figure~\ref{mkh8} shows  a composite picture of four cavities
 with different mixtures, obtained by numerically
tracking the contours of initially circular blobs for different
mixing protocols or initial positions. The colored circles
represent the initial position of the blobs. The upper, lower, and
right cavities represent the results of mixing with the same
periodic protocol with  $ H = $ 4 after 12 periods ($ M $ = 12).
The left cavity, in which the initial blob  is spread slightly
throughout the cavity region, demonstrates the result of mixing
for $ H = 2$. Despite the fact that for this specific case the
mixing
 process  was performed  for twice as long,
i.e., during 24 periods ($ M $ = 24) so that the
general work $ W = {\Theta} (a + b) M / {(\theta_0 a)} = 3HM $
for all the cavities is the same, the results of the mixing in
the four cases are
 different. The best mixing corresponds to the case
represented in the lower cavity, where the initial spot is divided
into four small round spots, placed in different parts of the
cavity. The upper picture corresponds to the mixing of one blob
placed initially around a single hyperbolic periodic point of
period-1 
for the protocol with $ H = 4 $ \cite{Krasnopolskaya1999,Krasnopolskaya2004}. Since the
selected periodic point is hyperbolic, the unstable
manifold passes through it, and points in its vicinity are characterized
by chaotic trajectories, the initial green blob becomes fairly uniformly
distributed throughout the cavity. The right cavity gives the
result of mixing with the same protocol  as in the upper  cavity,
but with a slightly different  initial position of the blob. In
this case, the spreading of the blob over the cavity is less
efficient. The left cavity shows the worst mixing result:
 the  initial position of the spot was chosen around
elliptic points of period-1 (for a single protocol $ H = 2 $).  Under this protocol there is one
elliptic period-1 point.
An initial blob placed around an elliptical point
results in very poor mixing. The areas around the elliptic points
are only slightly deformed during the mixing protocol, returning
all material in the area around the elliptic point \cite{Meleshko1996_2}.

\subsubsection{First-order measures of mixing}

\begin{figure*}[tb]
\centering
\includegraphics*[width=1.5\columnwidth]{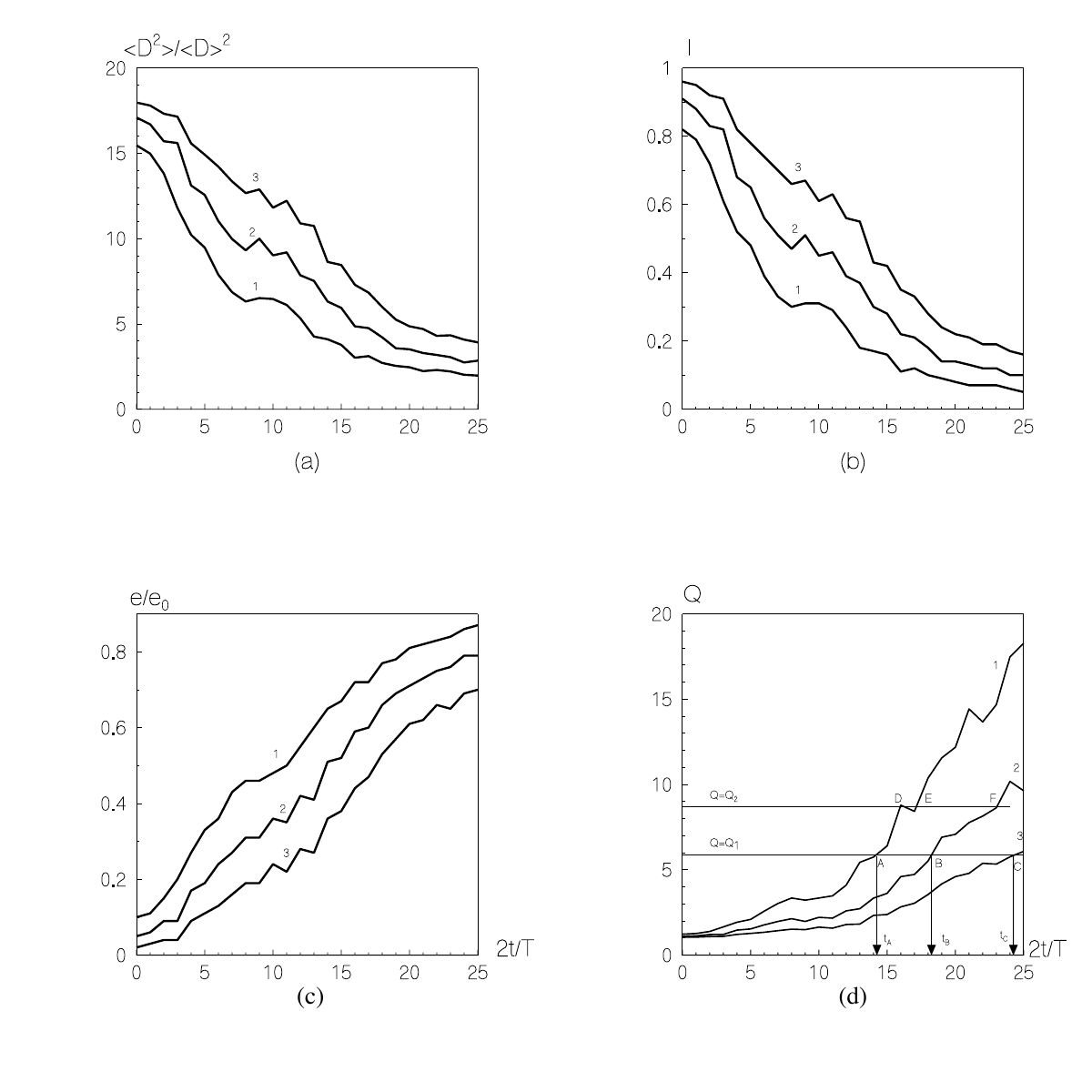}
\caption{Wedge cavity flow: Evolution of (a) the square density; (b) intensity
of segregation; (c) entropy; (d) quality of mixing. The initial
blob is  that shown as a green circle in the upper cavity of
Fig.~\ref{mkh8}. 
Reprinted from \textcite{Krasnopolskaya1999} with permission from Elsevier.
}
\label{mea-dyn}
\end{figure*}

For the case shown in the upper cavity of Fig.~\ref{mkh8}, i.e.
with the blob initially located around a hyperbolic point, we
consider the statistical measures of
the quality of the mixed state. For this purpose  the area of the
cavity is covered  with cells of  uniform size.
 The evolution  of the three criteria based on the coarse-grained
density $D$ is shown in Fig.~\ref{mea-dyn} for different cell
sizes. The mixing process is characterized by a decrease of
square density $\langle D^2 \rangle/\langle D \rangle^2$
 (Fig.~\ref{mea-dyn}(a)), by a decrease of
 the intensity of segregation $I$
(Fig.~\ref{mea-dyn}(b)), and by an increase of the  entropy
$e^2/e_0^2$ (Fig.~\ref{mea-dyn}(c)). The three curves in each
figure correspond to three different cell sizes, the labels 1, 2
and 3 corresponding to $\delta = 0.1a$, $\delta = 0.05a$ and
$\delta = 0.025a$, respectively. From Figs~\ref{mea-dyn}(a,b)
it follows that the intensity and the square density show the same
evolution but over different scales: the square density has values approximately in the range
between 20 and 2; and the intensity is in the range between 1 and 0.
As the intensity always lies in the same range (0,1),
using this criterion
 it is possible to compare  mixing processes for different values of the
 ratio
  $S_b/S$ and  to compare different  mixtures with the same ratio
$S_b/S$. For instance, it is easy to answer the  question after
how much time of mixing  the intensity of segregation will have
some given value for different box  sizes (which basically
represents the problem of scaling in mixing processes), by
drawing a horizontal line: $I=const$. By stating that a mixture
is uniform enough, for a given box size,  when $I$ is less than
some minimum value $I_{min}$, we know how long we have to proceed
with the mixing for other box sizes, i.e., for differently sized mixers. For
example, if $I_{min}=0.05$, then, for the box size $\delta=0.1a$
(curve 1,
 Fig.~\ref{mea-dyn}(b)) the mixture is uniform for $t/T>25$ according to this
requirement.

Alternatively, we may use the entropy
 (Fig.~\ref{mea-dyn}(c)) or the quality $Q$, defined as the reciprocal of
the intensity $I$ (see Fig.~\ref{mea-dyn}(d)), which both
increase during the mixing process. For
the same cell size $\delta$, the same value of the mixing quality
(or intensity) can be repeatedly reached. The quality
of a mixture can decrease for some time during the mixing process,
after which it increases again.

\begin{figure}[tb]
\centering
\includegraphics*[width=\columnwidth]{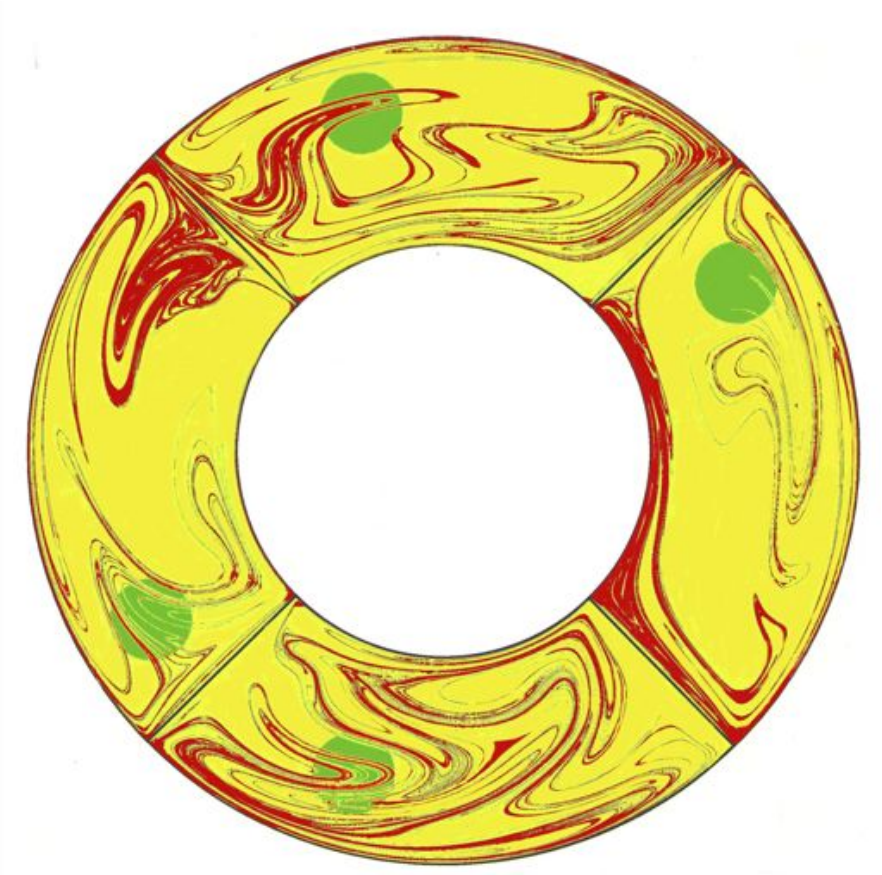}
\caption{Wedge cavity flow: Mixing patterns generated by the same  protocol for
different initial locations of the circular blobs. In the left and
right cavities the stretching is large; the upper cavity shows
large stretching and relatively good mixing; the bottom cavity
shows poor
 stretching combined with best mixing.
 Adapted from \textcite{Krasnopolskaya1999} with permission from Elsevier.
}
\label{mkh-stret}
\end{figure}

Figure~\ref{mkh-stret} shows the results of mixing for the same
protocol with $H = 4$  in four equal wedge cavities for
different initial locations of the blobs (green circular patches).
 The left and right cavities have  the initial blob  centered around a hyperbolic
 point of
period 2 and covering three hyperbolic points of period 6 and one
of period 4.  For this case the stretching of the contour (length
of interface) is the largest, while the distribution of the
deformed blob (indicated by the red color) over the cavity is
poorest. The graphs show the distribution of the blob material
after 18.5
 (right cavity) and 19 half-periods (left). The  length of the contour
 of the initial green circle  is extended 1760 (right) and 2010 (left) times, respectively.
Visual inspection shows, however, that large parts of the cavity,
in particular in the central area, are not covered by the red blob
material. The distribution of the colored fluid over the cavity
domain can be quantified, for example, by the intensity of
segregation $I$; for box size $\delta = 0.025a$ we find $I = 0.21$
(right) and $0.20$ (left), respectively. For the calculation shown
in the upper cavity, the initial position of the blob was chosen
around the other hyperbolic points of period 2. Stretching of the
contour line in this case was almost as large as in previous
cases:  the line stretched with a factor of 1850. However,
the quality of the mixture is much better, as  there are no large
uncovered parts. In this case the intensity of segregation  for
box size $ \delta = 0.025a $ has a value of $ I = 0.17 $.

The bottom cavity in Fig.~\ref{mkh-stret} shows a calculation
for
 the initial position of the blob around a
 hyperbolic point of period 1. Stretching of the contour line of
 the
 initial green blob
 is lower than in the previous cases with a factor of only
986.
  However, the
 mixture in the bottom  cavity
 shows that the red color was distributed over all the  subregions of the cavity,  having an
  intensity of segregation value of
$ I = 0.15 $.
The greatest stretching of the length of the initial contour  does not guarantee
the best quality of mixture.

\subsubsection{Second-order measures of mixing}

\begin{figure*}[tb]
\centering
\includegraphics*[width=1.5\columnwidth]{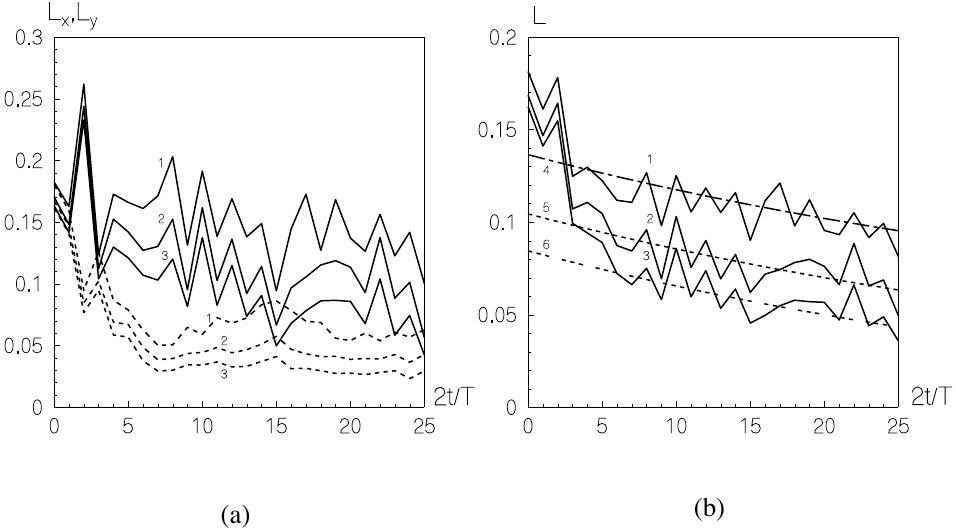}
\caption{Wedge cavity flow: Evolution of the scales of segregation (a) $L_x$ and
$L_y$, (b) the averaged scale  $L$ (solid) with the approximation
curves, 4 ($c_i=0.03$), 5 ($c_i=0$) and 6 ($c_i=-0.02$),
$c_1=1/50$ and $c_2=113$. Reprinted from \textcite{Krasnopolskaya1999} with permission from Elsevier.
}
\label{mkh-second}
\end{figure*}

Beyond first-order mixing measures there are second-order mixing measures that indicate the relative size of unmixed `rubbery' domains.
Criteria like those could be, in 
practical terms, more relevant than those of the first-order
statistics dealt with so far. In Fig.~\ref{mkh-second}(a) the
evolution of the scale of segregation $L$ in the two directions  $
x $ and $y$
 is shown for the same
mixing process as was used for calculations of the first-order
statistics (Fig.~\ref{mea-dyn}) for the same three box sizes
$\delta$. Initially, the scales in the    $ x $ direction (solid
lines) and $y$  direction (dashed lines) are almost equal and give
the approximate value of the radius of the initial blob
($R=0.2a$). For box counting with $\delta = 0.1a$ (curves 1 in
Fig.~\ref{mkh-second}(a)) the error of the value of $R$ is about
9\%; for cell size $\delta = 0.05a$ (curves 2 in
Fig.~\ref{mkh-second}(a)) it is slightly larger, just like for
$\delta = 0.025a$ (curves 3 in Fig.~\ref{mkh-second}(a)).
 In the course of time, owing to the anisotropy of the patterns (see
 Fig.~\ref{mkh-stret}), the scales of segregation in the   $ x $ direction
 $L_x= L$(\boldmath $x$\unboldmath)$/a$ and in the  $ y
$ direction
 $L_y= L$(\boldmath $ y$\unboldmath)$/a$ diverge. Nevertheless, both  have a tendency to
decrease, but not uniformly in time. In Fig.~\ref{mkh-second}(b)
the dependence of the averaged scale
 $L=(L_x+L_y)/2$
on the  number of half-periods $2t/T$ is presented for  the three
 box sizes. After the two first
periods of mixing, the curves can be approximated by exponential
functions of the form $c_i+ \exp(-c_1(2t/T+c_2))$.
 These functions are drawn in
Fig.~\ref{mkh-second}b as dashed lines 4, 5 and 6, respectively. 
Using these approximations we can roughly estimate after how many
 periods the averaged scale of segregation $L$ will be smaller than
some given value. For example, for  box size $\delta=0.05a$, $L$
becomes less than $\delta/2$ (the unmixed `rubbery' domain is
smaller than the area of the box)
 after approximately 35 periods. To find the value of periods $(t/T)$
we use the expression for curve 5 (for the chosen $\delta = 0.05a$), which
approximates the value of $L$, and subsequently solve the equation $\exp(-c_1(2t/T +c_2))< \delta /2$. This
shows that $2t/T$ should be larger than 70.

Thus, the existence and the evolution of the unmixed `rubbery'
domains in the mixture pattern can be determined from the
behavior of the scales of segregation. For one unmixed zone, as in
the initial situation with one circular blob, the scale $L$, as
defined in Eq.~(\ref{scale}), directly gives the size of the blob.
 For  well-mixed patterns of high quality,
 the non-zero $L$ values  indicate the existence of unmixed domains that
very slowly decline (or even
 stabilize, when the zone does not diminish in size at all)
 despite the continuing mixing.

The  evolution
of the first-order statistics measures (square density $\langle D^2 \rangle$, intensity of
segregation $I$, entropy $e$ and quality $Q$) are different from the dynamics of
the scale of segregation. The first ones  reflect the distribution
of filaments over the mixing domain (for the uniform mixture the coarse-grained density should be
less than 1 and larger than 0 all over domain, so filaments should be everywhere),
while  the latter
shows the behavior of the unmixed domains of the coherent structure
 \cite{Danckwerts1952}.
 They also change in  opposite ways when changing
 the box size.
 The smaller the box size $\delta$,
the worse is the mixing according to the first measures and the better
according to
the scale measures, which decrease with
decreasing cell size $\delta$.
Therefore, it is necessary to use both measures
 to judge how  well
or badly  materials are mixed.

\subsection{Mix-norms}\label{sec:mixnorms}

The issue with applying the above mixing measures has always been the arbitrariness of the  choice of scale.  But recent proposals for multi-scale mixing measures get around this issue, and in this section we discuss a multi-scale mixing measure originally introduced by \textcite{Mathew2005}.  While the multi-scale measure does not require diffusion to represent the amount of homogenization, it is convenient to begin the discussion by recalling the advection--diffusion equation 
$
\pd{C}{t} + \bu \cdot \bnabla C = \kappa \nabla^2 C,
$
where $C$ is a concentration field in a finite domain $\Omega$,
with no-net-flux boundary conditions. We assume without loss of
generality that
\begin{equation}
\int_{\Omega} C \intd \Omega = 0,
\end{equation}
and define the $L^2$-norm, or variance, as
\begin{equation}
\|C\|_2^2 = \int_{\Omega} C^2 \intd \Omega.
\end{equation}
The variance evolves according to
\begin{equation}
\label{eq:variance}
\td{}{t} \|C\|_2^2 = -2\kappa \|\bnabla C\|_2^2,
\end{equation}
and so decays in time as the system mixes. The variance indicates the extent
to which the concentration has homogenized and is thus a good measure of the
amount of mixing that has occurred.  However, a full computation of variance
requires knowledge of small scales in $C$, which we are not necessarily
interested in.  It would be better to use a measure that downplays the small
scales.  This is more in keeping with the definition of mixing in the sense of
ergodic theory~\cite{Lasota}.  We thus proceed to consider the pure advection
equation
\begin{equation}
\label{eq:advection}
\pd{C}{t} + \bu \cdot \bnabla C = 0.
\end{equation}
Note that in this case Eq.~\eqref{eq:variance} predicts that the variance
satisfies
\begin{equation}
\td{}{t} \|C\|_2^2 = 0,
\end{equation}
and cannot therefore be used as a measure of mixing.

The advection equation \eqref{eq:advection} returns us to the ergodic sense of mixing of Eq.~\eqref{eq:mixing}. Consider the advection due to the velocity field to be a time-dependent operator $S^t:\Omega \to \Omega$ that moves an initial patch of dye according to
\begin{equation}
C_0(\bx) \mapsto C(\bx,t) = S^t C_0(\bx).
\end{equation}
If we consider a region $A$ of uniform concentration defined by
\begin{equation}
C_0(\bx) = \left\{ \begin{array}{ll}
1 & \textrm{if $\bx \in A$}, \\
0 & \textrm{otherwise},
\end{array} \right.
\end{equation}
then the volume of the region $A$ remains constant in time by incompressibility, and can be associated with the Lebesgue measure $\mu(A)$. For a flow enjoying the property of strong mixing Eq.~\eqref{eq:mixing} is satisfied, with $f$ representing a stroboscopic map of the operator $S^t$. 

The intersection of the advected patch~$B$ with the reference patch
$A$, as in Fig.~\ref{fig:mixpatch} is analogous to projection onto~$L^2$ functions. This motivates
the following \emph{weak convergence} condition
\begin{equation}
\label{eq:weak_convergence}
\lim_{t\to\infty} \langle C(\bx,t), g \rangle = \bar g,
\end{equation}
for all functions $g \in L^2(\Omega)$ with spatial mean~$\bar g$, where the
inner product is defined by
\begin{equation}
\langle h, g \rangle = \int_{\Omega} h(\bx) g(\bx) \intd \Omega,
\end{equation}
and $h \in L^2(\Omega)$ if $\int_{\Omega} |h|^2 \intd \Omega < \infty$. Weak
convergence is equivalent to mixing as a consequence of the Riemann--Lebesgue
lemma. The equivalent conditions Eq.~\eqref{eq:mixing} and
Eq.~\eqref{eq:weak_convergence} require computing over all patches $A$ or
functions $g$, respectively. Thus, neither of these conditions, taken in isolation, is very useful
in practice. However, we now describe a theorem that shows that there is a
simpler way to determine whether or not weak convergence is satisfied.

\begin{figure}[tb]
  \centering
  \includegraphics[width=\columnwidth]{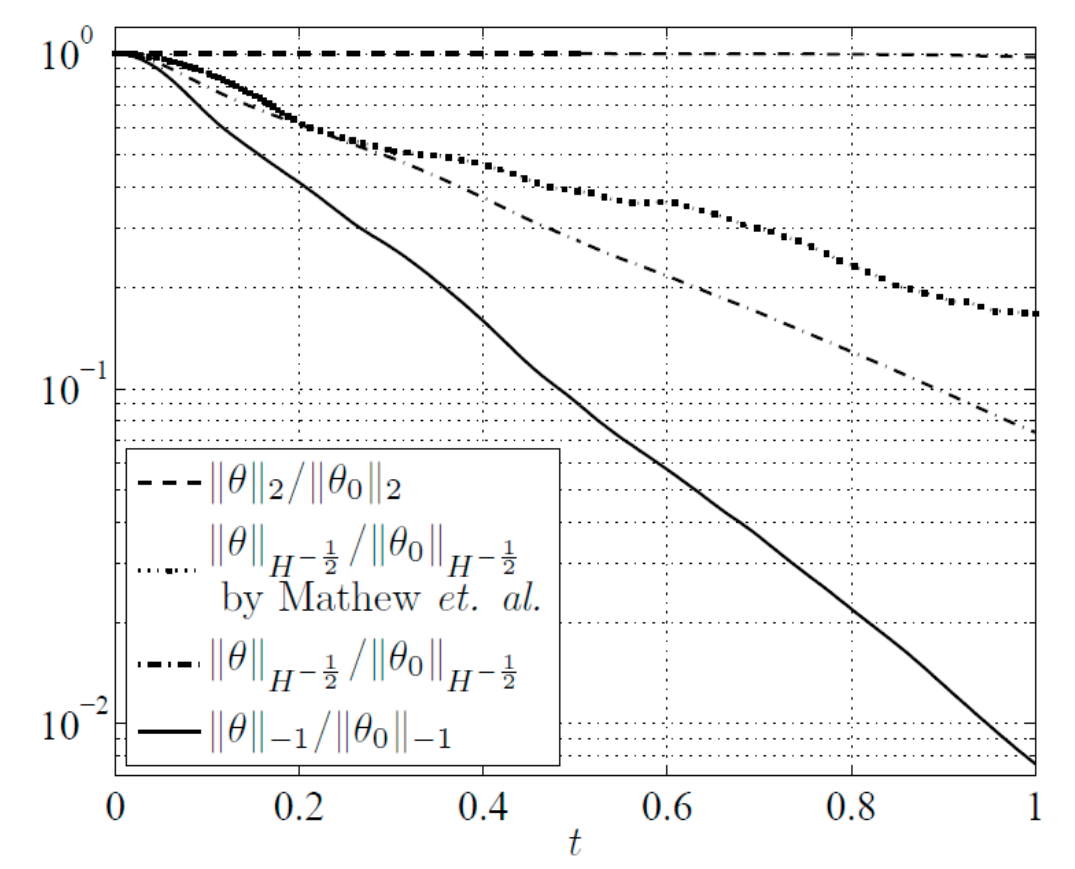}
  \caption{Mix-norms: Comparison for a flow optimized using the
    separate methods of optimal control and optimal instantaneous
    decay.     Reprinted with permission from \textcite{lin2011optimal}. Copyright (2011) Cambridge University Press.
}  \label{fig:mixnorms}
\end{figure}

\begin{figure}[tb]
  \centering
  \includegraphics[width=\columnwidth]{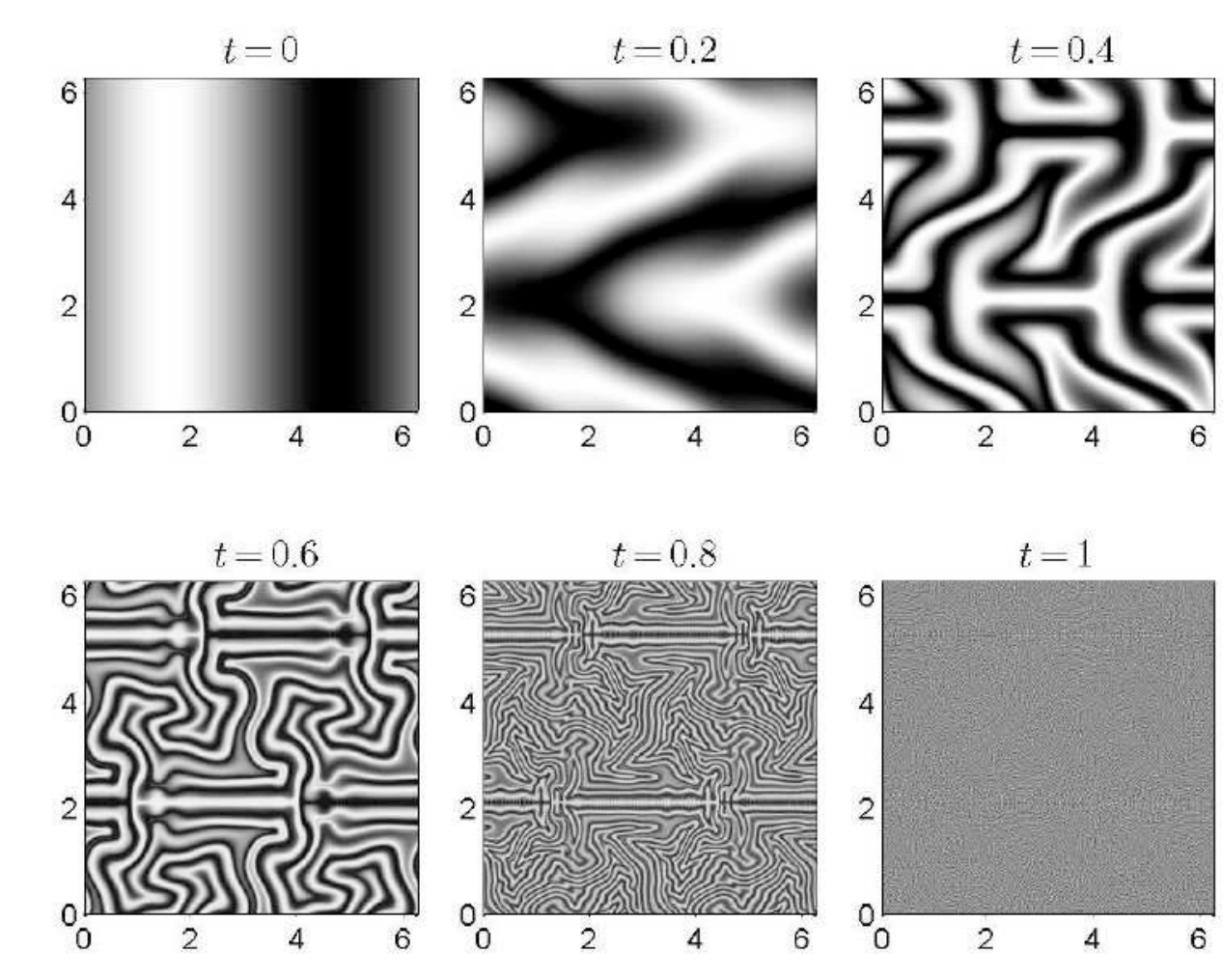}
  \caption{Evolution of the concentration field for the flow optimized
    in the case $q=-1$. 
    Reprinted with permission from \textcite{lin2011optimal}. Copyright (2011) Cambridge University Press.
}  \label{fig:optimized_flow}
\end{figure}

\textcite{Mathew2005} introduced the \emph{mix-norm},
which for mean-zero functions is equivalent to
\begin{equation}
\|C\|_{\dot{H}^{-1/2}} := \|\nabla^{-1/2} C\|_2\,.
\end{equation}
\textcite{DoeringThiffeault2006,lin2011optimal,Thiffeault2012} generalized the mix-norm to
\begin{equation}
\|C\|_{\dot{H}^q} := \|\nabla^q C\|_2,\qquad q<0,
\end{equation}
which formally is a negative homogeneous Sobolev pseudo-norm. This norm can be
interpreted for negative $q$ via eigenfunctions of the Laplacian
operator. For example, in a periodic domain, we have
\begin{equation}
\|C\|_{\dot{H}^q}^2 = \sum_{\bk} |\bk|^{2q} |\hat{C}_{\bk}|^2,
\end{equation}
from which we see that, for $q<0$, $\|C\|_{\dot{H}}^q$ smooths
$C$ before taking the $L^2$ norm. The theorem
\begin{equation}
\lim_{t\to\infty} \|C\|_{\dot{H}^q} = 0, \quad q < 0 \Longleftrightarrow C \textrm{ converges weakly to } 0,
\end{equation}
due to \textcite{Mathew2005,lin2011optimal,Thiffeault2012} shows that we can track any mix-norm
to determine whether a system is mixing. The existence of this quadratic norm
facilitates optimization of the velocity field to achieve good mixing. \textcite{Mathew2007} have used optimal
control to optimize the decay of the $q=-1/2$ mix-norm. \textcite{lin2011optimal} have optimized the instantaneous decay rate of the
$q=-1$ norm using the method of steepest descent, which is easier to compute
numerically but yields suboptimal, but nonetheless very effective, stirring
velocity fields. A comparison of the methods for optimized mixing is shown in
Fig.~\ref{fig:mixnorms}.  The solid line decays faster, but this is merely
because $\dot{H}^{-1}$ cannot be compared directly
with $\dot{H}^{-1/2}$. The corresponding evolution of the concentration field
for the case $q=-1$ from \textcite{lin2011optimal} is shown in Fig.~\ref{fig:optimized_flow}.  For a complete review, see \textcite{Thiffeault2012}.

\subsection{Rates of mixing}\label{rates}

Many applications may require the mixing of a particular scalar field, for example concentration or temperature. This problem leads to a natural reformulation of Eq.~(\ref{eq:mixing}) in functional form. Such an expression can then be used to ask  how quickly a mixing system achieves its goal. This procedure is a common one in the study of ergodic theory of dynamical systems, in abstract situations apparently unconnected with fluid flow. Nevertheless, rigorous results from this field can be appropriately applied to physical systems.

More formally, suppose a system has the strong-mixing property expressed by Eq.~(\ref{eq:mixing}). Observe first that $\mu (A) = \int_M \chi_A d\mu$, where $\chi_A$ is the characteristic indicator function for the set $A$. We can rewrite Eq.~(\ref{eq:mixing}) as
$$
\lim_{n \to \infty}  \int_M \chi_{f^n (A) \cap B d\mu} - \int_M \chi_A d\mu \int_M \chi_B d\mu = 0.
$$
Now $\chi_{f^n (A) \cap B} = \chi_B \chi_{f^n(A)} = \chi_B \chi_A \circ f^{-n}$, since $f^n(A)$ is the set of points that map into $A$ under $f^{-n}$. Hence the mixing condition can be written as:
$$
\lim_{n \to \infty} \int_M \chi_B \left(\chi_A \circ f^{-n} \right) \, d \mu - \int \chi_A \, d \mu \int \chi_B \, d \mu =0 .
$$
Replacing the characteristic indicator function $\chi$ with a pair of arbitrary scalar (observable) functions $\phi$, $\psi$ (typically chosen to possess some regularity properties) defines the correlation function:
\begin{equation}\label{eq:corr_decay}
\mathcal{C}_n (\phi, \psi) =\bigg| \int \phi \left(\psi \circ f^{-n} \right) \, d \mu - \int \phi \, d \mu \int \psi \, d \mu \bigg|.
\end{equation}
The decay of this correlation function then gives a measure of the rate of mixing of $f$. 

In principle, for a mixing chaotic system, the exponential separation of nearby initial conditions should cause $\mathcal{C}_n$ to decay to zero at exponential rate, that is,
\begin{equation}\label{eq:exp}
\mathcal{C}_n = \mathcal{O}(e^{-an})
\end{equation}
for some constant $a>0$. One can establish an exponential decay rate in Eq.~(\ref{eq:corr_decay}) for the cat map $\mathbf{A}$ (recall again the Arnol'd fast dynamo of  Section~\ref{turbulent}) and analytic observables relatively easily. Expanding $\varphi$ and $\psi$ as Fourier series (here also assuming without loss of generality that $\psi$ has zero mean),
$$
\varphi(\mathbf{x}) = \sum_{\mathbf{k} \in \mathbb{Z}^2} a_\mathbf{k} e^{i \mathbf{k}.\mathbf{x}}, \psi(\mathbf{x}) = \sum_{\mathbf{j} \in \mathbb{Z}^2} b_\mathbf{j} e^{i \mathbf{j}.\mathbf{x}},
$$
the analyticity assumption guarantees that the coefficients $a_k$ and $b_j$ decay exponentially quickly. Linearity and orthogonality mean that Eq.~(\ref{eq:corr_decay}) can be written as
\begin{eqnarray*}
\mathcal{C}_n(\varphi,\psi) &=& \int \sum_{\mathbf{k} \in \mathbb{Z}^2} a_\mathbf{k} e^{i \mathbf{k}.\mathbf{A}^n \mathbf{x}} \sum_{\mathbf{j} \in \mathbb{Z}^2} b_\mathbf{j} e^{i \mathbf{j}.\mathbf{x}} dx \\
&=& \sum_{\mathbf{k} \in \mathbb{Z}^2} a_{\mathbf{k}} b_{-\mathbf{k}\mathbf{A}^n}.
\end{eqnarray*}
Since $\mathbf{A}$ is a hyperbolic matrix, the exponential growth of $|\mathbf{k}\mathbf{A}^n|$ together with the exponential decay of Fourier coefficients  yield superexponential decay of $\mathcal{C}_n$ \cite{baladi2000positive}. Note that this calculation is only possible because the map is linear, with constant Jacobian.  The same properties also reveal why the cat map is a poor model for real mixing devices. It cannot represent any nonlinear, or non-uniform behavior. 

In practice it has been observed that the presence of boundaries can slow this rate.
This phenomenon can also be demonstrated in the measure-theoretic viewpoint by computation of the decay of correlations for linked twist maps \cite{sturman}. These form a class of maps which can be thought of as non-uniformly hyperbolic generalizations of the cat map, and which include boundary regions at which particular hydrodynamical boundary behavior can be modeled. Using a Young Tower \cite{young1998statistical} approach, \citet{springham2012polynomial} have demonstrated that in linked twist maps $\mathcal{C}_n$ decays at a polynomial rate for all choices of scalar observables $\varphi$ and $\psi$ (with H\"older regularity). Moreover, the contribution to $\mathcal{C}_n$ from the boundary regions can be computed explicitly, as in \citet{sturman2012rate}, suggesting that a linked twist map achieves its upper bound on mixing rate, that is, 
\begin{equation}\label{eq:poly}
\mathcal{C}_n = \mathcal{O}(n^{-b})
\end{equation}
for some constant $b>0$, and hence that the exponent in the underlying power law is determined entirely by the boundary conditions. 

Dynamical complexity of polynomial order is also a feature of another
class of system.  Interval exchange
transformations~\cite{keane1975interval}, and their generalizations
known as piecewise isometries~\cite{goetz2000dynamics} have no
hyperbolic behavior, and can be shown to have zero topological
entropy~\cite{buzzi2001piecewise}.  Instead, their complexity comes
from discontinuities in the system, and as such can create mixing by
cutting and shuffling, rather than by stretching and
folding \cite{sturman2012role}.  A natural place to find isometric
dynamics is in materials that exhibit highly localized, discontinuous
deformations such as slip surfaces and shear bands, e.g., colloidal suspensions, plastics, polymers, alloys and granular
matter.
Moreover, some of these classes of materials, along with valved fluid
systems, e.g., piping or vascular networks, multifunctional
microfluidic analysis chips, river networks with locks, or the heart,
can undergo a combination of cutting and shuffling along with
stretching and folding \cite{Smith_discontinuous_2016}.

\subsection{Visualization}\label{visualization}

\begin{figure}
\includegraphics[width=\columnwidth]{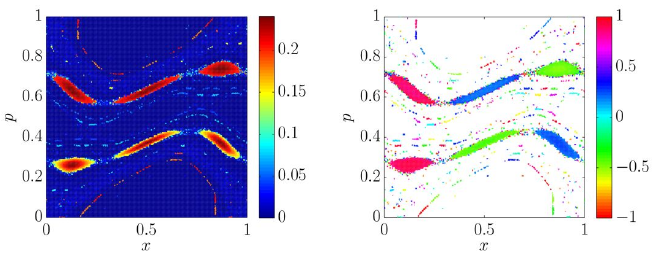}
\caption{
Filtering of invariant sets of a dynamical system using ergodic averages. The modulus (left) and phase (right) of a period-3 harmonic average of an observable along trajectories of the Chirikov standard map. Non-zero level sets of the modulus indicate period-3 sets (invariant sets of the map composed 3 times with itself). The phase of those sets indicates the order in which a trajectory visits subsets of the periodic sets. Reprinted from \textcite{Budisic2012} with the permission of AIP Publishing.
\label{visualization_fig}
}
\end{figure}

To inspect mixing in a fluid, it is reasonable to identify and visualize sets connected by fluid transport over a time period of interest. Any dynamics with mixing property would transport material between any two arbitrary sets, given a sufficiently long time window. Conversely, if there is a barrier separating the fluid into isolated islands, the fluid would not be well mixed globally; however, the fluid inside individual islands may still be mixed. Technically, the existence of material transport connecting all subsets over long time windows is ergodicity: a mixing flow is ergodic \emph{a fortiori}, but the two concepts are not equivalent, as we have earlier discussed.
Nevertheless, ergodicity can in practice be used as a first approximation to mixing. Construction of the ergodic partition, i.e., the collection of sets whose interiors contain ergodic dynamics, involves intersection of level sets of a sequence of static scalar fields averaged over Lagrangian particle paths \cite{Mezic1999}. This process is the basis for a numerical algorithm described in  \textcite{Budisic:2012woa}, which not only approximates the ergodic partition, but also aggregates smaller, dynamically-similar ergodic sets into larger invariant structures, allowing inspection of fluid features both on coarser and on finer scales; see Fig.~\ref{visualization_fig}.

The averaging algorithm for the ergodic partition can handle relatively sparse and restricted sets of initial points from which particles are launched. However, it requires time periods long enough for the particle paths to explore the mixing region. If only short bursts of dynamics are available, but the initial conditions can be sampled densely, an alternative is to form a Ulam approximation of the flow.  The Ulam method approximates the transfer operator of the flow, which describes how densities are advected, by a Markov chain transition matrix. To identify and visualize invariant and almost-invariant sets, eigenvectors of the transition matrix are used to color the state space, as described in \textcite{Dellnitz:2002wma,Froyland:2010jo,Froyland:2009ti}. 

In the ideal setting, the ergodic partition and the Ulam approximation are two sides of the same coin: they relate to the Koopman and Perron-Frobenius operators \cite{Budisic2012}, respectively, which form a dual pair. Practically, the two approaches have different numerical properties. To discover the true ergodic partition, the ergodic partition algorithm relies on long-time averages of a set of functions. Convergence of such averages can be expected in \(\mathcal O(T^{-1})\) for regular dynamics, and \(\mathcal O(T^{-1/2})\) for fully chaotic dynamics. For intermittent chaos, however, convergence can be much slower, although regions of intermittency are small in most problems, and do not cause problems, as explained in \textcite{Levnajic:2010gq}. The Ulam approximation, on the other hand, requires only short bursts of dynamics. To form the Ulam matrix, the state space is discretized into cells, which introduces a numerical stochastic element to the dynamics. This poses a problem in regions with many barriers to transport, since a poor choice of discretization can artificially connect otherwise independent regions. As the ergodic partition algorithm classifies trajectories directly, it does not suffer from the same problem.

\subsection{Summary of results on the structure of the mixed state}

We briefly summarize the main conclusions of this section.

Based upon the area-preservation property of a closed contour under topological transformation in an incompressible 2D flow, one suggestion for quantifying distributive mixing is to use the modified quantities of intensity of segregation and scale of segregation. All quantitative measures: square density, intensity and scale of segregation, and entropy reveal nonmonotonic behavior in time while approaching their limits for a uniform mixture. In order to judge properly how well or badly two substances are mixed, it is recommended to use both first-order measures (square density $\langle D^2\rangle$, intensity of segregation $I$, entropy $e$, quality $Q$) and second-order measures (e.g., the scale of segregation $L$) of the mixing quality.
Large (exponential) stretching rates do not always correspond to the best quality of mixture.
Many of these quantities have a dependence on the scale at which they are computed. The concept of mix-norm circumvents this problem.

A related point can be made in the context of decay of correlations in smooth ergodic theory. Equations~(\ref{eq:exp}) and (\ref{eq:poly}) represent exponential and polynomial decay of correlations respectively. Certainly, exponential decay will tend to zero at a faster rate than algebraic decay, but there is nothing in Eqs~(\ref{eq:exp}) and (\ref{eq:poly}) to suggest when this faster rate will occur. Although exponential mixing is a desirable property, it is quite possible that in finite time, a chaotic system with an exponential tail may be beaten by a system with a polynomial tail. (An alternative quantification of the rate of mixing has been proposed in \citet{MacKay_CCT2007}, namely the decay of the ``transportation" distance from an advected measure to the equilibrium measure.) 

We comment on the relationship between two different viewpoints on the dynamics. Topological ideas 
have been fruitful in providing bounds on rates of mixing via rates of stretching. In particular, \citet{boyland2000topological} and \citet{thiffeault2006topology} classify and optimize efficiency of stirring as measured by stretching of distinguished material lines in the fluid. The resulting {\em braid dynamics} imposes a minimum complexity on the flow, and the Thurston--Nielsen classification theorem reveals particular protocols of moving rods that achieve this complexity.  This topological approach leads to a {\em lower} bound for the {\em best} mixing behavior, while the measure-theoretic approach described above leads to an {\em upper} bound for the {\em worst} behavior. These two differing conclusions are the result of considering effectively the same dynamics in two different ways. An understanding of the gap (or perhaps overlap) between the topological bound and the measure-theoretic bound is likely to be a considerable mathematical challenge, but also promises to offer new insight into the details of mixing processes.

\section{Chaotic advection plus: Some illustrative applications}\label{advection_plus}

The general picture presented in this review shows the chaotic
advection of fluid as a consequence of the dynamical system
induced by the kinematic equation.  Fluid elements can take
regular paths, where regular regions are barriers to mixing and
transport, or chaotic paths, where chaotic regions have rapid mixing.
\textcite{aref1990} has termed the geometry created by all the paths that can
be taken by passive fluid particles the \emph{kinematic template} of the
flow.  But while the kinematic template is at the heart of all
applications of chaotic flow, few applications involve only the
kinematic template of chaotic fluid advection.  In applied chaos this
``pure'' form of motion is almost never encountered.  Applications
have multiple physical forces driving multiple modes of fluid or
particle motion in addition to the continuum stirred advection of a
carrier fluid \cite{Metcalfe_chaos_2010,Metcalfe_beyond_2012}.

\begin{figure}[tb]
\centering
\includegraphics[width=\columnwidth]{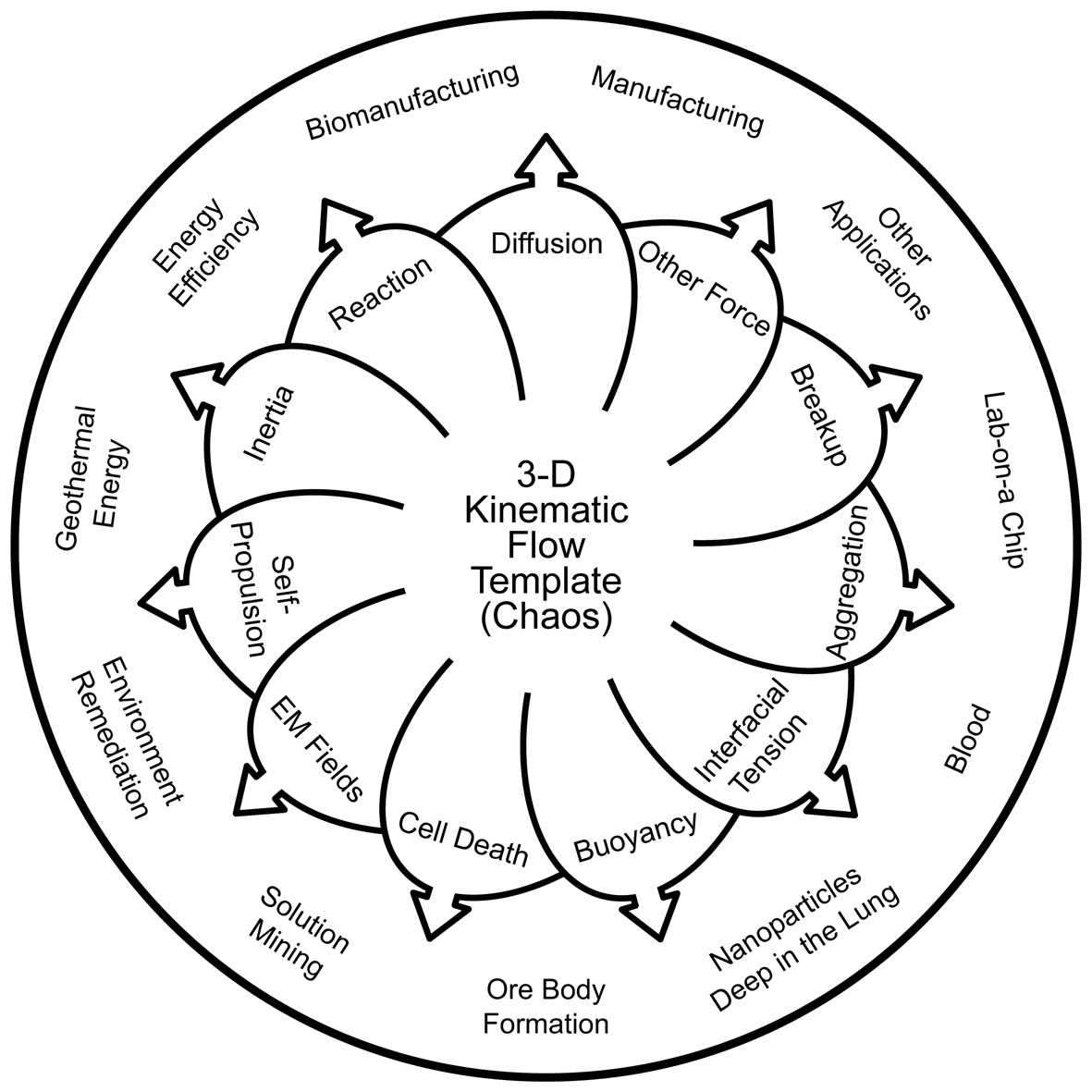}
\caption{The chaotic kinematic flow template (center) mediates the interaction
of all other physical processes to determine the outcome of
applications (outer ring).}
\label{fig:applications_template}
\end{figure}

The pathway to creative application lies in understanding the
interaction of the kinematic template with these other, multiple
application-dependent physical forces.
However, little systematic knowledge exists on the question of how
non-passive phenomena interact with the kinematic template to change
the organization of their own flow and transport properties.  For
instance, transport of elements that are not passive scalars (e.g., solid particles,
liquid droplets, biological cells, small self-propelling organisms) or
of materials that can react and diffuse (e.g., heat, chemical species) plays an
essential role in many industrial, geophysical, and biological
applications.  
The  interaction of
the chaotic flow template and these particular physical forces results in an expanded dynamical system with
the particles or scalars that is no longer conservative and is instead
dissipative (in the dynamical-systems sense of a contracting phase
space, not of fluid energy dissipation).  This leads to
attractors and repellors in particulate systems \cite{Babiano2000,Cartwright2002,Torney2007,Metcalfe_beyond_2012}
and strange eigenmodes with diffusive scalar fields 
\cite{Pierrehumbert1994,liu2004,Lester08a,Lester_RPM_2009, Lester_RPM_2010}. 
The kinematic template can be
identical in all these cases, but as the additional physics works
through and interactions are mediated by the advection, the transport
properties of the expanded systems can be quite different.

Examples occur in every domain in which fluid motion exists;
Fig.~\ref{fig:applications_template} shows schematically some selected
applications where mediation of physical forces by the kinematic
template is essential.  This section can only give a
flavor of the full range of applications possible with chaotic
advection; below we have chosen some that we find particularly interesting.

\subsection{Heat transfer}\label{sec:heat_transfer}

Control of heat transfer --- its augmentation or suppression --- is
pivotal for the efficiency of many processes
\cite{NAS_fluidsreport_2006}.  Understanding laminar-flow heat
transfer in foods, polymers, energy systems, the Earth's subsurface,
etc.\ is particularly problematic \cite{perugini03,perugini2012,ElOmari20151835,Petrelli2016425}.
Without turbulence to continually refresh material near heat transfer
boundaries, the temperature gradients at those boundaries lessen and
heat transfer slows \cite{leguer2012,Metcalfe_beyond_2012}.  Chaotic
advection can provide a means to deal with industrial applications
that involve highly viscous complex fluids and highly exothermic
reactions that are difficult to control.  In such cases, it is
important to control the rapid temperature increase resulting from
fast reactions in order to obtain the desired products. Heat removal
is then a crucial factor in controlling such fast reactions. This
problem can be tackled by using a large surface area per unit volume
ratio for the reactor (i.e., the case of microreactors), but this
solution is sometimes inappropriate, when large volumes of fluid are
considered, for example.  Another typical industrial solution is to
increase surface area by making the fluid transfer conduits very
narrow. However, this has a huge pumping energy penalty that increases
rapidly with the fluid viscosity. These problems are relevant in many
conventional industrial domains, in which the potentialities of
chaotic advection are not well recognized. In a large majority of
industrial reactors, heat transfer for either heating or cooling
processes is achieved through the walls of the reactor.  Furthermore,
many industrial fluids have non-Newtonian rheological behavior \cite{arratia2005}, and
their physical properties can also depend on the temperature.
Nowadays, highly energy-efficient heat-transfer devices have been
designed by considering chaotic flow coupled to the diffusing
temperature field
\cite{Lester_2007,Lester_nonNewtonian_2009,Metcalfe_foodRAM_2009}.

\subsubsection{Effect of the P{\'e}clet number}

As homogenizing the temperature field is equivalent to mixing a
diffusing passive scalar, the analysis proceeds via the strange
eigenmode solutions to the advection--diffusion equation, Eq.~\eqref{eqn:ADE}.
The solution to Eq.~\eqref{eqn:ADE} is given by a sum of the natural
modes $\varphi_k(\mathbf{x},t)$ of the advection--diffusion operator
\begin{equation}
\label{eqn:semodes}
\phi(\mathbf{x},t) =
\sum_{k=0}^{K} \alpha_k \, \varphi_k(\mathbf{x},t) \, e^{\lambda_k t},
\end{equation}
where the sum is ordered by the magnitude of the real parts of the
eigenvalues $\lambda_k$, with initial weights $\alpha_k$; and
physically the $\lambda_k$ have negative real parts
\cite{liu2004}.  

For the problem of material entering a device at one uniform
temperature and leaving at a different uniform temperature, only the
time to achieve thermal homogenization matters.  This time is given by
the most slowly decaying term in Eq.~\eqref{eqn:semodes}:
\begin{equation}
\phi(\mathbf{x},t) \rightarrow \alpha_0 \, \varphi_0(\mathbf{x},t) \, e^{\lambda_0 t}.
\end{equation}
The long-time transport rate is given by $Re(\lambda_0)$, and the
problem reduces to designing a device that minimizes $Re(\lambda_0)$
while keeping relatively open flow conduits, i.e., to simultaneously minimizing
the thermal homogenization time and the pumping energy.

\begin{figure}[tb]
\centering
\includegraphics[width=\columnwidth]{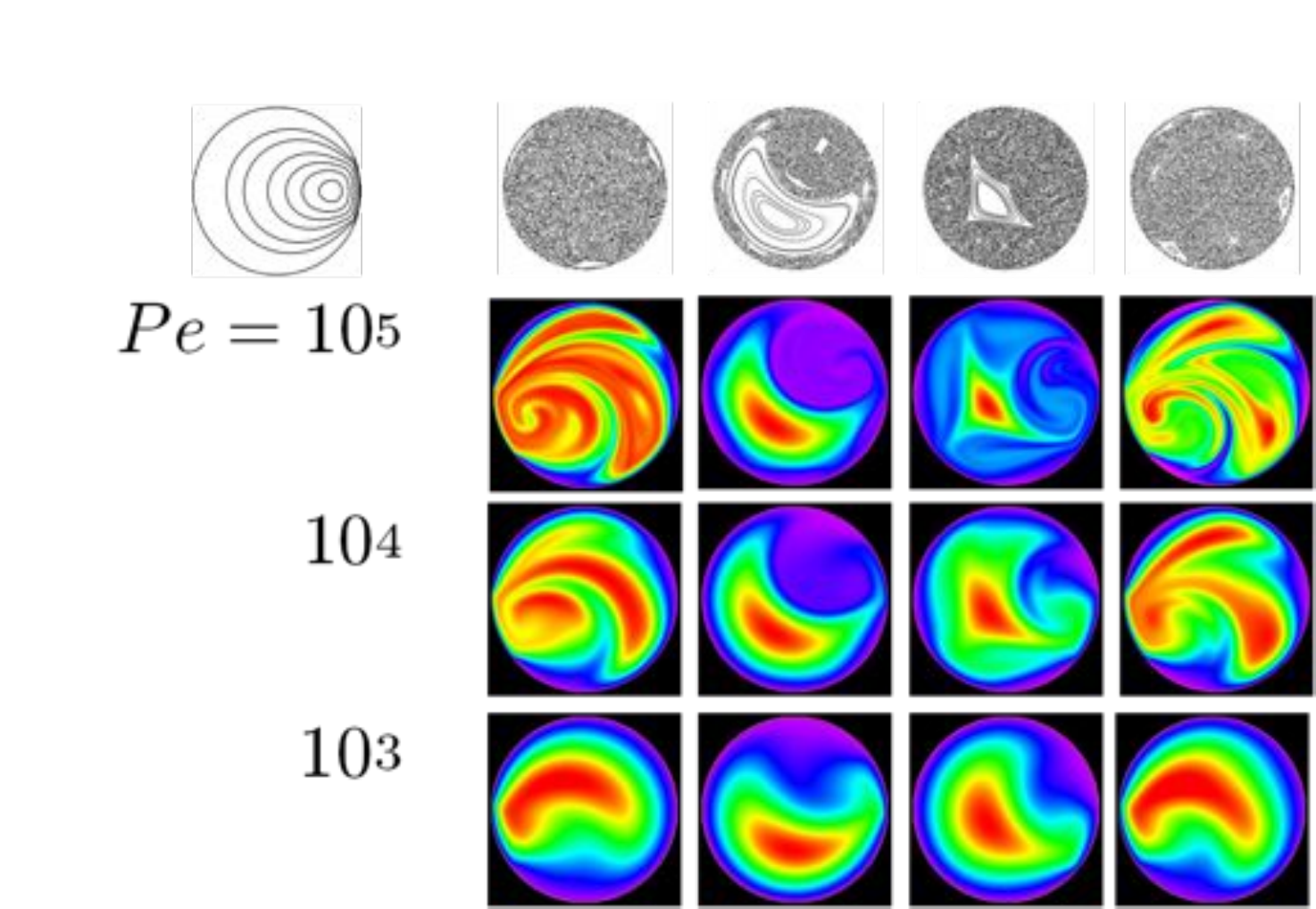} 
\caption{Dominant scalar concentration pattern in a RAM mixer for
  various reorientation parameter settings of the base flow in the top
  left, as defined in the text.  (top row) Poincar{\'e} sections
  (essentially infinite P\'eclet number results).  (columns) Below
  each Poincar{\'e} section are the patterns obtained at the indicated
  P\'eclet numbers.}
\label{fig:RAM_grid}
\end{figure}

\citet{Lester_2007,Lester08a,Lester_nonNewtonian_2009} have
used such a procedure to design a heat exchanger based on the rotated
arc mixing (RAM) flow \cite{Metcalfe_ram_2006}.  RAM flow
consists of an open tube through which fluid flows.  Along the length
of the tube portions of the boundary are in motion for a duration $\tau$, imparting
cross-sectional streamlines shown at the top left of
Fig.~\ref{fig:RAM_grid}.  This cross-sectional
flow is periodically reoriented by an angle $\Theta$ as material moves
down the tube.  The chaotic and symmetry properties of this flow have
been extensively studied
\cite{Speetjens_regime_2006,Speetjens_symmetry_2006,Metcalfe_chaos_2010}.
The top row in Fig.~\ref{fig:RAM_grid} shows Poincar\'e sections for
passive fluid advection at selected values of $\tau$ and $\Theta$
chosen to illustrate the connections between structures of various
sizes in the chaotic advection field and the resultant scalar
distribution when various levels of diffusion are added.  Each
column shows the eigenmode associated with the flow at the
top for the P{\'e}clet number indicated.  The initial condition is
a Gaussian blob of heat (red color) in the center of the
circle.  For low enough values of $Pe$ (bottom row) the asymptotic
pattern is a deformed version of the initial condition that varies
hardly at all even with large structural changes in the fluid
advection.  With increasing $Pe$ the scalar patterns can either be
dominated by large scale structures in the advection (middle two
columns), or they can seemingly decouple from the
advection field either when island structures in the advection field are
small enough (right-most column) or when there is global chaos
(left-most column).  However, other work suggests that the advection
and scalar fields are not really decoupled: it is just that Poincar\'e
sections do not show the relation because it occurs in the stretching
field.

\begin{figure}[tb]
\centering
\begin{tabular}{clcl}
\includegraphics[width=0.4\columnwidth]{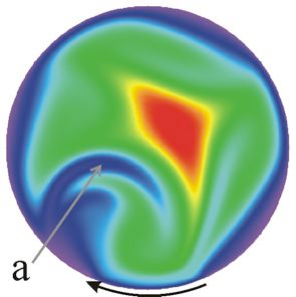} & 0 &
\includegraphics[width=0.4\columnwidth]{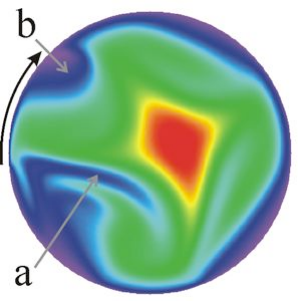} & $\tau/5$ \\
\includegraphics[width=0.4\columnwidth]{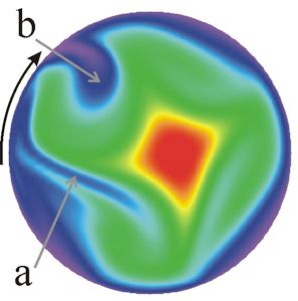} & $2\tau/5$ &
\includegraphics[width=0.4\columnwidth]{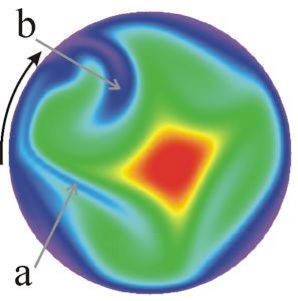} & $3\tau/5$ \\
\includegraphics[width=0.4\columnwidth]{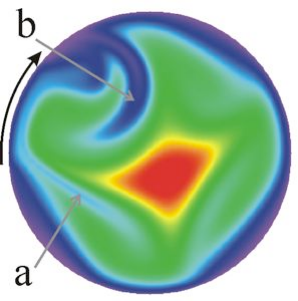} & $4\tau/5$ &
\includegraphics[width=0.4\columnwidth]{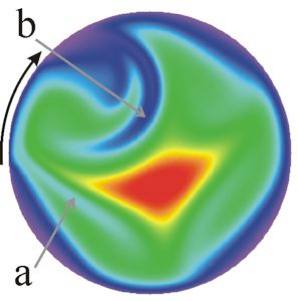} & $\tau$
\end{tabular}
\caption{How a scalar pattern evolves though a cycle of duration
  $\tau$ in a RAM flow.  At $0$ the moving boundary indicated by the
  arrow moves.  Region (a) has been stretched and folded; diffusion
  ``heals'' the pattern along folds.  Simultaneously advection
  stretches and makes new folds around the region (b).  The pattern at
  $\tau$ resumes its original shape, rotated by $\Theta$.
  Reprinted with permission from  \textcite{Lester_control_2014}.   Copyright (2014) the American Physical Society.
  }
\label{fig:heal}
\end{figure}

Figure~\ref{fig:heal} illustrates the cooperation between advection
and diffusion as the pattern $\varphi_0$ evolves through one
reorientation interval.  At $0$ the boundary arc indicated by the
arrow moves.  Region (a) has been stretched and folded during the
previous interval, and during the subsequent interval diffusion
``heals'' the pattern along this fold.  Simultaneously advection
stretches and makes new folds around the region (b).  The pattern at
$\tau$ resumes its original shape, rotated by $\Theta$.  Diffusion
heals the pattern wherever folds bring parts of the pattern close
together, and folding appears to play a larger role in sustaining the
pattern than is often credited.

\begin{figure}[tb]
\centering
\includegraphics[width=\columnwidth]{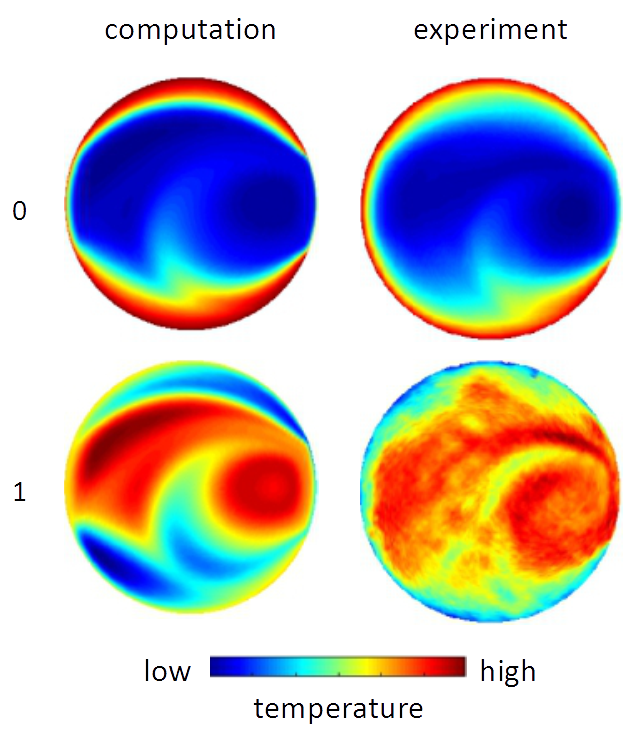}
\caption{The two leading eigenmodes (labeled 0 and 1) of the
  temperature evolution for a RAM flow at $Pe = 10^4$ from (left
  column) computation and (right column) experiment.    For
  numerical data $Re = 0$; for experimental data $Re = 0.1$.
  Reprinted from~\protect\citet{Baskan_modes_2015} with permission from Elsevier.
  }
\label{fig:thermal_expt}
\end{figure}

Experimental data for chaotic heat transfer are more difficult to
acquire, depending on what type of data are needed.  Overall heat
transfer measurements for an optimized RAM heat exchanger show that
the heat transfer rates are 4--10 times larger than for an open tube
with total pumping energy reductions of 60--80\% compared to an open
tube, depending on $Pe$ and fluid rheology.  Temperature pattern data
are scarce, but experiments are beginning to produce them:
Fig.~\ref{fig:thermal_expt} shows the measured temperature field
compared with a numerical computation of the same flow
\citep{Baskan_modes_2015}.

\subsubsection{Thermal boundary condition}

\begin{figure}[tb]
\begin{center}
\includegraphics[width=0.3\linewidth]{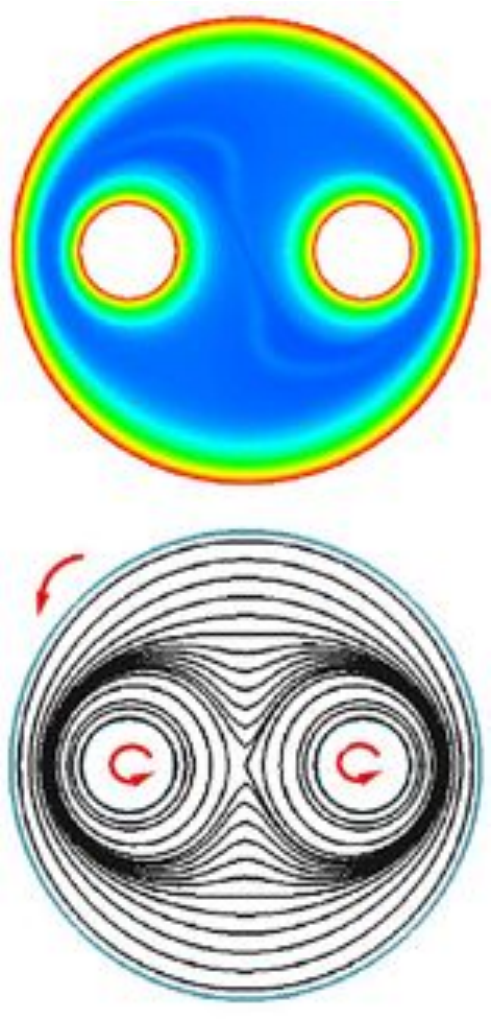}
\includegraphics[width=0.3\linewidth]{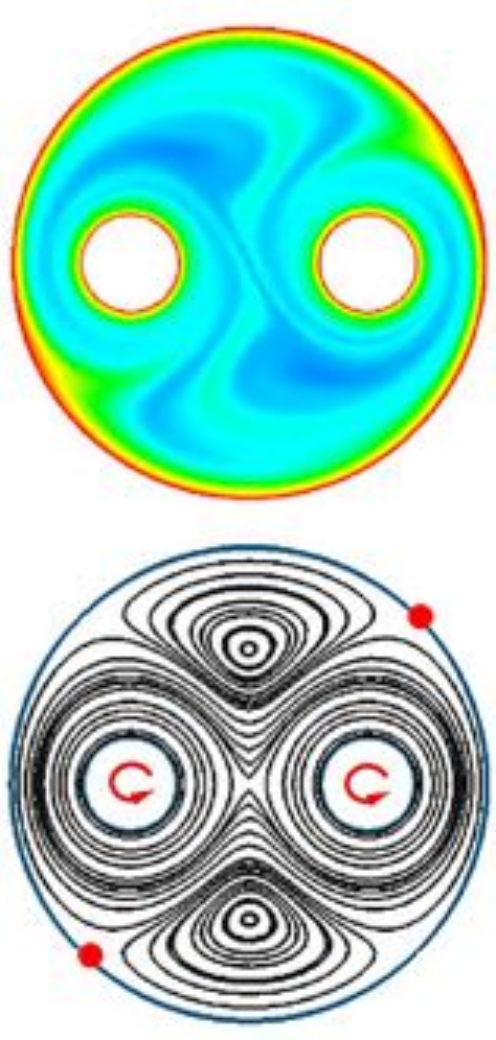}
\end{center}
 \caption{Two-rod mixer: Instantaneous temperature fields and streamlines after 4 periods of stirring. The rods and the tank rotate in the same direction. Continuously modulated stirring protocol (CM --- left) vs. alternated stirring protocol (ALT --- right). The dots at the wall indicate stagnation points, from which fluid is extracted towards the center of the mixer. Large zones of unheated fluid are observed in the CM case. Case of a constant wall temperature boundary condition. Reprinted from \citet{leguer2012} with permission from Elsevier.
}
 \label{fig:ALT_MC}
\end{figure}

 Consider a typical mixer composed of two circular rods inside a cylindrical tank. The tank and the rods are heated or cooled and can rotate around their respective axes. This two-rod mixer is suitable for obtaining global chaotic flow without large KAM regions \citep{elomari2009,elomari2010a}. Consider a highly viscous fluid of $Re \approx 1$ with a high Prandtl number $Pr=10^4$, so that it is difficult to mix.
Because of the combination of mixing with heat transfer, the wall boundary condition plays a particular role that is different from the mixing of a passive scalar considered without heating or cooling. The role of the wall has been discussed in previous sections, in Section~\ref{open} where the notion of a chaotic saddle in open flow is presented  and in Section~\ref{closed} where the case of a slip boundary is considered in a 2D bounded flow. 
\citet{elomari2010a, elomari2010b} have shown for both Newtonian and non-Newtonian fluids and for a constant wall-temperature boundary condition that the efficiency of thermal exchange is strongly dependent on the choice of the stirring protocol imposed on the walls. The main conclusion from these works is that maximizing heat transfer from the wall boundaries requires that the walls (i.e., the tank or rods here) move alternately. This way, one avoids the development of closed streamlines near the wall, which reduces advection from the wall zones and prevents the persistence of confined hot or cold fluid zones. Continuous modulation of wall movement is not sufficient to produce effective chaotic mixing and the existence of stagnation points on the static walls is necessary to create separatrices from which heteroclinic tangles are created (see Fig.~\ref{fig:ALT_MC}). Thus, thermal chaotic mixing is controlled by the topology of the flow near the moving or non-moving walls. However, the choice of a thermal wall boundary condition, either constant wall boundary condition or constant wall heat flux, gives rise to fundamental differences in the evolution of the fluid temperature and its homogenization.

\begin{figure}[tb]
\begin{center}
 \includegraphics[width=\linewidth]{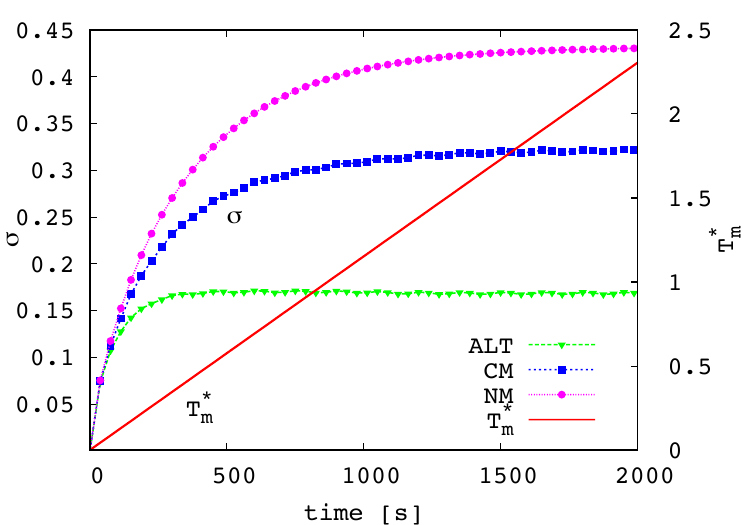}
\includegraphics[width=\linewidth]{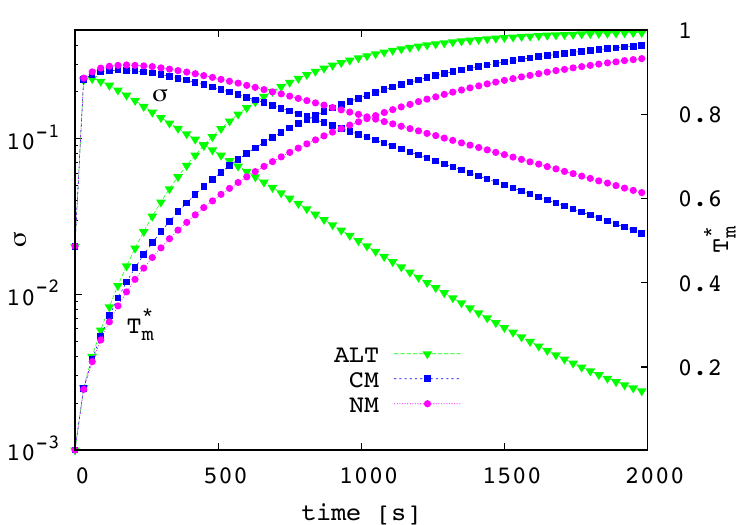}
\end{center}
 \caption{Two-rod mixer: Time evolution of the temperature standard deviation $\sigma$ for the non-modulated (NM), continuously modulated (CM) and alternated (ALT) stirring protocols. The non-dimensional mean temperature $T_m^*$ is also given. Modulation period is 30 s. Above: the imposed heat flux, below: the imposed temperature (note that in the latter case the $\sigma$ axis has a logarithmic scale). 
Reprinted from \textcite{elomari2012} with permission from SpringerNature.
}
 \label{fig:COMP_MC}
\end{figure}

For a constant wall temperature boundary condition (i.e., Dirichlet condition), the imposed wall temperature represents an asymptotic limit for the evolution of the mean fluid temperature, and the rapidity with which  this limit is reached is controlled by the efficiency of the mixing. In contrast, for a constant heat flux boundary condition (i.e., Neumann condition), there is no asymptotic limit for the evolution of the mean fluid temperature because its evolution is prescribed by the imposed heat flux density. In this case, the efficiency of mixing (of the stirring protocol) controls the heat homogenization only. As a consequence, the stirring strategy and the choice of the mixing measures must be selected in agreement with the type of the wall heating considered. This is illustrated in Fig.~\ref{fig:COMP_MC} for the two different boundary conditions and for three stirring protocols (non-modulated --- NM --- continuously modulated --- CM --- and alternated --- ALT). This figure displays the greater efficiency of the alternated stirring protocol for both thermal boundary conditions. Similarly to the non-modulated stirring protocols, the continuously modulated stirring protocols give rise to closed streamlines in the vicinity of the walls that prevent the radial transport of the temperature scalar inside the mixer. Thus, close to the rotating boundaries, radial heat transfer is achieved only by conduction across the streamlines; these streamlines act as an insulation medium \citep{Gouillart2010}.

\subsubsection{Effect of fluid rheology}

\begin{figure}[tb]
\begin{center}
 \includegraphics[width=0.6\linewidth]{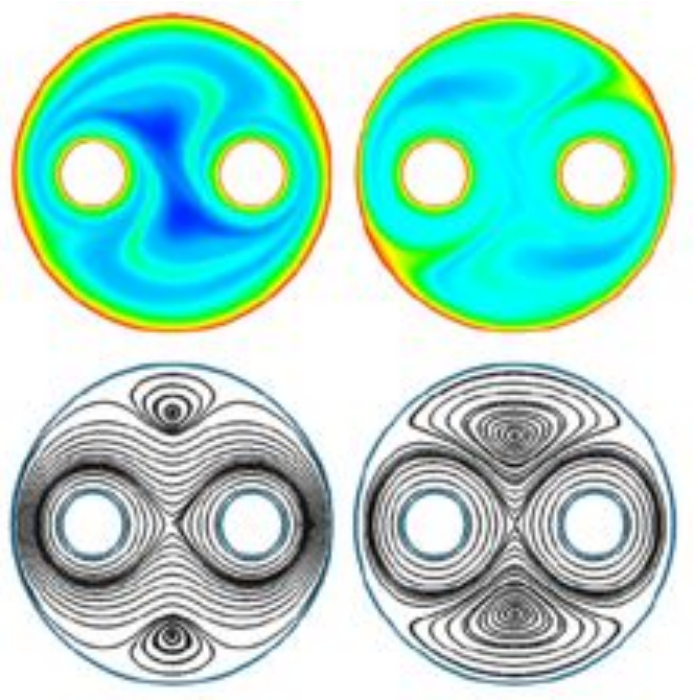}
\end{center}
 \caption{Two-rod mixer: Instantaneous temperature fields and streamlines after $4$ periods of stirring for shear-thinning (left) and shear-thickening (right)
rheological fluid behaviors for the alternated stirring protocol. Case of a constant wall temperature boundary condition. Reprinted from \citet{leguer2012} with permission from Elsevier.
}
 \label{fig:NN_ALT}
\end{figure}

\begin{figure}[tb]
 \centering
 \begin{center}
\includegraphics[width=0.6\linewidth]{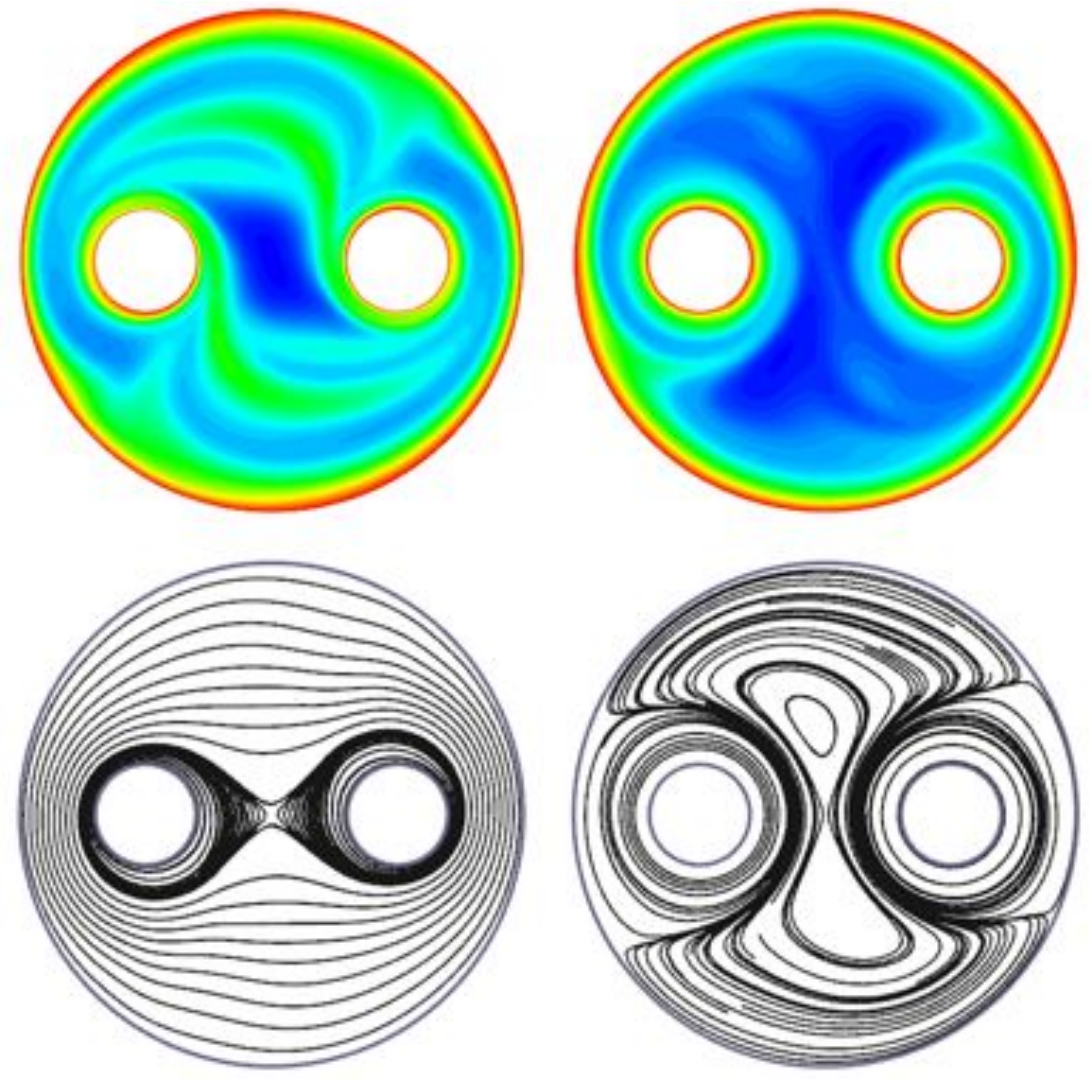}\\\textit{Heating}\hspace{1cm}\textit{Cooling} 
\end{center}
 \caption{Two-rod mixer: Temperature fields and streamlines after $4$ periods of stirring during a heating process (left) and a cooling process (right) for a Newtonian temperature-dependent fluid ($B=5$) and the alternated stirring protocol. Case of a constant wall temperature boundary condition. Reprinted from \citet{leguer2012} with permission from Elsevier.
 }
 \label{fig:TDepB5}
\end{figure}

\begin{figure}[tb]
 \centering
 \begin{center}
 \includegraphics[width=0.9\linewidth]{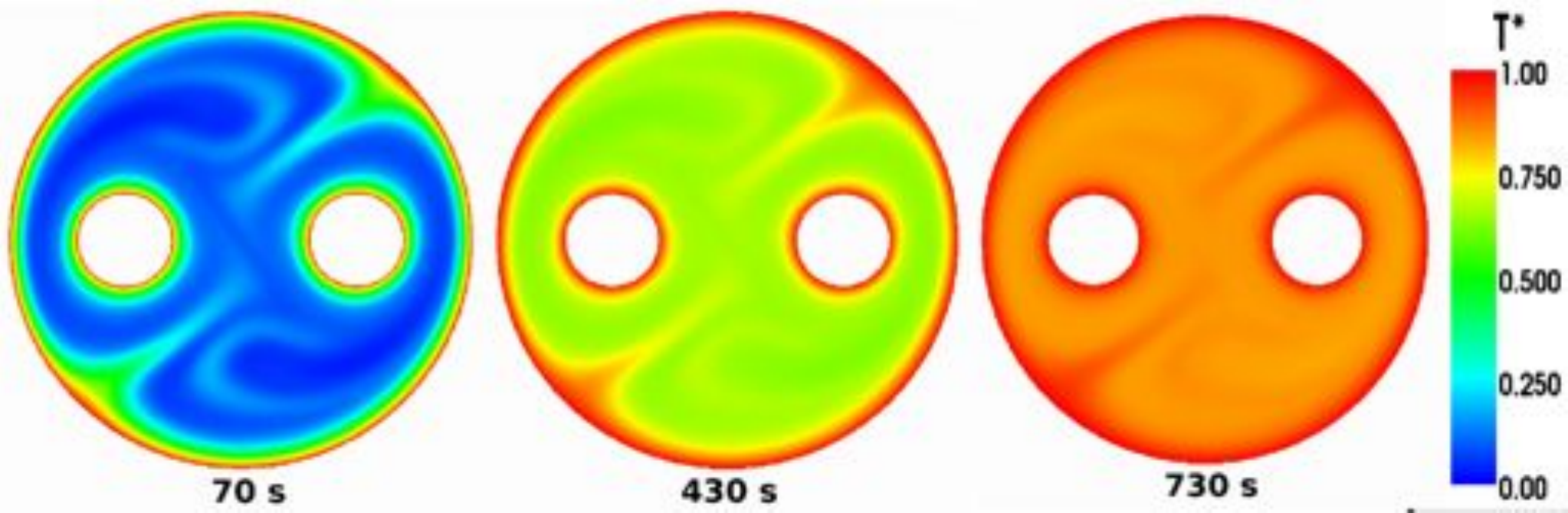}
  \includegraphics[width=0.85\linewidth]{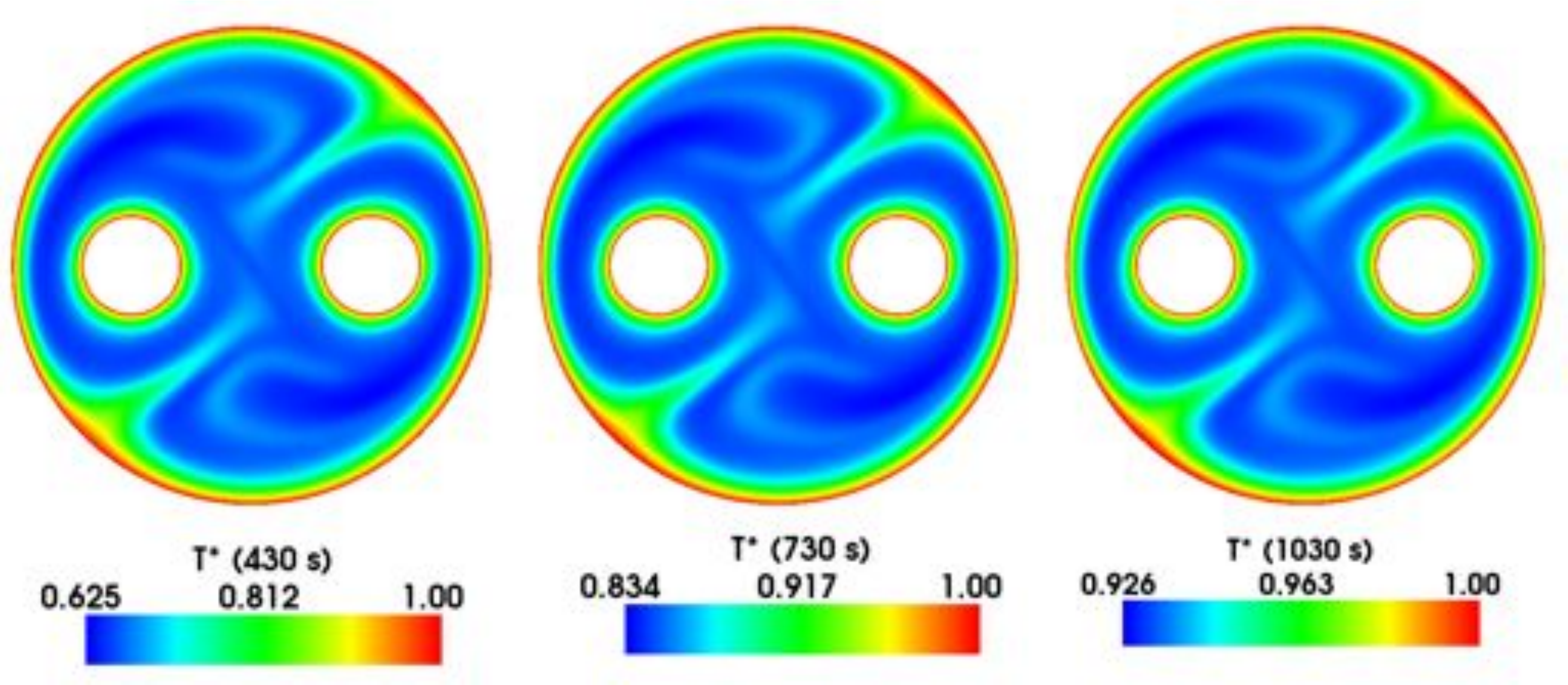}\mbox{\hspace{0.5cm}}
\end{center}
 \caption{Two-rod mixer: Snapshots at different periodic times show the evolution of the $T^*$ field towards a strange eigenmode (top row), and snapshots  rescaled between $T^*_{min}$ and $T^*_{max}$ (bottom row).  Reprinted from \citet{leguer2012} with permission from Elsevier.
 }
 \label{fig:carto_m_echel_CoCo30} 
\end{figure}

In order to show the impact of fluid rheological behavior on the pattern of the temperature field, we illustrate in Fig.~\ref{fig:NN_ALT} the temperature fields and the corresponding streamlines for shear-thinning (left) and shear-thickening (right) fluids in the case of the alternated stirring protocol. The Ostwald--de-Waele power-law model,  $\eta\left(\dot{\gamma}\right) = k \left(\dot{\gamma}\right)^{n-1}$, is chosen for the apparent viscosity. Two values of the behavior index $n$ are considered, $n=0.5$ and $n=1.5$, corresponding to shear-thinning and shear-thickening fluids, respectively. The values of the consistency index $k$ are adjusted to keep the generalized Reynolds number the same as the corresponding Newtonian case of Fig.~\ref{fig:ALT_MC}. Compared to Newtonian and shear-thinning fluids, the shear-thickening fluid is more driven by the wall due to its higher viscosity, resulting in larger vortices and more extended hot fluid streaks originating from the parietal parabolic points. This greater fluid driving also increases the folding of temperature striations. Thus, the mixing efficiency for this fluid is higher than that of the two other fluids. The vortices created in the shear-thinning fluid are smaller because of the viscosity reduction in the sheared fluid near the rotating walls. Fluid driving forces and mixing efficiency are then weak. The effects of the temperature dependence of viscosity on chaotic mixing are also important. As discussed above, fluid viscosity in the vicinity of the moving walls has a major influence on the fluid displacement and on the creation of large vortices accompanied by separatrices that help to promote mixing. When the temperature dependence is taken into account, the heating and cooling processes have different evolutions. This temperature dependence can be modeled by an exponential model as $\eta=\eta_0\exp{(- B T^*)}$  \citep{leguer2012}. The results obtained for Pearson number $B=5$ (relevant to melted polymers, for example) are shown in Fig.~\ref{fig:TDepB5} and may be compared to those of the non-temperature-dependent case ($B = 0$) of Fig.~\ref{fig:ALT_MC} (right).
In contrast to the non-temperature-dependent case, the velocity field for $B \neq 0$ is not reproducible from period to period owing to the continuous evolution of the viscosity, and therefore the mixing mechanisms are not exactly the same from the beginning of the mixing to its end.
The streamlines of Fig.~\ref{fig:TDepB5} show distinctly different patterns for heating and cooling and the both are different from the $B = 0$ case (Fig.~\ref{fig:ALT_MC} (right)). This difference is salient for the heating process, for which the viscosity is low in a thin fluid layer around the walls and higher elsewhere; thus, there is slippage between the fluid and the wall (i.e., the momentum diffusion is weak), and the fluid is not well driven. 
On the other hand, the lack of the fluid driving force of the moving walls is balanced by the ease of moving the less-viscous fluid at the static walls. 
The mixing mechanism of a temperature-dependent fluid is quite complex, with several superimposed effects, even for a Newtonian fluid. Also, flow morphologies evolve in time, and as a remarkable consequence, the strange eigenmodes existing in the non-temperature-dependent case (Fig.~\ref{fig:carto_m_echel_CoCo30}) are absent in the $B=5$ case because there are no recurrent patterns in the temperature field.

An important feature of thermal mixing in comparison with the mixing of a scalar substance is that the scalar source is located at the wall boundary. Thus the mode of heating (or cooling) of the boundary directly influences the characteristics of the mixing and heat transfer which are also significantly affected by the rheological behavior of the fluid through the shear rate dependence on viscosity. The thermal strange eigenmodes are always present for a non-Newtonian fluid (with a shear-thinning or shear-thickening behavior) but they disappear when the viscosity is temperature dependent because of the spatial modification of the flow field over time.

\subsection{Microfluidics}\label{Microfluidics}

In recent years, the transfer of lithographic techniques to the
microfluidics industry has led to a renascence of the chaotic advection paradigm for the
mixing of fluids at the lowest Reynolds numbers. Although microfluidic mixing is a key
process in a host of miniaturized analysis systems, with useful applications for chemical
reactions, crystallization, polymerization and organic synthesis, biological screening,
PCR (polymerase chain reaction) amplification and a great many other fields, it continues to pose challenges owing to
constraints associated with operating in an unfavorable laminar flow regime 
characterized by a combination of low Reynolds numbers and high
P{\'e}clet numbers. A wide variety of micromixing approaches have been
explored, most of which can be broadly classified as either active
(involving input of external energy) or passive (harnessing the
inherent hydrodynamic structure of specific flow fields to mix fluids
in the absence of external forces). Here we review some of the
most promising approaches.

\begin{figure*}[tb] \begin{center}
\includegraphics*[clip=true,width=0.95\textwidth]{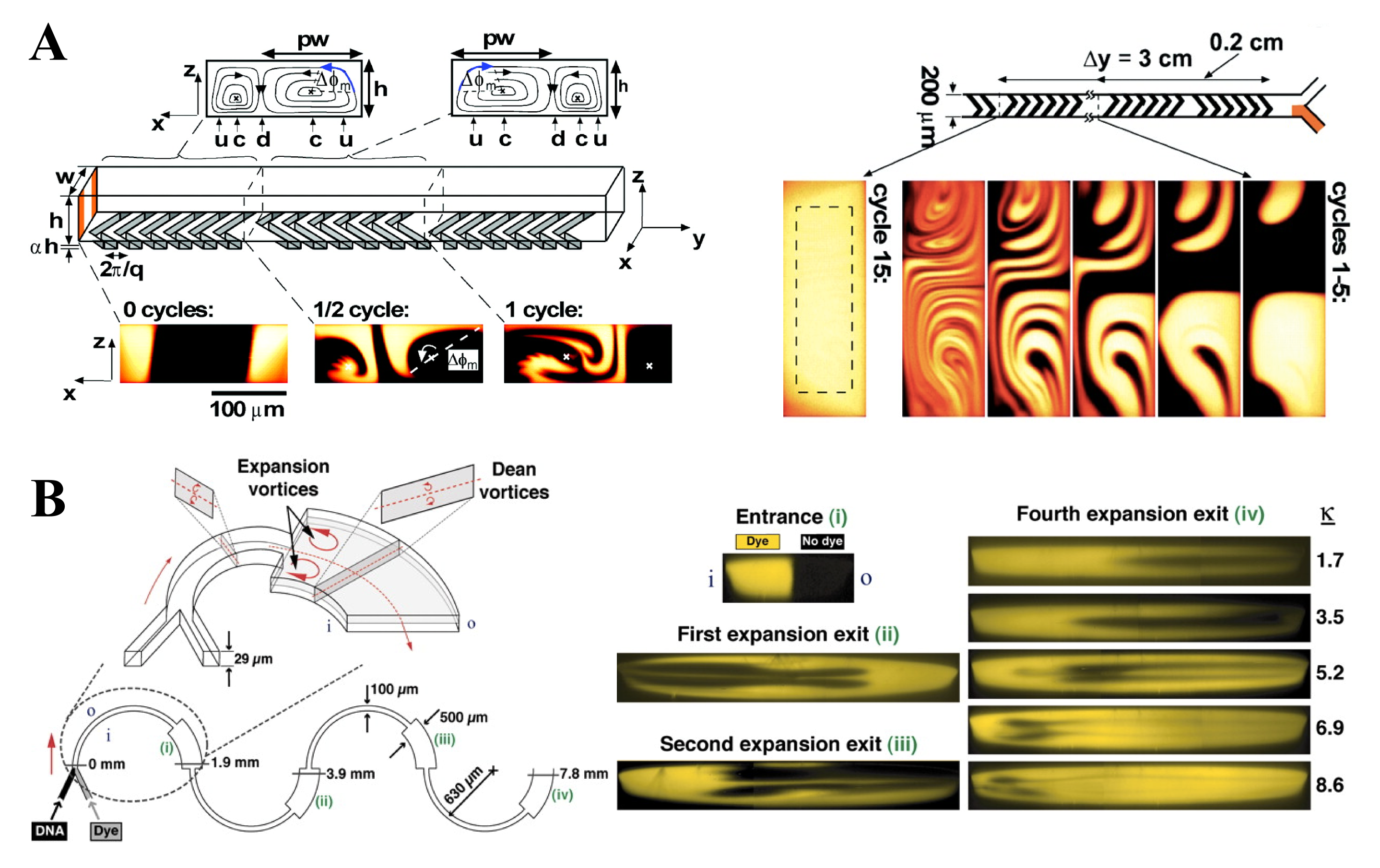} \end{center}
\caption{\label{microfluidics_passive} Passive micromixers: (a) The staggered-herringbone micromixer (SHM) has become a paradigm for passive mixing. A periodic sequence of micro-grooves imparts a passive perturbation to the primary channel flow to elicit chaotic advection and, as a consequence, rapid mixing. (b) More recent strategies include the use of finite $Re$ number effects in curved channels (e.g., a combination of Dean and expansion vortices).  (a) From  \textcite{Stroock2002}, reprinted with permission from AAAS; (b) reprinted with permission from \textcite{Sudarsan2006}, copyright (2006) National Academy of Sciences.} \end{figure*}

\subsubsection{Passive mixing} Passive designs are often desirable in applications involving
sensitive species (e.g., biological samples) because they do not impose strong mechanical,
electrical, or thermal agitation. The microchannel structures associated with these mixing
elements range from relatively simple topological features on one or more channel walls
--- ridges, grooves, or other protrusions that can, for example, be constructed by means of
multiple soft lithography, alignment, and bonding steps --- to intricate 3D flow networks.
Ultimately, it would be desirable to achieve gentle passive micromixing in the shortest
possible downstream distance by using simplified microchannel geometries (ideally, planar
2D smooth-walled) that can be easily constructed; ideally, in a single lithography step.

One of the best-known examples, the staggered-herringbone micromixer (SHM) shown in 
Fig.~\ref{microfluidics_passive}a \cite{Stroock2002}, subjects the fluid to a repeated sequence of rotational and extensional
local flow that, as result, produces a chaotic flow. The internal structures endow SHMs
with high mixing efficiency, allowing a short mixing length ($\sim$1--1.5~cm) at high
P\'eclet number ($Pe \sim 10^4$). A series of improved grooved patterned micromixers has been
proposed since then. More convoluted devices use a more faithful realization of
the Baker's map by a ``split-and-recombine'' strategy where the streams to be mixed are
divided or split into multiple channels and redirected along trajectories that allow them
to be reassembled subsequently as alternating lamellae \cite{Kim2005, Xia2005}. Another
interesting strategy for passive devices is the use of finite $Re$ effects to force the
fluid through an efficient chaotic transformation: examples include the use of Dean
vortices in curved channels depicted in Fig.~\ref{microfluidics_passive}b \cite{Sudarsan2006, Sudarsan2006a}, or the incorporation of
elastic passive elements in the microchannel to transform a simple steady laminar flow
into a forced oscillatory flow through hydroelastic instabilities \cite{Xia2012}.

\begin{figure*}[tb] \begin{center}
\includegraphics*[clip=true,width=0.95\textwidth]{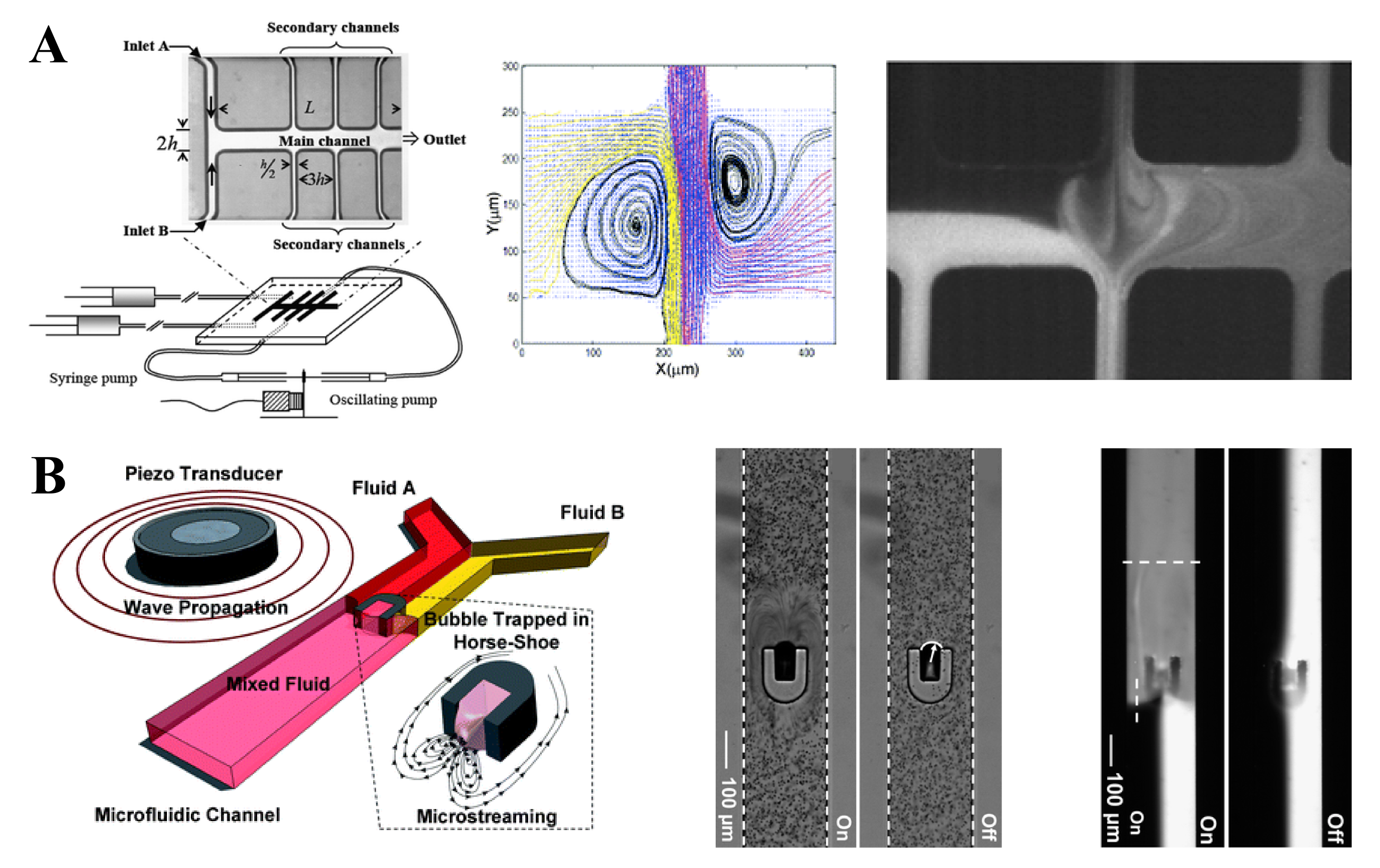} \end{center}
\caption{\label{microfluidics_active} Active micromixers: (a) Externally driven pressure actuators significantly
enhance mixing by imposing secondary oscillatory flows in an easily controlable fashion. In this example, efficient mixing is achieved within a single secondary channel width. (b) Active devices are nowadays developing new ways to decouple from bulky and complex connections to the outer world. In this example, an air bubble is trapped within the microchannel and external piezo transducers are used to excite remotely acoustic streaming. (a) Reproduced from \textcite{Bottausci2007} with permission of The Royal Society of Chemistry; (b) reproduced from \textcite{Ahmed2009} with permission of The Royal Society of Chemistry.} 
\end{figure*}

\subsubsection{Active mixing} In contrast to the passive types, active micromixers can keep their
functions under tighter control in most cases. External perturbation sources, such as
pressure, thermal, electrokinetic, magnetic and acoustic disturbances, can be designed to
induce chaotic advection. For instance, externally imposed oscillatory flows driven across
a main channel can stretch and fold fluid in the primary flow into itself, significantly
enhancing mixing. This secondary oscillatory flow can be generated by means of controlled
pressure, as in Fig.~\ref{microfluidics_active}a \cite{Glasgow2003, Bottausci2007, Okkels2004, Tabeling2004}, via low frequency
switching of transverse electroosmotic flows generated on integrated microelectrodes
\cite{Oddy2001, Song2010}, through ultrafast ion depletion and enrichment by
polyelectrolytic gel electrodes \cite{Chun2008}, or by trapping and exciting acoustic
streaming of air bubbles with external piezo-transducers, as illustrated in Fig.~\ref{microfluidics_active}b \cite{Liu2002, Ahmed2009,
Wang2011}.

\begin{figure*}[tb] \begin{center}
\includegraphics*[clip=true,width=0.95\textwidth]{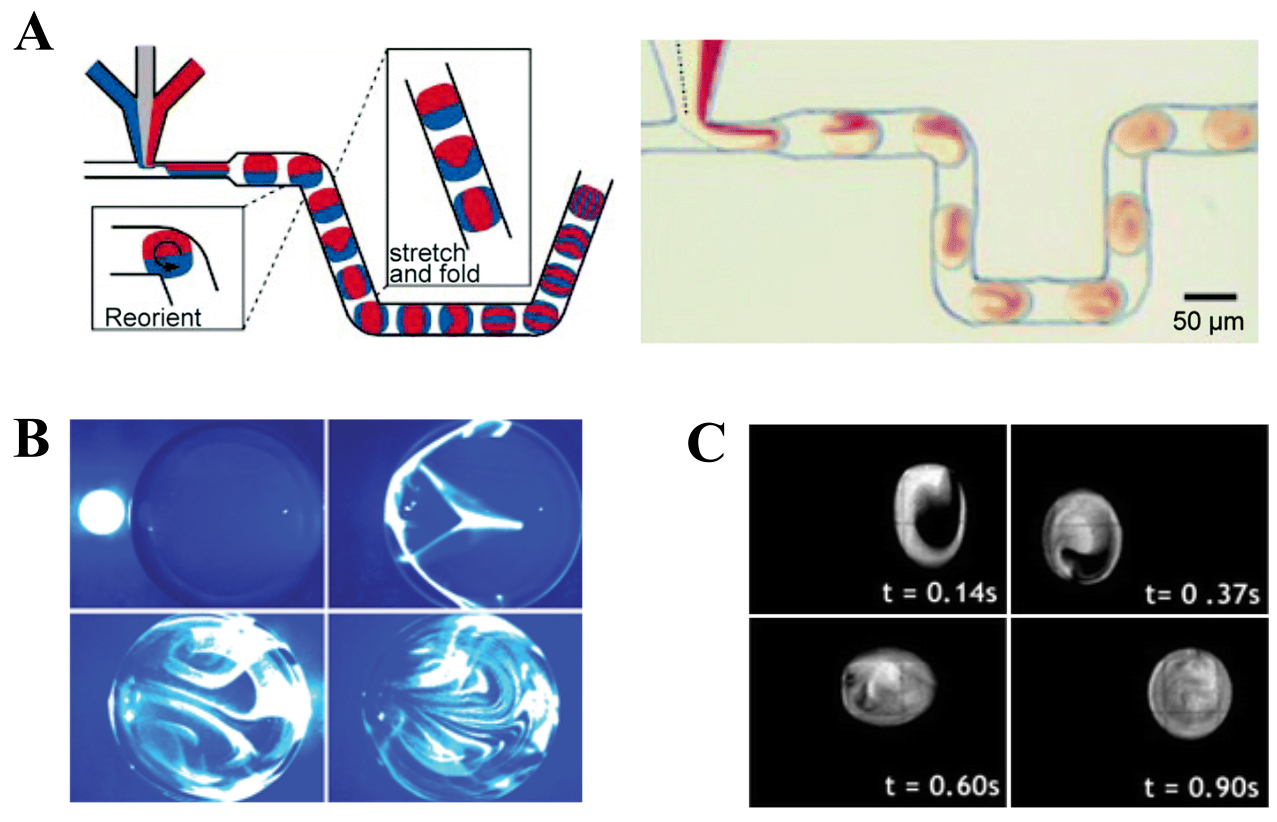} \end{center}
\caption{\label{microfluidics_droplets} Droplet based micromixers: Highly monodisperse emulsions have been shown recently to be easily controllable in microchannels, opening the door for an era of versatile high throughput use of microfluidic platforms. (a) The ``planar serpentine micromixer'' (PSM) is one of the best known devices that shows efficient mixing of individual microdroplets. Mixing is achieved by the introduction of turns and bends in the channel, which induces a periodic asymmetric flow perturbation within the drop. Active perturbations to individual droplets can also be achieved by external means, for instance: (b) by imposing surface-tension gradients through heating by a laser beam, or (c) by the fine tuning of surface electrowetting induced by microelectrodes embedded in the microchannels. (a) Reprinted from \textcite{Song03} with the permission of AIP Publishing; (b) 
reproduced from  \textcite{Grigoriev2006} with permission of The Royal Society of Chemistry; (c) reproduced from   \textcite{Paik2003} with permission of The Royal Society of Chemistry.
}
\end{figure*}

\subsubsection{Mixing in microdroplets} Droplet-based microfluidic systems, sometimes referred to
as ``digital microfluidics'', use two immiscible fluids such as water and oil to produce
highly monodisperse emulsions in microchannels with a small size variation. By varying the
viscosity of the two phases, the relative flow rates, or the channel dimension, it is
possible to tune the dimensions of the microdroplets produced. Such microdroplets can be
used to encapsulate, for instance, a number of chemical reagents that need to be mixed
quickly for a chemical reaction to take place. This compartmentalization is ideally
suited to a large number of applications, including the synthesis of biomolecules, drug
delivery, and diagnostic testing \cite{Teh2008}. The discrete nature of microdroplets and
the feasibility of individual control of their distinct volumes of fluids contrasts with
the continuum nature of other systems and emphasizes their use as a versatile 
high-throughput platform.

Interaction between the droplet surface and the channel walls can be used to generate
recirculating flows within the droplet \cite{Song2003, Tice2003}. The introduction of turns and bends in the channel induces an asymmetric flow pattern within the drop that can be used to perturb periodically the steady flow and, hence, induce mixing by chaotic advection. The best known device based on this principle is the ``planar serpentine micromixer'' (PSM) shown in Fig.~\ref{microfluidics_droplets}a \cite{Song03,Tice2004, Bringer2004a} in which the extent of mixing can be
controlled by altering the number and distribution of turns in the microchannel.
Conversely, the microdroplet surface can be perturbed by external forcing. Examples
include the generation of surface-tension gradients, for instance, generated by heating
from a laser beam \cite{Grigoriev2006}, or the fine tuning of surface electrowetting
induced by microelectrodes \cite{Paik2003, Paik2003a} illustrated in Figs.~\ref{microfluidics_droplets}b and \ref{microfluidics_droplets}c, respectively.

The above classification is by no means an exhaustive inventory of the available techniques,
as the field is rapidly evolving and new configurations and implementations for effective
mixing in microchannels based on the chaotic advection paradigm emerge every day. For
further reading on this topic see recent compendia of the field; e.g., \textcite{Squires2005, Tabeling2005, Lin2011, Nguyen2011} and references therein. The ubiquity
of the phenomenon in its diverse forms emphasizes the relevance of the physics of mixing
for the bulk of the miniaturization industry and, more particularly, for lab-on-a-chip
applications.

\subsection{Biology}

\subsubsection{Ciliary and flagellar chaotic advection}

\begin{figure*}[tb] \begin{center}
\includegraphics*[clip=true,width=0.95\textwidth]{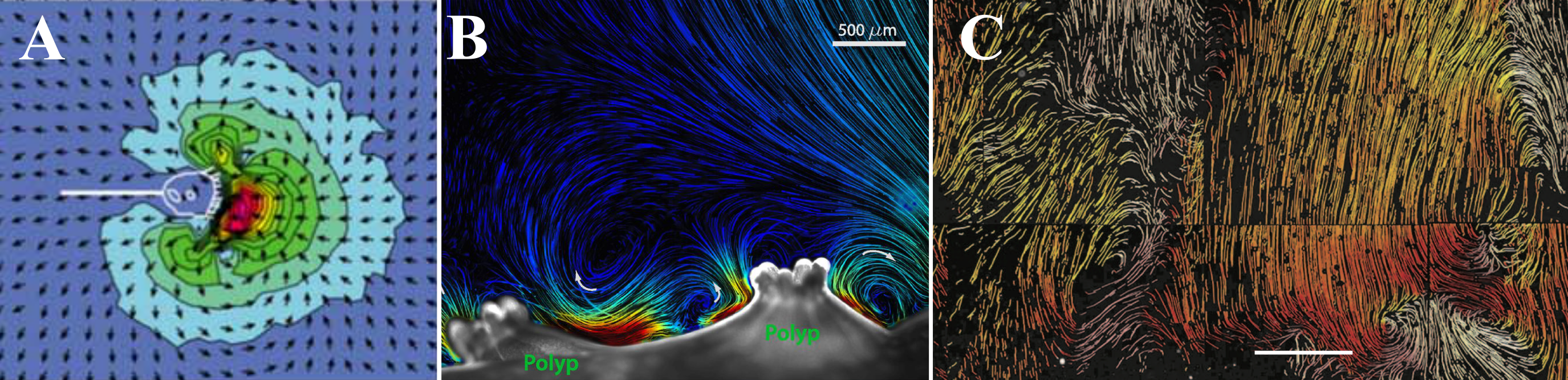} \end{center}
\caption{Ciliary and flagellar chaotic advection: (a) The fluid flow around \emph{Vorticella}. Experimental data showing a cell in the center overlain with the fluid velocity field. The \emph{Vorticella} body is outlined. (b) Cilia-driven vortical flows between two polyps on the surface of a small branch of \emph{ P. damicornis}. (c) The flow map of cerebrospinal fluid shows that it is highly organized and has cilia modules, separatrices, and whirls. Scale bar, 1mm. Color coding/shading shows the local flow orientation.  (a) 
Reprinted from \textcite{Nagai09}, with the permission of AIP Publishing; (b) from \textcite{Shapiro2014},
copyright (2014) National Academy of Sciences;  and (c) from \textcite{Faubel2016}, reprinted with permission from AAAS.}
\label{mixing_biology}
\end{figure*}

Flagella and cilia are microscopic organelles --- flexible rope-like driven  structures that form part of a single cell --- that act as ubiquitous biological stirrers.  These organelles are one of the most highly conserved structures in biology \cite{Gibbons1981, Gardiner2005, Marshall2006} and, as such, they have a wide range of biological fluid-mechanical functions: 
from propulsion in flagellated and ciliated micro-organisms \cite{Taylor1951, Lighthill1976, Brennen1977, Drescher2010} and the sperm and ovum in reproduction \cite{Fauci2006, Gaffney2011}, to feeding and filter feeding in sessile organisms \cite{Blake74b} and mollusks \cite{blake95}, and as a physical guide for diverse developmental processes \cite{Cartwright2009a, Freund2012}.

The flow induced by the beating of cilia and flagella, which falls in the regime of very low Reynolds numbers, has been modeled by means of fundamental singularity for Stokes flow: as single point forces (stokeslets) or torques (rotlets) \cite{Blake96, Cartwright2004, Niedermayer2008} or as a line distribution of these singular solutions \cite{Smith12} with, as expected from Stokes flows, boundaries playing a major role on the extent, topology and magnitude of the fluid flow \cite{Blake71a, Blake74b, Montenegro-Johnson12}. However, the role of cilia and flagella in biological fluid mixing has been much less studied. 

The sessile ciliated protozoan \emph{Vorticella}, which lives in freshwater ponds, has attracted considerable interest. The flow generated by the continual beating of its oral cilia (Fig.~\ref{mixing_biology}a) is well captured by a stokeslet near a flat rigid no-slip boundary, with the force acting normal to the boundary and, hence, generating a toroidal eddy \cite{Blake96,Nagai09,Pepper2010}. Reports on the sudden changes in the length of the stalk linking the cell and the substrate to which it is anchored led to a series of studies on mixing by a `blinking stokeslet' \cite{Otto_blinking_2001, Orme2001a, Orme2001b, blake2001}, a model inspired on the now classical `blinking vortex' model \cite{Aref1984, aref1986, meleshko1996_3} but in which the height of the point force above the no-slip boundary changes periodically. Furthermore, \emph{Vorticella's} bell changes its orientation erratically \cite{Pepper2013} which can be modeled with time-dependent variations in the direction of the applied point force. Both mechanisms might enhance the feeding process of \emph{Vorticella} by allowing nearby nutrients to be driven through filtering cilia on the cell surface.  More recently \emph{Vorticella} has even inspired a biomimetic solution for mixing in microfluidic chips \cite{Nagai2014}.

Other ciliated organisms (e.g., \emph{Opalina}, \emph{Paramecium} and \emph{Volvox}) and ciliated tissues (e.g. epithelia, corals) show a remarkable coordination of nearby beating cilia, including complete synchrony \cite{Goldstein2009, Goldstein2011} and metachronal waves (i.e., large-scale modulations of the beating phase) \cite{Gueron1997, Lenz2006, Brumley2012, Elgeti2013, Brumley2015}. The unsteady flow generated by their coordinated beating leads to stretching and folding and, over time, to mixing of the adjacent fluid \cite{Solari2006,Ding2014,Shapiro2014} (Fig.~\ref{mixing_biology}b). This mechanism has also been implemented in biomimetic solutions for lab-on-a-chip mixing using actuated artificial cilia-like protrusions  \cite{Toonder2008, Chen2013, Kongthon2011}. 

Cilia are also known to play a major role in the processes that shape biological vertebrate development \cite{Cartwright2009a, Freund2012}. For instance, they are involved in the development of the ear system of fish, a process known as otolith seeding \cite{Wu2011, Colantonio2009}; keep the cerebrospinal fluid flowing through the ventricular system of our brains (Fig.~\ref{mixing_biology}c) \cite{Guirao2010, Faubel2016}; and are fundamental for defining our vertebrate body plan by breaking the left--right symmetry of the early embryo \cite{Nonaka2002, Cartwright2004, Cartwright2007, Supatto2008, Smith12, Montenegro-Johnson12}. In all these cases, not only fluid transport but also mixing by cilia appears to be of possible importance. A few observations of flow patterns with widely divergent flow for initially nearby particles exist, although no detailed analysis has yet taken place to assess the relevance of fluid mixing in each case.

\subsubsection{Biological activity in chaotic oceanic flows}

The growth of marine organisms is closely related to ocean currents. Since hydrodynamic flows constitute an important physical factor for productivity in the ocean, the interplay between the physical environment and biological growth is an increasing field of research \cite{Mann-Lazier-91,Denman-Gargett-95,Peters-Marrase-00,Tel-et-al-05,Karolyi-et-al-00}. In particular, the growth of plankton, which is the basis of the food web in the ocean, is influenced by currents. Plankton species and their nutrients are only passively transported, while higher organisms (nekton) have the ability to swim actively and are therefore more independent of the flow. Starting with the work of \citet{Abraham-98} and \textcite{bees1998planktonic}, who first applied the concept of chaotic advection to the study of plankton patterns in the ocean, various authors have contributed to this subject \cite{Lopez-et-al-01a,Lopez-et-al-01b,Martin-et-al-02,Martin-03,Sandulescu-et-al-07}.

One of the major requirements for the growth of plankton species is the availability of nutrients, which depends crucially on the hydrodynamic flow patterns in the oceans. Of particular interest in the context of plankton growth are mesoscale flow patterns like fronts, jets, and vortices, which have a strong impact on plankton patterns. Though the full velocity field in the ocean is 3D, giving rise to horizontal and vertical transport of nutrients and plankton, a 2D field is often a rather good approximation for the study of plankton patterns. On one hand, the vertical velocity component is frequently much smaller than the horizontal ones; on the other, since phytoplankton --- being the plants of the ocean --- need sunlight for photosynthesis, their abundance is confined to the upper layer of the ocean. These features make the investigation of phytoplankton blooms, i.e., the emergence of high phytoplankton abundances, an ideal application field for chaotic advection. Though ocean circulation models are far more realistic, the study of simple flows leading to typical chaotic advection problems have revealed many insights into the biological--physical interactions. This is due to the much higher spatial resolution of the flow structures which can be achieved by simple, mostly analytically given flows compared to ocean flows. In particular, the emerging fine structure of filamentary plankton patterns is responsible for certain phenomena that cannot be explained by coarse-grained oceanic flows. Concepts of chaotic advection have led to fundamental contributions to the understanding of the mechanisms of plankton-bloom formation as well as of the coexistence of plankton species.

To investigate the impact of mesoscale hydrodynamic structures on plankton growth, a variety of different kinematic models for the flow have been considered. Taking into account the arguments mentioned above, 2D horizontal flow patterns have been studied in which the velocity field can be described
by a stream function. Three paradigmatic models exhibiting certain important features of an oceanic flow have been
used to elucidate the interplay between plankton dynamics and hydrodynamic
flows: (i) the blinking vortex flow \cite{Neufeld-et-al-02},
(ii) the flow in the wake of an obstacle \cite{Jung-et-al-93,Sandulescu-et-al-06}, and (iii) a jet flow \cite{Lopez-et-al-01b}. All of these
flows are periodically forced, which leads to chaotic advection of
passive tracers. Each of them focuses on particular properties of a
real flow. While (i) introduces a temporarily changing mixing
region, (ii) mimics the dynamics of a von K\'arm\'an vortex street in the
wake of an island located in an ocean current, while (iii) is a simple description of a meandering jet. Besides these idealized flow fields, 2D turbulence models are used in more realistic settings \cite{Martin-et-al-02}. These velocity fields are coupled to simplified nonlinear plankton growth models that often consist of two or three interacting species, e.g., nutrients, phytoplankton, and zooplankton. In some cases plankton blooms are modeled with excitable dynamics \cite{Neufeld-et-al-02}. The combination of these simplified plankton models and the above-mentioned kinematic flows in the corresponding reaction--diffusion--advection equations are the basis of the study of chaotic advection in plankton growth.

\begin{figure}[tb]
\includegraphics[width=\columnwidth]{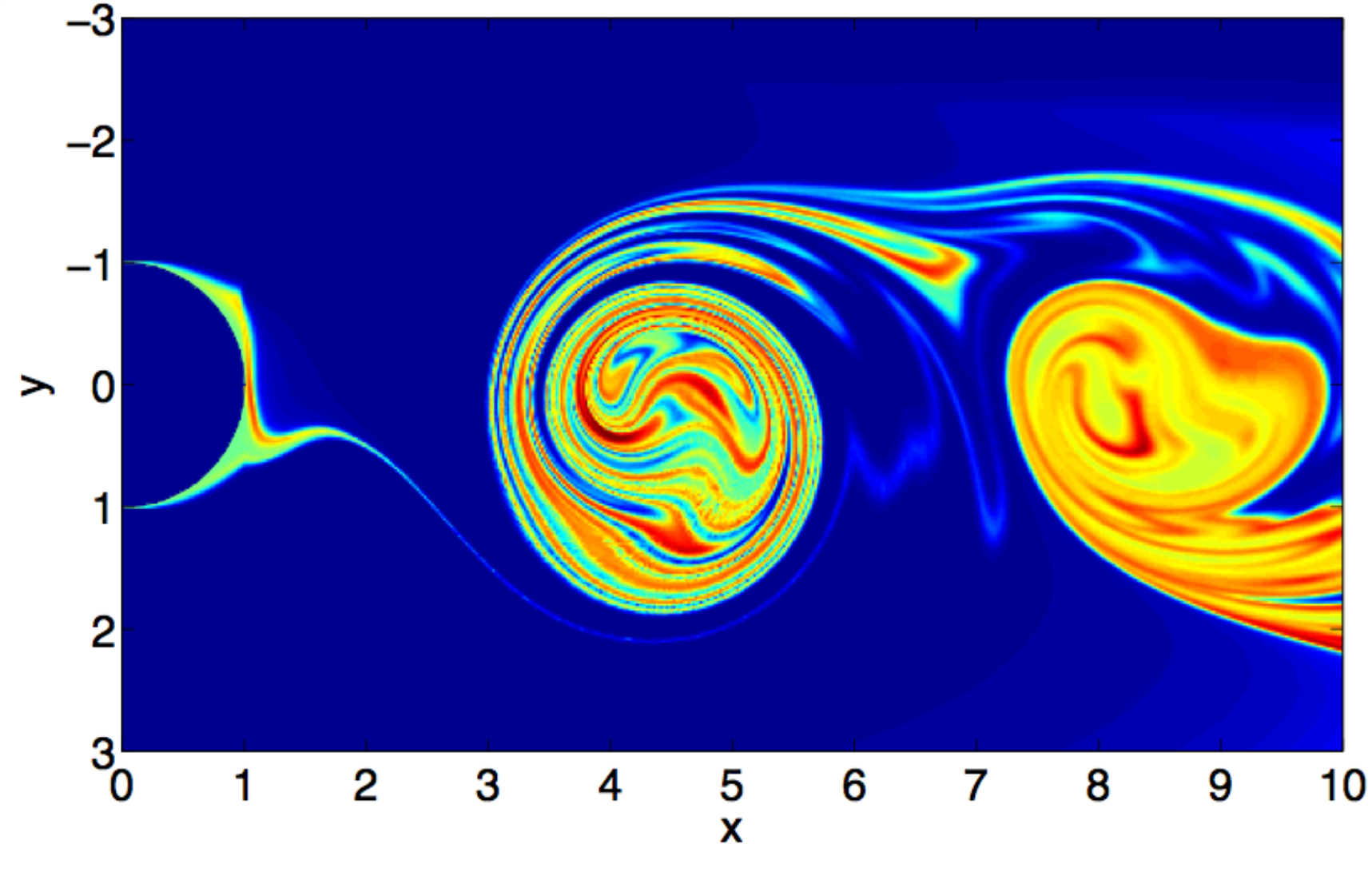}
\caption{Plankton bloom: Snapshot of the simulated phytoplankton concentration in the wake of an island (at left) with
  low inflow concentrations from the surrounding ocean (inflow from the left). Concentrations of phytoplankton
  from low to high are denoted by colors and shades. 
 }
\label{bloom}
\end{figure}

Using these approaches it has been shown that the redistribution of nutrients and plankton by horizontal stirring can enhance the growth of plankton \cite{Abraham-98,Lopez-et-al-01b,Hernandez-Garcia-et-al-02,Hernandez-Garcia-et-al-03,Martin-03} and even cause \cite{Karolyi-et-al-00} and sustain \cite{Hernandez-Garcia-Lopez-04} a plankton bloom. Moreover, mesoscale hydrodynamic vortices can serve as incubators of plankton blooms by keeping the plankton in the interior of the vortex without much exchange with its exterior \cite{Martin-03,Sandulescu-et-al-07}. In general the timescale of the hydrodynamic processes is much faster than the timescale of biological growth. However, the confinement of plankton within the vortex leads to a much larger residence time of the planktonic organisms in a certain area, so that growth processes can take place; see Fig.~\ref{bloom}.

An even simpler approach for growth processes has been studied by
considering autocatalytic reactions
\cite{Metcalfe_autocatalytic_1994,Toroczkai-et-al-98,Karoly-et-al-99,Tel-et-al-00}.
Planktonic organisms are considered as point tracers that are
advected by the flow and that undergo a reaction $A + B \rightarrow
2B$. Using this modeling technique it has been shown that the
biological activity is enhanced since the fractal filaments formed by
chaotic advection lead to an increase in the reaction surface and,
hence, of the reaction rate.  The fractal skeleton of the reaction is
characterized by its capacity dimension $D_0$ which in turn is an
essential parameter that can be used to estimate the scaling
properties of the reaction rate \cite{Tel-et-al-04}.
On the other hand, the enhancement of autocatalysis due to chaotic advection has been proposed as an explanation for deracemization during the chiral crystallization of achiral molecules \cite{Metcalfe_autocatalytic_1994}. In this case, an agent-based model constructed to describe the breakdown and subsequent reproduction of growing crystals submitted to shear stresses is immersed in a chaotic flow. As a result an unbalanced population of crystals with both chiralities is obtained \cite{Cartwright_chiral_2004}. When the agents are endowed also with a dynamics mimicking the Ostwald ripening phenomenon, one of the chiralities ends up being completely annihilated \cite{Cartwright_chiral_2007}.

The filamentary structures that develop in chaotic advection are not only advantageous for the growth of species but lead also to an increased separation of different species competing for the same nutrients \cite{Bracco-et-al-00,Karolyi-et-al-00,Scheuring-et-al-03,Bastine-Feudel-10}. As a consequence, competition is diminished, so that more species feeding upon the same resources can coexist. Weaker species with smaller abundance obtain more access to resources along the fractal filaments and are able to survive and compete with stronger competitors.  This effect is even stronger when the inertia of the organisms is taken into account \cite{Benczik-et-al-06}. In this way, chaotic advection can be considered as a possible mechanism to solve the \emph{paradox of plankton}
\cite{Hutchinson-61}, an intriguing problem in theoretical ecology.

\section{Perspectives}\label{perspectives}

For efficient mixing to be achieved, the velocity field must stir together different portions of the fluid to within a scale that is small enough for diffusion to take over and homogenize the concentrations of the advected quantities. One traditional way to do this has been through fully-developed turbulence, where the coexisting eddies of many different scales do the job of generating the large concentration gradients at small scales which molecular diffusion smoothes out. The other approach, which we review in this work, is to generate these small-scale structures through the repeated processes of stretching and folding that characterize deterministic chaos. Unlike turbulent advection, chaotic advection does not involve an energy cascade, and it works even in flows that are too viscous or too small-scale to have large Reynolds numbers, or are effectively 2D. This generality provokes questions, conjectures, remarks and reflections  from a wide range of applications and theory. In this section we raise some of the current challenges and opportunities in chaotic advection.

We do so under the assumption that an engineer faced with a mixing problem, whether from a traditional industry such as textiles, food or chemicals, or a more technologically sophisticated field such as micro- or nano-science, is already well aware of the potential advantages that chaotic advection can bring. The notion of mixing using chaotic dynamics is by now common-place. What then are the timely and relevant questions to ask the engineer, and which aspects of theory and practice can now be developed to answer questions posed by the engineer in return? 
Such questions may include: the difficulty of optimizing a mixing process, both in general and in particular circumstances; the effect of specific physical features of the flow, such as fluid inertia, boundaries and walls, and rheological behaviors; and how to describe, analyze and quantify chaotic advection in the most appropriate way. Answers to such questions require a synthesis between rigorous mathematics, experimental pragmatism, physical understanding and engineering intuition.

\subsection{How to choose the best stirring protocol for a given mixing problem?}

To maximize the performance of a mixing device there are typically many parameters, several of which are particular to that device. Some features are common to a range of generic devices, however, and as such suggest general schemes for optimization. 

For example, blinking flows and channel mixers have a natural temporal or spatial periodicity to which classical dynamical-systems theory can be applied. Less well understood is the influence of aperiodicity in stirring protocols \cite{Liu94}. Between the two extremes of a periodic stirring protocol, and a random one, one can introduce a small perturbation in a periodic sequence in order to study the effect of the degree of aperiodicity. The periodic length of a sequence in the stirring protocol is also a parameter to test. It has been shown that the choice of a globally aperiodic sequence is not appropriate for an efficient stirring protocol \cite{gibout2006coupling}. Moreover, in the case of fixed-time switching between horizontal and vertical shears on a periodic domain, the best mixing (in the sense of Kolmogorov--Sinai entropy) has been rigorously shown to be achieved by the simplest periodic protocol \cite{d'alessandroetal:1999}. This was later used to prove optimality of mixing protocols in \textcite{boyland2000topological}.

The geometrical details of a mixer can also be considered parameters for optimization. For example, is the angle between superimposed crossing streamlines taken at two different times, and their distribution over the fluid domains, a good criterion for the optimization of mixing? One attempt to investigate this can be found in \textcite{sturman2009eulerian}. Does good mixing always require a fairly uniform distribution of stretching rates within the flow? How does the movement of a mixer wall, or its radius of curvature affect the amplitude of the modulation, and in turn the mixing quality and efficiency? What is the relation between stretching fields and curvature fields, or the differences between cases of bounded and unbounded flows for the same stirring protocols? What is the influence of the phase shift between the waveforms applied to different rotating or moving walls. How does the response of the scalar field orientation to a modification of the stirring protocol impact on the quality of mixing? Many of these questions do not yet have full answers; see also \textcite{balasuriya2005optimal, balasuriya2005approach, balasuriya2010,cortelezzi2008feasibility}.

In applications, achieving a sufficiently good  mixing performance for the least energy input may be the chief motivation \cite{alvarez-hernandez2002}. Of the many existing mixing indicators and diagnostics discussed earlier, those which take into account the energy communicated to the fluid, or the energy saved, may be the most practical. \textcite{Mathew2007} coupled the two for the first time, using the mix-norm metric  supplemented by an energy measure. \textcite{lin2011optimal} advocated the use of an $H^{-1}$ norm, but $H^{-1/2}$ seems to have better  physical motivation, as it incorporates the idea of {\em equal distributions in equal volumes} for the perfect mixture. As always, the distinction between unsteady 2D flows and 3D steady flows is likely to be challenging, as is that between active and passive modes of generation of the advective flow. 

As the question of how to pick a stirring protocol only has a definitive theoretical answer in a specific model system, various computational approaches
have been developed to approximate an answer in 
applications.  For diffusive scalar transport the work of \textcite{Lester08a,Lester_lagrangian_2008,Lester_nonNewtonian_2009,Lester_RPM_2010,Lester_control_2014}
has produced a fast method to calculate the most slowly decaying (or
the first few) eigenmodes when the stirring flow consists of
one or a few basic flows (and symmetries of these base flows) that are
governed by a set of control parameters.  The general method
\cite{Lester08a} incorporates any symmetry, e.g., rotation,
reflection, scaling, superposition.  However, it still remains up to
the flow designer to determine the decomposition of fundamental flows
and symmetries.  

\begin{figure}[tb]
\centering
\includegraphics[width=\columnwidth]{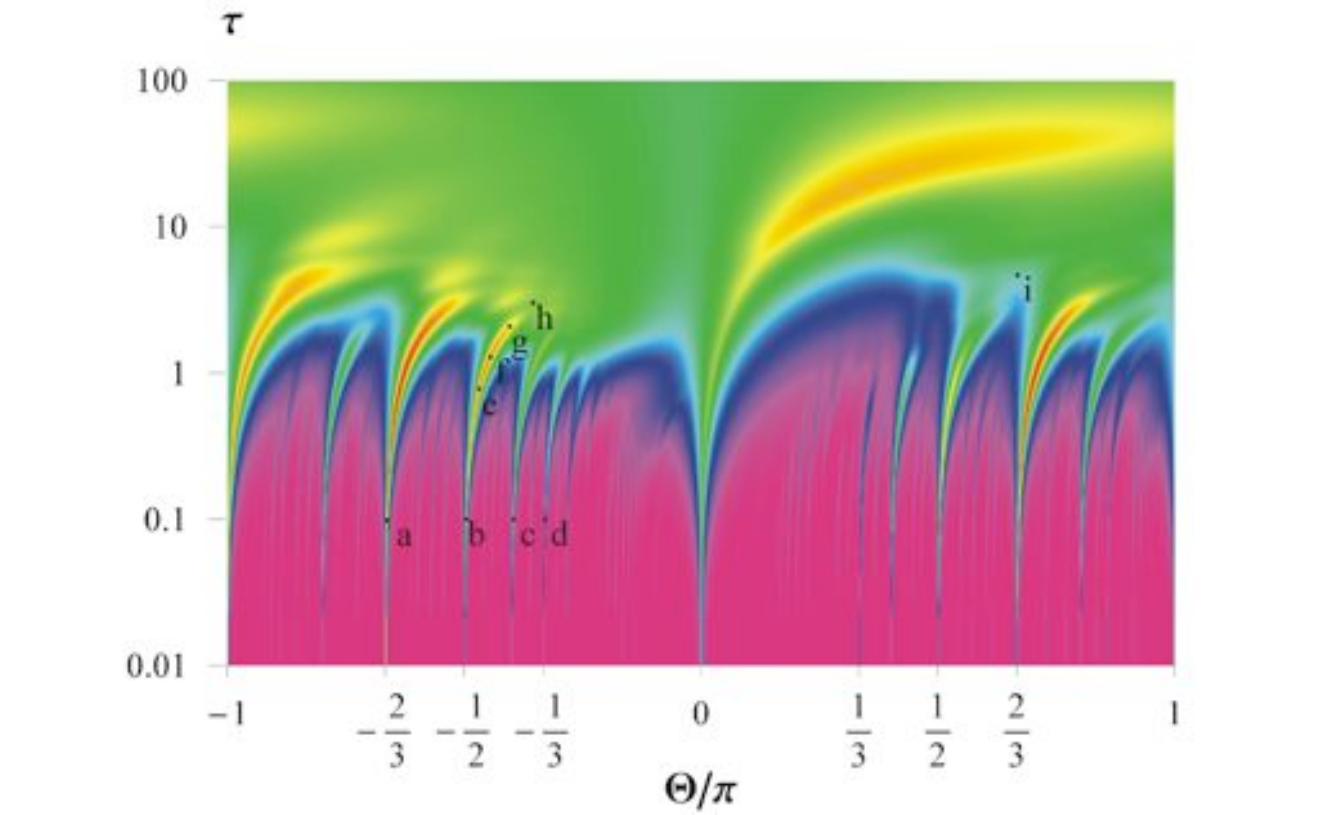}
\caption{Map of the asymptotic decay rate $Re(\lambda_0)$ scaled by
  the diffusion rate for Dirichlet boundary conditions,
  $\Delta={\pi}/{4}$, and $Pe=10^3$ for a periodically reoriented flow
  with $\tau$ the time between reorientation scaled by the flow
  circulation time and $\Theta$ the reorientation angle.  Note logarithmic scaling of
  $\tau$ axis and $\lambda_0$ contours.  Letters a--i are specific
  modes discussed in \protect\textcite{Lester_nonNewtonian_2009}.
  Reprinted from \textcite{Lester_nonNewtonian_2009} with permission from Elsevier.
 }
\label{fig:lambdacontour}
\end{figure}

For example, the RAM flow discussed in Section~\ref{sec:heat_transfer} has one
fundamental flow, from which all flows in the device are obtained by
rotation, and in the simplest case two control parameters, $\tau$, the
time any one flow operates, and $\Theta$, the fixed rotation angle of
the flow.  
Figure~\ref{fig:lambdacontour} is a contour plot of the asymptotic decay rate $Re(\lambda_0)$
calculated at $1.2 \times 10^5$ points over the $\tau$--$\Theta$ plane
for $Pe=10^3$ and homogeneous Dirichlet boundary conditions;
$\lambda_0$ is scaled by the diffusion rate (the most slowly decaying
eigenvalue of the diffusion operator in the disc).
This figure maps transport enhancement relative to
diffusion alone, and shows what can be
termed a complete parametric solution for scalar transport, in that
the optimum stirring protocol --- for the goal of thermal
homogenization --- can be read directly from the graph to the precision
of the computation.  \textcite{Lester08a} report that 
for symmetry-derived stirring
protocols, it is possible to obtain the complete parametric solution
about 6\,000 times faster.  The relative performance improves as
$Pe$ increases, but the absolute time to obtain the complete solution
also increases.

The structure of the scalar transport solution is worth commenting upon.  At low values of $\tau$ the enhancement distribution is fractal with many localized maxima, which can only be discerned with highly resolved solutions on the control parameter space.  The spiky regions originate from rational values of $\Theta/\pi$ at $\tau = 0$ and grow in width with increasing $\tau$.  Inside each region the scalar spatial distribution is locked into a symmetric pattern whose azimuthal wave number $m$ is a rational multiple of the forcing wave-number $k = 2\pi/\Theta$; these regions are symmetry-locked ``tongues'' similar to frequency-locked Arnol'd tongues \cite{Glazier_quasi_1988}.  This type of behavior was first observed in the context of micromixer design in \textcite{Mathew2004optimization} using the mix-norm as the measure of goodness of mixing. As $\tau$ becomes $\mathcal{O}(1)$ the spreading tongues interact to produce an order--disorder transition, where the symmetric spatial patterns change into asymmetric patterns.  At large values of $\tau > 10^3$ the enhancement ratio everywhere takes on the same value as that on the $\Theta = 0$ line, i.e., no enhancement.  For low $Pe$, such as $Pe = 10^3$ as in Fig.~\ref{fig:lambdacontour}, the maximum transport enhancement occurs in the $1/3$ resonance tongue at low values of $\tau$. This seems counter-intuitive because the no-diffusion mixing optimum \cite{Metcalfe_ram_2006,Speetjens_symmetry_2006}  occurs at $(\Theta, \tau) \approx (\pi/5, 15)$, and at low $\tau$ chaotic advection is, at best, weak.  As $Pe$ increases the maximum transport enhancement increases, and its location moves towards the fluid mixing optimum.

\subsection{Effect of fluid inertia}

In most chaotic-advection studies, a Stokes flow regime is envisaged
with low Reynolds numbers and a quasi-steady-state approximation of low
Strouhal number, $Sr$.  This implies that viscous forces dominate the
flow dynamics and leads the Navier--Stokes equations to be linear.
The linearity of the problem gives rise to reversibility of the flow
and a passive blob of tracer will return to its original place if the
effect of diffusion is weak or not amplified by chaotic advection.
Indeed, for chaotic-advection flows, the trajectories can be so
complex in space and time that a particle will lose the memory of its
initial starting position if we reverse the flow, due to the
accumulation of diffusion responsible for tiny jumps from streamline
to streamline, trapping events or L\'evy flights in hyperbolic regions
\citep{Solomon94,solomon2001}.  If fluid inertia is included (i.e.,
when $Re$ is increased but the flow is still laminar), the
nonlinearity of the Navier--Stokes equations introduces a new source
of transient dynamics that leads to breakdown of invariant manifolds
(see Section~\ref{Inertial} and the example of 3D unsteady flow in mergera
cylinder) and impacts strongly on the transport properties of the
flow.  How exactly the increase in $Re$ changes mixing properties is an
open problem at the intersection between Eulerian and Lagrangian
analyses.  There has been relatively little work on chaotic advection
that considers the effect of fluid inertia
\citep{dutta1995,mezic2001chaotic,horner2002,mezic2002extension,balasuriya2003weak,wang2009, Speetjens06b, MichelChaos,Pouransari2010}.
  Mezic and collaborators focus on possible nonmonotonic mixing created by  inertial effects, and specifically on the possibility of a decrease of mixing because of these. Follow-up experimental and numerical studies confirm these predictions \cite{lackey2006relationship,pratt2014chaotic}.
We should emphasize that inertia is not the only mechanism to induce chaos. It is also possible to obtain chaos from a breaking of time reversibility by the time derivative in the time-dependent Stokes equation, see \textcite{Hascoet2005}, or simply by minor imperfections in an experimental set-up, see \textcite{Wu2014}.

Inertial effects are also beginning to be used in microfluidic systems
\citep{dicarlo2009} where chaotic advection is now commonly
employed (see Section~\ref{Microfluidics}).  \textcite{horner2002}
introduced time dependence through transient
acceleration--deceleration of a flat wall driving the flow in a cavity
and, despite the fact that the streamline portrait
changes very little during the transition, they observed a significant
transport enhancement with the increase of the control
parameter  $Re Sr$.  In general, fluid inertia also introduces supplementary
secondary flows that bring new fixed parabolic or hyperbolic points
from which stable and unstable manifolds can emanate, thereby
improving the stirring process.  At the same time, inertial forces
influence chaotic advection by causing a distortion of the streamlines
as well as making them time-dependent \citep{wang2009}.

The effect of inertia is crucial for mixing in chaotic flows through a sequence of alternating curved ducts \citep{JTA89,castelain2001} for which it is necessary to have a large enough Dean number \citep{dean28} to produce a minimum transverse displacement of a passive scalar along a helical trajectory, otherwise the stirring will be ineffective. Such an arrangement of curved ducts is able to generate very complex flows with the presence of regular, partially or fully chaotic regions; they are used as mixers or reactors \citep{boesinger2005}. For a particular geometry, the coexistence of non-mixed islands and chaotic zones in the cross-section with the presence of the solid wall boundary depends highly on the flow rate and is a source of multimodality of the finite-time particle distributions along the pipe mixer, while a sole ergodic region in the transverse cross-section (globally chaotic flow) gives rise to a narrow distribution of residence times  \citep{mezic1999_2}.

Fluid inertia is also important in oscillating flows. \textcite{hydon1995} showed, for a pulsed flow in a curved pipe, that particles on certain trajectories are transported by resonant interaction between the secondary orbital motion and the longitudinal oscillation, inducing ballistic transport. For some trajectories trapped behind cantori, particles are transported intermittently by resonance; this mechanism leads to anomalous diffusion \citep{young1991}. An important physiological example is blood flow around a stenosis zone \citep{schelin2009,maiti2013}, for which chaotic advection is triggered by a combination of pulsating flow driven by the heart and constriction of the blood-vessel walls.

In the domain of geophysical flows, \textcite{pratt2014chaotic} have recently revisited the problem of chaotic advection within a 3D steady flow in a closed cylinder with a rotating bottom.  They show a non-monotonic variation of the stirring rate with the Reynolds number (more exactly with the Ekman number $Ek$, which is directly proportional to the inverse of the Reynolds number for fixed Rossby number). The bulk stirring rate, estimated from the tracer concentration variance, has a maximum at intermediate $Ek^{-1/2}$ and a complex dependency of the Lagrangian structures is noted, with thin or thick resonant layers sandwiched between KAM barriers; compare \textcite{mezic2001chaotic}.

The issue of fluid inertia also arises when addressing the question of
periodicity of the flow which does not necessarily match the
geometrical periodicity of the wall boundaries chosen for the
generation of chaotic advection \citep{blancher2014}.  This has direct
implication for the appropriate choice of the periodic velocity field
in order to map the flow as in the context of dynamical systems.

\begin{figure}[tb]
\centering
\includegraphics[width=0.9\columnwidth]{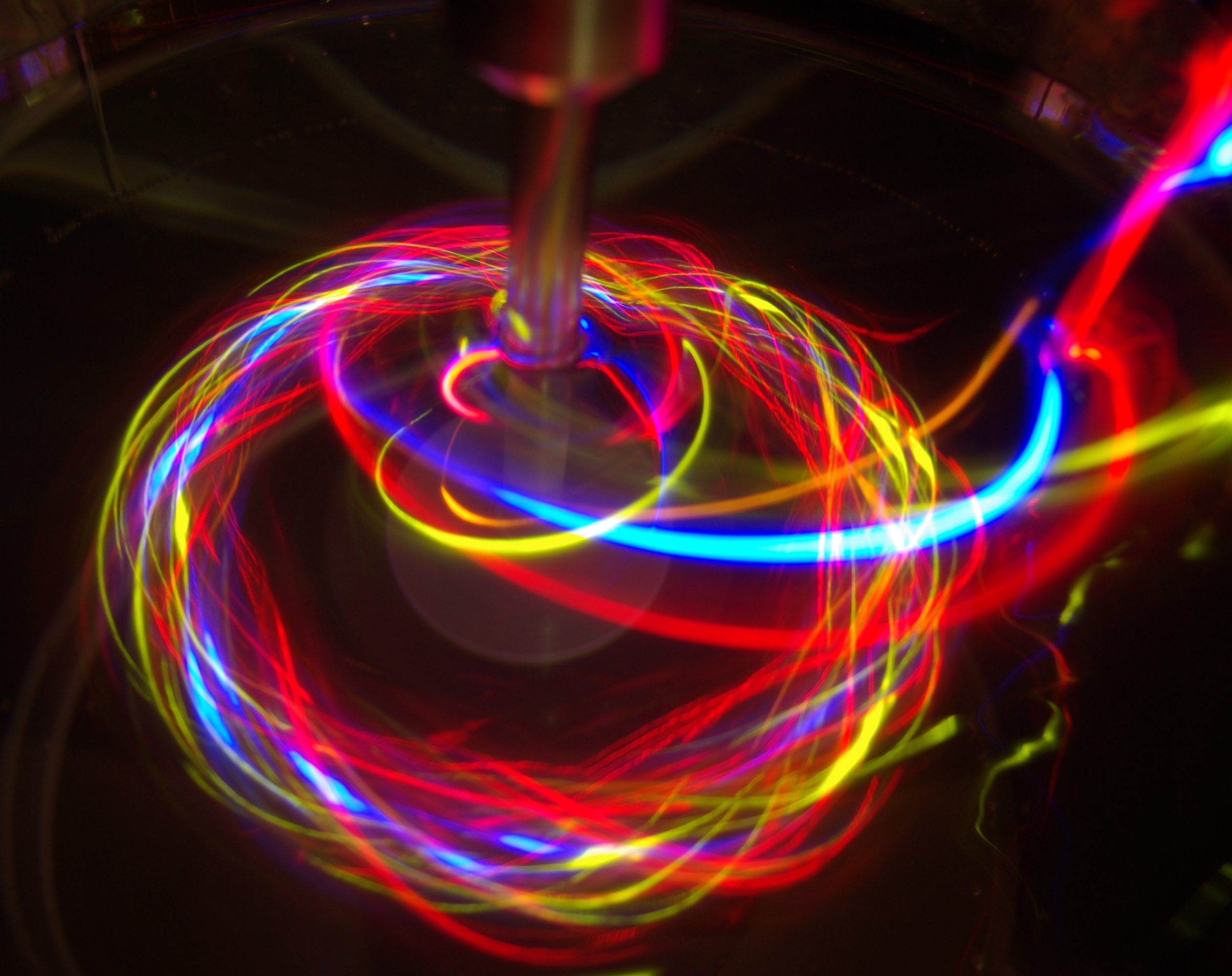}
\caption{Long exposure photograph of several finite Stokes
number particles moving in the laminar flow of a stirred tank.  All
particles start in a chaotic flow region but are eventually attracted
into a KAM tube where they execute helical trajectories.  Photo from
Wang, Stewart \& Metcalfe.}
\label{fig:light_streams}
\end{figure}

When
we consider non-passive scalar transport (i.e., diffusive or
non-buoyant scalars) such as finite-mass and -size particles, the
transient effects due to particle inertia and fluid inertia are
intimately linked owing to the presence of new forces.  Indeed, a non-passive particle responds to
intrinsic hydrodynamic forces, to externally imposed forces, or to
internally generated forces \citep{Metcalfe_beyond_2012}.  The
hydrodynamic forces include drag force, lift force, added mass force,
buoyancy force, Boussinesq--Basset history force \cite{Babiano2000} and also Brownian motion that
produces molecular diffusion \citep{Springer2010}.  The external forces are produced by an
external force field (i.e., an electric or magnetic field) and the
internal forces refer essentially to motile organisms or active
particles.  The dynamics of a finite-size particle in a chaotic flow
is thus nontrivial and can lead to very different and rich behaviors
because the particle's trajectory can diverge from the trajectory of
the fluid parcel, since a slip velocity exists between the particle
and the surrounding fluid whose velocity can eventually fluctuate in a
random manner.  The fact that the finite-size particle dynamics is
dissipative furthermore complicates the story by allowing the
existence of attractors in the phase space \citep{Springer2010}.
Figure~\ref{fig:light_streams} shows trajectories of
several finite Stokes number particles localizing onto an attracting
KAM tube in a 3D laminar flow. Predictions of what level of inertia
will cause spontaneous localization of finite inertia particles have
been quantified in recent experiments   \citep{Wang_separation_2014,Wang_visualization_2016}.  A direct
consequence is that invariant curves no longer exist for the case of
inertial finite-size particles.

\subsection{How to control mixing?}

The first question to address is: Why would we enhance mixing, and for what? Very often the answer is: to increase its quality (i.e., to obtain a complete homogenization) and its rate (i.e., the speed to achieve the uniform state).  For many engineering applications, however,  the problem is far from being so simple and the objectives of the mixing problem could be completely different depending on the application field (reactor and polymer engineering, heat exchangers, processes with particles or powders, combustion, etc). Heat and mass transfer, reactions, multiphase flows and various coupling mechanisms occur in these systems at a wide range of space and time scales. The geometry is also of primary importance: micro- or macro-fluidics, closed or open flow, static or moving boundaries, and so on. If the goal is carefully chosen for the mixing problem under consideration with all its constraints, the choice of the control parameter(s) can be envisaged. But immediately two other questions arise: what is the optimal manner to reach the goal function and which is the appropriate measure to adopt for the optimization of mixing?

This question of generating effective mixing involves solving an optimal control problem. Pioneering work was done by  \textcite{d'alessandroetal:1999} in this domain. They used mathematical tools from control and ergodic theory of dynamical systems to derive rigorously the mixing protocol that maximizes entropy among all the possible periodic sequences composed of two shear flows orthogonal to each other (i.e., the eggbeater flow). Other studies have followed in the same spirit \cite{vikhansky2002,vainchtein2004,stremler2006,Mathew2007,liu2008,gubanov2010,gubanov2012, lin2011optimal,lunasin2012}. 
Work of \textcite{aamo2003} has addressed the problem of enhancing mixing by means of active boundary feedback control in 2D channel flow.  Very efficient mixing was obtained by designing feedback control strategies for the stabilization of the parabolic equilibrium flow, then applying this feedback with the sign of the input reversed. \textcite{foures2014} examine different
norms for mixing efficiency and show  the impact of the mix-norm theory in higher-Reynolds-number flows.

Despite the fact that many theoretical ingredients are in place to address the problem of optimizing mixing, most of the studies to date concern idealized model flows. In recent work, \textcite{balasuriya2010} developed a theoretical tool for the determination of an optimal forcing frequency for a time-periodic mixing strategy in order to increase mixing across a fluid interface between two fluids in a microfluidic device. In engineering processes, the aforementioned strong coupling between different physical phenomena renders the theoretical design of an efficient mixing strategy very challenging. As an example, we can cite the interesting problem of optimization of the formation of an emulsion, a dispersion of droplets in a viscous fluid, in a laminar flow \cite{caubet2011} by controlling the droplet size distribution, which influences important properties of the emulsion for cosmetic or food applications such as rheology, texture, shelf life (stability), appearance or taste. For such an application we have contradictory targets: on one hand, we need to achieve an efficient mixing in order to avoid KAM regions that could trap large droplets, which implies imposing a modulation of the wall velocity of the stirring device. On the other hand, a continuous movement of the rotating elements is needed in order to achieve a continuous stretching of large droplets to promote their transformation into elongated threads and their rupture by the Rayleigh instability. Indeed, a decrease of the wall velocity will cause the relaxation of the stretched threads towards their initial spherical shape. Thus, in this case, a particular stirring strategy with the consideration of two contradictory objectives needs to be found. Another example of a complex mixing problem is the optimization of chaotic advection in the case of pulsatile switching flows in porous media, with applications to soil remediation or enhanced oil recovery \cite{trefry2012, lester2013}.

\subsection{Dynamics of the wall}

Since almost all flows of interest have solid boundaries, it is obviously important to establish what effects walls have on the dynamics of mixing. This topic has been surprisingly little explored, and although there are some important results described below, much remains to be done.

Walls have a dramatic effect on the rate of mixing because the no-slip boundary condition ensures that all points on (stationary) walls are parabolic fixed points. This manifold of non-isolated degenerate fixed points makes the resulting advective dynamical system non-hyperbolic. But this non-hyperbolicity is different from that arising from the presence of KAM islands, because the fractal hierarchy of cantori and inner islands is absent near walls. 

In closed flows (Section~\ref{closed}), the presence of walls slows down mixing, as should be expected from the parabolic fixed points on the wall: as reported in \textcite{Gouillart2007}, the variance of the concentration distribution of a passively advected scalar field decays in time following an algebraic law, as opposed to the exponential decay observed in the absence of boundaries. This is caused by the slow dynamics of the portions of the fluid in the vicinity of the walls. Interestingly, if the walls of a container are rotated as part of the mixing protocol, it has been demonstrated experimentally \cite{Gouillart2010} that the exponential decay of correlations is recovered. This is achieved because rotation of the walls creates a dynamical barrier surrounding a central chaotic region, and isolating it from the wall. This shows that moving boundaries can result in very different mixing dynamics, and there is currently no general theory capable of predicting mixing dynamics in the presence of multiple moving boundaries. 
In the classic system of fluid in a closed container (fixed walls) being mixed by moving rods (moving boundaries), in some circumstances one can prove that the advective dynamics has a strictly positive topological entropy, and is therefore chaotic \cite{Thiffeault2008}. Despite that, the variance of an advected scalar still decays exponentially because of the walls. This highlights the fact that there is no general understanding of the connection between dynamical invariance of the advective system and the mixing dynamics of an advected scalar with finite diffusion.

For open flows in a fluid transported from the inlet to the outlet of a pipe mixing device, it will be interesting to relate the form of the distribution of the residence times associated with elementary fluid particles to the chaotic behavior of the flow. It is known that a globally chaotic flow tightens the distribution of residence times, but how the existence of parabolic points along the wall boundaries and the fluid behavior at the vicinity of these hyperbolic points influence the shape of the distribution of residence time has not been studied. This will be particularly interesting for reactive problems in chemical engineering.

If a scalar is being input into the fluid by a continuous source, the
variance will not decay to zero in time, and can be expected to reach
a steady state or a periodic state, depending on the dynamics of the
source and of the flow.  This will almost always be the case for open
flows, for example a chemical reactor.  It can also be the case in
closed flows, for example if we expect ohmic heating of the
fluid in its volume, the heating or cooling of the fluid is done
through the wall boundaries, and thus temperature acts as a continuous
scalar source.  In these cases, the asymptotic scalar distribution
after the transient depends not only on the chaotic dynamics of the
flow, but also on details of the sources.  This is again poorly
understood at the moment, and it is clearly important in practical
applications of chaotic mixing.

Finally, chaos can be generated directly by the walls in more complex
duct networks that contain branchings and mergings of ducts, e.g., in
pore networks.  Many of the branch and merge locations will have
hyperbolic node points in the skin friction field.  If the branch and
merge are symmetric, the unstable and stable manifolds emanating from
the branching and merging hyperbolic points respectively will join
smoothly.  However, if the branch and merge pair are twisted, then the
hyperbolic manifolds must intersect transversely, generating chaotic
fluid trajectories.  \textcite{lester2013} developed this
argument and used it to prove that chaos is inherent to porous media
flows, where topological considerations guarentee an abundance of
hyperbolic points.  It turns out that chaos, through walls, has
considerable influence on the macroscopic transport properties of
natural and engineered porous media \cite{Lester_anomalous_2014}.
The dispersion of a scalar in an isotropic porous medium with a characteristic length over which neighboring fluid-element paths de-correlate is directly proportional to this length and to the interstitial fluid speed \cite{phillips1991}. Such a relation is responsible for the observed fast chemical transport, typically orders of magnitude larger than expected from molecular diffusion, in applications ranging from buoyancy-driven flows in soil \cite{Kal2005} to fluid motion over wavy precipitates in hydrothermal vents \cite{Ding2016}. 
There is definitely more to be understood about chaotic advection in porous media.

\subsection{Strange eigenmodes}

Strange eigenmodes --- first discussed by \textcite{Pierrehumbert1994} ---
characterize the mixing dynamics of advected fields in flows, and have been observed experimentally in both closed and open flows. 
Because of the non-self-adjoint nature of the advection--diffusion operator, completeness of the eigenfunctions is not guaranteed. Rigorous results on the existence and completeness have been obtained by  \textcite{liu2004}. Although strange eigenmodes have been observed experimentally in open flows, a corresponding rigorous result for open flows is missing. More generally, very little is known about strange eigenmodes in open flows. For example, questions about how the eigenmodes are related to the fractal structure of the chaotic saddle and its stable and unstable manifolds are still open.

\subsection{3D unsteady flows}

Great progress has been made on Lagrangian transport phenomena in 3D unsteady flows since the first pioneering studies from the mid-1980s \cite{Dombre1986,Feingold1987,Feingold1988}. However, many challenges remain. Firstly, 3D spaces admit far greater topological complexity of coherent structures compared to their counterparts in 2D \cite{Alexandroff1961}.  Secondly, the 3D equations of motion lack the well-defined Hamiltonian structure of 2D configurations. (3D steady systems under certain conditions admit representation as a 2D non-autonomous Hamiltonian systems (\textcite{Bajer1994,mezic1994thesis}; see Section~\ref{turbulent}). The latter in particular continues to be a major obstacle to advances in 3D Lagrangian transport phenomena.

Development of a comprehensive mathematical framework thus is imperative for a systematic description and analysis of 3D Lagrangian transport. Theoretical developments primarily expand on the classical Hamiltonian concept of action-angle variables by (local) representation of coherent structures as invariant surfaces and curves defined by constants of motion, denoted ``actions'' in Hamiltonian terminology (see e.g.,
\citet{Feingold1987,Feingold1988,Mackay1994,Mezic1994,Cartwright1996,Mezic2001}). A promising recent concept with Hamiltonian foundations devised specifically for mixing applications is found in the linked twist map \cite{sturman,meier07,sturman08}.

Establishment of a Hamiltonian-like formalism for 3D Lagrangian transport --- in particular response scenarios to perturbations and routes to chaos ---  is nonetheless in its infancy. The most important generalization of classical Hamiltonian mechanics to 3D systems is the 3D
counterpart to the KAM theorem, describing the fate of invariant tori under weak perturbations
\cite{cheng90,Mezic1994,broer96,vaidya2012existence}. However, similar universal response scenarios for coherent structures of different topology, most notably the important case of invariant spheroids, remain outstanding. Moreover, the scope must be widened to the effect of strong perturbations on the flow topology. Generically, isolated periodic points and periodic lines will emerge. 
However, the routes towards such states are largely unknown terrain, though bifurcations similar to those studied in \citet{Mullowney2005,Mullowney2008} are likely to play a pivotal role. Further reconciliation with concepts from mathematical physics \cite{Arnold1978,Arnold1998,Mezic1994,Haller1998,Bennet2006} and magnetohydrodynamics \cite{Biskamp1993,Moffatt1992,Moffatt2000} is essential in strengthening the current framework.

Theoretical developments largely concentrate on kinematic properties of divergence-free vector fields and volume-preserving maps and thus in essence account only for the role of continuity in the dynamics. However, the impact of momentum conservation as a facilitator or inhibitor of certain kinematic events in realistic fluid flows must be an integral part of scientific studies. A fundamental question is whether universal dynamic conditions, reminiscent of the Beltrami condition for 3D steady Euler flows, exist for 3D chaos and/or (degrees of) integrability in flows with significant viscous effects. A step towards this is in \textcite{mezic2002extension}.

Essentially 3D phenomena similar to those exemplified by the 3D cylinder flow in Section~\ref{3D} have been observed in a wide range of 3D unsteady systems. This includes generic 3D volume-preserving maps with non-toroidal invariant surfaces \cite{gomez02,Mullowney2005,Mullowney2008}, the 3D sphere-driven flow of \textcite{Moharana2013}, the 3D lid-driven cube of \textcite{anderson99,anderson2006} and 3D granular flows inside a spherical tumbler \cite{meier07,sturman08}.
The emergence of similar dynamics in this great diversity of systems reflects the universality of many of the observed phenomena.

Efforts to date overwhelmingly concern theoretical and computational investigations. Experimental studies remain scarce yet are essential to conclusively establish the physical validity, relevance and robustness of predicted phenomena. Moreover, experiments enable exposition and exploration of phenomena that are beyond present models and may thus contribute to progress in the field at a far deeper level than validation and verification alone.
Available techniques include laser-induced fluorescence (LIF) for visualization of coherent structures  \cite{Fountain1998,Fountain2000,sotiropoulos2002experiments,mezic2002ergodic,alvarez2002} and 3D particle-tracking velocimetry (3DPTV) for measurement of 3D Lagrangian fluid trajectories and 3D velocity fields 
enabling direct (quantitative) measurements on coherent structures
\cite{luethi}. Experimental studies have demonstrated the capabilities of 3DPTV for
such analyses \cite{Michel2004,Otto2008,Dore2009,Cheng2011,Znaien2012,Jilisen2012}.

A formidable challenge in 3D transport studies is the visualization and isolation of coherent structures in numerical and experimental data. Poincar\'{e} sectioning is the method of choice for time-periodic flows yet has the drawback that results depend critically on the initial tracer positions. The ergodic-partition method following \citet{mezic1994thesis,Budisic2012, mezic2013analysis}, reviewed in Section~\ref{visualization}, in essence generalizes this concept and --- in principle --- enables visualization of all relevant topological features with one generic ansatz. Further promising alternatives exist in the Lagrangian techniques by \citet{Haller2001,Branicki2009} developed specifically for 3D unsteady systems.

\subsection{Synthesis}

Why is this subject attractive? Firstly, it is rather interdisciplinary. There are typically at least three areas that are involved: (a) fluid mechanics; (b) dynamical-systems theory; and (c) the application area, which may be from other parts of physics or from engineering, biology, chemistry, etc. Such interdisciplinary problems are typically challenging and thus scientifically exciting. Secondly, analysis, experiment, and numerics are all possible and productive. Thirdly, the fundamentals are quite close to applications. Fourthly, 
mixing via chaotic fluid motion is a case where chaos is ``good''. Usually, chaos is associated with something ``bad'': undesirable vibrations in machinery, loss of particles in accelerators, difficulty in control problems, yet another type of noise, etc.  But in fluid mixing, chaos is good. We want as much of it as we can possibly get. In fact, we go to great lengths to eliminate or design away any vestiges of regularity. Fifthly, the concepts are applicable over a substantial range of length-scales ($\sim10^{-6}$ to $10^{5}$~m), comparable to the ten-decade range of length-scales found in turbulent flows in industrial applications and geophysical flows.   
Sixthly, there are still a number of interesting open problems. Finally, we must note that the data from experiments in this field are often aesthetically pleasing, as may be seen from the figures in this paper, which adds an extra attraction to chaotic advection.

The community of researchers working on the fundamentals and applications of chaotic advection has made great strides since the 1980s when the field was thus named.
Its impact has been felt in many fields of science and engineering and
the area of applications has grown enormously.
Yet, the theory is still incomplete and 
presents great challenges. Moreover, in spite of the hundreds of references cited covering a broad range of areas we are aware that important areas of interest to many possible readers are not covered in this review, such as those on non-Newtonian flows, transport in granular media, chemical reactions in chaotic flows, and so on. 

Key to advancing in this field is: developing a thorough understanding of laminar transport mechanisms in realistic fluid systems
as used in industry, including non-Newtonian fluids, chemical reactions, aggregation processes;
their rigorous experimental validation; further development of transport formalisms on the basis of principles from mathematical physics; their translation and integration into analysis and design strategies; and further development of numerical and experimental methods for transport studies.
As we write, new frontiers are being explored. 

\begin{quote}
The physics of mixing \\
is the subject to enter. \\
So say researchers \\
at the famed Lorentz Center.

It may sound quite inane \\
this exact predilection, \\
but there is nothing mundane \\
in chaotic advection.

Experiment rules \\
the precise mathematics \\
that eventually leads \\
to chaotic dynamics.

The flood gates now open \\
and every technique \\
is applied to the problem \\
however unique.

Poincar\'e to Koopman \\
and Ulam to Stokes; \\
all methods are chosen \\
by these different folks.

At the end of the workshop \\
one looks at each one, \\
but they still stir their tea \\
as they always have done!

So thanks to Herman, \\
GertJan and Michel. \\
We enjoyed being with you, \\
and with Julyan as well.

\emph{Hassan Aref}, \\
\emph{Lorentz Center, Leiden, January 2011.}
\end{quote}

\begin{acknowledgments}
This paper has grown out of the workshop ``Physics of Mixing'' organized by Cartwright, Clercx, van Heijst, and Speetjens, attended by the authors at the Lorentz Center, Leiden from 24--28 January 2011. We thank the Lorentz Center for its financial support, and the Lorentz Center staff for making the workshop a great success. 
We acknowledge funding from Netherlands (STW 11054; STW 10458; FOM 05PR2474) and Spain (FIS2013-48444-C2-1-P; FIS2013-48444-C2-2-P;  FIS2016-77692-C2-1-P; FIS2016-77692-C2-2-P).
We acknowledge the three of us who are no longer with us, Hassan Aref (1950--2011), John Blake (1947--2016), and Slava (Viatcheslav) Meleshko (1951--2011), with whom we planned and worked on this paper.
\end{acknowledgments}

\bibliographystyle{apsrmp4-1long}
\bibliography{frontiers}

\begin{thebibliography}{445}%
\makeatletter
\providecommand \@ifxundefined [1]{%
 \ifx #1\undefined \expandafter \@firstoftwo
 \else \expandafter \@secondoftwo
\fi
}%
\providecommand \@ifnum [1]{%
 \ifnum #1\expandafter \@firstoftwo
 \else \expandafter \@secondoftwo
\fi
}%
\providecommand \natexlab [1]{#1}%
\providecommand \enquote [1]{``#1''}%
\providecommand \bibnamefont  [1]{#1}%
\providecommand \bibfnamefont [1]{#1}%
\providecommand \citenamefont [1]{#1}%
\providecommand\href[0]{\@sanitize\@href}%
\providecommand\@href[1]{\endgroup\@@startlink{#1}\endgroup\@@href}%
\providecommand\@@href[1]{#1\@@endlink}%
\providecommand \@sanitize [0]{\begingroup\catcode`\&12\catcode`\#12\relax}%
\@ifxundefined \pdfoutput {\@firstoftwo}{%
 \@ifnum{\z@=\pdfoutput}{\@firstoftwo}{\@secondoftwo}%
}{%
 \providecommand\@@startlink[1]{\leavevmode\special{html:<a href="#1">}}%
 \providecommand\@@endlink[0]{\special{html:</a>}}%
}{%
 \providecommand\@@startlink[1]{%
  \leavevmode
  \pdfstartlink
   attr{/Border[0 0 1 ]/H/I/C[0 1 1]}%
   user{/Subtype/Link/A<</Type/Action/S/URI/URI(#1)>>}%
  \relax
 }%
 \providecommand\@@endlink[0]{\pdfendlink}%
}%
\providecommand \url  [0]{\begingroup\@sanitize \@url }%
\providecommand \@url [1]{\endgroup\@href {#1}{\urlprefix}}%
\providecommand \urlprefix [0]{URL }%
\providecommand \Eprint[0]{\href }%
\@ifxundefined \urlstyle {%
  \providecommand \doi [1]{doi:\discretionary{}{}{}#1}%
}{%
  \providecommand \doi [0]{doi:\discretionary{}{}{}\begingroup
  \urlstyle{rm}\Url }%
}%
\providecommand \doibase [0]{http://dx.doi.org/}%
\providecommand \Doi[1]{\href{\doibase#1}}%
\providecommand \bibAnnote [3]{%
  \BibitemShut{#1}%
  \begin{quotation}\noindent
    \textsc{Key:}\ #2\\\textsc{Annotation:}\ #3%
  \end{quotation}%
}%
\providecommand \bibAnnoteFile [2]{%
  \IfFileExists{#2}{\bibAnnote {#1} {#2} {\input{#2}}}{}%
}%
\providecommand \typeout [0]{\immediate \write \m@ne }%
\providecommand \selectlanguage [0]{\@gobble}%
\providecommand \bibinfo [0]{\@secondoftwo}%
\providecommand \bibfield [0]{\@secondoftwo}%
\providecommand \translation [1]{[#1]}%
\providecommand \BibitemOpen[0]{}%
\providecommand \bibitemStop [0]{}%
\providecommand \bibitemNoStop [0]{.\EOS\space}%
\providecommand \EOS [0]{\spacefactor3000\relax}%
\providecommand \BibitemShut [1]{\csname bibitem#1\endcsname}%
\bibitem[{\citenamefont{Aamo}\ \emph{et~al.}(2003)\citenamefont{Aamo},
  \citenamefont{Krsti\'{c}},\ and\ \citenamefont{Bewley}}]{aamo2003}%
  \BibitemOpen
  \bibfield{author}{%
  \bibinfo {author} {\bibnamefont{Aamo}, \bibfnamefont{O.M.}}, \bibinfo
  {author} {\bibfnamefont{M.}~\bibnamefont{Krsti\'{c}}},\ and\ \bibinfo
  {author} {\bibfnamefont{T.R.}\ \bibnamefont{Bewley}}}%
  , \bibinfo {year} {2003},\ \bibfield{title}{%
  \enquote{\bibinfo {title} {Control of mixing by boundary feedback in {2D}
  channel flow},}\ }%
  \bibfield{journal}{%
  \bibinfo {journal} {Automatica}\ }%
  \textbf{\bibinfo {volume} {39}},\ \bibinfo {pages} {1597--1606}%
  \bibAnnoteFile{NoStop}{aamo2003}%
\bibitem[{\citenamefont{Abraham}(1998)}]{Abraham-98}%
  \BibitemOpen
  \bibfield{author}{%
  \bibinfo {author} {\bibnamefont{Abraham}, \bibfnamefont{E.}}}%
  , \bibinfo {year} {1998},\ \bibfield{title}{%
  \enquote{\bibinfo {title} {The generation of plankton patchiness by turbulent
  stirring},}\ }%
  \bibfield{journal}{%
  \bibinfo {journal} {Nature}\ }%
  \textbf{\bibinfo {volume} {391}},\ \bibinfo {pages} {577--580}%
  \bibAnnoteFile{NoStop}{Abraham-98}%
\bibitem[{\citenamefont{Abraham}\ \emph{et~al.}(2000)\citenamefont{Abraham},
  \citenamefont{Law}, \citenamefont{Boyd}, \citenamefont{Lavender},
  \citenamefont{Maldonado},\ and\ \citenamefont{Bowie}}]{Abrah2}%
  \BibitemOpen
  \bibfield{author}{%
  \bibinfo {author} {\bibnamefont{Abraham}, \bibfnamefont{E.~R.}}, \bibinfo
  {author} {\bibfnamefont{C.~S.}\ \bibnamefont{Law}}, \bibinfo {author}
  {\bibfnamefont{P.~W.}\ \bibnamefont{Boyd}}, \bibinfo {author}
  {\bibfnamefont{S.~J.}\ \bibnamefont{Lavender}}, \bibinfo {author}
  {\bibfnamefont{M.~T.}\ \bibnamefont{Maldonado}},\ and\ \bibinfo {author}
  {\bibfnamefont{A.~R.}\ \bibnamefont{Bowie}}}%
  , \bibinfo {year} {2000},\ \bibfield{title}{%
  \enquote{\bibinfo {title} {Importance of stirring in the development of an
  iron-fertilized phytoplankton bloom},}\ }%
  \bibfield{journal}{%
  \bibinfo {journal} {Nature}\ }%
  \textbf{\bibinfo {volume} {407}},\ \bibinfo {pages} {727--730}%
  \bibAnnoteFile{NoStop}{Abrah2}%
\bibitem[{\citenamefont{Ahmed}\ \emph{et~al.}(2009)\citenamefont{Ahmed},
  \citenamefont{Mao}, \citenamefont{Shi}, \citenamefont{Juluri},\ and\
  \citenamefont{Huang}}]{Ahmed2009}%
  \BibitemOpen
  \bibfield{author}{%
  \bibinfo {author} {\bibnamefont{Ahmed}, \bibfnamefont{D.}}, \bibinfo {author}
  {\bibfnamefont{X.}~\bibnamefont{Mao}}, \bibinfo {author}
  {\bibfnamefont{J.}~\bibnamefont{Shi}}, \bibinfo {author}
  {\bibfnamefont{B.~K.}\ \bibnamefont{Juluri}},\ and\ \bibinfo {author}
  {\bibfnamefont{T.~J.}\ \bibnamefont{Huang}}}%
  , \bibinfo {year} {2009},\ \bibfield{title}{%
  \enquote{\bibinfo {title} {A millisecond micromixer via single-bubble-based
  acoustic streaming},}\ }%
  \bibfield{journal}{%
  \bibinfo {journal} {Lab on a chip}\ }%
  \textbf{\bibinfo {volume} {9}},\ \bibinfo {pages} {2738--2741}%
  \bibAnnoteFile{NoStop}{Ahmed2009}%
\bibitem[{\citenamefont{Alexandroff}(1961)}]{Alexandroff1961}%
  \BibitemOpen
  \bibfield{author}{%
  \bibinfo {author} {\bibnamefont{Alexandroff}, \bibfnamefont{P.}}}%
  , \bibinfo {year} {1961},\ \emph{\bibinfo {title} {Elementary Concepts of
  Topology}}\ (\bibinfo {publisher} {Dover},\ \bibinfo {address} {New York})%
  \bibAnnoteFile{NoStop}{Alexandroff1961}%
\bibitem[{\citenamefont{Allshouse}\ and\
  \citenamefont{Thiffeault}(2012)}]{Allshouse:2012kc}%
  \BibitemOpen
  \bibfield{author}{%
  \bibinfo {author} {\bibnamefont{Allshouse}, \bibfnamefont{M.~R.}},\ and\
  \bibinfo {author} {\bibfnamefont{{J.-L.}}\ \bibnamefont{Thiffeault}}}%
  , \bibinfo {year} {2012},\ \bibfield{title}{%
  \enquote{\bibinfo {title} {Detecting coherent structures using braids},}\ }%
  \bibinfo {journal} {Physica D},\ \bibinfo {pages} {95--105}%
  \bibAnnoteFile{NoStop}{Allshouse:2012kc}%
\bibitem[{\citenamefont{Alvarez}\ \emph{et~al.}(2002)\citenamefont{Alvarez},
  \citenamefont{Zalc}, \citenamefont{Shinbrot}, \citenamefont{Arratia},\ and\
  \citenamefont{Muzzio}}]{alvarez2002}%
  \BibitemOpen
\bibfield{journal}{%
    }%
  \bibfield{author}{%
  \bibinfo {author} {\bibnamefont{Alvarez}, \bibfnamefont{M.~M.}}, \bibinfo
  {author} {\bibfnamefont{J.~M.}\ \bibnamefont{Zalc}}, \bibinfo {author}
  {\bibfnamefont{T.}~\bibnamefont{Shinbrot}}, \bibinfo {author}
  {\bibfnamefont{P.~E.}\ \bibnamefont{Arratia}},\ and\ \bibinfo {author}
  {\bibfnamefont{F.~J.}\ \bibnamefont{Muzzio}}}%
  , \bibinfo {year} {2002},\ \bibfield{title}{%
  \enquote{\bibinfo {title} {Mechanisms of mixing and creation of structure in
  laminar stirred tanks},}\ }%
  \bibfield{journal}{%
  \bibinfo {journal} {AIChE J.}\ }%
  \textbf{\bibinfo {volume} {48}},\ \bibinfo {pages} {2135--2148}%
  \bibAnnoteFile{NoStop}{alvarez2002}%
\bibitem[{\citenamefont{Alvarez-Hern\'andez}\
  \emph{et~al.}(2002)\citenamefont{Alvarez-Hern\'andez},
  \citenamefont{Shinbrot}, \citenamefont{Zalc},\ and\
  \citenamefont{Muzzio}}]{alvarez-hernandez2002}%
  \BibitemOpen
  \bibfield{author}{%
  \bibinfo {author} {\bibnamefont{Alvarez-Hern\'andez}, \bibfnamefont{M.~M.}},
  \bibinfo {author} {\bibfnamefont{T.}~\bibnamefont{Shinbrot}}, \bibinfo
  {author} {\bibfnamefont{J.}~\bibnamefont{Zalc}},\ and\ \bibinfo {author}
  {\bibfnamefont{F.~J.}\ \bibnamefont{Muzzio}}}%
  , \bibinfo {year} {2002},\ \bibfield{title}{%
  \enquote{\bibinfo {title} {Practical chaotic mixing},}\ }%
  \bibfield{journal}{%
  \bibinfo {journal} {Chem. Eng. Sci.}\ }%
  \textbf{\bibinfo {volume} {57}},\ \bibinfo {pages} {3749--3753}%
  \bibAnnoteFile{NoStop}{alvarez-hernandez2002}%
\bibitem[{\citenamefont{Anderson}\ \emph{et~al.}(1999)\citenamefont{Anderson},
  \citenamefont{Galaktionov}, \citenamefont{Peters}, \citenamefont{Vosse},\
  and\ \citenamefont{Meijer}}]{anderson99}%
  \BibitemOpen
  \bibfield{author}{%
  \bibinfo {author} {\bibnamefont{Anderson}, \bibfnamefont{P.~D.}}, \bibinfo
  {author} {\bibfnamefont{O.~S.}\ \bibnamefont{Galaktionov}}, \bibinfo {author}
  {\bibfnamefont{G.~W.~M.}\ \bibnamefont{Peters}}, \bibinfo {author}
  {\bibfnamefont{F.~N. Van~De}\ \bibnamefont{Vosse}},\ and\ \bibinfo {author}
  {\bibfnamefont{H.~E.~H.}\ \bibnamefont{Meijer}}}%
  , \bibinfo {year} {1999},\ \bibfield{title}{%
  \enquote{\bibinfo {title} {Analysis of mixing in three-dimensional
  time-periodic cavity flows},}\ }%
  \bibfield{journal}{%
  \bibinfo {journal} {J.~Fluid Mech.}\ }%
  \textbf{\bibinfo {volume} {386}},\ \bibinfo {pages} {149--166}%
  \bibAnnoteFile{NoStop}{anderson99}%
\bibitem[{\citenamefont{Anderson}\ \emph{et~al.}(2006)\citenamefont{Anderson},
  \citenamefont{Ternet}, \citenamefont{Peters},\ and\
  \citenamefont{Meijer}}]{anderson2006}%
  \BibitemOpen
  \bibfield{author}{%
  \bibinfo {author} {\bibnamefont{Anderson}, \bibfnamefont{P.~D.}}, \bibinfo
  {author} {\bibfnamefont{D.~J.}\ \bibnamefont{Ternet}}, \bibinfo {author}
  {\bibfnamefont{G.~W.~M.}\ \bibnamefont{Peters}},\ and\ \bibinfo {author}
  {\bibfnamefont{H.~E.~H.}\ \bibnamefont{Meijer}}}%
  , \bibinfo {year} {2006},\ \bibfield{title}{%
  \enquote{\bibinfo {title} {Experimental/numerical analysis of chaotic
  advection in a three-dimensional cavity flow},}\ }%
  \bibfield{journal}{%
  \bibinfo {journal} {Int. Polymer Process.}\ }%
  \textbf{\bibinfo {volume} {4}},\ \bibinfo {pages} {412--420}%
  \bibAnnoteFile{NoStop}{anderson2006}%
\bibitem[{\citenamefont{Anosov}(1967)}]{Anosov1967}%
  \BibitemOpen
  \bibfield{author}{%
  \bibinfo {author} {\bibnamefont{Anosov}, \bibfnamefont{D.~V.}}}%
  , \bibinfo {year} {1967},\ \bibfield{title}{%
  \enquote{\bibinfo {title} {Geodesic flows on closed riemannian manifolds of
  negative curvature},}\ }%
  \bibfield{journal}{%
  \bibinfo {journal} {Trudy Mat. Inst. Steklov}\ }%
  \textbf{\bibinfo {volume} {90}},\ \bibinfo {pages} {211 pp.}%
  \bibAnnoteFile{Stop}{Anosov1967}%
\bibitem[{\citenamefont{Antonsen}\ \emph{et~al.}(1995)\citenamefont{Antonsen},
  \citenamefont{Fan},\ and\ \citenamefont{Ott}}]{Antonsen1995}%
  \BibitemOpen
  \bibfield{author}{%
  \bibinfo {author} {\bibnamefont{Antonsen}, \bibfnamefont{T.~M., Jr.}},
  \bibinfo {author} {\bibfnamefont{Z.}~\bibnamefont{Fan}},\ and\ \bibinfo
  {author} {\bibfnamefont{E.}~\bibnamefont{Ott}}}%
  , \bibinfo {year} {1995},\ \bibfield{title}{%
  \enquote{\bibinfo {title} {{$k$} spectrum of passive scalars in {L}agrangian
  chaotic fluid flows},}\ }%
  \bibfield{journal}{%
  \bibinfo {journal} {Phys. Rev. Lett.}\ }%
  \textbf{\bibinfo {volume} {75}},\ \bibinfo {pages} {1751--1754}%
  \bibAnnoteFile{NoStop}{Antonsen1995}%
\bibitem[{\citenamefont{Antonsen}\ \emph{et~al.}(1996)\citenamefont{Antonsen},
  \citenamefont{Fan}, \citenamefont{Ott},\ and\
  \citenamefont{Garcia-Lopez}}]{Antonsen1996}%
  \BibitemOpen
  \bibfield{author}{%
  \bibinfo {author} {\bibnamefont{Antonsen}, \bibfnamefont{T.~M., Jr.}},
  \bibinfo {author} {\bibfnamefont{Z.}~\bibnamefont{Fan}}, \bibinfo {author}
  {\bibfnamefont{E.}~\bibnamefont{Ott}},\ and\ \bibinfo {author}
  {\bibfnamefont{E.}~\bibnamefont{Garcia-Lopez}}}%
  , \bibinfo {year} {1996},\ \bibfield{title}{%
  \enquote{\bibinfo {title} {The role of chaotic orbits in the determination of
  power spectra},}\ }%
  \bibfield{journal}{%
  \bibinfo {journal} {Phys. Fluids}\ }%
  \textbf{\bibinfo {volume} {8}},\ \bibinfo {pages} {3094--3104}%
  \bibAnnoteFile{NoStop}{Antonsen1996}%
\bibitem[{\citenamefont{Aref}(1982)}]{Aref1982}%
  \BibitemOpen
  \bibfield{author}{%
  \bibinfo {author} {\bibnamefont{Aref}, \bibfnamefont{H.}}}%
  , \bibinfo {year} {1982},\ \enquote{\bibinfo {title} {An idealized model of
  stirring},}\ in\ \emph{\bibinfo {booktitle} {Woods Hole Oceanographic
  Institution Technical Report WHOI-82-45}},\ pp.\ \bibinfo {pages} {188--189}%
  \bibAnnoteFile{NoStop}{Aref1982}%
\bibitem[{\citenamefont{Aref}(1984)}]{Aref1984}%
  \BibitemOpen
  \bibfield{author}{%
  \bibinfo {author} {\bibnamefont{Aref}, \bibfnamefont{H.}}}%
  , \bibinfo {year} {1984},\ \bibfield{title}{%
  \enquote{\bibinfo {title} {Stirring by chaotic advection},}\ }%
  \bibfield{journal}{%
  \bibinfo {journal} {J. Fluid Mech.}\ }%
  \textbf{\bibinfo {volume} {143}},\ \bibinfo {pages} {1--21}%
  \bibAnnoteFile{NoStop}{Aref1984}%
\bibitem[{\citenamefont{Aref}(1990)}]{aref1990}%
  \BibitemOpen
  \bibfield{author}{%
  \bibinfo {author} {\bibnamefont{Aref}, \bibfnamefont{H.}}}%
  , \bibinfo {year} {1990},\ \bibfield{title}{%
  \enquote{\bibinfo {title} {Chaotic advection of fluid particles},}\ }%
  \bibfield{journal}{%
  \bibinfo {journal} {Phil. Trans. Roy. Soc. London A}\ }%
  \textbf{\bibinfo {volume} {333}},\ \bibinfo {pages} {273--288}%
  \bibAnnoteFile{NoStop}{aref1990}%
\bibitem[{\citenamefont{Aref}(2002)}]{Aref02}%
  \BibitemOpen
  \bibfield{author}{%
  \bibinfo {author} {\bibnamefont{Aref}, \bibfnamefont{H.}}}%
  , \bibinfo {year} {2002},\ \bibfield{title}{%
  \enquote{\bibinfo {title} {The development of chaotic advection},}\ }%
  \bibfield{journal}{%
  \bibinfo {journal} {Phys. Fluids}\ }%
  \textbf{\bibinfo {volume} {14}},\ \bibinfo {pages} {1315--1325}%
  \bibAnnoteFile{NoStop}{Aref02}%
\bibitem[{\citenamefont{Aref}\ and\
  \citenamefont{Balachandar}(1986)}]{aref1986}%
  \BibitemOpen
  \bibfield{author}{%
  \bibinfo {author} {\bibnamefont{Aref}, \bibfnamefont{H.}},\ and\ \bibinfo
  {author} {\bibfnamefont{S.}~\bibnamefont{Balachandar}}}%
  , \bibinfo {year} {1986},\ \bibfield{title}{%
  \enquote{\bibinfo {title} {Chaotic advection in {Stokes} flow},}\ }%
  \bibfield{journal}{%
  \bibinfo {journal} {Phys. Fluids}\ }%
  \textbf{\bibinfo {volume} {29}},\ \bibinfo {pages} {3515--3521}%
  \bibAnnoteFile{NoStop}{aref1986}%
\bibitem[{\citenamefont{Arnol'd}(1964)}]{Arnold1964}%
  \BibitemOpen
  \bibfield{author}{%
  \bibinfo {author} {\bibnamefont{Arnol'd}, \bibfnamefont{V.~I.}}}%
  , \bibinfo {year} {1964},\ \bibfield{title}{%
  \enquote{\bibinfo {title} {Instability of dynamical systems with several
  degrees of freedom},}\ }%
  \bibfield{journal}{%
  \bibinfo {journal} {Soviet Math. Doklady}\ }%
  \textbf{\bibinfo {volume} {5}},\ \bibinfo {pages} {581--585}%
  \bibAnnoteFile{NoStop}{Arnold1964}%
\bibitem[{\citenamefont{Arnol'd}(1965)}]{arnold1965}%
  \BibitemOpen
  \bibfield{author}{%
  \bibinfo {author} {\bibnamefont{Arnol'd}, \bibfnamefont{V.~I.}}}%
  , \bibinfo {year} {1965},\ \bibfield{title}{%
  \enquote{\bibinfo {title} {Sur la topologie des \'ecoulements stationnaires
  des fluides parfaits},}\ }%
  \bibfield{journal}{%
  \bibinfo {journal} {C. R. Acad. Sci. Paris A}\ }%
  \textbf{\bibinfo {volume} {261}},\ \bibinfo {pages} {17--20}%
  \bibAnnoteFile{NoStop}{arnold1965}%
\bibitem[{\citenamefont{Arnol'd}(1978)}]{Arnold1978}%
  \BibitemOpen
  \bibfield{author}{%
  \bibinfo {author} {\bibnamefont{Arnol'd}, \bibfnamefont{V.~I.}}}%
  , \bibinfo {year} {1978},\ \emph{\bibinfo {title} {Mathematical Methods of
  Classical Mechanics}}\ (\bibinfo {publisher} {Springer},\ \bibinfo {address}
  {New York})%
  \bibAnnoteFile{NoStop}{Arnold1978}%
\bibitem[{\citenamefont{Arnol'd}\ and\
  \citenamefont{Avez}(1968)}]{arnol?d1968ergodic}%
  \BibitemOpen
  \bibfield{author}{%
  \bibinfo {author} {\bibnamefont{Arnol'd}, \bibfnamefont{V.~I.}},\ and\
  \bibinfo {author} {\bibfnamefont{A.}~\bibnamefont{Avez}}}%
  , \bibinfo {year} {1968},\ \emph{\bibinfo {title} {Ergodic problems of
  classical mechanics}}\ (\bibinfo {publisher} {Benjamin, New York})%
  \bibAnnoteFile{NoStop}{arnol?d1968ergodic}%
\bibitem[{\citenamefont{Arnol'd}\ and\
  \citenamefont{Khesin}(1992)}]{arnold1992}%
  \BibitemOpen
  \bibfield{author}{%
  \bibinfo {author} {\bibnamefont{Arnol'd}, \bibfnamefont{V.~I.}},\ and\
  \bibinfo {author} {\bibfnamefont{B.~A.}\ \bibnamefont{Khesin}}}%
  , \bibinfo {year} {1992},\ \bibfield{title}{%
  \enquote{\bibinfo {title} {Topological methods in hydrodynamics},}\ }%
  \bibfield{journal}{%
  \bibinfo {journal} {Annu. Rev. Fluid Mech.}\ }%
  \textbf{\bibinfo {volume} {24}},\ \bibinfo {pages} {145--166}%
  \bibAnnoteFile{NoStop}{arnold1992}%
\bibitem[{\citenamefont{Arnol'd}\ and\
  \citenamefont{Khesin}(1998)}]{Arnold1998}%
  \BibitemOpen
  \bibfield{author}{%
  \bibinfo {author} {\bibnamefont{Arnol'd}, \bibfnamefont{V.~I.}},\ and\
  \bibinfo {author} {\bibfnamefont{B.~A.}\ \bibnamefont{Khesin}}}%
  , \bibinfo {year} {1998},\ \emph{\bibinfo {title} {Topological Methods in
  Hydrodynamics}}\ (\bibinfo {publisher} {Springer},\ \bibinfo {address} {New
  York})%
  \bibAnnoteFile{NoStop}{Arnold1998}%
\bibitem[{\citenamefont{Arratia}\ \emph{et~al.}(2005)\citenamefont{Arratia},
  \citenamefont{Shinbrot}, \citenamefont{Alvarez},\ and\
  \citenamefont{Muzzio}}]{arratia2005}%
  \BibitemOpen
  \bibfield{author}{%
  \bibinfo {author} {\bibnamefont{Arratia}, \bibfnamefont{P.~E.}}, \bibinfo
  {author} {\bibfnamefont{T.}~\bibnamefont{Shinbrot}}, \bibinfo {author}
  {\bibfnamefont{M.~M.}\ \bibnamefont{Alvarez}},\ and\ \bibinfo {author}
  {\bibfnamefont{F.~J.}\ \bibnamefont{Muzzio}}}%
  , \bibinfo {year} {2005},\ \bibfield{title}{%
  \enquote{\bibinfo {title} {Mixing of non-{Newtonian} fluids in steadily
  forced systems},}\ }%
  \bibfield{journal}{%
  \bibinfo {journal} {Phys. Rev. Lett.}\ }%
  \textbf{\bibinfo {volume} {94}},\ \bibinfo {pages} {084501}%
  \bibAnnoteFile{NoStop}{arratia2005}%
\bibitem[{\citenamefont{Arrieta}\ \emph{et~al.}(2015)\citenamefont{Arrieta},
  \citenamefont{Cartwright}, \citenamefont{Gouillart}, \citenamefont{Piro},
  \citenamefont{Piro},\ and\ \citenamefont{Tuval}}]{arrieta2015}%
  \BibitemOpen
  \bibfield{author}{%
  \bibinfo {author} {\bibnamefont{Arrieta}, \bibfnamefont{J.}}, \bibinfo
  {author} {\bibfnamefont{J.~H.~E.}\ \bibnamefont{Cartwright}}, \bibinfo
  {author} {\bibfnamefont{E.}~\bibnamefont{Gouillart}}, \bibinfo {author}
  {\bibfnamefont{N.}~\bibnamefont{Piro}}, \bibinfo {author}
  {\bibfnamefont{O.}~\bibnamefont{Piro}},\ and\ \bibinfo {author}
  {\bibfnamefont{I.}~\bibnamefont{Tuval}}}%
  , \bibinfo {year} {2015},\ \bibfield{title}{%
  \enquote{\bibinfo {title} {Geometric mixing, peristalsis, and the geometric
  phase of the stomach},}\ }%
  \bibfield{journal}{%
  \bibinfo {journal} {PlosOne}\ }%
  \textbf{\bibinfo {volume} {10}},\ \bibinfo {pages} {e0130735}%
  \bibAnnoteFile{NoStop}{arrieta2015}%
\bibitem[{\citenamefont{Arter}(1983)}]{arter1983}%
  \BibitemOpen
  \bibfield{author}{%
  \bibinfo {author} {\bibnamefont{Arter}, \bibfnamefont{W.}}}%
  , \bibinfo {year} {1983},\ \bibfield{title}{%
  \enquote{\bibinfo {title} {Ergodic streamlines in three-dimensional
  convection},}\ }%
  \bibfield{journal}{%
  \bibinfo {journal} {Phys. Lett.}\ }%
  \textbf{\bibinfo {volume} {97A}},\ \bibinfo {pages} {171--174}%
  \bibAnnoteFile{NoStop}{arter1983}%
\bibitem[{\citenamefont{Babiano}\ \emph{et~al.}(2000)\citenamefont{Babiano},
  \citenamefont{Cartwright}, \citenamefont{Piro},\ and\
  \citenamefont{Provenzale}}]{Babiano2000}%
  \BibitemOpen
  \bibfield{author}{%
  \bibinfo {author} {\bibnamefont{Babiano}, \bibfnamefont{A.}}, \bibinfo
  {author} {\bibfnamefont{J.~H.~E.}\ \bibnamefont{Cartwright}}, \bibinfo
  {author} {\bibfnamefont{O.}~\bibnamefont{Piro}},\ and\ \bibinfo {author}
  {\bibfnamefont{A.}~\bibnamefont{Provenzale}}}%
  , \bibinfo {year} {2000},\ \bibfield{title}{%
  \enquote{\bibinfo {title} {Dynamics of a small neutrally buoyant sphere in a
  fluid and targeting in {Hamiltonian} systems},}\ }%
  \bibfield{journal}{%
  \bibinfo {journal} {Phys. Rev. Lett.}\ }%
  \textbf{\bibinfo {volume} {84}},\ \bibinfo {pages} {5764--5767}%
  \bibAnnoteFile{NoStop}{Babiano2000}%
\bibitem[{\citenamefont{Bajer}(1994)}]{Bajer1994}%
  \BibitemOpen
  \bibfield{author}{%
  \bibinfo {author} {\bibnamefont{Bajer}, \bibfnamefont{K.}}}%
  , \bibinfo {year} {1994},\ \bibfield{title}{%
  \enquote{\bibinfo {title} {Hamiltonian formulation of the equations of
  streamlines in three-dimensional steady flows},}\ }%
  \bibfield{journal}{%
  \bibinfo {journal} {Chaos,~Solitons \& Fractals}\ }%
  \textbf{\bibinfo {volume} {4}},\ \bibinfo {pages} {895--911}%
  \bibAnnoteFile{NoStop}{Bajer1994}%
\bibitem[{\citenamefont{Bajer}\ and\ \citenamefont{Moffatt}(1990)}]{Bajer1990}%
  \BibitemOpen
  \bibfield{author}{%
  \bibinfo {author} {\bibnamefont{Bajer}, \bibfnamefont{K.}},\ and\ \bibinfo
  {author} {\bibfnamefont{H.~K.}\ \bibnamefont{Moffatt}}}%
  , \bibinfo {year} {1990},\ \bibfield{title}{%
  \enquote{\bibinfo {title} {On a class of steady confined {Stokes} flows with
  chaotic streamlines},}\ }%
  \bibfield{journal}{%
  \bibinfo {journal} {J. Fluid Mech.}\ }%
  \textbf{\bibinfo {volume} {212}},\ \bibinfo {pages} {337--363}%
  \bibAnnoteFile{NoStop}{Bajer1990}%
\bibitem[{\citenamefont{Bajer}\ and\ \citenamefont{Moffatt}(1992)}]{Bajer1992}%
  \BibitemOpen
  \bibfield{author}{%
  \bibinfo {author} {\bibnamefont{Bajer}, \bibfnamefont{K.}},\ and\ \bibinfo
  {author} {\bibfnamefont{H.~K.}\ \bibnamefont{Moffatt}}}%
  , \bibinfo {year} {1992},\ \enquote{\bibinfo {title} {Chaos associated with
  fluid inertia},}\ in\ \emph{\bibinfo {booktitle} {Topological Aspects of the
  Dynamics of Fluids and Plasmas}}\ (\bibinfo {publisher} {Kluwer Academic
  Publishers},\ \bibinfo {address} {Dordrecht})\ pp.\ \bibinfo {pages}
  {517--534}%
  \bibAnnoteFile{NoStop}{Bajer1992}%
\bibitem[{\citenamefont{Baladi}(2000)}]{baladi2000positive}%
  \BibitemOpen
  \bibfield{author}{%
  \bibinfo {author} {\bibnamefont{Baladi}, \bibfnamefont{V.}}}%
  , \bibinfo {year} {2000},\ \emph{\bibinfo {title} {Positive transfer
  operators and decay of correlations}},\ Vol.~\bibinfo {volume} {16}\
  (\bibinfo {publisher} {World Scientific})%
  \bibAnnoteFile{NoStop}{baladi2000positive}%
\bibitem[{\citenamefont{Balasuriya}(2005{\natexlab{a}})}]{balasuriya2005approach}%
  \BibitemOpen
  \bibfield{author}{%
  \bibinfo {author} {\bibnamefont{Balasuriya}, \bibfnamefont{S.}}}%
  , \bibinfo {year} {2005}{\natexlab{a}},\ \bibfield{title}{%
  \enquote{\bibinfo {title} {Approach for maximizing chaotic mixing in
  microfluidic devices},}\ }%
  \bibfield{journal}{%
  \bibinfo {journal} {Phys. Fluids}\ }%
  \textbf{\bibinfo {volume} {17}},\ \bibinfo {pages} {118103}%
  \bibAnnoteFile{NoStop}{balasuriya2005approach}%
\bibitem[{\citenamefont{Balasuriya}(2005{\natexlab{b}})}]{balasuriya2005optimal}%
  \BibitemOpen
  \bibfield{author}{%
  \bibinfo {author} {\bibnamefont{Balasuriya}, \bibfnamefont{S.}}}%
  , \bibinfo {year} {2005}{\natexlab{b}},\ \bibfield{title}{%
  \enquote{\bibinfo {title} {Optimal perturbation for enhanced chaotic
  transport},}\ }%
  \bibfield{journal}{%
  \bibinfo {journal} {Physica D}\ }%
  \textbf{\bibinfo {volume} {202}},\ \bibinfo {pages} {155--176}%
  \bibAnnoteFile{NoStop}{balasuriya2005optimal}%
\bibitem[{\citenamefont{Balasuriya}(2010)}]{balasuriya2010}%
  \BibitemOpen
  \bibfield{author}{%
  \bibinfo {author} {\bibnamefont{Balasuriya}, \bibfnamefont{S.}}}%
  , \bibinfo {year} {2010},\ \bibfield{title}{%
  \enquote{\bibinfo {title} {Optimal frequency for microfluidic mixing across a
  fluid interface},}\ }%
  \bibfield{journal}{%
  \bibinfo {journal} {Phys. Rev. Lett.}\ }%
  \textbf{\bibinfo {volume} {105}},\ \bibinfo {pages} {064501}%
  \bibAnnoteFile{NoStop}{balasuriya2010}%
\bibitem[{\citenamefont{Balasuriya}\
  \emph{et~al.}(2003)\citenamefont{Balasuriya}, \citenamefont{Mezi{\'c}},\ and\
  \citenamefont{Jones}}]{balasuriya2003weak}%
  \BibitemOpen
  \bibfield{author}{%
  \bibinfo {author} {\bibnamefont{Balasuriya}, \bibfnamefont{S.}}, \bibinfo
  {author} {\bibfnamefont{I.}~\bibnamefont{Mezi{\'c}}},\ and\ \bibinfo {author}
  {\bibfnamefont{C.~K. R.~T.}\ \bibnamefont{Jones}}}%
  , \bibinfo {year} {2003},\ \bibfield{title}{%
  \enquote{\bibinfo {title} {Weak finite-time {Melnikov} theory and {3D}
  viscous perturbations of {Euler} flows},}\ }%
  \bibfield{journal}{%
  \bibinfo {journal} {Physica D}\ }%
  \textbf{\bibinfo {volume} {176}},\ \bibinfo {pages} {82--106}%
  \bibAnnoteFile{NoStop}{balasuriya2003weak}%
\bibitem[{\citenamefont{Balkovsky}\ and\
  \citenamefont{Fouxon}(1999)}]{Balkovsky1999}%
  \BibitemOpen
  \bibfield{author}{%
  \bibinfo {author} {\bibnamefont{Balkovsky}, \bibfnamefont{E.}},\ and\
  \bibinfo {author} {\bibfnamefont{A.}~\bibnamefont{Fouxon}}}%
  , \bibinfo {year} {1999},\ \bibfield{title}{%
  \enquote{\bibinfo {title} {Universal long-time properties of {L}agrangian
  statistics in the {B}atchelor regime and their application to the passive
  scalar problem},}\ }%
  \bibfield{journal}{%
  \bibinfo {journal} {Phys. Rev. E}\ }%
  \textbf{\bibinfo {volume} {60}},\ \bibinfo {pages} {4164--4174}%
  \bibAnnoteFile{NoStop}{Balkovsky1999}%
\bibitem[{\citenamefont{Baskan}\ \emph{et~al.}(2015)\citenamefont{Baskan},
  \citenamefont{Speetjens}, \citenamefont{Metcalfe},\ and\
  \citenamefont{Clercx}}]{Baskan_modes_2015}%
  \BibitemOpen
  \bibfield{author}{%
  \bibinfo {author} {\bibnamefont{Baskan}, \bibfnamefont{{\"O}.}}, \bibinfo
  {author} {\bibfnamefont{M.~F.~M.}\ \bibnamefont{Speetjens}}, \bibinfo
  {author} {\bibfnamefont{G.}~\bibnamefont{Metcalfe}},\ and\ \bibinfo {author}
  {\bibfnamefont{H.~J.~H.}\ \bibnamefont{Clercx}}}%
  , \bibinfo {year} {2015},\ \bibfield{title}{%
  \enquote{\bibinfo {title} {Experimental and computational study of scalar
  modes in a periodic laminar flow},}\ }%
  \bibfield{journal}{%
  \bibinfo {journal} {Int. J. Thermal Sci.}\ }%
  \textbf{\bibinfo {volume} {96}},\ \bibinfo {pages} {102--118}%
  \bibAnnoteFile{NoStop}{Baskan_modes_2015}%
\bibitem[{\citenamefont{Bastine}\ and\
  \citenamefont{Feudel}(2010)}]{Bastine-Feudel-10}%
  \BibitemOpen
  \bibfield{author}{%
  \bibinfo {author} {\bibnamefont{Bastine}, \bibfnamefont{D.}},\ and\ \bibinfo
  {author} {\bibfnamefont{U.}~\bibnamefont{Feudel}}}%
  , \bibinfo {year} {2010},\ \bibfield{title}{%
  \enquote{\bibinfo {title} {Inhomogeneous dominance patterns of competing
  phytoplankton groups in the wake of an island},}\ }%
  \bibfield{journal}{%
  \bibinfo {journal} {Nonlin. Processes Geophys.}\ }%
  \textbf{\bibinfo {volume} {17}},\ \bibinfo {pages} {715--731}%
  \bibAnnoteFile{NoStop}{Bastine-Feudel-10}%
\bibitem[{\citenamefont{Batchelor}(1959)}]{Batchelor1959}%
  \BibitemOpen
  \bibfield{author}{%
  \bibinfo {author} {\bibnamefont{Batchelor}, \bibfnamefont{G.~K.}}}%
  , \bibinfo {year} {1959},\ \bibfield{title}{%
  \enquote{\bibinfo {title} {Small-scale variation of convected quantities like
  temperature in turbulent fluid: {P}art 1. {G}eneral discussion and the case
  of small conductivity},}\ }%
  \bibfield{journal}{%
  \bibinfo {journal} {J. Fluid Mech.}\ }%
  \textbf{\bibinfo {volume} {5}},\ \bibinfo {pages} {113--133}%
  \bibAnnoteFile{NoStop}{Batchelor1959}%
\bibitem[{\citenamefont{Bees}\ \emph{et~al.}(1998)\citenamefont{Bees},
  \citenamefont{Mezi\'{c}},\ and\ \citenamefont{McGlade}}]{bees1998planktonic}%
  \BibitemOpen
  \bibfield{author}{%
  \bibinfo {author} {\bibnamefont{Bees}, \bibfnamefont{M.~A.}}, \bibinfo
  {author} {\bibfnamefont{I.}~\bibnamefont{Mezi\'{c}}},\ and\ \bibinfo {author}
  {\bibfnamefont{J.}~\bibnamefont{McGlade}}}%
  , \bibinfo {year} {1998},\ \bibfield{title}{%
  \enquote{\bibinfo {title} {Planktonic interactions and chaotic advection in
  langmuir circulation},}\ }%
  \bibfield{journal}{%
  \bibinfo {journal} {Mathematics and Computers in Simulation}\ }%
  \textbf{\bibinfo {volume} {44}},\ \bibinfo {pages} {527--544}%
  \bibAnnoteFile{NoStop}{bees1998planktonic}%
\bibitem[{\citenamefont{Benczik}\ \emph{et~al.}(2006)\citenamefont{Benczik},
  \citenamefont{K{\'a}rolyi}, \citenamefont{Scheuring},\ and\
  \citenamefont{T{\'e}l}}]{Benczik-et-al-06}%
  \BibitemOpen
  \bibfield{author}{%
  \bibinfo {author} {\bibnamefont{Benczik}, \bibfnamefont{I.}}, \bibinfo
  {author} {\bibfnamefont{G.}~\bibnamefont{K{\'a}rolyi}}, \bibinfo {author}
  {\bibfnamefont{I.}~\bibnamefont{Scheuring}},\ and\ \bibinfo {author}
  {\bibfnamefont{T.}~\bibnamefont{T{\'e}l}}}%
  , \bibinfo {year} {2006},\ \bibfield{title}{%
  \enquote{\bibinfo {title} {Coexistence of inertial competitors in chaotic
  flows},}\ }%
  \bibfield{journal}{%
  \bibinfo {journal} {Chaos}\ }%
  \textbf{\bibinfo {volume} {16}},\ \bibinfo {pages} {043110}%
  \bibAnnoteFile{NoStop}{Benczik-et-al-06}%
\bibitem[{\citenamefont{Benczik}\ \emph{et~al.}(2002)\citenamefont{Benczik},
  \citenamefont{Toroczkai},\ and\ \citenamefont{T\'el}}]{Benczik2002}%
  \BibitemOpen
  \bibfield{author}{%
  \bibinfo {author} {\bibnamefont{Benczik}, \bibfnamefont{I.~J.}}, \bibinfo
  {author} {\bibfnamefont{Z.}~\bibnamefont{Toroczkai}},\ and\ \bibinfo {author}
  {\bibfnamefont{T.}~\bibnamefont{T\'el}}}%
  , \bibinfo {year} {2002},\ \bibfield{title}{%
  \enquote{\bibinfo {title} {Selective sensitivity of open chaotic flows on
  inertial tracer advection: catching particles with a stick},}\ }%
  \bibfield{journal}{%
  \bibinfo {journal} {Phys. Rev. Lett.}\ }%
  \textbf{\bibinfo {volume} {89}},\ \bibinfo {pages} {164501}%
  \bibAnnoteFile{NoStop}{Benczik2002}%
\bibitem[{\citenamefont{Bennet}(2006)}]{Bennet2006}%
  \BibitemOpen
  \bibfield{author}{%
  \bibinfo {author} {\bibnamefont{Bennet}, \bibfnamefont{A.}}}%
  , \bibinfo {year} {2006},\ \emph{\bibinfo {title} {Lagrangian Fluid
  Dynamics}}\ (\bibinfo {publisher} {Cambridge University Press},\ \bibinfo
  {address} {Cambridge})%
  \bibAnnoteFile{NoStop}{Bennet2006}%
\bibitem[{\citenamefont{{Beron-Vera}}\
  \emph{et~al.}(2013)\citenamefont{{Beron-Vera}}, \citenamefont{Wang},
  \citenamefont{Olascoaga}, \citenamefont{Goni},\ and\
  \citenamefont{Haller}}]{beron-vera_objective_2013}%
  \BibitemOpen
  \bibfield{author}{%
  \bibinfo {author} {\bibnamefont{{Beron-Vera}}, \bibfnamefont{F.~J.}},
  \bibinfo {author} {\bibfnamefont{Y.}~\bibnamefont{Wang}}, \bibinfo {author}
  {\bibfnamefont{M.~J.}\ \bibnamefont{Olascoaga}}, \bibinfo {author}
  {\bibfnamefont{G.~J.}\ \bibnamefont{Goni}},\ and\ \bibinfo {author}
  {\bibfnamefont{G.}~\bibnamefont{Haller}}}%
  , \bibinfo {year} {2013},\ \bibfield{title}{%
  \enquote{\bibinfo {title} {Objective detection of oceanic eddies and the
  agulhas leakage},}\ }%
  \bibfield{journal}{%
  \bibinfo {journal} {J. Phys. Oceanography}\ }%
  \textbf{\bibinfo {volume} {43}},\ \bibinfo {pages} {1426--1438}%
  \bibAnnoteFile{NoStop}{beron-vera_objective_2013}%
\bibitem[{\citenamefont{Bettencourt}\
  \emph{et~al.}(2012)\citenamefont{Bettencourt}, \citenamefont{Lopez},\ and\
  \citenamefont{{Hernandez-Garcia}}}]{Bettencourt:2012dv}%
  \BibitemOpen
  \bibfield{author}{%
  \bibinfo {author} {\bibnamefont{Bettencourt}, \bibfnamefont{J.~H.}}, \bibinfo
  {author} {\bibfnamefont{C.}~\bibnamefont{Lopez}},\ and\ \bibinfo {author}
  {\bibfnamefont{E.}~\bibnamefont{{Hernandez-Garcia}}}}%
  , \bibinfo {year} {2012},\ \bibfield{title}{%
  \enquote{\bibinfo {title} {Oceanic three-dimensional lagrangian coherent
  structures: A study of a mesoscale eddy in the benguela upwelling region},}\
  }%
  \bibfield{journal}{%
  \bibinfo {journal} {Ocean Modelling}\ }%
  \textbf{\bibinfo {volume} {51}},\ \bibinfo {pages} {73--83}%
  \bibAnnoteFile{NoStop}{Bettencourt:2012dv}%
\bibitem[{\citenamefont{Biemond}\ \emph{et~al.}(2008)\citenamefont{Biemond},
  \citenamefont{de~Moura}, \citenamefont{K\'arolyi}, \citenamefont{Grebogi},\
  and\ \citenamefont{Nijmeijer}}]{Biemond2008}%
  \BibitemOpen
  \bibfield{author}{%
  \bibinfo {author} {\bibnamefont{Biemond}, \bibfnamefont{J.~J.~B.}}, \bibinfo
  {author} {\bibfnamefont{A.~P.~S.}\ \bibnamefont{de~Moura}}, \bibinfo {author}
  {\bibfnamefont{G.}~\bibnamefont{K\'arolyi}}, \bibinfo {author}
  {\bibfnamefont{C.}~\bibnamefont{Grebogi}},\ and\ \bibinfo {author}
  {\bibfnamefont{H.}~\bibnamefont{Nijmeijer}}}%
  , \bibinfo {year} {2008},\ \bibfield{title}{%
  \enquote{\bibinfo {title} {Onset of chaotic advection in open flows},}\ }%
  \bibfield{journal}{%
  \bibinfo {journal} {Phys. Rev. E}\ }%
  \textbf{\bibinfo {volume} {78}},\ \bibinfo {pages} {016317}%
  \bibAnnoteFile{NoStop}{Biemond2008}%
\bibitem[{\citenamefont{Biskamp}(1993)}]{Biskamp1993}%
  \BibitemOpen
  \bibfield{author}{%
  \bibinfo {author} {\bibnamefont{Biskamp}, \bibfnamefont{D.}}}%
  , \bibinfo {year} {1993},\ \emph{\bibinfo {title} {Nonlinear
  Magnetohydrodynamics}}\ (\bibinfo {publisher} {Cambridge University Press},\
  \bibinfo {address} {Cambridge})%
  \bibAnnoteFile{NoStop}{Biskamp1993}%
\bibitem[{\citenamefont{Blake}(1971)}]{Blake71a}%
  \BibitemOpen
  \bibfield{author}{%
  \bibinfo {author} {\bibnamefont{Blake}, \bibfnamefont{J.~R}}}%
  , \bibinfo {year} {1971},\ \bibfield{title}{%
  \enquote{\bibinfo {title} {{A note on the image system for a Stokeslet in a
  no-slip boundary}},}\ }%
  \bibfield{journal}{%
  \bibinfo {journal} {Proc. Camb. Phil. Soc.}\ }%
  \textbf{\bibinfo {volume} {70}},\ \bibinfo {pages} {303--310}%
  \bibAnnoteFile{NoStop}{Blake71a}%
\bibitem[{\citenamefont{Blake}(2001)}]{blake2001}%
  \BibitemOpen
  \bibfield{author}{%
  \bibinfo {author} {\bibnamefont{Blake}, \bibfnamefont{J.~R.}}}%
  , \bibinfo {year} {2001},\ \enquote{\bibinfo {title} {Fluid mechanics of
  ciliary propulsion},}\ in\ \emph{\bibinfo {booktitle} {IMA Volume in
  Mathematics and its Applications: Computational Modelling in Biological Fluid
  Dynamics}},\ Vol.\ \bibinfo {volume} {124}\ (\bibinfo {publisher} {IMA})\
  pp.\ \bibinfo {pages} {1--51}%
  \bibAnnoteFile{NoStop}{blake2001}%
\bibitem[{\citenamefont{Blake}\ and\ \citenamefont{Fulford}(1995)}]{blake95}%
  \BibitemOpen
  \bibfield{author}{%
  \bibinfo {author} {\bibnamefont{Blake}, \bibfnamefont{J.~R.}},\ and\ \bibinfo
  {author} {\bibfnamefont{G.~R.}\ \bibnamefont{Fulford}}}%
  , \bibinfo {year} {1995},\ \bibfield{title}{%
  \enquote{\bibinfo {title} {Hydrodynamics of filter feeding},}\ }%
  \bibfield{journal}{%
  \bibinfo {journal} {Symp. Soc. Exp. Biol.}\ }%
  \textbf{\bibinfo {volume} {49}},\ \bibinfo {pages} {183--197}%
  \bibAnnoteFile{NoStop}{blake95}%
\bibitem[{\citenamefont{Blake}\ and\ \citenamefont{Otto}(1996)}]{Blake96}%
  \BibitemOpen
  \bibfield{author}{%
  \bibinfo {author} {\bibnamefont{Blake}, \bibfnamefont{J.~R.}},\ and\ \bibinfo
  {author} {\bibfnamefont{S.~R.}\ \bibnamefont{Otto}}}%
  , \bibinfo {year} {1996},\ \bibfield{title}{%
  \enquote{\bibinfo {title} {{Ciliary propulsion, chaotic filtration and a
  `blinking' stokeslet}},}\ }%
  \bibfield{journal}{%
  \bibinfo {journal} {J. Eng. Math.}\ }%
  \textbf{\bibinfo {volume} {30}},\ \bibinfo {pages} {151--168}%
  \bibAnnoteFile{NoStop}{Blake96}%
\bibitem[{\citenamefont{Blake}\ and\ \citenamefont{Sleigh}(1974)}]{Blake74b}%
  \BibitemOpen
  \bibfield{author}{%
  \bibinfo {author} {\bibnamefont{Blake}, \bibfnamefont{J.~R.}},\ and\ \bibinfo
  {author} {\bibfnamefont{M.~A.}\ \bibnamefont{Sleigh}}}%
  , \bibinfo {year} {1974},\ \bibfield{title}{%
  \enquote{\bibinfo {title} {{Mechanics of ciliary locomotion.}}.}\ }%
  \bibfield{journal}{%
  \bibinfo {journal} {Biol. Rev. Cambridge Philos. Soc.}\ }%
  \textbf{\bibinfo {volume} {49}},\ \bibinfo {pages} {85--125}%
  \bibAnnoteFile{NoStop}{Blake74b}%
\bibitem[{\citenamefont{Blancher}\ \emph{et~al.}(2014)\citenamefont{Blancher},
  \citenamefont{Le~Guer},\ and\ \citenamefont{El~Omari}}]{blancher2014}%
  \BibitemOpen
  \bibfield{author}{%
  \bibinfo {author} {\bibnamefont{Blancher}, \bibfnamefont{S.}}, \bibinfo
  {author} {\bibfnamefont{Y.}~\bibnamefont{Le~Guer}},\ and\ \bibinfo {author}
  {\bibfnamefont{K.}~\bibnamefont{El~Omari}}}%
  , \bibinfo {year} {2014},\ \bibfield{title}{%
  \enquote{\bibinfo {title} {Spatio-temporal structure of the fully developed
  transitional flow in a symmetric wavy channel. part. 1: Linear and weakly
  nonlinear stability analysis},}\ }%
  \bibfield{journal}{%
  \bibinfo {journal} {J. Fluid Mech.}\ }%
  \textbf{\bibinfo {volume} {764}},\ \bibinfo {pages} {250--276}%
  \bibAnnoteFile{NoStop}{blancher2014}%
\bibitem[{\citenamefont{Blazevski}\ and\
  \citenamefont{Haller}(2014)}]{Blazevski:2013ws}%
  \BibitemOpen
  \bibfield{author}{%
  \bibinfo {author} {\bibnamefont{Blazevski}, \bibfnamefont{D.}},\ and\
  \bibinfo {author} {\bibfnamefont{G.}~\bibnamefont{Haller}}}%
  , \bibinfo {year} {2014},\ \bibfield{title}{%
  \enquote{\bibinfo {title} {Hyperbolic and elliptic transport barriers in
  three-dimensional unsteady flows},}\ }%
  \bibfield{journal}{%
  \bibinfo {journal} {Physica D}\ }%
  \textbf{\bibinfo {volume} {273}},\ \bibinfo {pages} {46--62}%
  \bibAnnoteFile{NoStop}{Blazevski:2013ws}%
\bibitem[{\citenamefont{Boesinger}\
  \emph{et~al.}(2005)\citenamefont{Boesinger}, \citenamefont{Le~Guer},\ and\
  \citenamefont{Mory}}]{boesinger2005}%
  \BibitemOpen
  \bibfield{author}{%
  \bibinfo {author} {\bibnamefont{Boesinger}, \bibfnamefont{C.}}, \bibinfo
  {author} {\bibfnamefont{Y.}~\bibnamefont{Le~Guer}},\ and\ \bibinfo {author}
  {\bibfnamefont{M.}~\bibnamefont{Mory}}}%
  , \bibinfo {year} {2005},\ \bibfield{title}{%
  \enquote{\bibinfo {title} {Experimental study of reactive chaotic flows in
  tubular reactors},}\ }%
  \bibfield{journal}{%
  \bibinfo {journal} {AIChE J.}\ }%
  \textbf{\bibinfo {volume} {51}},\ \bibinfo {pages} {2122--2132}%
  \bibAnnoteFile{NoStop}{boesinger2005}%
\bibitem[{\citenamefont{Boffetta}\ \emph{et~al.}(2009)\citenamefont{Boffetta},
  \citenamefont{De~Lillo},\ and\ \citenamefont{Mazzino}}]{Boffetta2009}%
  \BibitemOpen
  \bibfield{author}{%
  \bibinfo {author} {\bibnamefont{Boffetta}, \bibfnamefont{G.}}, \bibinfo
  {author} {\bibfnamefont{F.}~\bibnamefont{De~Lillo}},\ and\ \bibinfo {author}
  {\bibfnamefont{A.}~\bibnamefont{Mazzino}}}%
  , \bibinfo {year} {2009},\ \bibfield{title}{%
  \enquote{\bibinfo {title} {{Peripheral mixing of passive scalar at small
  Reynolds number}},}\ }%
  \bibfield{journal}{%
  \bibinfo {journal} {J. Fluid Mech.}\ }%
  \textbf{\bibinfo {volume} {624}},\ \bibinfo {pages} {151--158}%
  \bibAnnoteFile{NoStop}{Boffetta2009}%
\bibitem[{\citenamefont{Bottausci}\
  \emph{et~al.}(2007)\citenamefont{Bottausci}, \citenamefont{Cardonne},
  \citenamefont{Meinhart},\ and\ \citenamefont{Mezi\'{c}}}]{Bottausci2007}%
  \BibitemOpen
  \bibfield{author}{%
  \bibinfo {author} {\bibnamefont{Bottausci}, \bibfnamefont{F.}}, \bibinfo
  {author} {\bibfnamefont{C.}~\bibnamefont{Cardonne}}, \bibinfo {author}
  {\bibfnamefont{C.}~\bibnamefont{Meinhart}},\ and\ \bibinfo {author}
  {\bibfnamefont{I.}~\bibnamefont{Mezi\'{c}}}}%
  , \bibinfo {year} {2007},\ \bibfield{title}{%
  \enquote{\bibinfo {title} {An ultrashort mixing length micromixer: the shear
  superposition micromixer},}\ }%
  \bibfield{journal}{%
  \bibinfo {journal} {Lab on a chip}\ }%
  \textbf{\bibinfo {volume} {7}},\ \bibinfo {pages} {396--398}%
  \bibAnnoteFile{NoStop}{Bottausci2007}%
\bibitem[{\citenamefont{Boyd}\ \emph{et~al.}(2000)\citenamefont{Boyd},
  \citenamefont{Watson}, \citenamefont{Law}, \citenamefont{Abraham},
  \citenamefont{Trull}, \citenamefont{Murdoch}, \citenamefont{Bakker},
  \citenamefont{Bowie}, \citenamefont{Buesseler},\ and\
  \citenamefont{Chang}}]{Abrah1}%
  \BibitemOpen
  \bibfield{author}{%
  \bibinfo {author} {\bibnamefont{Boyd}, \bibfnamefont{P.~W.}}, \bibinfo
  {author} {\bibfnamefont{A.~J.}\ \bibnamefont{Watson}}, \bibinfo {author}
  {\bibfnamefont{C.~S.}\ \bibnamefont{Law}}, \bibinfo {author}
  {\bibfnamefont{E.~R.}\ \bibnamefont{Abraham}}, \bibinfo {author}
  {\bibfnamefont{T.}~\bibnamefont{Trull}}, \bibinfo {author}
  {\bibfnamefont{R.}~\bibnamefont{Murdoch}}, \bibinfo {author}
  {\bibfnamefont{D.~C.~E.}\ \bibnamefont{Bakker}}, \bibinfo {author}
  {\bibfnamefont{A.~R.}\ \bibnamefont{Bowie}}, \bibinfo {author}
  {\bibfnamefont{K.~O.}\ \bibnamefont{Buesseler}},\ and\ \bibinfo {author}
  {\bibfnamefont{H.}~\bibnamefont{Chang}}}%
  , \bibinfo {year} {2000},\ \bibfield{title}{%
  \enquote{\bibinfo {title} {A mesoscale phytoplankton bloom in the polar
  {Southern Ocean} stimulated by iron fertilization},}\ }%
  \bibinfo {journal} {Nature},\ \bibinfo {pages} {695--702}%
  \bibAnnoteFile{NoStop}{Abrah1}%
\bibitem[{\citenamefont{Boyland}\ \emph{et~al.}(2000)\citenamefont{Boyland},
  \citenamefont{Aref},\ and\ \citenamefont{Stremler}}]{boyland2000topological}%
  \BibitemOpen
\bibfield{journal}{%
    }%
  \bibfield{author}{%
  \bibinfo {author} {\bibnamefont{Boyland}, \bibfnamefont{P.~L.}}, \bibinfo
  {author} {\bibfnamefont{H.}~\bibnamefont{Aref}},\ and\ \bibinfo {author}
  {\bibfnamefont{M.~A.}\ \bibnamefont{Stremler}}}%
  , \bibinfo {year} {2000},\ \bibfield{title}{%
  \enquote{\bibinfo {title} {Topological fluid mechanics of stirring},}\ }%
  \bibfield{journal}{%
  \bibinfo {journal} {J. Fluid Mech.}\ }%
  \textbf{\bibinfo {volume} {403}},\ \bibinfo {pages} {277--304}%
  \bibAnnoteFile{NoStop}{boyland2000topological}%
\bibitem[{\citenamefont{{BozorgMagham}}\
  \emph{et~al.}(2013)\citenamefont{{BozorgMagham}}, \citenamefont{{Ross}},\
  and\ \citenamefont{{Schmale}~{III}}}]{BozorgMagham2013}%
  \BibitemOpen
  \bibfield{author}{%
  \bibinfo {author} {\bibnamefont{{BozorgMagham}}, \bibfnamefont{{A.}~{E}.}},
  \bibinfo {author} {\bibfnamefont{{S.}~{D}.}\ \bibnamefont{{Ross}}},\ and\
  \bibinfo {author} {\bibfnamefont{{D.}~{G}.}\ \bibnamefont{{Schmale}~{III}}}}%
  , \bibinfo {year} {2013},\ \bibfield{title}{%
  \enquote{\bibinfo {title} {Real-time prediction of atmospheric {Lagrangian}
  coherent structures based on forecast data: {An} application and error
  analysis},}\ }%
  \bibfield{journal}{%
  \bibinfo {journal} {Physica {D}}\ }%
  \textbf{\bibinfo {volume} {258}},\ \bibinfo {pages} {47--60}%
  \bibAnnoteFile{NoStop}{BozorgMagham2013}%
\bibitem[{\citenamefont{Bracco}\ \emph{et~al.}(2000)\citenamefont{Bracco},
  \citenamefont{Provenzale},\ and\ \citenamefont{Scheuring}}]{Bracco-et-al-00}%
  \BibitemOpen
  \bibfield{author}{%
  \bibinfo {author} {\bibnamefont{Bracco}, \bibfnamefont{A.}}, \bibinfo
  {author} {\bibfnamefont{A.}~\bibnamefont{Provenzale}},\ and\ \bibinfo
  {author} {\bibfnamefont{I.}~\bibnamefont{Scheuring}}}%
  , \bibinfo {year} {2000},\ \bibfield{title}{%
  \enquote{\bibinfo {title} {Mesoscale vortices and the paradox of the
  plankton},}\ }%
  \bibfield{journal}{%
  \bibinfo {journal} {Proc. Roy. Soc. Lond. B}\ }%
  \textbf{\bibinfo {volume} {267}},\ \bibinfo {pages} {1795--1800}%
  \bibAnnoteFile{NoStop}{Bracco-et-al-00}%
\bibitem[{\citenamefont{Branicki}\ and\
  \citenamefont{Wiggins}(2009)}]{Branicki2009}%
  \BibitemOpen
  \bibfield{author}{%
  \bibinfo {author} {\bibnamefont{Branicki}, \bibfnamefont{M.}},\ and\ \bibinfo
  {author} {\bibfnamefont{S.}~\bibnamefont{Wiggins}}}%
  , \bibinfo {year} {2009},\ \bibfield{title}{%
  \enquote{\bibinfo {title} {An adaptive method for computing invariant
  manifolds in non-autonomous, three-dimensional dynamical systems},}\ }%
  \bibfield{journal}{%
  \bibinfo {journal} {Physica D}\ }%
  \textbf{\bibinfo {volume} {238}},\ \bibinfo {pages} {1625--1657}%
  \bibAnnoteFile{NoStop}{Branicki2009}%
\bibitem[{\citenamefont{Brennen}\ and\
  \citenamefont{Winet}(1977)}]{Brennen1977}%
  \BibitemOpen
  \bibfield{author}{%
  \bibinfo {author} {\bibnamefont{Brennen}, \bibfnamefont{C.}},\ and\ \bibinfo
  {author} {\bibfnamefont{H.}~\bibnamefont{Winet}}}%
  , \bibinfo {year} {1977},\ \bibfield{title}{%
  \enquote{\bibinfo {title} {{Fluid Mechanics of propulsion by cilia and
  flagella}},}\ }%
  \bibfield{journal}{%
  \bibinfo {journal} {Annu. Rev. Fluid Mech.}\ }%
  \textbf{\bibinfo {volume} {9}},\ \bibinfo {pages} {339--398}%
  \bibAnnoteFile{NoStop}{Brennen1977}%
\bibitem[{\citenamefont{Bringer}\ \emph{et~al.}(2004)\citenamefont{Bringer},
  \citenamefont{Gerdts}, \citenamefont{Song}, \citenamefont{Tice},\ and\
  \citenamefont{Ismagilov}}]{Bringer2004a}%
  \BibitemOpen
  \bibfield{author}{%
  \bibinfo {author} {\bibnamefont{Bringer}, \bibfnamefont{M.~R.}}, \bibinfo
  {author} {\bibfnamefont{C.~J.}\ \bibnamefont{Gerdts}}, \bibinfo {author}
  {\bibfnamefont{H.}~\bibnamefont{Song}}, \bibinfo {author}
  {\bibfnamefont{J.~D.}\ \bibnamefont{Tice}},\ and\ \bibinfo {author}
  {\bibfnamefont{R.~F.}\ \bibnamefont{Ismagilov}}}%
  , \bibinfo {year} {2004},\ \bibfield{title}{%
  \enquote{\bibinfo {title} {Microfluidic systems for chemical kinetics that
  rely on chaotic mixing in droplets},}\ }%
  \bibfield{journal}{%
  \bibinfo {journal} {Phil. Trans. Roy. Soc. London A}\ }%
  \textbf{\bibinfo {volume} {362}},\ \bibinfo {pages} {1087--1104}%
  \bibAnnoteFile{NoStop}{Bringer2004a}%
\bibitem[{\citenamefont{Broer}\ \emph{et~al.}(1996)\citenamefont{Broer},
  \citenamefont{Huitema},\ and\ \citenamefont{Sevryuk}}]{broer96}%
  \BibitemOpen
  \bibfield{author}{%
  \bibinfo {author} {\bibnamefont{Broer}, \bibfnamefont{H.~W.}}, \bibinfo
  {author} {\bibfnamefont{G.~B.}\ \bibnamefont{Huitema}},\ and\ \bibinfo
  {author} {\bibfnamefont{M.~B.}\ \bibnamefont{Sevryuk}}}%
  , \bibinfo {year} {1996},\ \enquote{\bibinfo {title} {Mathematical methods of
  classical mechanics},}\ in\ \emph{\bibinfo {booktitle} {Quasi-periodic
  Motions in Families of Dynamical Systems}}\ (\bibinfo {publisher}
  {Springer})%
  \bibAnnoteFile{NoStop}{broer96}%
\bibitem[{\citenamefont{Brumley}\ \emph{et~al.}(2012)\citenamefont{Brumley},
  \citenamefont{Polin}, \citenamefont{Pedley},\ and\
  \citenamefont{Goldstein}}]{Brumley2012}%
  \BibitemOpen
  \bibfield{author}{%
  \bibinfo {author} {\bibnamefont{Brumley}, \bibfnamefont{D.~R.}}, \bibinfo
  {author} {\bibfnamefont{M.}~\bibnamefont{Polin}}, \bibinfo {author}
  {\bibfnamefont{T.~J.}\ \bibnamefont{Pedley}},\ and\ \bibinfo {author}
  {\bibfnamefont{R.~E.}\ \bibnamefont{Goldstein}}}%
  , \bibinfo {year} {2012},\ \bibfield{title}{%
  \enquote{\bibinfo {title} {Hydrodynamic synchronization and metachronal waves
  on the surface of the colonial alga {Volvox carteri}},}\ }%
  \bibfield{journal}{%
  \bibinfo {journal} {Phys. Rev. Lett.}\ }%
  \textbf{\bibinfo {volume} {109}},\ \bibinfo {pages} {268102}%
  \bibAnnoteFile{NoStop}{Brumley2012}%
\bibitem[{\citenamefont{Brumley}\ \emph{et~al.}(2015)\citenamefont{Brumley},
  \citenamefont{Polin}, \citenamefont{Pedley},\ and\
  \citenamefont{Goldstein}}]{Brumley2015}%
  \BibitemOpen
  \bibfield{author}{%
  \bibinfo {author} {\bibnamefont{Brumley}, \bibfnamefont{D.~R.}}, \bibinfo
  {author} {\bibfnamefont{M.}~\bibnamefont{Polin}}, \bibinfo {author}
  {\bibfnamefont{T.~J.}\ \bibnamefont{Pedley}},\ and\ \bibinfo {author}
  {\bibfnamefont{R.~E.}\ \bibnamefont{Goldstein}}}%
  , \bibinfo {year} {2015},\ \bibfield{title}{%
  \enquote{\bibinfo {title} {{Metachronal waves in the flagellar beating of
  Volvox and their hydrodynamic origin}},}\ }%
  \bibfield{journal}{%
  \bibinfo {journal} {J. Roy. Soc. Interface}\ }%
  \textbf{\bibinfo {volume} {12}},\ \bibinfo {pages} {20141358}%
  \bibAnnoteFile{NoStop}{Brumley2015}%
\bibitem[{\citenamefont{Budi{\v s}i{\'c}}\ and\
  \citenamefont{Mezi{\'c}}(2012)}]{Budisic:2012woa}%
  \BibitemOpen
  \bibfield{author}{%
  \bibinfo {author} {\bibnamefont{Budi{\v s}i{\'c}}, \bibfnamefont{M.}},\ and\
  \bibinfo {author} {\bibfnamefont{I.}~\bibnamefont{Mezi{\'c}}}}%
  , \bibinfo {year} {2012},\ \bibfield{title}{%
  \enquote{\bibinfo {title} {Geometry of the ergodic quotient reveals coherent
  structures in flows},}\ }%
  \bibfield{journal}{%
  \bibinfo {journal} {Physica D}\ }%
  \textbf{\bibinfo {volume} {241}},\ \bibinfo {pages} {1255--1269}%
  \bibAnnoteFile{NoStop}{Budisic:2012woa}%
\bibitem[{\citenamefont{Budi\v{s}i\'c}\
  \emph{et~al.}(2012)\citenamefont{Budi\v{s}i\'c}, \citenamefont{Mohr},\ and\
  \citenamefont{Mezi\'{c}}}]{Budisic2012}%
  \BibitemOpen
  \bibfield{author}{%
  \bibinfo {author} {\bibnamefont{Budi\v{s}i\'c}, \bibfnamefont{M.}}, \bibinfo
  {author} {\bibfnamefont{R.}~\bibnamefont{Mohr}},\ and\ \bibinfo {author}
  {\bibfnamefont{I.}~\bibnamefont{Mezi\'{c}}}}%
  , \bibinfo {year} {2012},\ \bibfield{title}{%
  \enquote{\bibinfo {title} {Applied {Koopmanism}},}\ }%
  \bibfield{journal}{%
  \bibinfo {journal} {Chaos}\ }%
  \textbf{\bibinfo {volume} {22}},\ \bibinfo {pages} {047510}%
  \bibAnnoteFile{NoStop}{Budisic2012}%
\bibitem[{\citenamefont{Budi\v{s}i\'{c}}\
  \emph{et~al.}(2016)\citenamefont{Budi\v{s}i\'{c}}, \citenamefont{Siegmund},
  \citenamefont{Son},\ and\ \citenamefont{Mezi\'c}}]{budisic2016}%
  \BibitemOpen
  \bibfield{author}{%
  \bibinfo {author} {\bibnamefont{Budi\v{s}i\'{c}}, \bibfnamefont{M.}},
  \bibinfo {author} {\bibfnamefont{S.}~\bibnamefont{Siegmund}}, \bibinfo
  {author} {\bibfnamefont{D.~Thai}\ \bibnamefont{Son}},\ and\ \bibinfo {author}
  {\bibfnamefont{I.}~\bibnamefont{Mezi\'c}}}%
  , \bibinfo {year} {2016},\ \bibfield{title}{%
  \enquote{\bibinfo {title} {Mesochronic classification of trajectories in
  incompressible {3D} vector fields over finite times},}\ }%
  \bibfield{journal}{%
  \bibinfo {journal} {Discrete and Continuous Dyn. Sys. --- Series S}\ }%
  \textbf{\bibinfo {volume} {9}},\ \bibinfo {pages} {923--958}%
  \bibAnnoteFile{NoStop}{budisic2016}%
\bibitem[{\citenamefont{Buzzi}(2001)}]{buzzi2001piecewise}%
  \BibitemOpen
  \bibfield{author}{%
  \bibinfo {author} {\bibnamefont{Buzzi}, \bibfnamefont{J.}}}%
  , \bibinfo {year} {2001},\ \bibfield{title}{%
  \enquote{\bibinfo {title} {Piecewise isometries have zero topological
  entropy},}\ }%
  \bibfield{journal}{%
  \bibinfo {journal} {Ergod. Theor. Dyn. Sys.}\ }%
  \textbf{\bibinfo {volume} {21}},\ \bibinfo {pages} {1371--1377}%
  \bibAnnoteFile{NoStop}{buzzi2001piecewise}%
\bibitem[{\citenamefont{Cartwright}\
  \emph{et~al.}(1994)\citenamefont{Cartwright}, \citenamefont{Feingold},\ and\
  \citenamefont{Piro}}]{Cartwright1994}%
  \BibitemOpen
  \bibfield{author}{%
  \bibinfo {author} {\bibnamefont{Cartwright}, \bibfnamefont{J.~H.~E.}},
  \bibinfo {author} {\bibfnamefont{M.}~\bibnamefont{Feingold}},\ and\ \bibinfo
  {author} {\bibfnamefont{O.}~\bibnamefont{Piro}}}%
  , \bibinfo {year} {1994},\ \bibfield{title}{%
  \enquote{\bibinfo {title} {Passive scalars and three-dimensional
  {Liouvillian} maps},}\ }%
  \bibfield{journal}{%
  \bibinfo {journal} {Physica D}\ }%
  \textbf{\bibinfo {volume} {76}},\ \bibinfo {pages} {22--33}%
  \bibAnnoteFile{NoStop}{Cartwright1994}%
\bibitem[{\citenamefont{Cartwright}\
  \emph{et~al.}(1995)\citenamefont{Cartwright}, \citenamefont{Feingold},\ and\
  \citenamefont{Piro}}]{Cartwright1995}%
  \BibitemOpen
  \bibfield{author}{%
  \bibinfo {author} {\bibnamefont{Cartwright}, \bibfnamefont{J.~H.~E.}},
  \bibinfo {author} {\bibfnamefont{M.}~\bibnamefont{Feingold}},\ and\ \bibinfo
  {author} {\bibfnamefont{O.}~\bibnamefont{Piro}}}%
  , \bibinfo {year} {1995},\ \bibfield{title}{%
  \enquote{\bibinfo {title} {Global diffusion in a realistic three-dimensional
  time-dependent nonturbulent fluid flow},}\ }%
  \bibfield{journal}{%
  \bibinfo {journal} {Phys. Rev. Lett.}\ }%
  \textbf{\bibinfo {volume} {75}},\ \bibinfo {pages} {3669--3672}%
  \bibAnnoteFile{NoStop}{Cartwright1995}%
\bibitem[{\citenamefont{Cartwright}\
  \emph{et~al.}(1996)\citenamefont{Cartwright}, \citenamefont{Feingold},\ and\
  \citenamefont{Piro}}]{Cartwright1996}%
  \BibitemOpen
  \bibfield{author}{%
  \bibinfo {author} {\bibnamefont{Cartwright}, \bibfnamefont{J.~H.~E.}},
  \bibinfo {author} {\bibfnamefont{M.}~\bibnamefont{Feingold}},\ and\ \bibinfo
  {author} {\bibfnamefont{O.}~\bibnamefont{Piro}}}%
  , \bibinfo {year} {1996},\ \bibfield{title}{%
  \enquote{\bibinfo {title} {Chaotic advection in three-dimensional unsteady
  incompressible laminar flow},}\ }%
  \bibfield{journal}{%
  \bibinfo {journal} {J. Fluid Mech.}\ }%
  \textbf{\bibinfo {volume} {316}},\ \bibinfo {pages} {259--284}%
  \bibAnnoteFile{NoStop}{Cartwright1996}%
\bibitem[{\citenamefont{Cartwright}\
  \emph{et~al.}(2010)\citenamefont{Cartwright}, \citenamefont{Feudel},
  \citenamefont{K\'arolyi}, \citenamefont{Moura}, \citenamefont{Piro},\ and\
  \citenamefont{T\'el}}]{Springer2010}%
  \BibitemOpen
  \bibfield{author}{%
  \bibinfo {author} {\bibnamefont{Cartwright}, \bibfnamefont{J.~H.~E.}},
  \bibinfo {author} {\bibfnamefont{U.}~\bibnamefont{Feudel}}, \bibinfo {author}
  {\bibfnamefont{G.}~\bibnamefont{K\'arolyi}}, \bibinfo {author}
  {\bibfnamefont{A.}~\bibnamefont{Moura}}, \bibinfo {author}
  {\bibfnamefont{O.}~\bibnamefont{Piro}},\ and\ \bibinfo {author}
  {\bibfnamefont{T.}~\bibnamefont{T\'el}}}%
  , \bibinfo {year} {2010},\ \enquote{\bibinfo {title} {Dynamics of finite-size
  particles in chaotic fluid flows},}\ in\ \emph{\bibinfo {booktitle}
  {Dynamical Systems and Chaos: Advances and Perspectives}},\ \bibinfo {editor}
  {edited by\ \bibinfo {editor} {\bibfnamefont{M.}~\bibnamefont{Thiel}},
  \bibinfo {editor} {\bibfnamefont{J.}~\bibnamefont{Kurths}}, \bibinfo {editor}
  {\bibfnamefont{M.~C.}\ \bibnamefont{Romano}}, \bibinfo {editor}
  {\bibfnamefont{A.}~\bibnamefont{Moura}},\ and\ \bibinfo {editor}
  {\bibfnamefont{G.}~\bibnamefont{Karolyi}}}\ (\bibinfo {publisher}
  {Springer-Verlag})\ pp.\ \bibinfo {pages} {51--87}%
  \bibAnnoteFile{NoStop}{Springer2010}%
\bibitem[{\citenamefont{Cartwright}\
  \emph{et~al.}(2004{\natexlab{a}})\citenamefont{Cartwright},
  \citenamefont{Garc\'{\i}a-Ruiz}, \citenamefont{Piro},
  \citenamefont{Sainz-D\'{\i}az},\ and\
  \citenamefont{Tuval}}]{Cartwright_chiral_2004}%
  \BibitemOpen
  \bibfield{author}{%
  \bibinfo {author} {\bibnamefont{Cartwright}, \bibfnamefont{J.~H.~E.}},
  \bibinfo {author} {\bibfnamefont{J.~M.}\ \bibnamefont{Garc\'{\i}a-Ruiz}},
  \bibinfo {author} {\bibfnamefont{O.}~\bibnamefont{Piro}}, \bibinfo {author}
  {\bibfnamefont{C.~I.}\ \bibnamefont{Sainz-D\'{\i}az}},\ and\ \bibinfo
  {author} {\bibfnamefont{I.}~\bibnamefont{Tuval}}}%
  , \bibinfo {year} {2004}{\natexlab{a}},\ \bibfield{title}{%
  \enquote{\bibinfo {title} {Chiral symmetry breaking during crystallization:
  An advection-mediated nonlinear autocatalytic process},}\ }%
  \bibfield{journal}{%
  \bibinfo {journal} {Phys. Rev. Lett.}\ }%
  \textbf{\bibinfo {volume} {93}},\ \bibinfo {pages} {035502}%
  \bibAnnoteFile{NoStop}{Cartwright_chiral_2004}%
\bibitem[{\citenamefont{Cartwright}\
  \emph{et~al.}(2002{\natexlab{a}})\citenamefont{Cartwright},
  \citenamefont{Magnasco},\ and\ \citenamefont{Piro}}]{Cartwright2002}%
  \BibitemOpen
  \bibfield{author}{%
  \bibinfo {author} {\bibnamefont{Cartwright}, \bibfnamefont{J.~H.~E.}},
  \bibinfo {author} {\bibfnamefont{M.~O.}\ \bibnamefont{Magnasco}},\ and\
  \bibinfo {author} {\bibfnamefont{O.}~\bibnamefont{Piro}}}%
  , \bibinfo {year} {2002}{\natexlab{a}},\ \bibfield{title}{%
  \enquote{\bibinfo {title} {Noise- and inertia-induced inhomogeneity in the
  distribution of small particles in fluid flows},}\ }%
  \bibfield{journal}{%
  \bibinfo {journal} {Chaos}\ }%
  \textbf{\bibinfo {volume} {12}},\ \bibinfo {pages} {489--495}%
  \bibAnnoteFile{NoStop}{Cartwright2002}%
\bibitem[{\citenamefont{Cartwright}\
  \emph{et~al.}(2002{\natexlab{b}})\citenamefont{Cartwright},
  \citenamefont{Magnasco}, \citenamefont{Piro},\ and\
  \citenamefont{Tuval}}]{Cartwright2002_1}%
  \BibitemOpen
  \bibfield{author}{%
  \bibinfo {author} {\bibnamefont{Cartwright}, \bibfnamefont{J.~H.~E.}},
  \bibinfo {author} {\bibfnamefont{M.~O.}\ \bibnamefont{Magnasco}}, \bibinfo
  {author} {\bibfnamefont{O.}~\bibnamefont{Piro}},\ and\ \bibinfo {author}
  {\bibfnamefont{I.}~\bibnamefont{Tuval}}}%
  , \bibinfo {year} {2002}{\natexlab{b}},\ \bibfield{title}{%
  \enquote{\bibinfo {title} {Bailout embeddings and neutrally buoyant particles
  in three-dimensional flows},}\ }%
  \bibfield{journal}{%
  \bibinfo {journal} {Phys. Rev. Lett.}\ }%
  \textbf{\bibinfo {volume} {89}},\ \bibinfo {pages} {264501}%
  \bibAnnoteFile{NoStop}{Cartwright2002_1}%
\bibitem[{\citenamefont{Cartwright}\
  \emph{et~al.}(2007{\natexlab{a}})\citenamefont{Cartwright},
  \citenamefont{Piro}, \citenamefont{Piro},\ and\
  \citenamefont{Tuval}}]{Cartwright2007}%
  \BibitemOpen
  \bibfield{author}{%
  \bibinfo {author} {\bibnamefont{Cartwright}, \bibfnamefont{J.~H.~E.}},
  \bibinfo {author} {\bibfnamefont{N.}~\bibnamefont{Piro}}, \bibinfo {author}
  {\bibfnamefont{O.}~\bibnamefont{Piro}},\ and\ \bibinfo {author}
  {\bibfnamefont{I.}~\bibnamefont{Tuval}}}%
  , \bibinfo {year} {2007}{\natexlab{a}},\ \bibfield{title}{%
  \enquote{\bibinfo {title} {{Embryonic nodal flow and the dynamics of nodal
  vesicular parcels}},}\ }%
  \bibfield{journal}{%
  \bibinfo {journal} {J. Roy. Soc. Interface}\ }%
  \textbf{\bibinfo {volume} {4}},\ \bibinfo {pages} {49--55}%
  \bibAnnoteFile{NoStop}{Cartwright2007}%
\bibitem[{\citenamefont{Cartwright}\
  \emph{et~al.}(2004{\natexlab{b}})\citenamefont{Cartwright},
  \citenamefont{Piro},\ and\ \citenamefont{Tuval}}]{Cartwright2004}%
  \BibitemOpen
  \bibfield{author}{%
  \bibinfo {author} {\bibnamefont{Cartwright}, \bibfnamefont{J.~H.~E.}},
  \bibinfo {author} {\bibfnamefont{O.}~\bibnamefont{Piro}},\ and\ \bibinfo
  {author} {\bibfnamefont{I.}~\bibnamefont{Tuval}}}%
  , \bibinfo {year} {2004}{\natexlab{b}},\ \bibfield{title}{%
  \enquote{\bibinfo {title} {{Fluid-dynamical basis of the embryonic
  development of left-right asymmetry in vertebrates}},}\ }%
  \bibfield{journal}{%
  \bibinfo {journal} {Proc. Natl Acad. Sci. USA}\ }%
  \textbf{\bibinfo {volume} {101}},\ \bibinfo {pages} {7234--7239}%
  \bibAnnoteFile{NoStop}{Cartwright2004}%
\bibitem[{\citenamefont{Cartwright}\
  \emph{et~al.}(2007{\natexlab{b}})\citenamefont{Cartwright},
  \citenamefont{Piro},\ and\ \citenamefont{Tuval}}]{Cartwright_chiral_2007}%
  \BibitemOpen
  \bibfield{author}{%
  \bibinfo {author} {\bibnamefont{Cartwright}, \bibfnamefont{J.~H.~E.}},
  \bibinfo {author} {\bibfnamefont{O.}~\bibnamefont{Piro}},\ and\ \bibinfo
  {author} {\bibfnamefont{I.}~\bibnamefont{Tuval}}}%
  , \bibinfo {year} {2007}{\natexlab{b}},\ \bibfield{title}{%
  \enquote{\bibinfo {title} {Ostwald ripening, chiral crystallization, and the
  common-ancestor effect},}\ }%
  \bibfield{journal}{%
  \bibinfo {journal} {Phys. Rev. Lett.}\ }%
  \textbf{\bibinfo {volume} {98}},\ \bibinfo {pages} {165501}%
  \bibAnnoteFile{NoStop}{Cartwright_chiral_2007}%
\bibitem[{\citenamefont{Cartwright}\
  \emph{et~al.}(2009)\citenamefont{Cartwright}, \citenamefont{Piro},\ and\
  \citenamefont{Tuval}}]{Cartwright2009a}%
  \BibitemOpen
  \bibfield{author}{%
  \bibinfo {author} {\bibnamefont{Cartwright}, \bibfnamefont{J.~H.~E.}},
  \bibinfo {author} {\bibfnamefont{O.}~\bibnamefont{Piro}},\ and\ \bibinfo
  {author} {\bibfnamefont{I.}~\bibnamefont{Tuval}}}%
  , \bibinfo {year} {2009},\ \bibfield{title}{%
  \enquote{\bibinfo {title} {{Fluid dynamics in developmental biology: moving
  fluids that shape ontogeny}},}\ }%
  \bibfield{journal}{%
  \bibinfo {journal} {HFSP J.}\ }%
  \textbf{\bibinfo {volume} {3}},\ \bibinfo {pages} {77--93}%
  \bibAnnoteFile{NoStop}{Cartwright2009a}%
\bibitem[{\citenamefont{Castelain}\
  \emph{et~al.}(2001)\citenamefont{Castelain}, \citenamefont{Mokrani},
  \citenamefont{Le~Guer},\ and\ \citenamefont{Peerhossaini}}]{castelain2001}%
  \BibitemOpen
  \bibfield{author}{%
  \bibinfo {author} {\bibnamefont{Castelain}, \bibfnamefont{C.}}, \bibinfo
  {author} {\bibfnamefont{A.}~\bibnamefont{Mokrani}}, \bibinfo {author}
  {\bibfnamefont{Y.}~\bibnamefont{Le~Guer}},\ and\ \bibinfo {author}
  {\bibfnamefont{H.}~\bibnamefont{Peerhossaini}}}%
  , \bibinfo {year} {2001},\ \bibfield{title}{%
  \enquote{\bibinfo {title} {Experimental study of chaotic advection regime in
  a twisted duct flow},}\ }%
  \bibfield{journal}{%
  \bibinfo {journal} {Eur. J. Mech. B}\ }%
  \textbf{\bibinfo {volume} {20}},\ \bibinfo {pages} {205--232}%
  \bibAnnoteFile{NoStop}{castelain2001}%
\bibitem[{\citenamefont{Caubet}\ \emph{et~al.}(2011)\citenamefont{Caubet},
  \citenamefont{Le~Guer}, \citenamefont{Grassl}, \citenamefont{El~Omari},\ and\
  \citenamefont{Normandin}}]{caubet2011}%
  \BibitemOpen
  \bibfield{author}{%
  \bibinfo {author} {\bibnamefont{Caubet}, \bibfnamefont{S.}}, \bibinfo
  {author} {\bibfnamefont{Y.}~\bibnamefont{Le~Guer}}, \bibinfo {author}
  {\bibfnamefont{B.}~\bibnamefont{Grassl}}, \bibinfo {author}
  {\bibfnamefont{K.}~\bibnamefont{El~Omari}},\ and\ \bibinfo {author}
  {\bibfnamefont{E.}~\bibnamefont{Normandin}}}%
  , \bibinfo {year} {2011},\ \bibfield{title}{%
  \enquote{\bibinfo {title} {A low-energy emulsification batch mixer for
  concentrated oil-in-water emulsions},}\ }%
  \bibfield{journal}{%
  \bibinfo {journal} {AIChE J.}\ }%
  \textbf{\bibinfo {volume} {57}},\ \bibinfo {pages} {27--39}%
  \bibAnnoteFile{NoStop}{caubet2011}%
\bibitem[{\citenamefont{Cerbelli}\ and\
  \citenamefont{Giona}(2005)}]{Cerbelli2005}%
  \BibitemOpen
  \bibfield{author}{%
  \bibinfo {author} {\bibnamefont{Cerbelli}, \bibfnamefont{S.}},\ and\ \bibinfo
  {author} {\bibfnamefont{M.}~\bibnamefont{Giona}}}%
  , \bibinfo {year} {2005},\ \bibfield{title}{%
  \enquote{\bibinfo {title} {A continuous archetype of nonuniform chaos in
  area-preserving dynamical systems},}\ }%
  \bibfield{journal}{%
  \bibinfo {journal} {J. Nonlin. Sci.}\ }%
  \textbf{\bibinfo {volume} {15}},\ \bibinfo {pages} {387--421}%
  \bibAnnoteFile{NoStop}{Cerbelli2005}%
\bibitem[{\citenamefont{Chaiken}\ \emph{et~al.}(1986)\citenamefont{Chaiken},
  \citenamefont{Chevray}, \citenamefont{Tabor},\ and\
  \citenamefont{Tan}}]{chaiken}%
  \BibitemOpen
  \bibfield{author}{%
  \bibinfo {author} {\bibnamefont{Chaiken}, \bibfnamefont{J.}}, \bibinfo
  {author} {\bibfnamefont{R.}~\bibnamefont{Chevray}}, \bibinfo {author}
  {\bibfnamefont{M.}~\bibnamefont{Tabor}},\ and\ \bibinfo {author}
  {\bibfnamefont{Q.~M.}\ \bibnamefont{Tan}}}%
  , \bibinfo {year} {1986},\ \bibfield{title}{%
  \enquote{\bibinfo {title} {Experimental study of {Lagrangian} turbulence in a
  {Stokes} flow},}\ }%
  \bibfield{journal}{%
  \bibinfo {journal} {Proc. Roy. Soc. Lond. A}\ }%
  \textbf{\bibinfo {volume} {408}},\ \bibinfo {pages} {165--174}%
  \bibAnnoteFile{NoStop}{chaiken}%
\bibitem[{\citenamefont{Chen}\ \emph{et~al.}(2013)\citenamefont{Chen},
  \citenamefont{Chen}, \citenamefont{Lin},\ and\ \citenamefont{Hu}}]{Chen2013}%
  \BibitemOpen
  \bibfield{author}{%
  \bibinfo {author} {\bibnamefont{Chen}, \bibfnamefont{C.-Y.}}, \bibinfo
  {author} {\bibfnamefont{C.-Y.}\ \bibnamefont{Chen}}, \bibinfo {author}
  {\bibfnamefont{C.-Y.}\ \bibnamefont{Lin}},\ and\ \bibinfo {author}
  {\bibfnamefont{Y.-T.}\ \bibnamefont{Hu}}}%
  , \bibinfo {year} {2013},\ \bibfield{title}{%
  \enquote{\bibinfo {title} {{Magnetically actuated artificial cilia for
  optimum mixing performance in microfluidics}},}\ }%
  \bibfield{journal}{%
  \bibinfo {journal} {Lab on a chip}\ }%
  \textbf{\bibinfo {volume} {13}},\ \bibinfo {pages} {2834--2839}%
  \bibAnnoteFile{NoStop}{Chen2013}%
\bibitem[{\citenamefont{Cheng}\ and\
  \citenamefont{Sun}(1990{\natexlab{a}})}]{cheng90}%
  \BibitemOpen
  \bibfield{author}{%
  \bibinfo {author} {\bibnamefont{Cheng}, \bibfnamefont{C.~Q.}},\ and\ \bibinfo
  {author} {\bibfnamefont{Y.~S.}\ \bibnamefont{Sun}}}%
  , \bibinfo {year} {1990}{\natexlab{a}},\ \bibfield{title}{%
  \enquote{\bibinfo {title} {Existence of invariant tori in three-dimensional
  measure preserving mappings},}\ }%
  \bibfield{journal}{%
  \bibinfo {journal} {Celest. Mech.}\ }%
  \textbf{\bibinfo {volume} {47}},\ \bibinfo {pages} {275--292}%
  \bibAnnoteFile{NoStop}{cheng90}%
\bibitem[{\citenamefont{Cheng}\ and\
  \citenamefont{Sun}(1990{\natexlab{b}})}]{cheng90b}%
  \BibitemOpen
  \bibfield{author}{%
  \bibinfo {author} {\bibnamefont{Cheng}, \bibfnamefont{C.~Q.}},\ and\ \bibinfo
  {author} {\bibfnamefont{Y.~S.}\ \bibnamefont{Sun}}}%
  , \bibinfo {year} {1990}{\natexlab{b}},\ \bibfield{title}{%
  \enquote{\bibinfo {title} {Existence of periodically invariant curves in
  three-dimensional measure-preserving mappings},}\ }%
  \bibfield{journal}{%
  \bibinfo {journal} {Celest. Mech.}\ }%
  \textbf{\bibinfo {volume} {47}},\ \bibinfo {pages} {293�--303}%
  \bibAnnoteFile{NoStop}{cheng90b}%
\bibitem[{\citenamefont{Cheng}\ and\ \citenamefont{Diez}(2011)}]{Cheng2011}%
  \BibitemOpen
  \bibfield{author}{%
  \bibinfo {author} {\bibnamefont{Cheng}, \bibfnamefont{Y.}},\ and\ \bibinfo
  {author} {\bibfnamefont{F.~J.}\ \bibnamefont{Diez}}}%
  , \bibinfo {year} {2011},\ \bibfield{title}{%
  \enquote{\bibinfo {title} {A {4D} imaging tool for {Lagrangian} particle
  tracking in stirred tanks},}\ }%
  \bibfield{journal}{%
  \bibinfo {journal} {AIChE J.}\ }%
  \textbf{\bibinfo {volume} {57}},\ \bibinfo {pages} {1983--1996}%
  \bibAnnoteFile{NoStop}{Cheng2011}%
\bibitem[{\citenamefont{Chernykh}\ and\
  \citenamefont{Lebedev}(2008)}]{Chernykh2008}%
  \BibitemOpen
  \bibfield{author}{%
  \bibinfo {author} {\bibnamefont{Chernykh}, \bibfnamefont{A.}},\ and\ \bibinfo
  {author} {\bibfnamefont{V.}~\bibnamefont{Lebedev}}}%
  , \bibinfo {year} {2008},\ \bibfield{title}{%
  \enquote{\bibinfo {title} {Passive scalar structures in peripheral regions of
  random flows},}\ }%
  \bibfield{journal}{%
  \bibinfo {journal} {{JETP} Lett.}\ }%
  \textbf{\bibinfo {volume} {87}},\ \bibinfo {pages} {682--686}%
  \bibAnnoteFile{NoStop}{Chernykh2008}%
\bibitem[{\citenamefont{Chertkov}\ and\
  \citenamefont{Lebedev}(2003)}]{Chertkov2003b}%
  \BibitemOpen
  \bibfield{author}{%
  \bibinfo {author} {\bibnamefont{Chertkov}, \bibfnamefont{M.}},\ and\ \bibinfo
  {author} {\bibfnamefont{V.}~\bibnamefont{Lebedev}}}%
  , \bibinfo {year} {2003},\ \bibfield{title}{%
  \enquote{\bibinfo {title} {Boundary effects on chaotic advection-diffusion
  chemical reactions},}\ }%
  \bibfield{journal}{%
  \bibinfo {journal} {Phys. Rev Lett.}\ }%
  \textbf{\bibinfo {volume} {90}},\ \bibinfo {pages} {134501}%
  \bibAnnoteFile{NoStop}{Chertkov2003b}%
\bibitem[{\citenamefont{Chun}\ \emph{et~al.}(2008)\citenamefont{Chun},
  \citenamefont{Kim},\ and\ \citenamefont{Chung}}]{Chun2008}%
  \BibitemOpen
  \bibfield{author}{%
  \bibinfo {author} {\bibnamefont{Chun}, \bibfnamefont{H.}}, \bibinfo {author}
  {\bibfnamefont{H.~C.}\ \bibnamefont{Kim}},\ and\ \bibinfo {author}
  {\bibfnamefont{T.~D.}\ \bibnamefont{Chung}}}%
  , \bibinfo {year} {2008},\ \bibfield{title}{%
  \enquote{\bibinfo {title} {Ultrafast active mixer using polyelectrolytic ion
  extractor},}\ }%
  \bibfield{journal}{%
  \bibinfo {journal} {Lab on a chip}\ }%
  \textbf{\bibinfo {volume} {8}},\ \bibinfo {pages} {764--771}%
  \bibAnnoteFile{NoStop}{Chun2008}%
\bibitem[{\citenamefont{Colantonio}\
  \emph{et~al.}(2009)\citenamefont{Colantonio}, \citenamefont{Vermot},
  \citenamefont{Wu}, \citenamefont{Langenbacher}, \citenamefont{Fraser},
  \citenamefont{Chen},\ and\ \citenamefont{Hill}}]{Colantonio2009}%
  \BibitemOpen
  \bibfield{author}{%
  \bibinfo {author} {\bibnamefont{Colantonio}, \bibfnamefont{J.~R.}}, \bibinfo
  {author} {\bibfnamefont{J.}~\bibnamefont{Vermot}}, \bibinfo {author}
  {\bibfnamefont{D.}~\bibnamefont{Wu}}, \bibinfo {author}
  {\bibfnamefont{A.~D.}\ \bibnamefont{Langenbacher}}, \bibinfo {author}
  {\bibfnamefont{S.}~\bibnamefont{Fraser}}, \bibinfo {author}
  {\bibfnamefont{J.-N.}\ \bibnamefont{Chen}},\ and\ \bibinfo {author}
  {\bibfnamefont{K.~L.}\ \bibnamefont{Hill}}}%
  , \bibinfo {year} {2009},\ \bibfield{title}{%
  \enquote{\bibinfo {title} {{The dynein regulatory complex is required for
  ciliary motility and otolith biogenesis in the inner ear}},}\ }%
  \bibfield{journal}{%
  \bibinfo {journal} {Nature}\ }%
  \textbf{\bibinfo {volume} {457}},\ \bibinfo {pages} {205--209}%
  \bibAnnoteFile{NoStop}{Colantonio2009}%
\bibitem[{\citenamefont{Cortelezzi}\
  \emph{et~al.}(2008)\citenamefont{Cortelezzi}, \citenamefont{Adrover},\ and\
  \citenamefont{Giona}}]{cortelezzi2008feasibility}%
  \BibitemOpen
  \bibfield{author}{%
  \bibinfo {author} {\bibnamefont{Cortelezzi}, \bibfnamefont{L.}}, \bibinfo
  {author} {\bibfnamefont{A.}~\bibnamefont{Adrover}},\ and\ \bibinfo {author}
  {\bibfnamefont{M.}~\bibnamefont{Giona}}}%
  , \bibinfo {year} {2008},\ \bibfield{title}{%
  \enquote{\bibinfo {title} {Feasibility, efficiency and transportability of
  short-horizon optimal mixing protocols},}\ }%
  \bibfield{journal}{%
  \bibinfo {journal} {J. Fluid Mech.}\ }%
  \textbf{\bibinfo {volume} {597}},\ \bibinfo {pages} {199--231}%
  \bibAnnoteFile{NoStop}{cortelezzi2008feasibility}%
\bibitem[{\citenamefont{D'Alessandro}\
  \emph{et~al.}(1999)\citenamefont{D'Alessandro}, \citenamefont{Dahleh},\ and\
  \citenamefont{Mezi\'{c}}}]{d'alessandroetal:1999}%
  \BibitemOpen
  \bibfield{author}{%
  \bibinfo {author} {\bibnamefont{D'Alessandro}, \bibfnamefont{D.}}, \bibinfo
  {author} {\bibfnamefont{M.}~\bibnamefont{Dahleh}},\ and\ \bibinfo {author}
  {\bibfnamefont{I.}~\bibnamefont{Mezi\'{c}}}}%
  , \bibinfo {year} {1999},\ \bibfield{title}{%
  \enquote{\bibinfo {title} {Control of mixing in fluid flow: A maximum entropy
  approach},}\ }%
  \bibfield{journal}{%
  \bibinfo {journal} {IEEE Trans. Automatic Control}\ }%
  \textbf{\bibinfo {volume} {44}},\ \bibinfo {pages} {1852--1863}%
  \bibAnnoteFile{NoStop}{d'alessandroetal:1999}%
\bibitem[{\citenamefont{Danckwerts}(1952)}]{Danckwerts1952}%
  \BibitemOpen
  \bibfield{author}{%
  \bibinfo {author} {\bibnamefont{Danckwerts}, \bibfnamefont{P.~V.}}}%
  , \bibinfo {year} {1952},\ \bibfield{title}{%
  \enquote{\bibinfo {title} {The definition and measurement of some
  characteristics of mixtures},}\ }%
  \bibfield{journal}{%
  \bibinfo {journal} {Appl. Sci. Res.}\ }%
  \textbf{\bibinfo {volume} {A3}},\ \bibinfo {pages} {279--296}%
  \bibAnnoteFile{NoStop}{Danckwerts1952}%
\bibitem[{\citenamefont{De~Moura}(2011)}]{moura2011}%
  \BibitemOpen
  \bibfield{author}{%
  \bibinfo {author} {\bibnamefont{De~Moura}, \bibfnamefont{A.}}}%
  , \bibinfo {year} {2011},\ \bibfield{title}{%
  \enquote{\bibinfo {title} {Reacting particles in open chaotic flows},}\ }%
  \bibfield{journal}{%
  \bibinfo {journal} {Phys. Rev. Lett.}\ }%
  \textbf{\bibinfo {volume} {107}},\ \bibinfo {pages} {274501}%
  \bibAnnoteFile{NoStop}{moura2011}%
\bibitem[{\citenamefont{Dean}(1928)}]{dean28}%
  \BibitemOpen
  \bibfield{author}{%
  \bibinfo {author} {\bibnamefont{Dean}, \bibfnamefont{W.~R.}}}%
  , \bibinfo {year} {1928},\ \bibfield{title}{%
  \enquote{\bibinfo {title} {The streamline motion of fluid in curved pipes},}\
  }%
  \bibfield{journal}{%
  \bibinfo {journal} {Phil. Mag.}\ }%
  \textbf{\bibinfo {volume} {5}},\ \bibinfo {pages} {673--93}%
  \bibAnnoteFile{NoStop}{dean28}%
\bibitem[{\citenamefont{Dellnitz}\ and\
  \citenamefont{Junge}(1999)}]{Dellnitz:1999tr}%
  \BibitemOpen
  \bibfield{author}{%
  \bibinfo {author} {\bibnamefont{Dellnitz}, \bibfnamefont{M.}},\ and\ \bibinfo
  {author} {\bibfnamefont{O.}~\bibnamefont{Junge}}}%
  , \bibinfo {year} {1999},\ \bibfield{title}{%
  \enquote{\bibinfo {title} {On the approximation of complicated dynamical
  behavior},}\ }%
  \bibfield{journal}{%
  \bibinfo {journal} {{SIAM} Journal on Numerical Analysis}\ }%
  \textbf{\bibinfo {volume} {36}},\ \bibinfo {pages} {491--515}%
  \bibAnnoteFile{NoStop}{Dellnitz:1999tr}%
\bibitem[{\citenamefont{Dellnitz}\ and\
  \citenamefont{Junge}(2002)}]{Dellnitz:2002wma}%
  \BibitemOpen
  \bibfield{author}{%
  \bibinfo {author} {\bibnamefont{Dellnitz}, \bibfnamefont{M.}},\ and\ \bibinfo
  {author} {\bibfnamefont{O.}~\bibnamefont{Junge}}}%
  , \bibinfo {year} {2002},\ \enquote{\bibinfo {title} {Set oriented numerical
  methods for dynamical systems},}\ in\ \emph{\bibinfo {booktitle} {Handbook of
  dynamical systems, Vol. 2}}\ (\bibinfo {publisher} {{North-Holland}},\
  \bibinfo {address} {Amsterdam})\ pp.\ \bibinfo {pages} {221--264}%
  \bibAnnoteFile{NoStop}{Dellnitz:2002wma}%
\bibitem[{\citenamefont{Denman}\ and\
  \citenamefont{Gargett}(1995)}]{Denman-Gargett-95}%
  \BibitemOpen
  \bibfield{author}{%
  \bibinfo {author} {\bibnamefont{Denman}, \bibfnamefont{K.}},\ and\ \bibinfo
  {author} {\bibfnamefont{A.}~\bibnamefont{Gargett}}}%
  , \bibinfo {year} {1995},\ \bibfield{title}{%
  \enquote{\bibinfo {title} {Biological-physical interactions in the upper
  ocean: the role of vertical and small scale transport processes},}\ }%
  \bibfield{journal}{%
  \bibinfo {journal} {Annu. Rev. Fluid Mech.}\ }%
  \textbf{\bibinfo {volume} {27}},\ \bibinfo {pages} {225--255}%
  \bibAnnoteFile{NoStop}{Denman-Gargett-95}%
\bibitem[{\citenamefont{Di~Carlo}(2009)}]{dicarlo2009}%
  \BibitemOpen
  \bibfield{author}{%
  \bibinfo {author} {\bibnamefont{Di~Carlo}, \bibfnamefont{D.}}}%
  , \bibinfo {year} {2009},\ \bibfield{title}{%
  \enquote{\bibinfo {title} {Inertial microfluidics},}\ }%
  \bibfield{journal}{%
  \bibinfo {journal} {Lab on a chip}\ }%
  \textbf{\bibinfo {volume} {9}},\ \bibinfo {pages} {3038--3046}%
  \bibAnnoteFile{NoStop}{dicarlo2009}%
\bibitem[{\citenamefont{Ding}\ \emph{et~al.}(2014)\citenamefont{Ding},
  \citenamefont{Nawroth}, \citenamefont{McFall-Ngai},\ and\
  \citenamefont{Kanso}}]{Ding2014}%
  \BibitemOpen
  \bibfield{author}{%
  \bibinfo {author} {\bibnamefont{Ding}, \bibfnamefont{Y.}}, \bibinfo {author}
  {\bibfnamefont{J.~C.}\ \bibnamefont{Nawroth}}, \bibinfo {author}
  {\bibfnamefont{M.~J.}\ \bibnamefont{McFall-Ngai}},\ and\ \bibinfo {author}
  {\bibfnamefont{E.}~\bibnamefont{Kanso}}}%
  , \bibinfo {year} {2014},\ \bibfield{title}{%
  \enquote{\bibinfo {title} {{Mixing and transport by ciliary carpets: a
  numerical study}},}\ }%
  \bibfield{journal}{%
  \bibinfo {journal} {J. Fluid Mech.}\ }%
  \textbf{\bibinfo {volume} {743}}~(\bibinfo {number} {2007}),\ \bibinfo
  {pages} {124--140}%
  \bibAnnoteFile{NoStop}{Ding2014}%
\bibitem[{\citenamefont{Doering}\ and\
  \citenamefont{Thiffeault}(2006)}]{DoeringThiffeault2006}%
  \BibitemOpen
  \bibfield{author}{%
  \bibinfo {author} {\bibnamefont{Doering}, \bibfnamefont{C.~R.}},\ and\
  \bibinfo {author} {\bibfnamefont{J.-L.}\ \bibnamefont{Thiffeault}}}%
  , \bibinfo {year} {2006},\ \bibfield{title}{%
  \enquote{\bibinfo {title} {Multiscale mixing efficiencies for steady
  sources},}\ }%
  \bibfield{journal}{%
  \bibinfo {journal} {Phys. Rev. E}\ }%
  \textbf{\bibinfo {volume} {74}},\ \bibinfo {pages} {025301(R)}%
  \bibAnnoteFile{NoStop}{DoeringThiffeault2006}%
\bibitem[{\citenamefont{Dombre}\ \emph{et~al.}(1986)\citenamefont{Dombre},
  \citenamefont{Frisch}, \citenamefont{Greene}, \citenamefont{H\'{e}non},
  \citenamefont{Mehr},\ and\ \citenamefont{Soward}}]{Dombre1986}%
  \BibitemOpen
  \bibfield{author}{%
  \bibinfo {author} {\bibnamefont{Dombre}, \bibfnamefont{T.}}, \bibinfo
  {author} {\bibfnamefont{U.}~\bibnamefont{Frisch}}, \bibinfo {author}
  {\bibfnamefont{J.~M.}\ \bibnamefont{Greene}}, \bibinfo {author}
  {\bibfnamefont{M.}~\bibnamefont{H\'{e}non}}, \bibinfo {author}
  {\bibfnamefont{A.}~\bibnamefont{Mehr}},\ and\ \bibinfo {author}
  {\bibfnamefont{A.~M.}\ \bibnamefont{Soward}}}%
  , \bibinfo {year} {1986},\ \bibfield{title}{%
  \enquote{\bibinfo {title} {Chaotic streamlines in the abc flows},}\ }%
  \bibfield{journal}{%
  \bibinfo {journal} {J.~Fluid Mech.}\ }%
  \textbf{\bibinfo {volume} {167}},\ \bibinfo {pages} {353--391}%
  \bibAnnoteFile{NoStop}{Dombre1986}%
\bibitem[{\citenamefont{Domingos}\ and\
  \citenamefont{Cardoso}(2013)}]{Domingos2013}%
  \BibitemOpen
  \bibfield{author}{%
  \bibinfo {author} {\bibnamefont{Domingos}, \bibfnamefont{M.~G.}},\ and\
  \bibinfo {author} {\bibfnamefont{S.~S.~S}\ \bibnamefont{Cardoso}}}%
  , \bibinfo {year} {2013},\ \bibfield{title}{%
  \enquote{\bibinfo {title} {Turbulent two-phase plumes with bubble-size
  reduction owing to dissolution or chemical reaction},}\ }%
  \bibfield{journal}{%
  \bibinfo {journal} {J. Fluid Mech}\ }%
  \textbf{\bibinfo {volume} {716}},\ \bibinfo {pages} {120--136}%
  \bibAnnoteFile{NoStop}{Domingos2013}%
\bibitem[{\citenamefont{Domingos}\ and\
  \citenamefont{Cardoso}(2015)}]{Domingos2015}%
  \BibitemOpen
  \bibfield{author}{%
  \bibinfo {author} {\bibnamefont{Domingos}, \bibfnamefont{M.~G.}},\ and\
  \bibinfo {author} {\bibfnamefont{S.~S.~S}\ \bibnamefont{Cardoso}}}%
  , \bibinfo {year} {2015},\ \bibfield{title}{%
  \enquote{\bibinfo {title} {Turbulent thermals with chemical reaction},}\ }%
  \bibfield{journal}{%
  \bibinfo {journal} {J. Fluid Mech}\ }%
  \textbf{\bibinfo {volume} {784}},\ \bibinfo {pages} {5--29}%
  \bibAnnoteFile{NoStop}{Domingos2015}%
\bibitem[{\citenamefont{Dore}\ \emph{et~al.}(2009)\citenamefont{Dore},
  \citenamefont{Moroni}, \citenamefont{Menach},\ and\
  \citenamefont{Cenedese}}]{Dore2009}%
  \BibitemOpen
  \bibfield{author}{%
  \bibinfo {author} {\bibnamefont{Dore}, \bibfnamefont{V.}}, \bibinfo {author}
  {\bibfnamefont{M.}~\bibnamefont{Moroni}}, \bibinfo {author}
  {\bibfnamefont{M.~L.}\ \bibnamefont{Menach}},\ and\ \bibinfo {author}
  {\bibfnamefont{A.}~\bibnamefont{Cenedese}}}%
  , \bibinfo {year} {2009},\ \bibfield{title}{%
  \enquote{\bibinfo {title} {Investigation of penetrative convection in
  stratified fluids through {3D-PTV}},}\ }%
  \bibfield{journal}{%
  \bibinfo {journal} {Exp. Fluids}\ }%
  \textbf{\bibinfo {volume} {47}},\ \bibinfo {pages} {811--825}%
  \bibAnnoteFile{NoStop}{Dore2009}%
\bibitem[{\citenamefont{Drescher}\ \emph{et~al.}(2010)\citenamefont{Drescher},
  \citenamefont{Goldstein}, \citenamefont{Michel}, \citenamefont{Polin},\ and\
  \citenamefont{Tuval}}]{Drescher2010}%
  \BibitemOpen
  \bibfield{author}{%
  \bibinfo {author} {\bibnamefont{Drescher}, \bibfnamefont{K.}}, \bibinfo
  {author} {\bibfnamefont{R.~E.}\ \bibnamefont{Goldstein}}, \bibinfo {author}
  {\bibfnamefont{N.}~\bibnamefont{Michel}}, \bibinfo {author}
  {\bibfnamefont{M.}~\bibnamefont{Polin}},\ and\ \bibinfo {author}
  {\bibfnamefont{I.}~\bibnamefont{Tuval}}}%
  , \bibinfo {year} {2010},\ \bibfield{title}{%
  \enquote{\bibinfo {title} {Direct measurement of the flow field around
  swimming microorganisms},}\ }%
  \bibfield{journal}{%
  \bibinfo {journal} {Phys. Rev. Lett.}\ }%
  \textbf{\bibinfo {volume} {105}},\ \bibinfo {pages} {168101}%
  \bibAnnoteFile{NoStop}{Drescher2010}%
\bibitem[{\citenamefont{Dutta}\ and\ \citenamefont{Chevray}(1995)}]{dutta1995}%
  \BibitemOpen
  \bibfield{author}{%
  \bibinfo {author} {\bibnamefont{Dutta}, \bibfnamefont{P.}},\ and\ \bibinfo
  {author} {\bibfnamefont{R.}~\bibnamefont{Chevray}}}%
  , \bibinfo {year} {1995},\ \bibfield{title}{%
  \enquote{\bibinfo {title} {Inertial effects in chaotic mixing with
  diffusion},}\ }%
  \bibfield{journal}{%
  \bibinfo {journal} {J. Fluid Mech.}\ }%
  \textbf{\bibinfo {volume} {285}},\ \bibinfo {pages} {1--13}%
  \bibAnnoteFile{NoStop}{dutta1995}%
\bibitem[{\citenamefont{Eckart}(1948)}]{Eckart1948}%
  \BibitemOpen
  \bibfield{author}{%
  \bibinfo {author} {\bibnamefont{Eckart}, \bibfnamefont{C.}}}%
  , \bibinfo {year} {1948},\ \bibfield{title}{%
  \enquote{\bibinfo {title} {An analysis of the stirring and mixing processes
  in incompressible fluids},}\ }%
  \bibfield{journal}{%
  \bibinfo {journal} {J. Marine Res.}\ }%
  \textbf{\bibinfo {volume} {7}},\ \bibinfo {pages} {265--275}%
  \bibAnnoteFile{NoStop}{Eckart1948}%
\bibitem[{\citenamefont{Eckhardt}\ and\
  \citenamefont{Hasco\"et}(2005)}]{Hascoet2005}%
  \BibitemOpen
  \bibfield{author}{%
  \bibinfo {author} {\bibnamefont{Eckhardt}, \bibfnamefont{B.}},\ and\ \bibinfo
  {author} {\bibfnamefont{E.}~\bibnamefont{Hasco\"et}}}%
  , \bibinfo {year} {2005},\ \bibfield{title}{%
  \enquote{\bibinfo {title} {Breaking time reversal symmetry by viscous
  dephasing},}\ }%
  \bibfield{journal}{%
  \bibinfo {journal} {Phys. Rev. E}\ }%
  \textbf{\bibinfo {volume} {72}},\ \bibinfo {pages} {037301}%
  \bibAnnoteFile{NoStop}{Hascoet2005}%
\bibitem[{\citenamefont{El~Omari}\ and\
  \citenamefont{Le~Guer}(2009)}]{elomari2009}%
  \BibitemOpen
  \bibfield{author}{%
  \bibinfo {author} {\bibnamefont{El~Omari}, \bibfnamefont{K.}},\ and\ \bibinfo
  {author} {\bibfnamefont{Y.}~\bibnamefont{Le~Guer}}}%
  , \bibinfo {year} {2009},\ \bibfield{title}{%
  \enquote{\bibinfo {title} {{A numerical study of thermal chaotic mixing in a
  two rod rotating mixer}},}\ }%
  \bibfield{journal}{%
  \bibinfo {journal} {Comput. Thermal Sci.}\ }%
  \textbf{\bibinfo {volume} {1}},\ \bibinfo {pages} {55--73}%
  \bibAnnoteFile{NoStop}{elomari2009}%
\bibitem[{\citenamefont{El~Omari}\ and\
  \citenamefont{Le~Guer}(2010{\natexlab{a}})}]{elomari2010a}%
  \BibitemOpen
  \bibfield{author}{%
  \bibinfo {author} {\bibnamefont{El~Omari}, \bibfnamefont{K.}},\ and\ \bibinfo
  {author} {\bibfnamefont{Y.}~\bibnamefont{Le~Guer}}}%
  , \bibinfo {year} {2010}{\natexlab{a}},\ \bibfield{title}{%
  \enquote{\bibinfo {title} {Alternate rotating walls for thermal chaotic
  mixing},}\ }%
  \bibfield{journal}{%
  \bibinfo {journal} {Int. J. Heat Mass Transfer}\ }%
  \textbf{\bibinfo {volume} {53}},\ \bibinfo {pages} {123--134}%
  \bibAnnoteFile{NoStop}{elomari2010a}%
\bibitem[{\citenamefont{El~Omari}\ and\
  \citenamefont{Le~Guer}(2010{\natexlab{b}})}]{elomari2010b}%
  \BibitemOpen
  \bibfield{author}{%
  \bibinfo {author} {\bibnamefont{El~Omari}, \bibfnamefont{K.}},\ and\ \bibinfo
  {author} {\bibfnamefont{Y.}~\bibnamefont{Le~Guer}}}%
  , \bibinfo {year} {2010}{\natexlab{b}},\ \bibfield{title}{%
  \enquote{\bibinfo {title} {Thermal chaotic mixing of power law fluids in a
  mixer with alternately-rotating walls},}\ }%
  \bibfield{journal}{%
  \bibinfo {journal} {J. Non-Newtonian Fluid Mech.}\ }%
  \textbf{\bibinfo {volume} {165}},\ \bibinfo {pages} {641--651}%
  \bibAnnoteFile{NoStop}{elomari2010b}%
\bibitem[{\citenamefont{El~Omari}\ and\
  \citenamefont{Le~Guer}(2012)}]{elomari2012}%
  \BibitemOpen
  \bibfield{author}{%
  \bibinfo {author} {\bibnamefont{El~Omari}, \bibfnamefont{K.}},\ and\ \bibinfo
  {author} {\bibfnamefont{Y.}~\bibnamefont{Le~Guer}}}%
  , \bibinfo {year} {2012},\ \bibfield{title}{%
  \enquote{\bibinfo {title} {{Laminar mixing and heat transfer for constant
  heat flux boundary condition}},}\ }%
  \bibfield{journal}{%
  \bibinfo {journal} {Heat and Mass Transfer}\ }%
  \textbf{\bibinfo {volume} {48}},\ \bibinfo {pages} {1285--1296}%
  \bibAnnoteFile{NoStop}{elomari2012}%
\bibitem[{\citenamefont{El~Omari}\ \emph{et~al.}(2015)\citenamefont{El~Omari},
  \citenamefont{Le~Guer}, \citenamefont{Perugini},\ and\
  \citenamefont{Petrelli}}]{ElOmari20151835}%
  \BibitemOpen
  \bibfield{author}{%
  \bibinfo {author} {\bibnamefont{El~Omari}, \bibfnamefont{K.}}, \bibinfo
  {author} {\bibfnamefont{Y.}~\bibnamefont{Le~Guer}}, \bibinfo {author}
  {\bibfnamefont{D.}~\bibnamefont{Perugini}},\ and\ \bibinfo {author}
  {\bibfnamefont{M.}~\bibnamefont{Petrelli}}}%
  , \bibinfo {year} {2015},\ \bibfield{title}{%
  \enquote{\bibinfo {title} {Cooling of a magmatic system under thermal chaotic
  mixing},}\ }%
  \bibfield{journal}{%
  \bibinfo {journal} {Pure Appl. Geophys.}\ }%
  \textbf{\bibinfo {volume} {172}},\ \bibinfo {pages} {1835--1849}%
  \bibAnnoteFile{NoStop}{ElOmari20151835}%
\bibitem[{\citenamefont{Elgeti}\ and\
  \citenamefont{Gompper}(2013)}]{Elgeti2013}%
  \BibitemOpen
  \bibfield{author}{%
  \bibinfo {author} {\bibnamefont{Elgeti}, \bibfnamefont{J.}},\ and\ \bibinfo
  {author} {\bibfnamefont{G.}~\bibnamefont{Gompper}}}%
  , \bibinfo {year} {2013},\ \bibfield{title}{%
  \enquote{\bibinfo {title} {{Emergence of metachronal waves in cilia
  arrays}},}\ }%
  \bibfield{journal}{%
  \bibinfo {journal} {Proc. Natl Acad. Sci. USA}\ }%
  \textbf{\bibinfo {volume} {110}},\ \bibinfo {pages} {4470--4475}%
  \bibAnnoteFile{NoStop}{Elgeti2013}%
\bibitem[{\citenamefont{Falconer}(2003)}]{falconer}%
  \BibitemOpen
  \bibfield{author}{%
  \bibinfo {author} {\bibnamefont{Falconer}, \bibfnamefont{K.}}}%
  , \bibinfo {year} {2003},\ \emph{\bibinfo {title} {Fractal Geometry:
  Mathematical Foundations and Applications}},\ \bibinfo {edition} {2nd}\ ed.\
  (\bibinfo {publisher} {Wiley})%
  \bibAnnoteFile{NoStop}{falconer}%
\bibitem[{\citenamefont{Falkovich}\
  \emph{et~al.}(2001)\citenamefont{Falkovich}, \citenamefont{Gaw\c{e}dzki},\
  and\ \citenamefont{Vergassola}}]{Falkovich2001}%
  \BibitemOpen
  \bibfield{author}{%
  \bibinfo {author} {\bibnamefont{Falkovich}, \bibfnamefont{G.}}, \bibinfo
  {author} {\bibfnamefont{K.}~\bibnamefont{Gaw\c{e}dzki}},\ and\ \bibinfo
  {author} {\bibfnamefont{M.}~\bibnamefont{Vergassola}}}%
  , \bibinfo {year} {2001},\ \bibfield{title}{%
  \enquote{\bibinfo {title} {Particles and fields in turbulence},}\ }%
  \bibfield{journal}{%
  \bibinfo {journal} {Rev. Mod. Phys.}\ }%
  \textbf{\bibinfo {volume} {73}},\ \bibinfo {pages} {913--975}%
  \bibAnnoteFile{NoStop}{Falkovich2001}%
\bibitem[{\citenamefont{Faubel}\ \emph{et~al.}(2016)\citenamefont{Faubel},
  \citenamefont{Westendorf}, \citenamefont{Bodenschatz},\ and\
  \citenamefont{Eichele}}]{Faubel2016}%
  \BibitemOpen
  \bibfield{author}{%
  \bibinfo {author} {\bibnamefont{Faubel}, \bibfnamefont{R.}}, \bibinfo
  {author} {\bibfnamefont{C.}~\bibnamefont{Westendorf}}, \bibinfo {author}
  {\bibfnamefont{E.}~\bibnamefont{Bodenschatz}},\ and\ \bibinfo {author}
  {\bibfnamefont{G.}~\bibnamefont{Eichele}}}%
  , \bibinfo {year} {2016},\ \bibfield{title}{%
  \enquote{\bibinfo {title} {{Cilia-based flow network in the brain
  ventricles}},}\ }%
  \bibfield{journal}{%
  \bibinfo {journal} {Science}\ }%
  \textbf{\bibinfo {volume} {353}},\ \bibinfo {pages} {176--178}%
  \bibAnnoteFile{NoStop}{Faubel2016}%
\bibitem[{\citenamefont{Fauci}\ and\ \citenamefont{Dillon}(2006)}]{Fauci2006}%
  \BibitemOpen
  \bibfield{author}{%
  \bibinfo {author} {\bibnamefont{Fauci}, \bibfnamefont{L.~J.}},\ and\ \bibinfo
  {author} {\bibfnamefont{R.}~\bibnamefont{Dillon}}}%
  , \bibinfo {year} {2006},\ \bibfield{title}{%
  \enquote{\bibinfo {title} {Biofluidmechanics of reproduction},}\ }%
  \bibfield{journal}{%
  \bibinfo {journal} {Annu. Rev. Fluid Mech.}\ }%
  \textbf{\bibinfo {volume} {38}},\ \bibinfo {pages} {371--394}%
  \bibAnnoteFile{NoStop}{Fauci2006}%
\bibitem[{\citenamefont{Feingold}\ \emph{et~al.}(1987)\citenamefont{Feingold},
  \citenamefont{Kadanoff},\ and\ \citenamefont{Piro}}]{Feingold1987}%
  \BibitemOpen
  \bibfield{author}{%
  \bibinfo {author} {\bibnamefont{Feingold}, \bibfnamefont{M.}}, \bibinfo
  {author} {\bibfnamefont{L.~P.}\ \bibnamefont{Kadanoff}},\ and\ \bibinfo
  {author} {\bibfnamefont{O.}~\bibnamefont{Piro}}}%
  , \bibinfo {year} {1987},\ \enquote{\bibinfo {title} {A way to connect fluid
  dynamics to dynamical systems: passive scalars},}\ in\ \emph{\bibinfo
  {booktitle} {Fractal aspects of materials: disordered systems}},\ \bibinfo
  {editor} {edited by\ \bibinfo {editor} {\bibfnamefont{A.~J.}\
  \bibnamefont{Hurd}}, \bibinfo {editor} {\bibfnamefont{D.~A.}\
  \bibnamefont{Weitz}},\ and\ \bibinfo {editor} {\bibfnamefont{B.~B.}\
  \bibnamefont{Mandelbrot}}}\ (\bibinfo {publisher} {Materials Research
  Society})\ pp.\ \bibinfo {pages} {203--205}%
  \bibAnnoteFile{NoStop}{Feingold1987}%
\bibitem[{\citenamefont{Feingold}\
  \emph{et~al.}(1988{\natexlab{a}})\citenamefont{Feingold},
  \citenamefont{Kadanoff},\ and\ \citenamefont{Piro}}]{Feingold1988b}%
  \BibitemOpen
  \bibfield{author}{%
  \bibinfo {author} {\bibnamefont{Feingold}, \bibfnamefont{M.}}, \bibinfo
  {author} {\bibfnamefont{L.~P.}\ \bibnamefont{Kadanoff}},\ and\ \bibinfo
  {author} {\bibfnamefont{O.}~\bibnamefont{Piro}}}%
  , \bibinfo {year} {1988}{\natexlab{a}},\ \enquote{\bibinfo {title} {Diffusion
  of passive scalars in fluid flows: maps in three dimensions},}\ in\
  \emph{\bibinfo {booktitle} {Universalities in Condensed Matter}},\ \bibinfo
  {editor} {edited by\ \bibinfo {editor}
  {\bibfnamefont{R.}~\bibnamefont{Jullien}}, \bibinfo {editor}
  {\bibfnamefont{L.}~\bibnamefont{Peliti}}, \bibinfo {editor}
  {\bibfnamefont{R.}~\bibnamefont{Rammal}},\ and\ \bibinfo {editor}
  {\bibfnamefont{N.}~\bibnamefont{Boccara}}}\ (\bibinfo {publisher} {Springer
  Proc. Phys. (Springer, Berlin, Heidelberg)})%
  \bibAnnoteFile{NoStop}{Feingold1988b}%
\bibitem[{\citenamefont{Feingold}\
  \emph{et~al.}(1988{\natexlab{b}})\citenamefont{Feingold},
  \citenamefont{Kadanoff},\ and\ \citenamefont{Piro}}]{Feingold1988}%
  \BibitemOpen
  \bibfield{author}{%
  \bibinfo {author} {\bibnamefont{Feingold}, \bibfnamefont{M.}}, \bibinfo
  {author} {\bibfnamefont{L.~P.}\ \bibnamefont{Kadanoff}},\ and\ \bibinfo
  {author} {\bibfnamefont{O.}~\bibnamefont{Piro}}}%
  , \bibinfo {year} {1988}{\natexlab{b}},\ \bibfield{title}{%
  \enquote{\bibinfo {title} {Passive scalars, three-dimensional
  volume-preserving maps and chaos},}\ }%
  \bibfield{journal}{%
  \bibinfo {journal} {J. Stat. Phys.}\ }%
  \textbf{\bibinfo {volume} {50}},\ \bibinfo {pages} {529--565}%
  \bibAnnoteFile{NoStop}{Feingold1988}%
\bibitem[{\citenamefont{Feingold}\ \emph{et~al.}(1989)\citenamefont{Feingold},
  \citenamefont{Kadanoff},\ and\ \citenamefont{Piro}}]{Feingold1989}%
  \BibitemOpen
  \bibfield{author}{%
  \bibinfo {author} {\bibnamefont{Feingold}, \bibfnamefont{M.}}, \bibinfo
  {author} {\bibfnamefont{L.~P.}\ \bibnamefont{Kadanoff}},\ and\ \bibinfo
  {author} {\bibfnamefont{O.}~\bibnamefont{Piro}}}%
  , \bibinfo {year} {1989},\ \enquote{\bibinfo {title} {Transport of passive
  scalars: Kam surfaces and diffusion in three-dimensional {L}iouvillean
  maps},}\ in\ \emph{\bibinfo {booktitle} {Instabilities and Nonequilibrium
  Structures}},\ \bibinfo {editor} {edited by\ \bibinfo {editor}
  {\bibfnamefont{E.}~\bibnamefont{Tirapegui}}\ and\ \bibinfo {editor}
  {\bibfnamefont{D.}~\bibnamefont{Villarroel}}}\ (\bibinfo {publisher} {D.
  Reidel, Dordrecht})%
  \bibAnnoteFile{NoStop}{Feingold1989}%
\bibitem[{\citenamefont{Fereday}\ and\
  \citenamefont{Haynes}(2004)}]{Fereday2004}%
  \BibitemOpen
  \bibfield{author}{%
  \bibinfo {author} {\bibnamefont{Fereday}, \bibfnamefont{D.~R.}},\ and\
  \bibinfo {author} {\bibfnamefont{P.~H.}\ \bibnamefont{Haynes}}}%
  , \bibinfo {year} {2004},\ \bibfield{title}{%
  \enquote{\bibinfo {title} {Scalar decay in two-dimensional chaotic advection
  and {B}atchelor-regime turbulence},}\ }%
  \bibfield{journal}{%
  \bibinfo {journal} {Phys. Fluids}\ }%
  \textbf{\bibinfo {volume} {16}},\ \bibinfo {pages} {4359--4370}%
  \bibAnnoteFile{NoStop}{Fereday2004}%
\bibitem[{\citenamefont{Fereday}\ \emph{et~al.}(2002)\citenamefont{Fereday},
  \citenamefont{Haynes}, \citenamefont{Wonhas},\ and\
  \citenamefont{Vassilicos}}]{Fereday2002}%
  \BibitemOpen
  \bibfield{author}{%
  \bibinfo {author} {\bibnamefont{Fereday}, \bibfnamefont{D.~R.}}, \bibinfo
  {author} {\bibfnamefont{P.~H.}\ \bibnamefont{Haynes}}, \bibinfo {author}
  {\bibfnamefont{A.}~\bibnamefont{Wonhas}},\ and\ \bibinfo {author}
  {\bibfnamefont{J.~C.}\ \bibnamefont{Vassilicos}}}%
  , \bibinfo {year} {2002},\ \bibfield{title}{%
  \enquote{\bibinfo {title} {Scalar variance decay in chaotic advection and
  {B}atchelor-regime turbulence},}\ }%
  \bibfield{journal}{%
  \bibinfo {journal} {Phys. Rev. E}\ }%
  \textbf{\bibinfo {volume} {65}},\ \bibinfo {pages} {035301(R)}%
  \bibAnnoteFile{NoStop}{Fereday2002}%
\bibitem[{\citenamefont{Finn}\ and\
  \citenamefont{{Thiffeault}}(2011)}]{Finn2011}%
  \BibitemOpen
  \bibfield{author}{%
  \bibinfo {author} {\bibnamefont{Finn}, \bibfnamefont{{M}.}},\ and\ \bibinfo
  {author} {\bibfnamefont{{J}.}~\bibnamefont{{Thiffeault}}}}%
  , \bibinfo {year} {2011},\ \bibfield{title}{%
  \enquote{\bibinfo {title} {Topological {optimization} of {rod}-{stirring}
  {devices}},}\ }%
  \bibfield{journal}{%
  \bibinfo {journal} {{SIAM} {Rev.}}\ }%
  \textbf{\bibinfo {volume} {53}},\ \bibinfo {pages} {723--743}%
  \bibAnnoteFile{NoStop}{Finn2011}%
\bibitem[{\citenamefont{Fountain}\ \emph{et~al.}(2000)\citenamefont{Fountain},
  \citenamefont{Khakhar}, \citenamefont{Mezi\'{c}},\ and\
  \citenamefont{Ottino}}]{Fountain2000}%
  \BibitemOpen
  \bibfield{author}{%
  \bibinfo {author} {\bibnamefont{Fountain}, \bibfnamefont{G.~O.}}, \bibinfo
  {author} {\bibfnamefont{D.~V.}\ \bibnamefont{Khakhar}}, \bibinfo {author}
  {\bibfnamefont{I.}~\bibnamefont{Mezi\'{c}}},\ and\ \bibinfo {author}
  {\bibfnamefont{J.~M.}\ \bibnamefont{Ottino}}}%
  , \bibinfo {year} {2000},\ \bibfield{title}{%
  \enquote{\bibinfo {title} {Chaotic mixing in a bounded three-dimensional
  flow},}\ }%
  \bibfield{journal}{%
  \bibinfo {journal} {J.~Fluid Mech.}\ }%
  \textbf{\bibinfo {volume} {417}},\ \bibinfo {pages} {265--301}%
  \bibAnnoteFile{NoStop}{Fountain2000}%
\bibitem[{\citenamefont{Fountain}\ \emph{et~al.}(1998)\citenamefont{Fountain},
  \citenamefont{Khakhar},\ and\ \citenamefont{Ottino}}]{Fountain1998}%
  \BibitemOpen
  \bibfield{author}{%
  \bibinfo {author} {\bibnamefont{Fountain}, \bibfnamefont{G.~O.}}, \bibinfo
  {author} {\bibfnamefont{D.~V.}\ \bibnamefont{Khakhar}},\ and\ \bibinfo
  {author} {\bibfnamefont{J.~M.}\ \bibnamefont{Ottino}}}%
  , \bibinfo {year} {1998},\ \bibfield{title}{%
  \enquote{\bibinfo {title} {Chaotic mixing in a bounded three-dimensional
  flow},}\ }%
  \bibfield{journal}{%
  \bibinfo {journal} {Science}\ }%
  \textbf{\bibinfo {volume} {281}},\ \bibinfo {pages} {683--686}%
  \bibAnnoteFile{NoStop}{Fountain1998}%
\bibitem[{\citenamefont{Foures}\ \emph{et~al.}(2014)\citenamefont{Foures},
  \citenamefont{Caulfield},\ and\ \citenamefont{Schmid}}]{foures2014}%
  \BibitemOpen
  \bibfield{author}{%
  \bibinfo {author} {\bibnamefont{Foures}, \bibfnamefont{D.~P.~G.}}, \bibinfo
  {author} {\bibfnamefont{C.~P.}\ \bibnamefont{Caulfield}},\ and\ \bibinfo
  {author} {\bibfnamefont{P.~J.}\ \bibnamefont{Schmid}}}%
  , \bibinfo {year} {2014},\ \bibfield{title}{%
  \enquote{\bibinfo {title} {Optimal mixing in two-dimensional plane
  {Poiseuille} flow at finite {P\'eclet} number},}\ }%
  \bibfield{journal}{%
  \bibinfo {journal} {J. Fluid Mech.}\ }%
  \textbf{\bibinfo {volume} {748}},\ \bibinfo {pages} {241--277}%
  \bibAnnoteFile{NoStop}{foures2014}%
\bibitem[{\citenamefont{Franjione}\
  \emph{et~al.}(1989)\citenamefont{Franjione}, \citenamefont{Leong},\ and\
  \citenamefont{Ottino}}]{Franjione1989}%
  \BibitemOpen
  \bibfield{author}{%
  \bibinfo {author} {\bibnamefont{Franjione}, \bibfnamefont{J.~G.}}, \bibinfo
  {author} {\bibfnamefont{C-W.}\ \bibnamefont{Leong}},\ and\ \bibinfo {author}
  {\bibfnamefont{J.~M.}\ \bibnamefont{Ottino}}}%
  , \bibinfo {year} {1989},\ \bibfield{title}{%
  \enquote{\bibinfo {title} {Symmetries within chaos: A route to effective
  mixing},}\ }%
  \bibfield{journal}{%
  \bibinfo {journal} {Phys.\ Fluids \ A}\ }%
  \textbf{\bibinfo {volume} {11}},\ \bibinfo {pages} {1772--1783}%
  \bibAnnoteFile{NoStop}{Franjione1989}%
\bibitem[{\citenamefont{Freund}\ \emph{et~al.}(2012)\citenamefont{Freund},
  \citenamefont{Goetz}, \citenamefont{Hill},\ and\
  \citenamefont{Vermot}}]{Freund2012}%
  \BibitemOpen
  \bibfield{author}{%
  \bibinfo {author} {\bibnamefont{Freund}, \bibfnamefont{J.~B.}}, \bibinfo
  {author} {\bibfnamefont{J.~G.}\ \bibnamefont{Goetz}}, \bibinfo {author}
  {\bibfnamefont{K.~L.}\ \bibnamefont{Hill}},\ and\ \bibinfo {author}
  {\bibfnamefont{J.}~\bibnamefont{Vermot}}}%
  , \bibinfo {year} {2012},\ \bibfield{title}{%
  \enquote{\bibinfo {title} {{Fluid flows and forces in development: functions,
  features and biophysical principles}},}\ }%
  \bibfield{journal}{%
  \bibinfo {journal} {Development}\ }%
  \textbf{\bibinfo {volume} {139}},\ \bibinfo {pages} {1229--1245}%
  \bibAnnoteFile{NoStop}{Freund2012}%
\bibitem[{\citenamefont{Frisch}(1996)}]{frisch1996turbulence}%
  \BibitemOpen
  \bibfield{author}{%
  \bibinfo {author} {\bibnamefont{Frisch}, \bibfnamefont{U.}}}%
  , \bibinfo {year} {1996},\ \emph{\bibinfo {title} {Turbulence}}\ (\bibinfo
  {publisher} {Cambridge University Press, Cambridge})%
  \bibAnnoteFile{NoStop}{frisch1996turbulence}%
\bibitem[{\citenamefont{Froyland}\ and\
  \citenamefont{Dellnitz}(2003)}]{Froyland:2003jj}%
  \BibitemOpen
  \bibfield{author}{%
  \bibinfo {author} {\bibnamefont{Froyland}, \bibfnamefont{G.}},\ and\ \bibinfo
  {author} {\bibfnamefont{M.}~\bibnamefont{Dellnitz}}}%
  , \bibinfo {year} {2003},\ \bibfield{title}{%
  \enquote{\bibinfo {title} {Detecting and locating near-optimal
  almost-invariant sets and cycles},}\ }%
  \bibfield{journal}{%
  \bibinfo {journal} {{SIAM} Journal on Scientific Computing}\ }%
  \textbf{\bibinfo {volume} {24}},\ \bibinfo {pages} {1839--1863}%
  \bibAnnoteFile{NoStop}{Froyland:2003jj}%
\bibitem[{\citenamefont{Froyland}\ \emph{et~al.}(2012)\citenamefont{Froyland},
  \citenamefont{Horenkamp}, \citenamefont{Rossi},
  \citenamefont{Santitissadeekorn},\ and\
  \citenamefont{Gupta}}]{Froyland:2012fo}%
  \BibitemOpen
  \bibfield{author}{%
  \bibinfo {author} {\bibnamefont{Froyland}, \bibfnamefont{G.}}, \bibinfo
  {author} {\bibfnamefont{C.}~\bibnamefont{Horenkamp}}, \bibinfo {author}
  {\bibfnamefont{V.}~\bibnamefont{Rossi}}, \bibinfo {author}
  {\bibfnamefont{N.}~\bibnamefont{Santitissadeekorn}},\ and\ \bibinfo {author}
  {\bibfnamefont{A.~S.}\ \bibnamefont{Gupta}}}%
  , \bibinfo {year} {2012},\ \bibfield{title}{%
  \enquote{\bibinfo {title} {Three-dimensional characterization and tracking of
  an agulhas ring},}\ }%
  \bibfield{journal}{%
  \bibinfo {journal} {Ocean Modelling}\ }%
  \textbf{\bibinfo {volume} {52}},\ \bibinfo {pages} {69--75}%
  \bibAnnoteFile{NoStop}{Froyland:2012fo}%
\bibitem[{\citenamefont{Froyland}\
  \emph{et~al.}(2010{\natexlab{a}})\citenamefont{Froyland},
  \citenamefont{Lloyd},\ and\
  \citenamefont{Santitissadeekorn}}]{froyland_coherent_2010}%
  \BibitemOpen
  \bibfield{author}{%
  \bibinfo {author} {\bibnamefont{Froyland}, \bibfnamefont{G.}}, \bibinfo
  {author} {\bibfnamefont{S.}~\bibnamefont{Lloyd}},\ and\ \bibinfo {author}
  {\bibfnamefont{N.}~\bibnamefont{Santitissadeekorn}}}%
  , \bibinfo {year} {2010}{\natexlab{a}},\ \bibfield{title}{%
  \enquote{\bibinfo {title} {Coherent sets for nonautonomous dynamical
  systems},}\ }%
  \bibfield{journal}{%
  \bibinfo {journal} {Physica D}\ }%
  \textbf{\bibinfo {volume} {239}},\ \bibinfo {pages} {1527--1541}%
  \bibAnnoteFile{NoStop}{froyland_coherent_2010}%
\bibitem[{\citenamefont{Froyland}\ and\
  \citenamefont{Padberg}(2009)}]{Froyland:2009ti}%
  \BibitemOpen
  \bibfield{author}{%
  \bibinfo {author} {\bibnamefont{Froyland}, \bibfnamefont{G.}},\ and\ \bibinfo
  {author} {\bibfnamefont{K.}~\bibnamefont{Padberg}}}%
  , \bibinfo {year} {2009},\ \bibfield{title}{%
  \enquote{\bibinfo {title} {Almost-invariant sets and invariant manifolds ---
  connecting probabilistic and geometric descriptions of coherent structures in
  flows},}\ }%
  \bibfield{journal}{%
  \bibinfo {journal} {Physica D}\ }%
  \textbf{\bibinfo {volume} {238}},\ \bibinfo {pages} {1507--1523}%
  \bibAnnoteFile{NoStop}{Froyland:2009ti}%
\bibitem[{\citenamefont{Froyland}\ and\
  \citenamefont{{Padberg-Gehle}}(2012)}]{froyland_finite-time_2012}%
  \BibitemOpen
  \bibfield{author}{%
  \bibinfo {author} {\bibnamefont{Froyland}, \bibfnamefont{G.}},\ and\ \bibinfo
  {author} {\bibfnamefont{K.}~\bibnamefont{{Padberg-Gehle}}}}%
  , \bibinfo {year} {2012},\ \bibfield{title}{%
  \enquote{\bibinfo {title} {Finite-time entropy: A probabilistic approach for
  measuring nonlinear stretching},}\ }%
  \bibfield{journal}{%
  \bibinfo {journal} {Physica D}\ }%
  \textbf{\bibinfo {volume} {241}},\ \bibinfo {pages} {1612--1628}%
  \bibAnnoteFile{NoStop}{froyland_finite-time_2012}%
\bibitem[{\citenamefont{Froyland}\ and\
  \citenamefont{{Padberg-Gehle}}(2013)}]{Froyland:2013}%
  \BibitemOpen
  \bibfield{author}{%
  \bibinfo {author} {\bibnamefont{Froyland}, \bibfnamefont{G.}},\ and\ \bibinfo
  {author} {\bibfnamefont{K.}~\bibnamefont{{Padberg-Gehle}}}}%
  , \bibinfo {year} {2013},\ \enquote{\bibinfo {title} {Almost-invariant and
  finite-time coherent sets: directionality, duration, and diffusion},}\ in\
  \emph{\bibinfo {booktitle} {Ergodic Theory, Open Dynamics, and Coherent
  Structures}}\ (\bibinfo {publisher} {Springer})\ pp.\ \bibinfo {pages}
  {171--216}%
  \bibAnnoteFile{NoStop}{Froyland:2013}%
\bibitem[{\citenamefont{Froyland}\
  \emph{et~al.}(2010{\natexlab{b}})\citenamefont{Froyland},
  \citenamefont{Santitissadeekorn},\ and\
  \citenamefont{Monahan}}]{Froyland:2010jo}%
  \BibitemOpen
  \bibfield{author}{%
  \bibinfo {author} {\bibnamefont{Froyland}, \bibfnamefont{G.}}, \bibinfo
  {author} {\bibfnamefont{N.}~\bibnamefont{Santitissadeekorn}},\ and\ \bibinfo
  {author} {\bibfnamefont{A.}~\bibnamefont{Monahan}}}%
  , \bibinfo {year} {2010}{\natexlab{b}},\ \bibfield{title}{%
  \enquote{\bibinfo {title} {Transport in time-dependent dynamical systems:
  Finite-time coherent sets},}\ }%
  \bibfield{journal}{%
  \bibinfo {journal} {Chaos}\ }%
  \textbf{\bibinfo {volume} {20}},\ \bibinfo {pages} {043116}%
  \bibAnnoteFile{NoStop}{Froyland:2010jo}%
\bibitem[{\citenamefont{Gaffney}\ \emph{et~al.}(2011)\citenamefont{Gaffney},
  \citenamefont{Gadelha}, \citenamefont{Smith}, \citenamefont{Blake},\ and\
  \citenamefont{Kirkman-Brown}}]{Gaffney2011}%
  \BibitemOpen
  \bibfield{author}{%
  \bibinfo {author} {\bibnamefont{Gaffney}, \bibfnamefont{E.~A.}}, \bibinfo
  {author} {\bibfnamefont{H.}~\bibnamefont{Gadelha}}, \bibinfo {author}
  {\bibfnamefont{D.~J.}\ \bibnamefont{Smith}}, \bibinfo {author}
  {\bibfnamefont{J.~R.}\ \bibnamefont{Blake}},\ and\ \bibinfo {author}
  {\bibfnamefont{J.~C.}\ \bibnamefont{Kirkman-Brown}}}%
  , \bibinfo {year} {2011},\ \bibfield{title}{%
  \enquote{\bibinfo {title} {Mammalian sperm motility: observation and
  theory},}\ }%
  \bibfield{journal}{%
  \bibinfo {journal} {Annu. Rev. Fluid Mech.}\ }%
  \textbf{\bibinfo {volume} {43}},\ \bibinfo {pages} {501--528}%
  \bibAnnoteFile{NoStop}{Gaffney2011}%
\bibitem[{\citenamefont{Gardiner}(2005)}]{Gardiner2005}%
  \BibitemOpen
  \bibfield{author}{%
  \bibinfo {author} {\bibnamefont{Gardiner}, \bibfnamefont{M.~B.}}}%
  , \bibinfo {year} {2005},\ \bibfield{title}{%
  \enquote{\bibinfo {title} {The importance of being cilia},}\ }%
  \bibfield{journal}{%
  \bibinfo {journal} {Howard Hughes Medical Institute Bulletin}\ }%
  \textbf{\bibinfo {volume} {18}},\ \bibinfo {pages} {33--36}%
  \bibAnnoteFile{NoStop}{Gardiner2005}%
\bibitem[{\citenamefont{Gibbons}(1981)}]{Gibbons1981}%
  \BibitemOpen
  \bibfield{author}{%
  \bibinfo {author} {\bibnamefont{Gibbons}, \bibfnamefont{I.~R.}}}%
  , \bibinfo {year} {1981},\ \bibfield{title}{%
  \enquote{\bibinfo {title} {{Cilia and flagella of eukaryotes}},}\ }%
  \bibfield{journal}{%
  \bibinfo {journal} {J. Cell Biol.}\ }%
  \textbf{\bibinfo {volume} {91}},\ \bibinfo {pages} {107--124}%
  \bibAnnoteFile{NoStop}{Gibbons1981}%
\bibitem[{\citenamefont{Gibbs}(1902)}]{Gibbs1902}%
  \BibitemOpen
  \bibfield{author}{%
  \bibinfo {author} {\bibnamefont{Gibbs}, \bibfnamefont{J.~W.}}}%
  , \bibinfo {year} {1902},\ \emph{\bibinfo {title} {Elementary Principles in
  Statistical Mechanics}}\ (\bibinfo {publisher} {Scribner, New Haven})%
  \bibAnnoteFile{NoStop}{Gibbs1902}%
\bibitem[{\citenamefont{Gibout}\ \emph{et~al.}(2006)\citenamefont{Gibout},
  \citenamefont{Le~Guer},\ and\ \citenamefont{Schall}}]{gibout2006coupling}%
  \BibitemOpen
  \bibfield{author}{%
  \bibinfo {author} {\bibnamefont{Gibout}, \bibfnamefont{S.}}, \bibinfo
  {author} {\bibfnamefont{Y.}~\bibnamefont{Le~Guer}},\ and\ \bibinfo {author}
  {\bibfnamefont{E.}~\bibnamefont{Schall}}}%
  , \bibinfo {year} {2006},\ \bibfield{title}{%
  \enquote{\bibinfo {title} {Coupling of a mapping method and a genetic
  algorithm to optimize mixing efficiency in periodic chaotic flows},}\ }%
  \bibfield{journal}{%
  \bibinfo {journal} {Comm. Nonlin. Sci. Numer. Sim.}\ }%
  \textbf{\bibinfo {volume} {11}},\ \bibinfo {pages} {413--423}%
  \bibAnnoteFile{NoStop}{gibout2006coupling}%
\bibitem[{\citenamefont{Glasgow}\ and\
  \citenamefont{Aubry}(2003)}]{Glasgow2003}%
  \BibitemOpen
  \bibfield{author}{%
  \bibinfo {author} {\bibnamefont{Glasgow}, \bibfnamefont{I.}},\ and\ \bibinfo
  {author} {\bibfnamefont{N.}~\bibnamefont{Aubry}}}%
  , \bibinfo {year} {2003},\ \bibfield{title}{%
  \enquote{\bibinfo {title} {Enhancement of microfluidic mixing using time
  pulsing},}\ }%
  \bibfield{journal}{%
  \bibinfo {journal} {Lab on a chip}\ }%
  \textbf{\bibinfo {volume} {3}},\ \bibinfo {pages} {114--120}%
  \bibAnnoteFile{NoStop}{Glasgow2003}%
\bibitem[{\citenamefont{Glazier}\ and\
  \citenamefont{Libchaber}(1988)}]{Glazier_quasi_1988}%
  \BibitemOpen
  \bibfield{author}{%
  \bibinfo {author} {\bibnamefont{Glazier}, \bibfnamefont{J.~A.}},\ and\
  \bibinfo {author} {\bibfnamefont{A.}~\bibnamefont{Libchaber}}}%
  , \bibinfo {year} {1988},\ \bibfield{title}{%
  \enquote{\bibinfo {title} {Quasi-periodicity and dynamical systems: {A}n
  experimentalist's view},}\ }%
  \bibfield{journal}{%
  \bibinfo {journal} {{IEEE} Trans. Circuits Syst.}\ }%
  \textbf{\bibinfo {volume} {35}},\ \bibinfo {pages} {790--809}%
  \bibAnnoteFile{NoStop}{Glazier_quasi_1988}%
\bibitem[{\citenamefont{Goetz}(2000)}]{goetz2000dynamics}%
  \BibitemOpen
  \bibfield{author}{%
  \bibinfo {author} {\bibnamefont{Goetz}, \bibfnamefont{A.}}}%
  , \bibinfo {year} {2000},\ \bibfield{title}{%
  \enquote{\bibinfo {title} {Dynamics of piecewise isometries},}\ }%
  \bibfield{journal}{%
  \bibinfo {journal} {Illinois J. Math.}\ }%
  \textbf{\bibinfo {volume} {44}},\ \bibinfo {pages} {465--478}%
  \bibAnnoteFile{NoStop}{goetz2000dynamics}%
\bibitem[{\citenamefont{Goldstein}\
  \emph{et~al.}(2009)\citenamefont{Goldstein}, \citenamefont{Polin},\ and\
  \citenamefont{Tuval}}]{Goldstein2009}%
  \BibitemOpen
  \bibfield{author}{%
  \bibinfo {author} {\bibnamefont{Goldstein}, \bibfnamefont{R.~E.}}, \bibinfo
  {author} {\bibfnamefont{M.}~\bibnamefont{Polin}},\ and\ \bibinfo {author}
  {\bibfnamefont{I.}~\bibnamefont{Tuval}}}%
  , \bibinfo {year} {2009},\ \bibfield{title}{%
  \enquote{\bibinfo {title} {Noise and synchronization in pairs of beating
  eukaryotic flagella},}\ }%
  \bibfield{journal}{%
  \bibinfo {journal} {Phys. Rev. Lett.}\ }%
  \textbf{\bibinfo {volume} {103}},\ \bibinfo {pages} {168103}%
  \bibAnnoteFile{NoStop}{Goldstein2009}%
\bibitem[{\citenamefont{Goldstein}\
  \emph{et~al.}(2011)\citenamefont{Goldstein}, \citenamefont{Polin},\ and\
  \citenamefont{Tuval}}]{Goldstein2011}%
  \BibitemOpen
  \bibfield{author}{%
  \bibinfo {author} {\bibnamefont{Goldstein}, \bibfnamefont{R.~E.}}, \bibinfo
  {author} {\bibfnamefont{M.}~\bibnamefont{Polin}},\ and\ \bibinfo {author}
  {\bibfnamefont{I.}~\bibnamefont{Tuval}}}%
  , \bibinfo {year} {2011},\ \bibfield{title}{%
  \enquote{\bibinfo {title} {Emergence of synchronized beating during the
  regrowth of eukaryotic flagella},}\ }%
  \bibfield{journal}{%
  \bibinfo {journal} {Phys. Rev. Lett.}\ }%
  \textbf{\bibinfo {volume} {107}},\ \bibinfo {pages} {148103}%
  \bibAnnoteFile{NoStop}{Goldstein2011}%
\bibitem[{\citenamefont{Gollub}\ \emph{et~al.}(2006)\citenamefont{Gollub},
  \citenamefont{Fernando}, \citenamefont{Gharib}, \citenamefont{Kim},
  \citenamefont{Pope}, \citenamefont{Smits},\ and\
  \citenamefont{Stone}}]{NAS_fluidsreport_2006}%
  \BibitemOpen
  \bibfield{author}{%
  \bibinfo {author} {\bibnamefont{Gollub}, \bibfnamefont{J.}}, \bibinfo
  {author} {\bibfnamefont{H.}~\bibnamefont{Fernando}}, \bibinfo {author}
  {\bibfnamefont{M.}~\bibnamefont{Gharib}}, \bibinfo {author}
  {\bibfnamefont{J.}~\bibnamefont{Kim}}, \bibinfo {author}
  {\bibfnamefont{S.}~\bibnamefont{Pope}}, \bibinfo {author}
  {\bibfnamefont{A.}~\bibnamefont{Smits}},\ and\ \bibinfo {author}
  {\bibfnamefont{H.}~\bibnamefont{Stone}}}%
  , \bibinfo {year} {2006},\ \emph{\bibinfo {title} {Research in Fluid
  Dynamics: {M}eeting National Needs}},\ \bibinfo {type} {Tech. Rep.}\
  (\bibinfo {institution} {U.S.\/ National Committee on Theoretical and Applied
  Mechanics})%
  \bibAnnoteFile{NoStop}{NAS_fluidsreport_2006}%
\bibitem[{\citenamefont{G\'{o}mez}\ and\ \citenamefont{Meiss}(2002)}]{gomez02}%
  \BibitemOpen
  \bibfield{author}{%
  \bibinfo {author} {\bibnamefont{G\'{o}mez}, \bibfnamefont{A.}},\ and\
  \bibinfo {author} {\bibfnamefont{J.~D.}\ \bibnamefont{Meiss}}}%
  , \bibinfo {year} {2002},\ \bibfield{title}{%
  \enquote{\bibinfo {title} {Volume-preserving maps with an invariant},}\ }%
  \bibfield{journal}{%
  \bibinfo {journal} {Chaos}\ }%
  \textbf{\bibinfo {volume} {12}},\ \bibinfo {pages} {289--299}%
  \bibAnnoteFile{NoStop}{gomez02}%
\bibitem[{\citenamefont{Gouillart}\
  \emph{et~al.}(2008)\citenamefont{Gouillart}, \citenamefont{Dauchot},
  \citenamefont{Dubrulle}, \citenamefont{Roux},\ and\
  \citenamefont{Thiffeault}}]{Gouillart2008}%
  \BibitemOpen
  \bibfield{author}{%
  \bibinfo {author} {\bibnamefont{Gouillart}, \bibfnamefont{E.}}, \bibinfo
  {author} {\bibfnamefont{O.}~\bibnamefont{Dauchot}}, \bibinfo {author}
  {\bibfnamefont{B.}~\bibnamefont{Dubrulle}}, \bibinfo {author}
  {\bibfnamefont{S.}~\bibnamefont{Roux}},\ and\ \bibinfo {author}
  {\bibfnamefont{J.-L.}\ \bibnamefont{Thiffeault}}}%
  , \bibinfo {year} {2008},\ \bibfield{title}{%
  \enquote{\bibinfo {title} {Slow decay of concentration variance due to
  no-slip walls in chaotic mixing},}\ }%
  \bibfield{journal}{%
  \bibinfo {journal} {Phys. Rev. E}\ }%
  \textbf{\bibinfo {volume} {78}},\ \bibinfo {pages} {026211}%
  \bibAnnoteFile{NoStop}{Gouillart2008}%
\bibitem[{\citenamefont{Gouillart}\
  \emph{et~al.}(2010{\natexlab{a}})\citenamefont{Gouillart},
  \citenamefont{Dauchot},\ and\ \citenamefont{Thiffeault}}]{Gouillart2010b}%
  \BibitemOpen
  \bibfield{author}{%
  \bibinfo {author} {\bibnamefont{Gouillart}, \bibfnamefont{E.}}, \bibinfo
  {author} {\bibfnamefont{O.}~\bibnamefont{Dauchot}},\ and\ \bibinfo {author}
  {\bibfnamefont{J.-L.}\ \bibnamefont{Thiffeault}}}%
  , \bibinfo {year} {2010}{\natexlab{a}},\ \bibfield{title}{%
  \enquote{\bibinfo {title} {Measures of mixing quality in open flows with
  chaotic advection},}\ }%
  \bibfield{journal}{%
  \bibinfo {journal} {Phys. Fluids}\ }%
  \textbf{\bibinfo {volume} {23}},\ \bibinfo {pages} {013604}%
  \bibAnnoteFile{NoStop}{Gouillart2010b}%
\bibitem[{\citenamefont{Gouillart}\
  \emph{et~al.}(2009)\citenamefont{Gouillart}, \citenamefont{Dauchot},
  \citenamefont{Thiffeault},\ and\ \citenamefont{Roux}}]{Gouillart2009}%
  \BibitemOpen
  \bibfield{author}{%
  \bibinfo {author} {\bibnamefont{Gouillart}, \bibfnamefont{E.}}, \bibinfo
  {author} {\bibfnamefont{O.}~\bibnamefont{Dauchot}}, \bibinfo {author}
  {\bibfnamefont{J.-L.}\ \bibnamefont{Thiffeault}},\ and\ \bibinfo {author}
  {\bibfnamefont{S.}~\bibnamefont{Roux}}}%
  , \bibinfo {year} {2009},\ \bibfield{title}{%
  \enquote{\bibinfo {title} {Open-flow mixing: Experimental evidence for
  strange eigenmodes},}\ }%
  \bibfield{journal}{%
  \bibinfo {journal} {Phys. Fluids}\ }%
  \textbf{\bibinfo {volume} {21}},\ \bibinfo {pages} {022603}%
  \bibAnnoteFile{NoStop}{Gouillart2009}%
\bibitem[{\citenamefont{Gouillart}\
  \emph{et~al.}(2007)\citenamefont{Gouillart}, \citenamefont{Kuncio},
  \citenamefont{Dauchot}, \citenamefont{Dubrulle}, \citenamefont{Roux},\ and\
  \citenamefont{Thiffeault}}]{Gouillart2007}%
  \BibitemOpen
  \bibfield{author}{%
  \bibinfo {author} {\bibnamefont{Gouillart}, \bibfnamefont{E.}}, \bibinfo
  {author} {\bibfnamefont{N.}~\bibnamefont{Kuncio}}, \bibinfo {author}
  {\bibfnamefont{O.}~\bibnamefont{Dauchot}}, \bibinfo {author}
  {\bibfnamefont{B.}~\bibnamefont{Dubrulle}}, \bibinfo {author}
  {\bibfnamefont{S.}~\bibnamefont{Roux}},\ and\ \bibinfo {author}
  {\bibfnamefont{J.-L.}\ \bibnamefont{Thiffeault}}}%
  , \bibinfo {year} {2007},\ \bibfield{title}{%
  \enquote{\bibinfo {title} {Walls inhibit chaotic mixing},}\ }%
  \bibfield{journal}{%
  \bibinfo {journal} {Phys. Rev. Lett.}\ }%
  \textbf{\bibinfo {volume} {99}},\ \bibinfo {pages} {114501}%
  \bibAnnoteFile{NoStop}{Gouillart2007}%
\bibitem[{\citenamefont{Gouillart}\
  \emph{et~al.}(2010{\natexlab{b}})\citenamefont{Gouillart},
  \citenamefont{Thiffeault},\ and\ \citenamefont{Dauchot}}]{Gouillart2010}%
  \BibitemOpen
  \bibfield{author}{%
  \bibinfo {author} {\bibnamefont{Gouillart}, \bibfnamefont{E.}}, \bibinfo
  {author} {\bibfnamefont{J.-L.}\ \bibnamefont{Thiffeault}},\ and\ \bibinfo
  {author} {\bibfnamefont{O.}~\bibnamefont{Dauchot}}}%
  , \bibinfo {year} {2010}{\natexlab{b}},\ \bibfield{title}{%
  \enquote{\bibinfo {title} {Rotation shields chaotic mixing regions from
  no-slip walls},}\ }%
  \bibfield{journal}{%
  \bibinfo {journal} {Phys. Rev. Lett.}\ }%
  \textbf{\bibinfo {volume} {104}},\ \bibinfo {pages} {204502}%
  \bibAnnoteFile{NoStop}{Gouillart2010}%
\bibitem[{\citenamefont{Gouillart}\
  \emph{et~al.}(2006)\citenamefont{Gouillart}, \citenamefont{Thiffeault},\ and\
  \citenamefont{Finn}}]{Gouillart2006}%
  \BibitemOpen
  \bibfield{author}{%
  \bibinfo {author} {\bibnamefont{Gouillart}, \bibfnamefont{E.}}, \bibinfo
  {author} {\bibfnamefont{J.-L.}\ \bibnamefont{Thiffeault}},\ and\ \bibinfo
  {author} {\bibfnamefont{M.~D.}\ \bibnamefont{Finn}}}%
  , \bibinfo {year} {2006},\ \bibfield{title}{%
  \enquote{\bibinfo {title} {Topological mixing with ghosts rods},}\ }%
  \bibfield{journal}{%
  \bibinfo {journal} {Phys. Rev. E}\ }%
  \textbf{\bibinfo {volume} {73}},\ \bibinfo {pages} {036311}%
  \bibAnnoteFile{NoStop}{Gouillart2006}%
\bibitem[{\citenamefont{Grebogi}\ \emph{et~al.}(1983)\citenamefont{Grebogi},
  \citenamefont{McDonald}, \citenamefont{Ott},\ and\
  \citenamefont{Yorke}}]{Grebogi1983}%
  \BibitemOpen
  \bibfield{author}{%
  \bibinfo {author} {\bibnamefont{Grebogi}, \bibfnamefont{C.}}, \bibinfo
  {author} {\bibfnamefont{S.~W.}\ \bibnamefont{McDonald}}, \bibinfo {author}
  {\bibfnamefont{E.}~\bibnamefont{Ott}},\ and\ \bibinfo {author}
  {\bibfnamefont{J.~A.}\ \bibnamefont{Yorke}}}%
  , \bibinfo {year} {1983},\ \bibfield{title}{%
  \enquote{\bibinfo {title} {Final-state sensitivity --- an obstruction to
  predictability},}\ }%
  \bibfield{journal}{%
  \bibinfo {journal} {Phys. Lett. A}\ }%
  \textbf{\bibinfo {volume} {99}},\ \bibinfo {pages} {415--418}%
  \bibAnnoteFile{NoStop}{Grebogi1983}%
\bibitem[{\citenamefont{Greene}(1968)}]{Greene:1968ua}%
  \BibitemOpen
  \bibfield{author}{%
  \bibinfo {author} {\bibnamefont{Greene}, \bibfnamefont{J.~M.}}}%
  , \bibinfo {year} {1968},\ \bibfield{title}{%
  \enquote{\bibinfo {title} {Two-dimensional measure-preserving mappings},}\ }%
  \bibfield{journal}{%
  \bibinfo {journal} {J. Math. Phys.}\ }%
  \textbf{\bibinfo {volume} {9}},\ \bibinfo {pages} {760--768}%
  \bibAnnoteFile{NoStop}{Greene:1968ua}%
\bibitem[{\citenamefont{Greene}(1979)}]{Greene:1979jr}%
  \BibitemOpen
  \bibfield{author}{%
  \bibinfo {author} {\bibnamefont{Greene}, \bibfnamefont{J.~M.}}}%
  , \bibinfo {year} {1979},\ \bibfield{title}{%
  \enquote{\bibinfo {title} {Method for determining a stochastic transition},}\
  }%
  \bibfield{journal}{%
  \bibinfo {journal} {J. Math. Phys.}\ }%
  \textbf{\bibinfo {volume} {20}},\ \bibinfo {pages} {1183--1201}%
  \bibAnnoteFile{NoStop}{Greene:1979jr}%
\bibitem[{\citenamefont{Grigoriev}\
  \emph{et~al.}(2006)\citenamefont{Grigoriev}, \citenamefont{Schatz},\ and\
  \citenamefont{Sharma}}]{Grigoriev2006}%
  \BibitemOpen
  \bibfield{author}{%
  \bibinfo {author} {\bibnamefont{Grigoriev}, \bibfnamefont{R.~O.}}, \bibinfo
  {author} {\bibfnamefont{M.~F.}\ \bibnamefont{Schatz}},\ and\ \bibinfo
  {author} {\bibfnamefont{V.}~\bibnamefont{Sharma}}}%
  , \bibinfo {year} {2006},\ \bibfield{title}{%
  \enquote{\bibinfo {title} {Chaotic mixing in microdroplets},}\ }%
  \bibfield{journal}{%
  \bibinfo {journal} {Lab on a chip}\ }%
  \textbf{\bibinfo {volume} {6}},\ \bibinfo {pages} {1369--1372}%
  \bibAnnoteFile{NoStop}{Grigoriev2006}%
\bibitem[{\citenamefont{Grover}\ \emph{et~al.}(2012)\citenamefont{Grover},
  \citenamefont{Ross}, \citenamefont{Stremler},\ and\
  \citenamefont{Kumar}}]{Grover:2012iy}%
  \BibitemOpen
  \bibfield{author}{%
  \bibinfo {author} {\bibnamefont{Grover}, \bibfnamefont{P.}}, \bibinfo
  {author} {\bibfnamefont{S.~D.}\ \bibnamefont{Ross}}, \bibinfo {author}
  {\bibfnamefont{M.~A.}\ \bibnamefont{Stremler}},\ and\ \bibinfo {author}
  {\bibfnamefont{P.}~\bibnamefont{Kumar}}}%
  , \bibinfo {year} {2012},\ \bibfield{title}{%
  \enquote{\bibinfo {title} {Topological chaos, braiding and bifurcation of
  almost-cyclic sets},}\ }%
  \bibfield{journal}{%
  \bibinfo {journal} {Chaos}\ }%
  \textbf{\bibinfo {volume} {22}},\ \bibinfo {pages} {043135}%
  \bibAnnoteFile{NoStop}{Grover:2012iy}%
\bibitem[{\citenamefont{Gubanov}\ and\
  \citenamefont{Cortelezzi}(2010)}]{gubanov2012}%
  \BibitemOpen
  \bibfield{author}{%
  \bibinfo {author} {\bibnamefont{Gubanov}, \bibfnamefont{O.}},\ and\ \bibinfo
  {author} {\bibfnamefont{L.}~\bibnamefont{Cortelezzi}}}%
  , \bibinfo {year} {2010},\ \bibfield{title}{%
  \enquote{\bibinfo {title} {Toward the design of an optimal mixer},}\ }%
  \bibfield{journal}{%
  \bibinfo {journal} {J. Fluid Mech.}\ }%
  \textbf{\bibinfo {volume} {651}},\ \bibinfo {pages} {27--53}%
  \bibAnnoteFile{NoStop}{gubanov2012}%
\bibitem[{\citenamefont{Gubanov}\ and\
  \citenamefont{Cortelezzi}(2012)}]{gubanov2010}%
  \BibitemOpen
  \bibfield{author}{%
  \bibinfo {author} {\bibnamefont{Gubanov}, \bibfnamefont{O.}},\ and\ \bibinfo
  {author} {\bibfnamefont{L.}~\bibnamefont{Cortelezzi}}}%
  , \bibinfo {year} {2012},\ \bibfield{title}{%
  \enquote{\bibinfo {title} {On the cost efficiency of mixing optimization},}\
  }%
  \bibfield{journal}{%
  \bibinfo {journal} {J. Fluid Mech.}\ }%
  \textbf{\bibinfo {volume} {692}},\ \bibinfo {pages} {112--136}%
  \bibAnnoteFile{NoStop}{gubanov2010}%
\bibitem[{\citenamefont{Guckenheimer}\ and\
  \citenamefont{Holmes}(1983)}]{GuckenheimerHolmes1983}%
  \BibitemOpen
  \bibfield{author}{%
  \bibinfo {author} {\bibnamefont{Guckenheimer}, \bibfnamefont{J.}},\ and\
  \bibinfo {author} {\bibfnamefont{P.}~\bibnamefont{Holmes}}}%
  , \bibinfo {year} {1983},\ \emph{\bibinfo {title} {Nonlinear Oscillations,
  Dynamical Systems and Bifurcations of Vector Fields}}\ (\bibinfo {publisher}
  {Springer},\ \bibinfo {address} {New York})%
  \bibAnnoteFile{NoStop}{GuckenheimerHolmes1983}%
\bibitem[{\citenamefont{Gueron}\ \emph{et~al.}(1997)\citenamefont{Gueron},
  \citenamefont{Levit-Gurevich}, \citenamefont{Liron},\ and\
  \citenamefont{Blum}}]{Gueron1997}%
  \BibitemOpen
  \bibfield{author}{%
  \bibinfo {author} {\bibnamefont{Gueron}, \bibfnamefont{S.}}, \bibinfo
  {author} {\bibfnamefont{K.}~\bibnamefont{Levit-Gurevich}}, \bibinfo {author}
  {\bibfnamefont{N.}~\bibnamefont{Liron}},\ and\ \bibinfo {author}
  {\bibfnamefont{J.~J.}\ \bibnamefont{Blum}}}%
  , \bibinfo {year} {1997},\ \bibfield{title}{%
  \enquote{\bibinfo {title} {{Cilia internal mechanism and metachronal
  coordination as the result of hydrodynamical coupling}},}\ }%
  \bibfield{journal}{%
  \bibinfo {journal} {Proc. Natl Acad. Sci. USA}\ }%
  \textbf{\bibinfo {volume} {94}},\ \bibinfo {pages} {6001--6006}%
  \bibAnnoteFile{NoStop}{Gueron1997}%
\bibitem[{\citenamefont{Guirao}\ \emph{et~al.}(2010)\citenamefont{Guirao},
  \citenamefont{Meunier}, \citenamefont{Mortaud}, \citenamefont{Aguilar},
  \citenamefont{Corsi}, \citenamefont{Strehl}, \citenamefont{Hirota},
  \citenamefont{Desoeuvre}, \citenamefont{Boutin}, \citenamefont{Han},
  \citenamefont{Mirzadeh}, \citenamefont{Cremer}, \citenamefont{Montcouquiol},
  \citenamefont{Sawamoto},\ and\ \citenamefont{Spassky}}]{Guirao2010}%
  \BibitemOpen
  \bibfield{author}{%
  \bibinfo {author} {\bibnamefont{Guirao}, \bibfnamefont{B.}}, \bibinfo
  {author} {\bibfnamefont{A.}~\bibnamefont{Meunier}}, \bibinfo {author}
  {\bibfnamefont{S.}~\bibnamefont{Mortaud}}, \bibinfo {author}
  {\bibfnamefont{A.}~\bibnamefont{Aguilar}}, \bibinfo {author}
  {\bibfnamefont{J.-M.}\ \bibnamefont{Corsi}}, \bibinfo {author}
  {\bibfnamefont{L.}~\bibnamefont{Strehl}}, \bibinfo {author}
  {\bibfnamefont{Y.}~\bibnamefont{Hirota}}, \bibinfo {author}
  {\bibfnamefont{A.}~\bibnamefont{Desoeuvre}}, \bibinfo {author}
  {\bibfnamefont{C.}~\bibnamefont{Boutin}}, \bibinfo {author}
  {\bibfnamefont{Y.-G.}\ \bibnamefont{Han}}, \bibinfo {author}
  {\bibfnamefont{Z.}~\bibnamefont{Mirzadeh}}, \bibinfo {author}
  {\bibfnamefont{H.}~\bibnamefont{Cremer}}, \bibinfo {author}
  {\bibfnamefont{M.}~\bibnamefont{Montcouquiol}}, \bibinfo {author}
  {\bibfnamefont{K.}~\bibnamefont{Sawamoto}},\ and\ \bibinfo {author}
  {\bibfnamefont{N.}~\bibnamefont{Spassky}}}%
  , \bibinfo {year} {2010},\ \bibfield{title}{%
  \enquote{\bibinfo {title} {{Coupling between hydrodynamic forces and planar
  cell polarity orients mammalian motile cilia}},}\ }%
  \bibfield{journal}{%
  \bibinfo {journal} {Nature Cell Biol.}\ }%
  \textbf{\bibinfo {volume} {12}},\ \bibinfo {pages} {341--350}%
  \bibAnnoteFile{NoStop}{Guirao2010}%
\bibitem[{\citenamefont{Haller}(2000)}]{haller_finding_2000}%
  \BibitemOpen
  \bibfield{author}{%
  \bibinfo {author} {\bibnamefont{Haller}, \bibfnamefont{G.}}}%
  , \bibinfo {year} {2000},\ \bibfield{title}{%
  \enquote{\bibinfo {title} {Finding finite-time invariant manifolds in
  two-dimensional velocity fields},}\ }%
  \bibfield{journal}{%
  \bibinfo {journal} {Chaos}\ }%
  \textbf{\bibinfo {volume} {10}},\ \bibinfo {pages} {99--108}%
  \bibAnnoteFile{NoStop}{haller_finding_2000}%
\bibitem[{\citenamefont{Haller}(2001{\natexlab{a}})}]{Haller2001}%
  \BibitemOpen
  \bibfield{author}{%
  \bibinfo {author} {\bibnamefont{Haller}, \bibfnamefont{G.}}}%
  , \bibinfo {year} {2001}{\natexlab{a}},\ \bibfield{title}{%
  \enquote{\bibinfo {title} {Distinguished material surfaces and coherent
  structures in three-dimensional fluid flows},}\ }%
  \bibfield{journal}{%
  \bibinfo {journal} {Physica D}\ }%
  \textbf{\bibinfo {volume} {149}},\ \bibinfo {pages} {248--277}%
  \bibAnnoteFile{NoStop}{Haller2001}%
\bibitem[{\citenamefont{Haller}(2001{\natexlab{b}})}]{Haller:2001ed}%
  \BibitemOpen
  \bibfield{author}{%
  \bibinfo {author} {\bibnamefont{Haller}, \bibfnamefont{G.}}}%
  , \bibinfo {year} {2001}{\natexlab{b}},\ \bibfield{title}{%
  \enquote{\bibinfo {title} {Lagrangian structures and the rate of strain in a
  partition of two-dimensional turbulence},}\ }%
  \bibfield{journal}{%
  \bibinfo {journal} {Phys. Fluids}\ }%
  \textbf{\bibinfo {volume} {13}},\ \bibinfo {pages} {3365--3385}%
  \bibAnnoteFile{NoStop}{Haller:2001ed}%
\bibitem[{\citenamefont{Haller}(2002)}]{Haller:2002bf}%
  \BibitemOpen
  \bibfield{author}{%
  \bibinfo {author} {\bibnamefont{Haller}, \bibfnamefont{G.}}}%
  , \bibinfo {year} {2002},\ \bibfield{title}{%
  \enquote{\bibinfo {title} {Lagrangian coherent structures from approximate
  velocity data},}\ }%
  \bibfield{journal}{%
  \bibinfo {journal} {Phys. Fluids}\ }%
  \textbf{\bibinfo {volume} {14}},\ \bibinfo {pages} {1851--1861}%
  \bibAnnoteFile{NoStop}{Haller:2002bf}%
\bibitem[{\citenamefont{Haller}(2011)}]{Haller:2011kr}%
  \BibitemOpen
  \bibfield{author}{%
  \bibinfo {author} {\bibnamefont{Haller}, \bibfnamefont{G.}}}%
  , \bibinfo {year} {2011},\ \bibfield{title}{%
  \enquote{\bibinfo {title} {A variational theory of hyperbolic lagrangian
  coherent structures},}\ }%
  \bibfield{journal}{%
  \bibinfo {journal} {Physica D}\ }%
  \textbf{\bibinfo {volume} {240}},\ \bibinfo {pages} {574--598}%
  \bibAnnoteFile{NoStop}{Haller:2011kr}%
\bibitem[{\citenamefont{Haller}(2015)}]{Haller2015}%
  \BibitemOpen
  \bibfield{author}{%
  \bibinfo {author} {\bibnamefont{Haller}, \bibfnamefont{{G.}}}}%
  , \bibinfo {year} {2015},\ \bibfield{title}{%
  \enquote{\bibinfo {title} {Lagrangian {coherent} {structures}},}\ }%
  \bibfield{journal}{%
  \bibinfo {journal} {Annu. {Rev.} {Fluid} {Mech.}}\ }%
  \textbf{\bibinfo {volume} {47}},\ \bibinfo {pages} {137--162}%
  \bibAnnoteFile{NoStop}{Haller2015}%
\bibitem[{\citenamefont{{Haller}}\ and\
  \citenamefont{{Beron-Vera}}(2012)}]{haller2012}%
  \BibitemOpen
  \bibfield{author}{%
  \bibinfo {author} {\bibnamefont{{Haller}}, \bibfnamefont{G.}},\ and\ \bibinfo
  {author} {\bibfnamefont{F.~J.}\ \bibnamefont{{Beron-Vera}}}}%
  , \bibinfo {year} {2012},\ \bibfield{title}{%
  \enquote{\bibinfo {title} {Geodesic theory of transport barriers in
  two-dimensional flows},}\ }%
  \bibfield{journal}{%
  \bibinfo {journal} {Physica D}\ }%
  \textbf{\bibinfo {volume} {241}},\ \bibinfo {pages} {1680--1702}%
  \bibAnnoteFile{NoStop}{haller2012}%
\bibitem[{\citenamefont{Haller}\ and\
  \citenamefont{Iacono}(2003)}]{Haller:2003cf}%
  \BibitemOpen
  \bibfield{author}{%
  \bibinfo {author} {\bibnamefont{Haller}, \bibfnamefont{G.}},\ and\ \bibinfo
  {author} {\bibfnamefont{R.}~\bibnamefont{Iacono}}}%
  , \bibinfo {year} {2003},\ \bibfield{title}{%
  \enquote{\bibinfo {title} {Stretching, alignment, and shear in slowly varying
  velocity fields},}\ }%
  \bibfield{journal}{%
  \bibinfo {journal} {Phys. Rev. E}\ }%
  \textbf{\bibinfo {volume} {68}},\ \bibinfo {pages} {056304}%
  \bibAnnoteFile{NoStop}{Haller:2003cf}%
\bibitem[{\citenamefont{Haller}\ and\
  \citenamefont{Mezi\'{c}}(1998)}]{Haller1998}%
  \BibitemOpen
  \bibfield{author}{%
  \bibinfo {author} {\bibnamefont{Haller}, \bibfnamefont{G.}},\ and\ \bibinfo
  {author} {\bibfnamefont{I.}~\bibnamefont{Mezi\'{c}}}}%
  , \bibinfo {year} {1998},\ \bibfield{title}{%
  \enquote{\bibinfo {title} {Reduction of three-dimensional, volume-preserving
  flows by symmetry},}\ }%
  \bibfield{journal}{%
  \bibinfo {journal} {Nonlinearity}\ }%
  \textbf{\bibinfo {volume} {11}},\ \bibinfo {pages} {319--339}%
  \bibAnnoteFile{NoStop}{Haller1998}%
\bibitem[{\citenamefont{Haller}\ and\
  \citenamefont{Poje}(1998)}]{Haller:1998cx}%
  \BibitemOpen
  \bibfield{author}{%
  \bibinfo {author} {\bibnamefont{Haller}, \bibfnamefont{G.}},\ and\ \bibinfo
  {author} {\bibfnamefont{A.~C.}\ \bibnamefont{Poje}}}%
  , \bibinfo {year} {1998},\ \bibfield{title}{%
  \enquote{\bibinfo {title} {Finite time transport in aperiodic flows},}\ }%
  \bibfield{journal}{%
  \bibinfo {journal} {Physica D}\ }%
  \textbf{\bibinfo {volume} {119}},\ \bibinfo {pages} {352--380}%
  \bibAnnoteFile{NoStop}{Haller:1998cx}%
\bibitem[{\citenamefont{Haller}\ and\
  \citenamefont{Sapsis}(2011)}]{haller_lagrangian_2011}%
  \BibitemOpen
  \bibfield{author}{%
  \bibinfo {author} {\bibnamefont{Haller}, \bibfnamefont{G.}},\ and\ \bibinfo
  {author} {\bibfnamefont{T.}~\bibnamefont{Sapsis}}}%
  , \bibinfo {year} {2011},\ \bibfield{title}{%
  \enquote{\bibinfo {title} {Lagrangian coherent structures and the smallest
  finite-time lyapunov exponent},}\ }%
  \bibfield{journal}{%
  \bibinfo {journal} {Chaos}\ }%
  \textbf{\bibinfo {volume} {21}},\ \bibinfo {pages} {023115}%
  \bibAnnoteFile{NoStop}{haller_lagrangian_2011}%
\bibitem[{\citenamefont{Haller}\ and\
  \citenamefont{Yuan}(2000)}]{Haller:2000us}%
  \BibitemOpen
  \bibfield{author}{%
  \bibinfo {author} {\bibnamefont{Haller}, \bibfnamefont{G.}},\ and\ \bibinfo
  {author} {\bibfnamefont{G.}~\bibnamefont{Yuan}}}%
  , \bibinfo {year} {2000},\ \bibfield{title}{%
  \enquote{\bibinfo {title} {Lagrangian coherent structures and mixing in
  two-dimensional turbulence},}\ }%
  \bibfield{journal}{%
  \bibinfo {journal} {Physica D}\ }%
  \textbf{\bibinfo {volume} {147}},\ \bibinfo {pages} {352--370}%
  \bibAnnoteFile{NoStop}{Haller:2000us}%
\bibitem[{\citenamefont{Haynes}\ and\
  \citenamefont{Vanneste}(2005)}]{Haynes2005}%
  \BibitemOpen
  \bibfield{author}{%
  \bibinfo {author} {\bibnamefont{Haynes}, \bibfnamefont{P.~H.}},\ and\
  \bibinfo {author} {\bibfnamefont{J.}~\bibnamefont{Vanneste}}}%
  , \bibinfo {year} {2005},\ \bibfield{title}{%
  \enquote{\bibinfo {title} {What controls the decay of passive scalars in
  smooth flows?}.}\ }%
  \bibfield{journal}{%
  \bibinfo {journal} {Phys. Fluids}\ }%
  \textbf{\bibinfo {volume} {17}},\ \bibinfo {pages} {097103}%
  \bibAnnoteFile{NoStop}{Haynes2005}%
\bibitem[{\citenamefont{Heller}(1960)}]{heller}%
  \BibitemOpen
  \bibfield{author}{%
  \bibinfo {author} {\bibnamefont{Heller}, \bibfnamefont{J.~P.}}}%
  , \bibinfo {year} {1960},\ \bibfield{title}{%
  \enquote{\bibinfo {title} {{An unmixing demonstration}},}\ }%
  \bibfield{journal}{%
  \bibinfo {journal} {Am. J. Phys.}\ }%
  \textbf{\bibinfo {volume} {28}},\ \bibinfo {pages} {348--351}%
  \bibAnnoteFile{NoStop}{heller}%
\bibitem[{\citenamefont{H{\'{e}}non}(1966)}]{henon1966}%
  \BibitemOpen
  \bibfield{author}{%
  \bibinfo {author} {\bibnamefont{H{\'{e}}non}, \bibfnamefont{M.}}}%
  , \bibinfo {year} {1966},\ \bibfield{title}{%
  \enquote{\bibinfo {title} {Sur la topologie des lignes de courant dans un cas
  particulier},}\ }%
  \bibfield{journal}{%
  \bibinfo {journal} {C. R. Acad. Sci. Paris A}\ }%
  \textbf{\bibinfo {volume} {262}},\ \bibinfo {pages} {312--314}%
  \bibAnnoteFile{NoStop}{henon1966}%
\bibitem[{\citenamefont{Hern\'andez-Garc\'ia}\ and\
  \citenamefont{L\'opez}(2004)}]{Hernandez-Garcia-Lopez-04}%
  \BibitemOpen
  \bibfield{author}{%
  \bibinfo {author} {\bibnamefont{Hern\'andez-Garc\'ia}, \bibfnamefont{E.}},\
  and\ \bibinfo {author} {\bibfnamefont{C.}~\bibnamefont{L\'opez}}}%
  , \bibinfo {year} {2004},\ \bibfield{title}{%
  \enquote{\bibinfo {title} {Sustained plankton blooms under open chaotic
  flows},}\ }%
  \bibfield{journal}{%
  \bibinfo {journal} {Ecol. Complex.}\ }%
  \textbf{\bibinfo {volume} {1}},\ \bibinfo {pages} {253--259}%
  \bibAnnoteFile{NoStop}{Hernandez-Garcia-Lopez-04}%
\bibitem[{\citenamefont{Hern\'andez-Garc\'ia}\
  \emph{et~al.}(2002)\citenamefont{Hern\'andez-Garc\'ia},
  \citenamefont{L\'opez},\ and\
  \citenamefont{Neufeld}}]{Hernandez-Garcia-et-al-02}%
  \BibitemOpen
  \bibfield{author}{%
  \bibinfo {author} {\bibnamefont{Hern\'andez-Garc\'ia}, \bibfnamefont{E.}},
  \bibinfo {author} {\bibfnamefont{C.}~\bibnamefont{L\'opez}},\ and\ \bibinfo
  {author} {\bibfnamefont{Z.}~\bibnamefont{Neufeld}}}%
  , \bibinfo {year} {2002},\ \bibfield{title}{%
  \enquote{\bibinfo {title} {Small-scale structure of nonlinearly interacting
  species advected by chaotic flows},}\ }%
  \bibfield{journal}{%
  \bibinfo {journal} {Chaos}\ }%
  \textbf{\bibinfo {volume} {12}},\ \bibinfo {pages} {470--480}%
  \bibAnnoteFile{NoStop}{Hernandez-Garcia-et-al-02}%
\bibitem[{\citenamefont{Hern\'andez-Garc\'ia}\
  \emph{et~al.}(2003)\citenamefont{Hern\'andez-Garc\'ia},
  \citenamefont{L\'opez},\ and\
  \citenamefont{Neufeld}}]{Hernandez-Garcia-et-al-03}%
  \BibitemOpen
  \bibfield{author}{%
  \bibinfo {author} {\bibnamefont{Hern\'andez-Garc\'ia}, \bibfnamefont{E.}},
  \bibinfo {author} {\bibfnamefont{C.}~\bibnamefont{L\'opez}},\ and\ \bibinfo
  {author} {\bibfnamefont{Z.}~\bibnamefont{Neufeld}}}%
  , \bibinfo {year} {2003},\ \enquote{\bibinfo {title} {Spatial patterns in
  chemically and biologically reacting flows},}\ in\ \emph{\bibinfo {booktitle}
  {Chaos in geophysical flows}},\ \bibinfo {editor} {edited by\ \bibinfo
  {editor} {\bibfnamefont{G.}~\bibnamefont{Bofetta}}, \bibinfo {editor}
  {\bibfnamefont{G.}~\bibnamefont{Lacorata}}, \bibinfo {editor}
  {\bibfnamefont{G.}~\bibnamefont{Visconti}},\ and\ \bibinfo {editor}
  {\bibfnamefont{A.}~\bibnamefont{Vulpiani}}}\ (\bibinfo {publisher} {OTTO
  Editore},\ \bibinfo {address} {Torino})\ pp.\ \bibinfo {pages} {35--61}%
  \bibAnnoteFile{NoStop}{Hernandez-Garcia-et-al-03}%
\bibitem[{\citenamefont{Horner}\ \emph{et~al.}(2002)\citenamefont{Horner},
  \citenamefont{Metcalfe}, \citenamefont{Wiggins},\ and\
  \citenamefont{Ottino}}]{horner2002}%
  \BibitemOpen
  \bibfield{author}{%
  \bibinfo {author} {\bibnamefont{Horner}, \bibfnamefont{M.}}, \bibinfo
  {author} {\bibfnamefont{G.}~\bibnamefont{Metcalfe}}, \bibinfo {author}
  {\bibfnamefont{S.}~\bibnamefont{Wiggins}},\ and\ \bibinfo {author}
  {\bibfnamefont{J.~M.}\ \bibnamefont{Ottino}}}%
  , \bibinfo {year} {2002},\ \bibfield{title}{%
  \enquote{\bibinfo {title} {Transport enhancement mechanisms in open
  cavities},}\ }%
  \bibfield{journal}{%
  \bibinfo {journal} {J. Fluid Mech.}\ }%
  \textbf{\bibinfo {volume} {452}},\ \bibinfo {pages} {199--229}%
  \bibAnnoteFile{NoStop}{horner2002}%
\bibitem[{\citenamefont{Hutchinson}(1961)}]{Hutchinson-61}%
  \BibitemOpen
  \bibfield{author}{%
  \bibinfo {author} {\bibnamefont{Hutchinson}, \bibfnamefont{G.~E.}}}%
  , \bibinfo {year} {1961},\ \bibfield{title}{%
  \enquote{\bibinfo {title} {The paradox of the plankton},}\ }%
  \bibfield{journal}{%
  \bibinfo {journal} {Am. Nat.}\ }%
  \textbf{\bibinfo {volume} {95}},\ \bibinfo {pages} {137--145}%
  \bibAnnoteFile{NoStop}{Hutchinson-61}%
\bibitem[{\citenamefont{Hydon}(1995)}]{hydon1995}%
  \BibitemOpen
  \bibfield{author}{%
  \bibinfo {author} {\bibnamefont{Hydon}, \bibfnamefont{P.E.}}}%
  , \bibinfo {year} {1995},\ \bibfield{title}{%
  \enquote{\bibinfo {title} {Resonant and chaotic advection in a curved
  pipe},}\ }%
  \bibfield{journal}{%
  \bibinfo {journal} {Chaos, Solitons and Fractals}\ }%
  \textbf{\bibinfo {volume} {4}},\ \bibinfo {pages} {197--210}%
  \bibAnnoteFile{NoStop}{hydon1995}%
\bibitem[{\citenamefont{Jilisen}\ \emph{et~al.}(2013)\citenamefont{Jilisen},
  \citenamefont{Bloemen},\ and\ \citenamefont{Speetjens}}]{Jilisen2012}%
  \BibitemOpen
  \bibfield{author}{%
  \bibinfo {author} {\bibnamefont{Jilisen}, \bibfnamefont{R.~T.~M.}}, \bibinfo
  {author} {\bibfnamefont{P.~R.}\ \bibnamefont{Bloemen}},\ and\ \bibinfo
  {author} {\bibfnamefont{M.~F.~M.}\ \bibnamefont{Speetjens}}}%
  , \bibinfo {year} {2013},\ \bibfield{title}{%
  \enquote{\bibinfo {title} {Three-dimensional flow measurements in a static
  mixer},}\ }%
  \bibfield{journal}{%
  \bibinfo {journal} {AIChE J.}\ }%
  \textbf{\bibinfo {volume} {159}},\ \bibinfo {pages} {1746--1761}%
  \bibAnnoteFile{NoStop}{Jilisen2012}%
\bibitem[{\citenamefont{Jones}\ \emph{et~al.}(1989)\citenamefont{Jones},
  \citenamefont{Thomas},\ and\ \citenamefont{Aref}}]{JTA89}%
  \BibitemOpen
  \bibfield{author}{%
  \bibinfo {author} {\bibnamefont{Jones}, \bibfnamefont{S.~W.}}, \bibinfo
  {author} {\bibfnamefont{O.~M.}\ \bibnamefont{Thomas}},\ and\ \bibinfo
  {author} {\bibfnamefont{H.}~\bibnamefont{Aref}}}%
  , \bibinfo {year} {1989},\ \bibfield{title}{%
  \enquote{\bibinfo {title} {Chaotic advection by laminar flow in a twisted
  pipe},}\ }%
  \bibfield{journal}{%
  \bibinfo {journal} {J. Fluid Mech.}\ }%
  \textbf{\bibinfo {volume} {209}},\ \bibinfo {pages} {335--357}%
  \bibAnnoteFile{NoStop}{JTA89}%
\bibitem[{\citenamefont{Jones}\ and\ \citenamefont{Young}(1994)}]{Jones1994}%
  \BibitemOpen
  \bibfield{author}{%
  \bibinfo {author} {\bibnamefont{Jones}, \bibfnamefont{S.~W.}},\ and\ \bibinfo
  {author} {\bibfnamefont{W.~R.}\ \bibnamefont{Young}}}%
  , \bibinfo {year} {1994},\ \bibfield{title}{%
  \enquote{\bibinfo {title} {Shear dispersion and anomalous diffusion by
  chaotic advection},}\ }%
  \bibfield{journal}{%
  \bibinfo {journal} {J. Fluid Mech.}\ }%
  \textbf{\bibinfo {volume} {280}},\ \bibinfo {pages} {149--172}%
  \bibAnnoteFile{NoStop}{Jones1994}%
\bibitem[{\citenamefont{Joseph}\ and\
  \citenamefont{Legras}(2002)}]{Joseph:2002vi}%
  \BibitemOpen
  \bibfield{author}{%
  \bibinfo {author} {\bibnamefont{Joseph}, \bibfnamefont{B.}},\ and\ \bibinfo
  {author} {\bibfnamefont{B.}~\bibnamefont{Legras}}}%
  , \bibinfo {year} {2002},\ \bibfield{title}{%
  \enquote{\bibinfo {title} {Relation between kinematic boundaries, stirring,
  and barriers for the antarctic polar vortex},}\ }%
  \bibfield{journal}{%
  \bibinfo {journal} {J. Atmos. Sci.}\ }%
  \textbf{\bibinfo {volume} {59}},\ \bibinfo {pages} {1198--1212}%
  \bibAnnoteFile{NoStop}{Joseph:2002vi}%
\bibitem[{\citenamefont{Jung}\ \emph{et~al.}(1993)\citenamefont{Jung},
  \citenamefont{T\'el},\ and\ \citenamefont{Ziemniak}}]{Jung-et-al-93}%
  \BibitemOpen
  \bibfield{author}{%
  \bibinfo {author} {\bibnamefont{Jung}, \bibfnamefont{C.}}, \bibinfo {author}
  {\bibfnamefont{T.}~\bibnamefont{T\'el}},\ and\ \bibinfo {author}
  {\bibfnamefont{E.}~\bibnamefont{Ziemniak}}}%
  , \bibinfo {year} {1993},\ \bibfield{title}{%
  \enquote{\bibinfo {title} {Application of scattering chaos to particle
  transport in a hydrodynamical flow},}\ }%
  \bibfield{journal}{%
  \bibinfo {journal} {Chaos}\ }%
  \textbf{\bibinfo {volume} {3}},\ \bibinfo {pages} {555--568}%
  \bibAnnoteFile{NoStop}{Jung-et-al-93}%
\bibitem[{\citenamefont{Kalejaiye}\ and\
  \citenamefont{Cardoso}(2005)}]{Kal2005}%
  \BibitemOpen
  \bibfield{author}{%
  \bibinfo {author} {\bibnamefont{Kalejaiye}, \bibfnamefont{B.~O.}},\ and\
  \bibinfo {author} {\bibfnamefont{S.~S.~S.}\ \bibnamefont{Cardoso}}}%
  , \bibinfo {year} {2005},\ \bibfield{title}{%
  \enquote{\bibinfo {title} {Specification of the dispersion coefficient in the
  modeling of gravity-driven flow in porous media},}\ }%
  \bibfield{journal}{%
  \bibinfo {journal} {Water Resources Res.}\ }%
  \textbf{\bibinfo {volume} {41}},\ \bibinfo {pages} {W1047}%
  \bibAnnoteFile{NoStop}{Kal2005}%
\bibitem[{\citenamefont{Kantz}\ and\
  \citenamefont{Grassberger}(1985)}]{Grassberger1985}%
  \BibitemOpen
  \bibfield{author}{%
  \bibinfo {author} {\bibnamefont{Kantz}, \bibfnamefont{H.}},\ and\ \bibinfo
  {author} {\bibfnamefont{P.}~\bibnamefont{Grassberger}}}%
  , \bibinfo {year} {1985},\ \bibfield{title}{%
  \enquote{\bibinfo {title} {Repellers, semi-attractors, and long-lived chaotic
  transients},}\ }%
  \bibfield{journal}{%
  \bibinfo {journal} {Physica D}\ }%
  \textbf{\bibinfo {volume} {17}},\ \bibinfo {pages} {75--86}%
  \bibAnnoteFile{NoStop}{Grassberger1985}%
\bibitem[{\citenamefont{K\'arolyi}\
  \emph{et~al.}(2000)\citenamefont{K\'arolyi}, \citenamefont{P\'entek},
  \citenamefont{Scheuring}, \citenamefont{T\'el},\ and\
  \citenamefont{Toroczkai}}]{Karolyi-et-al-00}%
  \BibitemOpen
  \bibfield{author}{%
  \bibinfo {author} {\bibnamefont{K\'arolyi}, \bibfnamefont{G.}}, \bibinfo
  {author} {\bibfnamefont{A.}~\bibnamefont{P\'entek}}, \bibinfo {author}
  {\bibfnamefont{I.}~\bibnamefont{Scheuring}}, \bibinfo {author}
  {\bibfnamefont{T.}~\bibnamefont{T\'el}},\ and\ \bibinfo {author}
  {\bibfnamefont{Z.}~\bibnamefont{Toroczkai}}}%
  , \bibinfo {year} {2000},\ \bibfield{title}{%
  \enquote{\bibinfo {title} {Chaotic flow: the physics of species
  coexistence},}\ }%
  \bibfield{journal}{%
  \bibinfo {journal} {Proc. Natl. Acad. Sci. USA}\ }%
  \textbf{\bibinfo {volume} {97}},\ \bibinfo {pages} {13661--13665}%
  \bibAnnoteFile{NoStop}{Karolyi-et-al-00}%
\bibitem[{\citenamefont{K{\'a}rolyi}\
  \emph{et~al.}(1999)\citenamefont{K{\'a}rolyi}, \citenamefont{P{\'e}ntek},
  \citenamefont{Toroczkai}, \citenamefont{T{\'e}l},\ and\
  \citenamefont{Grebogi}}]{Karoly-et-al-99}%
  \BibitemOpen
  \bibfield{author}{%
  \bibinfo {author} {\bibnamefont{K{\'a}rolyi}, \bibfnamefont{G.}}, \bibinfo
  {author} {\bibfnamefont{{\'A}.}~\bibnamefont{P{\'e}ntek}}, \bibinfo {author}
  {\bibfnamefont{Z.}~\bibnamefont{Toroczkai}}, \bibinfo {author}
  {\bibfnamefont{T.}~\bibnamefont{T{\'e}l}},\ and\ \bibinfo {author}
  {\bibfnamefont{C.}~\bibnamefont{Grebogi}}}%
  , \bibinfo {year} {1999},\ \bibfield{title}{%
  \enquote{\bibinfo {title} {Chemical and biological activity in open chaotic
  flows},}\ }%
  \bibfield{journal}{%
  \bibinfo {journal} {Phys. Rev. E}\ }%
  \textbf{\bibinfo {volume} {59}},\ \bibinfo {pages} {5468--5481}%
  \bibAnnoteFile{NoStop}{Karoly-et-al-99}%
\bibitem[{\citenamefont{K\'arolyi}\ and\ \citenamefont{T\'el}(1997)}]{PHR}%
  \BibitemOpen
  \bibfield{author}{%
  \bibinfo {author} {\bibnamefont{K\'arolyi}, \bibfnamefont{G.}},\ and\
  \bibinfo {author} {\bibfnamefont{T.}~\bibnamefont{T\'el}}}%
  , \bibinfo {year} {1997},\ \bibfield{title}{%
  \enquote{\bibinfo {title} {Chaotic tracer scattering and fractal basin
  boundaries in a blinking vortex-sink system},}\ }%
  \bibfield{journal}{%
  \bibinfo {journal} {Phys. Rep.}\ }%
  \textbf{\bibinfo {volume} {290}},\ \bibinfo {pages} {125--147}%
  \bibAnnoteFile{NoStop}{PHR}%
\bibitem[{\citenamefont{K\'arolyi}\
  \emph{et~al.}(2004)\citenamefont{K\'arolyi}, \citenamefont{T\'el},
  \citenamefont{de~Moura},\ and\ \citenamefont{Grebogi}}]{Karolyi2004}%
  \BibitemOpen
  \bibfield{author}{%
  \bibinfo {author} {\bibnamefont{K\'arolyi}, \bibfnamefont{G.}}, \bibinfo
  {author} {\bibfnamefont{T.}~\bibnamefont{T\'el}}, \bibinfo {author}
  {\bibfnamefont{A.~P.~S.}\ \bibnamefont{de~Moura}},\ and\ \bibinfo {author}
  {\bibfnamefont{C.}~\bibnamefont{Grebogi}}}%
  , \bibinfo {year} {2004},\ \bibfield{title}{%
  \enquote{\bibinfo {title} {Reactive particles in random flows},}\ }%
  \bibfield{journal}{%
  \bibinfo {journal} {Phys. Rev. Lett.}\ }%
  \textbf{\bibinfo {volume} {92}},\ \bibinfo {pages} {174101}%
  \bibAnnoteFile{NoStop}{Karolyi2004}%
\bibitem[{\citenamefont{Karrasch}\ and\
  \citenamefont{Haller}(2013)}]{karrasch_finite-size_2013}%
  \BibitemOpen
  \bibfield{author}{%
  \bibinfo {author} {\bibnamefont{Karrasch}, \bibfnamefont{D.}},\ and\ \bibinfo
  {author} {\bibfnamefont{G.}~\bibnamefont{Haller}}}%
  , \bibinfo {year} {2013},\ \bibfield{title}{%
  \enquote{\bibinfo {title} {Do {Finite-Size} lyapunov exponents detect
  coherent structures?}.}\ }%
  \bibfield{journal}{%
  \bibinfo {journal} {Chaos}\ }%
  \textbf{\bibinfo {volume} {23}},\ \bibinfo {pages} {043126}%
  \bibAnnoteFile{NoStop}{karrasch_finite-size_2013}%
\bibitem[{\citenamefont{Keane}(1975)}]{keane1975interval}%
  \BibitemOpen
  \bibfield{author}{%
  \bibinfo {author} {\bibnamefont{Keane}, \bibfnamefont{M.}}}%
  , \bibinfo {year} {1975},\ \bibfield{title}{%
  \enquote{\bibinfo {title} {Interval exchange transformations},}\ }%
  \bibfield{journal}{%
  \bibinfo {journal} {Math. Z.}\ }%
  \textbf{\bibinfo {volume} {141}},\ \bibinfo {pages} {25--31}%
  \bibAnnoteFile{NoStop}{keane1975interval}%
\bibitem[{\citenamefont{Kim}\ \emph{et~al.}(2005)\citenamefont{Kim},
  \citenamefont{Lee}, \citenamefont{Kwon},\ and\ \citenamefont{Ahn}}]{Kim2005}%
  \BibitemOpen
  \bibfield{author}{%
  \bibinfo {author} {\bibnamefont{Kim}, \bibfnamefont{D.~S.}}, \bibinfo
  {author} {\bibfnamefont{S.~H.}\ \bibnamefont{Lee}}, \bibinfo {author}
  {\bibfnamefont{T.~H.}\ \bibnamefont{Kwon}},\ and\ \bibinfo {author}
  {\bibfnamefont{C.~H.}\ \bibnamefont{Ahn}}}%
  , \bibinfo {year} {2005},\ \bibfield{title}{%
  \enquote{\bibinfo {title} {A serpentine laminating micromixer combining
  splitting/recombination and advection},}\ }%
  \bibfield{journal}{%
  \bibinfo {journal} {Lab on a chip}\ }%
  \textbf{\bibinfo {volume} {5}},\ \bibinfo {pages} {739--747}%
  \bibAnnoteFile{NoStop}{Kim2005}%
\bibitem[{\citenamefont{Koh}\ and\ \citenamefont{Legras}(2002)}]{vortex_atm}%
  \BibitemOpen
  \bibfield{author}{%
  \bibinfo {author} {\bibnamefont{Koh}, \bibfnamefont{T.-Y.}},\ and\ \bibinfo
  {author} {\bibfnamefont{B.}~\bibnamefont{Legras}}}%
  , \bibinfo {year} {2002},\ \bibfield{title}{%
  \enquote{\bibinfo {title} {Hyperbolic lines and the stratospheric vortex},}\
  }%
  \bibfield{journal}{%
  \bibinfo {journal} {Chaos}\ }%
  \textbf{\bibinfo {volume} {12}},\ \bibinfo {pages} {382--394}%
  \bibAnnoteFile{NoStop}{vortex_atm}%
\bibitem[{\citenamefont{Kolmogorov}(1941{\natexlab{a}})}]{k41a}%
  \BibitemOpen
  \bibfield{author}{%
  \bibinfo {author} {\bibnamefont{Kolmogorov}, \bibfnamefont{A.~N.}}}%
  , \bibinfo {year} {1941}{\natexlab{a}},\ \bibfield{title}{%
  \enquote{\bibinfo {title} {The local structure of turbulence in
  incompressible viscous fluid for very large reynolds number},}\ }%
  \bibfield{journal}{%
  \bibinfo {journal} {Dokl. Akad. Nauk. SSSR}\ }%
  \textbf{\bibinfo {volume} {30}},\ \bibinfo {pages} {9--13}%
  \bibAnnoteFile{NoStop}{k41a}%
\bibitem[{\citenamefont{Kolmogorov}(1941{\natexlab{b}})}]{k41b}%
  \BibitemOpen
  \bibfield{author}{%
  \bibinfo {author} {\bibnamefont{Kolmogorov}, \bibfnamefont{A.~N.}}}%
  , \bibinfo {year} {1941}{\natexlab{b}},\ \bibfield{title}{%
  \enquote{\bibinfo {title} {On degeneration (decay) of isotropic turbulence in
  an incompressible viscous liquid},}\ }%
  \bibfield{journal}{%
  \bibinfo {journal} {Dokl. Akad. Nauk. SSSR}\ }%
  \textbf{\bibinfo {volume} {31}},\ \bibinfo {pages} {538--540}%
  \bibAnnoteFile{NoStop}{k41b}%
\bibitem[{\citenamefont{Kongthon}\ \emph{et~al.}(2011)\citenamefont{Kongthon},
  \citenamefont{Chung}, \citenamefont{Riley},\ and\
  \citenamefont{Devasia}}]{Kongthon2011}%
  \BibitemOpen
  \bibfield{author}{%
  \bibinfo {author} {\bibnamefont{Kongthon}, \bibfnamefont{J.}}, \bibinfo
  {author} {\bibfnamefont{J.-H.}\ \bibnamefont{Chung}}, \bibinfo {author}
  {\bibfnamefont{J.~J.}\ \bibnamefont{Riley}},\ and\ \bibinfo {author}
  {\bibfnamefont{S.}~\bibnamefont{Devasia}}}%
  , \bibinfo {year} {2011},\ \bibfield{title}{%
  \enquote{\bibinfo {title} {Dynamics of cilia-based microfluidic devices},}\
  }%
  \bibfield{journal}{%
  \bibinfo {journal} {Journal of Dynamic Systems, Measurement, and Control}\ }%
  \textbf{\bibinfo {volume} {133}},\ \bibinfo {pages} {051012}%
  \bibAnnoteFile{NoStop}{Kongthon2011}%
\bibitem[{\citenamefont{Kozlov}(1993)}]{Kozlov1993}%
  \BibitemOpen
  \bibfield{author}{%
  \bibinfo {author} {\bibnamefont{Kozlov}, \bibfnamefont{V.~V.}}}%
  , \bibinfo {year} {1993},\ \bibfield{title}{%
  \enquote{\bibinfo {title} {Dynamical systems determined by the
  {Navier}--{Stokes} equations},}\ }%
  \bibfield{journal}{%
  \bibinfo {journal} {Russ. J. Math. Phys.}\ }%
  \textbf{\bibinfo {volume} {1}},\ \bibinfo {pages} {57--69}%
  \bibAnnoteFile{NoStop}{Kozlov1993}%
\bibitem[{\citenamefont{Krasnopolskaya}\ and\
  \citenamefont{Meleshko}(2004)}]{Krasnopolskaya2004}%
  \BibitemOpen
  \bibfield{author}{%
  \bibinfo {author} {\bibnamefont{Krasnopolskaya}, \bibfnamefont{T.~S.}},\ and\
  \bibinfo {author} {\bibfnamefont{V.~V.}\ \bibnamefont{Meleshko}}}%
  , \bibinfo {year} {2004},\ \bibfield{title}{%
  \enquote{\bibinfo {title} {Laminar stirring of fluids. {Part} 1.
  {M}ethodology aspects},}\ }%
  \bibfield{journal}{%
  \bibinfo {journal} {Appl. Hydromechanics}\ }%
  \textbf{\bibinfo {volume} {6}},\ \bibinfo {pages} {28--40}%
  \bibAnnoteFile{NoStop}{Krasnopolskaya2004}%
\bibitem[{\citenamefont{Krasnopolskaya}\ and\
  \citenamefont{Meleshko}(2009)}]{Krasnopolskaya2009}%
  \BibitemOpen
  \bibfield{author}{%
  \bibinfo {author} {\bibnamefont{Krasnopolskaya}, \bibfnamefont{T.~S.}},\ and\
  \bibinfo {author} {\bibfnamefont{V.~V.}\ \bibnamefont{Meleshko}}}%
  , \bibinfo {year} {2009},\ \enquote{\bibinfo {title} {Quality measures and
  transport properties},}\ in\ \emph{\bibinfo {booktitle} {Analysis and Control
  of Mixing with an Application to Micro and Macro Flow Processes}},\ \bibinfo
  {editor} {edited by\ \bibinfo {editor}
  {\bibfnamefont{L.}~\bibnamefont{Cortelezzi}}\ and\ \bibinfo {editor}
  {\bibfnamefont{I.}~\bibnamefont{Mezi\'{c}}}}\ (\bibinfo {publisher}
  {Springer})\ pp.\ \bibinfo {pages} {291--306}%
  \bibAnnoteFile{NoStop}{Krasnopolskaya2009}%
\bibitem[{\citenamefont{Krasnopolskaya}\
  \emph{et~al.}(1996)\citenamefont{Krasnopolskaya}, \citenamefont{Meleshko},
  \citenamefont{Peters},\ and\ \citenamefont{Meijer}}]{Krasnopolskaya1996}%
  \BibitemOpen
  \bibfield{author}{%
  \bibinfo {author} {\bibnamefont{Krasnopolskaya}, \bibfnamefont{T.~S.}},
  \bibinfo {author} {\bibfnamefont{V.~V.}\ \bibnamefont{Meleshko}}, \bibinfo
  {author} {\bibfnamefont{G.~W.~M.}\ \bibnamefont{Peters}},\ and\ \bibinfo
  {author} {\bibfnamefont{H.~E.~H.}\ \bibnamefont{Meijer}}}%
  , \bibinfo {year} {1996},\ \bibfield{title}{%
  \enquote{\bibinfo {title} {Steady {Stokes} flow in an annular cavity},}\ }%
  \bibfield{journal}{%
  \bibinfo {journal} {Quart. J. Mech. Appl. Math.}\ }%
  \textbf{\bibinfo {volume} {49}},\ \bibinfo {pages} {593--619}%
  \bibAnnoteFile{NoStop}{Krasnopolskaya1996}%
\bibitem[{\citenamefont{Krasnopolskaya}\
  \emph{et~al.}(1999)\citenamefont{Krasnopolskaya}, \citenamefont{Meleshko},
  \citenamefont{Peters},\ and\ \citenamefont{Meijer}}]{Krasnopolskaya1999}%
  \BibitemOpen
  \bibfield{author}{%
  \bibinfo {author} {\bibnamefont{Krasnopolskaya}, \bibfnamefont{T.~S.}},
  \bibinfo {author} {\bibfnamefont{V.~V.}\ \bibnamefont{Meleshko}}, \bibinfo
  {author} {\bibfnamefont{G.~W.~M.}\ \bibnamefont{Peters}},\ and\ \bibinfo
  {author} {\bibfnamefont{H.~E.~H.}\ \bibnamefont{Meijer}}}%
  , \bibinfo {year} {1999},\ \bibfield{title}{%
  \enquote{\bibinfo {title} {Mixing in {Stokes} flow in an annular wedge
  cavity},}\ }%
  \bibfield{journal}{%
  \bibinfo {journal} {Eur. J. Mech. B/Fluids}\ }%
  \textbf{\bibinfo {volume} {18}},\ \bibinfo {pages} {793--822}%
  \bibAnnoteFile{NoStop}{Krasnopolskaya1999}%
\bibitem[{\citenamefont{Kundu}\ and\ \citenamefont{Cohen}(2008)}]{kundu}%
  \BibitemOpen
  \bibfield{author}{%
  \bibinfo {author} {\bibnamefont{Kundu}, \bibfnamefont{P.~K.}},\ and\ \bibinfo
  {author} {\bibfnamefont{I.~M.}\ \bibnamefont{Cohen}}}%
  , \bibinfo {year} {2008},\ \emph{\bibinfo {title} {Fluid Mechanics}},\
  \bibinfo {edition} {4th}\ ed.\ (\bibinfo {publisher} {Academic, New York})%
  \bibAnnoteFile{NoStop}{kundu}%
\bibitem[{\citenamefont{Lackey}\ and\
  \citenamefont{Sotiropoulos}(2006)}]{lackey2006relationship}%
  \BibitemOpen
  \bibfield{author}{%
  \bibinfo {author} {\bibnamefont{Lackey}, \bibfnamefont{T.~C.}},\ and\
  \bibinfo {author} {\bibfnamefont{F.}~\bibnamefont{Sotiropoulos}}}%
  , \bibinfo {year} {2006},\ \bibfield{title}{%
  \enquote{\bibinfo {title} {Relationship between stirring rate and reynolds
  number in the chaotically advected steady flow in a container with exactly
  counter-rotating lids},}\ }%
  \bibfield{journal}{%
  \bibinfo {journal} {Phys. Fluids}\ }%
  \textbf{\bibinfo {volume} {18}},\ \bibinfo {pages} {053601}%
  \bibAnnoteFile{NoStop}{lackey2006relationship}%
\bibitem[{\citenamefont{Lai}\ and\ \citenamefont{T\'el}(2011)}]{cscatbook}%
  \BibitemOpen
  \bibfield{author}{%
  \bibinfo {author} {\bibnamefont{Lai}, \bibfnamefont{Y.-C.}},\ and\ \bibinfo
  {author} {\bibfnamefont{T.}~\bibnamefont{T\'el}}}%
  , \bibinfo {year} {2011},\ \emph{\bibinfo {title} {Transient Chaos}}\
  (\bibinfo {publisher} {Springer Verlag},\ \bibinfo {address} {New York})%
  \bibAnnoteFile{NoStop}{cscatbook}%
\bibitem[{\citenamefont{Lasota}\ and\ \citenamefont{Mackey}(1994)}]{Lasota}%
  \BibitemOpen
  \bibfield{author}{%
  \bibinfo {author} {\bibnamefont{Lasota}, \bibfnamefont{A.}},\ and\ \bibinfo
  {author} {\bibfnamefont{M.~C.}\ \bibnamefont{Mackey}}}%
  , \bibinfo {year} {1994},\ \emph{\bibinfo {title} {Chaos, Fractals, and
  Noise}}\ (\bibinfo {publisher} {Springer Verlag},\ \bibinfo {address} {New
  York})%
  \bibAnnoteFile{NoStop}{Lasota}%
\bibitem[{\citenamefont{Lau}\ \emph{et~al.}(1991)\citenamefont{Lau},
  \citenamefont{Finn},\ and\ \citenamefont{Ott}}]{Lau}%
  \BibitemOpen
  \bibfield{author}{%
  \bibinfo {author} {\bibnamefont{Lau}, \bibfnamefont{Y.-T.}}, \bibinfo
  {author} {\bibfnamefont{J.~M.}\ \bibnamefont{Finn}},\ and\ \bibinfo {author}
  {\bibfnamefont{E.}~\bibnamefont{Ott}}}%
  , \bibinfo {year} {1991},\ \bibfield{title}{%
  \enquote{\bibinfo {title} {Fractal dimension in nonhyperbolic chaotic
  scattering},}\ }%
  \bibfield{journal}{%
  \bibinfo {journal} {Phys. Rev. Lett.}\ }%
  \textbf{\bibinfo {volume} {66}},\ \bibinfo {pages} {978--981}%
  \bibAnnoteFile{NoStop}{Lau}%
\bibitem[{\citenamefont{Le~Guer}\ and\
  \citenamefont{El~Omari}(2012)}]{leguer2012}%
  \BibitemOpen
  \bibfield{author}{%
  \bibinfo {author} {\bibnamefont{Le~Guer}, \bibfnamefont{Y.}},\ and\ \bibinfo
  {author} {\bibfnamefont{K.}~\bibnamefont{El~Omari}}}%
  , \bibinfo {year} {2012},\ \bibfield{title}{%
  \enquote{\bibinfo {title} {Chaotic advection for thermal mixing},}\ }%
  \bibfield{journal}{%
  \bibinfo {journal} {Adv. Appl. Mech.}\ }%
  \textbf{\bibinfo {volume} {45}},\ \bibinfo {pages} {189--237}%
  \bibAnnoteFile{NoStop}{leguer2012}%
\bibitem[{\citenamefont{Lebedev}\ and\
  \citenamefont{Turitsyn}(2004)}]{Lebedev2004}%
  \BibitemOpen
  \bibfield{author}{%
  \bibinfo {author} {\bibnamefont{Lebedev}, \bibfnamefont{V.~V.}},\ and\
  \bibinfo {author} {\bibfnamefont{K.~S.}\ \bibnamefont{Turitsyn}}}%
  , \bibinfo {year} {2004},\ \bibfield{title}{%
  \enquote{\bibinfo {title} {Passive scalar evolution in peripheral regions},}\
  }%
  \bibfield{journal}{%
  \bibinfo {journal} {Phys. Rev. E}\ }%
  \textbf{\bibinfo {volume} {69}},\ \bibinfo {pages} {036301}%
  \bibAnnoteFile{NoStop}{Lebedev2004}%
\bibitem[{\citenamefont{Lenz}\ and\ \citenamefont{Ryskin}(2006)}]{Lenz2006}%
  \BibitemOpen
  \bibfield{author}{%
  \bibinfo {author} {\bibnamefont{Lenz}, \bibfnamefont{P.}},\ and\ \bibinfo
  {author} {\bibfnamefont{A.}~\bibnamefont{Ryskin}}}%
  , \bibinfo {year} {2006},\ \bibfield{title}{%
  \enquote{\bibinfo {title} {{Collective effects in ciliar arrays}},}\ }%
  \bibfield{journal}{%
  \bibinfo {journal} {Phys. Biol.}\ }%
  \textbf{\bibinfo {volume} {3}},\ \bibinfo {pages} {285--294}%
  \bibAnnoteFile{NoStop}{Lenz2006}%
\bibitem[{\citenamefont{Lester}\ and\
  \citenamefont{Metcalfe}(2009)}]{Lester_lagrangian_2008}%
  \BibitemOpen
  \bibfield{author}{%
  \bibinfo {author} {\bibnamefont{Lester}, \bibfnamefont{D.}},\ and\ \bibinfo
  {author} {\bibfnamefont{G.}~\bibnamefont{Metcalfe}}}%
  , \bibinfo {year} {2009},\ \enquote{\bibinfo {title} {Lagrangian topology of
  reoriented potential flows},}\ in\ \emph{\bibinfo {booktitle} {Biomedical
  Applications of Micro- and Nanoengineering IV and Complex Systems}},\ Vol.\
  \bibinfo {volume} {7270},\ \bibinfo {editor} {edited by\ \bibinfo {editor}
  {\bibfnamefont{D.~V.}\ \bibnamefont{Nicolau}}\ and\ \bibinfo {editor}
  {\bibfnamefont{G.}~\bibnamefont{Metcalfe}}}\ (\bibinfo {publisher} {SPIE})\
  p.\ \bibinfo {pages} {727013}%
  \bibAnnoteFile{NoStop}{Lester_lagrangian_2008}%
\bibitem[{\citenamefont{Lester}\ \emph{et~al.}(2007)\citenamefont{Lester},
  \citenamefont{Metcalfe},\ and\ \citenamefont{Rudman}}]{Lester_2007}%
  \BibitemOpen
  \bibfield{author}{%
  \bibinfo {author} {\bibnamefont{Lester}, \bibfnamefont{D.}}, \bibinfo
  {author} {\bibfnamefont{G.}~\bibnamefont{Metcalfe}},\ and\ \bibinfo {author}
  {\bibfnamefont{M.}~\bibnamefont{Rudman}}}%
  , \bibinfo {year} {2007},\ \enquote{\bibinfo {title} {Complete parametric
  scalar dispersion},}\ in\ \emph{\bibinfo {booktitle} {Proceedings of SPIE,
  Microelectronics, {MEMS}, and Nanotechnology, Complex Systems II}},\ Vol.\
  \bibinfo {volume} {6802}%
  \bibAnnoteFile{NoStop}{Lester_2007}%
\bibitem[{\citenamefont{Lester}\
  \emph{et~al.}(2014{\natexlab{a}})\citenamefont{Lester},
  \citenamefont{Metcalfe},\ and\ \citenamefont{Rudman}}]{Lester_control_2014}%
  \BibitemOpen
  \bibfield{author}{%
  \bibinfo {author} {\bibnamefont{Lester}, \bibfnamefont{D.}}, \bibinfo
  {author} {\bibfnamefont{G.}~\bibnamefont{Metcalfe}},\ and\ \bibinfo {author}
  {\bibfnamefont{M.}~\bibnamefont{Rudman}}}%
  , \bibinfo {year} {2014}{\natexlab{a}},\ \bibfield{title}{%
  \enquote{\bibinfo {title} {Control mechanisms for the global structure of
  scalar dispersion in chaotic flows},}\ }%
  \bibfield{journal}{%
  \bibinfo {journal} {Phys. Rev. E}\ }%
  \textbf{\bibinfo {volume} {90}},\ \bibinfo {pages} {022908}%
  \bibAnnoteFile{NoStop}{Lester_control_2014}%
\bibitem[{\citenamefont{Lester}\ \emph{et~al.}(2010)\citenamefont{Lester},
  \citenamefont{Metcalfe}, \citenamefont{Rudman}, \citenamefont{Ord},\ and\
  \citenamefont{Hobbs}}]{Lester_RPM_2010}%
  \BibitemOpen
  \bibfield{author}{%
  \bibinfo {author} {\bibnamefont{Lester}, \bibfnamefont{D.}}, \bibinfo
  {author} {\bibfnamefont{G.}~\bibnamefont{Metcalfe}}, \bibinfo {author}
  {\bibfnamefont{M.}~\bibnamefont{Rudman}}, \bibinfo {author}
  {\bibfnamefont{A.}~\bibnamefont{Ord}},\ and\ \bibinfo {author}
  {\bibfnamefont{B.}~\bibnamefont{Hobbs}}}%
  , \bibinfo {year} {2010},\ \bibfield{title}{%
  \enquote{\bibinfo {title} {Scalar dispersion in a periodically reoriented
  potential flow: {A}cceleration via {L}agrangian chaos},}\ }%
  \bibfield{journal}{%
  \bibinfo {journal} {Phys. Rev. E}\ }%
  \textbf{\bibinfo {volume} {81}},\ \bibinfo {pages} {046319}%
  \bibAnnoteFile{NoStop}{Lester_RPM_2010}%
\bibitem[{\citenamefont{Lester}\
  \emph{et~al.}(2009{\natexlab{a}})\citenamefont{Lester},
  \citenamefont{Metcalfe}, \citenamefont{Trefry}, \citenamefont{Ord},
  \citenamefont{Hobbs},\ and\ \citenamefont{Rudman}}]{Lester_RPM_2009}%
  \BibitemOpen
  \bibfield{author}{%
  \bibinfo {author} {\bibnamefont{Lester}, \bibfnamefont{D.}}, \bibinfo
  {author} {\bibfnamefont{G.}~\bibnamefont{Metcalfe}}, \bibinfo {author}
  {\bibfnamefont{M.}~\bibnamefont{Trefry}}, \bibinfo {author}
  {\bibfnamefont{A.}~\bibnamefont{Ord}}, \bibinfo {author}
  {\bibfnamefont{B.}~\bibnamefont{Hobbs}},\ and\ \bibinfo {author}
  {\bibfnamefont{M.}~\bibnamefont{Rudman}}}%
  , \bibinfo {year} {2009}{\natexlab{a}},\ \bibfield{title}{%
  \enquote{\bibinfo {title} {{L}agrangian topology of a periodically reoriented
  potential flow: {S}ymmetry, optimization and mixing},}\ }%
  \bibfield{journal}{%
  \bibinfo {journal} {Phys. Rev. E}\ }%
  \textbf{\bibinfo {volume} {80}},\ \bibinfo {pages} {036108}%
  \bibAnnoteFile{NoStop}{Lester_RPM_2009}%
\bibitem[{\citenamefont{Lester}\ \emph{et~al.}(2013)\citenamefont{Lester},
  \citenamefont{Metcalfe},\ and\ \citenamefont{Trefry}}]{lester2013}%
  \BibitemOpen
  \bibfield{author}{%
  \bibinfo {author} {\bibnamefont{Lester}, \bibfnamefont{D.~R.}}, \bibinfo
  {author} {\bibfnamefont{G.}~\bibnamefont{Metcalfe}},\ and\ \bibinfo {author}
  {\bibfnamefont{M.~G.}\ \bibnamefont{Trefry}}}%
  , \bibinfo {year} {2013},\ \bibfield{title}{%
  \enquote{\bibinfo {title} {Is chaotic advection inherent to porous media
  flow?}.}\ }%
  \bibfield{journal}{%
  \bibinfo {journal} {Phys. Rev. Lett.}\ }%
  \textbf{\bibinfo {volume} {111}},\ \bibinfo {pages} {174101}%
  \bibAnnoteFile{NoStop}{lester2013}%
\bibitem[{\citenamefont{Lester}\
  \emph{et~al.}(2014{\natexlab{b}})\citenamefont{Lester},
  \citenamefont{Metcalfe},\ and\
  \citenamefont{Trefry}}]{Lester_anomalous_2014}%
  \BibitemOpen
  \bibfield{author}{%
  \bibinfo {author} {\bibnamefont{Lester}, \bibfnamefont{D.~R.}}, \bibinfo
  {author} {\bibfnamefont{G.}~\bibnamefont{Metcalfe}},\ and\ \bibinfo {author}
  {\bibfnamefont{M.~G.}\ \bibnamefont{Trefry}}}%
  , \bibinfo {year} {2014}{\natexlab{b}},\ \bibfield{title}{%
  \enquote{\bibinfo {title} {Anomalous transport and chaotic advection in
  homogeneous porous media},}\ }%
  \bibfield{journal}{%
  \bibinfo {journal} {Phys. Rev. E}\ }%
  \textbf{\bibinfo {volume} {90}},\ \bibinfo {pages} {063012}%
  \bibAnnoteFile{NoStop}{Lester_anomalous_2014}%
\bibitem[{\citenamefont{Lester}\
  \emph{et~al.}(2009{\natexlab{b}})\citenamefont{Lester},
  \citenamefont{Rudman},\ and\
  \citenamefont{Metcalfe}}]{Lester_nonNewtonian_2009}%
  \BibitemOpen
  \bibfield{author}{%
  \bibinfo {author} {\bibnamefont{Lester}, \bibfnamefont{D.~R.}}, \bibinfo
  {author} {\bibfnamefont{M.}~\bibnamefont{Rudman}},\ and\ \bibinfo {author}
  {\bibfnamefont{G.}~\bibnamefont{Metcalfe}}}%
  , \bibinfo {year} {2009}{\natexlab{b}},\ \bibfield{title}{%
  \enquote{\bibinfo {title} {Low {R}eynolds number scalar transport enhancement
  in viscous and non-{N}ewtonian fluids},}\ }%
  \bibfield{journal}{%
  \bibinfo {journal} {Int. J. Heat Mass Transfer}\ }%
  \textbf{\bibinfo {volume} {52}},\ \bibinfo {pages} {655--664}%
  \bibAnnoteFile{NoStop}{Lester_nonNewtonian_2009}%
\bibitem[{\citenamefont{Lester}\ \emph{et~al.}(2008)\citenamefont{Lester},
  \citenamefont{Rudman}, \citenamefont{Metcalfe},\ and\
  \citenamefont{Blackburn}}]{Lester08a}%
  \BibitemOpen
  \bibfield{author}{%
  \bibinfo {author} {\bibnamefont{Lester}, \bibfnamefont{D.~R.}}, \bibinfo
  {author} {\bibfnamefont{M.}~\bibnamefont{Rudman}}, \bibinfo {author}
  {\bibfnamefont{G.}~\bibnamefont{Metcalfe}},\ and\ \bibinfo {author}
  {\bibfnamefont{H.~M.}\ \bibnamefont{Blackburn}}}%
  , \bibinfo {year} {2008},\ \bibfield{title}{%
  \enquote{\bibinfo {title} {Global parametric solutions of scalar
  transport},}\ }%
  \bibfield{journal}{%
  \bibinfo {journal} {J. Comput. Phys.}\ }%
  \textbf{\bibinfo {volume} {227}},\ \bibinfo {pages} {3032--3057}%
  \bibAnnoteFile{NoStop}{Lester08a}%
\bibitem[{\citenamefont{Levnaji{\'c}}\ and\
  \citenamefont{Mezi{\'c}}(2010)}]{Levnajic:2010gq}%
  \BibitemOpen
  \bibfield{author}{%
  \bibinfo {author} {\bibnamefont{Levnaji{\'c}}, \bibfnamefont{Z.}},\ and\
  \bibinfo {author} {\bibfnamefont{I.}~\bibnamefont{Mezi{\'c}}}}%
  , \bibinfo {year} {2010},\ \bibfield{title}{%
  \enquote{\bibinfo {title} {{Ergodic theory and visualization. I. Mesochronic
  plots for visualization of ergodic partition and invariant sets}},}\ }%
  \bibfield{journal}{%
  \bibinfo {journal} {Chaos}\ }%
  \textbf{\bibinfo {volume} {20}},\ \bibinfo {pages} {033114}%
  \bibAnnoteFile{NoStop}{Levnajic:2010gq}%
\bibitem[{\citenamefont{Lighthill}(1976)}]{Lighthill1976}%
  \BibitemOpen
  \bibfield{author}{%
  \bibinfo {author} {\bibnamefont{Lighthill}, \bibfnamefont{J.}}}%
  , \bibinfo {year} {1976},\ \bibfield{title}{%
  \enquote{\bibinfo {title} {{Flagellar Hydrodynamics}},}\ }%
  \bibfield{journal}{%
  \bibinfo {journal} {SIAM Rev.}\ }%
  \textbf{\bibinfo {volume} {18}},\ \bibinfo {pages} {161--230}%
  \bibAnnoteFile{NoStop}{Lighthill1976}%
\bibitem[{\citenamefont{Lin}\ and\ \citenamefont{Basuray}(2011)}]{Lin2011}%
  \BibitemOpen
  \bibfield{author}{%
  \bibinfo {author} {\bibnamefont{Lin}, \bibfnamefont{B.}},\ and\ \bibinfo
  {author} {\bibfnamefont{S.}~\bibnamefont{Basuray}}}%
  , \bibinfo {year} {2011},\ \emph{\bibinfo {title} {Microfluidics:
  Technologies and Applications}},\ Topics in Current Chemistry\ (\bibinfo
  {publisher} {Springer})%
  \bibAnnoteFile{NoStop}{Lin2011}%
\bibitem[{\citenamefont{Lin}\ \emph{et~al.}(2011)\citenamefont{Lin},
  \citenamefont{Thiffeault},\ and\ \citenamefont{Doering}}]{lin2011optimal}%
  \BibitemOpen
  \bibfield{author}{%
  \bibinfo {author} {\bibnamefont{Lin}, \bibfnamefont{Z.}}, \bibinfo {author}
  {\bibfnamefont{J.-L.}\ \bibnamefont{Thiffeault}},\ and\ \bibinfo {author}
  {\bibfnamefont{C.~R.}\ \bibnamefont{Doering}}}%
  , \bibinfo {year} {2011},\ \bibfield{title}{%
  \enquote{\bibinfo {title} {Optimal stirring strategies for passive scalar
  mixing},}\ }%
  \bibfield{journal}{%
  \bibinfo {journal} {J. Fluid Mech.}\ }%
  \textbf{\bibinfo {volume} {675}},\ \bibinfo {pages} {465--476}%
  \bibAnnoteFile{NoStop}{lin2011optimal}%
\bibitem[{\citenamefont{Litvak-Hinenzon}\ and\
  \citenamefont{Rom-Kedar}(2002)}]{Litvak2002}%
  \BibitemOpen
  \bibfield{author}{%
  \bibinfo {author} {\bibnamefont{Litvak-Hinenzon}, \bibfnamefont{A.}},\ and\
  \bibinfo {author} {\bibfnamefont{V.}~\bibnamefont{Rom-Kedar}}}%
  , \bibinfo {year} {2002},\ \bibfield{title}{%
  \enquote{\bibinfo {title} {Parabolic resonances in 3 degree of freedom
  near-integrable {Hamiltonian} systems},}\ }%
  \bibfield{journal}{%
  \bibinfo {journal} {Physica D}\ }%
  \textbf{\bibinfo {volume} {164}},\ \bibinfo {pages} {213--250}%
  \bibAnnoteFile{NoStop}{Litvak2002}%
\bibitem[{\citenamefont{Liu}\ \emph{et~al.}(1994)\citenamefont{Liu},
  \citenamefont{Muzzio},\ and\ \citenamefont{Peskin}}]{Liu94}%
  \BibitemOpen
  \bibfield{author}{%
  \bibinfo {author} {\bibnamefont{Liu}, \bibfnamefont{M.}}, \bibinfo {author}
  {\bibfnamefont{F.~J.}\ \bibnamefont{Muzzio}},\ and\ \bibinfo {author}
  {\bibfnamefont{R.~L.}\ \bibnamefont{Peskin}}}%
  , \bibinfo {year} {1994},\ \bibfield{title}{%
  \enquote{\bibinfo {title} {Quantification of mixing in aperiodic chaotic
  flows},}\ }%
  \bibfield{journal}{%
  \bibinfo {journal} {Chaos, Solitons and Fractals}\ }%
  \textbf{\bibinfo {volume} {4}},\ \bibinfo {pages} {869--893}%
  \bibAnnoteFile{NoStop}{Liu94}%
\bibitem[{\citenamefont{Liu}\ \emph{et~al.}(2002)\citenamefont{Liu},
  \citenamefont{Yang}, \citenamefont{Pindera}, \citenamefont{Athavale},\ and\
  \citenamefont{Grodzinski}}]{Liu2002}%
  \BibitemOpen
  \bibfield{author}{%
  \bibinfo {author} {\bibnamefont{Liu}, \bibfnamefont{R.~H.}}, \bibinfo
  {author} {\bibfnamefont{J.}~\bibnamefont{Yang}}, \bibinfo {author}
  {\bibfnamefont{M.~Z.}\ \bibnamefont{Pindera}}, \bibinfo {author}
  {\bibfnamefont{M.}~\bibnamefont{Athavale}},\ and\ \bibinfo {author}
  {\bibfnamefont{P.}~\bibnamefont{Grodzinski}}}%
  , \bibinfo {year} {2002},\ \bibfield{title}{%
  \enquote{\bibinfo {title} {Bubble-induced acoustic micromixing},}\ }%
  \bibfield{journal}{%
  \bibinfo {journal} {Lab on a chip}\ }%
  \textbf{\bibinfo {volume} {2}},\ \bibinfo {pages} {151--157}%
  \bibAnnoteFile{NoStop}{Liu2002}%
\bibitem[{\citenamefont{Liu}(2008)}]{liu2008}%
  \BibitemOpen
  \bibfield{author}{%
  \bibinfo {author} {\bibnamefont{Liu}, \bibfnamefont{W.}}}%
  , \bibinfo {year} {2008},\ \bibfield{title}{%
  \enquote{\bibinfo {title} {Mixing enhancement by optimal flow advection},}\
  }%
  \bibfield{journal}{%
  \bibinfo {journal} {SIAM J. Control Optim.}\ }%
  \textbf{\bibinfo {volume} {47}},\ \bibinfo {pages} {624--638}%
  \bibAnnoteFile{NoStop}{liu2008}%
\bibitem[{\citenamefont{Liu}\ and\ \citenamefont{Haller}(2004)}]{liu2004}%
  \BibitemOpen
  \bibfield{author}{%
  \bibinfo {author} {\bibnamefont{Liu}, \bibfnamefont{W.}},\ and\ \bibinfo
  {author} {\bibfnamefont{G.}~\bibnamefont{Haller}}}%
  , \bibinfo {year} {2004},\ \bibfield{title}{%
  \enquote{\bibinfo {title} {{Strange eigenmodes and decay of variance in the
  mixing of diffusive tracers}},}\ }%
  \bibfield{journal}{%
  \bibinfo {journal} {Physica D}\ }%
  \textbf{\bibinfo {volume} {188}},\ \bibinfo {pages} {1--39}%
  \bibAnnoteFile{NoStop}{liu2004}%
\bibitem[{\citenamefont{Lomeli}\ and\ \citenamefont{Meiss}(2009)}]{Lomeli2009}%
  \BibitemOpen
  \bibfield{author}{%
  \bibinfo {author} {\bibnamefont{Lomeli}, \bibfnamefont{H.~E.}},\ and\
  \bibinfo {author} {\bibfnamefont{J.~D.}\ \bibnamefont{Meiss}}}%
  , \bibinfo {year} {2009},\ \bibfield{title}{%
  \enquote{\bibinfo {title} {Resonance zones and lobe volumes for exact
  volume-preserving maps},}\ }%
  \bibfield{journal}{%
  \bibinfo {journal} {Nonlinearity}\ }%
  \textbf{\bibinfo {volume} {22}},\ \bibinfo {pages} {1761--1789}%
  \bibAnnoteFile{NoStop}{Lomeli2009}%
\bibitem[{\citenamefont{L\'opez}\
  \emph{et~al.}(2001{\natexlab{a}})\citenamefont{L\'opez},
  \citenamefont{Hern\'andez-Garc\'ia}, \citenamefont{Piro},
  \citenamefont{Vulpiani},\ and\ \citenamefont{Zambianchi}}]{Lopez-et-al-01a}%
  \BibitemOpen
  \bibfield{author}{%
  \bibinfo {author} {\bibnamefont{L\'opez}, \bibfnamefont{C.}}, \bibinfo
  {author} {\bibfnamefont{E.}~\bibnamefont{Hern\'andez-Garc\'ia}}, \bibinfo
  {author} {\bibfnamefont{O.}~\bibnamefont{Piro}}, \bibinfo {author}
  {\bibfnamefont{A.}~\bibnamefont{Vulpiani}},\ and\ \bibinfo {author}
  {\bibfnamefont{E.}~\bibnamefont{Zambianchi}}}%
  , \bibinfo {year} {2001}{\natexlab{a}},\ \bibfield{title}{%
  \enquote{\bibinfo {title} {Population dynamics advected by chaotic flows: A
  discrete-time map approach},}\ }%
  \bibfield{journal}{%
  \bibinfo {journal} {Chaos}\ }%
  \textbf{\bibinfo {volume} {11}},\ \bibinfo {pages} {397--403}%
  \bibAnnoteFile{NoStop}{Lopez-et-al-01a}%
\bibitem[{\citenamefont{L\'opez}\
  \emph{et~al.}(2001{\natexlab{b}})\citenamefont{L\'opez},
  \citenamefont{Neufeld}, \citenamefont{Hern\'andez-Garc\'ia},\ and\
  \citenamefont{Haynes}}]{Lopez-et-al-01b}%
  \BibitemOpen
  \bibfield{author}{%
  \bibinfo {author} {\bibnamefont{L\'opez}, \bibfnamefont{C.}}, \bibinfo
  {author} {\bibfnamefont{Z.}~\bibnamefont{Neufeld}}, \bibinfo {author}
  {\bibfnamefont{E.}~\bibnamefont{Hern\'andez-Garc\'ia}},\ and\ \bibinfo
  {author} {\bibfnamefont{P.}~\bibnamefont{Haynes}}}%
  , \bibinfo {year} {2001}{\natexlab{b}},\ \bibfield{title}{%
  \enquote{\bibinfo {title} {Chaotic advection of reacting substances: Plankton
  dynamics on a meandering jet},}\ }%
  \bibfield{journal}{%
  \bibinfo {journal} {Phys. Chem. Earth, B}\ }%
  \textbf{\bibinfo {volume} {26}},\ \bibinfo {pages} {313--317}%
  \bibAnnoteFile{NoStop}{Lopez-et-al-01b}%
\bibitem[{\citenamefont{Luethi}\ \emph{et~al.}(2005)\citenamefont{Luethi},
  \citenamefont{Tsinober},\ and\ \citenamefont{Kinzelbach}}]{luethi}%
  \BibitemOpen
  \bibfield{author}{%
  \bibinfo {author} {\bibnamefont{Luethi}, \bibfnamefont{B.}}, \bibinfo
  {author} {\bibfnamefont{A.}~\bibnamefont{Tsinober}},\ and\ \bibinfo {author}
  {\bibfnamefont{W.}~\bibnamefont{Kinzelbach}}}%
  , \bibinfo {year} {2005},\ \bibfield{title}{%
  \enquote{\bibinfo {title} {Lagrangian measurement of vorticity dynamics in
  turbulent flow},}\ }%
  \bibfield{journal}{%
  \bibinfo {journal} {J. Fluid Mech.}\ }%
  \textbf{\bibinfo {volume} {528}},\ \bibinfo {pages} {87--118}%
  \bibAnnoteFile{NoStop}{luethi}%
\bibitem[{\citenamefont{Lukens}\ \emph{et~al.}(2010)\citenamefont{Lukens},
  \citenamefont{Yang},\ and\ \citenamefont{Fauci}}]{lukens_using_2010}%
  \BibitemOpen
  \bibfield{author}{%
  \bibinfo {author} {\bibnamefont{Lukens}, \bibfnamefont{S.}}, \bibinfo
  {author} {\bibfnamefont{X.}~\bibnamefont{Yang}},\ and\ \bibinfo {author}
  {\bibfnamefont{L.}~\bibnamefont{Fauci}}}%
  , \bibinfo {year} {2010},\ \bibfield{title}{%
  \enquote{\bibinfo {title} {Using {Lagrangian} coherent structures to analyze
  fluid mixing by cilia},}\ }%
  \bibfield{journal}{%
  \bibinfo {journal} {Chaos}\ }%
  \textbf{\bibinfo {volume} {20}},\ \bibinfo {pages} {017511}%
  \bibAnnoteFile{NoStop}{lukens_using_2010}%
\bibitem[{\citenamefont{Lunasin}\ \emph{et~al.}(2012)\citenamefont{Lunasin},
  \citenamefont{Lin}, \citenamefont{Novikov}, \citenamefont{Mazzucato},\ and\
  \citenamefont{Doering}}]{lunasin2012}%
  \BibitemOpen
  \bibfield{author}{%
  \bibinfo {author} {\bibnamefont{Lunasin}, \bibfnamefont{E.}}, \bibinfo
  {author} {\bibfnamefont{Z.}~\bibnamefont{Lin}}, \bibinfo {author}
  {\bibfnamefont{A.}~\bibnamefont{Novikov}}, \bibinfo {author}
  {\bibfnamefont{A.}~\bibnamefont{Mazzucato}},\ and\ \bibinfo {author}
  {\bibfnamefont{C.~R.}\ \bibnamefont{Doering}}}%
  , \bibinfo {year} {2012},\ \bibfield{title}{%
  \enquote{\bibinfo {title} {Optimal mixing and optimal stirring for fixed
  energy, fixed power, or fixed palenstrophy flows},}\ }%
  \bibfield{journal}{%
  \bibinfo {journal} {J. Math. Phys.}\ }%
  \textbf{\bibinfo {volume} {53}},\ \bibinfo {pages} {115611}%
  \bibAnnoteFile{NoStop}{lunasin2012}%
\bibitem[{\citenamefont{MacKay}(1990)}]{mackay1990_knot}%
  \BibitemOpen
  \bibfield{author}{%
  \bibinfo {author} {\bibnamefont{MacKay}, \bibfnamefont{R.~S.}}}%
  , \bibinfo {year} {1990},\ \enquote{\bibinfo {title} {Postscript: Knot types
  for {3-D} vector fields},}\ in\ \emph{\bibinfo {booktitle} {Topological Fluid
  Mechanics}},\ \bibinfo {series and number} {IUTAM Conf. Proc.},\ \bibinfo
  {editor} {edited by\ \bibinfo {editor} {\bibfnamefont{Moffatt~H.}\
  \bibnamefont{K.}}\ and\ \bibinfo {editor} {\bibfnamefont{Tsinober}\
  \bibnamefont{A.}}}\ (\bibinfo {publisher} {Cambridge University Press})\ p.\
  \bibinfo {pages} {787}%
  \bibAnnoteFile{NoStop}{mackay1990_knot}%
\bibitem[{\citenamefont{MacKay}(1994)}]{Mackay1994}%
  \BibitemOpen
  \bibfield{author}{%
  \bibinfo {author} {\bibnamefont{MacKay}, \bibfnamefont{R.~S.}}}%
  , \bibinfo {year} {1994},\ \bibfield{title}{%
  \enquote{\bibinfo {title} {Transport in {3D} volume-preserving flows},}\ }%
  \bibfield{journal}{%
  \bibinfo {journal} {J. Nonlinear Sci.}\ }%
  \textbf{\bibinfo {volume} {4}},\ \bibinfo {pages} {329--354}%
  \bibAnnoteFile{NoStop}{Mackay1994}%
\bibitem[{\citenamefont{MacKay}(2001)}]{M01}%
  \BibitemOpen
  \bibfield{author}{%
  \bibinfo {author} {\bibnamefont{MacKay}, \bibfnamefont{R.~S.}}}%
  , \bibinfo {year} {2001},\ \bibfield{title}{%
  \enquote{\bibinfo {title} {Complicated dynamics from simple topological
  hypotheses},}\ }%
  \bibfield{journal}{%
  \bibinfo {journal} {Proc. Roy. Soc. A}\ }%
  \textbf{\bibinfo {volume} {359}},\ \bibinfo {pages} {1479--1496}%
  \bibAnnoteFile{NoStop}{M01}%
\bibitem[{\citenamefont{MacKay}(2006)}]{MacKay2006}%
  \BibitemOpen
  \bibfield{author}{%
  \bibinfo {author} {\bibnamefont{MacKay}, \bibfnamefont{R.~S.}}}%
  , \bibinfo {year} {2006},\ \bibfield{title}{%
  \enquote{\bibinfo {title} {{Cerbelli} and {Giona}'s map is pseudo-{Anosov}
  and nine consequences},}\ }%
  \bibfield{journal}{%
  \bibinfo {journal} {J Nonlin. Sci.}\ }%
  \textbf{\bibinfo {volume} {16}},\ \bibinfo {pages} {415--434}%
  \bibAnnoteFile{NoStop}{MacKay2006}%
\bibitem[{\citenamefont{MacKay}(2008)}]{MacKay_CCT2007}%
  \BibitemOpen
  \bibfield{author}{%
  \bibinfo {author} {\bibnamefont{MacKay}, \bibfnamefont{R.~S.}}}%
  , \bibinfo {year} {2008},\ \enquote{\bibinfo {title} {A steady mixing flow
  with no-slip boundaries},}\ in\ \emph{\bibinfo {booktitle} {Chaos,
  Complexity, and Transport: Theory and Applications}},\ \bibinfo {editor}
  {edited by\ \bibinfo {editor} {\bibfnamefont{C.}~\bibnamefont{Chandre}},
  \bibinfo {editor} {\bibfnamefont{X.}~\bibnamefont{Leoncini}},\ and\ \bibinfo
  {editor} {\bibfnamefont{G.}~\bibnamefont{Zaslavsky}}}\ (\bibinfo {publisher}
  {World Scientific},\ \bibinfo {address} {Singapore})\ pp.\ \bibinfo {pages}
  {55--68}%
  \bibAnnoteFile{NoStop}{MacKay_CCT2007}%
\bibitem[{\citenamefont{MacKay}\ and\ \citenamefont{Meiss}(1987)}]{Ham_chaos}%
  \BibitemOpen
  \bibfield{author}{%
  \bibinfo {author} {\bibnamefont{MacKay}, \bibfnamefont{R.~S.}},\ and\
  \bibinfo {author} {\bibfnamefont{J.~D.}\ \bibnamefont{Meiss}}}%
  , \bibinfo {year} {1987},\ \emph{\bibinfo {title} {Hamiltonian dynamical
  systems}}\ (\bibinfo {publisher} {Institute of Physics Publishing},\ \bibinfo
  {address} {London})%
  \bibAnnoteFile{NoStop}{Ham_chaos}%
\bibitem[{\citenamefont{MacKay}\ \emph{et~al.}(1984)\citenamefont{MacKay},
  \citenamefont{Meiss},\ and\ \citenamefont{Percival}}]{MacKay:1984bz}%
  \BibitemOpen
  \bibfield{author}{%
  \bibinfo {author} {\bibnamefont{MacKay}, \bibfnamefont{R.~S.}}, \bibinfo
  {author} {\bibfnamefont{J.~D.}\ \bibnamefont{Meiss}},\ and\ \bibinfo {author}
  {\bibfnamefont{I.~C.}\ \bibnamefont{Percival}}}%
  , \bibinfo {year} {1984},\ \bibfield{title}{%
  \enquote{\bibinfo {title} {Transport in {Hamiltonian} systems},}\ }%
  \bibfield{journal}{%
  \bibinfo {journal} {Physica D}\ }%
  \textbf{\bibinfo {volume} {13}},\ \bibinfo {pages} {55--81}%
  \bibAnnoteFile{NoStop}{MacKay:1984bz}%
\bibitem[{\citenamefont{Madrid}\ and\
  \citenamefont{Mancho}(2009)}]{madrid_distinguished_2009}%
  \BibitemOpen
  \bibfield{author}{%
  \bibinfo {author} {\bibnamefont{Madrid}, \bibfnamefont{J.~A.~Jim{\'e}nez}},\
  and\ \bibinfo {author} {\bibfnamefont{A.~M.}\ \bibnamefont{Mancho}}}%
  , \bibinfo {year} {2009},\ \bibfield{title}{%
  \enquote{\bibinfo {title} {Distinguished trajectories in time dependent
  vector fields},}\ }%
  \bibfield{journal}{%
  \bibinfo {journal} {Chaos}\ }%
  \textbf{\bibinfo {volume} {19}},\ \bibinfo {pages} {013111}%
  \bibAnnoteFile{NoStop}{madrid_distinguished_2009}%
\bibitem[{\citenamefont{Maiti}\ \emph{et~al.}(2013)\citenamefont{Maiti},
  \citenamefont{Chaudhury}, \citenamefont{DasGupta},\ and\
  \citenamefont{Chakraborty}}]{maiti2013}%
  \BibitemOpen
  \bibfield{author}{%
  \bibinfo {author} {\bibnamefont{Maiti}, \bibfnamefont{S.}}, \bibinfo {author}
  {\bibfnamefont{K.}~\bibnamefont{Chaudhury}}, \bibinfo {author}
  {\bibfnamefont{D.}~\bibnamefont{DasGupta}},\ and\ \bibinfo {author}
  {\bibfnamefont{S.}~\bibnamefont{Chakraborty}}}%
  , \bibinfo {year} {2013},\ \bibfield{title}{%
  \enquote{\bibinfo {title} {Alteration of chaotic advection in blood flow
  around particle blockage zone: role of hematocrit concentration},}\ }%
  \bibfield{journal}{%
  \bibinfo {journal} {J. Apply. Phys.}\ }%
  \textbf{\bibinfo {volume} {113}},\ \bibinfo {pages} {034701}%
  \bibAnnoteFile{NoStop}{maiti2013}%
\bibitem[{\citenamefont{Malhotra}\ \emph{et~al.}(1998)\citenamefont{Malhotra},
  \citenamefont{Mezi{\'c}},\ and\ \citenamefont{Wiggins}}]{Malhotra:1998tk}%
  \BibitemOpen
  \bibfield{author}{%
  \bibinfo {author} {\bibnamefont{Malhotra}, \bibfnamefont{N.}}, \bibinfo
  {author} {\bibfnamefont{I.}~\bibnamefont{Mezi{\'c}}},\ and\ \bibinfo {author}
  {\bibfnamefont{S.}~\bibnamefont{Wiggins}}}%
  , \bibinfo {year} {1998},\ \bibfield{title}{%
  \enquote{\bibinfo {title} {Patchiness: A new diagnostic for {Lagrangian}
  trajectory analysis in time-dependent fluid flows},}\ }%
  \bibfield{journal}{%
  \bibinfo {journal} {Int. J. Bifurcation and Chaos}\ }%
  \textbf{\bibinfo {volume} {8}},\ \bibinfo {pages} {1053--1093}%
  \bibAnnoteFile{NoStop}{Malhotra:1998tk}%
\bibitem[{\citenamefont{Malyuga}\ \emph{et~al.}(2002)\citenamefont{Malyuga},
  \citenamefont{Meleshko}, \citenamefont{Speetjens}, \citenamefont{Clercx},\
  and\ \citenamefont{van Heijst}}]{Malyuga2002}%
  \BibitemOpen
  \bibfield{author}{%
  \bibinfo {author} {\bibnamefont{Malyuga}, \bibfnamefont{V.~S.}}, \bibinfo
  {author} {\bibfnamefont{V.~V.}\ \bibnamefont{Meleshko}}, \bibinfo {author}
  {\bibfnamefont{M.~F.~M.}\ \bibnamefont{Speetjens}}, \bibinfo {author}
  {\bibfnamefont{H.~J.~H.}\ \bibnamefont{Clercx}},\ and\ \bibinfo {author}
  {\bibfnamefont{G.~J.~F.}\ \bibnamefont{van Heijst}}}%
  , \bibinfo {year} {2002},\ \bibfield{title}{%
  \enquote{\bibinfo {title} {Mixing in the {Stokes} flow in a cylindrical
  container},}\ }%
  \bibfield{journal}{%
  \bibinfo {journal} {Proc.\ R.\ Soc.\ A}\ }%
  \textbf{\bibinfo {volume} {458}},\ \bibinfo {pages} {1867--1885}%
  \bibAnnoteFile{NoStop}{Malyuga2002}%
\bibitem[{\citenamefont{Mancho}\ \emph{et~al.}(2013)\citenamefont{Mancho},
  \citenamefont{Wiggins}, \citenamefont{Curbelo},\ and\
  \citenamefont{Mendoza}}]{mancho_lagrangian_2013}%
  \BibitemOpen
  \bibfield{author}{%
  \bibinfo {author} {\bibnamefont{Mancho}, \bibfnamefont{A.~M.}}, \bibinfo
  {author} {\bibfnamefont{S.}~\bibnamefont{Wiggins}}, \bibinfo {author}
  {\bibfnamefont{J.}~\bibnamefont{Curbelo}},\ and\ \bibinfo {author}
  {\bibfnamefont{C.}~\bibnamefont{Mendoza}}}%
  , \bibinfo {year} {2013},\ \bibfield{title}{%
  \enquote{\bibinfo {title} {Lagrangian descriptors: A method for revealing
  phase space structures of general time dependent dynamical systems},}\ }%
  \bibfield{journal}{%
  \bibinfo {journal} {Communications in Nonlinear Science and Numerical
  Simulation}\ }%
  \textbf{\bibinfo {volume} {18}},\ \bibinfo {pages} {3530--3557}%
  \bibAnnoteFile{NoStop}{mancho_lagrangian_2013}%
\bibitem[{\citenamefont{Mann}\ and\
  \citenamefont{Lazier}(1991)}]{Mann-Lazier-91}%
  \BibitemOpen
  \bibfield{author}{%
  \bibinfo {author} {\bibnamefont{Mann}, \bibfnamefont{K.}},\ and\ \bibinfo
  {author} {\bibfnamefont{J.}~\bibnamefont{Lazier}}}%
  , \bibinfo {year} {1991},\ \emph{\bibinfo {title} {Dynamics of marine
  ecosystems, Biological-physical interactions in the oceans}}\ (\bibinfo
  {publisher} {Blackwell Scientific Publications},\ \bibinfo {address}
  {Boston})%
  \bibAnnoteFile{NoStop}{Mann-Lazier-91}%
\bibitem[{\citenamefont{Marshall}\ and\
  \citenamefont{Nonaka}(2006)}]{Marshall2006}%
  \BibitemOpen
  \bibfield{author}{%
  \bibinfo {author} {\bibnamefont{Marshall}, \bibfnamefont{W.~F.}},\ and\
  \bibinfo {author} {\bibfnamefont{S.}~\bibnamefont{Nonaka}}}%
  , \bibinfo {year} {2006},\ \bibfield{title}{%
  \enquote{\bibinfo {title} {{Cilia: tuning in to the cell's antenna}},}\ }%
  \bibfield{journal}{%
  \bibinfo {journal} {Curr. Biol.}\ }%
  \textbf{\bibinfo {volume} {16}},\ \bibinfo {pages} {R604--614}%
  \bibAnnoteFile{NoStop}{Marshall2006}%
\bibitem[{\citenamefont{Martin}(2003)}]{Martin-03}%
  \BibitemOpen
  \bibfield{author}{%
  \bibinfo {author} {\bibnamefont{Martin}, \bibfnamefont{A.}}}%
  , \bibinfo {year} {2003},\ \bibfield{title}{%
  \enquote{\bibinfo {title} {Phytoplankton patchiness: the role of lateral
  stirring and mixing},}\ }%
  \bibfield{journal}{%
  \bibinfo {journal} {Prog. Oceanogr.}\ }%
  \textbf{\bibinfo {volume} {57}},\ \bibinfo {pages} {125--174}%
  \bibAnnoteFile{NoStop}{Martin-03}%
\bibitem[{\citenamefont{Martin}\ \emph{et~al.}(2002)\citenamefont{Martin},
  \citenamefont{Richards}, \citenamefont{Bracco},\ and\
  \citenamefont{Provenzale}}]{Martin-et-al-02}%
  \BibitemOpen
  \bibfield{author}{%
  \bibinfo {author} {\bibnamefont{Martin}, \bibfnamefont{A.}}, \bibinfo
  {author} {\bibfnamefont{K.}~\bibnamefont{Richards}}, \bibinfo {author}
  {\bibfnamefont{A.}~\bibnamefont{Bracco}},\ and\ \bibinfo {author}
  {\bibfnamefont{A.}~\bibnamefont{Provenzale}}}%
  , \bibinfo {year} {2002},\ \bibfield{title}{%
  \enquote{\bibinfo {title} {Patchy productivity in the open ocean},}\ }%
  \bibfield{journal}{%
  \bibinfo {journal} {Global Biogeochem. Cy.}\ }%
  \textbf{\bibinfo {volume} {16}},\ \bibinfo {pages} {1025}%
  \bibAnnoteFile{NoStop}{Martin-et-al-02}%
\bibitem[{\citenamefont{Mathew}\ \emph{et~al.}(2007)\citenamefont{Mathew},
  \citenamefont{Mezi{\'c}}, \citenamefont{Grivopoulos}, \citenamefont{Vaidya},\
  and\ \citenamefont{Petzold}}]{Mathew2007}%
  \BibitemOpen
  \bibfield{author}{%
  \bibinfo {author} {\bibnamefont{Mathew}, \bibfnamefont{G.}}, \bibinfo
  {author} {\bibfnamefont{I.}~\bibnamefont{Mezi{\'c}}}, \bibinfo {author}
  {\bibfnamefont{S.}~\bibnamefont{Grivopoulos}}, \bibinfo {author}
  {\bibfnamefont{U.}~\bibnamefont{Vaidya}},\ and\ \bibinfo {author}
  {\bibfnamefont{L.}~\bibnamefont{Petzold}}}%
  , \bibinfo {year} {2007},\ \bibfield{title}{%
  \enquote{\bibinfo {title} {Optimal control of mixing in {S}tokes fluid
  flows},}\ }%
  \bibfield{journal}{%
  \bibinfo {journal} {J. Fluid Mech.}\ }%
  \textbf{\bibinfo {volume} {580}},\ \bibinfo {pages} {261--281}%
  \bibAnnoteFile{NoStop}{Mathew2007}%
\bibitem[{\citenamefont{Mathew}\ \emph{et~al.}(2005)\citenamefont{Mathew},
  \citenamefont{Mezi{\'c}},\ and\ \citenamefont{Petzold}}]{Mathew2005}%
  \BibitemOpen
  \bibfield{author}{%
  \bibinfo {author} {\bibnamefont{Mathew}, \bibfnamefont{G.}}, \bibinfo
  {author} {\bibfnamefont{I.}~\bibnamefont{Mezi{\'c}}},\ and\ \bibinfo {author}
  {\bibfnamefont{L.}~\bibnamefont{Petzold}}}%
  , \bibinfo {year} {2005},\ \bibfield{title}{%
  \enquote{\bibinfo {title} {A multiscale measure for mixing},}\ }%
  \bibfield{journal}{%
  \bibinfo {journal} {Physica D}\ }%
  \textbf{\bibinfo {volume} {211}},\ \bibinfo {pages} {23--46}%
  \bibAnnoteFile{NoStop}{Mathew2005}%
\bibitem[{\citenamefont{Mathew}\ \emph{et~al.}(2004)\citenamefont{Mathew},
  \citenamefont{Mezi\'{c}}, \citenamefont{Serban},\ and\
  \citenamefont{Petzold}}]{Mathew2004optimization}%
  \BibitemOpen
  \bibfield{author}{%
  \bibinfo {author} {\bibnamefont{Mathew}, \bibfnamefont{G.}}, \bibinfo
  {author} {\bibfnamefont{I.}~\bibnamefont{Mezi\'{c}}}, \bibinfo {author}
  {\bibfnamefont{R.}~\bibnamefont{Serban}},\ and\ \bibinfo {author}
  {\bibfnamefont{L.}~\bibnamefont{Petzold}}}%
  , \bibinfo {year} {2004},\ \enquote{\bibinfo {title} {Optimization of mixing
  in an active micromixing device},}\ in\ \emph{\bibinfo {booktitle} {Technical
  Proc. 2004 NSTI Nanotechnology Conf.\ Trade Show, Boston, MA}},\
  Vol.~\bibinfo {volume} {1},\ pp.\ \bibinfo {pages} {300--303}%
  \bibAnnoteFile{NoStop}{Mathew2004optimization}%
\bibitem[{\citenamefont{Meier}\ \emph{et~al.}(2007)\citenamefont{Meier},
  \citenamefont{Lueptow},\ and\ \citenamefont{Ottino}}]{meier07}%
  \BibitemOpen
  \bibfield{author}{%
  \bibinfo {author} {\bibnamefont{Meier}, \bibfnamefont{S.~W.}}, \bibinfo
  {author} {\bibfnamefont{R.~M.}\ \bibnamefont{Lueptow}},\ and\ \bibinfo
  {author} {\bibfnamefont{J.~M.}\ \bibnamefont{Ottino}}}%
  , \bibinfo {year} {2007},\ \bibfield{title}{%
  \enquote{\bibinfo {title} {A dynamical systems approach to mixing and
  segregation of granular materials in tumblers},}\ }%
  \bibfield{journal}{%
  \bibinfo {journal} {Adv. Phys.}\ }%
  \textbf{\bibinfo {volume} {56}},\ \bibinfo {pages} {757--827}%
  \bibAnnoteFile{NoStop}{meier07}%
\bibitem[{\citenamefont{Meiss}(2012)}]{Meiss2012}%
  \BibitemOpen
  \bibfield{author}{%
  \bibinfo {author} {\bibnamefont{Meiss}, \bibfnamefont{J.~D.}}}%
  , \bibinfo {year} {2012},\ \bibfield{title}{%
  \enquote{\bibinfo {title} {The destruction of tori in volume-preserving
  maps},}\ }%
  \bibfield{journal}{%
  \bibinfo {journal} {Comm. Nonlin. Sci. Num. Sim.}\ }%
  \textbf{\bibinfo {volume} {17}},\ \bibinfo {pages} {2108--2121}%
  \bibAnnoteFile{NoStop}{Meiss2012}%
\bibitem[{\citenamefont{Meiss}\ and\ \citenamefont{Ott}(1985)}]{Meiss1985}%
  \BibitemOpen
  \bibfield{author}{%
  \bibinfo {author} {\bibnamefont{Meiss}, \bibfnamefont{J.~D.}},\ and\ \bibinfo
  {author} {\bibfnamefont{E.}~\bibnamefont{Ott}}}%
  , \bibinfo {year} {1985},\ \bibfield{title}{%
  \enquote{\bibinfo {title} {Markov-tree model of intrinsic transport in
  {Hamiltonian} systems},}\ }%
  \bibfield{journal}{%
  \bibinfo {journal} {Phys. Rev. Lett.}\ }%
  \textbf{\bibinfo {volume} {55}},\ \bibinfo {pages} {2741--2744}%
  \bibAnnoteFile{NoStop}{Meiss1985}%
\bibitem[{\citenamefont{Meleshko}\ and\
  \citenamefont{Aref.}(1996)}]{meleshko1996_3}%
  \BibitemOpen
  \bibfield{author}{%
  \bibinfo {author} {\bibnamefont{Meleshko}, \bibfnamefont{V.~V.}},\ and\
  \bibinfo {author} {\bibfnamefont{H.}~\bibnamefont{Aref.}}}%
  , \bibinfo {year} {1996},\ \bibfield{title}{%
  \enquote{\bibinfo {title} {A blinking rotlet model for chaotic advection},}\
  }%
  \bibfield{journal}{%
  \bibinfo {journal} {Phys. Fluids}\ }%
  \textbf{\bibinfo {volume} {A8}},\ \bibinfo {pages} {3215--3217}%
  \bibAnnoteFile{NoStop}{meleshko1996_3}%
\bibitem[{\citenamefont{Meleshko}\ \emph{et~al.}(1996)\citenamefont{Meleshko},
  \citenamefont{Krasnopolskaya}, \citenamefont{Peters},\ and\
  \citenamefont{Meijer}}]{Meleshko1996_2}%
  \BibitemOpen
  \bibfield{author}{%
  \bibinfo {author} {\bibnamefont{Meleshko}, \bibfnamefont{V.~V.}}, \bibinfo
  {author} {\bibfnamefont{T.~S.}\ \bibnamefont{Krasnopolskaya}}, \bibinfo
  {author} {\bibfnamefont{G.~W.~M.}\ \bibnamefont{Peters}},\ and\ \bibinfo
  {author} {\bibfnamefont{H.~E.~H.}\ \bibnamefont{Meijer}}}%
  , \bibinfo {year} {1996},\ \enquote{\bibinfo {title} {Coherent structures and
  scales of {Lagrangian} turbulence},}\ in\ \emph{\bibinfo {booktitle}
  {Advances in Turbulence VI}},\ \bibinfo {editor} {edited by\ \bibinfo
  {editor} {\bibfnamefont{S.}~\bibnamefont{Gavrilakis}}, \bibinfo {editor}
  {\bibfnamefont{L.}~\bibnamefont{Machiels}},\ and\ \bibinfo {editor}
  {\bibfnamefont{A.~Monkewitz}\ \bibnamefont{P.}}}\ (\bibinfo {publisher}
  {Kluwer, Dordrecht})\ pp.\ \bibinfo {pages} {601--604}%
  \bibAnnoteFile{NoStop}{Meleshko1996_2}%
\bibitem[{\citenamefont{Meleshko}\ and\
  \citenamefont{Peters}(1996)}]{Meleshko1996}%
  \BibitemOpen
  \bibfield{author}{%
  \bibinfo {author} {\bibnamefont{Meleshko}, \bibfnamefont{V.~V.}},\ and\
  \bibinfo {author} {\bibfnamefont{G.~W.~M.}\ \bibnamefont{Peters}}}%
  , \bibinfo {year} {1996},\ \bibfield{title}{%
  \enquote{\bibinfo {title} {Periodic points for two-dimensional {Stokes} flow
  in a rectangular cavity},}\ }%
  \bibfield{journal}{%
  \bibinfo {journal} {Phys. Lett. A}\ }%
  \textbf{\bibinfo {volume} {216}},\ \bibinfo {pages} {87--96}%
  \bibAnnoteFile{NoStop}{Meleshko1996}%
\bibitem[{\citenamefont{Metcalfe}(2010)}]{Metcalfe_chaos_2010}%
  \BibitemOpen
  \bibfield{author}{%
  \bibinfo {author} {\bibnamefont{Metcalfe}, \bibfnamefont{G.}}}%
  , \bibinfo {year} {2010},\ \enquote{\bibinfo {title} {Complex physical,
  biophysical, and econophysical systems},}\ Chap.\ \bibinfo {chapter} {Applied
  Fluid Chaos: {D}esigning Advection with Periodically Reoriented Flows for
  Micro to Geophysical Mixing and Transport Enhancement}\ (\bibinfo {publisher}
  {World Scientific})\ pp.\ \bibinfo {pages} {189--242}%
  \bibAnnoteFile{NoStop}{Metcalfe_chaos_2010}%
\bibitem[{\citenamefont{Metcalfe}\
  \emph{et~al.}(2010{\natexlab{a}})\citenamefont{Metcalfe},
  \citenamefont{Lester}, \citenamefont{Ord}, \citenamefont{Kulkarni},
  \citenamefont{Rudman}, \citenamefont{Trefry}, \citenamefont{Hobbs},
  \citenamefont{Regenaur-Lieb},\ and\
  \citenamefont{Morris}}]{Metcalfe_ECC_2010}%
  \BibitemOpen
  \bibfield{author}{%
  \bibinfo {author} {\bibnamefont{Metcalfe}, \bibfnamefont{G.}}, \bibinfo
  {author} {\bibfnamefont{D.}~\bibnamefont{Lester}}, \bibinfo {author}
  {\bibfnamefont{A.}~\bibnamefont{Ord}}, \bibinfo {author}
  {\bibfnamefont{P.}~\bibnamefont{Kulkarni}}, \bibinfo {author}
  {\bibfnamefont{M.}~\bibnamefont{Rudman}}, \bibinfo {author}
  {\bibfnamefont{M.}~\bibnamefont{Trefry}}, \bibinfo {author}
  {\bibfnamefont{B.}~\bibnamefont{Hobbs}}, \bibinfo {author}
  {\bibfnamefont{K.}~\bibnamefont{Regenaur-Lieb}},\ and\ \bibinfo {author}
  {\bibfnamefont{J.}~\bibnamefont{Morris}}}%
  , \bibinfo {year} {2010}{\natexlab{a}},\ \bibfield{title}{%
  \enquote{\bibinfo {title} {An experimental and theoretical study of the
  mixing characteristics of a periodically reoriented irrotational flow},}\ }%
  \bibfield{journal}{%
  \bibinfo {journal} {Phil. Trans. Roy. Soc. A}\ }%
  \textbf{\bibinfo {volume} {368}},\ \bibinfo {pages} {2147--2162}%
  \bibAnnoteFile{NoStop}{Metcalfe_ECC_2010}%
\bibitem[{\citenamefont{Metcalfe}\
  \emph{et~al.}(2010{\natexlab{b}})\citenamefont{Metcalfe},
  \citenamefont{Lester}, \citenamefont{Ord}, \citenamefont{Kulkarni},
  \citenamefont{Trefry}, \citenamefont{Hobbs}, \citenamefont{Regenaur-Lieb},\
  and\ \citenamefont{Morris}}]{Metcalfe_PiP_2010}%
  \BibitemOpen
  \bibfield{author}{%
  \bibinfo {author} {\bibnamefont{Metcalfe}, \bibfnamefont{G.}}, \bibinfo
  {author} {\bibfnamefont{D.}~\bibnamefont{Lester}}, \bibinfo {author}
  {\bibfnamefont{A.}~\bibnamefont{Ord}}, \bibinfo {author}
  {\bibfnamefont{P.}~\bibnamefont{Kulkarni}}, \bibinfo {author}
  {\bibfnamefont{M.}~\bibnamefont{Trefry}}, \bibinfo {author}
  {\bibfnamefont{B.}~\bibnamefont{Hobbs}}, \bibinfo {author}
  {\bibfnamefont{K.}~\bibnamefont{Regenaur-Lieb}},\ and\ \bibinfo {author}
  {\bibfnamefont{J.}~\bibnamefont{Morris}}}%
  , \bibinfo {year} {2010}{\natexlab{b}},\ \bibfield{title}{%
  \enquote{\bibinfo {title} {A partially open porous media flow with chaotic
  advection: {T}owards a model of coupled fields},}\ }%
  \bibfield{journal}{%
  \bibinfo {journal} {Phil. Trans. Roy. Soc. A}\ }%
  \textbf{\bibinfo {volume} {368}},\ \bibinfo {pages} {217--230}%
  \bibAnnoteFile{NoStop}{Metcalfe_PiP_2010}%
\bibitem[{\citenamefont{Metcalfe}\ and\
  \citenamefont{Lester}(2009)}]{Metcalfe_foodRAM_2009}%
  \BibitemOpen
  \bibfield{author}{%
  \bibinfo {author} {\bibnamefont{Metcalfe}, \bibfnamefont{G.}},\ and\ \bibinfo
  {author} {\bibfnamefont{D.~R.}\ \bibnamefont{Lester}}}%
  , \bibinfo {year} {2009},\ \bibfield{title}{%
  \enquote{\bibinfo {title} {Mixing and heat transfer of highly viscous food
  products with a continuous chaotic duct flow},}\ }%
  \bibfield{journal}{%
  \bibinfo {journal} {J. Food Eng.}\ }%
  \textbf{\bibinfo {volume} {95}},\ \bibinfo {pages} {21--29}%
  \bibAnnoteFile{NoStop}{Metcalfe_foodRAM_2009}%
\bibitem[{\citenamefont{Metcalfe}\ and\
  \citenamefont{Ottino}(1994)}]{Metcalfe_autocatalytic_1994}%
  \BibitemOpen
  \bibfield{author}{%
  \bibinfo {author} {\bibnamefont{Metcalfe}, \bibfnamefont{G.}},\ and\ \bibinfo
  {author} {\bibfnamefont{J.~M.}\ \bibnamefont{Ottino}}}%
  , \bibinfo {year} {1994},\ \bibfield{title}{%
  \enquote{\bibinfo {title} {Autocatalytic processes in mixing flows},}\ }%
  \bibfield{journal}{%
  \bibinfo {journal} {Phys. Rev. Lett.}\ }%
  \textbf{\bibinfo {volume} {72}},\ \bibinfo {pages} {2875--2878},\ \bibinfo
  {note} {erratum Phys. Rev. Lett. {\bf 73} 212 (1994).}%
  \bibAnnoteFile{Stop}{Metcalfe_autocatalytic_1994}%
\bibitem[{\citenamefont{Metcalfe}\ \emph{et~al.}(2006)\citenamefont{Metcalfe},
  \citenamefont{Rudman}, \citenamefont{Brydon}, \citenamefont{Graham},\ and\
  \citenamefont{Hamilton}}]{Metcalfe_ram_2006}%
  \BibitemOpen
  \bibfield{author}{%
  \bibinfo {author} {\bibnamefont{Metcalfe}, \bibfnamefont{G.}}, \bibinfo
  {author} {\bibfnamefont{M.}~\bibnamefont{Rudman}}, \bibinfo {author}
  {\bibfnamefont{A.}~\bibnamefont{Brydon}}, \bibinfo {author}
  {\bibfnamefont{L.~J.~W.}\ \bibnamefont{Graham}},\ and\ \bibinfo {author}
  {\bibfnamefont{R.}~\bibnamefont{Hamilton}}}%
  , \bibinfo {year} {2006},\ \bibfield{title}{%
  \enquote{\bibinfo {title} {Composing chaos: An experimental and numerical
  study of an open duct mixing flow},}\ }%
  \bibfield{journal}{%
  \bibinfo {journal} {Am. Inst. Chem. Eng. J.}\ }%
  \textbf{\bibinfo {volume} {52}},\ \bibinfo {pages} {9--28}%
  \bibAnnoteFile{NoStop}{Metcalfe_ram_2006}%
\bibitem[{\citenamefont{Metcalfe}\ \emph{et~al.}(2012)\citenamefont{Metcalfe},
  \citenamefont{Speetjens}, \citenamefont{Lester},\ and\
  \citenamefont{Clercx}}]{Metcalfe_beyond_2012}%
  \BibitemOpen
  \bibfield{author}{%
  \bibinfo {author} {\bibnamefont{Metcalfe}, \bibfnamefont{G.}}, \bibinfo
  {author} {\bibfnamefont{M.~F.~M.}\ \bibnamefont{Speetjens}}, \bibinfo
  {author} {\bibfnamefont{D.~R.}\ \bibnamefont{Lester}},\ and\ \bibinfo
  {author} {\bibfnamefont{H.~J.~H.}\ \bibnamefont{Clercx}}}%
  , \bibinfo {year} {2012},\ \bibfield{title}{%
  \enquote{\bibinfo {title} {Beyond passive: {C}haotic transport in stirred
  fluids},}\ }%
  \bibfield{journal}{%
  \bibinfo {journal} {Adv. Appl. Mech.}\ }%
  \textbf{\bibinfo {volume} {45}},\ \bibinfo {pages} {109--188}%
  \bibAnnoteFile{NoStop}{Metcalfe_beyond_2012}%
\bibitem[{\citenamefont{Mezi\'{c}}(1994)}]{mezic1994thesis}%
  \BibitemOpen
  \bibfield{author}{%
  \bibinfo {author} {\bibnamefont{Mezi\'{c}}, \bibfnamefont{I.}}}%
  , \bibinfo {year} {1994},\ \emph{\bibinfo {title} {On the geometrical and
  statistical properties of dynamical systems: theory and applications}},\
  Ph.D. thesis\ (\bibinfo {school} {California Institute of Technology})%
  \bibAnnoteFile{NoStop}{mezic1994thesis}%
\bibitem[{\citenamefont{Mezi\'{c}}(2001{\natexlab{a}})}]{Mezic2001}%
  \BibitemOpen
  \bibfield{author}{%
  \bibinfo {author} {\bibnamefont{Mezi\'{c}}, \bibfnamefont{I.}}}%
  , \bibinfo {year} {2001}{\natexlab{a}},\ \bibfield{title}{%
  \enquote{\bibinfo {title} {Break-up of invariant surfaces in
  action-angle-angle maps and flows},}\ }%
  \bibfield{journal}{%
  \bibinfo {journal} {Physica D}\ }%
  \textbf{\bibinfo {volume} {154}},\ \bibinfo {pages} {51}%
  \bibAnnoteFile{NoStop}{Mezic2001}%
\bibitem[{\citenamefont{Mezi\'{c}}(2001{\natexlab{b}})}]{mezic2001chaotic}%
  \BibitemOpen
  \bibfield{author}{%
  \bibinfo {author} {\bibnamefont{Mezi\'{c}}, \bibfnamefont{I.}}}%
  , \bibinfo {year} {2001}{\natexlab{b}},\ \bibfield{title}{%
  \enquote{\bibinfo {title} {Chaotic advection in bounded navier--stokes
  flows},}\ }%
  \bibfield{journal}{%
  \bibinfo {journal} {J. Fluid Mech.}\ }%
  \textbf{\bibinfo {volume} {431}},\ \bibinfo {pages} {347--370}%
  \bibAnnoteFile{NoStop}{mezic2001chaotic}%
\bibitem[{\citenamefont{Mezi{\'c}}(2002)}]{mezic2002extension}%
  \BibitemOpen
  \bibfield{author}{%
  \bibinfo {author} {\bibnamefont{Mezi{\'c}}, \bibfnamefont{I.}}}%
  , \bibinfo {year} {2002},\ \bibfield{title}{%
  \enquote{\bibinfo {title} {An extension of {Prandtl--Batchelor} theory and
  consequences for chaotic advection},}\ }%
  \bibfield{journal}{%
  \bibinfo {journal} {Phys. Fluids}\ }%
  \textbf{\bibinfo {volume} {14}},\ \bibinfo {pages} {L61--L64}%
  \bibAnnoteFile{NoStop}{mezic2002extension}%
\bibitem[{\citenamefont{Mezi\'{c}}(2013)}]{mezic2013analysis}%
  \BibitemOpen
  \bibfield{author}{%
  \bibinfo {author} {\bibnamefont{Mezi\'{c}}, \bibfnamefont{I.}}}%
  , \bibinfo {year} {2013},\ \bibfield{title}{%
  \enquote{\bibinfo {title} {Analysis of fluid flows via spectral properties of
  the {Koopman} operator},}\ }%
  \bibfield{journal}{%
  \bibinfo {journal} {Annu. Rev. Fluid Mech.}\ }%
  \textbf{\bibinfo {volume} {45}},\ \bibinfo {pages} {357--378}%
  \bibAnnoteFile{NoStop}{mezic2013analysis}%
\bibitem[{\citenamefont{Mezi{\'c}}\
  \emph{et~al.}(2010)\citenamefont{Mezi{\'c}}, \citenamefont{Loire},
  \citenamefont{Fonoberov},\ and\ \citenamefont{Hogan}}]{Mezic:2010kh}%
  \BibitemOpen
  \bibfield{author}{%
  \bibinfo {author} {\bibnamefont{Mezi{\'c}}, \bibfnamefont{I.}}, \bibinfo
  {author} {\bibfnamefont{S.}~\bibnamefont{Loire}}, \bibinfo {author}
  {\bibfnamefont{V.~A.}\ \bibnamefont{Fonoberov}},\ and\ \bibinfo {author}
  {\bibfnamefont{P.~J.}\ \bibnamefont{Hogan}}}%
  , \bibinfo {year} {2010},\ \bibfield{title}{%
  \enquote{\bibinfo {title} {A new mixing diagnostic and {Gulf} oil spill
  movement},}\ }%
  \bibfield{journal}{%
  \bibinfo {journal} {Science}\ }%
  \textbf{\bibinfo {volume} {330}},\ \bibinfo {pages} {486--489}%
  \bibAnnoteFile{NoStop}{Mezic:2010kh}%
\bibitem[{\citenamefont{Mezi{\'c}}\ and\
  \citenamefont{Sotiropoulos}(2002)}]{mezic2002ergodic}%
  \BibitemOpen
  \bibfield{author}{%
  \bibinfo {author} {\bibnamefont{Mezi{\'c}}, \bibfnamefont{I.}},\ and\
  \bibinfo {author} {\bibfnamefont{F.}~\bibnamefont{Sotiropoulos}}}%
  , \bibinfo {year} {2002},\ \bibfield{title}{%
  \enquote{\bibinfo {title} {Ergodic theory and experimental visualization of
  invariant sets in chaotically advected flows},}\ }%
  \bibfield{journal}{%
  \bibinfo {journal} {Phys. Fluids}\ }%
  \textbf{\bibinfo {volume} {14}},\ \bibinfo {pages} {2235--2243}%
  \bibAnnoteFile{NoStop}{mezic2002ergodic}%
\bibitem[{\citenamefont{Mezi\'{c}}\ and\
  \citenamefont{Wiggins}(1994)}]{Mezic1994}%
  \BibitemOpen
  \bibfield{author}{%
  \bibinfo {author} {\bibnamefont{Mezi\'{c}}, \bibfnamefont{I.}},\ and\
  \bibinfo {author} {\bibfnamefont{S.}~\bibnamefont{Wiggins}}}%
  , \bibinfo {year} {1994},\ \bibfield{title}{%
  \enquote{\bibinfo {title} {On the integrability and perturbation of
  three-dimensional fluid flows with symmetry},}\ }%
  \bibfield{journal}{%
  \bibinfo {journal} {J. Nonlinear Sci.}\ }%
  \textbf{\bibinfo {volume} {4}},\ \bibinfo {pages} {157}%
  \bibAnnoteFile{NoStop}{Mezic1994}%
\bibitem[{\citenamefont{Mezi\'{c}}\ and\
  \citenamefont{Wiggins}(1999)}]{Mezic1999}%
  \BibitemOpen
  \bibfield{author}{%
  \bibinfo {author} {\bibnamefont{Mezi\'{c}}, \bibfnamefont{I.}},\ and\
  \bibinfo {author} {\bibfnamefont{S.}~\bibnamefont{Wiggins}}}%
  , \bibinfo {year} {1999},\ \bibfield{title}{%
  \enquote{\bibinfo {title} {A method for visualisation of invariant sets of
  dynamical systems based on ergodic partition},}\ }%
  \bibfield{journal}{%
  \bibinfo {journal} {Chaos}\ }%
  \textbf{\bibinfo {volume} {9}},\ \bibinfo {pages} {213}%
  \bibAnnoteFile{NoStop}{Mezic1999}%
\bibitem[{\citenamefont{Mezi\'{c}}\
  \emph{et~al.}(1999)\citenamefont{Mezi\'{c}}, \citenamefont{Wiggins},\ and\
  \citenamefont{Betz}}]{mezic1999_2}%
  \BibitemOpen
  \bibfield{author}{%
  \bibinfo {author} {\bibnamefont{Mezi\'{c}}, \bibfnamefont{I.}}, \bibinfo
  {author} {\bibfnamefont{S.}~\bibnamefont{Wiggins}},\ and\ \bibinfo {author}
  {\bibfnamefont{D.}~\bibnamefont{Betz}}}%
  , \bibinfo {year} {1999},\ \bibfield{title}{%
  \enquote{\bibinfo {title} {Residence-time distributions for chaotic flows in
  pipes},}\ }%
  \bibfield{journal}{%
  \bibinfo {journal} {Chaos}\ }%
  \textbf{\bibinfo {volume} {9}},\ \bibinfo {pages} {173--182}%
  \bibAnnoteFile{NoStop}{mezic1999_2}%
\bibitem[{\citenamefont{Moffatt}(2000)}]{Moffatt2000}%
  \BibitemOpen
  \bibfield{author}{%
  \bibinfo {author} {\bibnamefont{Moffatt}, \bibfnamefont{H.~K.}}}%
  , \bibinfo {year} {2000},\ \enquote{\bibinfo {title} {Vortex and
  magneto-dynamics --- a topological perspective},}\ in\ \emph{\bibinfo
  {booktitle} {Mathematical Physics 2000}},\ \bibinfo {editor} {edited by\
  \bibinfo {editor} {\bibfnamefont{A}~\bibnamefont{Fokas}}, \bibinfo {editor}
  {\bibfnamefont{A.}~\bibnamefont{Grigoryan}}, \bibinfo {editor}
  {\bibfnamefont{T.}~\bibnamefont{Kibble}},\ and\ \bibinfo {editor}
  {\bibfnamefont{B.}~\bibnamefont{Zegarlinski}}}\ (\bibinfo {publisher}
  {Imperial College Press})\ pp.\ \bibinfo {pages} {170--182}%
  \bibAnnoteFile{NoStop}{Moffatt2000}%
\bibitem[{\citenamefont{Moffatt}\ \emph{et~al.}(1992)\citenamefont{Moffatt},
  \citenamefont{Zaslavsky}, \citenamefont{Comte},\ and\
  \citenamefont{Tabor}}]{Moffatt1992}%
  \BibitemOpen
  \bibfield{author}{%
  \bibinfo {author} {\bibnamefont{Moffatt}, \bibfnamefont{H.~K.}}, \bibinfo
  {author} {\bibfnamefont{G.~M.}\ \bibnamefont{Zaslavsky}}, \bibinfo {author}
  {\bibfnamefont{P.}~\bibnamefont{Comte}},\ and\ \bibinfo {author}
  {\bibfnamefont{M.}~\bibnamefont{Tabor}}}%
  , \bibinfo {year} {1992},\ \emph{\bibinfo {title} {Topological Aspects of the
  Dynamics of Fluids and Plasmas}}\ (\bibinfo {publisher} {Kluwer Academic
  Publishers},\ \bibinfo {address} {Dordrecht})%
  \bibAnnoteFile{NoStop}{Moffatt1992}%
\bibitem[{\citenamefont{Moharana}\ \emph{et~al.}(2013)\citenamefont{Moharana},
  \citenamefont{Speetjens}, \citenamefont{Trieling},\ and\
  \citenamefont{Clercx}}]{Moharana2013}%
  \BibitemOpen
  \bibfield{author}{%
  \bibinfo {author} {\bibnamefont{Moharana}, \bibfnamefont{N.~R.}}, \bibinfo
  {author} {\bibfnamefont{M.~F.~M.}\ \bibnamefont{Speetjens}}, \bibinfo
  {author} {\bibfnamefont{R.~R.}\ \bibnamefont{Trieling}},\ and\ \bibinfo
  {author} {\bibfnamefont{H.~J.~H.}\ \bibnamefont{Clercx}}}%
  , \bibinfo {year} {2013},\ \bibfield{title}{%
  \enquote{\bibinfo {title} {Three-dimensional lagrangian transport phenomena
  in unsteady laminar flows driven by a rotating sphere},}\ }%
  \bibfield{journal}{%
  \bibinfo {journal} {Phys. Fluids}\ }%
  \textbf{\bibinfo {volume} {25}},\ \bibinfo {pages} {093602}%
  \bibAnnoteFile{NoStop}{Moharana2013}%
\bibitem[{\citenamefont{Montenegro-Johnson}\
  \emph{et~al.}(2012)\citenamefont{Montenegro-Johnson}, \citenamefont{Smith},
  \citenamefont{Smith}, \citenamefont{Loghin},\ and\
  \citenamefont{Blake}}]{Montenegro-Johnson12}%
  \BibitemOpen
  \bibfield{author}{%
  \bibinfo {author} {\bibnamefont{Montenegro-Johnson}, \bibfnamefont{T.~D.}},
  \bibinfo {author} {\bibfnamefont{A.~A.}\ \bibnamefont{Smith}}, \bibinfo
  {author} {\bibfnamefont{D.~J.}\ \bibnamefont{Smith}}, \bibinfo {author}
  {\bibfnamefont{D.}~\bibnamefont{Loghin}},\ and\ \bibinfo {author}
  {\bibfnamefont{J.~R.}\ \bibnamefont{Blake}}}%
  , \bibinfo {year} {2012},\ \bibfield{title}{%
  \enquote{\bibinfo {title} {Modelling the fluid mechanics of cilia and
  flagella in reproduction and development},}\ }%
  \bibfield{journal}{%
  \bibinfo {journal} {Eur. Phys. J. E}\ }%
  \textbf{\bibinfo {volume} {35}},\ \bibinfo {pages} {1--17}%
  \bibAnnoteFile{NoStop}{Montenegro-Johnson12}%
\bibitem[{\citenamefont{Motter}\ \emph{et~al.}(2003)\citenamefont{Motter},
  \citenamefont{Lai},\ and\ \citenamefont{Grebogi}}]{Motter2003}%
  \BibitemOpen
  \bibfield{author}{%
  \bibinfo {author} {\bibnamefont{Motter}, \bibfnamefont{A.~E.}}, \bibinfo
  {author} {\bibfnamefont{Y.-C.}\ \bibnamefont{Lai}},\ and\ \bibinfo {author}
  {\bibfnamefont{C.}~\bibnamefont{Grebogi}}}%
  , \bibinfo {year} {2003},\ \bibfield{title}{%
  \enquote{\bibinfo {title} {Reactive dynamics of inertial particles in
  nonhyperbolic chaotic flows},}\ }%
  \bibfield{journal}{%
  \bibinfo {journal} {Phys. Rev. E}\ }%
  \textbf{\bibinfo {volume} {68}},\ \bibinfo {pages} {056307}%
  \bibAnnoteFile{NoStop}{Motter2003}%
\bibitem[{\citenamefont{Motter}\ \emph{et~al.}(2005)\citenamefont{Motter},
  \citenamefont{de~Moura}, \citenamefont{Grebogi},\ and\
  \citenamefont{Kantz}}]{Motter2005}%
  \BibitemOpen
  \bibfield{author}{%
  \bibinfo {author} {\bibnamefont{Motter}, \bibfnamefont{A.~E}}, \bibinfo
  {author} {\bibfnamefont{A.~P.~S.}\ \bibnamefont{de~Moura}}, \bibinfo {author}
  {\bibfnamefont{C.}~\bibnamefont{Grebogi}},\ and\ \bibinfo {author}
  {\bibfnamefont{H.}~\bibnamefont{Kantz}}}%
  , \bibinfo {year} {2005},\ \bibfield{title}{%
  \enquote{\bibinfo {title} {Effective dynamics in {Hamiltonian} systems with
  mixed phase space.}.}\ }%
  \bibfield{journal}{%
  \bibinfo {journal} {Phys. Rev. E}\ }%
  \textbf{\bibinfo {volume} {71}},\ \bibinfo {pages} {036215}%
  \bibAnnoteFile{NoStop}{Motter2005}%
\bibitem[{\citenamefont{de~Moura}\ and\
  \citenamefont{Grebogi}(2004{\natexlab{a}})}]{Moura2004b}%
  \BibitemOpen
  \bibfield{author}{%
  \bibinfo {author} {\bibnamefont{de~Moura}, \bibfnamefont{A.~P.~S.}},\ and\
  \bibinfo {author} {\bibfnamefont{C.}~\bibnamefont{Grebogi}}}%
  , \bibinfo {year} {2004}{\natexlab{a}},\ \bibfield{title}{%
  \enquote{\bibinfo {title} {Chemical and biological activity in
  three-dimensional flows.}.}\ }%
  \bibfield{journal}{%
  \bibinfo {journal} {Phys. Rev. E}\ }%
  \textbf{\bibinfo {volume} {70}},\ \bibinfo {pages} {026218}%
  \bibAnnoteFile{NoStop}{Moura2004b}%
\bibitem[{\citenamefont{de~Moura}\ and\
  \citenamefont{Grebogi}(2004{\natexlab{b}})}]{Moura2004}%
  \BibitemOpen
  \bibfield{author}{%
  \bibinfo {author} {\bibnamefont{de~Moura}, \bibfnamefont{A.~P.~S.}},\ and\
  \bibinfo {author} {\bibfnamefont{C.}~\bibnamefont{Grebogi}}}%
  , \bibinfo {year} {2004}{\natexlab{b}},\ \bibfield{title}{%
  \enquote{\bibinfo {title} {Reactions in flows with nonhyperbolic
  dynamics.}.}\ }%
  \bibfield{journal}{%
  \bibinfo {journal} {Phys. Rev. E}\ }%
  \textbf{\bibinfo {volume} {70}},\ \bibinfo {pages} {036216}%
  \bibAnnoteFile{NoStop}{Moura2004}%
\bibitem[{\citenamefont{Mullowney}\
  \emph{et~al.}(2008{\natexlab{a}})\citenamefont{Mullowney},
  \citenamefont{Julien},\ and\ \citenamefont{Meiss}}]{Mullowney2005}%
  \BibitemOpen
  \bibfield{author}{%
  \bibinfo {author} {\bibnamefont{Mullowney}, \bibfnamefont{P.}}, \bibinfo
  {author} {\bibfnamefont{K.}~\bibnamefont{Julien}},\ and\ \bibinfo {author}
  {\bibfnamefont{J.~D.}\ \bibnamefont{Meiss}}}%
  , \bibinfo {year} {2008}{\natexlab{a}},\ \bibfield{title}{%
  \enquote{\bibinfo {title} {Blinking rolls: Chaotic advection in a
  three-dimensional flow with an invariant},}\ }%
  \bibfield{journal}{%
  \bibinfo {journal} {SIAM J. Appl. Dyn. Sys.}\ }%
  \textbf{\bibinfo {volume} {4}},\ \bibinfo {pages} {159--186}%
  \bibAnnoteFile{NoStop}{Mullowney2005}%
\bibitem[{\citenamefont{Mullowney}\
  \emph{et~al.}(2008{\natexlab{b}})\citenamefont{Mullowney},
  \citenamefont{Julien},\ and\ \citenamefont{Meiss}}]{Mullowney2008}%
  \BibitemOpen
  \bibfield{author}{%
  \bibinfo {author} {\bibnamefont{Mullowney}, \bibfnamefont{P.}}, \bibinfo
  {author} {\bibfnamefont{K.}~\bibnamefont{Julien}},\ and\ \bibinfo {author}
  {\bibfnamefont{J.~D.}\ \bibnamefont{Meiss}}}%
  , \bibinfo {year} {2008}{\natexlab{b}},\ \bibfield{title}{%
  \enquote{\bibinfo {title} {Chaotic advection and the emergence of tori in the
  {K}\"{u}ppers--{L}ortz state},}\ }%
  \bibfield{journal}{%
  \bibinfo {journal} {Chaos}\ }%
  \textbf{\bibinfo {volume} {18}},\ \bibinfo {pages} {033104}%
  \bibAnnoteFile{NoStop}{Mullowney2008}%
\bibitem[{\citenamefont{Nagai}\ \emph{et~al.}(2014)\citenamefont{Nagai},
  \citenamefont{Hayasaka}, \citenamefont{Kawashima},\ and\
  \citenamefont{Shibata}}]{Nagai2014}%
  \BibitemOpen
  \bibfield{author}{%
  \bibinfo {author} {\bibnamefont{Nagai}, \bibfnamefont{M.}}, \bibinfo {author}
  {\bibfnamefont{Y.}~\bibnamefont{Hayasaka}}, \bibinfo {author}
  {\bibfnamefont{T.}~\bibnamefont{Kawashima}},\ and\ \bibinfo {author}
  {\bibfnamefont{T.}~\bibnamefont{Shibata}}}%
  , \bibinfo {year} {2014},\ \bibfield{title}{%
  \enquote{\bibinfo {title} {{Active mixing in microchamber using cilia of
  Vorticella convallaria}},}\ }%
  \bibfield{journal}{%
  \bibinfo {journal} {IEEJ Trans. Elect. Electron. Eng.}\ }%
  \textbf{\bibinfo {volume} {9}},\ \bibinfo {pages} {575--576}%
  \bibAnnoteFile{NoStop}{Nagai2014}%
\bibitem[{\citenamefont{Nagai}\ \emph{et~al.}(2009)\citenamefont{Nagai},
  \citenamefont{Oishi}, \citenamefont{Oshima}, \citenamefont{Asai},\ and\
  \citenamefont{Fujita}}]{Nagai09}%
  \BibitemOpen
  \bibfield{author}{%
  \bibinfo {author} {\bibnamefont{Nagai}, \bibfnamefont{M.}}, \bibinfo {author}
  {\bibfnamefont{M.}~\bibnamefont{Oishi}}, \bibinfo {author}
  {\bibfnamefont{M.}~\bibnamefont{Oshima}}, \bibinfo {author}
  {\bibfnamefont{H.}~\bibnamefont{Asai}},\ and\ \bibinfo {author}
  {\bibfnamefont{H.}~\bibnamefont{Fujita}}}%
  , \bibinfo {year} {2009},\ \bibfield{title}{%
  \enquote{\bibinfo {title} {Three-dimensional two-component velocity
  measurement of the flow field induced by the \emph{Vorticella} picta
  microorganism using a confocal microparticle image velocimetry technique},}\
  }%
  \bibfield{journal}{%
  \bibinfo {journal} {Biomicrofluidics}\ }%
  \textbf{\bibinfo {volume} {3}},\ \bibinfo {pages} {014105}%
  \bibAnnoteFile{NoStop}{Nagai09}%
\bibitem[{\citenamefont{Neufeld}\ \emph{et~al.}(2002)\citenamefont{Neufeld},
  \citenamefont{L\'opez}, \citenamefont{Hern\'andez-Garc\'ia},\ and\
  \citenamefont{Piro}}]{Neufeld-et-al-02}%
  \BibitemOpen
  \bibfield{author}{%
  \bibinfo {author} {\bibnamefont{Neufeld}, \bibfnamefont{Z.}}, \bibinfo
  {author} {\bibfnamefont{C.}~\bibnamefont{L\'opez}}, \bibinfo {author}
  {\bibfnamefont{E.}~\bibnamefont{Hern\'andez-Garc\'ia}},\ and\ \bibinfo
  {author} {\bibfnamefont{O.}~\bibnamefont{Piro}}}%
  , \bibinfo {year} {2002},\ \bibfield{title}{%
  \enquote{\bibinfo {title} {Excitable media in open and closed chaotic
  flows},}\ }%
  \bibfield{journal}{%
  \bibinfo {journal} {Phys. Rev. E}\ }%
  \textbf{\bibinfo {volume} {66}},\ \bibinfo {pages} {066208}%
  \bibAnnoteFile{NoStop}{Neufeld-et-al-02}%
\bibitem[{\citenamefont{Nguyen}(2011)}]{Nguyen2011}%
  \BibitemOpen
  \bibfield{author}{%
  \bibinfo {author} {\bibnamefont{Nguyen}, \bibfnamefont{N.T.}}}%
  , \bibinfo {year} {2011},\ \emph{\bibinfo {title} {Micromixers: Fundamentals,
  Design and Fabrication}},\ Micro \& nano technologies series\ (\bibinfo
  {publisher} {Elsevier/William Andrew})%
  \bibAnnoteFile{NoStop}{Nguyen2011}%
\bibitem[{\citenamefont{Niedermayer}\
  \emph{et~al.}(2008)\citenamefont{Niedermayer}, \citenamefont{Eckhardt},\ and\
  \citenamefont{Lenz}}]{Niedermayer2008}%
  \BibitemOpen
  \bibfield{author}{%
  \bibinfo {author} {\bibnamefont{Niedermayer}, \bibfnamefont{T.}}, \bibinfo
  {author} {\bibfnamefont{B.}~\bibnamefont{Eckhardt}},\ and\ \bibinfo {author}
  {\bibfnamefont{P.}~\bibnamefont{Lenz}}}%
  , \bibinfo {year} {2008},\ \bibfield{title}{%
  \enquote{\bibinfo {title} {{Synchronization, phase locking, and metachronal
  wave formation in ciliary chains}},}\ }%
  \bibfield{journal}{%
  \bibinfo {journal} {Chaos}\ }%
  \textbf{\bibinfo {volume} {18}},\ \bibinfo {pages} {037128}%
  \bibAnnoteFile{NoStop}{Niedermayer2008}%
\bibitem[{\citenamefont{Nonaka}\ \emph{et~al.}(2002)\citenamefont{Nonaka},
  \citenamefont{Shiratori}, \citenamefont{Saijoh},\ and\
  \citenamefont{Hamada}}]{Nonaka2002}%
  \BibitemOpen
  \bibfield{author}{%
  \bibinfo {author} {\bibnamefont{Nonaka}, \bibfnamefont{S.}}, \bibinfo
  {author} {\bibfnamefont{H.}~\bibnamefont{Shiratori}}, \bibinfo {author}
  {\bibfnamefont{Y.}~\bibnamefont{Saijoh}},\ and\ \bibinfo {author}
  {\bibfnamefont{H.}~\bibnamefont{Hamada}}}%
  , \bibinfo {year} {2002},\ \bibfield{title}{%
  \enquote{\bibinfo {title} {{Determination of left--right patterning of the
  mouse embryo by artificial nodal flow}},}\ }%
  \bibfield{journal}{%
  \bibinfo {journal} {Nature}\ }%
  \textbf{\bibinfo {volume} {418}},\ \bibinfo {pages} {96--99}%
  \bibAnnoteFile{NoStop}{Nonaka2002}%
\bibitem[{\citenamefont{Oddy}\ \emph{et~al.}(2001)\citenamefont{Oddy},
  \citenamefont{Santiago},\ and\ \citenamefont{Mikkelsen}}]{Oddy2001}%
  \BibitemOpen
  \bibfield{author}{%
  \bibinfo {author} {\bibnamefont{Oddy}, \bibfnamefont{M.~H.}}, \bibinfo
  {author} {\bibfnamefont{J.~G.}\ \bibnamefont{Santiago}},\ and\ \bibinfo
  {author} {\bibfnamefont{J.~C.}\ \bibnamefont{Mikkelsen}}}%
  , \bibinfo {year} {2001},\ \bibfield{title}{%
  \enquote{\bibinfo {title} {Electrokinetic instability micromixing},}\ }%
  \bibfield{journal}{%
  \bibinfo {journal} {Analytical Chem.}\ }%
  \textbf{\bibinfo {volume} {73}},\ \bibinfo {pages} {5822--5832}%
  \bibAnnoteFile{NoStop}{Oddy2001}%
\bibitem[{\citenamefont{Okkels}\ and\
  \citenamefont{Tabeling}(2004)}]{Okkels2004}%
  \BibitemOpen
  \bibfield{author}{%
  \bibinfo {author} {\bibnamefont{Okkels}, \bibfnamefont{F.}},\ and\ \bibinfo
  {author} {\bibfnamefont{P.}~\bibnamefont{Tabeling}}}%
  , \bibinfo {year} {2004},\ \bibfield{title}{%
  \enquote{\bibinfo {title} {Spatiotemporal resonances in mixing of open
  viscous fluids},}\ }%
  \bibfield{journal}{%
  \bibinfo {journal} {Phys. Rev Lett.}\ }%
  \textbf{\bibinfo {volume} {92}},\ \bibinfo {pages} {038301}%
  \bibAnnoteFile{NoStop}{Okkels2004}%
\bibitem[{\citenamefont{Okubo}(1970)}]{Okubo:1970tr}%
  \BibitemOpen
  \bibfield{author}{%
  \bibinfo {author} {\bibnamefont{Okubo}, \bibfnamefont{A.}}}%
  , \bibinfo {year} {1970},\ \bibfield{title}{%
  \enquote{\bibinfo {title} {Horizontal dispersion of floatable particles in
  vicinity of velocity singularities such as convergences},}\ }%
  \bibfield{journal}{%
  \bibinfo {journal} {Deep-Sea Res.}\ }%
  \textbf{\bibinfo {volume} {17}},\ \bibinfo {pages} {445--454}%
  \bibAnnoteFile{NoStop}{Okubo:1970tr}%
\bibitem[{\citenamefont{Orme}\
  \emph{et~al.}(2001{\natexlab{a}})\citenamefont{Orme}, \citenamefont{Otto},\
  and\ \citenamefont{Blake}}]{Orme2001b}%
  \BibitemOpen
  \bibfield{author}{%
  \bibinfo {author} {\bibnamefont{Orme}, \bibfnamefont{B.~A.~A.}}, \bibinfo
  {author} {\bibfnamefont{S.~R.}\ \bibnamefont{Otto}},\ and\ \bibinfo {author}
  {\bibfnamefont{J.~R.}\ \bibnamefont{Blake}}}%
  , \bibinfo {year} {2001}{\natexlab{a}},\ \bibfield{title}{%
  \enquote{\bibinfo {title} {Chaos and mixing in micro-biological fluid
  dynamics:blinking stokeslets},}\ }%
  \bibfield{journal}{%
  \bibinfo {journal} {Math. Meth. Appl. Sci.}\ }%
  \textbf{\bibinfo {volume} {24}},\ \bibinfo {pages} {1337--1349}%
  \bibAnnoteFile{NoStop}{Orme2001b}%
\bibitem[{\citenamefont{Orme}\
  \emph{et~al.}(2001{\natexlab{b}})\citenamefont{Orme}, \citenamefont{Otto},\
  and\ \citenamefont{Blake}}]{Orme2001a}%
  \BibitemOpen
  \bibfield{author}{%
  \bibinfo {author} {\bibnamefont{Orme}, \bibfnamefont{B.~A.~A.}}, \bibinfo
  {author} {\bibfnamefont{S.~R.}\ \bibnamefont{Otto}},\ and\ \bibinfo {author}
  {\bibfnamefont{J.~R.}\ \bibnamefont{Blake}}}%
  , \bibinfo {year} {2001}{\natexlab{b}},\ \bibfield{title}{%
  \enquote{\bibinfo {title} {Enhanced efficiency of feeding and mixing due to
  chaotic flow patterns around choanoflagellates},}\ }%
  \bibfield{journal}{%
  \bibinfo {journal} {IMA J. Math Applied to Medicine and Biology}\ }%
  \textbf{\bibinfo {volume} {18}},\ \bibinfo {pages} {293--325}%
  \bibAnnoteFile{NoStop}{Orme2001a}%
\bibitem[{\citenamefont{Ott}(1993)}]{Ott}%
  \BibitemOpen
  \bibfield{author}{%
  \bibinfo {author} {\bibnamefont{Ott}, \bibfnamefont{E.}}}%
  , \bibinfo {year} {1993},\ \emph{\bibinfo {title} {Chaos in dynamical
  systems}}\ (\bibinfo {publisher} {Cambridge University Press, Cambdridge})%
  \bibAnnoteFile{NoStop}{Ott}%
\bibitem[{\citenamefont{Ottino}(1989)}]{Ottino1989}%
  \BibitemOpen
  \bibfield{author}{%
  \bibinfo {author} {\bibnamefont{Ottino}, \bibfnamefont{J.~M.}}}%
  , \bibinfo {year} {1989},\ \emph{\bibinfo {title} {The kinematics of mixing:
  Stretching, chaos and transport}}\ (\bibinfo {publisher} {Cambridge
  University Press},\ \bibinfo {address} {Cambridge})%
  \bibAnnoteFile{NoStop}{Ottino1989}%
\bibitem[{\citenamefont{Ottino}(1990)}]{Ottino90}%
  \BibitemOpen
  \bibfield{author}{%
  \bibinfo {author} {\bibnamefont{Ottino}, \bibfnamefont{J~M}}}%
  , \bibinfo {year} {1990},\ \bibfield{title}{%
  \enquote{\bibinfo {title} {Mixing, chaotic advection and turbulence},}\ }%
  \bibfield{journal}{%
  \bibinfo {journal} {Annu. Rev. Fluid Mech.}\ }%
  \textbf{\bibinfo {volume} {22}},\ \bibinfo {pages} {207--253}%
  \bibAnnoteFile{NoStop}{Ottino90}%
\bibitem[{\citenamefont{Ottino}\ \emph{et~al.}(1994)\citenamefont{Ottino},
  \citenamefont{Jana},\ and\ \citenamefont{Chakravarthy}}]{Ottino1994}%
  \BibitemOpen
  \bibfield{author}{%
  \bibinfo {author} {\bibnamefont{Ottino}, \bibfnamefont{J.~M.}}, \bibinfo
  {author} {\bibfnamefont{S.~C.}\ \bibnamefont{Jana}},\ and\ \bibinfo {author}
  {\bibfnamefont{V.~S.}\ \bibnamefont{Chakravarthy}}}%
  , \bibinfo {year} {1994},\ \bibfield{title}{%
  \enquote{\bibinfo {title} {{From Reynold's stretching and folding to mixing
  studies using horseshoe maps}},}\ }%
  \bibfield{journal}{%
  \bibinfo {journal} {Phys.\ Fluids}\ }%
  \textbf{\bibinfo {volume} {6}},\ \bibinfo {pages} {685--699}%
  \bibAnnoteFile{NoStop}{Ottino1994}%
\bibitem[{\citenamefont{Ottino}\ and\
  \citenamefont{Khakhar}(2000)}]{Ottino2000}%
  \BibitemOpen
  \bibfield{author}{%
  \bibinfo {author} {\bibnamefont{Ottino}, \bibfnamefont{J.~M.}},\ and\
  \bibinfo {author} {\bibfnamefont{D.~V.}\ \bibnamefont{Khakhar}}}%
  , \bibinfo {year} {2000},\ \bibfield{title}{%
  \enquote{\bibinfo {title} {Mixing and segregation of granular materials},}\
  }%
  \bibfield{journal}{%
  \bibinfo {journal} {Annu. Rev. Fluid Mech.}\ }%
  \textbf{\bibinfo {volume} {32}},\ \bibinfo {pages} {55--91}%
  \bibAnnoteFile{NoStop}{Ottino2000}%
\bibitem[{\citenamefont{Otto}\ \emph{et~al.}(2008)\citenamefont{Otto},
  \citenamefont{Riegler},\ and\ \citenamefont{Voth}}]{Otto2008}%
  \BibitemOpen
  \bibfield{author}{%
  \bibinfo {author} {\bibnamefont{Otto}, \bibfnamefont{F.}}, \bibinfo {author}
  {\bibfnamefont{E.}~\bibnamefont{Riegler}},\ and\ \bibinfo {author}
  {\bibfnamefont{G.}~\bibnamefont{Voth}}}%
  , \bibinfo {year} {2008},\ \bibfield{title}{%
  \enquote{\bibinfo {title} {Measurements of the steady streaming flow around
  oscillating spheres using three dimensional particle tracking velocimetry},}\
  }%
  \bibfield{journal}{%
  \bibinfo {journal} {Phys. Fluids}\ }%
  \textbf{\bibinfo {volume} {20}},\ \bibinfo {pages} {093304}%
  \bibAnnoteFile{NoStop}{Otto2008}%
\bibitem[{\citenamefont{Otto}\ \emph{et~al.}(2001)\citenamefont{Otto},
  \citenamefont{Yannacopoulos},\ and\
  \citenamefont{Blake}}]{Otto_blinking_2001}%
  \BibitemOpen
  \bibfield{author}{%
  \bibinfo {author} {\bibnamefont{Otto}, \bibfnamefont{S.~R.}}, \bibinfo
  {author} {\bibfnamefont{A.~N.}\ \bibnamefont{Yannacopoulos}},\ and\ \bibinfo
  {author} {\bibfnamefont{J.~R.}\ \bibnamefont{Blake}}}%
  , \bibinfo {year} {2001},\ \bibfield{title}{%
  \enquote{\bibinfo {title} {Transport and mixing in {Stokes} flow: the effect
  of chaotic dynamics on the blinking stokeslet},}\ }%
  \bibfield{journal}{%
  \bibinfo {journal} {J. Fluid Mech.}\ }%
  \textbf{\bibinfo {volume} {430}},\ \bibinfo {pages} {1--26}%
  \bibAnnoteFile{NoStop}{Otto_blinking_2001}%
\bibitem[{\citenamefont{Paik}\
  \emph{et~al.}(2003{\natexlab{a}})\citenamefont{Paik}, \citenamefont{Pamula},\
  and\ \citenamefont{Fair}}]{Paik2003}%
  \BibitemOpen
  \bibfield{author}{%
  \bibinfo {author} {\bibnamefont{Paik}, \bibfnamefont{P.}}, \bibinfo {author}
  {\bibfnamefont{V.~K.}\ \bibnamefont{Pamula}},\ and\ \bibinfo {author}
  {\bibfnamefont{R.~B.}\ \bibnamefont{Fair}}}%
  , \bibinfo {year} {2003}{\natexlab{a}},\ \bibfield{title}{%
  \enquote{\bibinfo {title} {Rapid droplet mixers for digital microfluidic
  systems},}\ }%
  \bibfield{journal}{%
  \bibinfo {journal} {Lab on a chip}\ }%
  \textbf{\bibinfo {volume} {3}},\ \bibinfo {pages} {253--239}%
  \bibAnnoteFile{NoStop}{Paik2003}%
\bibitem[{\citenamefont{Paik}\
  \emph{et~al.}(2003{\natexlab{b}})\citenamefont{Paik}, \citenamefont{Pamula},
  \citenamefont{Pollack},\ and\ \citenamefont{Fair}}]{Paik2003a}%
  \BibitemOpen
  \bibfield{author}{%
  \bibinfo {author} {\bibnamefont{Paik}, \bibfnamefont{P.}}, \bibinfo {author}
  {\bibfnamefont{V.~K.}\ \bibnamefont{Pamula}}, \bibinfo {author}
  {\bibfnamefont{M.~G.}\ \bibnamefont{Pollack}},\ and\ \bibinfo {author}
  {\bibfnamefont{R.~B.}\ \bibnamefont{Fair}}}%
  , \bibinfo {year} {2003}{\natexlab{b}},\ \bibfield{title}{%
  \enquote{\bibinfo {title} {Electrowetting-based droplet mixers for
  microfluidic systems},}\ }%
  \bibfield{journal}{%
  \bibinfo {journal} {Lab on a chip}\ }%
  \textbf{\bibinfo {volume} {3}},\ \bibinfo {pages} {28--33}%
  \bibAnnoteFile{NoStop}{Paik2003a}%
\bibitem[{\citenamefont{Pepper}\ \emph{et~al.}(2010)\citenamefont{Pepper},
  \citenamefont{Roper}, \citenamefont{Ryu}, \citenamefont{Matsudaira},\ and\
  \citenamefont{Stone}}]{Pepper2010}%
  \BibitemOpen
  \bibfield{author}{%
  \bibinfo {author} {\bibnamefont{Pepper}, \bibfnamefont{R.~E.}}, \bibinfo
  {author} {\bibfnamefont{M.}~\bibnamefont{Roper}}, \bibinfo {author}
  {\bibfnamefont{S.}~\bibnamefont{Ryu}}, \bibinfo {author}
  {\bibfnamefont{P.}~\bibnamefont{Matsudaira}},\ and\ \bibinfo {author}
  {\bibfnamefont{H.~A.}\ \bibnamefont{Stone}}}%
  , \bibinfo {year} {2010},\ \bibfield{title}{%
  \enquote{\bibinfo {title} {{Nearby boundaries create eddies near microscopic
  filter feeders}},}\ }%
  \bibfield{journal}{%
  \bibinfo {journal} {J. Roy. Soc. Interface}\ }%
  \textbf{\bibinfo {volume} {7}},\ \bibinfo {pages} {851--862}%
  \bibAnnoteFile{NoStop}{Pepper2010}%
\bibitem[{\citenamefont{Pepper}\ \emph{et~al.}(2013)\citenamefont{Pepper},
  \citenamefont{Roper}, \citenamefont{Ryu}, \citenamefont{Matsumoto},
  \citenamefont{Nagai},\ and\ \citenamefont{Stone}}]{Pepper2013}%
  \BibitemOpen
  \bibfield{author}{%
  \bibinfo {author} {\bibnamefont{Pepper}, \bibfnamefont{R.~E.}}, \bibinfo
  {author} {\bibfnamefont{M.}~\bibnamefont{Roper}}, \bibinfo {author}
  {\bibfnamefont{S.}~\bibnamefont{Ryu}}, \bibinfo {author}
  {\bibfnamefont{N.}~\bibnamefont{Matsumoto}}, \bibinfo {author}
  {\bibfnamefont{M.}~\bibnamefont{Nagai}},\ and\ \bibinfo {author}
  {\bibfnamefont{H.~A.}\ \bibnamefont{Stone}}}%
  , \bibinfo {year} {2013},\ \bibfield{title}{%
  \enquote{\bibinfo {title} {{A new angle on microscopic suspension feeders
  near boundaries}},}\ }%
  \bibfield{journal}{%
  \bibinfo {journal} {Biophys. J.}\ }%
  \textbf{\bibinfo {volume} {105}},\ \bibinfo {pages} {1796--1804}%
  \bibAnnoteFile{NoStop}{Pepper2013}%
\bibitem[{\citenamefont{Perugini}\ \emph{et~al.}(2012)\citenamefont{Perugini},
  \citenamefont{Campos}, \citenamefont{Ertel-Ingrisch},\ and\
  \citenamefont{Dingwell}}]{perugini2012}%
  \BibitemOpen
  \bibfield{author}{%
  \bibinfo {author} {\bibnamefont{Perugini}, \bibfnamefont{D.}}, \bibinfo
  {author} {\bibfnamefont{C.~P.~De}\ \bibnamefont{Campos}}, \bibinfo {author}
  {\bibfnamefont{W.}~\bibnamefont{Ertel-Ingrisch}},\ and\ \bibinfo {author}
  {\bibfnamefont{D.~B.}\ \bibnamefont{Dingwell}}}%
  , \bibinfo {year} {2012},\ \bibfield{title}{%
  \enquote{\bibinfo {title} {The space and time complexity of chaotic mixing of
  silicate melts: Implications for igneous petrology},}\ }%
  \bibfield{journal}{%
  \bibinfo {journal} {LITHOS}\ }%
  \textbf{\bibinfo {volume} {155}},\ \bibinfo {pages} {326--340}%
  \bibAnnoteFile{NoStop}{perugini2012}%
\bibitem[{\citenamefont{Perugini}\ \emph{et~al.}(2003)\citenamefont{Perugini},
  \citenamefont{Poli},\ and\ \citenamefont{Mazzuoli}}]{perugini03}%
  \BibitemOpen
  \bibfield{author}{%
  \bibinfo {author} {\bibnamefont{Perugini}, \bibfnamefont{D.}}, \bibinfo
  {author} {\bibfnamefont{G.}~\bibnamefont{Poli}},\ and\ \bibinfo {author}
  {\bibfnamefont{R.}~\bibnamefont{Mazzuoli}}}%
  , \bibinfo {year} {2003},\ \bibfield{title}{%
  \enquote{\bibinfo {title} {Chaotic advection, fractals and diffusion during
  mixing of magmas: evidence from lava flows},}\ }%
  \bibfield{journal}{%
  \bibinfo {journal} {J. Volcanol. Geotherm. Res.}\ }%
  \textbf{\bibinfo {volume} {2615}},\ \bibinfo {pages} {1--25}%
  \bibAnnoteFile{NoStop}{perugini03}%
\bibitem[{\citenamefont{Peters}\ and\
  \citenamefont{Marras\'e}(2000)}]{Peters-Marrase-00}%
  \BibitemOpen
  \bibfield{author}{%
  \bibinfo {author} {\bibnamefont{Peters}, \bibfnamefont{F.}},\ and\ \bibinfo
  {author} {\bibfnamefont{C.}~\bibnamefont{Marras\'e}}}%
  , \bibinfo {year} {2000},\ \bibfield{title}{%
  \enquote{\bibinfo {title} {Effects of turbulence on plankton: an overview of
  experimental evidence and some theoretical considerations},}\ }%
  \bibfield{journal}{%
  \bibinfo {journal} {Mar. Ecol. Prog. Ser.}\ }%
  \textbf{\bibinfo {volume} {205}},\ \bibinfo {pages} {291--306}%
  \bibAnnoteFile{NoStop}{Peters-Marrase-00}%
\bibitem[{\citenamefont{Petrelli}\ \emph{et~al.}(2016)\citenamefont{Petrelli},
  \citenamefont{El~Omari}, \citenamefont{Le~Guer},\ and\
  \citenamefont{Perugini}}]{Petrelli2016425}%
  \BibitemOpen
  \bibfield{author}{%
  \bibinfo {author} {\bibnamefont{Petrelli}, \bibfnamefont{M.}}, \bibinfo
  {author} {\bibfnamefont{K.}~\bibnamefont{El~Omari}}, \bibinfo {author}
  {\bibfnamefont{Y.}~\bibnamefont{Le~Guer}},\ and\ \bibinfo {author}
  {\bibfnamefont{D.}~\bibnamefont{Perugini}}}%
  , \bibinfo {year} {2016},\ \bibfield{title}{%
  \enquote{\bibinfo {title} {Effects of chaotic advection on the timescales of
  cooling and crystallization of magma bodies at mid crustal levels},}\ }%
  \bibfield{journal}{%
  \bibinfo {journal} {Geochem. Geophys, Geosys.}\ }%
  \textbf{\bibinfo {volume} {17}},\ \bibinfo {pages} {425--441}%
  \bibAnnoteFile{NoStop}{Petrelli2016425}%
\bibitem[{\citenamefont{Phillips}(1991)}]{phillips1991}%
  \BibitemOpen
  \bibfield{author}{%
  \bibinfo {author} {\bibnamefont{Phillips}, \bibfnamefont{O.~M.}}}%
  , \bibinfo {year} {1991},\ \emph{\bibinfo {title} {Flow and Reactions in
  Permeable Rocks}}\ (\bibinfo {publisher} {Cambridge University Press})%
  \bibAnnoteFile{NoStop}{phillips1991}%
\bibitem[{\citenamefont{Pierrehumbert}(1994)}]{Pierrehumbert1994}%
  \BibitemOpen
  \bibfield{author}{%
  \bibinfo {author} {\bibnamefont{Pierrehumbert}, \bibfnamefont{R.~T.}}}%
  , \bibinfo {year} {1994},\ \bibfield{title}{%
  \enquote{\bibinfo {title} {Tracer microstructure in the large-eddy dominated
  regime},}\ }%
  \bibfield{journal}{%
  \bibinfo {journal} {Chaos Solitons Fractals}\ }%
  \textbf{\bibinfo {volume} {4}},\ \bibinfo {pages} {1091--1110}%
  \bibAnnoteFile{NoStop}{Pierrehumbert1994}%
\bibitem[{\citenamefont{Pikovsky}\ and\
  \citenamefont{Popovych}(2003)}]{Pikovsky2003}%
  \BibitemOpen
  \bibfield{author}{%
  \bibinfo {author} {\bibnamefont{Pikovsky}, \bibfnamefont{A.}},\ and\ \bibinfo
  {author} {\bibfnamefont{O.}~\bibnamefont{Popovych}}}%
  , \bibinfo {year} {2003},\ \bibfield{title}{%
  \enquote{\bibinfo {title} {Persistent patterns in deterministic mixing
  flows},}\ }%
  \bibfield{journal}{%
  \bibinfo {journal} {Europhys. Lett.}\ }%
  \textbf{\bibinfo {volume} {61}},\ \bibinfo {pages} {625--631}%
  \bibAnnoteFile{NoStop}{Pikovsky2003}%
\bibitem[{\citenamefont{Piro}\ and\ \citenamefont{Feingold}(1988)}]{Piro1988}%
  \BibitemOpen
  \bibfield{author}{%
  \bibinfo {author} {\bibnamefont{Piro}, \bibfnamefont{O.}},\ and\ \bibinfo
  {author} {\bibfnamefont{M.}~\bibnamefont{Feingold}}}%
  , \bibinfo {year} {1988},\ \bibfield{title}{%
  \enquote{\bibinfo {title} {Diffusion in three-dimensional {L}iouvillian
  maps},}\ }%
  \bibfield{journal}{%
  \bibinfo {journal} {Phys. Rev. Lett.}\ }%
  \textbf{\bibinfo {volume} {61}},\ \bibinfo {pages} {1799--1802}%
  \bibAnnoteFile{NoStop}{Piro1988}%
\bibitem[{\citenamefont{Poje}\ \emph{et~al.}(1999)\citenamefont{Poje},
  \citenamefont{Haller},\ and\ \citenamefont{Mezi{\'c}}}]{Poje:1999wf}%
  \BibitemOpen
  \bibfield{author}{%
  \bibinfo {author} {\bibnamefont{Poje}, \bibfnamefont{A.}}, \bibinfo {author}
  {\bibfnamefont{G.}~\bibnamefont{Haller}},\ and\ \bibinfo {author}
  {\bibfnamefont{I.}~\bibnamefont{Mezi{\'c}}}}%
  , \bibinfo {year} {1999},\ \bibfield{title}{%
  \enquote{\bibinfo {title} {The geometry and statistics of mixing in aperiodic
  flows},}\ }%
  \bibfield{journal}{%
  \bibinfo {journal} {Phys. Fluids}\ }%
  \textbf{\bibinfo {volume} {11}},\ \bibinfo {pages} {2963--2968}%
  \bibAnnoteFile{NoStop}{Poje:1999wf}%
\bibitem[{\citenamefont{Popovych}\ \emph{et~al.}(2007)\citenamefont{Popovych},
  \citenamefont{Pikovsky},\ and\ \citenamefont{Eckhardt}}]{Popovych2007}%
  \BibitemOpen
  \bibfield{author}{%
  \bibinfo {author} {\bibnamefont{Popovych}, \bibfnamefont{O.~V.}}, \bibinfo
  {author} {\bibfnamefont{A.}~\bibnamefont{Pikovsky}},\ and\ \bibinfo {author}
  {\bibfnamefont{B.}~\bibnamefont{Eckhardt}}}%
  , \bibinfo {year} {2007},\ \bibfield{title}{%
  \enquote{\bibinfo {title} {Abnormal mixing of passive scalars in chaotic
  flows},}\ }%
  \bibfield{journal}{%
  \bibinfo {journal} {Phys. Rev. E}\ }%
  \textbf{\bibinfo {volume} {75}},\ \bibinfo {pages} {036308}%
  \bibAnnoteFile{NoStop}{Popovych2007}%
\bibitem[{\citenamefont{Pouransari}\
  \emph{et~al.}(2010)\citenamefont{Pouransari}, \citenamefont{Speetjens},\ and\
  \citenamefont{Clercx}}]{Pouransari2010}%
  \BibitemOpen
  \bibfield{author}{%
  \bibinfo {author} {\bibnamefont{Pouransari}, \bibfnamefont{Z.}}, \bibinfo
  {author} {\bibfnamefont{M.~F.~M.}\ \bibnamefont{Speetjens}},\ and\ \bibinfo
  {author} {\bibfnamefont{H.~J.~H.}\ \bibnamefont{Clercx}}}%
  , \bibinfo {year} {2010},\ \bibfield{title}{%
  \enquote{\bibinfo {title} {Formation of coherent structures by fluid inertia
  in three-dimensional laminar flows},}\ }%
  \bibfield{journal}{%
  \bibinfo {journal} {J.~Fluid Mech.}\ }%
  \textbf{\bibinfo {volume} {654}},\ \bibinfo {pages} {5--34}%
  \bibAnnoteFile{NoStop}{Pouransari2010}%
\bibitem[{\citenamefont{Pratt}\ \emph{et~al.}(2014)\citenamefont{Pratt},
  \citenamefont{Rypina}, \citenamefont{{\"O}zg{\"o}kmen}, \citenamefont{Wang},
  \citenamefont{Childs},\ and\ \citenamefont{Bebieva}}]{pratt2014chaotic}%
  \BibitemOpen
  \bibfield{author}{%
  \bibinfo {author} {\bibnamefont{Pratt}, \bibfnamefont{L.~J.}}, \bibinfo
  {author} {\bibfnamefont{I.~I.}\ \bibnamefont{Rypina}}, \bibinfo {author}
  {\bibfnamefont{T.~M.}\ \bibnamefont{{\"O}zg{\"o}kmen}}, \bibinfo {author}
  {\bibfnamefont{P.}~\bibnamefont{Wang}}, \bibinfo {author}
  {\bibfnamefont{H.}~\bibnamefont{Childs}},\ and\ \bibinfo {author}
  {\bibfnamefont{Y.}~\bibnamefont{Bebieva}}}%
  , \bibinfo {year} {2014},\ \bibfield{title}{%
  \enquote{\bibinfo {title} {Chaotic advection in a steady, three-dimensional,
  ekman-driven eddy},}\ }%
  \bibfield{journal}{%
  \bibinfo {journal} {J. Fluid Mech.}\ }%
  \textbf{\bibinfo {volume} {738}},\ \bibinfo {pages} {143--183}%
  \bibAnnoteFile{NoStop}{pratt2014chaotic}%
\bibitem[{\citenamefont{Raben}\ \emph{et~al.}(2013)\citenamefont{Raben},
  \citenamefont{{Ross}},\ and\ \citenamefont{{Vlachos}}}]{Raben2013}%
  \BibitemOpen
  \bibfield{author}{%
  \bibinfo {author} {\bibnamefont{Raben}, \bibfnamefont{{S.}~{G}.}}, \bibinfo
  {author} {\bibfnamefont{{S.}~{D}.}\ \bibnamefont{{Ross}}},\ and\ \bibinfo
  {author} {\bibfnamefont{{P.}~{P}.}\ \bibnamefont{{Vlachos}}}}%
  , \bibinfo {year} {2013},\ \bibfield{title}{%
  \enquote{\bibinfo {title} {Computation of finite-time {Lyapunov} exponents
  from time-resolved particle image velocimetry data},}\ }%
  \bibfield{journal}{%
  \bibinfo {journal} {Exper. {Fluids}}\ }%
  \textbf{\bibinfo {volume} {55}},\ \bibinfo {pages} {1--14}%
  \bibAnnoteFile{NoStop}{Raben2013}%
\bibitem[{\citenamefont{Rodrigues}\
  \emph{et~al.}(2010)\citenamefont{Rodrigues}, \citenamefont{de~Moura},\ and\
  \citenamefont{Grebogi}}]{Rodrigues2010}%
  \BibitemOpen
  \bibfield{author}{%
  \bibinfo {author} {\bibnamefont{Rodrigues}, \bibfnamefont{C.~S.}}, \bibinfo
  {author} {\bibfnamefont{A.~P.~S.}\ \bibnamefont{de~Moura}},\ and\ \bibinfo
  {author} {\bibfnamefont{C.}~\bibnamefont{Grebogi}}}%
  , \bibinfo {year} {2010},\ \bibfield{title}{%
  \enquote{\bibinfo {title} {Random fluctuation leads to forbidden escape of
  particles},}\ }%
  \bibfield{journal}{%
  \bibinfo {journal} {Phys. Rev. E}\ }%
  \textbf{\bibinfo {volume} {82}},\ \bibinfo {pages} {026211}%
  \bibAnnoteFile{NoStop}{Rodrigues2010}%
\bibitem[{\citenamefont{Rokhlin}(1960)}]{rokhlin1960new}%
  \BibitemOpen
  \bibfield{author}{%
  \bibinfo {author} {\bibnamefont{Rokhlin}, \bibfnamefont{V.~A.}}}%
  , \bibinfo {year} {1960},\ \bibfield{title}{%
  \enquote{\bibinfo {title} {New progress in the theory of transformations with
  invariant measure},}\ }%
  \bibfield{journal}{%
  \bibinfo {journal} {Russ. Math. Surv.}\ }%
  \textbf{\bibinfo {volume} {15}},\ \bibinfo {pages} {1--22}%
  \bibAnnoteFile{NoStop}{rokhlin1960new}%
\bibitem[{\citenamefont{Rypina}\ \emph{et~al.}(2007)\citenamefont{Rypina},
  \citenamefont{Brown}, \citenamefont{{Beron-Vera}}, \citenamefont{Ko{\c c}ak},
  \citenamefont{Olascoaga},\ and\
  \citenamefont{Udovydchenkov}}]{Rypina:2007ev}%
  \BibitemOpen
  \bibfield{author}{%
  \bibinfo {author} {\bibnamefont{Rypina}, \bibfnamefont{I.}}, \bibinfo
  {author} {\bibfnamefont{M.~G.}\ \bibnamefont{Brown}}, \bibinfo {author}
  {\bibfnamefont{F.~J.}\ \bibnamefont{{Beron-Vera}}}, \bibinfo {author}
  {\bibfnamefont{H.}~\bibnamefont{Ko{\c c}ak}}, \bibinfo {author}
  {\bibfnamefont{M.~J.}\ \bibnamefont{Olascoaga}},\ and\ \bibinfo {author}
  {\bibfnamefont{I.~A.}\ \bibnamefont{Udovydchenkov}}}%
  , \bibinfo {year} {2007},\ \bibfield{title}{%
  \enquote{\bibinfo {title} {On the lagrangian dynamics of atmospheric zonal
  jets and the permeability of the stratospheric polar vortex},}\ }%
  \bibfield{journal}{%
  \bibinfo {journal} {J. Atmos. Sci.}\ }%
  \textbf{\bibinfo {volume} {64}},\ \bibinfo {pages} {3595--3610}%
  \bibAnnoteFile{NoStop}{Rypina:2007ev}%
\bibitem[{\citenamefont{Salman}\ and\
  \citenamefont{Haynes}(2007)}]{Salman2007}%
  \BibitemOpen
  \bibfield{author}{%
  \bibinfo {author} {\bibnamefont{Salman}, \bibfnamefont{H.}},\ and\ \bibinfo
  {author} {\bibfnamefont{P.~H.}\ \bibnamefont{Haynes}}}%
  , \bibinfo {year} {2007},\ \bibfield{title}{%
  \enquote{\bibinfo {title} {A numerical study of passive scalar evolution in
  peripheral regions},}\ }%
  \bibfield{journal}{%
  \bibinfo {journal} {Phys. Fluids}\ }%
  \textbf{\bibinfo {volume} {19}},\ \bibinfo {pages} {067101}%
  \bibAnnoteFile{NoStop}{Salman2007}%
\bibitem[{\citenamefont{Samelson}(2013)}]{samelson_lagrangian_2013}%
  \BibitemOpen
  \bibfield{author}{%
  \bibinfo {author} {\bibnamefont{Samelson}, \bibfnamefont{R.~M.}}}%
  , \bibinfo {year} {2013},\ \bibfield{title}{%
  \enquote{\bibinfo {title} {Lagrangian motion, coherent structures, and lines
  of persistent material strain},}\ }%
  \bibfield{journal}{%
  \bibinfo {journal} {Annual Review of Marine Science}\ }%
  \textbf{\bibinfo {volume} {5}},\ \bibinfo {pages} {137--163}%
  \bibAnnoteFile{NoStop}{samelson_lagrangian_2013}%
\bibitem[{\citenamefont{Sandulescu}\
  \emph{et~al.}(2006)\citenamefont{Sandulescu},
  \citenamefont{Hern\'andez-Garc\'ia}, \citenamefont{L\'opez},\ and\
  \citenamefont{Feudel}}]{Sandulescu-et-al-06}%
  \BibitemOpen
  \bibfield{author}{%
  \bibinfo {author} {\bibnamefont{Sandulescu}, \bibfnamefont{M.}}, \bibinfo
  {author} {\bibfnamefont{E.}~\bibnamefont{Hern\'andez-Garc\'ia}}, \bibinfo
  {author} {\bibfnamefont{C.}~\bibnamefont{L\'opez}},\ and\ \bibinfo {author}
  {\bibfnamefont{U.}~\bibnamefont{Feudel}}}%
  , \bibinfo {year} {2006},\ \bibfield{title}{%
  \enquote{\bibinfo {title} {Kinematic studies of transport across an island
  wake, with application to the {Canary} islands},}\ }%
  \bibfield{journal}{%
  \bibinfo {journal} {Tellus A}\ }%
  \textbf{\bibinfo {volume} {58}},\ \bibinfo {pages} {605--615}%
  \bibAnnoteFile{NoStop}{Sandulescu-et-al-06}%
\bibitem[{\citenamefont{Sandulescu}\
  \emph{et~al.}(2007)\citenamefont{Sandulescu},
  \citenamefont{Hern\'andez-Garc\'ia}, \citenamefont{L\'opez},\ and\
  \citenamefont{Feudel}}]{Sandulescu-et-al-07}%
  \BibitemOpen
  \bibfield{author}{%
  \bibinfo {author} {\bibnamefont{Sandulescu}, \bibfnamefont{M.}}, \bibinfo
  {author} {\bibfnamefont{E.}~\bibnamefont{Hern\'andez-Garc\'ia}}, \bibinfo
  {author} {\bibfnamefont{C.}~\bibnamefont{L\'opez}},\ and\ \bibinfo {author}
  {\bibfnamefont{U.}~\bibnamefont{Feudel}}}%
  , \bibinfo {year} {2007},\ \bibfield{title}{%
  \enquote{\bibinfo {title} {Plankton blooms in vortices: the role of
  biological and hydrodynamic timescales},}\ }%
  \bibfield{journal}{%
  \bibinfo {journal} {Nonlin. Processes Geophys.}\ }%
  \textbf{\bibinfo {volume} {14}},\ \bibinfo {pages} {443--454}%
  \bibAnnoteFile{NoStop}{Sandulescu-et-al-07}%
\bibitem[{\citenamefont{Santitissadeekorn}\
  \emph{et~al.}(2010)\citenamefont{Santitissadeekorn},
  \citenamefont{Froyland},\ and\
  \citenamefont{Monahan}}]{santitissadeekorn_optimally_2010}%
  \BibitemOpen
  \bibfield{author}{%
  \bibinfo {author} {\bibnamefont{Santitissadeekorn}, \bibfnamefont{N.}},
  \bibinfo {author} {\bibfnamefont{G.}~\bibnamefont{Froyland}},\ and\ \bibinfo
  {author} {\bibfnamefont{A.}~\bibnamefont{Monahan}}}%
  , \bibinfo {year} {2010},\ \bibfield{title}{%
  \enquote{\bibinfo {title} {Optimally coherent sets in geophysical flows: A
  transfer-operator approach to delimiting the stratospheric polar vortex},}\
  }%
  \bibfield{journal}{%
  \bibinfo {journal} {Phy. Rev. E}\ }%
  \textbf{\bibinfo {volume} {82}},\ \bibinfo {pages} {056311}%
  \bibAnnoteFile{NoStop}{santitissadeekorn_optimally_2010}%
\bibitem[{\citenamefont{Schekochihin}\
  \emph{et~al.}(2004)\citenamefont{Schekochihin}, \citenamefont{Haynes},\ and\
  \citenamefont{Cowley}}]{Schekochihin2004}%
  \BibitemOpen
  \bibfield{author}{%
  \bibinfo {author} {\bibnamefont{Schekochihin}, \bibfnamefont{A.~A.}},
  \bibinfo {author} {\bibfnamefont{P.~H.}\ \bibnamefont{Haynes}},\ and\
  \bibinfo {author} {\bibfnamefont{S.~C.}\ \bibnamefont{Cowley}}}%
  , \bibinfo {year} {2004},\ \bibfield{title}{%
  \enquote{\bibinfo {title} {Diffusion of passive scalar in a finite-scale
  random flow},}\ }%
  \bibfield{journal}{%
  \bibinfo {journal} {Phys. Rev. E}\ }%
  \textbf{\bibinfo {volume} {70}},\ \bibinfo {pages} {046304}%
  \bibAnnoteFile{NoStop}{Schekochihin2004}%
\bibitem[{\citenamefont{Schelin}\ \emph{et~al.}(2009)\citenamefont{Schelin},
  \citenamefont{Karolyi}, \citenamefont{de~Moura}, \citenamefont{Booth},\ and\
  \citenamefont{Grebogi}}]{schelin2009}%
  \BibitemOpen
  \bibfield{author}{%
  \bibinfo {author} {\bibnamefont{Schelin}, \bibfnamefont{A.~B.}}, \bibinfo
  {author} {\bibfnamefont{G.}~\bibnamefont{Karolyi}}, \bibinfo {author}
  {\bibfnamefont{A.~P.~S.}\ \bibnamefont{de~Moura}}, \bibinfo {author}
  {\bibfnamefont{N.~A.}\ \bibnamefont{Booth}},\ and\ \bibinfo {author}
  {\bibfnamefont{C.}~\bibnamefont{Grebogi}}}%
  , \bibinfo {year} {2009},\ \bibfield{title}{%
  \enquote{\bibinfo {title} {Chaotic advection in blood flow},}\ }%
  \bibfield{journal}{%
  \bibinfo {journal} {Phys. Rev. E.}\ }%
  \textbf{\bibinfo {volume} {80}},\ \bibinfo {pages} {016213}%
  \bibAnnoteFile{NoStop}{schelin2009}%
\bibitem[{\citenamefont{Scheuring}\
  \emph{et~al.}(2003)\citenamefont{Scheuring}, \citenamefont{K\'arolyi},
  \citenamefont{Toroczkai}, \citenamefont{T\'el},\ and\
  \citenamefont{P\'entek}}]{Scheuring-et-al-03}%
  \BibitemOpen
  \bibfield{author}{%
  \bibinfo {author} {\bibnamefont{Scheuring}, \bibfnamefont{I.}}, \bibinfo
  {author} {\bibfnamefont{G.}~\bibnamefont{K\'arolyi}}, \bibinfo {author}
  {\bibfnamefont{Z.}~\bibnamefont{Toroczkai}}, \bibinfo {author}
  {\bibfnamefont{T.}~\bibnamefont{T\'el}},\ and\ \bibinfo {author}
  {\bibfnamefont{\'A.}\ \bibnamefont{P\'entek}}}%
  , \bibinfo {year} {2003},\ \bibfield{title}{%
  \enquote{\bibinfo {title} {Competing populations in flows with chaotic
  mixing},}\ }%
  \bibfield{journal}{%
  \bibinfo {journal} {Theor. Pop. Biol.}\ }%
  \textbf{\bibinfo {volume} {63}},\ \bibinfo {pages} {77--90}%
  \bibAnnoteFile{NoStop}{Scheuring-et-al-03}%
\bibitem[{\citenamefont{Sebille}\ \emph{et~al.}(2012)\citenamefont{Sebille},
  \citenamefont{{England}},\ and\ \citenamefont{{Froyland}}}]{Sebille2012}%
  \BibitemOpen
  \bibfield{author}{%
  \bibinfo {author} {\bibnamefont{Sebille}, \bibfnamefont{{E.}~van}}, \bibinfo
  {author} {\bibfnamefont{{M.}~{H}.}\ \bibnamefont{{England}}},\ and\ \bibinfo
  {author} {\bibfnamefont{{G.}}~\bibnamefont{{Froyland}}}}%
  , \bibinfo {year} {2012},\ \bibfield{title}{%
  \enquote{\bibinfo {title} {Origin, dynamics and evolution of ocean garbage
  patches from observed surface drifters},}\ }%
  \bibfield{journal}{%
  \bibinfo {journal} {Environ. {Res.} {Lett.}}\ }%
  \textbf{\bibinfo {volume} {7}},\ \bibinfo {pages} {044040}%
  \bibAnnoteFile{NoStop}{Sebille2012}%
\bibitem[{\citenamefont{Shadden}\ \emph{et~al.}(2005)\citenamefont{Shadden},
  \citenamefont{Lekien},\ and\ \citenamefont{Marsden}}]{Shadden:2005vn}%
  \BibitemOpen
  \bibfield{author}{%
  \bibinfo {author} {\bibnamefont{Shadden}, \bibfnamefont{S.~C.}}, \bibinfo
  {author} {\bibfnamefont{F.}~\bibnamefont{Lekien}},\ and\ \bibinfo {author}
  {\bibfnamefont{J.~E.}\ \bibnamefont{Marsden}}}%
  , \bibinfo {year} {2005},\ \bibfield{title}{%
  \enquote{\bibinfo {title} {Definition and properties of lagrangian coherent
  structures from finite-time {Lyapunov} exponents in two-dimensional aperiodic
  flows},}\ }%
  \bibfield{journal}{%
  \bibinfo {journal} {Physica D}\ }%
  \textbf{\bibinfo {volume} {212}},\ \bibinfo {pages} {271--304}%
  \bibAnnoteFile{NoStop}{Shadden:2005vn}%
\bibitem[{\citenamefont{Shadden}\ \emph{et~al.}(2009)\citenamefont{Shadden},
  \citenamefont{Lekien}, \citenamefont{Paduan}, \citenamefont{Chavez},\ and\
  \citenamefont{Marsden}}]{Shadden:2009cn}%
  \BibitemOpen
  \bibfield{author}{%
  \bibinfo {author} {\bibnamefont{Shadden}, \bibfnamefont{S.~C.}}, \bibinfo
  {author} {\bibfnamefont{F.}~\bibnamefont{Lekien}}, \bibinfo {author}
  {\bibfnamefont{J.~D.}\ \bibnamefont{Paduan}}, \bibinfo {author}
  {\bibfnamefont{F.~P.}\ \bibnamefont{Chavez}},\ and\ \bibinfo {author}
  {\bibfnamefont{J.~E.}\ \bibnamefont{Marsden}}}%
  , \bibinfo {year} {2009},\ \bibfield{title}{%
  \enquote{\bibinfo {title} {The correlation between surface drifters and
  coherent structures based on high-frequency radar data in monterey bay},}\ }%
  \bibfield{journal}{%
  \bibinfo {journal} {{Deep-Sea} Res. II}\ }%
  \textbf{\bibinfo {volume} {56}},\ \bibinfo {pages} {161--172}%
  \bibAnnoteFile{NoStop}{Shadden:2009cn}%
\bibitem[{\citenamefont{Shankar}(1997)}]{Shankar1997}%
  \BibitemOpen
  \bibfield{author}{%
  \bibinfo {author} {\bibnamefont{Shankar}, \bibfnamefont{P.~N.}}}%
  , \bibinfo {year} {1997},\ \bibfield{title}{%
  \enquote{\bibinfo {title} {Three-dimensional eddy structure in a cylindrical
  container},}\ }%
  \bibfield{journal}{%
  \bibinfo {journal} {J. Fluid Mech.}\ }%
  \textbf{\bibinfo {volume} {342}},\ \bibinfo {pages} {97--118}%
  \bibAnnoteFile{NoStop}{Shankar1997}%
\bibitem[{\citenamefont{Shankar}(1998)}]{Shankar1998}%
  \BibitemOpen
  \bibfield{author}{%
  \bibinfo {author} {\bibnamefont{Shankar}, \bibfnamefont{P.~N.}}}%
  , \bibinfo {year} {1998},\ \bibfield{title}{%
  \enquote{\bibinfo {title} {Three-dimensional {Stokes} flow in a cylindrical
  container},}\ }%
  \bibfield{journal}{%
  \bibinfo {journal} {Phys. Fluids}\ }%
  \textbf{\bibinfo {volume} {10}},\ \bibinfo {pages} {540--549}%
  \bibAnnoteFile{NoStop}{Shankar1998}%
\bibitem[{\citenamefont{Shankar}\ and\
  \citenamefont{Deshpande}(2000)}]{Shankar2000}%
  \BibitemOpen
  \bibfield{author}{%
  \bibinfo {author} {\bibnamefont{Shankar}, \bibfnamefont{P.~N.}},\ and\
  \bibinfo {author} {\bibfnamefont{M.~D.}\ \bibnamefont{Deshpande}}}%
  , \bibinfo {year} {2000},\ \bibfield{title}{%
  \enquote{\bibinfo {title} {Fluid mechanics in the driven cavity},}\ }%
  \bibfield{journal}{%
  \bibinfo {journal} {Annu. Rev. Fluid Mech.}\ }%
  \textbf{\bibinfo {volume} {32}},\ \bibinfo {pages} {93--136}%
  \bibAnnoteFile{NoStop}{Shankar2000}%
\bibitem[{\citenamefont{Shapere}\ and\
  \citenamefont{Wilczek}(1989{\natexlab{a}})}]{shapere}%
  \BibitemOpen
  \bibfield{author}{%
  \bibinfo {author} {\bibnamefont{Shapere}, \bibfnamefont{A.}},\ and\ \bibinfo
  {author} {\bibnamefont{Wilczek}}}%
  , \bibinfo {year} {1989}{\natexlab{a}},\ \emph{\bibinfo {title} {{Geometric
  phases in physics}}}\ (\bibinfo {publisher} {World Scientific})%
  \bibAnnoteFile{NoStop}{shapere}%
\bibitem[{\citenamefont{Shapere}\ and\
  \citenamefont{Wilczek}(1989{\natexlab{b}})}]{shapere2}%
  \BibitemOpen
  \bibfield{author}{%
  \bibinfo {author} {\bibnamefont{Shapere}, \bibfnamefont{A.}},\ and\ \bibinfo
  {author} {\bibfnamefont{F.}~\bibnamefont{Wilczek}}}%
  , \bibinfo {year} {1989}{\natexlab{b}},\ \bibfield{title}{%
  \enquote{\bibinfo {title} {Geometry of self-propulsion at low {Reynolds}
  number},}\ }%
  \bibfield{journal}{%
  \bibinfo {journal} {J. Fluid Mech.}\ }%
  \textbf{\bibinfo {volume} {198}},\ \bibinfo {pages} {557--585}%
  \bibAnnoteFile{NoStop}{shapere2}%
\bibitem[{\citenamefont{Shapiro}\ \emph{et~al.}(2014)\citenamefont{Shapiro},
  \citenamefont{Fernandez}, \citenamefont{Garren}, \citenamefont{Guasto},
  \citenamefont{Debaillon-Vesque}, \citenamefont{Kramarsky-Winter},
  \citenamefont{Vardi},\ and\ \citenamefont{Stocker}}]{Shapiro2014}%
  \BibitemOpen
  \bibfield{author}{%
  \bibinfo {author} {\bibnamefont{Shapiro}, \bibfnamefont{O.~H.}}, \bibinfo
  {author} {\bibfnamefont{V.~I.}\ \bibnamefont{Fernandez}}, \bibinfo {author}
  {\bibfnamefont{M.}~\bibnamefont{Garren}}, \bibinfo {author}
  {\bibfnamefont{J.~S.}\ \bibnamefont{Guasto}}, \bibinfo {author}
  {\bibfnamefont{F.~P.}\ \bibnamefont{Debaillon-Vesque}}, \bibinfo {author}
  {\bibfnamefont{E.}~\bibnamefont{Kramarsky-Winter}}, \bibinfo {author}
  {\bibfnamefont{A.}~\bibnamefont{Vardi}},\ and\ \bibinfo {author}
  {\bibfnamefont{R.}~\bibnamefont{Stocker}}}%
  , \bibinfo {year} {2014},\ \bibfield{title}{%
  \enquote{\bibinfo {title} {Vortical ciliary flows actively enhance mass
  transport in reef corals},}\ }%
  \bibfield{journal}{%
  \bibinfo {journal} {Proc. Natl Acad. Sci. USA}\ }%
  \textbf{\bibinfo {volume} {111}},\ \bibinfo {pages} {13391--13396}%
  \bibAnnoteFile{NoStop}{Shapiro2014}%
\bibitem[{\citenamefont{Shub}(2005)}]{shub2005}%
  \BibitemOpen
  \bibfield{author}{%
  \bibinfo {author} {\bibnamefont{Shub}, \bibfnamefont{M.}}}%
  , \bibinfo {year} {2005},\ \bibfield{title}{%
  \enquote{\bibinfo {title} {What is a horseshoe?}.}\ }%
  \bibfield{journal}{%
  \bibinfo {journal} {Notices of the AMS}\ }%
  \textbf{\bibinfo {volume} {52}},\ \bibinfo {pages} {516--517}%
  \bibAnnoteFile{NoStop}{shub2005}%
\bibitem[{\citenamefont{Smale}(1967)}]{smale1967}%
  \BibitemOpen
  \bibfield{author}{%
  \bibinfo {author} {\bibnamefont{Smale}, \bibfnamefont{S.}}}%
  , \bibinfo {year} {1967},\ \bibfield{title}{%
  \enquote{\bibinfo {title} {Differentiable dynamical systems},}\ }%
  \bibfield{journal}{%
  \bibinfo {journal} {Bull. Amer. Math. Soc.}\ }%
  \textbf{\bibinfo {volume} {73}},\ \bibinfo {pages} {747--817}%
  \bibAnnoteFile{NoStop}{smale1967}%
\bibitem[{\citenamefont{Smale}(1998)}]{smale1998}%
  \BibitemOpen
  \bibfield{author}{%
  \bibinfo {author} {\bibnamefont{Smale}, \bibfnamefont{S.}}}%
  , \bibinfo {year} {1998},\ \bibfield{title}{%
  \enquote{\bibinfo {title} {Finding a horseshoe on the beaches of {Rio}},}\ }%
  \bibfield{journal}{%
  \bibinfo {journal} {Math. Intelligencer}\ }%
  \textbf{\bibinfo {volume} {20}},\ \bibinfo {pages} {39--44}%
  \bibAnnoteFile{NoStop}{smale1998}%
\bibitem[{\citenamefont{Smith}\ \emph{et~al.}(2012)\citenamefont{Smith},
  \citenamefont{Johnson}, \citenamefont{Smith},\ and\
  \citenamefont{Blake}}]{Smith12}%
  \BibitemOpen
  \bibfield{author}{%
  \bibinfo {author} {\bibnamefont{Smith}, \bibfnamefont{A.~A.}}, \bibinfo
  {author} {\bibfnamefont{T.~D.}\ \bibnamefont{Johnson}}, \bibinfo {author}
  {\bibfnamefont{D.~J.}\ \bibnamefont{Smith}},\ and\ \bibinfo {author}
  {\bibfnamefont{J.~R.}\ \bibnamefont{Blake}}}%
  , \bibinfo {year} {2012},\ \bibfield{title}{%
  \enquote{\bibinfo {title} {Symmetry-breaking cilia driven flow in the
  zebrafish embryo},}\ }%
  \bibfield{journal}{%
  \bibinfo {journal} {J. Fluid Mech.}\ }%
  \textbf{\bibinfo {volume} {705}},\ \bibinfo {pages} {26--45}%
  \bibAnnoteFile{NoStop}{Smith12}%
\bibitem[{\citenamefont{Smith}\ \emph{et~al.}(2016)\citenamefont{Smith},
  \citenamefont{Rudman}, \citenamefont{Lester},\ and\
  \citenamefont{Metcalfe}}]{Smith_discontinuous_2016}%
  \BibitemOpen
  \bibfield{author}{%
  \bibinfo {author} {\bibnamefont{Smith}, \bibfnamefont{L.~D.}}, \bibinfo
  {author} {\bibfnamefont{M.}~\bibnamefont{Rudman}}, \bibinfo {author}
  {\bibfnamefont{D.~R.}\ \bibnamefont{Lester}},\ and\ \bibinfo {author}
  {\bibfnamefont{G.}~\bibnamefont{Metcalfe}}}%
  , \bibinfo {year} {2016},\ \bibfield{title}{%
  \enquote{\bibinfo {title} {Mixing of discontinuously deforming media},}\ }%
  \bibfield{journal}{%
  \bibinfo {journal} {Chaos}\ }%
  \textbf{\bibinfo {volume} {26}},\ \bibinfo {pages} {023113}%
  \bibAnnoteFile{NoStop}{Smith_discontinuous_2016}%
\bibitem[{\citenamefont{Solari}\ \emph{et~al.}(2006)\citenamefont{Solari},
  \citenamefont{Ganguly}, \citenamefont{Kessler}, \citenamefont{Michod},\ and\
  \citenamefont{Goldstein}}]{Solari2006}%
  \BibitemOpen
  \bibfield{author}{%
  \bibinfo {author} {\bibnamefont{Solari}, \bibfnamefont{C.~A.}}, \bibinfo
  {author} {\bibfnamefont{S.}~\bibnamefont{Ganguly}}, \bibinfo {author}
  {\bibfnamefont{J.~O.}\ \bibnamefont{Kessler}}, \bibinfo {author}
  {\bibfnamefont{R.~E.}\ \bibnamefont{Michod}},\ and\ \bibinfo {author}
  {\bibfnamefont{R.~E.}\ \bibnamefont{Goldstein}}}%
  , \bibinfo {year} {2006},\ \bibfield{title}{%
  \enquote{\bibinfo {title} {Multicellularity and the functional
  interdependence of motility and molecular transport},}\ }%
  \bibfield{journal}{%
  \bibinfo {journal} {Proc. Natl Acad. Sci. USA}\ }%
  \textbf{\bibinfo {volume} {103}},\ \bibinfo {pages} {1353--1358}%
  \bibAnnoteFile{NoStop}{Solari2006}%
\bibitem[{\citenamefont{Solomon}\ and\
  \citenamefont{Mezi\'{c}}(2003)}]{Solomon2003}%
  \BibitemOpen
  \bibfield{author}{%
  \bibinfo {author} {\bibnamefont{Solomon}, \bibfnamefont{T.~H.}},\ and\
  \bibinfo {author} {\bibfnamefont{I.}~\bibnamefont{Mezi\'{c}}}}%
  , \bibinfo {year} {2003},\ \bibfield{title}{%
  \enquote{\bibinfo {title} {Uniform resonant chaotic mixing in fluid flows},}\
  }%
  \bibfield{journal}{%
  \bibinfo {journal} {Nature}\ }%
  \textbf{\bibinfo {volume} {425}},\ \bibinfo {pages} {376--380}%
  \bibAnnoteFile{NoStop}{Solomon2003}%
\bibitem[{\citenamefont{Solomon}\ \emph{et~al.}(1994)\citenamefont{Solomon},
  \citenamefont{Weeks},\ and\ \citenamefont{Swinney}}]{Solomon94}%
  \BibitemOpen
  \bibfield{author}{%
  \bibinfo {author} {\bibnamefont{Solomon}, \bibfnamefont{T.~H.}}, \bibinfo
  {author} {\bibfnamefont{E.~R.}\ \bibnamefont{Weeks}},\ and\ \bibinfo {author}
  {\bibfnamefont{H.~L.}\ \bibnamefont{Swinney}}}%
  , \bibinfo {year} {1994},\ \bibfield{title}{%
  \enquote{\bibinfo {title} {Chaotic advection in a two-dimensional flow:
  L{\'e}vy flights and anomalous diffusion},}\ }%
  \bibinfo {journal} {Physica D},\ \bibinfo {pages} {70--84}%
  \bibAnnoteFile{NoStop}{Solomon94}%
\bibitem[{\citenamefont{Solomon}\ \emph{et~al.}(2009)\citenamefont{Solomon},
  \citenamefont{Lee},\ and\ \citenamefont{Fogleman}}]{solomon2001}%
  \BibitemOpen
\bibfield{journal}{%
    }%
  \bibfield{author}{%
  \bibinfo {author} {\bibnamefont{Solomon}, \bibfnamefont{T.H.}}, \bibinfo
  {author} {\bibfnamefont{A.T.}\ \bibnamefont{Lee}},\ and\ \bibinfo {author}
  {\bibfnamefont{M.A.}\ \bibnamefont{Fogleman}}}%
  , \bibinfo {year} {2009},\ \bibfield{title}{%
  \enquote{\bibinfo {title} {Resonant flights and transient superdiffusion in a
  time-periodic, two-dimensional flow},}\ }%
  \bibfield{journal}{%
  \bibinfo {journal} {Physica D}\ }%
  \textbf{\bibinfo {volume} {157}},\ \bibinfo {pages} {40--53}%
  \bibAnnoteFile{NoStop}{solomon2001}%
\bibitem[{\citenamefont{Sommerer}\ \emph{et~al.}(1996)\citenamefont{Sommerer},
  \citenamefont{Ku},\ and\ \citenamefont{Gilreath}}]{sommerer}%
  \BibitemOpen
  \bibfield{author}{%
  \bibinfo {author} {\bibnamefont{Sommerer}, \bibfnamefont{J.~C.}}, \bibinfo
  {author} {\bibfnamefont{H.~C.}\ \bibnamefont{Ku}},\ and\ \bibinfo {author}
  {\bibfnamefont{H.~E.}\ \bibnamefont{Gilreath}}}%
  , \bibinfo {year} {1996},\ \bibfield{title}{%
  \enquote{\bibinfo {title} {Experimental evidence for chaotic scattering in a
  fluid wake},}\ }%
  \bibfield{journal}{%
  \bibinfo {journal} {Phys. Rev. Lett.}\ }%
  \textbf{\bibinfo {volume} {77}},\ \bibinfo {pages} {5055--5058}%
  \bibAnnoteFile{NoStop}{sommerer}%
\bibitem[{\citenamefont{Song}\
  \emph{et~al.}(2003{\natexlab{a}})\citenamefont{Song}, \citenamefont{Bringer},
  \citenamefont{D.}, \citenamefont{Gerdts},\ and\
  \citenamefont{Ismagilov}}]{Song03}%
  \BibitemOpen
  \bibfield{author}{%
  \bibinfo {author} {\bibnamefont{Song}, \bibfnamefont{H.}}, \bibinfo {author}
  {\bibfnamefont{M.~R.}\ \bibnamefont{Bringer}}, \bibinfo {author}
  {\bibfnamefont{Tice~J.}\ \bibnamefont{D.}}, \bibinfo {author}
  {\bibfnamefont{C.~J.}\ \bibnamefont{Gerdts}},\ and\ \bibinfo {author}
  {\bibfnamefont{R.~F.}\ \bibnamefont{Ismagilov}}}%
  , \bibinfo {year} {2003}{\natexlab{a}},\ \bibfield{title}{%
  \enquote{\bibinfo {title} {Experimental test of scaling of mixing by chaotic
  advection in droplets moving through microfluidic channels},}\ }%
  \bibfield{journal}{%
  \bibinfo {journal} {Appl. Phys. Lett.}\ }%
  \textbf{\bibinfo {volume} {83}},\ \bibinfo {pages} {4664--4666}%
  \bibAnnoteFile{NoStop}{Song03}%
\bibitem[{\citenamefont{Song}\
  \emph{et~al.}(2003{\natexlab{b}})\citenamefont{Song}, \citenamefont{Bringer},
  \citenamefont{Tice}, \citenamefont{Gerdts},\ and\
  \citenamefont{Ismagilov}}]{Song2003}%
  \BibitemOpen
  \bibfield{author}{%
  \bibinfo {author} {\bibnamefont{Song}, \bibfnamefont{H.}}, \bibinfo {author}
  {\bibfnamefont{M.~R.}\ \bibnamefont{Bringer}}, \bibinfo {author}
  {\bibfnamefont{J.~D.}\ \bibnamefont{Tice}}, \bibinfo {author}
  {\bibfnamefont{C.~J.}\ \bibnamefont{Gerdts}},\ and\ \bibinfo {author}
  {\bibfnamefont{R.~F.}\ \bibnamefont{Ismagilov}}}%
  , \bibinfo {year} {2003}{\natexlab{b}},\ \bibfield{title}{%
  \enquote{\bibinfo {title} {Experimental test of scaling of mixing by chaotic
  advection in droplets moving through microfluidic channels},}\ }%
  \bibfield{journal}{%
  \bibinfo {journal} {Appl. Phys. Lett.}\ }%
  \textbf{\bibinfo {volume} {83}},\ \bibinfo {pages} {4664--4666}%
  \bibAnnoteFile{NoStop}{Song2003}%
\bibitem[{\citenamefont{Song}\ \emph{et~al.}(2010)\citenamefont{Song},
  \citenamefont{Cai}, \citenamefont{Noh},\ and\
  \citenamefont{Bennett}}]{Song2010}%
  \BibitemOpen
  \bibfield{author}{%
  \bibinfo {author} {\bibnamefont{Song}, \bibfnamefont{H.}}, \bibinfo {author}
  {\bibfnamefont{Z.}~\bibnamefont{Cai}}, \bibinfo {author}
  {\bibfnamefont{H.~M.}\ \bibnamefont{Noh}},\ and\ \bibinfo {author}
  {\bibfnamefont{D.~J.}\ \bibnamefont{Bennett}}}%
  , \bibinfo {year} {2010},\ \bibfield{title}{%
  \enquote{\bibinfo {title} {Chaotic mixing in microchannels via low frequency
  switching transverse electroosmotic flow generated on integrated
  microelectrodes},}\ }%
  \bibfield{journal}{%
  \bibinfo {journal} {Lab on a chip}\ }%
  \textbf{\bibinfo {volume} {10}},\ \bibinfo {pages} {734--740}%
  \bibAnnoteFile{NoStop}{Song2010}%
\bibitem[{\citenamefont{Sotiropoulos}\
  \emph{et~al.}(2002)\citenamefont{Sotiropoulos}, \citenamefont{Webster},\ and\
  \citenamefont{Lackey}}]{sotiropoulos2002experiments}%
  \BibitemOpen
  \bibfield{author}{%
  \bibinfo {author} {\bibnamefont{Sotiropoulos}, \bibfnamefont{F.}}, \bibinfo
  {author} {\bibfnamefont{D.~R.}\ \bibnamefont{Webster}},\ and\ \bibinfo
  {author} {\bibfnamefont{T.~C.}\ \bibnamefont{Lackey}}}%
  , \bibinfo {year} {2002},\ \bibfield{title}{%
  \enquote{\bibinfo {title} {Experiments on lagrangian transport in steady
  vortex-breakdown bubbles in a confined swirling flow},}\ }%
  \bibfield{journal}{%
  \bibinfo {journal} {J. Fluid Mech.}\ }%
  \textbf{\bibinfo {volume} {466}},\ \bibinfo {pages} {215--248}%
  \bibAnnoteFile{NoStop}{sotiropoulos2002experiments}%
\bibitem[{\citenamefont{Speetjens}\
  \emph{et~al.}(2006{\natexlab{a}})\citenamefont{Speetjens},
  \citenamefont{Metcalfe},\ and\
  \citenamefont{Rudman}}]{Speetjens_symmetry_2006}%
  \BibitemOpen
  \bibfield{author}{%
  \bibinfo {author} {\bibnamefont{Speetjens}, \bibfnamefont{M.}}, \bibinfo
  {author} {\bibfnamefont{G.}~\bibnamefont{Metcalfe}},\ and\ \bibinfo {author}
  {\bibfnamefont{M.}~\bibnamefont{Rudman}}}%
  , \bibinfo {year} {2006}{\natexlab{a}},\ \bibfield{title}{%
  \enquote{\bibinfo {title} {Topological mixing study of non-{N}ewtonian duct
  flows},}\ }%
  \bibfield{journal}{%
  \bibinfo {journal} {Phys. Fluids}\ }%
  \textbf{\bibinfo {volume} {18}},\ \bibinfo {pages} {103103}%
  \bibAnnoteFile{NoStop}{Speetjens_symmetry_2006}%
\bibitem[{\citenamefont{Speetjens}\
  \emph{et~al.}(2006{\natexlab{b}})\citenamefont{Speetjens},
  \citenamefont{Rudman},\ and\
  \citenamefont{Metcalfe}}]{Speetjens_regime_2006}%
  \BibitemOpen
  \bibfield{author}{%
  \bibinfo {author} {\bibnamefont{Speetjens}, \bibfnamefont{M.}}, \bibinfo
  {author} {\bibfnamefont{M.}~\bibnamefont{Rudman}},\ and\ \bibinfo {author}
  {\bibfnamefont{G.}~\bibnamefont{Metcalfe}}}%
  , \bibinfo {year} {2006}{\natexlab{b}},\ \bibfield{title}{%
  \enquote{\bibinfo {title} {Flow regime analysis of non-{N}ewtonian duct
  flows},}\ }%
  \bibfield{journal}{%
  \bibinfo {journal} {Phys. Fluids}\ }%
  \textbf{\bibinfo {volume} {18}},\ \bibinfo {pages} {013101}%
  \bibAnnoteFile{NoStop}{Speetjens_regime_2006}%
\bibitem[{\citenamefont{Speetjens}(2001)}]{michelthesis}%
  \BibitemOpen
  \bibfield{author}{%
  \bibinfo {author} {\bibnamefont{Speetjens}, \bibfnamefont{M.~F.~M.}}}%
  , \bibinfo {year} {2001},\ \emph{\bibinfo {title} {Three-dimensional Chaotic
  Advection in a Cylindrical Domain}},\ Ph.D. thesis\ (\bibinfo {school}
  {Eindhoven University of Technology})%
  \bibAnnoteFile{NoStop}{michelthesis}%
\bibitem[{\citenamefont{Speetjens}\ and\
  \citenamefont{Clercx}(2013)}]{Speetjens2013}%
  \BibitemOpen
  \bibfield{author}{%
  \bibinfo {author} {\bibnamefont{Speetjens}, \bibfnamefont{M.~F.~M.}},\ and\
  \bibinfo {author} {\bibfnamefont{H.~J.~H.}\ \bibnamefont{Clercx}}}%
  , \bibinfo {year} {2013},\ \enquote{\bibinfo {title} {Formation of coherent
  structures in a class of realistic {3D} unsteady flows},}\ in\ \emph{\bibinfo
  {booktitle} {Fluid Dynamics in Physics, Engineering and Environmental
  Applications, Part 1}},\ \bibinfo {editor} {edited by\ \bibinfo {editor}
  {\bibfnamefont{J.}~\bibnamefont{Klapp}}, \bibinfo {editor}
  {\bibfnamefont{A.}~\bibnamefont{Medina}}, \bibinfo {editor}
  {\bibfnamefont{A.}~\bibnamefont{Cros}},\ and\ \bibinfo {editor}
  {\bibfnamefont{C.}~\bibnamefont{Vargas}}}\ (\bibinfo {publisher} {Springer})\
  pp.\ \bibinfo {pages} {139--157}%
  \bibAnnoteFile{NoStop}{Speetjens2013}%
\bibitem[{\citenamefont{Speetjens}\
  \emph{et~al.}(2004)\citenamefont{Speetjens}, \citenamefont{Clercx},\ and\
  \citenamefont{van Heijst}}]{Michel2004}%
  \BibitemOpen
  \bibfield{author}{%
  \bibinfo {author} {\bibnamefont{Speetjens}, \bibfnamefont{M.~F.~M.}},
  \bibinfo {author} {\bibfnamefont{H.~J.~H.}\ \bibnamefont{Clercx}},\ and\
  \bibinfo {author} {\bibfnamefont{G.~J.~F.}\ \bibnamefont{van Heijst}}}%
  , \bibinfo {year} {2004},\ \bibfield{title}{%
  \enquote{\bibinfo {title} {A numerical and experimental study on advection in
  three-dimensional {Stokes} flows},}\ }%
  \bibfield{journal}{%
  \bibinfo {journal} {J.~Fluid Mech.}\ }%
  \textbf{\bibinfo {volume} {514}},\ \bibinfo {pages} {77--105}%
  \bibAnnoteFile{NoStop}{Michel2004}%
\bibitem[{\citenamefont{Speetjens}\
  \emph{et~al.}(2006{\natexlab{c}})\citenamefont{Speetjens},
  \citenamefont{Clercx},\ and\ \citenamefont{van Heijst}}]{Speetjens06b}%
  \BibitemOpen
  \bibfield{author}{%
  \bibinfo {author} {\bibnamefont{Speetjens}, \bibfnamefont{M.~F.~M.}},
  \bibinfo {author} {\bibfnamefont{H.~J.~H.}\ \bibnamefont{Clercx}},\ and\
  \bibinfo {author} {\bibfnamefont{G.~J.~F.}\ \bibnamefont{van Heijst}}}%
  , \bibinfo {year} {2006}{\natexlab{c}},\ \bibfield{title}{%
  \enquote{\bibinfo {title} {Inertia-induced coherent structures in a
  time-periodic viscous mixing flow},}\ }%
  \bibfield{journal}{%
  \bibinfo {journal} {Phys. Fluids}\ }%
  \textbf{\bibinfo {volume} {18}},\ \bibinfo {pages} {083603}%
  \bibAnnoteFile{NoStop}{Speetjens06b}%
\bibitem[{\citenamefont{Speetjens}\
  \emph{et~al.}(2006{\natexlab{d}})\citenamefont{Speetjens},
  \citenamefont{Clercx},\ and\ \citenamefont{van Heijst}}]{MichelChaos}%
  \BibitemOpen
  \bibfield{author}{%
  \bibinfo {author} {\bibnamefont{Speetjens}, \bibfnamefont{M.~F.~M.}},
  \bibinfo {author} {\bibfnamefont{H.~J.~H.}\ \bibnamefont{Clercx}},\ and\
  \bibinfo {author} {\bibfnamefont{G.~J.~F.}\ \bibnamefont{van Heijst}}}%
  , \bibinfo {year} {2006}{\natexlab{d}},\ \bibfield{title}{%
  \enquote{\bibinfo {title} {Merger of coherent structures in time-periodic
  viscous flows},}\ }%
  \bibfield{journal}{%
  \bibinfo {journal} {Chaos}\ }%
  \textbf{\bibinfo {volume} {16}},\ \bibinfo {pages} {0431104}%
  \bibAnnoteFile{NoStop}{MichelChaos}%
\bibitem[{\citenamefont{Springham}\ and\
  \citenamefont{Sturman}(2014)}]{springham2012polynomial}%
  \BibitemOpen
  \bibfield{author}{%
  \bibinfo {author} {\bibnamefont{Springham}, \bibfnamefont{J.}},\ and\
  \bibinfo {author} {\bibfnamefont{R.}~\bibnamefont{Sturman}}}%
  , \bibinfo {year} {2014},\ \bibfield{title}{%
  \enquote{\bibinfo {title} {Polynomial decay of correlations in linked--twist
  maps},}\ }%
  \bibfield{journal}{%
  \bibinfo {journal} {Ergod. Theory Dyn. Sys.}\ }%
  \textbf{\bibinfo {volume} {34}},\ \bibinfo {pages} {1724--1746}%
  \bibAnnoteFile{NoStop}{springham2012polynomial}%
\bibitem[{\citenamefont{Squires}\ and\
  \citenamefont{Quake}(2005)}]{Squires2005}%
  \BibitemOpen
  \bibfield{author}{%
  \bibinfo {author} {\bibnamefont{Squires}, \bibfnamefont{T.}},\ and\ \bibinfo
  {author} {\bibfnamefont{S.}~\bibnamefont{Quake}}}%
  , \bibinfo {year} {2005},\ \bibfield{title}{%
  \enquote{\bibinfo {title} {Microfluidics: Fluid physics at the nanoliter
  scale},}\ }%
  \bibfield{journal}{%
  \bibinfo {journal} {Rev. Mod. Phys.}\ }%
  \textbf{\bibinfo {volume} {77}},\ \bibinfo {pages} {977--1026}%
  \bibAnnoteFile{NoStop}{Squires2005}%
\bibitem[{\citenamefont{Stremler}\ and\
  \citenamefont{Cola}(2006)}]{stremler2006}%
  \BibitemOpen
  \bibfield{author}{%
  \bibinfo {author} {\bibnamefont{Stremler}, \bibfnamefont{M.~A.}},\ and\
  \bibinfo {author} {\bibfnamefont{B.~A.}\ \bibnamefont{Cola}}}%
  , \bibinfo {year} {2006},\ \bibfield{title}{%
  \enquote{\bibinfo {title} {A maximum entropy approach to optimal mixing in a
  pulsed source-sink flow},}\ }%
  \bibfield{journal}{%
  \bibinfo {journal} {Phys. Fluids}\ }%
  \textbf{\bibinfo {volume} {18}},\ \bibinfo {pages} {011701}%
  \bibAnnoteFile{NoStop}{stremler2006}%
\bibitem[{\citenamefont{Stremler}\ \emph{et~al.}(2011)\citenamefont{Stremler},
  \citenamefont{Ross}, \citenamefont{Grover},\ and\
  \citenamefont{Kumar}}]{Stremler:2011hu}%
  \BibitemOpen
  \bibfield{author}{%
  \bibinfo {author} {\bibnamefont{Stremler}, \bibfnamefont{M.~A.}}, \bibinfo
  {author} {\bibfnamefont{S.~D.}\ \bibnamefont{Ross}}, \bibinfo {author}
  {\bibfnamefont{P.}~\bibnamefont{Grover}},\ and\ \bibinfo {author}
  {\bibfnamefont{P.}~\bibnamefont{Kumar}}}%
  , \bibinfo {year} {2011},\ \bibfield{title}{%
  \enquote{\bibinfo {title} {Topological chaos and periodic braiding of
  {Almost-Cyclic} sets},}\ }%
  \bibfield{journal}{%
  \bibinfo {journal} {Phys. Rev. Lett.}\ }%
  \textbf{\bibinfo {volume} {106}},\ \bibinfo {pages} {114101}%
  \bibAnnoteFile{NoStop}{Stremler:2011hu}%
\bibitem[{\citenamefont{Stroock}\ \emph{et~al.}(2002)\citenamefont{Stroock},
  \citenamefont{Dertinger}, \citenamefont{Ajdari}, \citenamefont{Mezi\'{c}},
  \citenamefont{Stone},\ and\ \citenamefont{Whitesides}}]{Stroock2002}%
  \BibitemOpen
  \bibfield{author}{%
  \bibinfo {author} {\bibnamefont{Stroock}, \bibfnamefont{A.~D.}}, \bibinfo
  {author} {\bibfnamefont{S.~K.~W.}\ \bibnamefont{Dertinger}}, \bibinfo
  {author} {\bibfnamefont{A.}~\bibnamefont{Ajdari}}, \bibinfo {author}
  {\bibfnamefont{I.}~\bibnamefont{Mezi\'{c}}}, \bibinfo {author}
  {\bibfnamefont{H.~A.}\ \bibnamefont{Stone}},\ and\ \bibinfo {author}
  {\bibfnamefont{G.~M.}\ \bibnamefont{Whitesides}}}%
  , \bibinfo {year} {2002},\ \bibfield{title}{%
  \enquote{\bibinfo {title} {Chaotic mixer for microchannels},}\ }%
  \bibfield{journal}{%
  \bibinfo {journal} {Science}\ }%
  \textbf{\bibinfo {volume} {295}},\ \bibinfo {pages} {647--651}%
  \bibAnnoteFile{NoStop}{Stroock2002}%
\bibitem[{\citenamefont{Sturman}(2012)}]{sturman2012role}%
  \BibitemOpen
  \bibfield{author}{%
  \bibinfo {author} {\bibnamefont{Sturman}, \bibfnamefont{R.}}}%
  , \bibinfo {year} {2012},\ \bibfield{title}{%
  \enquote{\bibinfo {title} {The role of discontinuities in mixing},}\ }%
  \bibfield{journal}{%
  \bibinfo {journal} {Adv. Appl. Mech.}\ }%
  \textbf{\bibinfo {volume} {45}},\ \bibinfo {pages} {51--90}%
  \bibAnnoteFile{NoStop}{sturman2012role}%
\bibitem[{\citenamefont{Sturman}\ \emph{et~al.}(2008)\citenamefont{Sturman},
  \citenamefont{Meier}, \citenamefont{Ottino},\ and\
  \citenamefont{Wiggins}}]{sturman08}%
  \BibitemOpen
  \bibfield{author}{%
  \bibinfo {author} {\bibnamefont{Sturman}, \bibfnamefont{R.}}, \bibinfo
  {author} {\bibfnamefont{S.~W.}\ \bibnamefont{Meier}}, \bibinfo {author}
  {\bibfnamefont{J.~M.}\ \bibnamefont{Ottino}},\ and\ \bibinfo {author}
  {\bibfnamefont{S.}~\bibnamefont{Wiggins}}}%
  , \bibinfo {year} {2008},\ \bibfield{title}{%
  \enquote{\bibinfo {title} {Linked twist map formalism in two and three
  dimensions applied to mixing in tumbled granular flows},}\ }%
  \bibfield{journal}{%
  \bibinfo {journal} {J. Fluid Mech.}\ }%
  \textbf{\bibinfo {volume} {602}},\ \bibinfo {pages} {129--174}%
  \bibAnnoteFile{NoStop}{sturman08}%
\bibitem[{\citenamefont{Sturman}\ \emph{et~al.}(2006)\citenamefont{Sturman},
  \citenamefont{Ottino},\ and\ \citenamefont{Wiggins}}]{sturman}%
  \BibitemOpen
  \bibfield{author}{%
  \bibinfo {author} {\bibnamefont{Sturman}, \bibfnamefont{R.}}, \bibinfo
  {author} {\bibfnamefont{J.~M.}\ \bibnamefont{Ottino}},\ and\ \bibinfo
  {author} {\bibfnamefont{S.}~\bibnamefont{Wiggins}}}%
  , \bibinfo {year} {2006},\ \emph{\bibinfo {title} {The mathematical
  foundation of mixing}}\ (\bibinfo {publisher} {Cambridge University Press},\
  \bibinfo {address} {Cambridge})%
  \bibAnnoteFile{NoStop}{sturman}%
\bibitem[{\citenamefont{Sturman}\ and\
  \citenamefont{Springham}(2013)}]{sturman2012rate}%
  \BibitemOpen
  \bibfield{author}{%
  \bibinfo {author} {\bibnamefont{Sturman}, \bibfnamefont{R.}},\ and\ \bibinfo
  {author} {\bibfnamefont{J.}~\bibnamefont{Springham}}}%
  , \bibinfo {year} {2013},\ \bibfield{title}{%
  \enquote{\bibinfo {title} {Rate of chaotic mixing and boundary behavior},}\
  }%
  \bibfield{journal}{%
  \bibinfo {journal} {Phys. Rev. E}\ }%
  \textbf{\bibinfo {volume} {87}},\ \bibinfo {pages} {012906}%
  \bibAnnoteFile{NoStop}{sturman2012rate}%
\bibitem[{\citenamefont{Sturman}\ and\
  \citenamefont{Wiggins}(2009)}]{sturman2009eulerian}%
  \BibitemOpen
  \bibfield{author}{%
  \bibinfo {author} {\bibnamefont{Sturman}, \bibfnamefont{R.}},\ and\ \bibinfo
  {author} {\bibfnamefont{S.}~\bibnamefont{Wiggins}}}%
  , \bibinfo {year} {2009},\ \bibfield{title}{%
  \enquote{\bibinfo {title} {Eulerian indicators for predicting and optimizing
  mixing quality},}\ }%
  \bibfield{journal}{%
  \bibinfo {journal} {New J. Phys.}\ }%
  \textbf{\bibinfo {volume} {11}},\ \bibinfo {pages} {075031}%
  \bibAnnoteFile{NoStop}{sturman2009eulerian}%
\bibitem[{\citenamefont{Sudarsan}\ and\
  \citenamefont{Ugaz}(2006{\natexlab{a}})}]{Sudarsan2006a}%
  \BibitemOpen
  \bibfield{author}{%
  \bibinfo {author} {\bibnamefont{Sudarsan}, \bibfnamefont{A.~P.}},\ and\
  \bibinfo {author} {\bibfnamefont{V.~M.}\ \bibnamefont{Ugaz}}}%
  , \bibinfo {year} {2006}{\natexlab{a}},\ \bibfield{title}{%
  \enquote{\bibinfo {title} {Fluid mixing in planar spiral microchannels},}\ }%
  \bibfield{journal}{%
  \bibinfo {journal} {Lab on a chip}\ }%
  \textbf{\bibinfo {volume} {6}},\ \bibinfo {pages} {74--82}%
  \bibAnnoteFile{NoStop}{Sudarsan2006a}%
\bibitem[{\citenamefont{Sudarsan}\ and\
  \citenamefont{Ugaz}(2006{\natexlab{b}})}]{Sudarsan2006}%
  \BibitemOpen
  \bibfield{author}{%
  \bibinfo {author} {\bibnamefont{Sudarsan}, \bibfnamefont{A.~P.}},\ and\
  \bibinfo {author} {\bibfnamefont{V.~M.}\ \bibnamefont{Ugaz}}}%
  , \bibinfo {year} {2006}{\natexlab{b}},\ \bibfield{title}{%
  \enquote{\bibinfo {title} {Multivortex micromixing},}\ }%
  \bibfield{journal}{%
  \bibinfo {journal} {Proc. Natl Acad. Sci. USA}\ }%
  \textbf{\bibinfo {volume} {103}},\ \bibinfo {pages} {7228--7233}%
  \bibAnnoteFile{NoStop}{Sudarsan2006}%
\bibitem[{\citenamefont{Supatto}\ \emph{et~al.}(2008)\citenamefont{Supatto},
  \citenamefont{Fraser},\ and\ \citenamefont{Vermot}}]{Supatto2008}%
  \BibitemOpen
  \bibfield{author}{%
  \bibinfo {author} {\bibnamefont{Supatto}, \bibfnamefont{W.}}, \bibinfo
  {author} {\bibfnamefont{S.~E.}\ \bibnamefont{Fraser}},\ and\ \bibinfo
  {author} {\bibfnamefont{J.}~\bibnamefont{Vermot}}}%
  , \bibinfo {year} {2008},\ \bibfield{title}{%
  \enquote{\bibinfo {title} {An all-optical approach for probing microscopic
  flows in living embryos},}\ }%
  \bibfield{journal}{%
  \bibinfo {journal} {Biophys. J.}\ }%
  \textbf{\bibinfo {volume} {95}},\ \bibinfo {pages} {L29--31}%
  \bibAnnoteFile{NoStop}{Supatto2008}%
\bibitem[{\citenamefont{Tabeling}\ \emph{et~al.}(2004)\citenamefont{Tabeling},
  \citenamefont{Chabert}, \citenamefont{Dodge}, \citenamefont{Jullien},\ and\
  \citenamefont{Okkels}}]{Tabeling2004}%
  \BibitemOpen
  \bibfield{author}{%
  \bibinfo {author} {\bibnamefont{Tabeling}, \bibfnamefont{P.}}, \bibinfo
  {author} {\bibfnamefont{M.}~\bibnamefont{Chabert}}, \bibinfo {author}
  {\bibfnamefont{A.}~\bibnamefont{Dodge}}, \bibinfo {author}
  {\bibfnamefont{C.}~\bibnamefont{Jullien}},\ and\ \bibinfo {author}
  {\bibfnamefont{F.}~\bibnamefont{Okkels}}}%
  , \bibinfo {year} {2004},\ \bibfield{title}{%
  \enquote{\bibinfo {title} {Chaotic mixing in cross-channel micromixers},}\ }%
  \bibfield{journal}{%
  \bibinfo {journal} {Phil. Trans. R. Soc. London A}\ }%
  \textbf{\bibinfo {volume} {362}},\ \bibinfo {pages} {987--1000}%
  \bibAnnoteFile{NoStop}{Tabeling2004}%
\bibitem[{\citenamefont{Tabeling}\ and\
  \citenamefont{Chen}(2005)}]{Tabeling2005}%
  \BibitemOpen
  \bibfield{author}{%
  \bibinfo {author} {\bibnamefont{Tabeling}, \bibfnamefont{P.}},\ and\ \bibinfo
  {author} {\bibfnamefont{S.}~\bibnamefont{Chen}}}%
  , \bibinfo {year} {2005},\ \emph{\bibinfo {title} {Introduction to
  Microfluidics}}\ (\bibinfo {publisher} {OUP, Oxford})%
  \bibAnnoteFile{NoStop}{Tabeling2005}%
\bibitem[{\citenamefont{Tabor}(1989)}]{tabor}%
  \BibitemOpen
  \bibfield{author}{%
  \bibinfo {author} {\bibnamefont{Tabor}, \bibfnamefont{M.}}}%
  , \bibinfo {year} {1989},\ \emph{\bibinfo {title} {{Chaos and integrability
  in nonlinear dynamics: an introduction}}}\ (\bibinfo {publisher} {Wiley})%
  \bibAnnoteFile{NoStop}{tabor}%
\bibitem[{\citenamefont{Tallapragada}\ and\
  \citenamefont{Ross}(2013)}]{Tallapragada:2013bh}%
  \BibitemOpen
  \bibfield{author}{%
  \bibinfo {author} {\bibnamefont{Tallapragada}, \bibfnamefont{P.}},\ and\
  \bibinfo {author} {\bibfnamefont{S.~D.}\ \bibnamefont{Ross}}}%
  , \bibinfo {year} {2013},\ \bibfield{title}{%
  \enquote{\bibinfo {title} {A set oriented definition of finite-time
  {Lyapunov} exponents and coherent sets},}\ }%
  \bibfield{journal}{%
  \bibinfo {journal} {Commun. Nonlin. Sci. Numer. Simul.}\ }%
  \textbf{\bibinfo {volume} {18}},\ \bibinfo {pages} {1106--1126}%
  \bibAnnoteFile{NoStop}{Tallapragada:2013bh}%
\bibitem[{\citenamefont{Taylor}(1951)}]{Taylor1951}%
  \BibitemOpen
  \bibfield{author}{%
  \bibinfo {author} {\bibnamefont{Taylor}, \bibfnamefont{G.~I.}}}%
  , \bibinfo {year} {1951},\ \bibfield{title}{%
  \enquote{\bibinfo {title} {{Analysis of the swimming of microscopic
  organisms}},}\ }%
  \bibfield{journal}{%
  \bibinfo {journal} {Proc. Roy. Soc. A}\ }%
  \textbf{\bibinfo {volume} {209}},\ \bibinfo {pages} {447--461}%
  \bibAnnoteFile{NoStop}{Taylor1951}%
\bibitem[{\citenamefont{Taylor}(1954)}]{Taylor1954}%
  \BibitemOpen
  \bibfield{author}{%
  \bibinfo {author} {\bibnamefont{Taylor}, \bibfnamefont{G.~I.}}}%
  , \bibinfo {year} {1954},\ \bibfield{title}{%
  \enquote{\bibinfo {title} {The dispersion of matter in turbulent flow through
  a pipe},}\ }%
  \bibfield{journal}{%
  \bibinfo {journal} {Proc. Roy. Soc. A}\ }%
  \textbf{\bibinfo {volume} {223}},\ \bibinfo {pages} {446--468}%
  \bibAnnoteFile{NoStop}{Taylor1954}%
\bibitem[{\citenamefont{Taylor}(1960)}]{taylor_film}%
  \BibitemOpen
  \bibfield{author}{%
  \bibinfo {author} {\bibnamefont{Taylor}, \bibfnamefont{G.~I.}}}%
  , \bibinfo {year} {1960},\ \enquote{\bibinfo {title} {{Low Reynolds number
  flow}},}\ \bibinfo {howpublished} {video},\ \bibinfo {note} {educational
  Services Incorporated, 16 mm film}%
  \bibAnnoteFile{NoStop}{taylor_film}%
\bibitem[{\citenamefont{Teh}\ \emph{et~al.}(2008)\citenamefont{Teh},
  \citenamefont{Lin}, \citenamefont{Hung},\ and\ \citenamefont{Lee}}]{Teh2008}%
  \BibitemOpen
  \bibfield{author}{%
  \bibinfo {author} {\bibnamefont{Teh}, \bibfnamefont{S.-Y.}}, \bibinfo
  {author} {\bibfnamefont{R.}~\bibnamefont{Lin}}, \bibinfo {author}
  {\bibfnamefont{L.-H.}\ \bibnamefont{Hung}},\ and\ \bibinfo {author}
  {\bibfnamefont{A.~P.}\ \bibnamefont{Lee}}}%
  , \bibinfo {year} {2008},\ \bibfield{title}{%
  \enquote{\bibinfo {title} {Droplet microfluidics},}\ }%
  \bibfield{journal}{%
  \bibinfo {journal} {Lab on a chip}\ }%
  \textbf{\bibinfo {volume} {8}},\ \bibinfo {pages} {198--220}%
  \bibAnnoteFile{NoStop}{Teh2008}%
\bibitem[{\citenamefont{T{\'e}l}\ \emph{et~al.}(2000)\citenamefont{T{\'e}l},
  \citenamefont{K{\'a}rolyi}, \citenamefont{P{\'e}ntek},
  \citenamefont{Scheuring}, \citenamefont{Toroczkai}, \citenamefont{Grebogi},\
  and\ \citenamefont{Kadtke}}]{Tel-et-al-00}%
  \BibitemOpen
  \bibfield{author}{%
  \bibinfo {author} {\bibnamefont{T{\'e}l}, \bibfnamefont{T.}}, \bibinfo
  {author} {\bibfnamefont{G.}~\bibnamefont{K{\'a}rolyi}}, \bibinfo {author}
  {\bibfnamefont{{\'A}.}~\bibnamefont{P{\'e}ntek}}, \bibinfo {author}
  {\bibfnamefont{I.}~\bibnamefont{Scheuring}}, \bibinfo {author}
  {\bibfnamefont{Z.}~\bibnamefont{Toroczkai}}, \bibinfo {author}
  {\bibfnamefont{C.}~\bibnamefont{Grebogi}},\ and\ \bibinfo {author}
  {\bibfnamefont{J.}~\bibnamefont{Kadtke}}}%
  , \bibinfo {year} {2000},\ \bibfield{title}{%
  \enquote{\bibinfo {title} {Chaotic advection, diffusion, and reactions in
  open flows},}\ }%
  \bibfield{journal}{%
  \bibinfo {journal} {Chaos}\ }%
  \textbf{\bibinfo {volume} {10}},\ \bibinfo {pages} {89--98}%
  \bibAnnoteFile{NoStop}{Tel-et-al-00}%
\bibitem[{\citenamefont{T\'el}\ \emph{et~al.}(2005)\citenamefont{T\'el},
  \citenamefont{de~Moura}, \citenamefont{Grebogi},\ and\
  \citenamefont{K\'arolyi}}]{Tel-et-al-05}%
  \BibitemOpen
  \bibfield{author}{%
  \bibinfo {author} {\bibnamefont{T\'el}, \bibfnamefont{T.}}, \bibinfo {author}
  {\bibfnamefont{A.}~\bibnamefont{de~Moura}}, \bibinfo {author}
  {\bibfnamefont{C.}~\bibnamefont{Grebogi}},\ and\ \bibinfo {author}
  {\bibfnamefont{G.}~\bibnamefont{K\'arolyi}}}%
  , \bibinfo {year} {2005},\ \bibfield{title}{%
  \enquote{\bibinfo {title} {Chemical and biological activity in open flows: a
  dynamical systems approach},}\ }%
  \bibfield{journal}{%
  \bibinfo {journal} {Phys. Rep.}\ }%
  \textbf{\bibinfo {volume} {413}},\ \bibinfo {pages} {91--196}%
  \bibAnnoteFile{NoStop}{Tel-et-al-05}%
\bibitem[{\citenamefont{T{\'e}l}\ \emph{et~al.}(2004)\citenamefont{T{\'e}l},
  \citenamefont{Nishikawa}, \citenamefont{Motter}, \citenamefont{Grebogi},\
  and\ \citenamefont{Toroczkai}}]{Tel-et-al-04}%
  \BibitemOpen
  \bibfield{author}{%
  \bibinfo {author} {\bibnamefont{T{\'e}l}, \bibfnamefont{T.}}, \bibinfo
  {author} {\bibfnamefont{T.}~\bibnamefont{Nishikawa}}, \bibinfo {author}
  {\bibfnamefont{A.~E.}\ \bibnamefont{Motter}}, \bibinfo {author}
  {\bibfnamefont{C.}~\bibnamefont{Grebogi}},\ and\ \bibinfo {author}
  {\bibfnamefont{Z.}~\bibnamefont{Toroczkai}}}%
  , \bibinfo {year} {2004},\ \bibfield{title}{%
  \enquote{\bibinfo {title} {Universality in active chaos},}\ }%
  \bibfield{journal}{%
  \bibinfo {journal} {Chaos}\ }%
  \textbf{\bibinfo {volume} {14}},\ \bibinfo {pages} {72--78}%
  \bibAnnoteFile{NoStop}{Tel-et-al-04}%
\bibitem[{\citenamefont{Thiffeault}(2005)}]{Thiffeault2005}%
  \BibitemOpen
  \bibfield{author}{%
  \bibinfo {author} {\bibnamefont{Thiffeault}, \bibfnamefont{{J.}-{L.}}}}%
  , \bibinfo {year} {2005},\ \bibfield{title}{%
  \enquote{\bibinfo {title} {Measuring {topological} {chaos}},}\ }%
  \bibfield{journal}{%
  \bibinfo {journal} {Phys. {Rev.} {Lett.}}\ }%
  \textbf{\bibinfo {volume} {94}},\ \bibinfo {pages} {084502}%
  \bibAnnoteFile{NoStop}{Thiffeault2005}%
\bibitem[{\citenamefont{Thiffeault}(2008)}]{ThiffeaultAosta2004}%
  \BibitemOpen
  \bibfield{author}{%
  \bibinfo {author} {\bibnamefont{Thiffeault}, \bibfnamefont{J.-L.}}}%
  , \bibinfo {year} {2008},\ \enquote{\bibinfo {title} {Scalar decay in chaotic
  mixing},}\ in\ \emph{\bibinfo {booktitle} {Transport and Mixing in
  Geophysical Flows}},\ \bibinfo {series} {Lecture Notes in Physics}, Vol.\
  \bibinfo {volume} {744},\ \bibinfo {editor} {edited by\ \bibinfo {editor}
  {\bibfnamefont{J.~B.}\ \bibnamefont{Weiss}}\ and\ \bibinfo {editor}
  {\bibfnamefont{A.}~\bibnamefont{Provenzale}}}\ (\bibinfo {publisher}
  {Springer},\ \bibinfo {address} {Berlin})\ pp.\ \bibinfo {pages} {3--35}%
  \bibAnnoteFile{NoStop}{ThiffeaultAosta2004}%
\bibitem[{\citenamefont{Thiffeault}(2010)}]{Thiffeault2010}%
  \BibitemOpen
  \bibfield{author}{%
  \bibinfo {author} {\bibnamefont{Thiffeault}, \bibfnamefont{{J.}-{L.}}}}%
  , \bibinfo {year} {2010},\ \bibfield{title}{%
  \enquote{\bibinfo {title} {Braids of entangled particle trajectories},}\ }%
  \bibfield{journal}{%
  \bibinfo {journal} {Chaos}\ }%
  \textbf{\bibinfo {volume} {20}},\ \bibinfo {pages} {017516--017514}%
  \bibAnnoteFile{NoStop}{Thiffeault2010}%
\bibitem[{\citenamefont{Thiffeault}(2012)}]{Thiffeault2012}%
  \BibitemOpen
  \bibfield{author}{%
  \bibinfo {author} {\bibnamefont{Thiffeault}, \bibfnamefont{J.-L.}}}%
  , \bibinfo {year} {2012},\ \bibfield{title}{%
  \enquote{\bibinfo {title} {Using multiscale norms to quantify mixing and
  transport},}\ }%
  \bibfield{journal}{%
  \bibinfo {journal} {Nonlinearity}\ }%
  \textbf{\bibinfo {volume} {25}},\ \bibinfo {pages} {R1--R44}%
  \bibAnnoteFile{NoStop}{Thiffeault2012}%
\bibitem[{\citenamefont{Thiffeault}\ and\ \citenamefont{{Budi}{\v
  s}i{\'c}}(2014)}]{Thiffeault2014v3}%
  \BibitemOpen
  \bibfield{author}{%
  \bibinfo {author} {\bibnamefont{Thiffeault}, \bibfnamefont{{J.}-{L.}}},\ and\
  \bibinfo {author} {\bibfnamefont{{M.}}~\bibnamefont{{Budi}{\v s}i{\'c}}}}%
  , \bibinfo {year} {2014},\ \bibfield{title}{%
  \enquote{\bibinfo {title} {Braidlab: {A} {Software} {Package} for {Braids}
  and {Loops} (v.3.1)},}\ }%
  \bibinfo {journal} {{arXiv}:1410.0849v3 {[}math{]}}%
  \bibAnnoteFile{NoStop}{Thiffeault2014v3}%
\bibitem[{\citenamefont{Thiffeault}\ and\
  \citenamefont{Childress}(2003)}]{Thiffeault2003d}%
  \BibitemOpen
\bibfield{journal}{%
    }%
  \bibfield{author}{%
  \bibinfo {author} {\bibnamefont{Thiffeault}, \bibfnamefont{J.-L.}},\ and\
  \bibinfo {author} {\bibfnamefont{S.}~\bibnamefont{Childress}}}%
  , \bibinfo {year} {2003},\ \bibfield{title}{%
  \enquote{\bibinfo {title} {Chaotic mixing in a torus map},}\ }%
  \bibfield{journal}{%
  \bibinfo {journal} {Chaos}\ }%
  \textbf{\bibinfo {volume} {13}},\ \bibinfo {pages} {502--507}%
  \bibAnnoteFile{NoStop}{Thiffeault2003d}%
\bibitem[{\citenamefont{Thiffeault}\ and\
  \citenamefont{Finn}(2006)}]{thiffeault2006topology}%
  \BibitemOpen
  \bibfield{author}{%
  \bibinfo {author} {\bibnamefont{Thiffeault}, \bibfnamefont{J.-L.}},\ and\
  \bibinfo {author} {\bibfnamefont{M.~D.}\ \bibnamefont{Finn}}}%
  , \bibinfo {year} {2006},\ \bibfield{title}{%
  \enquote{\bibinfo {title} {Topology, braids and mixing in fluids},}\ }%
  \bibfield{journal}{%
  \bibinfo {journal} {Phil. Trans. Roy. Soc. A}\ }%
  \textbf{\bibinfo {volume} {364}},\ \bibinfo {pages} {3251--3266}%
  \bibAnnoteFile{NoStop}{thiffeault2006topology}%
\bibitem[{\citenamefont{Thiffeault}\
  \emph{et~al.}(2008)\citenamefont{Thiffeault}, \citenamefont{Finn},
  \citenamefont{Gouillart},\ and\ \citenamefont{Hall}}]{Thiffeault2008}%
  \BibitemOpen
  \bibfield{author}{%
  \bibinfo {author} {\bibnamefont{Thiffeault}, \bibfnamefont{J.-L.}}, \bibinfo
  {author} {\bibfnamefont{M.~D.}\ \bibnamefont{Finn}}, \bibinfo {author}
  {\bibfnamefont{E.}~\bibnamefont{Gouillart}},\ and\ \bibinfo {author}
  {\bibfnamefont{T.}~\bibnamefont{Hall}}}%
  , \bibinfo {year} {2008},\ \bibfield{title}{%
  \enquote{\bibinfo {title} {Topology of chaotic mixing patterns},}\ }%
  \bibfield{journal}{%
  \bibinfo {journal} {Chaos}\ }%
  \textbf{\bibinfo {volume} {18}},\ \bibinfo {pages} {033123}%
  \bibAnnoteFile{NoStop}{Thiffeault2008}%
\bibitem[{\citenamefont{Thiffeault}\
  \emph{et~al.}(2011)\citenamefont{Thiffeault}, \citenamefont{Gouillart},\ and\
  \citenamefont{Dauchot}}]{Thiffeault2011c}%
  \BibitemOpen
  \bibfield{author}{%
  \bibinfo {author} {\bibnamefont{Thiffeault}, \bibfnamefont{J.-L.}}, \bibinfo
  {author} {\bibfnamefont{E.}~\bibnamefont{Gouillart}},\ and\ \bibinfo {author}
  {\bibfnamefont{O.}~\bibnamefont{Dauchot}}}%
  , \bibinfo {year} {2011},\ \bibfield{title}{%
  \enquote{\bibinfo {title} {Moving walls accelerate mixing},}\ }%
  \bibfield{journal}{%
  \bibinfo {journal} {Phys. Rev. E}\ }%
  \textbf{\bibinfo {volume} {84}},\ \bibinfo {pages} {036313}%
  \bibAnnoteFile{NoStop}{Thiffeault2011c}%
\bibitem[{\citenamefont{Tice}\ \emph{et~al.}(2004)\citenamefont{Tice},
  \citenamefont{Lyon},\ and\ \citenamefont{Ismagilov}}]{Tice2004}%
  \BibitemOpen
  \bibfield{author}{%
  \bibinfo {author} {\bibnamefont{Tice}, \bibfnamefont{J.~D.}}, \bibinfo
  {author} {\bibfnamefont{A.~D.}\ \bibnamefont{Lyon}},\ and\ \bibinfo {author}
  {\bibfnamefont{R.~F.}\ \bibnamefont{Ismagilov}}}%
  , \bibinfo {year} {2004},\ \bibfield{title}{%
  \enquote{\bibinfo {title} {Effects of viscosity on droplet formation and
  mixing in microfluidic channels},}\ }%
  \bibfield{journal}{%
  \bibinfo {journal} {Analytica Chimica Acta}\ }%
  \textbf{\bibinfo {volume} {507}},\ \bibinfo {pages} {73--77}%
  \bibAnnoteFile{NoStop}{Tice2004}%
\bibitem[{\citenamefont{Tice}\ \emph{et~al.}(2003)\citenamefont{Tice},
  \citenamefont{Song}, \citenamefont{Lyon},\ and\
  \citenamefont{Ismagilov}}]{Tice2003}%
  \BibitemOpen
  \bibfield{author}{%
  \bibinfo {author} {\bibnamefont{Tice}, \bibfnamefont{J.~D}}, \bibinfo
  {author} {\bibfnamefont{H.}~\bibnamefont{Song}}, \bibinfo {author}
  {\bibfnamefont{A.~D.}\ \bibnamefont{Lyon}},\ and\ \bibinfo {author}
  {\bibfnamefont{R.~F.}\ \bibnamefont{Ismagilov}}}%
  , \bibinfo {year} {2003},\ \bibfield{title}{%
  \enquote{\bibinfo {title} {Formation of droplets and mixing in multiphase
  microfluidics at low values of the {Reynolds} and the capillary numbers},}\
  }%
  \bibfield{journal}{%
  \bibinfo {journal} {Langmuir}\ }%
  \textbf{\bibinfo {volume} {19}},\ \bibinfo {pages} {9127--9133}%
  \bibAnnoteFile{NoStop}{Tice2003}%
\bibitem[{\citenamefont{den Toonder}\ \emph{et~al.}(2008)\citenamefont{den
  Toonder}, \citenamefont{Bos}, \citenamefont{Broer}, \citenamefont{Filippini},
  \citenamefont{Gillies}, \citenamefont{de~Goede}, \citenamefont{Mol},
  \citenamefont{Reijme}, \citenamefont{Talen}, \citenamefont{Wilderbeek},
  \citenamefont{Khatavkar},\ and\ \citenamefont{Anderson}}]{Toonder2008}%
  \BibitemOpen
  \bibfield{author}{%
  \bibinfo {author} {\bibnamefont{den Toonder}, \bibfnamefont{J.}}, \bibinfo
  {author} {\bibfnamefont{F.}~\bibnamefont{Bos}}, \bibinfo {author}
  {\bibfnamefont{D.}~\bibnamefont{Broer}}, \bibinfo {author}
  {\bibfnamefont{L.}~\bibnamefont{Filippini}}, \bibinfo {author}
  {\bibfnamefont{M.}~\bibnamefont{Gillies}}, \bibinfo {author}
  {\bibfnamefont{J.}~\bibnamefont{de~Goede}}, \bibinfo {author}
  {\bibfnamefont{T.}~\bibnamefont{Mol}}, \bibinfo {author}
  {\bibfnamefont{M.}~\bibnamefont{Reijme}}, \bibinfo {author}
  {\bibfnamefont{W.}~\bibnamefont{Talen}}, \bibinfo {author}
  {\bibfnamefont{H.}~\bibnamefont{Wilderbeek}}, \bibinfo {author}
  {\bibfnamefont{V.}~\bibnamefont{Khatavkar}},\ and\ \bibinfo {author}
  {\bibfnamefont{P.}~\bibnamefont{Anderson}}}%
  , \bibinfo {year} {2008},\ \bibfield{title}{%
  \enquote{\bibinfo {title} {Artificial cilia for active micro-fluidic
  mixing},}\ }%
  \bibfield{journal}{%
  \bibinfo {journal} {Lab on a chip}\ }%
  \textbf{\bibinfo {volume} {8}},\ \bibinfo {pages} {533--541}%
  \bibAnnoteFile{NoStop}{Toonder2008}%
\bibitem[{\citenamefont{Torney}\ and\
  \citenamefont{Neufeld}(2007)}]{Torney2007}%
  \BibitemOpen
  \bibfield{author}{%
  \bibinfo {author} {\bibnamefont{Torney}, \bibfnamefont{C.}},\ and\ \bibinfo
  {author} {\bibfnamefont{Z.}~\bibnamefont{Neufeld}}}%
  , \bibinfo {year} {2007},\ \bibfield{title}{%
  \enquote{\bibinfo {title} {Transport and aggregation of self-propelled
  particles in fluid flows},}\ }%
  \bibfield{journal}{%
  \bibinfo {journal} {Phys. Rev. Lett.}\ }%
  \textbf{\bibinfo {volume} {99}},\ \bibinfo {pages} {078101}%
  \bibAnnoteFile{NoStop}{Torney2007}%
\bibitem[{\citenamefont{Toroczkai}\
  \emph{et~al.}(1998)\citenamefont{Toroczkai}, \citenamefont{K{\'a}rolyi},
  \citenamefont{P{\'e}ntek}, \citenamefont{T{\'e}l},\ and\
  \citenamefont{Grebogi}}]{Toroczkai-et-al-98}%
  \BibitemOpen
  \bibfield{author}{%
  \bibinfo {author} {\bibnamefont{Toroczkai}, \bibfnamefont{Z.}}, \bibinfo
  {author} {\bibfnamefont{G.}~\bibnamefont{K{\'a}rolyi}}, \bibinfo {author}
  {\bibfnamefont{{\'A}.}~\bibnamefont{P{\'e}ntek}}, \bibinfo {author}
  {\bibfnamefont{T.}~\bibnamefont{T{\'e}l}},\ and\ \bibinfo {author}
  {\bibfnamefont{C.}~\bibnamefont{Grebogi}}}%
  , \bibinfo {year} {1998},\ \bibfield{title}{%
  \enquote{\bibinfo {title} {Advection of active particles in open chaotic
  flows},}\ }%
  \bibfield{journal}{%
  \bibinfo {journal} {Phys. Rev. Lett.}\ }%
  \textbf{\bibinfo {volume} {80}},\ \bibinfo {pages} {500--503}%
  \bibAnnoteFile{NoStop}{Toroczkai-et-al-98}%
\bibitem[{\citenamefont{Trefry}\ \emph{et~al.}(2012)\citenamefont{Trefry},
  \citenamefont{Lester}, \citenamefont{Metcalfe}, \citenamefont{Ord},\ and\
  \citenamefont{Regenauer-Lieb}}]{trefry2012}%
  \BibitemOpen
  \bibfield{author}{%
  \bibinfo {author} {\bibnamefont{Trefry}, \bibfnamefont{M.~G.}}, \bibinfo
  {author} {\bibfnamefont{D.~R.}\ \bibnamefont{Lester}}, \bibinfo {author}
  {\bibfnamefont{G.}~\bibnamefont{Metcalfe}}, \bibinfo {author}
  {\bibfnamefont{A.}~\bibnamefont{Ord}},\ and\ \bibinfo {author}
  {\bibfnamefont{K.}~\bibnamefont{Regenauer-Lieb}}}%
  , \bibinfo {year} {2012},\ \bibfield{title}{%
  \enquote{\bibinfo {title} {Toward enhanced subsurface intervention methods
  using chaotic advection},}\ }%
  \bibfield{journal}{%
  \bibinfo {journal} {J. Contaminant Hydrology}\ }%
  \textbf{\bibinfo {volume} {127}},\ \bibinfo {pages} {15--29}%
  \bibAnnoteFile{NoStop}{trefry2012}%
\bibitem[{\citenamefont{Tritton}(1988)}]{tritton}%
  \BibitemOpen
  \bibfield{author}{%
  \bibinfo {author} {\bibnamefont{Tritton}, \bibfnamefont{D.~J.}}}%
  , \bibinfo {year} {1988},\ \emph{\bibinfo {title} {Physical Fluid
  Dynamics}},\ \bibinfo {edition} {2nd}\ ed.\ (\bibinfo {publisher} {Oxford
  University Press})%
  \bibAnnoteFile{NoStop}{tritton}%
\bibitem[{\citenamefont{Tumasz}\ and\
  \citenamefont{{Thiffeault}}(2013)}]{Tumasz2013}%
  \BibitemOpen
  \bibfield{author}{%
  \bibinfo {author} {\bibnamefont{Tumasz}, \bibfnamefont{{S.}~{E.}}},\ and\
  \bibinfo {author} {\bibfnamefont{{J.}-{L.}}\ \bibnamefont{{Thiffeault}}}}%
  , \bibinfo {year} {2013},\ \bibfield{title}{%
  \enquote{\bibinfo {title} {Estimating {topological} {entropy} from the
  {motion} of {stirring} {rods}},}\ }%
  \bibfield{journal}{%
  \bibinfo {journal} {Procedia {IUTAM}}\ }%
  \textbf{\bibinfo {volume} {7}},\ \bibinfo {pages} {117--126}%
  \bibAnnoteFile{NoStop}{Tumasz2013}%
\bibitem[{\citenamefont{Tuval}\ \emph{et~al.}(2004)\citenamefont{Tuval},
  \citenamefont{Schneider}, \citenamefont{Piro},\ and\
  \citenamefont{T\'el}}]{Tuval2004}%
  \BibitemOpen
  \bibfield{author}{%
  \bibinfo {author} {\bibnamefont{Tuval}, \bibfnamefont{I.}}, \bibinfo {author}
  {\bibfnamefont{J.}~\bibnamefont{Schneider}}, \bibinfo {author}
  {\bibfnamefont{O.}~\bibnamefont{Piro}},\ and\ \bibinfo {author}
  {\bibfnamefont{T.}~\bibnamefont{T\'el}}}%
  , \bibinfo {year} {2004},\ \bibfield{title}{%
  \enquote{\bibinfo {title} {Opening up fractal structures of three-dimensional
  flows via leaking},}\ }%
  \bibfield{journal}{%
  \bibinfo {journal} {Europhys. Lett.}\ }%
  \textbf{\bibinfo {volume} {65}},\ \bibinfo {pages} {633--639}%
  \bibAnnoteFile{NoStop}{Tuval2004}%
\bibitem[{\citenamefont{Vaidya}\ and\
  \citenamefont{Mezi{\'c}}(2012)}]{vaidya2012existence}%
  \BibitemOpen
  \bibfield{author}{%
  \bibinfo {author} {\bibnamefont{Vaidya}, \bibfnamefont{U.}},\ and\ \bibinfo
  {author} {\bibfnamefont{I.}~\bibnamefont{Mezi{\'c}}}}%
  , \bibinfo {year} {2012},\ \bibfield{title}{%
  \enquote{\bibinfo {title} {Existence of invariant tori in three dimensional
  maps with degeneracy},}\ }%
  \bibfield{journal}{%
  \bibinfo {journal} {Physica D}\ }%
  \textbf{\bibinfo {volume} {241}},\ \bibinfo {pages} {1136--1145}%
  \bibAnnoteFile{NoStop}{vaidya2012existence}%
\bibitem[{\citenamefont{Vainchtein}\ and\
  \citenamefont{Mezi\'{c}}(2004)}]{vainchtein2004}%
  \BibitemOpen
  \bibfield{author}{%
  \bibinfo {author} {\bibnamefont{Vainchtein}, \bibfnamefont{D.}},\ and\
  \bibinfo {author} {\bibfnamefont{I.}~\bibnamefont{Mezi\'{c}}}}%
  , \bibinfo {year} {2004},\ \bibfield{title}{%
  \enquote{\bibinfo {title} {Optimal control of a co-rotating vortex pair:
  averaging and impulsive control},}\ }%
  \bibfield{journal}{%
  \bibinfo {journal} {Physica D}\ }%
  \textbf{\bibinfo {volume} {192}},\ \bibinfo {pages} {63--82}%
  \bibAnnoteFile{NoStop}{vainchtein2004}%
\bibitem[{\citenamefont{Vainchtein}\
  \emph{et~al.}(2006)\citenamefont{Vainchtein}, \citenamefont{Neishtadt},\ and\
  \citenamefont{Mezi\'{c}}}]{Vainchtein2006}%
  \BibitemOpen
  \bibfield{author}{%
  \bibinfo {author} {\bibnamefont{Vainchtein}, \bibfnamefont{D.~L.}}, \bibinfo
  {author} {\bibfnamefont{A.~I.}\ \bibnamefont{Neishtadt}},\ and\ \bibinfo
  {author} {\bibfnamefont{I.}~\bibnamefont{Mezi\'{c}}}}%
  , \bibinfo {year} {2006},\ \bibfield{title}{%
  \enquote{\bibinfo {title} {On passage through resonances in volume-preserving
  systems},}\ }%
  \bibfield{journal}{%
  \bibinfo {journal} {Chaos}\ }%
  \textbf{\bibinfo {volume} {16}},\ \bibinfo {pages} {043123}%
  \bibAnnoteFile{NoStop}{Vainchtein2006}%
\bibitem[{\citenamefont{Vainchtein}\
  \emph{et~al.}(2007)\citenamefont{Vainchtein}, \citenamefont{Widloski},\ and\
  \citenamefont{Grigoriev}}]{Vainchtein2007}%
  \BibitemOpen
  \bibfield{author}{%
  \bibinfo {author} {\bibnamefont{Vainchtein}, \bibfnamefont{D.~L.}}, \bibinfo
  {author} {\bibfnamefont{J.}~\bibnamefont{Widloski}},\ and\ \bibinfo {author}
  {\bibfnamefont{R.~O.}\ \bibnamefont{Grigoriev}}}%
  , \bibinfo {year} {2007},\ \bibfield{title}{%
  \enquote{\bibinfo {title} {Resonant chaotic mixing in a cellular flow},}\ }%
  \bibfield{journal}{%
  \bibinfo {journal} {Phys. Rev. Lett.}\ }%
  \textbf{\bibinfo {volume} {99}},\ \bibinfo {pages} {094501}%
  \bibAnnoteFile{NoStop}{Vainchtein2007}%
\bibitem[{\citenamefont{{Valentine}}\
  \emph{et~al.}(2012)\citenamefont{{Valentine}}, \citenamefont{{Mezi\'{c}}},
  \citenamefont{{Macesic}}, \citenamefont{{Crnjaric-Zic}},
  \citenamefont{{Ivic}}, \citenamefont{{Hogan}}, \citenamefont{{Fonoberov}},\
  and\ \citenamefont{{Loire}}}]{valentine2012}%
  \BibitemOpen
  \bibfield{author}{%
  \bibinfo {author} {\bibnamefont{{Valentine}}, \bibfnamefont{D.~L.}}, \bibinfo
  {author} {\bibfnamefont{I.}~\bibnamefont{{Mezi\'{c}}}}, \bibinfo {author}
  {\bibfnamefont{S.}~\bibnamefont{{Macesic}}}, \bibinfo {author}
  {\bibfnamefont{N.}~\bibnamefont{{Crnjaric-Zic}}}, \bibinfo {author}
  {\bibfnamefont{S.}~\bibnamefont{{Ivic}}}, \bibinfo {author}
  {\bibfnamefont{P.~J.}\ \bibnamefont{{Hogan}}}, \bibinfo {author}
  {\bibfnamefont{V.~A.}\ \bibnamefont{{Fonoberov}}},\ and\ \bibinfo {author}
  {\bibfnamefont{S.}~\bibnamefont{{Loire}}}}%
  , \bibinfo {year} {2012},\ \bibfield{title}{%
  \enquote{\bibinfo {title} {Dynamic autoinoculation and the microbial ecology
  of a deep water hydrocarbon irruption},}\ }%
  \bibfield{journal}{%
  \bibinfo {journal} {Proc. Natl Acad. Sci. USA}\ }%
  \textbf{\bibinfo {volume} {109}},\ \bibinfo {pages} {20286--20291}%
  \bibAnnoteFile{NoStop}{valentine2012}%
\bibitem[{\citenamefont{Vikhansky}(2002)}]{vikhansky2002}%
  \BibitemOpen
  \bibfield{author}{%
  \bibinfo {author} {\bibnamefont{Vikhansky}, \bibfnamefont{A.}}}%
  , \bibinfo {year} {2002},\ \bibfield{title}{%
  \enquote{\bibinfo {title} {Enhancement of laminar mixing by optimal control
  methods},}\ }%
  \bibfield{journal}{%
  \bibinfo {journal} {Chem. Eng. Sci.}\ }%
  \textbf{\bibinfo {volume} {57}},\ \bibinfo {pages} {2719--2725}%
  \bibAnnoteFile{NoStop}{vikhansky2002}%
\bibitem[{\citenamefont{Vilela}\ \emph{et~al.}(2006)\citenamefont{Vilela},
  \citenamefont{de~Moura},\ and\ \citenamefont{Grebogi}}]{Vilela2006}%
  \BibitemOpen
  \bibfield{author}{%
  \bibinfo {author} {\bibnamefont{Vilela}, \bibfnamefont{R.~D}}, \bibinfo
  {author} {\bibfnamefont{A.~P.~S.}\ \bibnamefont{de~Moura}},\ and\ \bibinfo
  {author} {\bibfnamefont{C.}~\bibnamefont{Grebogi}}}%
  , \bibinfo {year} {2006},\ \bibfield{title}{%
  \enquote{\bibinfo {title} {Finite-size effects on open chaotic advection},}\
  }%
  \bibfield{journal}{%
  \bibinfo {journal} {Phys. Rev. E}\ }%
  \textbf{\bibinfo {volume} {73}},\ \bibinfo {pages} {026302}%
  \bibAnnoteFile{NoStop}{Vilela2006}%
\bibitem[{\citenamefont{Vilela}\ \emph{et~al.}(2007)\citenamefont{Vilela},
  \citenamefont{T\'el}, \citenamefont{de~Moura},\ and\
  \citenamefont{Grebogi}}]{Vilela2007}%
  \BibitemOpen
  \bibfield{author}{%
  \bibinfo {author} {\bibnamefont{Vilela}, \bibfnamefont{R.~D}}, \bibinfo
  {author} {\bibfnamefont{T.}~\bibnamefont{T\'el}}, \bibinfo {author}
  {\bibfnamefont{A.~P.~S.}\ \bibnamefont{de~Moura}},\ and\ \bibinfo {author}
  {\bibfnamefont{C.}~\bibnamefont{Grebogi}}}%
  , \bibinfo {year} {2007},\ \bibfield{title}{%
  \enquote{\bibinfo {title} {Signatures of fractal clustering of aerosols
  advected under gravity},}\ }%
  \bibfield{journal}{%
  \bibinfo {journal} {Phys. Rev. E}\ }%
  \textbf{\bibinfo {volume} {75}},\ \bibinfo {pages} {065203}%
  \bibAnnoteFile{NoStop}{Vilela2007}%
\bibitem[{\citenamefont{Wang}\ \emph{et~al.}(2009)\citenamefont{Wang},
  \citenamefont{Feng}, \citenamefont{Ottino},\ and\
  \citenamefont{R.}}]{wang2009}%
  \BibitemOpen
  \bibfield{author}{%
  \bibinfo {author} {\bibnamefont{Wang}, \bibfnamefont{J.}}, \bibinfo {author}
  {\bibfnamefont{L.}~\bibnamefont{Feng}}, \bibinfo {author}
  {\bibfnamefont{J.M.}\ \bibnamefont{Ottino}},\ and\ \bibinfo {author}
  {\bibfnamefont{Lueptow}\ \bibnamefont{R.}}}%
  , \bibinfo {year} {2009},\ \bibfield{title}{%
  \enquote{\bibinfo {title} {Inertial effects on chaotic advection and mixing
  in a {2D} cavity flow},}\ }%
  \bibfield{journal}{%
  \bibinfo {journal} {Ind. Eng. Chem. Res.}\ }%
  \textbf{\bibinfo {volume} {48}},\ \bibinfo {pages} {2436--2442}%
  \bibAnnoteFile{NoStop}{wang2009}%
\bibitem[{\citenamefont{Wang}\ \emph{et~al.}(2011)\citenamefont{Wang},
  \citenamefont{Huang},\ and\ \citenamefont{Yang}}]{Wang2011}%
  \BibitemOpen
  \bibfield{author}{%
  \bibinfo {author} {\bibnamefont{Wang}, \bibfnamefont{S.}}, \bibinfo {author}
  {\bibfnamefont{X.}~\bibnamefont{Huang}},\ and\ \bibinfo {author}
  {\bibfnamefont{C.}~\bibnamefont{Yang}}}%
  , \bibinfo {year} {2011},\ \bibfield{title}{%
  \enquote{\bibinfo {title} {Mixing enhancement for high viscous fluids in a
  microfluidic chamber},}\ }%
  \bibfield{journal}{%
  \bibinfo {journal} {Lab on a chip}\ }%
  \textbf{\bibinfo {volume} {11}},\ \bibinfo {pages} {2081--2087}%
  \bibAnnoteFile{NoStop}{Wang2011}%
\bibitem[{\citenamefont{Wang}\ \emph{et~al.}(2014)\citenamefont{Wang},
  \citenamefont{Metcalfe}, \citenamefont{Stewart}, \citenamefont{Wu},
  \citenamefont{Ohmura}, \citenamefont{Feng},\ and\
  \citenamefont{Yang}}]{Wang_separation_2014}%
  \BibitemOpen
  \bibfield{author}{%
  \bibinfo {author} {\bibnamefont{Wang}, \bibfnamefont{S.}}, \bibinfo {author}
  {\bibfnamefont{G.}~\bibnamefont{Metcalfe}}, \bibinfo {author}
  {\bibfnamefont{R.~L.}\ \bibnamefont{Stewart}}, \bibinfo {author}
  {\bibfnamefont{J.}~\bibnamefont{Wu}}, \bibinfo {author}
  {\bibfnamefont{N.}~\bibnamefont{Ohmura}}, \bibinfo {author}
  {\bibfnamefont{X.}~\bibnamefont{Feng}},\ and\ \bibinfo {author}
  {\bibfnamefont{C.}~\bibnamefont{Yang}}}%
  , \bibinfo {year} {2014},\ \bibfield{title}{%
  \enquote{\bibinfo {title} {Solid-liquid separation by
  particle-flow-instability},}\ }%
  \bibfield{journal}{%
  \bibinfo {journal} {Energy \& Environmental Sci.}\ }%
  \textbf{\bibinfo {volume} {7}},\ \bibinfo {pages} {3982--3988}%
  \bibAnnoteFile{NoStop}{Wang_separation_2014}%
\bibitem[{\citenamefont{Wang}\ \emph{et~al.}(2016)\citenamefont{Wang},
  \citenamefont{Stewart},\ and\
  \citenamefont{Metcalfe}}]{Wang_visualization_2016}%
  \BibitemOpen
  \bibfield{author}{%
  \bibinfo {author} {\bibnamefont{Wang}, \bibfnamefont{S.}}, \bibinfo {author}
  {\bibfnamefont{R.}~\bibnamefont{Stewart}},\ and\ \bibinfo {author}
  {\bibfnamefont{G.}~\bibnamefont{Metcalfe}}}%
  , \bibinfo {year} {2016},\ \bibfield{title}{%
  \enquote{\bibinfo {title} {Visualization of the trapping of inertial
  particles in a laminar mixing tank},}\ }%
  \bibfield{journal}{%
  \bibinfo {journal} {Chem. Eng. Sci.}\ }%
  \textbf{\bibinfo {volume} {143}},\ \bibinfo {pages} {99--104}%
  \bibAnnoteFile{NoStop}{Wang_visualization_2016}%
\bibitem[{\citenamefont{Weiss}(1991)}]{Weiss:1991kv}%
  \BibitemOpen
  \bibfield{author}{%
  \bibinfo {author} {\bibnamefont{Weiss}, \bibfnamefont{J.}}}%
  , \bibinfo {year} {1991},\ \bibfield{title}{%
  \enquote{\bibinfo {title} {The dynamics of enstrophy transfer in
  two-dimensional hydrodynamics},}\ }%
  \bibfield{journal}{%
  \bibinfo {journal} {Physica D}\ }%
  \textbf{\bibinfo {volume} {48}},\ \bibinfo {pages} {273--294}%
  \bibAnnoteFile{NoStop}{Weiss:1991kv}%
\bibitem[{\citenamefont{Welander}(1954)}]{Welander1954}%
  \BibitemOpen
  \bibfield{author}{%
  \bibinfo {author} {\bibnamefont{Welander}, \bibfnamefont{P.}}}%
  , \bibinfo {year} {1954},\ \bibfield{title}{%
  \enquote{\bibinfo {title} {On the temperature jump in a rarefied gas},}\ }%
  \bibfield{journal}{%
  \bibinfo {journal} {Ark. f. Fysik}\ }%
  \textbf{\bibinfo {volume} {7}},\ \bibinfo {pages} {507--553}%
  \bibAnnoteFile{NoStop}{Welander1954}%
\bibitem[{\citenamefont{Welander}(1955)}]{Welander1955}%
  \BibitemOpen
  \bibfield{author}{%
  \bibinfo {author} {\bibnamefont{Welander}, \bibfnamefont{P.}}}%
  , \bibinfo {year} {1955},\ \bibfield{title}{%
  \enquote{\bibinfo {title} {Studies of the general development of motion in a
  two-dimensional ideal fluid},}\ }%
  \bibfield{journal}{%
  \bibinfo {journal} {Tellus}\ }%
  \textbf{\bibinfo {volume} {7}},\ \bibinfo {pages} {141--156}%
  \bibAnnoteFile{NoStop}{Welander1955}%
\bibitem[{\citenamefont{Wonhas}\ and\
  \citenamefont{Vassilicos}(2002)}]{Wonhas2002}%
  \BibitemOpen
  \bibfield{author}{%
  \bibinfo {author} {\bibnamefont{Wonhas}, \bibfnamefont{A.}},\ and\ \bibinfo
  {author} {\bibfnamefont{J.~C.}\ \bibnamefont{Vassilicos}}}%
  , \bibinfo {year} {2002},\ \bibfield{title}{%
  \enquote{\bibinfo {title} {Mixing in fully chaotic flows},}\ }%
  \bibfield{journal}{%
  \bibinfo {journal} {Phys. Rev. E}\ }%
  \textbf{\bibinfo {volume} {66}},\ \bibinfo {pages} {051205}%
  \bibAnnoteFile{NoStop}{Wonhas2002}%
\bibitem[{\citenamefont{Wu}\ \emph{et~al.}(2011)\citenamefont{Wu},
  \citenamefont{Freund}, \citenamefont{Fraser},\ and\
  \citenamefont{Vermot}}]{Wu2011}%
  \BibitemOpen
  \bibfield{author}{%
  \bibinfo {author} {\bibnamefont{Wu}, \bibfnamefont{D.}}, \bibinfo {author}
  {\bibfnamefont{J.~B.}\ \bibnamefont{Freund}}, \bibinfo {author}
  {\bibfnamefont{S.~E.}\ \bibnamefont{Fraser}},\ and\ \bibinfo {author}
  {\bibfnamefont{J.}~\bibnamefont{Vermot}}}%
  , \bibinfo {year} {2011},\ \bibfield{title}{%
  \enquote{\bibinfo {title} {Mechanistic basis of otolith formation during
  teleost inner ear development},}\ }%
  \bibfield{journal}{%
  \bibinfo {journal} {Dev. Cell}\ }%
  \textbf{\bibinfo {volume} {20}},\ \bibinfo {pages} {271--278}%
  \bibAnnoteFile{NoStop}{Wu2011}%
\bibitem[{\citenamefont{Wu}\ \emph{et~al.}(2014)\citenamefont{Wu},
  \citenamefont{Speetjens}, \citenamefont{Vainchtein},
  \citenamefont{Trieling},\ and\ \citenamefont{Clercx}}]{Wu2014}%
  \BibitemOpen
  \bibfield{author}{%
  \bibinfo {author} {\bibnamefont{Wu}, \bibfnamefont{F.}}, \bibinfo {author}
  {\bibfnamefont{M.~F.~M.}\ \bibnamefont{Speetjens}}, \bibinfo {author}
  {\bibfnamefont{D.~L.}\ \bibnamefont{Vainchtein}}, \bibinfo {author}
  {\bibfnamefont{R.~R.}\ \bibnamefont{Trieling}},\ and\ \bibinfo {author}
  {\bibfnamefont{H.~J.~H.}\ \bibnamefont{Clercx}}}%
  , \bibinfo {year} {2014},\ \bibfield{title}{%
  \enquote{\bibinfo {title} {Comparative numerical-experimental analysis of the
  universal impact of arbitrary perturbations on transport in three-dimensional
  unsteady flows},}\ }%
  \bibfield{journal}{%
  \bibinfo {journal} {Phys. Rev. E}\ }%
  \textbf{\bibinfo {volume} {90}},\ \bibinfo {pages} {063002}%
  \bibAnnoteFile{NoStop}{Wu2014}%
\bibitem[{\citenamefont{Xia}\ \emph{et~al.}(2005)\citenamefont{Xia},
  \citenamefont{Wan}, \citenamefont{Shu},\ and\ \citenamefont{Chew}}]{Xia2005}%
  \BibitemOpen
  \bibfield{author}{%
  \bibinfo {author} {\bibnamefont{Xia}, \bibfnamefont{H.~M.}}, \bibinfo
  {author} {\bibfnamefont{S.~Y.~M.}\ \bibnamefont{Wan}}, \bibinfo {author}
  {\bibfnamefont{C.}~\bibnamefont{Shu}},\ and\ \bibinfo {author}
  {\bibfnamefont{Y.~T.}\ \bibnamefont{Chew}}}%
  , \bibinfo {year} {2005},\ \bibfield{title}{%
  \enquote{\bibinfo {title} {Chaotic micromixers using two-layer crossing
  channels to exhibit fast mixing at low {Reynolds} numbers},}\ }%
  \bibfield{journal}{%
  \bibinfo {journal} {Lab on a chip}\ }%
  \textbf{\bibinfo {volume} {5}},\ \bibinfo {pages} {748--755}%
  \bibAnnoteFile{NoStop}{Xia2005}%
\bibitem[{\citenamefont{Xia}\ \emph{et~al.}(2012)\citenamefont{Xia},
  \citenamefont{Wang}, \citenamefont{Fan}, \citenamefont{Wijaya},
  \citenamefont{Wang},\ and\ \citenamefont{Wang}}]{Xia2012}%
  \BibitemOpen
  \bibfield{author}{%
  \bibinfo {author} {\bibnamefont{Xia}, \bibfnamefont{H.~M.}}, \bibinfo
  {author} {\bibfnamefont{Z.~P.}\ \bibnamefont{Wang}}, \bibinfo {author}
  {\bibfnamefont{W.}~\bibnamefont{Fan}}, \bibinfo {author}
  {\bibfnamefont{A.}~\bibnamefont{Wijaya}}, \bibinfo {author}
  {\bibfnamefont{W.}~\bibnamefont{Wang}},\ and\ \bibinfo {author}
  {\bibfnamefont{Z.~F.}\ \bibnamefont{Wang}}}%
  , \bibinfo {year} {2012},\ \bibfield{title}{%
  \enquote{\bibinfo {title} {Converting steady laminar flow to oscillatory flow
  through a hydroelasticity approach at microscales},}\ }%
  \bibfield{journal}{%
  \bibinfo {journal} {Lab on a chip}\ }%
  \textbf{\bibinfo {volume} {12}},\ \bibinfo {pages} {60--64}%
  \bibAnnoteFile{NoStop}{Xia2012}%
\bibitem[{\citenamefont{Yang}\ \emph{et~al.}(2016)\citenamefont{Yang},
  \citenamefont{Batista}, \citenamefont{Steinbock}, \citenamefont{Cartwright},\
  and\ \citenamefont{Cardoso}}]{Ding2016}%
  \BibitemOpen
  \bibfield{author}{%
  \bibinfo {author} {\bibnamefont{Yang}, \bibfnamefont{D.}}, \bibinfo {author}
  {\bibfnamefont{B.}~\bibnamefont{Batista}}, \bibinfo {author}
  {\bibfnamefont{O.}~\bibnamefont{Steinbock}}, \bibinfo {author}
  {\bibfnamefont{J~.H.~E.}\ \bibnamefont{Cartwright}},\ and\ \bibinfo {author}
  {\bibfnamefont{S.~S.~S.}\ \bibnamefont{Cardoso}}}%
  , \bibinfo {year} {2016},\ \bibfield{title}{%
  \enquote{\bibinfo {title} {Wavy membranes and the growth rate of a planar
  chemical garden: Enhanced diffusion and bioenergetics},}\ }%
  \bibfield{journal}{%
  \bibinfo {journal} {Proc. Natl Acad. Sci. USA}\ }%
  \textbf{\bibinfo {volume} {113}},\ \bibinfo {pages} {9182--9186}%
  \bibAnnoteFile{NoStop}{Ding2016}%
\bibitem[{\citenamefont{Yannacopoulos}\
  \emph{et~al.}(1998)\citenamefont{Yannacopoulos}, \citenamefont{Mezi\'{c}},
  \citenamefont{Rowlands},\ and\ \citenamefont{King}}]{Yanna1998}%
  \BibitemOpen
  \bibfield{author}{%
  \bibinfo {author} {\bibnamefont{Yannacopoulos}, \bibfnamefont{A.~N.}},
  \bibinfo {author} {\bibfnamefont{I.}~\bibnamefont{Mezi\'{c}}}, \bibinfo
  {author} {\bibfnamefont{G.}~\bibnamefont{Rowlands}},\ and\ \bibinfo {author}
  {\bibfnamefont{G.~P.}\ \bibnamefont{King}}}%
  , \bibinfo {year} {1998},\ \bibfield{title}{%
  \enquote{\bibinfo {title} {Eulerian diagnostics for {Lagrangian} chaos in
  three-dimensional {Navier--Stokes} flows},}\ }%
  \bibfield{journal}{%
  \bibinfo {journal} {Phys. Rev. E}\ }%
  \textbf{\bibinfo {volume} {57}},\ \bibinfo {pages} {482--490}%
  \bibAnnoteFile{NoStop}{Yanna1998}%
\bibitem[{\citenamefont{Young}(1998)}]{young1998statistical}%
  \BibitemOpen
  \bibfield{author}{%
  \bibinfo {author} {\bibnamefont{Young}, \bibfnamefont{L.~S.}}}%
  , \bibinfo {year} {1998},\ \bibfield{title}{%
  \enquote{\bibinfo {title} {Statistical properties of dynamical systems with
  some hyperbolicity},}\ }%
  \bibfield{journal}{%
  \bibinfo {journal} {Annals Math.}\ }%
  \textbf{\bibinfo {volume} {147}},\ \bibinfo {pages} {585--650}%
  \bibAnnoteFile{NoStop}{young1998statistical}%
\bibitem[{\citenamefont{Young}\ and\ \citenamefont{Jones}(1991)}]{young1991}%
  \BibitemOpen
  \bibfield{author}{%
  \bibinfo {author} {\bibnamefont{Young}, \bibfnamefont{W.~R.}},\ and\ \bibinfo
  {author} {\bibfnamefont{S.~W.}\ \bibnamefont{Jones}}}%
  , \bibinfo {year} {1991},\ \bibfield{title}{%
  \enquote{\bibinfo {title} {Shear dispersion},}\ }%
  \bibinfo {journal} {Phys. Fluids A3},\ \bibinfo {pages} {1087--1101}%
  \bibAnnoteFile{NoStop}{young1991}%
\bibitem[{\citenamefont{Zaggout}\ and\
  \citenamefont{Gilbert}(2012)}]{Zaggout2012}%
  \BibitemOpen
\bibfield{journal}{%
    }%
  \bibfield{author}{%
  \bibinfo {author} {\bibnamefont{Zaggout}, \bibfnamefont{F.~A.}},\ and\
  \bibinfo {author} {\bibfnamefont{A.~D.}\ \bibnamefont{Gilbert}}}%
  , \bibinfo {year} {2012},\ \bibfield{title}{%
  \enquote{\bibinfo {title} {Passive scalar decay in chaotic flows with
  boundaries},}\ }%
  \bibfield{journal}{%
  \bibinfo {journal} {Fluid Dyn. Res.}\ }%
  \textbf{\bibinfo {volume} {44}},\ \bibinfo {pages} {025504}%
  \bibAnnoteFile{NoStop}{Zaggout2012}%
\bibitem[{\citenamefont{Zeidler}(2012)}]{Zeidler2012}%
  \BibitemOpen
  \bibfield{author}{%
  \bibinfo {author} {\bibnamefont{Zeidler}, \bibfnamefont{E.}}}%
  , \bibinfo {year} {2012},\ \emph{\bibinfo {title} {Applied Functional
  Analysis: Applications to Mathematical Physics}}\ (\bibinfo {publisher}
  {Springer})%
  \bibAnnoteFile{NoStop}{Zeidler2012}%
\bibitem[{\citenamefont{Znaien}\ \emph{et~al.}(2012)\citenamefont{Znaien},
  \citenamefont{Speetjens}, \citenamefont{Trieling},\ and\
  \citenamefont{Clercx}}]{Znaien2012}%
  \BibitemOpen
  \bibfield{author}{%
  \bibinfo {author} {\bibnamefont{Znaien}, \bibfnamefont{J.~G.}}, \bibinfo
  {author} {\bibfnamefont{M.~F.~M.}\ \bibnamefont{Speetjens}}, \bibinfo
  {author} {\bibfnamefont{R.~R.}\ \bibnamefont{Trieling}},\ and\ \bibinfo
  {author} {\bibfnamefont{H.~J.~H.}\ \bibnamefont{Clercx}}}%
  , \bibinfo {year} {2012},\ \bibfield{title}{%
  \enquote{\bibinfo {title} {Observability of periodic lines in
  three-dimensional lid-driven cylindrical cavity flows},}\ }%
  \bibfield{journal}{%
  \bibinfo {journal} {Phys. Rev. E}\ }%
  \textbf{\bibinfo {volume} {85}},\ \bibinfo {pages} {066320}%
  \bibAnnoteFile{NoStop}{Znaien2012}%
\end{thebibliography}%

\end{document}